\RequirePackage[hyphens]{url}
\newif\ifcolor
\colortrue
\ifcolor
\else

 \PassOptionsToPackage{monochrome}{xcolor}

\fi

\documentclass[a4paper]{tufte-book}

\ifcolor
\else

 \definecolor{black}{RGB}{0,0,0}
 \definecolor{blue}{RGB}{0,0,0}
 \definecolor{brown}{RGB}{0,0,0}
 \definecolor{cyan}{RGB}{0,0,0}
 \definecolor{darkgray}{RGB}{0,0,0}
 \definecolor{gray}{RGB}{0,0,0}
 \definecolor{green}{RGB}{0,0,0}
 \definecolor{lightgray}{RGB}{0,0,0}
 \definecolor{lime}{RGB}{0,0,0}
 \definecolor{magenta}{RGB}{0,0,0}
 \definecolor{olive}{RGB}{0,0,0}
 \definecolor{orange}{RGB}{0,0,0}
 \definecolor{pink}{RGB}{0,0,0}
 \definecolor{purple}{RGB}{0,0,0}
 \definecolor{red}{RGB}{0,0,0}
 \definecolor{teal}{RGB}{0,0,0}
 \definecolor{violet}{RGB}{0,0,0}
 \definecolor{yellow}{RGB}{0,0,0}
 \definecolor{Apricot}{RGB}{0,0,0}
 \definecolor{Aquamarine}{RGB}{0,0,0}
 \definecolor{Bittersweet}{RGB}{0,0,0}
 \definecolor{Black}{RGB}{0,0,0}
 \definecolor{Blue}{RGB}{0,0,0}
 \definecolor{BlueGreen}{RGB}{0,0,0}
 \definecolor{BlueViolet}{RGB}{0,0,0}
 \definecolor{BrickRed}{RGB}{0,0,0}
 \definecolor{Brown}{RGB}{0,0,0}
 \definecolor{BurntOrange}{RGB}{0,0,0}
 \definecolor{CadetBlue}{RGB}{0,0,0}
 \definecolor{CarnationPink}{RGB}{0,0,0}
 \definecolor{Cerulean}{RGB}{0,0,0}
 \definecolor{CornflowerBlue}{RGB}{0,0,0}
 \definecolor{Cyan}{RGB}{0,0,0}
 \definecolor{Dandelion}{RGB}{0,0,0}
 \definecolor{DarkOrchid}{RGB}{0,0,0}
 \definecolor{Emerald}{RGB}{0,0,0}
 \definecolor{ForestGreen}{RGB}{0,0,0}
 \definecolor{Fuchsia}{RGB}{0,0,0}
 \definecolor{Goldenrod}{RGB}{0,0,0}
 \definecolor{Gray}{RGB}{0,0,0}
 \definecolor{Green}{RGB}{0,0,0}
 \definecolor{GreenYellow}{RGB}{0,0,0}
 \definecolor{JungleGreen}{RGB}{0,0,0}
 \definecolor{Lavender}{RGB}{0,0,0}
 \definecolor{LimeGreen}{RGB}{0,0,0}
 \definecolor{Magenta}{RGB}{0,0,0}
 \definecolor{Mahogany}{RGB}{0,0,0}
 \definecolor{Maroon}{RGB}{0,0,0}
 \definecolor{Melon}{RGB}{0,0,0}
 \definecolor{MidnightBlue}{RGB}{0,0,0}
 \definecolor{Mulberry}{RGB}{0,0,0}
 \definecolor{NavyBlue}{RGB}{0,0,0}
 \definecolor{OliveGreen}{RGB}{0,0,0}
 \definecolor{Orange}{RGB}{0,0,0}
 \definecolor{OrangeRed}{RGB}{0,0,0}
 \definecolor{Orchid}{RGB}{0,0,0}
 \definecolor{Peach}{RGB}{0,0,0}
 \definecolor{Periwinkle}{RGB}{0,0,0}
 \definecolor{PineGreen}{RGB}{0,0,0}
 \definecolor{Plum}{RGB}{0,0,0}
 \definecolor{ProcessBlue}{RGB}{0,0,0}
 \definecolor{Purple}{RGB}{0,0,0}
 \definecolor{RawSienna}{RGB}{0,0,0}
 \definecolor{Red}{RGB}{0,0,0}
 \definecolor{RedOrange}{RGB}{0,0,0}
 \definecolor{RedViolet}{RGB}{0,0,0}
 \definecolor{Rhodamine}{RGB}{0,0,0}
 \definecolor{RoyalBlue}{RGB}{0,0,0}
 \definecolor{RoyalPurple}{RGB}{0,0,0}
 \definecolor{RubineRed}{RGB}{0,0,0}
 \definecolor{Salmon}{RGB}{0,0,0}
 \definecolor{SeaGreen}{RGB}{0,0,0}
 \definecolor{Sepia}{RGB}{0,0,0}
 \definecolor{SkyBlue}{RGB}{0,0,0}
 \definecolor{SpringGreen}{RGB}{0,0,0}
 \definecolor{Tan}{RGB}{0,0,0}
 \definecolor{TealBlue}{RGB}{0,0,0}
 \definecolor{Thistle}{RGB}{0,0,0}
 \definecolor{Turquoise}{RGB}{0,0,0}
 \definecolor{Violet}{RGB}{0,0,0}
 \definecolor{VioletRed}{RGB}{0,0,0}
 \definecolor{WildStrawberry}{RGB}{0,0,0}
 \definecolor{Yellow}{RGB}{0,0,0}
 \definecolor{YellowGreen}{RGB}{0,0,0}
 \definecolor{YellowOrange}{RGB}{0,0,0}

\fi

\makeatletter
\DeclareRobustCommand\natcite{%
  \begingroup\let\NAT@ctype\z@\NAT@partrue\NAT@swatrue
    \@ifstar{\NAT@fulltrue\NAT@cites}{\NAT@fullfalse\NAT@cites}%
}

\renewcommand{\@tufte@infootnote@cite}[1]{%
  \natcite{#1}
  \@tufte@add@citation{#1}%
}

\renewcommand\@tufte@add@citation[1]{\relax
  \ifx\@tufte@citations\@empty\else
    \g@addto@macro\@tufte@citations{,}
  \fi
  \g@addto@macro\@tufte@citations{#1}
}
\makeatother
\hypersetup{colorlinks}
\title{Mathematical Methods of Theoretical Physics\thanks{Based on the Vienna University of Technology course {\em Mathematische Methoden der Physik}.}}
\author[Karl Svozil]{Karl Svozil}
\publisher{Edition Funzl}

\renewcommand{\smallcaps}[1]{\color{Grey} \sffamily #1}

\newcommand{\chapterlabel}{\textcolor{BrickRed}{\thechapter}}%
\titleformat{\chapter}[display]
 {\begin{fullwidth}\raggedright}
 {{\fontsize{60}{68} \selectfont   \chapterlabel}}
 {-2\baselineskip}
 {\vspace{2\baselineskip}\vspace{1ex}\Huge\color{BrickRed}}
 [\end{fullwidth}\vspace{1ex}]

\usepackage{titletoc}

\usepackage{tikz}
\usetikzlibrary{positioning}
\tikzset{face/.style={shape=circle,minimum size=4ex,shading=radial,outer sep=0pt,
        inner color=white!50!yellow,outer color= yellow!70!orange}}

\usepackage{pgfplots}

\newcommand{\emoticon}[1][]{%
  \node[face,#1] (emoticon) {};
  \draw[fill=white] (-1ex,0ex) ..controls (-0.5ex,0.2ex)and(0.5ex,0.2ex)..
        (1ex,0.0ex) ..controls ( 1.5ex,1.5ex)and( 0.2ex,1.7ex)..
        (0ex,0.4ex) ..controls (-0.2ex,1.7ex)and(-1.5ex,1.5ex)..
        (-1ex,0ex)--cycle;}
\newcommand{\pupils}{
  \fill[shift={(0.5ex,0.5ex)},rotate=80]
       (0,0) ellipse (0.3ex and 0.15ex);
  \fill[shift={(-0.5ex,0.5ex)},rotate=100]
       (0,0) ellipse (0.3ex and 0.15ex);}

\ifcolor

\contentsmargin{0pt}
\titlecontents{section}[1.8pc]
  {\addvspace{3pt}\itshape}
  {\color{OliveGreen}\contentslabel[\thecontentslabel   ]{1.8pc}}
  {}
  {\quad\thecontentspage}

\titlecontents*{subsection}[1.8pc]
  {\small \itshape}
  {\color{Grey}\thecontentslabel~$\;$}
  {}
  {, \thecontentspage}
  [. --- ][.]

\else

\contentsmargin{0pt}
  \titlecontents{part}[2.8pc]
  {\addvspace{0.1in}\begin{fullwidth}\bfseries\normalsize\itshape}
  {\color{OliveGreen}\hspace*{4em}\contentslabel[\thecontentslabel   ]{2.2pc}}
  {}
  { \dotfill\rm\normalsize\itshape\thecontentspage}
  [\end{fullwidth}]

\titlecontents{chapter}[5cm]
  {\begin{fullwidth}\bfseries\normalsize\itshape}
  {\color{OliveGreen}\hspace*{2.4em}\contentslabel[\thecontentslabel   ]{2pc}}
  {}
  {\dotfill\rm\normalsize\itshape\thecontentspage}
  [\end{fullwidth}]

\titlecontents{section}[2pc]
  {\vspace*{-2pt}\begin{fullwidth}\normalsize\itshape}
  {\color{OliveGreen}\hspace*{5.5em}\contentslabel[\thecontentslabel   ]{2.2pc}}
  {}
  {\dotfill\normalsize\itshape\thecontentspage}
  [\end{fullwidth}]

\titlecontents{subsection}[2pc]
  {\vspace*{-4pt}\begin{fullwidth}\normalsize\itshape}
  {\color{Grey}\hspace*{8.7em}\contentslabel[\thecontentslabel   ]{2.4pc}}
  {}
  {\dotfill\thecontentspage}
  [\end{fullwidth}]

\fi

 \fancypagestyle{plain}{}

 \usepackage{microtype}

\usepackage{booktabs}

\usepackage{amssymb}
\usepackage[tbtags]{amsmath}
\usepackage{curves}
\usepackage{wasysym}
\usepackage{fourier}
\newcommand{\bexample}{ }
\newcommand{\eexample}{ }
\newcommand{\bproof}{ }
\newcommand{\eproof}{ }

\definecolor{lightgrey}{rgb}{0.95,0.95,0.95}

\usepackage{graphicx}
\setkeys{Gin}{width=\linewidth,totalheight=\textheight,keepaspectratio}
\graphicspath{{graphics/}}

\usepackage{fancyvrb}
\fvset{fontsize=\normalsize}

\usepackage{xspace}

\newcommand{\monthyear}{%
  \ifcase\month\or January\or February\or March\or April\or May\or June\or
  July\or August\or September\or October\or November\or
  December\fi\space\number\year
}

\renewcommand{\frak}{\cal}

\usepackage{stmaryrd}
\usepackage{units}

\newcommand{\hlred}[1]{\textcolor{Maroon}{#1}}
\newcommand{\hangleft}[1]{\makebox[0pt][r]{#1}}

\providecommand{\XeLaTeX}{X\lower.5ex\hbox{\kern-0.15em\reflectbox{E}}\kern-0.1em\LaTeX}

\newcommand{\tuftebs}{\symbol{'134}}
\newcommand{\doccmddef}[2][]{%
  \hlred{\texttt{\tuftebs#2}}\label{cmd:#2}%
  \ifthenelse{\isempty{#1}}%
    {
      \index{#2 command@\protect\hangleft{\texttt{\tuftebs}}\texttt{#2}}
    }%
    {
      \index{#2 command@\protect\hangleft{\texttt{\tuftebs}}\texttt{#2} (\texttt{#1} package)}
      \index{#1 package@\texttt{#1} package}\index{packages!#1@\texttt{#1}}
    }%
}
\newcommand{\doccmd}[2][]{%
  \texttt{\tuftebs#2}%
  \ifthenelse{\isempty{#1}}%
    {
      \index{#2 command@\protect\hangleft{\texttt{\tuftebs}}\texttt{#2}}
    }%
    {
      \index{#2 command@\protect\hangleft{\texttt{\tuftebs}}\texttt{#2} (\texttt{#1} package)}
      \index{#1 package@\texttt{#1} package}\index{packages!#1@\texttt{#1}}
    }%
}

\usepackage{mathbbol} 

\usepackage{makeidx}
\makeindex

\begin{document}
\sloppy




\maketitle

\newpage
\begin{fullwidth}
~\vfill
\thispagestyle{empty}
\setlength{\parindent}{0pt}
\setlength{\parskip}{\baselineskip}
\ifcolor
Copyright \copyright\ \the\year\ \thanklessauthor

\par\smallcaps{Published by \thanklesspublisher}

\par For academic use only. You may not reproduce or distribute without permission of the author.
\index{license}

\par\textit{First Edition, October 2011}
\par\textit{Second Edition, October 2013}
\par\textit{Third Edition, October 2014}
\par\textit{Fourth Edition, October 2016}
\par\textit{Fifth Edition, October 2018}
\par\textit{Sixth Edition, \monthyear}
\fi
\end{fullwidth}


\tableofcontents




\mainmatter
\thispagestyle{fancy}

\chapter*{Why mathematics?}
\addcontentsline{toc}{chapter}{Why mathematics?}
\markboth{Why mathematics?}{Why mathematics?}


\newthought{Nobody knows} why the application of mathematics is effective in physics
and the sciences in general.
Indeed, some greater (mathematical) minds have found this so mind-boggling they have called it unreasonable\cite[-65mm]{wigner}: {\em
``$\ldots$~the enormous usefulness of mathematics in the natural sciences is something
bordering on the mysterious and~$\ldots$ there is no rational explanation for it.''}

{A rather straightforward way} of getting rid of this issue (and probably too much more) entirely
would be to consider it a metaphysical sophism\cite[-60mm]{Hume-Enquiry,Hahn1930,Carnap-1931-engl}
-- a  pseudo-statement devoid of any empirical and operational or logical substance whatsoever.
Nevertheless, it might be amusing to contemplate two extremely speculative positions pertinent to the topic.

{A Pythagorean scenario} would be to identify Nature with mathematics.
In particular, suppose we are embedded minds inhabiting
a ``calculating space''\cite[-18mm]{zuse-70} --
some sort of virtual reality, or clockwork universe, rendered by some computing machinery ``located'' in the beyond
``out of our immediate reach.''
Our accessible gaming environment may exist autonomous (without intervention); or it may be interconnected
to some external universe by some interfaces
which appear as immanent indeterminates or gaps in the laws of physics\cite[-32mm]{frank,franke}
without violating these laws.

{Another, converse, scenario} postulates totally chaotic, stochastic processes
at the lowest, foundational, level of description\cite[-13mm]{Exner-1908,Stoeltzner-1999,svozil-2018-was}.
In this line of thought, long before humans created mathematics the following hierarchy evolved:
the primordial chaos has ``expressed'' itself in some form of physical laws,
like the law of large numbers or the ones encountered in Ramsey theory.
The physical laws have expressed themselves in matter and biological ``stuff'' like genes.
The genes, in turn, have expressed themselves in individual minds,
and those minds create ideas about their surroundings\cite{berkeley}.

{In any case}
mathematics might have evolved by abductive inference and adaption --
as a collection of emergent cognitive concepts to ``understand,'' or at least predict and manipulate,
the human environment.
Thereby, {\em mathematics provides intrinsic, embedded means and ways by which the universe contemplates itself.}
Its {\em instrument art thou}\sidenote{Krishna in {\it The Bhagavad-Gita.} Chapter XI.}. 

This makes mathematics an endeavor both glorious and prone to deficiencies.
What a pathetic yet sobering perspective!
In its humility
it may point to an existential freedom\cite{camus-mos} in creating and using mathematical entities.
And it might offer some consolation when
encountering inconsistencies in the formalism,
and the sometimes pragmatic (if not outright ignorant) ways to cope with them.

For instance, Hilbert's reaction with regards to employing
Cantor's (inspiring yet inconsistent) ``na\"ive'' set theory was enthusiastic\cite{hilbert-26}:
{\em ``from the paradise, that Cantor created for us, no-one shall be able to expel us.''}
Another example is the inconsistency arising from insisting on Bohr's measurement concept
-- which effectively amounts to a many-to-one process --
in lieu of the uniform unitary state evolution -- essentially a one-to-one function
and nesting.
Or take  Heaviside's not uncontroversial stance\cite{heaviside-EMT}:
\begin{quote}
{\em
I suppose all workers
in mathematical physics have noticed how the mathematics
seems made for the physics, the latter suggesting the former, and
that practical ways of working arise naturally. $\ldots$ But then the
rigorous logic of the matter is not plain! Well, what of that?
Shall I refuse my dinner because I do not fully understand the
process of digestion? No, not if I am satisfied with the result.
Now a physicist may in like manner employ unrigorous processes with satisfaction and usefulness if he, by the application
of tests, satisfies himself of the accuracy of his results. At
the same time he may be fully aware of his want of infallibility,
and that his investigations are largely of an experimental character, and maybe repellent to unsympathetically
constituted mathematicians accustomed to a different kind
of work.~[p.~9, \S~225]
}
\label{2013-m-ch-intro-cooking}
\end{quote}
\begin{marginfigure}
\begin{center}
\includegraphics[width=3cm]{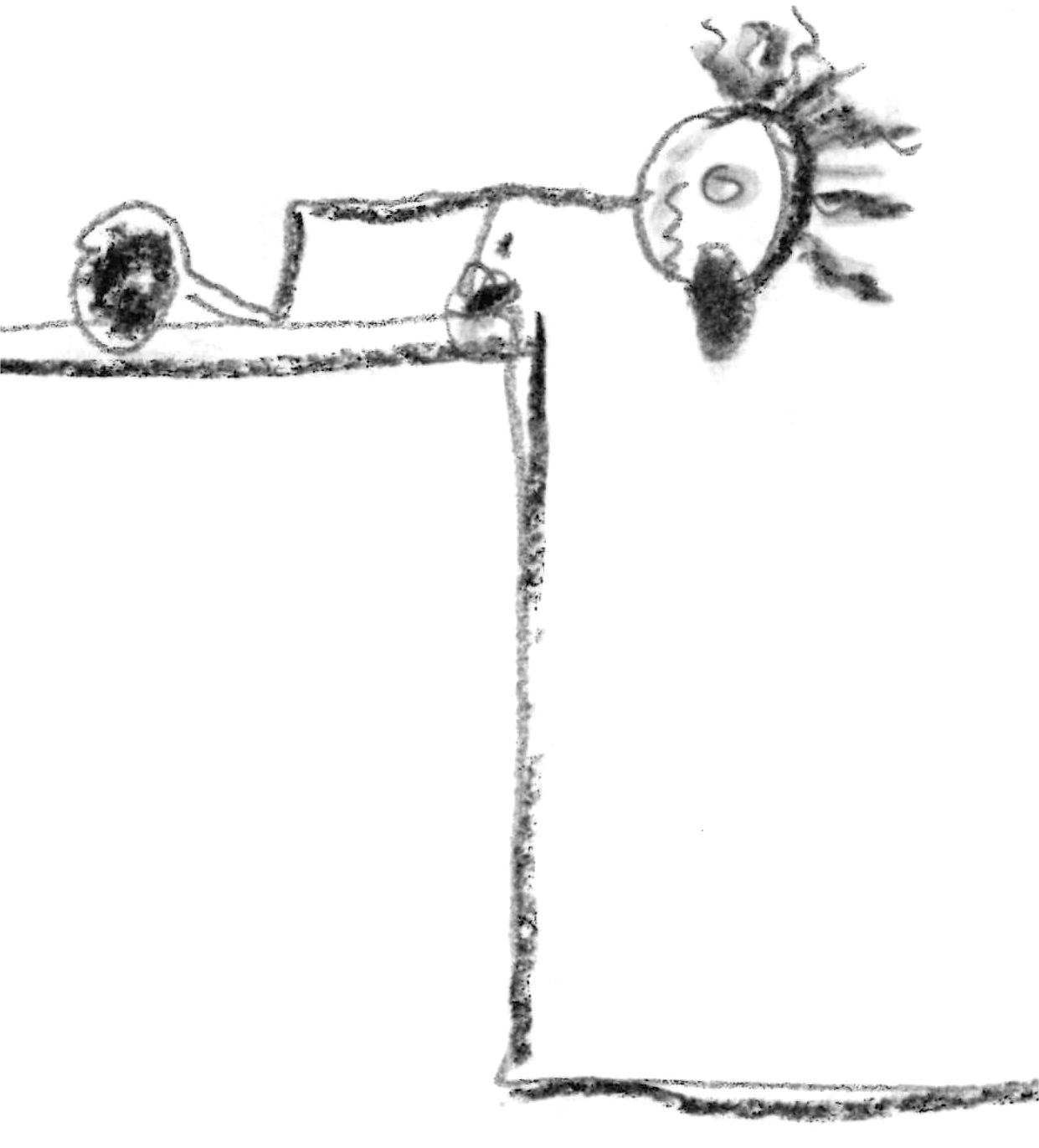}
\caption{Contemporary mathematicians may have perceived the introduction of Heaviside's unit step function with some concern.
It is good in the modeling of, say, switching on and off electric currents, but it is nonsmooth and nondifferentiable.}
\label{2018-m-cartoon-Heaviside1}
\end{center}
\end{marginfigure}
Mathematicians finally succeeded in (what they currently consider)
properly coping with such sort of entities, as reviewed in Chapter~\ref{2011-m-ch:gf}; but it took a while.
Currently we are experiencing interest in another challinging field, still {\it in statu nascendi} and
exposed in Chapter~\ref{2011-m-ch-ds}, the  asymptotic expansion of divergent series:
for some finite number of terms these series ``converge'' towards a meaningful value, only to resurge \index{resurgence} later;
a phenomenon encountered in perturbation theory, approximating solutions of differential equations by series expansions.

Dietrich K\"uchemann,
the ingenious German-British aerodynamicist and
one of the main contributors to the wing design of the {\em Concord} supersonic civil aircraft, tells us
\cite[5mm]{Kuchemann}
\begin{quote}
{\em
[Again,] the most drastic simplifying assumptions must be made before we can even think about
the flow of gases and arrive at equations which are amenable to treatment. Our whole
science lives on highly-idealized concepts and ingenious abstractions and approximations.
We should remember this in all modesty at all times, especially when somebody claims to
have obtained ``the right answer'' or ``the exact solution''.
At the same time, we must acknowledge and admire the intuitive art of those scientists
to whom we owe the many useful concepts and approximations with which we work~[page 23].
}
\end{quote}

{The relationship between physics and formalism}, in particular, has been debated by
Bridgman\cite[-30mm]{bridgman},
Feynman\cite[-15mm]{feynman-computation},
and  Landauer\cite{landauer},
among many others.
It has many twists, anecdotes, and opinions.
Already Zeno of Elea and Parmenides wondered how there can be motion if
our universe is either infinitely divisible or discrete.
Because in the dense case (between any two points there is another point),
the slightest finite move would require an infinity of actions.
Likewise, in the discrete case,
how can there be motion if everything is not moving at all times\cite[-10mm]{zeno}?

{The question arises:} to what extent should we take the formalism as a mere convenience?
Or should we take it very seriously and literally,
using it as a guide to new territories, which might even appear absurd, inconsistent and mind-boggling?
Should we expect that all the wild things
formally imaginable,  such as, for instance, the Banach-Tarski paradox\cite[-30mm]{wagon1},  have a physical realization?

{It might be prudent} to
adopt a contemplative strategy of {\em evenly-suspended attention}
outlined by  Freud\cite[-15mm]{Freud-1912}, who admonishes analysts to be aware of the dangers
caused by {\em ``temptations to project,
what  [the analyst]  in dull self-perception recognizes as the peculiarities of his own personality,
as generally valid theory into science.''}
Nature is thereby treated as a  client-patient,  and whatever findings come up are accepted  as is  without any
immediate emphasis or judgment.
This also alleviates the dangers of becoming embittered with the reactions of ``the peers,''
a problem sometimes encountered when ``surfing on the edge'' of contemporary knowledge; such as, for
example, Everett's case\cite[-15mm]{everett-collw}.

{I am calling} for more tolerance and greater unity in physics;
as well as for greater esteem on ``both sides of the same effort;''
I am also opting for more pragmatism;
one that acknowledges the mutual benefits and oneness of
theoretical and empirical physical world perceptions.
Schr\"odinger\cite[-10mm]{schroed:natgr}
cites  Democritus with arguing against a too great separation of the  intellect ($\delta \iota {\alpha}\nu o \iota \alpha$, dianoia) and the senses
($\alpha \iota \sigma \theta {\eta} \sigma \epsilon \iota \varsigma$, aitheseis).
In fragment D 125 from Galen\cite{Diels-fdv}, p. 408, footnote 125 , the intellect claims
``ostensibly there is color, ostensibly sweetness, ostensibly bitterness, actually only atoms and the void;''
to which the senses retort:
``Poor intellect, do you hope to defeat us while from us you borrow your evidence? Your victory is your defeat.''

Jaynes has warned us of the {\em ``Mind Projection Fallacy''}\cite{jaynes-89,jaynes-90}, pointing out that
{\em ``we are all under an ego-driven temptation to project our private
thoughts out onto the real world, by supposing that the creations of one's own imagination are real
properties of Nature, or that one's own ignorance signifies some kind of indecision on the part of
Nature.''}

It is also important to emphasize that, in order to absorb formalisims one needs not only talent but,
in particular, a high degree of resilience.
Mathematics (at least to me) turns out to be humbling; a training in tolerance and modesty:
most of us experience no difficulties in
finding very personal challenges by excessive demands.
And oftentimes this may even amount to (temporary) defeat.
Nevertheless, I am inclined to quote {\it Rocky Balboa},
{\em ``$\ldots$~it's about how hard you can get hit and keep moving forward; how much you can take
and keep moving forward~$\ldots$''.}

And yet, despite all aforementioned {\it provisos}, formalized science finally succeeded to do what the alchemists sought for so long:
it transmuted mercury into gold\cite{PhysRev.60.473}.

Let me close this informal rant by contemplating the question: ``what is truth?''
If one sticks to empirical truth then one is reminded of Hannah Arendt's\cite{Arendt-1967-truth}:
``we may call truth what we cannot change; metaphorically,
it is the ground on which we stand and the sky that stretches above us.''
That is very poetic but not easily transferable to science.
For instance, does a click in a detector from a particle prepared in a complementary (relative to the detector) quantum state
represent or correspond to the ``true state'' of the particle in the detector frame? With Niels Bohr, I believe not.

Let us recall what Heinrich Hertz\cite{hertz-94e} wrote about physical theory:
``The most direct, and in a sense the most important, problem
which our conscious knowledge of nature should enable us to
solve is the anticipation of future events, so that we may
arrange our present affairs in accordance with such anticipation.
As a basis for the solution of this problem we always
make use of our knowledge of events which have already
occurred, obtained by chance observation or by prearranged
experiment. In endeavouring thus to draw inferences as to
the future from the past, we always adopt the following process.
We form for ourselves images or symbols of external objects;
and the form which we give them is such that the necessary
consequents of the images in thought are always the images of
the necessary consequents in nature of the things pictured. In
order that this requirement may be satisfied, there must be a
certain conformity between nature and our thought. Experience
teaches us that the requirement can be satisfied, and hence that
such a conformity does in fact exist. When from our accumulated
previous experience wre have once succeeded in deducing
images of the desired nature, we can then in a short time
develop by means of them, as by means of models, the
consequences which in the external world only arise in a comparatively
long time, or as the result of our own interposition.
We are thus enabled to be in advance of the facts, and to
decide as to present affairs in accordance with the insight so
obtained. The images which we here speak of are our conceptions
of things. With the things themselves they are in
conformity in one important respect, namely, in satisfying the
above - mentioned requirement. For our purpose it is not
necessary that they should be in conformity with the things in
any other respect whatever. As a matter of fact, we do not
know, nor have we any means of knowing, whether our conceptions
of things conform with them in any other
than this one fundamental respect.''

It is my conviction that, very much in the spirit of Hertz, a careful investigation into ``scientific truth'',
in particular, when it comes to formalizations,
suggests a subjective, individualistic answer:
``a person's belief, an image, suspended in free thought, that is often consistent with empirical corroborations.''

Relative to mild side assumptions, such as consistency, the general induction problem is provable unsolvable.
A little bit more formally,
the general rule inference problem---one machine figuring out the working of another machine---can be reduced to the halting problem.
The term reduction here means that to solve the general rule inference problem one would need to be able to solve the halting problem.
This latter problem, like many metamathematical problems such as G\"odel's incompleteness theorems, turns out to be unsolvable
within the framework of any ``sufficiently (allowing Peano arithmetic) strong'' formalism.

Besides formal logic and mathematics this has consequences for physics:
while it may be possible to guess physical theories even by methods of machine learning, there will never be a systematic way of
figuring out if the world is lawful, and what laws there are.

Confronted with this situation several contemporary philosophers of science have suggested more or less pragmatic criteria
for theory formation.
For instance,
Karl Raimund Popper~\cite{popper-en} suggested falsification as a demarcation criterion, separating useless ideology, ``blablabla'' as he
called it, from useful science: the standard on which a judgment or decision may be based is a theoretical prediction that can be tested.
The emphasis is not so much on corroboration than on falsification.

Imre Lakatos\cite{lakatosch} has criticized Popper's demarcation criterion because such a test of the core of a research program,
its main idea or metaphor,
may depend on so many side assumptions, and may involve so many historic issues that it renders falsification practically useless.
As a result, contemporary scientists are incapable to differentiate between progressive research programs and degenerative ones.
Lakatosh also points out that there is no straightforward semantic  convergence of research programs: he
quotes gravity and points out that the Ptolemaic model of epicycles, a purely geometric model putting Earth in the middle of the Universe,
was so sophisticated that it outperformed the heliocentric Copernican model initially.
The heliocentric model, supported by Newton's force model of gravity, a long-range interaction, eventually superseded the Ptolemaian geometric model.
In another turn of science history, the general theory of relativity, resolving Newtonian forces of gravity into
space-time curvature, brought back a geometrical model---so, from geometry to force, and back to geometry!
It will be an interesting challenge of what comes next,
given the amazing  maneuvers of unidentified areal vehicles (if they exist)
that defy inertial motion.

Thomas Kuhn~\cite[-30mm]{kuhn} has observed that often science progresses in terms of revolutions,
followed by longer periods of working out the consequences thereof.
There are long periods of consolidation, interrupted by short periods of
iconoclastic upheaval.

I have attended lectures of the late Paul Feyerabend\cite[-20mm]{feyerabend} in Berkeley in which he suggested that, because of
all of these issues it might be best to distribute scientific resources through a system of lay judges, very similar to
existing courts of lay assessors.
He is supported by Swizz investigations into what the experts considered progressive research areas in which to invest resources,
that turned out to be anticorrelated to what happened later\cite[-30mm]{swizz-science}.

Let me close this short review of truth with encouragement by Immanuel Kant
that has given me both strength and resilience in my personal pursuit.
This dictum of the enlightenment might guide the reader as well\cite[-30mm]{kant-Aufklaerung}:
{\it ``sapere aude!''}---``Have the courage to make use of thy own understanding!''
And, one may add, do not get distracted by absorbing bullshit~\cite[-10mm]{Frankfurt-OnBullshit}.
\marginnote{This is an enumeration of wrong proof methods (in German): \url{http://kamelopedia.net/wiki/Beweis}}

\begin{center}\color{black}
$\widetilde{\qquad \qquad }$
$\widetilde{\qquad \qquad}$
$\widetilde{\qquad \qquad }$
\end{center}



\newpage

\newthought{This is an ongoing attempt}
to provide some written material of a course in mathematical methods of theoretical physics.
Who knows (see Ref.\cite{Aquinas} part one, question 14, article 13; and  also Ref.\cite{specker-60}, p. 243)
if I have succeeded?
I kindly ask the perplexed to please be patient, do not panic under any circumstances,
and do not allow themselves to be too upset with mistakes, omissions \& other problems of this text.
At the end of the day, everything will be fine, and in the long run, we will be dead anyway.
Or, to quote Karl Kraus, {\em ``it is not enough to have no concept,
one must also be capable of expressing it.''}
\marginnote{
From the German original in {\em Karl Kraus, {\em Die Fackel} {\bf 697}, 60 (1925)}:
{\em ``Es gen\"ugt nicht, keinen Gedanken zu haben: man muss ihn auch ausdr\"ucken k\"onnen.''
}}

{The problem}
with all such presentations is to present the material in sufficient depth while at the same time not to get buried by the formalism.
As every individual has his or her own mode of comprehension there is no canonical answer to this challenge.

{So not all that is presented here} will be acceptable to everybody; for various reasons.
Some people will claim that I am too confused and utterly formalistic, others will claim my arguments are in desperate need of rigor.
Many formally fascinated readers will demand to go deeper into the meaning of the subjects;
others may want some easy-to-identify pragmatic, syntactic rules of deriving results.
I apologize to both groups from the outset.
This is the best I can do; from certain different perspectives, others, maybe even some tutors or students, might perform much better.

{In 1987 in his} {\it Abschiedsvorlesung} professor Ernst Specker
at the {\it Eidgen\"ossische Hochschule Z\"urich}
remarked that
the many books authored by David Hilbert carry his name first,
and the name(s) of his co-author(s) second,
although the subsequent author(s) had actually written these books;
the only exception of this rule being Courant and Hilbert's 1924 book
{\em Methoden der mathematischen Physik},
comprising around 1000 densely packed pages,
which allegedly none of these authors had actually written.
It appears to be some sort of collective effort of scholars from the University of G\"ottingen.

I most humbly present my own version of what is important for standard courses of contemporary physics.
Thereby, I am quite aware that, not dissimilar with some attempts of that sort undertaken so far, I might fail miserably.
Because even if I manage to induce some interest, affection, passion, and understanding in the audience --
as Danny Greenberger put it,
inevitably
four hundred years from now, all our present physical theories of today will appear transient\cite[-40mm]{lakatosch}, if not laughable.
And thus, in the long run, my efforts will be forgotten (although, I do hope, not totally futile); and some other brave, courageous guy
will continue attempting to (re)present the most important mathematical methods in theoretical physics.
{\it Per aspera ad astra}\sidenote[][-20mm]{Quoted from {\em Hercules Furens} by Lucius Annaeus Seneca (c. 4~BC -- AD~65), line 437, spoken by Megara, Hercules' wife:
{\it ``non est ad astra mollis e terris via''} (``there is no easy way from the earth to the stars.'')}!

\newpage

I would like to gratefully acknowledge the input, corrections and encouragements by numerous (former) students and colleagues,
in particular also professors Hans Havlicek, Jose Maria Isidro San Juan, Thomas Sommer and Reinhard Winkler.
I also would kindly like to thank the publisher, and, in particular, the Editor Nur Syarfeena Binte Mohd Fauzi
for her patience with numerous preliminary versions, and the kind care dedicated to this volume.
Needless to say, all remaining errors and misrepresentations
are my own fault. I am grateful for any correction and suggestion for an improvement of this text.

\begin{center}
{\color{lightgray}   \Huge
\aldine
}
\end{center}

\cleardoublepage
\setcounter{page}{1}
\pagenumbering{arabic}
\chapter*{\color{BurntOrange}\thispagestyle{empty} {\fontsize{40}{168} \selectfont   Part I} \\ {\fontsize{30}{68} \selectfont Linear vector spaces}
\addcontentsline{toc}{part}{Part I:  Linear vector spaces}
{\newpage \thispagestyle{empty}   $\;$ \vskip 9 true cm \begin{center}\includegraphics[width=0.7\textwidth, angle=-20]{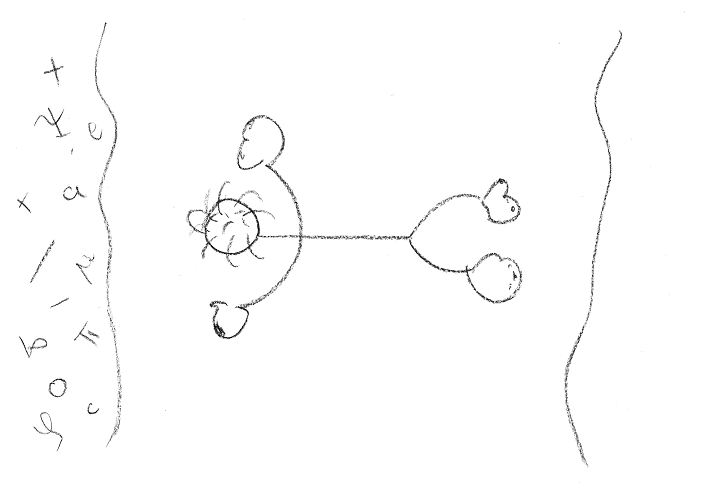}\end{center}}}
\chapter{Finite-dimensional vector spaces and linear algebra}
\label{ch:lvs}

\begin{quote}
{\it ``I would have written a shorter letter, but I did not have the time.''}
(Literally: {\it ``I made this [letter] very long because I did not have the leisure to make it shorter.''})
Blaise Pascal, {\it Provincial Letters: Letter XVI (English Translation)}\\
{\it ``Perhaps if I had spent more time I should have been able to make a shorter report~$\ldots$''}
James Clerk Maxwell~\cite[-20mm]{garber}, Document~15,  p.~426
\end{quote}

\newthought{Vector Spaces} are prevalent in physics;
they are essential for an understanding
of mechanics, relativity theory, quantum mechanics, and statistical physics.

\section{Conventions and basic definitions}

This presentation is greatly inspired
by Halmos' compact yet comprehensive treatment
``Finite-Dimensional Vector Spaces''.\cite[-40mm]{halmos-vs}
I greatly encourage the reader to have a look into that book.
Of course, there exist zillions of other very nice presentations, among them
Greub's ``Linear algebra,'' and
Strang's ``Introduction to Linear Algebra,''
among many others, even freely downloadable  ones
\cite[-30mm]{Greub75,Strang:2009:ILA,Homes-rorres,lipschutz:schaul-la,Hefferon}
competing for your attention.

Unless stated differently, only
finite-dimensional vector spaces will be considered.

In what follows
the overline sign stands for complex conjugation; that is,
if
${a}= \Re a +i\Im a $ is a complex number, then
$\overline{a}= \Re a -i\Im a$.
Very often vector and other coordinates will be real- or complex-valued scalars, which are elements of a field (see Section~\ref{2017-m-ch-fdvs-fields}).

A superscript ``$\intercal$'' means transposition.

The physically oriented notation in Mermin's book on
quantum information theory\cite[-7mm]{mermin-04} is adopted.
Vectors are either typed in boldface, or in Dirac's ``bra-ket'' notation.\cite{dirac}
Both notations will be used simultaneously and equivalently;
not to confuse or obfuscate, but to make  the reader familiar with the bra-ket
notation used in quantum physics.

Thereby,
the vector ${\bf x}$ is identified with the ``ket vector'' $\vert {\bf x}\rangle$.
Ket vectors will be represented by column vectors, that is, by vertically arranged tuples of scalars,
or, equivalently, as $n \times 1$ matrices; that is,
\begin{equation}
{\bf x}
\equiv
\vert {\bf x}\rangle
\equiv
\begin{pmatrix}
x_1\\
x_2\\
\vdots \\
x_n
\end{pmatrix}
=\begin{pmatrix}
x_1,
x_2,
\ldots ,
x_n
\end{pmatrix}^\intercal
.
\end{equation}

A vector ${\bf x}^\ast$ with an asterisk symbol ``$\ast$'' in its superscript denotes an element of the dual space (see later, Section \ref{2011-m-dvs} on page \pageref{2011-m-dvs}).
It is also identified with the ``bra vector'' $\langle {\bf x}\vert$.
\index{ket vector}
\index{bra vector}
Bra vectors will be represented by row vectors, that is, by horizontally arranged tuples of scalars,
or, equivalently, as $1 \times n$ matrices; that is,
\begin{equation}
{\bf x}^\ast
\equiv
\langle {\bf x}   \vert
\equiv
\begin{pmatrix}
x_1,
x_2,
\ldots ,
x_n
\end{pmatrix}
.
\end{equation}

\index{inner product}
\index{scalar product}
Dot (scalar or inner) products between two vectors ${\bf x}$ and ${\bf y}$   in Euclidean space are then
denoted by ``$\langle \textrm{bra} \vert  (\textrm{c}) \vert \textrm{ket}  \rangle$''  form;
that is, by $\langle {\bf x} \vert  {\bf y}  \rangle$.

For an $n \times m$ matrix $\textsf{\textbf{A}}\equiv a_{ij}$ we shall use the following {\em index notation}:
suppose the (column) index $j$  indicates their column number in a matrix-like object $a_{i{\color{red} j}}$ ``runs {\em horizontally},'' that is, from left to right.
The (row) index $i$  indicates their row number in a matrix-like object  $a_{{\color{red}  i}j}$ ``runs {\em vertically},''
so that, with
$1 \le i \le n$ and
$1 \le j \le m$,
\index{index notation}
\begin{equation}
\textsf{\textbf{A}}
\equiv
\begin{pmatrix}
a_{11}&a_{12}& \cdots & a_{1m}\\
a_{21}&a_{22}& \cdots & a_{2m}\\
\vdots &\vdots & \ddots & \vdots \\
a_{n1}&a_{n2}& \cdots & a_{nm}
\end{pmatrix}
\equiv a_{ij}
.
\label{2016-m-ch-fdvs-matrixind}
\end{equation}
Stated differently, $a_{ij}$ is the element  of the table representing $\textsf{\textbf{A}}$ which is in the $i$th row and in the $j$th column.

A {\em matrix multiplication}
\index{matrix multiplication} (written with or without dot)
$\textsf{\textbf{A}} \cdot \textsf{\textbf{B}} = \textsf{\textbf{A}}  \textsf{\textbf{B}}$
of an $n \times m$ matrix $\textsf{\textbf{A}}\equiv a_{ij}$
with an $m \times l$ matrix $\textsf{\textbf{B}}\equiv b_{pq}$
can then be written as an $n \times l$ matrix
$\textsf{\textbf{A}} \cdot \textsf{\textbf{B}} \equiv a_{ij}b_{jk}$,
$1\le i\le n$,
$1\le j\le m$,
$1\le k\le l$.
Here the {\em Einstein summation convention}
\index{Einstein summation convention}
$a_{ij}b_{jk} = \sum_j a_{ij}b_{jk}$
has been used,
which requires that, when an index variable appears twice in a single term, one has to
sum over all of the possible index values.
Stated differently, if $\textsf{\textbf{A}}$ is an $n \times m$ matrix and $\textsf{\textbf{B}}$ is an $m \times l$ matrix,
their matrix product $\textsf{\textbf{A}}\textsf{\textbf{B}}$ is an $n \times l$ matrix, in which the $m$
entries across the rows of $\textsf{\textbf{A}}$ are multiplied with the $m$ entries down the columns of $\textsf{\textbf{B}}$.

As stated earlier ket and bra vectors (from the original or the dual vector space; exact definitions will be given later)
will be encoded -- with respect to a basis or coordinate system -- as an $n$-tuple of numbers;
which are arranged either in $n \times 1$ matrices (column vectors),
or in $1 \times n$ matrices (row vectors), respectively.
We can then write certain terms very compactly (alas often misleadingly).
Suppose, for instance, that
$\vert {\bf x} \rangle \equiv {\bf x}\equiv \begin{pmatrix}x_1, x_2,\ldots ,x_n\end{pmatrix}^\intercal $
and
$\vert {\bf y} \rangle \equiv {\bf y}\equiv \begin{pmatrix}y_1, y_2,\ldots ,y_n\end{pmatrix}^\intercal $
are two (column) vectors (with respect to a given basis).
Then, $x_iy_j a_{ij}$ can (somewhat superficially) be represented as a matrix multiplication ${\bf x}^\intercal  \textsf{\textbf{A}} {\bf y}$ of
a row vector with a matrix and a column vector  yielding a scalar; which in turn can be interpreted as a $1 \times 1$ matrix.
Note that, as ``$\intercal$'' indicates transposition\marginnote{Note that double transposition yields the identity.}
$\left({\bf y}^\intercal \right)^\intercal \equiv \left[ \begin{pmatrix}
y_1,
y_2,
\ldots ,
y_n
\end{pmatrix}^\intercal \right]^\intercal
=
\begin{pmatrix}
y_1,
y_2,
\ldots ,
y_n
\end{pmatrix}
$
represents a row vector, whose components or coordinates with respect to a particular (here undisclosed) basis are the scalars -- that is, an element of a field which will mostly be real or complex numbers -- $y_i$.

\subsection{Fields of real and complex numbers}
\label{2017-m-ch-fdvs-fields}

In physics, scalars occur either as real or complex numbers.
Thus we shall restrict our attention to these cases.

A {\em field}  $\langle  {\Bbb F} , +, \cdot , -, ^{-1}, 0, 1\rangle$
\index{field}
is a set together with two operations,
usually called {\em addition} and {\em multiplication},   denoted by ``$+$'' and ``$\cdot$''
(often  ``$a\cdot b$'' is identified with the expression ``$ab$'' without the center dot)
respectively, such that the following conditions (or, stated differently, axioms) hold:
\begin{itemize}
\item[(i)]
{\bf closure} of ${\Bbb F}$ with respect to addition and multiplication:
for all $a, b \in {\Bbb F}$, both $a + b$ as well as $a   b$ are in ${\Bbb F}$;
\item[(ii)]
{\bf associativity} of addition and multiplication:
for all $a$, $b$, and $c$ in ${\Bbb F}$,
the following equalities hold: $a + (b + c) = (a + b) + c$,
and
$a (b c) = (a  b) c$;
\item[(iii)]
{\bf commutativity} of addition and multiplication:
for all $a$ and $b$ in ${\Bbb F}$, the following equalities hold: $a + b = b + a$ and $a b = b  a$;
\item[(iv)]
additive and multiplicative {\bf identities}:
there exists an element of ${\Bbb F}$,
called the additive identity element and denoted by $0$, such that for all $a$ in ${\Bbb F}$,
$a + 0 = a$.
Likewise, there is an element, called the multiplicative identity element and denoted by $1$,
such that for all $a$ in ${\Bbb F}$, $1 \cdot a  = a$.
(To exclude the trivial ring, the additive identity and the multiplicative
identity are required to be distinct.)
\item[(v)]
additive and multiplicative {\bf inverses}:
for every $a$ in ${\Bbb F}$, there exists an element $-a$ in ${\Bbb F}$, such that $a + (-a) = 0$.
Similarly, for any $a$ in ${\Bbb F}$ other than $0$, there exists an element $a^{-1}$ in ${\Bbb F}$,
such that $a \cdot a^{-1} = 1$.
(The elements $+ (-a)$ and  $a^{-1}$
are also denoted $-a$ and $\frac{1}{a}$, respectively.)
Stated differently: subtraction and division operations exist.
\item[(vi)]
{\bf Distributivity} of multiplication over addition:
For all $a$, $b$ and $c$ in ${\Bbb F}$, the following equality holds:
$a (b + c) = (a  b) + (a  c)$.
\end{itemize}

\subsection{Vectors and vector space}
\marginnote[-7mm]{For proofs and additional information see  {\S}2   in~\bibentry{halmos-vs}.}

Vector spaces are structures or sets  allowing the
summation (addition,  ``coherent superposition'') of objects called ``vectors,'' and the multiplication
of these objects by scalars --  thereby remaining in these structures or sets, and hence satisfying a closure property.
\index{closure}
That is, for instance, the ``coherent superposition''  ${\bf a} + {\bf b}\equiv \vert {\bf a} + {\bf b} \rangle $
\index{coherent superposition}
\index{superposition}
of two vectors ${\bf a} \equiv \vert {\bf a} \rangle $ and ${\bf b} \equiv \vert {\bf b} \rangle $ can be guaranteed to
be a vector.
\sidenote{In order to define {\em length}, we have to engage an additional structure, namely the {\em norm}
$\|  {\bf a} \|$ of a vector ${\bf a}$.
\index{length}
And in order to define relative {\em direction} and {\em orientation}, and, in particular,
{\em orthogonality} and {\em collinearity} we have to define the {\em scalar product}
$\langle {\bf a} \vert {\bf b} \rangle$ of two vectors ${\bf a}$ and ${\bf b}$.
\index{direction}
\index{scalar product}
}
At this stage, little can be said about the length or relative direction or orientation of these ``vectors.''
Algebraically, ``vectors'' are elements of vector spaces.
Geometrically a vector may be interpreted as ``a quantity which is usefully represented by an arrow''.\cite{Weinreich}

A {\em linear vector space}      $\langle  \frak V , +, \cdot , -,  0, 1\rangle$
\index{linear vector space}
is a set $\frak V$ of elements called {\em vectors},
\index{vector}
here denoted by  bold face  symbols such as
$
{\bf a},
{\bf x},
{\bf v},
{\bf w},
\ldots
$,
or, equivalently, denoted by
$
\vert {\bf a}\rangle,
\vert {\bf x}\rangle,
\vert {\bf v}\rangle,
\vert {\bf w}\rangle,
\ldots
$,
satisfying certain conditions (or, stated differently, axioms); among them,
with respect to addition of vectors:
\begin{itemize}
\item[(i)]
{\bf commutativity}, that is, $\vert {\bf x}\rangle + \vert {\bf y}\rangle   = \vert {\bf y}\rangle + \vert {\bf x}\rangle$;
\item[(ii)]
{\bf associativity}, that is, $(\vert {\bf x}\rangle + \vert {\bf y}\rangle ) +  \vert {\bf z}\rangle = \vert {\bf x}\rangle + (\vert {\bf y}\rangle +  \vert {\bf z}\rangle )$;
\item[(iii)]
the uniqueness of the origin or {\bf null vector} $0$;
as well as
\item[(iv)]
the uniqueness of  the {\bf negative vector};
\item[ ]
with respect to multiplication of vectors with scalars:
\item[(v)]
the existence of an {\bf identity} or unit factor $1$; and
\item[(vi)]
{\bf distributivity} with respect to scalar and vector additions; that is,
\begin{equation}
\begin{split}
(\alpha +\beta ){\bf x} = \alpha {\bf x}+\beta  {\bf x}, \\
\alpha ({\bf x} +{\bf y}) = \alpha {\bf x}+\alpha {\bf y},
\end{split}
\end{equation}
with ${\bf x}, {\bf y} \in \frak V$ and scalars $\alpha ,\beta \in  {\Bbb F}$,
respectively.
\end{itemize}

{
\color{blue}
\bexample
Examples of vector spaces are:
\begin{itemize}
\item[(i)]
The set ${\Bbb C}$ of complex numbers: ${\Bbb C}$  can be interpreted as a complex vector space by  interpreting as vector addition and scalar multiplication
as the usual addition and multiplication of complex numbers, and with $0$ as the null vector;
\item[(ii)]
The set ${\Bbb C}^n$, $n \in {\Bbb N}$ of $n$-tuples of complex numbers:
Let
${\bf x}=
(x_1,\ldots , x_n)$
and
${\bf y}=
(y_1,\ldots , y_n)$.
 ${\Bbb C}^n$  can be interpreted as a complex vector space by  interpreting
the ordinary addition  $ {\bf x} + {\bf y} =  (x_1+y_1,\ldots , x_n+y_n)$
and the multiplication $\alpha {\bf x}=
(\alpha  x_1,\ldots ,\alpha  x_n)$ by a complex number $\alpha$ as vector addition
 and scalar multiplication, respectively;
the null tuple $0 =
(0,\ldots ,0)$ is the neutral element of vector addition;
\item[(iii)]
The set ${\frak P}$
 of all polynomials with complex coefficients in a variable~$t$:
${\frak P}$  can be interpreted as a complex vector space by  interpreting
the ordinary addition of polynomials and the multiplication of a polynomial by a complex number as vector addition and scalar multiplication,
respectively;
the null polynomial is the neutral element of vector addition.
\end{itemize}
\eexample
}

\section{Linear independence}

A set ${\frak S}=\{
{\bf x}_1,
{\bf x}_2,
\ldots ,
{\bf x}_k\} \subset {\frak V}$
of vectors ${\bf x}_i$ in a linear vector space
is {\em linearly independent}
\index{linear independence}
if ${\bf x}_i\neq 0\,\forall 1\le i\le k$,
and additionally, if either $k=1$,
or if no vector in ${\frak S}$ can be written as a linear combination of other vectors in this set ${\frak S}$;
that is, there are no scalars $\alpha_j$ satisfying
 ${\bf x}_i=\sum_{1\le j \le k, \; j\neq i}\alpha_j {\bf x}_j$.

Equivalently, if $\sum_{i=1}^k \alpha_i {\bf x}_i = 0$
 implies $\alpha_i =0$ for each $i$, then the set
${\frak S}=\{
{\bf x}_1,
{\bf x}_2,
\ldots ,
{\bf x}_k\}  $ is linearly independent.

Note that the vectors of a basis are linear independent and ``maximal''
insofar as any inclusion of an additional vector results in a linearly dependent set;
that is, this additional vector can be expressed in terms of a linear combination of the
existing basis vectors; see also Section~\ref{2012-m-ch-fdvs-Basis} on page \pageref{2012-m-ch-fdvs-Basis}.

\section{Subspace}
\label{2011-m-subspace}
\marginnote{For proofs and additional information see {\S}10 in~\bibentry{halmos-vs}.}
A nonempty subset ${\frak M}$ of a vector space is a {\em subspace}
\index{subspace}
or, used synonymously,
a {\em linear manifold}\index{linear manifold},
 if, along with every pair of vectors ${\bf x}$   and  ${\bf y}$
 contained in  ${\frak M}$,
 every linear combination
$\alpha {\bf x} + \beta {\bf y}$ is also contained in  ${\frak M}$.

If
${\frak U}$
and
${\frak V}$
are two subspaces of a vector space,
then
${\frak U}+{\frak V}$
is the subspace spanned by
${\frak U}$
and
${\frak V}$;
that is,
it contains all vectors
${\bf z}={\bf x}+{\bf y}$, with
${\bf x}\in {\frak U}$  and
${\bf y}\in {\frak V}$.

${\frak M}$ is the {\em linear span}
\index{linear span}
\index{span}
\begin{equation}
{\frak M}
= \textrm{span}({\frak U},{\frak V})
= \textrm{span}({\bf x},{\bf y}) =
\{\alpha {\bf x} +\beta {\bf y}\mid \alpha ,\beta \in {\Bbb F}, {\bf x} \in {\frak U},
{\bf y} \in {\frak V}\}.
\end{equation}

A generalization to more than two vectors and more than two subspaces is straightforward.

For every vector space ${\frak V}$, the vector space containing only the null vector,
 and the vector space ${\frak V}$ itself are subspaces of ${\frak V}$.

\subsection{Scalar or inner product}
\marginnote{For proofs and additional information see {\S}61 in~\bibentry{halmos-vs}.}
\index{scalar product}
\index{inner product}
\label{2011-m-scalarproduct}

A {\em scalar} or {\em inner} product presents some form of measure of ``distance'' or ``apartness''
of two vectors in a linear vector space.
It should not be confused with the bilinear functionals (introduced on page \pageref{2011-m-dvs}) that connect a vector space with its dual vector space,
although for real Euclidean vector spaces these may coincide,
and although the scalar product is also bilinear in its arguments.
It should also not be confused with the tensor product introduced in Section~\ref{2011-m-tensorp} on page \pageref{2011-m-tensorp}.

An inner product space is a vector space $\frak V$,
together with an inner product; that is, with a map
 $\langle \cdot \vert \cdot \rangle :  \frak V  \times  \frak V  \longrightarrow {\Bbb F}$
 (usually ${\Bbb F}={\Bbb C}$ or ${\Bbb F}={\Bbb R}$)
 that satisfies the following three conditions (or, stated differently, axioms) for all vectors  and all scalars:
\begin{itemize}
\item[(i)]
{\bf Conjugate (Hermitian) symmetry}:
\index{conjugate symmetry}
\index{Hermitian symmetry}
$
\langle {\bf x}\vert {\bf y}\rangle
=
\overline{\langle {\bf y}\vert {\bf x}\rangle }$;
\marginnote{For real, Euclidean vector spaces, this function is symmetric; that is
$
\langle {\bf x}\vert {\bf y}\rangle
=
{\langle {\bf y}\vert {\bf x}\rangle}$.
}
\item[(ii)]
{\bf linearity} in the second argument:
\marginnote{This definition and nomenclature is different from Halmos' axiom which defines linearity in the  {\em first} argument.
We chose linearity in the second argument because this is usually assumed in physics textbooks, and because Thomas Sommer strongly insisted.}
$$
\langle {\bf z}  \vert \alpha{\bf x} + \beta {\bf y} \rangle
=
\alpha {\langle {\bf z}\vert {\bf x}\rangle}
+
\beta {\langle {\bf z}\vert {\bf y}\rangle}
;
$$
\item[(iii)]
\label{2016-m-ch-fdvs-pd}
{\bf positive-definiteness}:
$
\langle {\bf x}\vert {\bf x}\rangle
\ge
0$;  with equality if and only if ${\bf x} = 0$.
\end{itemize}

Note that from the first two properties, it follows that the inner product is
{\em antilinear}, or synonymously,
{\em conjugate-linear}, in its {\em first} argument (note that $\overline{(uv)}= (\overline{u}) {\;} (\overline{v}) $ for all $u,v\in {\Bbb C}$):
\begin{equation}
\begin{split}
 \langle \alpha {\bf x} + \beta {\bf y}   \vert {\bf z}\rangle
 =
 \overline{\langle   {\bf z} \vert \alpha {\bf x} + \beta {\bf y}\rangle }
 =
 \overline{ \alpha}\overline{\langle {\bf z} \vert {\bf x} \rangle}
 + \overline{\beta }\overline{\langle {\bf z} \vert {\bf y} \rangle}
 =
 \overline{\alpha} {\langle {\bf x}\vert {\bf z}\rangle}
 +\overline{\beta} {\langle {\bf y}\vert {\bf z}\rangle}
.
\end{split}
\end{equation}

{
\color{blue}
\bexample
One example of an inner product is  the
{\em dot product}
\index{dot product}
\begin{equation}
\langle {{\bf x}} \vert  {\bf y} \rangle
=
\sum_{i=1}^n \overline{x_i}y_i
\end{equation}
of two vectors ${\bf x}=
(x_1,\ldots , x_n)$
and
${\bf y}=
(y_1,\ldots , y_n)$ in ${\Bbb C}^n$,
which, for real Euclidean space,  reduces to the well-known dot product
$\langle  {\bf x} \vert {\bf y} \rangle
=
{x_1}y_1 + \cdots + {x_n}y_n  = \| {\bf x}\| \| {\bf y}\| \cos \angle ({\bf x},{\bf y})$.

It is mentioned without proof that the most general form of an inner product in ${\Bbb C}^n$
is
$\langle  {\bf x} \vert {\bf y} \rangle
=  {\bf y}^\dagger \textsf{\textbf{A}} {\bf x}= \overline{ {\bf x}^\dagger \textsf{\textbf{A}} {\bf y}}$,
where the symbol ``$\dagger$'' stands for the {\em conjugate transpose} (also denoted as
{\em Hermitian conjugate} or {\em Hermitian adjoint}),
\index{conjugate transpose}
\index{Hermitian conjugate}
\index{Hermitian adjoint}
and $ \textsf{\textbf{A}} $ is a positive definite Hermitian matrix (all of its eigenvalues are positive).
\eexample
}

The {\em norm} of a vector ${\bf x}$
\index{norm}
is defined by
\begin{equation}
\|
{\bf x}
\|
=\sqrt{\langle {\bf x}\vert {\bf x}\rangle }
.
\label{2018-mm-ch-vn}
\end{equation}

Conversely, the {\em polarization identity}
\index{polarization identity}
expresses the inner product of two vectors in terms of the norm of their differences; that is,

\begin{equation}
\langle {\bf x}\vert {\bf y}\rangle
=
\frac{1}{4}\left[
\|  {\bf x}+ {\bf y} \|^2
-
\|  {\bf x}- {\bf y} \|^2
+ i
\left(
 \|  {\bf x}- i{\bf y} \|^2
-
\|  {\bf x}+ i{\bf y} \|^2
\right)
\right]
.
\label{2015-m-ch-polidc}
\end{equation}

{\color{OliveGreen}
\bproof
In complex vector space,  a direct but tedious calculation
-- with conjugate-linearity (antilinearity)
in the first argument and linearity in the second argument of the inner product -- yields
\begin{equation}
\begin{split}
\frac{1}{4}\left(
\|  {\bf x}+ {\bf y} \|^2
-
\|  {\bf x}- {\bf y} \|^2
+ i
\|  {\bf x}- i{\bf y} \|^2
- i
\|  {\bf x}+ i{\bf y} \|^2
\right)
\\
=
\frac{1}{4}\left(
\langle {\bf x}+ {\bf y} \vert {\bf x}+ {\bf y}\rangle
-
\langle {\bf x}- {\bf y} \vert {\bf x}- {\bf y}\rangle
+ i
 \langle {\bf x}- i{\bf y} \vert {\bf x} -i {\bf y}\rangle
- i   \langle {\bf x}+ i{\bf y} \vert {\bf x}+ i{\bf y}\rangle
\right)
\\
=
\frac{1}{4}\left[
(
\langle {\bf x} \vert {\bf x}\rangle +
\langle {\bf x} \vert {\bf y}\rangle +
\langle {\bf y} \vert  {\bf x}\rangle +
\langle {\bf y} \vert {\bf y}\rangle
)
-
(
 \langle {\bf x} \vert {\bf x}\rangle    -
\langle {\bf x} \vert {\bf y}\rangle    -
\langle {\bf y} \vert {\bf x}\rangle    +
\langle {\bf y} \vert {\bf y}\rangle
)
\right.
\\
\left.
+ i
(
\langle {\bf x} \vert {\bf x}\rangle -
\langle {\bf x} \vert i {\bf y}\rangle  -
\langle i{\bf y} \vert {\bf x}\rangle  +
\langle  i{\bf y} \vert i {\bf y}\rangle
)
- i
(
\langle {\bf x} \vert {\bf x}\rangle +
\langle {\bf x} \vert  i{\bf y}\rangle  +
\langle i{\bf y} \vert {\bf x}\rangle   +
\langle i{\bf y} \vert i{\bf y}\rangle
)
\right]
\\
=
\frac{1}{4}\left[
\langle {\bf x} \vert {\bf x}\rangle +
\langle {\bf x} \vert {\bf y}\rangle +
\langle {\bf y} \vert  {\bf x}\rangle +
\langle {\bf y} \vert {\bf y}\rangle
-
\langle {\bf x} \vert {\bf x}\rangle    +
\langle {\bf x} \vert {\bf y}\rangle    +
\langle {\bf y} \vert {\bf x}\rangle    -
\langle {\bf y} \vert {\bf y}\rangle
)
\right.
\\
\left.
+
i \langle {\bf x} \vert {\bf x}\rangle -
i\langle {\bf x} \vert  i{\bf y}\rangle  -
i\langle i{\bf y} \vert {\bf x}\rangle   +
i\langle i{\bf y} \vert i{\bf y}\rangle
-
i\langle {\bf x} \vert {\bf x}\rangle -
i\langle {\bf x} \vert i {\bf y}\rangle -
i\langle i{\bf y} \vert {\bf x}\rangle  -
i\langle  i{\bf y} \vert i {\bf y}\rangle
)
\right]
\\
=
\frac{1}{4}\left[
2 (
\langle {\bf x} \vert {\bf y}\rangle +
\langle {\bf y} \vert  {\bf x}\rangle
)
- 2 i (
\langle {\bf x} \vert  i{\bf y}\rangle  +
\langle i{\bf y} \vert {\bf x}\rangle
)
\right]
\\
=
\frac{1}{2}\left[
(
\langle {\bf x} \vert {\bf y}\rangle +
\langle {\bf y} \vert  {\bf x}\rangle
)
- i (
i\langle {\bf x} \vert  {\bf y}\rangle     -
i\langle {\bf y} \vert {\bf x}\rangle
)
\right]
\\
=
\frac{1}{2}\left[
\langle {\bf x} \vert {\bf y}\rangle +
\langle {\bf y} \vert  {\bf x}\rangle
+
\langle {\bf x} \vert  {\bf y}\rangle
-\langle {\bf y} \vert {\bf x}\rangle
\right]
=
\langle {\bf x} \vert {\bf y}\rangle
.
\end{split}
\label{2015-m-ch-polidc2}
\end{equation}
\eproof
}

For any real vector space the imaginary terms in (\ref{2015-m-ch-polidc}) are absent, and (\ref{2015-m-ch-polidc}) reduces to
\begin{equation}
\langle {\bf x}\vert {\bf y}\rangle
=
\frac{1}{4}\left(
\langle {\bf x}+ {\bf y} \vert {\bf x}+ {\bf y}\rangle
-
\langle {\bf x}- {\bf y} \vert {\bf x}- {\bf y}\rangle
\right)
=
\frac{1}{4}\left(
\|  {\bf x}+ {\bf y} \|^2
-
\|  {\bf x}- {\bf y} \|^2
\right).
\label{2015-m-ch-polidr}
\end{equation}

Two nonzero vectors $  {\bf x} , {\bf y} \in {\frak V}$,  $  {\bf x},   {\bf y}\neq 0$
are {\em orthogonal}, denoted by ``${\bf x} \perp {\bf y}$''
\index{othogonality}
if their scalar product vanishes; that is, if
\begin{equation}
\langle  {\bf x} \vert {\bf y} \rangle   = 0.
\end{equation}

Let ${\frak E}$ be any set of vectors in an inner product space ${\frak V}$.
The symbol
\begin{equation}
{\frak E}^\perp  = \left\{ {\bf x}\mid  \langle  {\bf x} \vert {\bf y} \rangle=0,  {\bf x} \in {\frak V},
\forall {\bf y} \in {\frak E} \right\}
\end{equation}
 denotes the set of all vectors in  ${\frak V}$ that are
orthogonal to every vector in  ${\frak E}$.

Note that, regardless of whether or not ${\frak E}$ is a subspace,
\marginnote{See page \pageref {2011-m-subspace} for a definition of subspace.}
${\frak E}^\perp$ is a subspace.
Furthermore,
${\frak E}$  is contained in $({\frak E}^\perp)^\perp= {\frak E}^{\perp\perp}$.
In case ${\frak E}$ is a subspace, we call ${\frak E}^\perp$
the {\em orthogonal complement}
\index{orthogonal complement}
of ${\frak E}$.

The following {\em projection theorem}
\index{projection theorem}
is mentioned without proof.
If ${\frak M}$ is any subspace of a finite-dimensional inner product space ${\frak V}$,
then ${\frak V}$ is the direct sum of ${\frak M}$ and ${\frak M}^\perp$;
that is, ${\frak M}^{\perp \perp}={\frak M}$.

{
\color{blue}
\bexample
For the sake of an example, suppose ${\frak V}={\Bbb R}^2$,
and take ${\frak E}$ to be the set of all vectors spanned by the vector $(1,0)$;
then ${\frak E}^\perp$ is the set of all vectors spanned by $(0,1)$.
\eexample
}

\subsection{Hilbert space}

A (quantum mechanical) {\em Hilbert space}  is a linear
\index{Hilbert space}
vector space ${\frak V}$ over the field ${\Bbb C}$ of complex numbers
(sometimes only ${\Bbb R}$ is used)
equipped with vector addition, scalar multiplication, and some inner (scalar) product.
Furthermore, {\em completeness}
by the Cauchy criterion for sequences
is an additional requirement,
but nobody has made operational sense of that so far:
If ${\bf x}_n\in {\frak V}$, $n=1,2,\ldots$, and if $\lim_{n,m\rightarrow
\infty} ({\bf x}_n-{\bf x}_m,{\bf x}_n-{\bf x}_m)=0$,
then there exists an ${\bf x}\in {\frak V}$ with
$\lim_{n\rightarrow \infty} ({\bf x}_n-{\bf x},{\bf x}_n-{\bf x})=0$.

Infinite dimensional vector spaces and continuous spectra are nontrivial
extensions of the finite
dimensional Hilbert space treatment. As a heuristic rule -- which is not
always correct -- it might be
stated that the sums become integrals, and the Kronecker delta function
$\delta_{ij}$ defined by
\index{Kronecker delta function}
\begin{equation}
\delta_{ij} =\begin{cases}
0  &\text{ for }i\neq j , \\
1  &\text{ for }i = j.
\end{cases}
\end{equation}
becomes the Dirac delta function $\delta (x-y)$, which is a
generalized function in the continuous variables $x,y$.
In the Dirac bra-ket notation, the resolution of the identity operator,
sometimes also referred to as {\em completeness}, is given by
\index{resolution of the identity}
\index{completeness}
$\mathbb{1} = \int_{-\infty}^{+\infty} \vert x\rangle \langle  x\vert \, dx$.
For a careful treatment, see, for instance,
the books by
Reed and Simon,\cite{reed-sim1,reed-sim2}
or wait for Chapter~\ref{2011-m-ch:gf}, page~\pageref{2011-m-ch:gf}.

\section{Basis}
\label{2012-m-ch-fdvs-Basis}
\marginnote{For proofs and additional information see {\S}7 in~\bibentry{halmos-vs}.}

We shall use bases of vector spaces to formally represent vectors (elements) therein.

A (linear) {\em basis}
\index{basis}
[or a {\em coordinate system}, or a {\em frame (of reference)}]
\index{coordinate system}
\index{frame}
is a set    $\frak B$
of linearly independent vectors
such that every vector
in $\frak V$ is a linear combination of the vectors in the basis; hence
$\frak B$ spans $\frak V$.

What particular basis should one choose?
{\em A priori} no basis is privileged over the other.
Yet, in view of certain (mutual) properties of elements of some bases (such as orthogonality or orthonormality)
we shall prefer some or one over others.

Note that a vector is some directed entity with a particular length,
oriented in some (vector) ``space.''
It is ``laid out there'' in front of our eyes, as it is: some directed entity.
{\it A priori}, this space, in its most primitive form,
is not equipped with a basis, or synonymously, a frame of reference, or reference frame.
Insofar it is not yet coordinatized.
In order to formalize the notion of a vector, we have to encode this vector by ``coordinates''
or ``components'' which are the coefficients with respect to a (de)composition into basis elements.
Therefore, just as for numbers (e.g., by different numeral bases, or by prime decomposition),
there exist many ``competing'' ways to encode a vector.

Some of these ways appear to be rather straightforward,
such as, in particular, the {\em Cartesian basis},
also synonymously called the  {\em standard basis}.
\index{Cartesian basis}
\index{standard basis}
It is, however, not in any way {\it a priori}
``evident'' or ``necessary'' what should be specified to be ``the Cartesian basis.''
Actually, specification of a ``Cartesian basis'' seems to be mainly motivated by
physical inertial motion --
and thus identified with some inertial frame of reference --
``without any friction and
forces,'' resulting in a ``straight line motion at constant speed.''
(This sentence is  cyclic because heuristically any such absence of ``friction and
force''  can only be operationalized by testing if the motion is a
``straight line motion at constant speed.'')
If we grant that in this way straight lines can be defined, then
Cartesian bases in Euclidean vector spaces can be characterized by
orthogonal (orthogonality is defined {\it via} vanishing scalar products between nonzero vectors)
straight lines spanning the entire space.
In this way, we arrive, say for a planar situation, at the coordinates
characterized by some basis $\{(0,1),(1,0)\}$,
where, for instance, the basis vector ``$(1,0)$'' literally and physically
means ``a unit arrow pointing in some particular, specified direction.''

Alas, if we would prefer, say, cyclic motion in the plane,
we might want to call a frame based on the polar coordinates $r$ and $\theta$ ``Cartesian,''
resulting in some ``Cartesian basis'' $\{(0,1),(1,0)\}$;
but this ``Cartesian basis'' would be very different from the Cartesian
basis mentioned earlier,
as ``$(1,0)$'' would refer to some specific unit radius,
and ``$(0,1)$'' would refer to some specific unit angle (with respect to a specific zero angle).
In terms of the ``straight'' coordinates (with respect to ``the usual Cartesian basis'')
$x,y$, the polar coordinates are $r = \sqrt{x^2+y^2}$ and $\theta = \mathrm{tan}^{-1} (y/x)$.
We obtain the original ``straight'' coordinates (with respect to ``the usual Cartesian basis'')
back if we take
$x=r\cos \theta$
and
$y=r\sin \theta$.

Other bases than the ``Cartesian'' one may be less suggestive at first; alas it may be ``economical'' or pragmatical to use them;
mostly to cope with, and adapt to, the {\em symmetry} of a physical configuration:
if the physical situation at hand is, for instance, rotationally invariant,
we might want to use rotationally invariant bases --
such as, for instance, polar coordinates in two dimensions, or spherical coordinates in three dimensions --
to represent a vector, or, more generally, to encode any given representation of a physical entity
(e.g., tensors, operators) by such bases.

\section{Dimension}
\marginnote{For proofs and additional information see {\S}8 in~\bibentry{halmos-vs}.}
The {\em dimension}
\index{dimension}
of $\frak V$ is the number of elements in $\frak B$.

All bases $\frak B$ of $\frak V$ contain the same number of elements.

A vector space is finite dimensional if its bases are finite; that is, its bases
contain a finite number of elements.

{\color{Purple}
In quantum physics, the dimension of a quantized system is associated with
the {\em number of mutually exclusive measurement outcomes}.
For a spin state measurement of an electron
along with a particular direction,
as well as for a measurement of the linear polarization
of a photon in a particular direction,
the dimension is two, since both measurements
may yield two distinct outcomes
which we can
interpret as vectors in two-dimensional Hilbert space,
which, in Dirac's bra-ket notation,\cite{dirac} can be written as
$
\vert \uparrow \rangle$ and $\vert \downarrow \rangle$,
or $\vert + \rangle$ and $
\vert - \rangle
$,
or
$
\vert H \rangle $ and $
\vert V \rangle
$,
or
$
\vert 0 \rangle $ and $
\vert 1 \rangle
$,
or
$
\bigg|$\raisebox{-0.4\height}{\begin{tikzpicture}
    \emoticon\pupils
    \draw[very thick,red,line cap=round] (-1ex,-1ex)
               ..controls (-0.5ex,-1.5ex)and(0.5ex,-1.5ex)..(1ex,-1ex);
    \draw (0.60ex,1.20ex)--(0.60ex,1.60ex)
          (0.85ex,1.25ex)--(0.95ex,1.45ex)
          (1.00ex,1.00ex)--(1.20ex,1.10ex)
          (0.35ex,1.15ex)--(0.25ex,1.35ex)
          (-0.60ex,1.20ex)--(-0.60ex,1.60ex)
          (-0.85ex,1.25ex)--(-0.95ex,1.45ex)
          (-1.00ex,1.00ex)--(-1.20ex,1.10ex)
          (-0.35ex,1.15ex)--(-0.25ex,1.35ex);
\end{tikzpicture}}$\bigg\rangle $ and $
\bigg|$\raisebox{-0.4\height}{\begin{tikzpicture}
\emoticon\pupils
    \fill[shift={( 0.5ex,0.5ex)},rotate=90] (0,0) ellipse (0.3ex and 0.15ex);
    \fill[shift={(-0.5ex,0.5ex)},rotate=90] (0,0) ellipse (0.3ex and 0.15ex);
    \draw[thick] (-1ex,-1ex)..controls
        (-0.5ex,-0.5ex)and(0.5ex,-0.5ex)..(1ex,-1ex);
    \draw[thick] (0.2ex,1.15ex)--(0.5ex,1.6ex)(-0.2ex,1.15ex)--(-0.5ex,1.6ex);
\end{tikzpicture}}$\bigg\rangle
$,
respectively.
}

\section{Vector coordinates or components}
\marginnote{For proofs and additional information see {\S}46 in~\bibentry{halmos-vs}.}
The coordinates or components of a vector with respect to some basis
represent the coding of that vector in that particular basis.
It is important to realize that, as bases change, so do coordinates.
Indeed, the changes in coordinates have to ``compensate'' for the bases change,
because the same coordinates in a different basis would render an altogether different
vector.
Thus it is often said that, in order to represent one and the same vector,
if the base vectors {\em vary}, the corresponding components or coordinates have to {\em contra-vary.}
Figure~\ref{2011-m-bases} presents some geometrical demonstration of
these thoughts, for your contemplation.

\begin{figure}[ht]
\caption{Coordinazation of vectors:
(a) some primitive vector;
(b)  some primitive vectors, laid out in some space, denoted by dotted lines
(c) vector coordinates $x_1$ and $x_2$ of the vector  ${\bf x} =  (x_1,x_2) = x_1{\bf e}_1 +  x_2{\bf e}_2$ in a standard basis;
(d) vector coordinates $x_1'$ and $x_2'$ of the vector  ${\bf x} = (x_1',x_2') =  x_1'{\bf e}_1' +  x_2'{\bf e}_2'$  in some nonorthogonal basis.
\label{2011-m-bases}}
\begin{center}
\begin{tabular}{cc}
\begin{tikzpicture}[ scale=0.8,]

\tikzstyle{every path}=[line width=2pt]

\begin{axis}[
axis equal,
ymin=-6cm,
ymax=6cm,
xmin=-6cm,
xmax=6cm,
height=6cm,
width=6cm,
axis line style={draw=none},
tick style={draw=none},
xticklabels={,,},
yticklabels={,,},
]

\draw[orange,line width=3pt,->] (axis cs:-3cm,-3cm) -- (axis cs:3cm,3cm) node[above right]{${\bf x}$};

\end{axis}
\end{tikzpicture}
&
\begin{tikzpicture}[ scale=0.8,]

\tikzstyle{every path}=[line width=2pt]

\begin{axis}[
axis equal,
draw=gray!80,
ymin=-6cm,
ymax=6cm,
xmin=-6cm,
xmax=6cm,
height=6cm,
width=6cm,
axis line style={dotted},
tick style={draw=none},
xticklabels={,,},
yticklabels={,,},
]

\draw[orange,line width=3pt,->] (axis cs:-3cm,-3cm) -- (axis cs:3cm,3cm) node[above right]{${\bf x}$};

\end{axis}
\end{tikzpicture}
\\
(a)&(b)\\
$\;$\\
\begin{tikzpicture}[ scale=0.8,]

\tikzstyle{every path}=[line width=2pt]

\begin{axis}[
axis equal,
draw=gray!80,
ymin=-6cm,
ymax=6cm,
xmin=-6cm,
xmax=6cm,
height=6cm,
width=6cm,
axis line style={dotted},
tick style={draw=none},
xticklabels={,,},
yticklabels={,,},
]

\draw[blue,line width=2pt,->] (axis cs:-3cm,-3cm) -- (axis cs:4cm,-3cm) node[right]{${\bf e}_1$};
\draw[blue,line width=2pt,->] (axis cs:-3cm,-3cm) -- (axis cs:-3cm,4cm) node[above]{${\bf e}_2$};

\draw[blue,line width=1pt,dashed] (axis cs:3cm,-3cm) node[below]{$x_1$} -- (axis cs:3cm,3cm);
\draw[blue,line width=1pt,dashed] (axis cs:-3cm,3cm) node[left]{$x_2$} -- (axis cs:3cm,3cm);

\draw[orange,line width=3pt,->] (axis cs:-3cm,-3cm) -- (axis cs:3cm,3cm) node[above right]{${\bf x}$};

\end{axis}

\end{tikzpicture}
&
\begin{tikzpicture}[ scale=0.8,]

\tikzstyle{every path}=[line width=2pt]

\begin{axis}[
axis equal,
draw=gray!80,
ymin=-6cm,
ymax=6cm,
xmin=-6cm,
xmax=6cm,
height=6cm,
width=6cm,
axis line style={dotted},
tick style={draw=none},
xticklabels={,,},
yticklabels={,,},
]

\draw[blue,line width=2pt,->] (axis cs:-3cm,-3cm) -- (axis cs:3cm,-0cm) node[right]{${\bf e}_1$};
\draw[blue,line width=2pt,->] (axis cs:-3cm,-3cm) -- (axis cs:0cm,3cm) node[above]{${\bf e}_2$};

\draw[blue,line width=1pt,dashed] (axis cs:1cm,-1cm) node[below]{$x_1$} -- (axis cs:3cm,3cm);
\draw[blue,line width=1pt,dashed] (axis cs:-1cm,1cm) node[left]{$x_2$} -- (axis cs:3cm,3cm);

\draw[orange,line width=3pt,->] (axis cs:-3cm,-3cm) -- (axis cs:3cm,3cm) node[above right]{${\bf x}$};

\end{axis}

\end{tikzpicture}
\\
(c)&(d)\\
\end{tabular}
\end{center}
\end{figure}

Elementary high school tutorials often condition students into believing that the components of the vector
``is'' the vector, rather than emphasizing that these components {\em represent} or {\em encode}
the vector with respect to some (mostly implicitly assumed) basis.
A similar situation occurs in many introductions to quantum theory,
where the span
(i.e., the one-dimensional linear subspace spanned by that vector)
\index{span}
$\{
{\bf y}
\mid
{\bf y} = \alpha {\bf x}, \alpha \in {\Bbb C}
\}$, or, equivalently,  for orthogonal projections,
the {\em projection} (i.e., the projection operator; see also page \pageref{2011-m-projec})
\index{projection}
$\textsf{\textbf{E}}_{\bf x} \equiv {\bf x} \otimes {\bf x}^\dagger  \equiv \vert {\bf x} \rangle \langle {\bf x}\vert$
corresponding to a unit (of length $1$) vector ${\bf x}$
often is identified with that vector.
In many instances, this is a great help and,
if administered properly, is consistent and fine (at least for all practical purposes).

The  Cartesian  standard basis in $n$-dimensional complex space ${\Bbb C}^n$
\index{Cartesian basis}
\index{standard basis}
is the set of (usually ``straight'')
vectors $x_i, i=1, \ldots , n$, of ``unit length''
-- the unit is conventional and thus needs to be fixed as operationally precisely as possible,
such as in the {\em International System of Units (SI)}
\marginnote{In the {\em International System of Units (SI)}
\index{International System of Units}
the ``second'' as the unit of time is defined to be the duration of 9 192 631 770 periods of the radiation corresponding to the transition
between the two hyperfine levels of the ground state of the cesium 133 atom.
The `` meter'' as the unit of length is defined to be the length of the path traveled by light in vacuum during a time interval
of 1/299 792 458 of a second
--
or, equivalently,
as light travels 299 792 458 meters per second,
a duration in which 9 192 631 770 transitions between two orthogonal quantum states of a cesium 133 atom occur
--
during
9 192 631 770/299 792 458 $\approx 31$ transitions of two orthogonal quantum states of a cesium 133 atom.
Thereby, the speed of light in the vacuum is fixed at exactly 299 792 458 meters per second; see also \bibentry{peres-84}.}
--
represented by $n$-tuples,
defined by the condition that the $i$'th coordinate of the $j$'th basis vector
${\bf e}_j$ is given by $\delta_{ij}$.
Likewise, $\delta_{ij}$ can be interpreted as the  $j$'th coordinate of the $i$'th basis vector.
Thereby $\delta_{ij}$ is the Kronecker delta function
\index{Kronecker delta function}
\begin{equation}
\delta_{ij} =\delta_{ji} =\begin{cases}
0  &\text{ for }i\neq j , \\
1  &\text{ for }i = j.
\end{cases}
\end{equation}
Thus we can represent the basis vectors by
\begin{equation}
\begin{split}
\vert {\bf e}_1 \rangle \equiv {\bf e}_1 \equiv \begin{pmatrix}1\\ 0\\ \vdots\\ 0\end{pmatrix},\quad
\vert {\bf e}_2 \rangle \equiv {\bf e}_2 \equiv \begin{pmatrix}0\\ 1\\ \vdots\\ 0\end{pmatrix},\quad
\ldots\quad
\vert {\bf e}_n \rangle \equiv {\bf e}_n \equiv \begin{pmatrix}0\\ 0\\ \vdots\\ 1\end{pmatrix}.
\end{split}
\label{2016-m-fdvs-csb}
\end{equation}

In terms of these standard base vectors, every vector ${\bf x}$
can be written as a linear combination -- in quantum physics, this is called
{\em coherent superposition}
\index{coherent superposition}
\index{superposition}
\begin{equation}
\vert {\bf x} \rangle \equiv {\bf x} = \sum_{i=1}^n x_i{\bf e}_i \equiv  \sum_{i=1}^n x_i \vert {\bf e}_i \rangle
\equiv
\begin{pmatrix}x_1\\x_2\\ \vdots \\ x_n\end{pmatrix}
\label{2016-m-fdvs-rv0}
\end{equation}
with respect  to the basis
${\frak B}= \{ {\bf e}_1 ,
  {\bf e}_2   ,
\ldots ,
 {\bf e}_n
\}$.

With the notation\marginnote{For reasons demonstrated later in Equation~(\ref{2015-m-ch-fdlvs-uniascolv})
$\textsf{\textbf{U}}$ is a unitary matrix, that is,
$\textsf{\textbf{U}}^{-1}=\textsf{\textbf{U}}^\dagger= \overline{\textsf{\textbf{U}}}^\intercal $,
where the overline stands for complex conjugation $\overline{u}_{ij}$ of the entries $u_{ij}$ of $\textsf{\textbf{U}}$, and
the superscript ``$\intercal$'' indicates transposition; that is, $\textsf{\textbf{U}}^\intercal $ has entries $u_{ji}$.
}
defined by
\begin{equation}
\begin{split} X = \begin{pmatrix}
x_1, x_2, \ldots , x_n
\end{pmatrix}^\dagger
\textrm{, and }
\\
\textsf{\textbf{U}} =
\begin{pmatrix}{\bf e}_1,{\bf e}_2, \ldots , {\bf e}_n\end{pmatrix}
\equiv
\begin{pmatrix} \vert {\bf e}_1\rangle ,\vert {\bf e}_2 \rangle ,  \ldots , \vert {\bf e}_n\rangle \end{pmatrix}
,
\label{2016-m-fdvs-not}
\end{split}
\end{equation}
such that  $u_{ij} = e_{i,j}$ is the $j$th component of the $i$th vector,
Equation~(\ref{2016-m-fdvs-rv0}) can be written
in ``Euclidean dot product notation,''
that is,
``column times row''
and
``row times column'' (the dot is usually omitted)
\begin{equation}
\begin{split}
\vert {\bf x} \rangle \equiv {\bf x} =
\begin{pmatrix}{\bf e}_1,{\bf e}_2, \ldots , {\bf e}_n\end{pmatrix}
\begin{pmatrix} x_1\\x_2\\ \vdots \\ x_n \end{pmatrix}
\equiv
\begin{pmatrix} \vert {\bf e}_1\rangle ,\vert {\bf e}_2 \rangle ,  \ldots , \vert {\bf e}_n\rangle \end{pmatrix}
\begin{pmatrix} x_1\\x_2\\ \vdots \\ x_n \end{pmatrix} \equiv
\\
\equiv
\begin{pmatrix}
{\bf e}_{1,1} &   {\bf e}_{2,1}  &  \cdots &  {\bf e}_{n,1}\\
{\bf e}_{1,2} &   {\bf e}_{2,2}  &  \cdots &  {\bf e}_{n,2}\\
\cdots  & \cdots  &  \ddots &  \cdots \\
{\bf e}_{1,n} &   {\bf e}_{2,n}  &  \cdots &  {\bf e}_{n,n}\\
\end{pmatrix}
\begin{pmatrix} x_1\\x_2\\ \vdots \\ x_n \end{pmatrix}
\equiv
\textsf{\textbf{U}} X
 .
\label{2016-m-fdvs-rv}
\end{split}
\end{equation}
Of course, with the Cartesian standard basis (\ref{2016-m-fdvs-csb}), $\textsf{\textbf{U}} = \mathbb{1}_n$, but
(\ref{2016-m-fdvs-rv}) remains valid for general bases.

In (\ref{2016-m-fdvs-rv}) the identification of the tuple
$ X = \begin{pmatrix}
x_1, x_2, \ldots , x_n
\end{pmatrix}^\intercal
$
containing the vector components $x_i$
with the vector $\vert {\bf x} \rangle \equiv {\bf x}$
really means
``coded {\em with respect}, or {\em relative},  to the basis ${\frak B}=\{{\bf e}_1,{\bf e}_2, \ldots , {\bf e}_n\}$.''
Thus in what follows, we shall often identify the column vector
$
\begin{pmatrix}
x_1, x_2, \ldots , x_n
\end{pmatrix}^\intercal
$
containing the coordinates of the vector
with the vector ${\bf x}\equiv \vert {\bf x} \rangle$, but we always need to keep in mind that
the tuples of coordinates are defined only with respect to a particular basis
$\{ {\bf e}_1,{\bf e}_2, \ldots , {\bf e}_n \}$; otherwise these numbers lack any meaning whatsoever.

Indeed, with respect to some arbitrary  basis ${\frak B}=\{
{\bf f}_1, \ldots , {\bf f}_n\}$ of some $n$-dimensional vector space ${\frak V}$
with the base vectors ${\bf f}_i$, $1\le i\le n$, every vector ${\bf x}$ in  ${\frak V}$
can be written as a unique linear combination
\begin{equation}
 \vert {\bf x} \rangle \equiv {\bf x} = \sum_{i=1}^n x_i{\bf f}_i
\equiv
\sum_{i=1}^n x_i \vert {\bf f}_i \rangle
\equiv \begin{pmatrix} x_1\\x_2\\ \vdots \\ x_n \end{pmatrix}
\end{equation}
with respect to the basis  ${\frak B}=\{
{\bf f}_1, \ldots , {\bf f}_n\}$.

{\color{OliveGreen}
\bproof
The uniqueness of the coordinates is proven indirectly by {\em reductio ad absurdum:}
Suppose there is another decomposition
${\bf x} = \sum_{i=1}^n y_i{\bf f}_i = (y_1,y_2, \ldots , y_n) $;
then by subtraction, $0 = \sum_{i=1}^n (x_i-y_i) {\bf f}_i = (0,0, \ldots , 0)$.
Since the basis vectors ${\bf f}_i$ are linearly independent,
this can only be valid if all coefficients in the summation  vanish;
thus $x_i-y_i=0$ for all $1\le i\le n$; hence finally  $x_i=y_i$ for all $1\le i\le n$.
This is in contradiction with our assumption that the coordinates $x_i$ and $y_i$
(or at least some of them) are different.
Hence the only consistent alternative is the assumption that, with respect to a given basis, the coordinates are uniquely determined.
\eproof
}

A  set    ${\frak B} = \{ {\bf a}_1, \ldots , {\bf a}_n\}$
of  vectors   of the inner product space $\frak V$
is {\em orthonormal}
\index{orthonormal}
if, for all
 ${\bf a}_i\in\frak B$ and
 ${\bf a}_j\in\frak B$,
it follows that
\begin{equation}
\langle {\bf a}_i \mid {\bf a}_j \rangle =\delta_{ij}.
\label{2013-m-ch-fdvs-orthonorm}
\end{equation}
Any such set is called {\em complete}
\index{completeness}
if it is not a subset of any larger orthonormal set of vectors of $\frak V$.
Any complete set is a basis.
If, instead of Equation~(\ref{2013-m-ch-fdvs-orthonorm}),
$\langle {\bf a}_i \mid {\bf a}_j \rangle = \alpha_i \delta_{ij}$
with nonzero factors $\alpha_i$, the set is called {\em orthogonal}.

\section{Finding orthogonal bases from nonorthogonal ones}
\label{2019-mm-ch-fdvs-GS}

A {\em Gram-Schmidt process}\cite{Leon-MR3054730}
or {\em Householder orthonormalization}
is a systematic method for orthonormalising a set of vectors
\index{Gram-Schmidt process}
\index{Householder orthonormalization}
\index{scalar product}
\index{inner product}
in a space equipped with a {\em scalar product,}
or by a synonym preferred in mathematics, {\em inner product.}

The Gram-Schmidt process or Householder
orthonormalization\marginnote{The Householder orthonormalization will be dealt with in Section~\ref{2021-m-ch-hposholder} on page~\pageref{2021-m-ch-hposholder}.}
takes a finite, linearly independent set
of base vectors
and generates an orthonormal basis that spans the same (sub)space as the original set.

The general method of the Gram-Schmidt process  is to start with the original basis,
say,  \\
$\{
{\bf x}_1,
{\bf x}_2,
{\bf x}_3,
\ldots ,
{\bf x}_n
\}$,
and generate a new orthogonal basis
by
\begin{equation}
\begin{split}
{\bf y}_1={\bf x}_1,\\
{\bf y}_2={\bf x}_2 - P_{{\bf y}_1}({\bf x}_2),\\
{\bf y}_3={\bf x}_3 - P_{{\bf y}_1}({\bf x}_3)- P_{{\bf y}_2}({\bf x}_3),\\
 \vdots \\
{\bf y}_n={\bf x}_n -\sum_{i=1}^{n-1} P_{{\bf y}_i}({\bf x}_n),
\end{split}
\end{equation}
where
\marginnote{The scalar or inner product
$\langle {\bf x}\vert {\bf y} \rangle$ of two vectors
${\bf x}$ and ${\bf y}$ is defined on page \pageref{2011-m-scalarproduct}.
In Euclidean  space such as ${\Bbb R}^n$,
one often identifies the ``dot product''
${\bf x}\cdot {\bf y} =x_1y_1+ \cdots +x_ny_n$
of two vectors ${\bf x}$ and $ {\bf y}$ with their scalar or inner product.}
$\{
{\bf y}_1,
{\bf y}_2,
{\bf y}_3,
\ldots ,
{\bf y}_n
\}$
\begin{equation}
P_{{\bf y}}({\bf x}) =
\frac{\langle {\bf y} \vert  {\bf x}\rangle }
{\langle {\bf y}\vert {\bf y} \rangle }
{\bf y}
,\textrm{ and }
P_{{\bf y}}^\perp ({\bf x}) = {\bf x} -
\frac{\langle {\bf y} \vert  {\bf x}\rangle }
{\langle {\bf y}\vert {\bf y} \rangle }
{\bf y}
\end{equation}
are the orthogonal projections of ${\bf x}$ onto ${\bf y}$ and ${\bf y}^\perp$, respectively
(the latter is mentioned for the sake of completeness and is not required here).
\label{2011-m-gsp}
Note that these orthogonal projections are idempotent
\index{idempotence}
and mutually orthogonal; that is,
\begin{equation}
\begin{split}
P_{{\bf y}}^2({\bf x})  = P_{{\bf y}}(P_{{\bf y}}({\bf x}) ) =
\frac{\langle {\bf y}\vert {\bf y} \rangle }{\langle {\bf y}\vert {\bf y} \rangle }
\frac{\langle {\bf y} \vert  {\bf x}\rangle }{\langle {\bf y}\vert {\bf y} \rangle }
{\bf y} =P_{{\bf y}}({\bf x}),  \\
(P_{{\bf y}}^\perp)^2({\bf x})  = P_{{\bf y}}^\perp(P_{{\bf y}}^\perp({\bf x}) ) =
{\bf x}- \frac{\langle {\bf y} \vert  {\bf x}\rangle }{\langle {\bf y}\vert {\bf y} \rangle }{\bf y}
-\left(
\frac{\langle {\bf y} \vert  {\bf x}\rangle }{\langle {\bf y}\vert {\bf y} \rangle }
-
\frac{\langle {\bf y}\vert {\bf y} \rangle \langle {\bf y} \vert  {\bf x}\rangle
}{\langle {\bf y}\vert {\bf y} \rangle^2 }
\right)
{\bf y}
=P_{{\bf y}}^\perp({\bf x}),  \\
P_{{\bf y}}(P_{{\bf y}}^\perp({\bf x}) ) =  P_{{\bf y}}^\perp(P_{{\bf y}}({\bf x}) ) =
\frac{\langle {\bf y} \vert  {\bf x}\rangle}{\langle {\bf y}\vert {\bf y} \rangle }{\bf y}
-
\frac{\langle {\bf y}\vert {\bf y} \rangle \langle {\bf y} \vert  {\bf x}\rangle }{\langle {\bf y}\vert {\bf y} \rangle^2 }
{\bf y}
=0.
\end{split}
\end{equation}
For a more general discussion of projections, see also page \pageref{2011-m-projec}.

Subsequently, in order to obtain an orthonormal basis,
one can divide every basis vector by its length.

{\color{OliveGreen}
\bproof
The idea of the proof is as follows (see also Section~7.9 of Ref.\cite{Greub75}).
In order to generate an orthogonal basis from a nonorthogonal one,
the first vector of the old basis is identified with the first vector of the new basis;
that is ${\bf y}_1={\bf x}_1$.
Then, as depicted in Figure~\ref{2012-m-fdvs-ideaofGS}, the second vector of the new basis is obtained by
taking the second vector of the old basis and
subtracting its projection on the first vector of the new basis.
\begin{marginfigure}%
{\color{black}
\begin{center}%
\begin{tikzpicture}[ scale=0.6]

\draw[black,line width=0.5pt,dashed] (0cm,3cm) -- (4cm,3cm);

\draw[black,line width=0.5pt,dashed] (4cm,0cm) -- (4cm,3cm);

\draw[blue,line width=1pt,->] (0cm,0cm) -- (5.5cm,0cm) node[right]{${\bf x}_1= {\color{orange}{\bf y}_1}$};

\draw[blue,line width=1pt,->] (0cm,0cm) -- (4cm,3cm) node[above right]{${\bf x}_2$};

\draw[black,line width=1pt,->] (0cm,0cm) -- (4cm,0cm) node[below]{$P_{{\bf y}_1}({\bf x}_2)$};

\draw[orange,line width=1pt,->] (0cm,0cm) -- (0cm,3cm) node[above]{${\bf y}_2= {\bf x}_2 -P_{{\bf y}_1}({\bf x}_2)$};

\end{tikzpicture}
\end{center}%
\caption{\label{2012-m-fdvs-ideaofGS}Gram-Schmidt construction for two nonorthogonal vectors ${\bf x}_1$ and ${\bf x}_2$,
yielding two  orthogonal vectors ${\bf y}_1$ and ${\bf y}_2$.}
}
\end{marginfigure}
More precisely, take the Ansatz
\begin{equation}
{\bf y}_2={\bf x}_2 + \lambda  {\bf y}_1,
\end{equation}
thereby determining the arbitrary scalar $\lambda$ such that
${\bf y}_1$
and
${\bf y}_2$
are orthogonal; that is,
$\langle{\bf y}_2 \vert  {\bf y}_1\rangle =0$.
This yields
\begin{equation}
\langle {\bf y}_1\vert  {\bf y}_2 \rangle
=\langle {\bf y}_1 \vert {\bf x}_2 \rangle
+ \lambda
\langle {\bf y}_1\vert  {\bf y}_1\rangle =0,
\end{equation}
and thus, since ${\bf y}_1 \neq 0$,
\begin{equation}
\lambda =
-
\frac{\langle {\bf y}_1 \vert {\bf x}_2 \rangle}
{\langle {\bf y}_1\vert {\bf y}_1 \rangle} .
\end{equation}
To obtain the third vector ${\bf y}_3$ of the new basis,
take the Ansatz
\begin{equation}
{\bf y}_3={\bf x}_3 + \mu  {\bf y}_1  + \nu  {\bf y}_2,
\label{2012-m-ch-gs1}
\end{equation}
and require that it is orthogonal to the two previous orthogonal basis vectors
${\bf y}_1$
and
${\bf y}_2$;
that is
$\langle {\bf y}_1\vert {\bf y}_3 \rangle =\langle  {\bf y}_2\vert{\bf y}_3  \rangle =0$.
We already know that $\langle {\bf y}_1\vert {\bf y}_2 \rangle = 0$.
Consider the scalar products of ${\bf y}_1$
and ${\bf y}_2$
with the {\it Ansatz} for ${\bf y}_3$ in Equation~(\ref{2012-m-ch-gs1}); that is,
\begin{equation}
\begin{split}
\langle {\bf y}_1\vert {\bf y}_3\rangle
=
\langle {\bf y}_1 \vert {\bf x}_3 \rangle + \mu  \langle {\bf y}_1\vert {\bf y}_1 \rangle  + \nu   \underbrace{\langle {\bf y}_1\vert {\bf y}_2\rangle}_{=0}
 =0,
\end{split}
\end{equation}
and
\begin{equation}
\begin{split}
\langle{\bf y}_2 \vert {\bf y}_3\rangle =\langle {\bf y}_2\vert  {\bf x}_3\rangle + \mu \underbrace{ \langle {\bf y}_2\vert {\bf y}_1 \rangle}_{=0}   + \nu \langle {\bf y}_2\vert {\bf y}_2\rangle
  =0.
\end{split}
\end{equation}
As a result,
\begin{equation}
\mu = -  \frac{\langle {\bf y}_1 \vert {\bf x}_3 \rangle}
{\langle {\bf y}_1\vert {\bf y}_1 \rangle},\quad
\nu =- \frac{\langle {\bf y}_2\vert  {\bf x}_3\rangle}
{\langle {\bf y}_2 \vert {\bf y}_2 \rangle}.
\end{equation}
A generalization of this construction
for all the other new base vectors
${\bf y}_3, \ldots ,  {\bf y}_n$, and thus a proof by complete induction,
proceeds by a generalized construction.
\eproof
}

{\color{blue}
\bexample
Consider, as an example, the standard Euclidean scalar product denoted by ``$\cdot$''
and the basis
$\left\{\begin{pmatrix}0\\1\end{pmatrix},\begin{pmatrix}1\\1\end{pmatrix}\right\}$.
Then two orthogonal bases are obtained by taking
\begin{itemize}
\item[(i)]
either the basis vector
$\begin{pmatrix}0\\1\end{pmatrix}$, together with
$
\begin{pmatrix}1\\1\end{pmatrix} -
\frac{\begin{pmatrix}1\\1\end{pmatrix}\cdot \begin{pmatrix}0\\1\end{pmatrix}}{\begin{pmatrix}0\\1\end{pmatrix}\cdot \begin{pmatrix}0\\1\end{pmatrix}} \begin{pmatrix}0\\1\end{pmatrix} = \begin{pmatrix}1\\0\end{pmatrix},
$
\item[(ii)]
or the basis vector
$\begin{pmatrix}1\\1\end{pmatrix}$, together with
$
\begin{pmatrix}0\\1\end{pmatrix} -
\frac{\begin{pmatrix}0\\1\end{pmatrix}\cdot \begin{pmatrix}1\\1\end{pmatrix}}{\begin{pmatrix}1\\1\end{pmatrix}\cdot \begin{pmatrix}1\\1\end{pmatrix}} \begin{pmatrix}1\\1\end{pmatrix} = \frac{1}{2}\begin{pmatrix}-1\\1\end{pmatrix}. \textrm{\eexample}
$
\end{itemize}
}

\section{Dual space}
\label{2011-m-dvs}
\marginnote{For proofs and additional information see {\S}13--15 in~\bibentry{halmos-vs}.}

Every vector space ${\frak V}$
has a corresponding {\em dual vector space}
\index{dual vector space}
\index{dual space}
(or just {\em dual space})   ${\frak V}^\ast$
consisting of all linear functionals on ${\frak V}$.

A {\em linear functional}
\index{linear functional}
on a vector space ${\frak V}$ is a scalar-valued linear function ${\bf y}$
defined for every vector   ${\bf x} \in {\frak V}$, with the linear property that\marginnote{
Although
the linear functional ${\bf y}$ is written in vector notation,
elements of its codomain or set of destination or outputs are scalars
(note also that elements of its  domain or set of departure or inputs are vectors of a vector space).
The vector notation has been chosen because every such linear functional ${\bf y}$
can be represented as a vector in a linear vector space
spanned by the dual basis
defined in Equation~(\ref{2011-m-Dualbasis-e1}).
An example (with scalar domain and codomain) are polynomials $x^l$ or Legendre polynomials $P_l$ (cf. Section~\ref{2013-m-sf-lp})
\index{Legendre polynomial}
with $i\in \mathbb{N}_0$, spanning an infinite dimensional vector space.
}
\begin{equation}
{\bf y} (\alpha_1 {\bf x}_1 +\alpha_2 {\bf x}_2)
=
\alpha_1 {\bf y} ({\bf x}_1) +\alpha_2 {\bf y} ({\bf x}_2) .
\end{equation}

{\color{blue}
\bexample
For example,
let ${\bf x} = (x_1,\ldots , x_n)$, and
take
${\bf y} ({\bf x}) =x_1$.

For another example,
let again ${\bf x} = (x_1,\ldots , x_n)$, and
let $\alpha_1,\ldots , \alpha_n \in {\Bbb C}$ be scalars; and
take
${\bf y} ({\bf x}) =\alpha_1 x_1 + \cdots +\alpha_n x_n$.

The following supermarket example has been
communicated to me by Hans Havlicek:\cite{havlicek-priv3}
suppose you visit a supermarket, with a variety of products therein.
Suppose further that you select some items and collect them in a cart or trolley.
Suppose further that, in order to complete your purchase, you finally go to the cash desk,
where the sum total of your purchase is computed from the price-per-product information stored
in the memory of the cash register.

In this example, the vector space can be identified with all conceivable configurations of products in a cart or trolley.
Its dimension is determined by the number of different, mutually distinct products in the supermarket.
Its ``base vectors'' can be identified with the mutually distinct products in the supermarket.
The respective functional is the computation of the price of any such purchase.
It is based on a particular price information.
Every such price information contains one price per item for all mutually distinct products.
The dual space consists of all conceivable price details.
The number of respective basis vectors---encoded as price-per-product---needs to be the same as the number of products.
Therefore, the dimensions of both all product configurations ``spanning'' the vector space,
as well as of all possible prices rendered, needs to be the same.
\eexample
}

We adopt a doublesquare bracket notation ``$\llbracket \cdot , \cdot \rrbracket$''
for the functional

\begin{equation}
{\bf y} ({\bf x})
=
\llbracket {\bf x},{\bf y}\rrbracket .
\end{equation}

The set of linear functionals is closed with respect to
the addition of two or more of such functionals, as well as multiplication of scalars with a functional;
that is,
\begin{equation}
(a {\bf y} + b {\bf z}) ({\bf x})
=
  a {\bf y} ({\bf x}) + b {\bf z} ({\bf x})
.
\end{equation}
Together with the ``zero functional''
(mapping every argument to zero), as well as other algebrac properties,
this induces a kind of linear vector space structure, where the ``vectors''
are identified with the linear functionals.
This vector space will be called {\em dual space} ${\frak V}^\ast $.
\index{dual space}

As a result, this ``bracket'' functional is
{\em bilinear} in its two arguments; that is,
\begin{equation}
\llbracket  \alpha_1 {\bf x}_1 +\alpha_2 {\bf x}_2, {\bf y}\rrbracket
=
\alpha_1 \llbracket {\bf x}_1 ,{\bf y}\rrbracket   +\alpha_2  \llbracket {\bf x}_2,{\bf y}\rrbracket ,
\end{equation}
and
\begin{equation}
\llbracket
{\bf x}, \alpha_1 {\bf y}_1 +\alpha_2 {\bf y}_2
\rrbracket
=
\alpha_1
\llbracket {\bf x},{\bf y}_1 \rrbracket
+
\alpha_2
\llbracket {\bf x},{\bf y}_2\rrbracket .
\end{equation}
\marginnote{The square bracket can be identified with the scalar dot product
$\llbracket  {\bf x},{\bf y} \rrbracket  = \langle {\bf x}\mid {\bf y}\rangle$
only for Euclidean
vector spaces ${\Bbb R}^n$, since for complex spaces this would no longer be positive definite.
That is, for Euclidean
vector spaces ${\Bbb R}^n$ the inner or scalar product is bilinear.
}

Because of linearity, we can completely characterize an arbitrary linear functional
${\bf y} \in {\frak V}^\ast $ by its values of the vectors of some basis of ${\frak V}$:
If we know the functional value on the basis vectors in ${\frak B}$, we know the functional
on all elements of the vector space ${\frak V}$.
If ${\frak V}$ is an $n$-dimensional vector space, and if ${\frak B} = \{{\bf f}_1,\ldots , {\bf f}_n\}$
is a basis of  ${\frak V}$, and if
$\{\alpha_1, \ldots ,\alpha_n\}$  is any set of $n$ scalars, then there is
a unique linear functional ${\bf y}$  on  ${\frak V}$ such that
$ \llbracket  {\bf f}_i, {\bf y}\rrbracket  = \alpha_i $ for all $0\le i \le n$.

{\color{OliveGreen}
\bproof
A constructive proof  of this theorem can be given as follows:
Because every ${\bf x}\in {\frak V}$
can be written as a linear combination $ {\bf x} = x_1 {\bf f}_1 +\cdots + x_n {\bf f}_n$
of the basis vectors of ${\frak B} = \{{\bf f}_1,\ldots , {\bf f}_n\}$
in one and only one (unique) way, we obtain for any arbitrary linear functional ${\bf y} \in {\frak V}^\ast $  a unique decomposition
in terms of the basis vectors of  ${\frak B} = \{{\bf f}_1,\ldots , {\bf f}_n\}$; that is,
\begin{equation}
\llbracket {\bf x},{\bf y}\rrbracket
=
x_1 \llbracket {\bf f}_1 ,{\bf y}\rrbracket  +\cdots + x_n \llbracket {\bf f}_n ,{\bf y}\rrbracket  .
\end{equation}
By identifying  $\llbracket {\bf f}_i ,{\bf y}\rrbracket =\alpha_i$ we obtain
\begin{equation}
\llbracket {\bf x},{\bf y}\rrbracket
=
x_1 \alpha_1 +\cdots + x_n \alpha_n .
\end{equation}
\eproof
}

Conversely, if we {\em define} ${\bf y}$ by $\llbracket {\bf x},{\bf y}\rrbracket  =\alpha_1  x_1+ \cdots +\alpha_n x_n$, then ${\bf y}$
can be interpreted as a linear functional in ${\frak V}^\ast$ with $\llbracket {\bf f}_i,{\bf y}\rrbracket  = \alpha_i$.

If we introduce a {\em dual basis}
by requiring that $\llbracket {\bf f}_i,  {\bf f}_j^\ast \rrbracket =\delta_{ij}$ [cf. Equation~(\ref{2011-m-Dualbasis-e1})],
then the coefficients $\llbracket {\bf f}_i ,{\bf y}\rrbracket  = \alpha_i$,
$1\le i \le n$, can be interpreted
as the {\em coordinates} of the linear functional ${\bf y}$ with respect to the dual
basis ${\frak B}^\ast$, such that or,
relative to the dual basis defined in the next Section~\ref{2011-m-Dualbasis},
${\bf y}=\sum_i \alpha_i {\bf f}_i^\ast$, or, with respect to the dual basis,
${\bf y}=(\alpha_1,\alpha_2,\ldots , \alpha_n)$.

Likewise, as will be shown in (\ref{2014-m-ch-fdlvs-kju}),
$
x_i =
 \llbracket {\bf x},{\bf f}_i^\ast \rrbracket
$; that is, the vector coordinates can be represented by the functionals of the elements of the dual basis.

The number of such basis vectors---and thus the dimension of the dual space---needs to be the same
as the dimension of the original vector space: because of linearity it is necessary and sufficient to know all functional values on the
vectors of the basis of the original space.

{\color{blue}
\bexample
Let us explicitly construct an example of a linear functional $\varphi ({\bf x})\equiv \llbracket {\bf x},\varphi\rrbracket $ that is defined
on all vectors ${\bf x}=
\alpha {\bf e}_1
+
\beta {\bf e}_2
$
of a two-dimensional vector space with the basis $\{{\bf e}_1, {\bf e}_2 \}$
by enumerating its ``performance on the basis vectors''  ${\bf e}_1=\begin{pmatrix}1,0\end{pmatrix}^\intercal$
and ${\bf e}_2=\begin{pmatrix}0,1\end{pmatrix}^\intercal$;
more explicitly,  say, for an example's sake,
$\varphi ({\bf e}_1 ) \equiv \llbracket  {\bf e}_1,\varphi \rrbracket  = 2$ and
$\varphi ({\bf e}_2 ) \equiv \llbracket  {\bf e}_2,\varphi \rrbracket  = 3$.
Therefore, for example for the vector $\begin{pmatrix}5,7\end{pmatrix}^\intercal$,
$\varphi \left(\begin{pmatrix}5,7\end{pmatrix}^\intercal \right)
\equiv \left\llbracket \begin{pmatrix}5,7\end{pmatrix}^\intercal,
\varphi \right\rrbracket  = 5 \llbracket {\bf e}_1, \varphi \rrbracket  + 7 \llbracket  {\bf e}_2,\varphi \rrbracket
=10+21=31$.

In general the performance of the linear function on just one vector renders insufficient
information to uniquely define a linear functional of vectors of dimension two or higher:
one needs as many values on mutually
linear independent vectors as there are dimensions for a complete specification of the linear functional.
Take, for example, just one value of $\varphi$ on a single vector, say
${\bf x}=\begin{pmatrix} 5,7 \end{pmatrix}^\intercal$; that is,
$\varphi \left( {\bf x} \right)=31$. If one does not
know the linear functional beforehand,
all one can do is to write $\varphi$ in terms of its components (with respect to the dual basis)
$\varphi = \begin{pmatrix} \varphi_1 ,\varphi_2 \end{pmatrix}$
and evaluate
$\begin{pmatrix} \varphi_1 ,\varphi_2 \end{pmatrix} \cdot \begin{pmatrix} 5,7 \end{pmatrix}^\intercal
= 5\varphi_1 +7\varphi_2 = 31$, which just yields one component of $\varphi$ in terms of the other;
that is, $\varphi_1  = (31-7\varphi_2)/5$.
The components of $\varphi$ (with respect to the dual basis) are uniquely fixed
only by presentation of another value, say
$\varphi \left( {\bf y} \right)=13$,
on another vector
${\bf y}=\begin{pmatrix} 2,3 \end{pmatrix}^\intercal$ not collinear to the first vector ${\bf x}$.
Then
$\begin{pmatrix} \varphi_1 ,\varphi_2 \end{pmatrix} \cdot \begin{pmatrix} 2,3 \end{pmatrix}^\intercal
= 2\varphi_1 +3\varphi_2 = 13$  yields $\varphi_1  = (13-3\varphi_2)/2$.
Equating those two equations for $\varphi_1$ yields
$(31-7\varphi_2)/5=(13-3\varphi_2)/2$ and thus $\varphi_2=3$ and therefore $\varphi_1=2$.

\eexample
}

\subsection{Dual basis}
\label{2011-m-Dualbasis}

We now can define a {\em dual basis}, or, used synonymously, a {\em reciprocal} or {\em contravariant} basis.
\index{dual basis}
\index{reciprocal basis}
\index{contravariant basis}
If ${\frak V}$ is an $n$-dimensional vector space, and if
${\frak B} = \{{\bf f}_1,\ldots , {\bf f}_n\}$
is a basis of  ${\frak V}$,
then there is a unique {\em dual basis}
${\frak B}^\ast
=\{{\bf f}_1^\ast ,\ldots , {\bf f}_n^\ast \}$ in the dual vector space ${\frak V}^\ast $
defined by
\begin{equation}
{\bf f}_j^\ast ({\bf f}_i) =  \llbracket {\bf f}_i,  {\bf f}_j^\ast \rrbracket =\delta_{ij},
\label{2011-m-Dualbasis-e1}
\end{equation}
where  $\delta_{ij}$
is the Kronecker delta function.
The dual space  ${\frak V}^\ast $ spanned by the dual basis ${\frak B}^\ast $ is $n$-dimensional.

In a different notation involving subscripts (lower indices) for (basis) vectors of the base vector space,
and superscripts (upper indices) ${\bf f}^j = {\bf f}_j^\ast $,
for (basis) vectors of the dual vector space,
Equation~(\ref{2011-m-Dualbasis-e1}) can be written as
\begin{equation}
{\bf f}^j ( {\bf f}_i ) = \llbracket {\bf f}_i,{\bf f}^j\rrbracket =\delta_{ij}.
\label{2011-m-Dualbasis-e2}
\end{equation}
Suppose
$g$ is a {\em metric},
\index{metric}
facilitating the translation from vectors of the base vectors into vectors of the dual space and {\it vice versa}
(cf. Section~\ref{2011-m-metrict} on page~\pageref{2011-m-metrict} for a definition and more details),
in particular, ${\bf f}_i =  g_{il}{\bf f}^l$
as well as  ${\bf f}_j^\ast  = {\bf f}^j = g^{jk}{\bf f}_k$.
Then Eqs.~(\ref{2011-m-Dualbasis-e1}) and (\ref{2011-m-Dualbasis-e2}) can be rewritten as
\begin{equation}
\llbracket g_{il} {\bf f}^l, {\bf f}^j\rrbracket     = \llbracket {\bf f}_i, g^{jk} {\bf f}_k\rrbracket   = \delta_{ij}.
\label{2011-m-Dualbasis-e3}
\end{equation}

Note that the vectors ${\bf f}^\ast_i = {\bf f}^i$ of the dual basis can be used to ``retrieve''
or ``extract'' the components of arbitrary vectors
${\bf x} = \sum_j x_j {\bf f}_j$  through
\begin{equation}
{\bf f}^\ast_i ( {\bf x} ) =
{\bf f}^\ast_i \left( \sum_j x_j {\bf f}_j \right) =
\sum_j  x_j {\bf f}^\ast_i \left({\bf f}_j \right) =
\sum_j  x_j \delta_{ij} =
x_i.
\label{2016-m-fdlvs-recoverc}
\end{equation}
Likewise, the basis vectors ${\bf f}_i$ can be used to extract the respective coordinates of any dual vector.

In terms of the inner products of the base vector space and its dual vector space the representation
of the metric
may be defined by
$g_{ij}=g({\bf f}_i,{\bf f}_j)=\langle {\bf f}_i\mid {\bf f}_j\rangle$,
as well as
$g^{ij}=g({\bf f}^i,{\bf f}^j) =\langle {\bf f}^i\mid {\bf f}^j\rangle$, respectively.
Note, however, that the coordinates $g_{ij}$ of
the metric $g$ need not necessarily be positive definite.
For example,  special relativity uses the ``pseudo-Euclidean'' metric
 $g={\rm diag}(+1,+1,+1,-1)$ (or just $g={\rm diag}(+,+,+,-)$), where ``${\rm diag}$''
stands for the {\em diagonal matrix}
\index{diagonal matrix}
with the arguments in the diagonal.
\marginnote{The metric tensor $g_{ij}$ represents a {\em bilinear functional}
$g({\bf x},{\bf y}) =x^iy^j g_{ij}$ that is {\em symmetric}; that is,
$g({\bf x},{\bf y}) = g({\bf y},{\bf x})$
and {\em nondegenerate}; that is, for any nonzero vector ${\bf x}\in {\frak V}$,   ${\bf x}\neq 0$,
there is some  vector  ${\bf y}\in {\frak V}$, so that  $g({\bf x},{\bf y}) \neq 0$.
$g$ also satisfies the triangle
inequality
$\vert\vert {\bf x} -{\bf z} \vert\vert  \le \vert\vert {\bf x} - {\bf y} \vert\vert  + \vert\vert  {\bf y} - {\bf z} \vert\vert $.
}

In a real Euclidean vector space ${\Bbb R}^n$
with the dot product as the scalar product,
the dual basis of an orthogonal basis  is also orthogonal, and contains vectors with the same directions,
although with {\em reciprocal length} (thereby explaining the wording ``reciprocal basis'').
Moreover, for an orthonormal basis, the basis vectors are uniquely identifiable by
${\bf e}_i \longrightarrow {\bf e}_i^* = {\bf e}_i^\intercal $.
This identification can only be made for orthonormal bases; it is {\em not} true for nonorthonormal bases.


A ``reverse construction'' of the elements ${\bf f}_j^\ast $ of the dual basis ${\frak B}^\ast $
-- thereby using the definition ``$\llbracket {\bf f}_i,{\bf y}\rrbracket  = \alpha_i$
for all $1 \le i \le n$''
for any element ${\bf y}$ in ${\frak V}^\ast $ introduced earlier
--
can be given as follows:
for every $1\le j \le n$,
we can {\em define} a vector ${\bf f}_j^\ast $ in the dual basis ${\frak B}^\ast $
by the {\em requirement}   $\llbracket {\bf f}_i,{\bf f}_j^\ast \rrbracket  = \delta_{ij}$.
That is, in words:
the dual basis element, when applied to the elements of the original $n$-dimensional basis,
yields one if and only if  it corresponds to the respective equally indexed basis element;
for all the other $n-1$ basis elements it yields zero.

What remains to be proven is the conjecture that
${\frak B}^\ast  = \{{\bf f}_1^\ast ,\ldots , {\bf f}_n^\ast \}$
is a basis of ${\frak V}^\ast $; that is, that the vectors in ${\frak B}^\ast $ are linear independent,
and that they span ${\frak V}^\ast $.

First observe that ${\frak B}^\ast $ is a set of linear independent vectors,
for if
$ \alpha_1 {\bf f}_1^\ast  + \cdots + \alpha_n {\bf f}_n^\ast =0$, then also
\begin{equation}
 \llbracket {\bf x},\alpha_1 {\bf f}_1^\ast  + \cdots + \alpha_n {\bf f}_n^\ast \rrbracket =
 \alpha_1 \llbracket {\bf x},{\bf f}_1^\ast \rrbracket  + \cdots + \alpha_n \llbracket {\bf x},{\bf f}_n^\ast \rrbracket =
0
\end{equation}
for arbitrary ${\bf x}\in {\frak V}$.
In particular, by identifying ${\bf x}$ with ${\bf f}_i \in {\frak B}$, for $1 \le i \le n$,
\begin{equation}
 \alpha_1 \llbracket {\bf f}_i,{\bf f}_1^\ast \rrbracket  + \cdots + \alpha_n \llbracket {\bf f}_i,{\bf f}_n^\ast \rrbracket = \alpha_j \llbracket {\bf f}_i,{\bf f}_j^\ast \rrbracket  = \alpha_j \delta_{ij} = \alpha_i=
0.
\end{equation}

Second,  every ${\bf y} \in {\frak V}^\ast $ is a linear combination of elements in
${\frak B}^\ast  = \{{\bf f}_1^\ast ,\ldots , {\bf f}_n^\ast \}$, because by
starting from
$\llbracket {\bf f}_i,  {\bf y}\rrbracket =\alpha_i$,
with
$ {\bf x} = x_1 {\bf f}_1 +\cdots + x_n {\bf f}_n$
we obtain
\begin{equation}
 \llbracket {\bf x},{\bf y}\rrbracket
= x_1 \llbracket {\bf f}_1,{\bf y}\rrbracket  +\cdots + x_n \llbracket {\bf f}_n,{\bf y}\rrbracket
= x_1 \alpha_1 +\cdots + x_n \alpha_n.
\label{2014-m-fdvs-ba}
\end{equation}
Note that , for arbitrary  ${\bf x}\in {\frak V}$,
\begin{equation}
 \llbracket {\bf x},{\bf f}_i^\ast \rrbracket
= x_1 \llbracket {\bf f}_1,{\bf f}_i^\ast \rrbracket  +\cdots + x_n \llbracket {\bf f}_n,{\bf f}_i^\ast \rrbracket =  x_j \llbracket {\bf f}_j,{\bf f}_i^\ast \rrbracket =  x_j \delta_{ji}=  x_i,
\label{2014-m-ch-fdlvs-kju}
\end{equation}
and by substituting $\llbracket {\bf x},{\bf f}_i\rrbracket $ for $x_i$ in Equation~(\ref{2014-m-fdvs-ba}) we obtain
\begin{equation}
\begin{split}
 \llbracket {\bf x},{\bf y}\rrbracket  =
 x_1\alpha_1 +\cdots + x_n \alpha_n \\
= \llbracket {\bf x},{\bf f}_1\rrbracket  \alpha_1 +\cdots + \llbracket {\bf x},{\bf f}_n\rrbracket  \alpha_n  \\
= \llbracket {\bf x},\alpha_1 {\bf f}_1 +\cdots + \alpha_n{\bf f}_n\rrbracket  ,
\label{2014-m-fdvs-ba1}
\end{split}
\end{equation}
and therefore ${\bf y} =\alpha_1 {\bf f}_1 +\cdots + \alpha_n{\bf f}_n=\alpha_i{\bf f}_i$.


How can one determine the dual basis from a given,
not necessarily orthogonal, basis?
For the rest of this section, suppose that the metric is identical to the Euclidean metric ${\rm diag}(+,+,\cdots ,+)$ representable as the usual ``dot product.''
The tuples of {\em column vectors} of the basis ${\frak B} = \{{\bf f}_1,\ldots , {\bf f}_n\}$
can be arranged into a $n \times n$ matrix
\begin{equation}
\textsf{\textbf{B}}
\equiv
\begin{pmatrix} \vert {\bf f}_{1} \rangle , \vert {\bf f}_{2}\rangle , \cdots ,\vert  {\bf f}_{n}\rangle \end{pmatrix}
\equiv
\begin{pmatrix} {\bf f}_{1} , {\bf f}_{2}, \cdots , {\bf f}_{n} \end{pmatrix}
=
\begin{pmatrix}
{\bf f}_{1,1}&\cdots & {\bf f}_{n,1}\\
{\bf f}_{1,2}&\cdots & {\bf f}_{n,2}\\
\vdots&\vdots & \vdots \\
{\bf f}_{1,n}&\cdots & {\bf f}_{n,n}
\end{pmatrix}
.
\end{equation}
Then take the
{\em inverse matrix}
$\textsf{\textbf{B}}^{-1}$,
and interpret the
{\em row vectors} ${\bf f}_{i}^\ast$ of
\begin{equation}
\begin{split}
\textsf{\textbf{B}}^{\ast}=\textsf{\textbf{B}}^{-1}
\equiv
\begin{pmatrix}
\langle {\bf f}_{1} \vert \\
\langle {\bf f}_{2} \vert \\
                      \vdots  \\
\langle {\bf f}_{n} \vert
\end{pmatrix}
\equiv
\begin{pmatrix}
{\bf f}_{1}^\ast\\
{\bf f}_{2}^\ast\\
\vdots  \\
{\bf f}_{n}^\ast
\end{pmatrix}
=
\begin{pmatrix}
{\bf f}_{1,1}^\ast&\cdots & {\bf f}_{1,n}^\ast\\
{\bf f}_{2,1}^\ast&\cdots & {\bf f}_{2,n}^\ast\\
\vdots&\vdots & \vdots \\
{\bf f}_{n,1}^\ast&\cdots & {\bf f}_{n,n}^\ast
\end{pmatrix}
\end{split}
\end{equation}
as the tuples of elements of the dual basis of ${\frak B}^\ast $.

For orthogonal but not orthonormal bases, the term {\em reciprocal} basis
can be easily explained by the fact that the norm (or length) of each vector in the {\em reciprocal basis}
is just the {\em inverse} of the length of the original vector.

{\color{OliveGreen}
\bproof
For a direct proof  consider $\textsf{\textbf{B}}\cdot \textsf{\textbf{B}}^{-1} =\mathbb{1}_n$.
\eproof
}

{\color{blue}
\bexample
\begin{itemize}
\item[(i)]
For example,
if
\begin{equation}
{\frak B}
\equiv
\{ \vert {\bf e}_1 \rangle ,\vert  {\bf e}_2\rangle ,\ldots ,\vert {\bf e}_n\rangle  \}
\equiv
\{{\bf e}_1, {\bf e}_2,\ldots ,{\bf e}_n\}
\equiv
\left\{
\begin{pmatrix}
1\\
0\\
\vdots \\
0
\end{pmatrix}
,
\begin{pmatrix}
0\\
1\\
\vdots \\
0
\end{pmatrix}
,
\ldots ,
\begin{pmatrix}
0\\
0\\
\vdots \\
1
\end{pmatrix}
\right\}
\end{equation}
is the standard basis in $n$-dimensional vector space containing unit vectors of norm (or length) one,
then
\begin{equation}
\begin{split}
{\frak B}^\ast
\equiv
\{ \langle {\bf e}_1 \vert , \langle {\bf e}_2 \vert , \ldots ,\langle {\bf e}_n \vert \}
\\
\equiv
\{{\bf e}_1^\ast  ,{\bf e}_2^\ast , \ldots ,{\bf e}_n^\ast  \}
\equiv
\left\{
(1,0,\ldots,0),
(0,1,\ldots,0),
\ldots,
(0,0,\ldots,1)
\right\}
\end{split}
\end{equation}
has elements with identical components,
but those tuples are the transposed ones.

\item[(ii)]
If
\begin{equation}
\begin{split}
{\frak X}
\equiv
\{ \alpha_1 \vert {\bf e}_1 \rangle , \alpha_2 \vert  {\bf e}_2 \rangle ,\ldots ,\alpha_n \vert {\bf e}_n \rangle  \}
\equiv
\{\alpha_1 {\bf e}_1, \alpha_2 {\bf e}_2,\ldots ,\alpha_n {\bf e}_n\}
\\
\equiv
\left\{
\begin{pmatrix}
\alpha_1\\
0\\
\vdots \\
0
\end{pmatrix}
,
\begin{pmatrix}
0\\
\alpha_2\\
\vdots \\
0
\end{pmatrix}
,
\ldots ,
\begin{pmatrix}
0\\
0\\
\vdots \\
\alpha_n
\end{pmatrix}
\right\}
,
\end{split}
\end{equation}
with nonzero $\alpha_1,\alpha_2,\ldots ,\alpha_n \in {\Bbb R}$,
is a ``dilated'' basis in $n$-dimensional vector space containing vectors of norm (or length) $\alpha_i$,
then
\begin{equation}
\begin{split}
{\frak X}^\ast
\equiv
\left\{  \frac{1}{\alpha_1 }\langle {\bf e}_1 \vert ,\frac{1}{\alpha_2 } \langle  {\bf e}_2 \vert , \ldots ,\frac{1}{\alpha_n } \langle {\bf e}_n \vert \right\}
\\
\equiv
\left \{\frac{1}{\alpha_1 }{\bf e}_1^\ast  ,\frac{1}{\alpha_2 }{\bf e}_2^\ast , \ldots ,\frac{1}{\alpha_n }{\bf e}_n^\ast  \right\}
\\
\equiv
\left\{
\begin{pmatrix}\frac{1}{\alpha_1 },0,\ldots,0\end{pmatrix},
\begin{pmatrix}0,\frac{1}{\alpha_2 },\ldots,0\end{pmatrix},
\ldots,
\begin{pmatrix}0,0,\ldots,\frac{1}{\alpha_n }\end{pmatrix}\right\}
\end{split}
\end{equation}
has elements with identical components of inverse length $\frac{1}{\alpha_i }$,
and again those tuples are the transposed tuples.

\item[(iii)]
Consider the nonorthogonal basis
${\frak B} =
\left\{\begin{pmatrix}1\\3\end{pmatrix}, \begin{pmatrix}2\\ 4\end{pmatrix}\right\}$.
The associated column matrix is
\begin{equation}
\textsf{\textbf{B}}
=
\begin{pmatrix}
1&2\\
3&4
\end{pmatrix}
.
\end{equation}
The inverse matrix is
\begin{equation}
\textsf{\textbf{B}}^{-1}
=
\begin{pmatrix}
-2&1\\
\frac{3}{2}&-\frac{1}{2}
\end{pmatrix}
,
\end{equation}
and the associated dual basis is obtained from the rows of $\textsf{\textbf{B}}^{-1} $ by
\begin{equation}
{\frak B}^\ast  =
\left\{
\begin{pmatrix}
-2, 1
\end{pmatrix},
\begin{pmatrix}
\frac{3}{2},
-\frac{1}{2}
\end{pmatrix}
\right\} = \frac{1}{2}
\left\{
\begin{pmatrix}
-4,
2
\end{pmatrix},
\begin{pmatrix}
3,
-1
\end{pmatrix}
\right\}
.
\label{2011-m-cenobdb}
\end{equation}
\eexample
\end{itemize}
}

\subsection{Dual coordinates}

With respect to a given basis,
the components of a vector are often written as tuples of ordered
(``$x_i$ is written before $x_{i+1}$'' -- not ``$x_i < x_{i+1}$'')
scalars  as {\em column vectors}
\begin{equation}
\vert {\bf x} \rangle
\equiv
{\bf x}
\equiv
\begin{pmatrix}
x_1,x_2,
\cdots ,
x_n
\end{pmatrix}^\intercal
,
\end{equation}
whereas the components of vectors in dual spaces are often written in terms of
tuples of ordered
scalars  as {\em row vectors}
\begin{equation}
\langle {\bf x} \vert
\equiv
{\bf x}^\ast \equiv
\begin{pmatrix}
x_1^\ast ,x_2^\ast ,\ldots , x_n^\ast
\end{pmatrix}
.
\end{equation}
The coordinates  of  vectors $\vert {\bf x} \rangle
\equiv
{\bf x}
$
of the base vector space ${\frak V}$
-- and by definition (or rather, declaration) the vectors $\vert {\bf x} \rangle
\equiv
{\bf x}
$ themselves --
are called
{\em contravariant}:
\index{contravariance}
\index{contravariant vector}
because in order to compensate for scale changes of the reference axes (the basis vectors)
$\vert {\bf e}_1 \rangle ,\vert  {\bf e}_2\rangle ,\ldots ,\vert {\bf e}_n\rangle
\equiv
{\bf e}_1, {\bf e}_2,\ldots ,{\bf e}_n$
these coordinates have to contra-vary (inversely vary)  with respect to any such change.

In contradistinction the coordinates of dual vectors,
that is, vectors of the dual vector space ${\frak V}^\ast$,
$\langle {\bf x} \vert
\equiv
{\bf x}^\ast $ -- and by definition (or rather, declaration) the vectors $\langle {\bf x} \vert
\equiv
{\bf x}^\ast $ themselves --
are called
{\em covariant}.
\index{covariant coordinates}
\index{covariant vectors}

Alternatively covariant coordinates could be denoted by subscripts (lower indices),
and contravariant coordinates can be denoted by superscripts (upper indices); that is
(see also
Havlicek\cite{havlicek-laftm}, Section 11.4),
\begin{equation}
\begin{split}
{\bf x}  \equiv
\vert {\bf x}  \rangle \equiv
\begin{pmatrix}
x^1,
x^2,
\cdots ,
x^n
\end{pmatrix}^\intercal
\textrm{, and }
\\
 {\bf x}^\ast \equiv
\langle {\bf x}  \vert \equiv
(x_1^\ast ,x_2^\ast ,\ldots , x_n^\ast  ) \equiv   (x_1,x_2,\ldots , x_n )
.
\end{split}
\end{equation}
This notation will be used in the chapter~\ref{ch:t} on tensors.
Note again that the covariant and contravariant components
$x_k$ and $x^k$ are not absolute, but always defined {\em with respect to}
a particular (dual) basis.

Note that, for orthormal bases it is possible to interchange contravariant and covariant coordinates by
taking the conjugate transpose; that is,
\index{conjugate transpose}
\index{Hermitian conjugate}
\index{Hermitian adjoint}
\begin{equation}
\left(\langle {\bf x} \vert \right)^\dagger = \vert {\bf x}  \rangle
\textrm {, and }
\left(\vert {\bf x}  \rangle \right)^\dagger = \langle {\bf x} \vert
.
\label{2015-m-ch-fdlvs-inter}
\end{equation}

Note also that the {\em Einstein summation convention}
\index{Einstein summation convention}
requires that, when an index variable appears twice in a single term, one has to
sum over all of the possible index values.
This saves us from drawing the sum sign ``$\sum_i$'' for the index $i$;
for instance $x_iy_i =\sum_{i}x_iy_i$.

In the particular context of covariant and contravariant components
--
made necessary by nonorthogonal bases whose associated dual bases are {\em not} identical
--
the summation always is between some superscript (upper index) and some subscript (lower index);
e.g., $x_iy^i$.

Note again that for orthonormal basis,
$x^i=x_i$.

\subsection{Representation of a functional by inner product}
\label{2011-m-corr-bil-ip}
\marginnote{For proofs and additional information see {\S}67 in~\bibentry{halmos-vs}.}
The following representation theorem,
often called
{\em Riesz representation theorem}
\index{Riesz representation theorem}
\index{Fr\'echet-Riesz representation theorem}
(sometimes also called the {\em Fr\'echet-Riesz theorem}),
is about the connection between any functional
in a vector space and its inner product:
To any linear functional ${\bf z}$
on a finite-dimensional inner product space ${\frak V}$
there corresponds a unique vector   ${\bf y}\in {\frak V}$,
such that
\begin{equation}
{\bf z} ({\bf x}) \equiv \llbracket {\bf x}, {\bf z}\rrbracket = \langle {\bf y} \mid {\bf x} \rangle
\label{2015-m-ch-fdlvs-e1}
\end{equation}
for all ${\bf x}\in {\frak V}$.\marginnote{See Theorem~4.12  in~\bibentry{Rudin-RaCA}.}

{\color{OliveGreen}
\bproof
One way of constructing the vector ${\bf y}\in {\frak V}$ is by noticing that,
by assumption, the linear functional  ${\bf z} \in {\frak V}^\ast$ is linear.
Thus it suffices to know its values $a_i={\bf z} ({\bf e}_i)$, $i=1,\ldots , n$
on all vectors of some orthonormal basis
${\frak B}= \{ {\bf e}_1 ,   {\bf e}_2   , \ldots ,  {\bf e}_n \}$.
With respect to that basis, because of antilinearity of the scalar product in the complex case,
the components of the vector ${\bf y}$
associated with the linear functional ${\bf z}$ will be $\overline{a}_i$.
That is,
${\bf y} = \overline{a}_i {\bf e}_i \equiv \begin{pmatrix} \overline{a}_1,\overline{a}_2,\ldots , \overline{a}_n\end{pmatrix}^\intercal $.
In that way an arbitrary vector ${\bf x} = x_j {\bf e}_j \in {\frak V}$ is mapped by the
scalar product as
$\langle {\bf y} \mid {\bf x} \rangle = \langle \overline{a}_i {\bf e}_i  \mid x_j {\bf e}_j  \rangle
=  a_i x_j \underbrace{\langle  {\bf e}_i  \mid {\bf e}_j  \rangle}_{=\delta_{ij}}
=a_i x_i = x_i a_i =  x_i {\bf z} ({\bf e}_i) = {\bf z} (x_i{\bf e}_i)={\bf z} ({\bf x})$.

Another constructive proof
provides a method to compute the vector ${\bf y} \in {\frak V}$
given the linear functional
${\bf z} \in {\frak V}^\ast$.
The proof idea is to ``go back'' to the
target vector ${\bf y}$ from
the original vector ${\bf z}$
by formation of the ``orthogonal'' subspace twice
-- the first time defining a kind of ``orthogonality''
between a functional ${\bf z} \in {\frak V}^\ast$ and vectors ${\bf z} \in {\frak V}$
by ${\bf z} ({\bf x})=0$.

Let us first consider the case
of ${\bf z} =0$, for which we can {\it ad hoc} identify
the zero vector with ${\bf y}$; that is, ${\bf y}=0$.

For any nonzero ${\bf z}({\bf x}) \neq 0$  on some ${\bf x}$ we first need to locate the subspace
\begin{equation}
{\frak M}= \Big\{ {\bf x} \Big| {\bf z}({\bf x}) = 0, {\bf x} \in {\frak V} \Big\}
\end{equation}
consisting of all vectors ${\bf x}$ for which ${\bf z}({\bf x})$ vanishes.

In a second step consider ${\frak M}^\perp$, the orthogonal complement of ${\frak M}$ with respect to ${\frak V}$.
${\frak M}^\perp$ consists of all vectors  orthogonal to all vectors in  ${\frak M}$,
such that $\langle {\bf x}\mid {\bf w} \rangle =0$ for ${\bf x}\in {\frak M}$ and ${\bf w}\in {\frak M}^\perp$.

The assumption ${\bf z}({\bf x}) \neq 0$  on some ${\bf x}$
guarantees that  ${\frak M}^\perp$ does not consist of the zero vector $0$ alone.
That is, ${\frak M}^\perp$ must contain a nonzero unit vector ${\bf y}_0\in {\frak M}^\perp$.
(It turns out that ${\frak M}^\perp$  is one-dimensional and spanned by ${\bf y}_0$;
that is, up to a multiplicative constant ${\bf y}_0$ is proportional to the vector ${\bf y}$.)

In a next step define the vector
\begin{equation}
 {\bf u} = {\bf z}({\bf x}) {\bf y}_0 - {\bf z}({\bf y}_0) {\bf x}
\label{2018-mm-ch-fdlvs-rrtd0}
\end{equation}
for which, due to linearity of ${\bf z}$,
\begin{equation}
 {\bf z}({\bf u}) = {\bf z}\Big[{\bf z}({\bf x}) {\bf y}_0 - {\bf z}({\bf y}_0)   {\bf x}\Big]
= {\bf z}({\bf x}) {\bf z}({\bf y}_0) - {\bf z}({\bf y}_0) {\bf z}({\bf x})
=0
.
\label{2018-mm-ch-fdlvs-rrtd1}
\end{equation}
Thus ${\bf u} \in {\frak M}$, and therefore also $\langle {\bf u} \mid {\bf y}_0 \rangle =0$.
Insertion of ${\bf u}$ from~(\ref{2018-mm-ch-fdlvs-rrtd0})
and antilinearity in the first argument and linearity in the second argument  of the inner product yields
\begin{equation}
\begin{split}
 \langle {\bf z}({\bf x}) {\bf y}_0  - {\bf z}({\bf y}_0) {\bf x} \mid  {\bf y}_0 \rangle=0,\\
\overline{{\bf z}({\bf x})} \underbrace{\langle {\bf y}_0 \mid  {\bf y}_0 \rangle}_{=1}
               - \overline{{\bf z}({\bf y}_0)} \langle {\bf x} \mid  {\bf y}_0 \rangle=0,\\
{\bf z}({\bf x}) = {\bf z}({\bf y}_0) \langle    {\bf y}_0\mid  {\bf x}\rangle\
=  \langle   \overline{{\bf z}({\bf y}_0)} {\bf y}_0\mid  {\bf x}\rangle
.
\end{split}
\label{2018-mm-ch-fdlvs-rrtd2}
\end{equation}
Thus we can identify the ``target'' vector
\begin{equation}
{\bf y} =  \overline{{\bf z}({\bf y}_0)} {\bf y}_0
\label{2018-mm-ch-fdlvs-rrtdtv}
\end{equation}
associated with the functional ${\bf z}$.

The proof of uniqueness is by (wrongly) assuming
that there exist two (presumably different) ${\bf y}_1$ and ${\bf y}_2$
such that
$
\langle {\bf x} \vert {\bf y}_1 \rangle
=
\langle {\bf x} \vert {\bf y}_2 \rangle
$ for all ${\bf x} \in {\frak V}$.
Due to linearity of the scalar product,
$
\langle  {\bf x}\vert {\bf y}_1-{\bf y}_2\rangle =0
$;
in particular, if we identify
${\bf x}={\bf y}_1-{\bf y}_2$,
then
$
\langle {\bf y}_1-{\bf y}_2 \vert {\bf y}_1-{\bf y}_2\rangle =0
$
and thus ${\bf y}_1={\bf y}_2$.

This proof is constructive in the sense that it yields ${\bf y}$, given ${\bf z}$.
Note that, because of uniqueness,
${\frak M}^\perp$ has to be a one dimensional subspace of ${\frak V}$  spanned by the unit vector ${\bf y}_0$.

Another, more direct, proof
is a straightforward construction of the ``target'' vector
${\bf y} \in {\frak V}$
associated with the linear functional
${\bf z} \in {\frak V}^\ast$
in terms of some orthonormal basis
${\frak B} = \{ {\bf e}_1, \ldots , {\bf e}_n\}$
of ${\frak V}$:
We obtain the components (coordinates) $y_i$, $1\le i\le n$  of
${\bf y} =
\sum_{j=1}^n y_j{\bf e}_j
\equiv
\begin{pmatrix}
y_1,
\cdots ,
y_n
\end{pmatrix}^\intercal  $ with respect to the orthonormal basis (coordinate system) ${\frak B}$
by  evaluating the ``performance'' of
${\bf z} $
on all vectors of the basis ${\bf e}_i$, $1\le i\le n$ in that basis:
\begin{equation}
{\bf z}({\bf e}_i)
= \langle  {\bf y} \mid  {\bf e}_i \rangle
= \big\langle \sum_{j=1}^n y_j{\bf e}_j  \big\vert  {\bf e}_i \big\rangle
= \sum_{j=1}^n \overline{y_j}\underbrace{\langle {\bf e}_j  \mid  {\bf e}_i}_{\delta{ij}} \rangle
= \overline{y_i}
.
\label{2018-mm-ch-fdlvs-rrtddirect}
\end{equation}
Hence, the ``target'' vector can be written as
\begin{equation}
{\bf y}   =  \sum_{j=1}^n  \overline{{\bf z}({\bf e}_j)}  {\bf e}_j
.
\label{2018-mm-ch-fdlvs-rrtddirect2}
\end{equation}
Both proofs yield the same ``target'' vector ${\bf y}$ associated with ${\bf z}$, as   insertion into~(\ref{2018-mm-ch-fdlvs-rrtdtv}) and~(\ref{2018-mm-ch-fdlvs-rrtddirect}) results in
\marginnote{Einstein's summation convention is used here.}
\index{Einstein summation convention}
\begin{equation}
\begin{split}
{\bf y}
=  \overline{{\bf z}({\bf y}_0)} {\bf y}_0
=  \overline{{\bf z}\left( \frac{y_i}{\sqrt{\langle{\bf y}\mid{\bf y}\rangle}}{\bf e}_i \right)}  \frac{y_j}{\sqrt{\langle{\bf y}\mid{\bf y}\rangle}}{\bf e}_j  \\
=  \frac{\overline{y_i}}{ \langle{\bf y}\mid{\bf y}\rangle }\underbrace{\overline{{\bf z}\left( {\bf e}_i \right)}}_{y_i}   y_j {\bf e}_j
=  \frac{\overline{y_i}y_i}{\langle{\bf y}\mid{\bf y}\rangle }   y_j{\bf e}_j  = y_j{\bf e}_j
.
\end{split}
\label{2018-mm-ch-fdlvs-rrtddirect21}
\end{equation}

\eproof
}

{\color{blue}
\bexample
In the Babylonian
tradition\sidenote[][-13mm]{The Babylonians ``proved'' arithmetical statements
by inserting ``large numbers'' in the respective conjectures;
cf. Chapter V of \bibentry{neugeb}}
and for the sake of an example
consider the Cartesian standard basis of ${\frak V} = \mathbb{R}^2$;
with the two basis vectors
${\bf e}_1 =\begin{pmatrix} 1,0\end{pmatrix}^\intercal $ and
${\bf e}_2 =\begin{pmatrix} 0,1\end{pmatrix}^\intercal $.
Suppose further that the linear functional ${\bf z}$ is defined by its ``behavior'' on these basis elements
${\bf e}_1$  and
${\bf e}_2$ as follows:
\begin{equation}
 {\bf z}({\bf e}_1) = 1 ,
\; {\bf z}({\bf e}_2) = 2.
\end{equation}

In a first step, let us construct
${\frak M}=\{ {\bf x} \mid {\bf z}({\bf x}) = 0, {\bf x} \in  \mathbb{R}^2\}$.
Consider an arbitrary vector ${\bf x} = x_1 {\bf e}_1 + x_2 {\bf e}_2 \in {\frak M}$.
Then,
\begin{equation}
\begin{split}
{\bf z}  ( {\bf x} )
=  {\bf z}  ( x_1 {\bf e}_1 + x_2 {\bf e}_2 )
=  x_1 {\bf z}  ( {\bf e}_1) + x_2 {\bf z}  ( {\bf e}_2 )
=  x_1  + 2 x_2 = 0
,
\end{split}
\end{equation}
and therefore $x_1 =  - 2 x_2 $.
The normalized vector spanning ${\frak M}$ thus is
$\frac{1}{\sqrt{5}} \begin{pmatrix} -2,1\end{pmatrix}^\intercal $.

In the second step, a normalized vector  ${\bf y}_0 \in  {\frak N} = {\frak M}^\perp$ orthogonal to ${\frak M}$ is constructed by
$\frac{1}{\sqrt{5}} \begin{pmatrix} -2,1\end{pmatrix}^\intercal  \cdot {\bf y}_0=0$,
resulting in
${\bf y}_0=\frac{1}{\sqrt{5}} \begin{pmatrix} 1,2\end{pmatrix}^\intercal  =
\frac{1}{\sqrt{5}} \left( {\bf e}_1 +2 {\bf e}_2 \right) $.

In the third and final step ${\bf y}$ is constructed through
\begin{equation}
\begin{split}
{\bf y}
=  {\bf z}  ( {\bf y}_0 ){\bf y}_0
=  {\bf z}  \left( \frac{1}{\sqrt{5}}\left( {\bf e}_1 +2 {\bf e}_2 \right) \right)\frac{1}{\sqrt{5}} \begin{pmatrix} 1,2\end{pmatrix}^\intercal   \\
= \frac{1}{5}  \left[ {\bf z} ( {\bf e}_1) + 2 {\bf z}  ( {\bf e}_2 ) \right] \begin{pmatrix} 1,2\end{pmatrix}^\intercal
= \frac{1}{5}  \left[ 1 + 4 \right] \begin{pmatrix} 1,2\end{pmatrix}^\intercal
=    \begin{pmatrix} 1,2\end{pmatrix}^\intercal
.
\end{split}
\end{equation}

It is always prudent -- and in the ``Babylonian spirit'' -- to check this out by inserting ``large numbers'' (maybe even primes):
suppose ${\bf x} =\begin{pmatrix} 11,13\end{pmatrix}^\intercal $; then
${\bf z}  ( {\bf x} ) = 11 + 26 = 37$; whereas, according to Equation~(\ref{2015-m-ch-fdlvs-e1}),
$\langle {\bf y}\mid {\bf x}\rangle = \begin{pmatrix} 1,2\end{pmatrix}^\intercal   \cdot      \begin{pmatrix} 11,13\end{pmatrix}^\intercal
= 37$.

\eexample
}

Note that  in  real or complex vector space ${\Bbb R}^n$ or ${\Bbb C}^n$, and with the dot product,  ${\bf y}^\dagger \equiv {\bf z}$.
Indeed, this construction induces a ``conjugate'' (in the complex case, referring to the conjugate symmetry of the scalar product
in Equation~(\ref{2015-m-ch-fdlvs-e1}),
which is {\em conjugate-linear} in its second argument) isomorphisms between a vector space
${\frak V}$ and its dual space ${\frak V}^\ast$.

Note also that every inner product
$\langle {\bf y}\mid {\bf x} \rangle = \phi_y(x)$ defines a linear
functional $\phi_y(x)$ for all ${\bf x}\in {\frak V}$.

{\color{Purple}
In quantum mechanics,
this representation of a functional by the inner product suggests
the (unique) existence of
the bra vector $\langle \psi \vert \in {\frak V}^\ast $
associated with every ket vector $\vert \psi \rangle \in {\frak V}$.

It also suggests a ``natural'' duality between
propositions and states
--
that is, between (i)
dichotomic (yes/no, or 1/0) observables
represented by projections $\textsf{\textbf{E}}_{\bf x}= \vert {\bf x} \rangle \langle {\bf x} \vert$
and their associated linear subspaces  spanned by unit vectors  $\vert {\bf x} \rangle $
on the one hand,
and (ii) pure states, which are also represented by projections $\boldsymbol{\rho}_{\psi}= \vert \psi \rangle \langle \psi \vert$
and their associated subspaces spanned by unit vectors  $ \vert {\psi} \rangle$
on the other hand
--
{\em via} the scalar product ``$\langle \cdot \vert \cdot \rangle$.''
In particular,\cite{hamhalter-book}
\begin{equation}
{{\psi}} ({\bf x})  = \langle \psi \mid  {\bf x} \rangle
\end{equation}
represents the {\em probability amplitude.}
By the {\em Born rule}
\index{Born rule}
for pure states,
the absolute square $\vert \langle {\bf x} \mid \psi \rangle \vert^2$
of this probability amplitude is identified with the probability of the occurrence of the proposition
$\textsf{\textbf{E}}_{\bf x}$,
given the state  $ \vert {\psi} \rangle$.

More general,  due to linearity and the spectral theorem
(cf. Section \ref{2012-m-ch-Spectraltheorem} on page \pageref{2012-m-ch-Spectraltheorem}),
the statistical expectation for a Hermitian (normal) operator $\textsf{\textbf{A}}=
\sum_{i=0}^k   \lambda_i \textsf{\textbf{E}}_i$
and a quantized system prepared in pure state
\index{pure state}
(cf. Section~\ref{2011-m-projec})
$\boldsymbol{\rho}_\psi = \vert {\psi}\rangle \langle \psi \vert$ for some unit vector $\vert {\psi}\rangle$
is given by the {\em Born rule}
\index{Born rule}
\begin{equation}
\begin{split}
\langle \textsf{\textbf{A}}\rangle_{\psi} = \text{Tr} (\boldsymbol{\rho}_\psi \textsf{\textbf{A}})
=
\text{Tr}  \left[\boldsymbol{\rho}_\psi  \left(\sum_{i=0}^k   \lambda_i  \textsf{\textbf{E}}_i  \right)\right] =
\text{Tr}  \left(\sum_{i=0}^k   \lambda_i \boldsymbol{\rho}_\psi  \textsf{\textbf{E}}_i  \right)\\
=
 \text{Tr} \left(\sum_{i=0}^k   \lambda_i (\vert \psi \rangle \langle \psi \vert )( \vert {\bf x}_i \rangle \langle {\bf x}_i \vert )\right)
=
 \text{Tr} \left(\sum_{i=0}^k   \lambda_i  \vert \psi \rangle \langle \psi \vert   {\bf x}_i \rangle \langle {\bf x}_i \vert  \right)\\
=
\sum_{j=0}^k \langle {\bf x}_j \vert
\left(  \sum_{i=0}^k   \lambda_i  \vert \psi \rangle  \langle \psi   \vert {\bf x}_i \rangle   \langle {\bf x}_i \vert \right)   \vert{\bf x}_j \rangle    \\
=
\sum_{j=0}^k
   \sum_{i=0}^k   \lambda_i  \langle {\bf x}_j \vert \psi \rangle   \langle \psi   \vert {\bf x}_i \rangle
\underbrace{ \langle {\bf x}_i \vert    {\bf x}_j \rangle }_{\delta_{ij}}  \\
=
\sum_{i=0}^k   \lambda_i  \langle {\bf x}_i \vert \psi \rangle \langle \psi   \vert {\bf x}_i \rangle
=
\sum_{i=0}^k   \lambda_i \vert \langle {\bf x}_i \vert \psi \rangle \vert^2,
\end{split}
\label{2015-m-ch-fdlvs-bornr}
\end{equation}
where $\text{Tr}$ stands for the trace (cf. Section \ref{2013-ch-fdvs-trace} on page \pageref{2013-ch-fdvs-trace}),
and we have used the spectral decomposition $\textsf{\textbf{A}}= \sum_{i=0}^k   \lambda_i \textsf{\textbf{E}}_i$
(cf. Section \ref{2012-m-ch-Spectraltheorem} on page \pageref{2012-m-ch-Spectraltheorem}).
}

\subsection{Double dual space}
\index{double dual space}
\label{2012-m-dds}

In the following, we strictly limit the discussion to finite dimensional vector spaces.

Because to every vector  space ${\frak V}$
there exists a dual vector  space ${\frak V}^\ast$ ``spanned''
by all linear functionals on ${\frak V}$,
there exists also a dual vector space $({\frak V}^{\ast})^{\ast}={\frak V}^{\ast \ast}$ to the dual  vector  space ${\frak V}^\ast$
 ``spanned'' by all linear functionals on ${\frak V}^\ast$.
This construction can be iterated and is the basis of a constructively definable ``succession''
of spaces of ever increasing duality.

At the same time, by a sort of ``inversion''
of the linear functional
(or by exchanging the corresponding arguments of the inner product)
every vector in ${\frak V}$
can be thought of as a linear functional on ${\frak V}^\ast$:
just define
${\bf x}({\bf y})\stackrel{{\tiny \textrm{ def }}}{\equiv}{\bf y}({\bf x})$
for ${\bf x}\in {\frak V}$ and ${\bf y}\in {\frak V}^\ast$,
thereby rendering an element in ${\frak V}^{\ast \ast}$.
So is there some sort of ``connection'' between
a vector space and its double dual space?

\marginnote{For proofs and additional information see {\S}16 in~\bibentry{halmos-vs}.}
We state without proof that indeed
there is a canonical identification
\index{canonical identification}
between ${\frak V}$ and ${\frak V}^{\ast \ast}$:
corresponding to every linear functional
${\bf z}\in  {\frak V}^{\ast \ast}$ on
the dual space  ${\frak V}^\ast$ of   ${\frak V}$
there exists a vector ${\bf x} \in  {\frak V}$
such that ${\bf z}({\bf y}) = {\bf y}({\bf x})$
for every ${\bf y} \in  {\frak V}^\ast$.
Thereby this ${\bf x}-{\bf z}$ correspondence
${\frak V} \equiv {\frak V}^{\ast \ast}$ between
${\frak V}$ and ${\frak V}^{\ast \ast}$
is an isomorphism; that is, a structure preserving map
which is one-to-one and onto.

With this in mind, we obtain
\begin{equation}
\begin{split}
{\frak V}\equiv {\frak V}^{\ast \ast},\\
{\frak V}^\ast \equiv {\frak V}^{\ast \ast \ast},\\
{\frak V}^{\ast \ast} \equiv {\frak V}^{\ast \ast \ast \ast}\equiv {\frak V},\\
{\frak V}^{\ast  \ast \ast} \equiv {\frak V}^{\ast \ast \ast  \ast \ast}\equiv {\frak V}^\ast ,\\
\qquad \vdots
\end{split}
\end{equation}

\section{Tensor product}
\label{2011-m-tensorp}
\marginnote{For proofs and additional information see {\S}24 in~\bibentry{halmos-vs}.}

\index{tensor product}

\subsection{Sloppy definition}

Informally speaking the {\em tensor product}
\index{tensor product}
 ${\frak V} \otimes {\frak U}$
of two linear vector spaces  ${\frak V}$ and  ${\frak U}$
should be such that,
to every
${\bf x} \in  {\frak V}$
and every
${\bf y} \in  {\frak U}$
there corresponds a tensor product ${\bf z} = {\bf x} \otimes {\bf y}
\in {\frak V} \otimes {\frak U}$
which is bilinear; that is, linear in both factors.

A generalization to more factors appears to present no further conceptual difficulties.

\subsection{Definition}

A more rigorous definition uses ``double-duality'' by
considering the dual space of all bilinear functionals as follows:
The {\em tensor product} ${\frak V} \otimes {\frak U}$ of two vector spaces ${\frak V}$
and
${\frak U}$ (over the same field, say ${\Bbb R}$ or ${\Bbb C}$)
is the dual vector space of all bilinear forms on ${\frak V}$ and ${\frak U}$.

That is, for each pair of vectors ${\bf x} \in  {\frak V}$
and
${\bf y} \in  {\frak U}$   the tensor product ${\bf z} = {\bf x} \otimes {\bf y}$
is identified (in the sense of double-duality) with the element of ${\frak V} \otimes {\frak U}$
such that ${\bf z}({\bf w})={\bf w}({\bf x},{\bf y})$ for every bilinear form
${\bf w}$.

Alternatively and more concretely
we could define the {\em tensor product}  as the coherent superpositions (aka linear combination)
\marginnote{The terms ``coherent superposition'' and ``linear combination''
will be used synonymously; the former being much used in quantum mechanics, the latter in mathematics.}
of products ${\bf e}_i \otimes {\bf f}_j$
\index{coherent superposition}
\index{superposition}
of all basis vectors
${\bf e}_i \in  {\frak V}$, with $1\le i \le n$,
and
${\bf f}_j \in  {\frak U}$, with $1\le j \le m$ as follows.
First we note without proof that if ${\frak A} =\{{\bf e}_1,\ldots , {\bf e}_n\}$ and
${\frak B} =\{{\bf f}_1,\ldots , {\bf f}_m\}$
are bases of  $n$- and $m$-
dimensional vector spaces ${\frak V}$ and  ${\frak U}$, respectively,
then the set
of vectors ${\bf e}_i \otimes {\bf f}_j$
with $i=1,\ldots n$ and $j=1,\ldots m$
 is a basis of the { tensor product}
 ${\frak V} \otimes {\frak U}$.
Then an arbitrary tensor product can be written as the coherent superposition of all
\index{coherent superposition}
\index{superposition}
its basis vectors ${\bf e}_i \otimes {\bf f}_j$
with
${\bf e}_i \in  {\frak V}$, with $1\le i \le n$,
and
${\bf f}_j \in  {\frak U}$, with $1\le j \le m$; that is,
\begin{equation}
{\bf z}=\sum_{i,j} c_{ij} \; {\bf e}_i \otimes {\bf f}_j
\equiv \sum_{i,j} c_{ij} \; \vert {\bf e}_i \rangle \otimes  \vert {\bf f}_j\rangle
\equiv \sum_{i,j} c_{ij} \; \vert {\bf e}_i \rangle   \vert {\bf f}_j\rangle
\equiv \sum_{i,j} c_{ij} \; \vert {\bf e}_i  {\bf f}_j\rangle
.
\label{2014-m-ch-fdvs-lsqv}
\end{equation}

We state without proof that the dimension
of ${\frak V} \otimes {\frak U}$ of an $n$-dimensional vector space ${\frak V}$
and an $m$-dimensional vector space
${\frak U}$
is multiplicative,
that is, the dimension of  ${\frak V} \otimes {\frak U}$ is $nm$.
Informally, this is evident from the number of basis pairs ${\bf e}_i \otimes {\bf f}_j$.

\subsection{Representation}
\index{dyadic product}
\index{outer product}

A tensor (dyadic, outer) product ${\bf z} = {\bf x} \otimes {\bf y}$ of two vectors ${\bf x}$ and ${\bf y}$
has three equivalent notations or representations:
\begin{itemize}
\item[(i)]
as the scalar coordinates $x_iy_j$ with respect to the basis in which the vectors ${\bf  x}$ and ${\bf y}$ have been defined and encoded;
\item[(ii)]
as a quasi-matrix $z_{ij}  =x_iy_j$,
whose components $z_{ij}$ are  defined with respect to the basis in which the vectors ${\bf  x}$ and ${\bf y}$
have been defined and encoded;
\item[(iii)]
as a list, or quasi-vector, or ``flattened matrix'' defined by the Kronecker product
${\bf z} = ({ x}_1  {\bf y}, { x}_2  {\bf y}, \ldots , { x}_n  {\bf y})^\intercal =
({ x}_1  { y}_1, { x}_1  { y}_2, \ldots , { x}_n  { y}_n)^\intercal
$. Again, the scalar coordinates $x_iy_j$ are defined
with respect to the basis in which the vectors ${\bf  x}$ and ${\bf y}$ have been defined and encoded.
\index{Kronecker product}
\end{itemize}
In all three cases, the pairs $x_i y_j$  are properly represented by distinct mathematical entities.

{\color{blue}
\bexample
Take, for example,
${\bf x}=(2,3)^\intercal $
and
${\bf y}=(5,7,11)^\intercal $.
Then ${\bf z} = {\bf x} \otimes {\bf y}$  can be represented by
(i) the four scalars
$x_1y_1=10$,
$x_1y_2=14$,
$x_1y_3=22$,
$x_2y_1=15$,
$x_2y_2=21$,
$x_2y_3=33$,
or by
(ii) a $2 \times 3$ matrix
$
\begin{pmatrix}
10&14&22\\
15&21&33
\end{pmatrix}
$,
or by
(iii) a $2 \times 3 = 6$-tuple
$
\begin{pmatrix}  10,14,22,15,21,33\end{pmatrix} ^\intercal
$.
\eexample
}

Note, however, that this kind of quasi-matrix or quasi-vector representation of vector products
can be misleading insofar as
it (wrongly) suggests that all vectors in the tensor product space are accessible (representable) as quasi-vectors
-- they are, however, accessible by {\em coherent superpositions} (\ref{2014-m-ch-fdvs-lsqv})
\index{coherent superposition}
\index{superposition}
of such quasi-vectors. \label{2012-m-c-fdvs-entanglement}
\marginnote{In quantum mechanics this amounts to the fact that not all pure two-particle states can be
written in terms of (tensor) products of single-particle states; see also Section 1.5 of~\bibentry{mermin-07}.}
For instance, take the arbitrary form of a (quasi-)vector in ${\Bbb C}^4$, which can be parameterized by
\begin{equation}
\begin{pmatrix}\alpha_1,\alpha_2,\alpha_3,\alpha_4\end{pmatrix}^\intercal , \textrm{ with } \alpha_1,\alpha_3,\alpha_3,\alpha_4 \in {\Bbb C},
\label{2012-m-ch-fdvs-dectp-gf}
\end{equation}
and compare (\ref{2012-m-ch-fdvs-dectp-gf}) with the general form of a tensor product of two quasi-vectors in  ${\Bbb C}^2$
\begin{equation}
\begin{pmatrix}a_1,a_2\end{pmatrix}^\intercal \otimes \begin{pmatrix}b_1,b_2\end{pmatrix}^\intercal
\equiv \begin{pmatrix}a_1b_1, a_1 b_2,a_2b_1,a_2b_2\end{pmatrix}^\intercal , \textrm{ with } a_1,a_2,b_1,b_2\in {\Bbb C}.
\label{2012-m-ch-fdvs-dectp-gftp}
\end{equation}
A comparison of the coordinates in
(\ref{2012-m-ch-fdvs-dectp-gf})
and
(\ref{2012-m-ch-fdvs-dectp-gftp})
yields
\begin{equation}
\begin{split}
\alpha_1=a_1b_1,\quad
\alpha_2=a_1b_2,\quad
\alpha_3=a_2b_1,\quad
\alpha_4=a_2b_2.
\end{split}
\label{2012-m-ch-fdvs-dectp-gftp-a}
\end{equation}
By taking the product of both sides of (i) the first and the last
equations, as well as (ii) the second and the third equations
one obtains because of commutativity
\begin{equation}
\begin{split}
{\alpha_1}{\alpha_4}= (a_1b_1) (a_2b_2) = (a_1b_2) (a_2b_1)
={\alpha_2}{\alpha_3},
\end{split}
\label{2012-m-ch-fdvs-dectp-gftp-fr}
\end{equation}
which amounts to a condition for the four coordinates  $\alpha_1,\alpha_2,\alpha_3,\alpha_4$
in order for this four-dimensional vector to be decomposable into a tensor product of two two-dimensional quasi-vectors.
In quantum mechanics, pure states which are not decomposable into a  product of single-particle states
are called {\em entangled}.
\index{entanglement}

{\color{blue}
\bexample
\label{bellstate1}
A typical example of an entangled state is the
{\em Bell state}, \index{Bell state}  $\vert \Psi^- \rangle$
or, more generally, states in the Bell basis  \index{Bell basis}:
with the notation $
{\bf a}
\otimes
{\bf b}
\equiv
{\bf a}
{\bf b}
\equiv
\vert {\bf a}\rangle
\otimes
\vert {\bf b}\rangle
\equiv
\vert {\bf a}\rangle
\vert {\bf b}\rangle
\equiv
\vert {\bf a} {\bf b}\rangle
$
and the identifications
$\vert 0 \rangle \equiv  \begin{pmatrix}1,0\end{pmatrix}^\intercal$ and
$\vert 1 \rangle \equiv  \begin{pmatrix}0,1\end{pmatrix}^\intercal$
\begin{equation}
\begin{split}
\vert \Psi^\pm \rangle
 = \frac{1}{\sqrt{2}}\left(\vert 0 \rangle \vert 1 \rangle \pm \vert 1 \rangle \vert 0 \rangle  \right)
\equiv \frac{1}{\sqrt{2}}\left(\vert 0   1 \rangle \pm \vert 1   0 \rangle  \right)
\qquad \qquad
\\
\equiv
\frac{1}{\sqrt{2}}
\left[
\begin{pmatrix}1,0\end{pmatrix}^\intercal
\begin{pmatrix}0,1\end{pmatrix}^\intercal
\pm
\begin{pmatrix}0,1\end{pmatrix}^\intercal
\begin{pmatrix}1,0\end{pmatrix}^\intercal
\right]
=
\frac{1}{\sqrt{2}}
\begin{pmatrix}0,1,\pm 1,0\end{pmatrix}^\intercal
,\\
\vert \Phi^\pm \rangle
 = \frac{1}{\sqrt{2}}\left(\vert 0 \rangle \vert 0 \rangle \pm \vert 1 \rangle \vert 1 \rangle  \right)
 \equiv \frac{1}{\sqrt{2}}\left(\vert 0   0 \rangle \pm \vert 1   1 \rangle  \right)
\qquad \qquad
\\
\equiv
\frac{1}{\sqrt{2}}
\left[
\begin{pmatrix}1,0\end{pmatrix}^\intercal
\begin{pmatrix}1,0\end{pmatrix}^\intercal
\pm
\begin{pmatrix}0,1\end{pmatrix}^\intercal
\begin{pmatrix}0,1\end{pmatrix}^\intercal
\right]
=
\frac{1}{\sqrt{2}}
\begin{pmatrix}1,0,0,\pm 1\end{pmatrix}^\intercal.
\end{split}
\label{2014-m-ch-fdvs-bellbasis}
\end{equation}

For instance, in the case of $\vert \Psi^- \rangle$ a comparison of coefficient yields
\begin{equation}
\begin{split}
\alpha_1=a_1b_1=0=a_2b_2=\alpha_4,      \\
\alpha_2=a_1b_2=\frac{1}{\sqrt{2}}=-a_2b_1=-\alpha_3;
\end{split}
\label{2012-m-ch-fdvs-BellSCC}
\end{equation}
and thus
the entanglement, since
\begin{equation}
{\alpha_1}{\alpha_4}=0 \neq {\alpha_2}{\alpha_3}=-\frac{1}{2}.
\end{equation}
This shows that  $\vert \Psi^- \rangle$ cannot be considered as a two particle product state.
Indeed, the state can only be characterized by considering the {\em relative properties}
of the two particles --
in the case of  $\vert \Psi^- \rangle$ they are associated with the statements:\cite{zeil-99}
``the quantum numbers (in this case ``$0$'' and ``$1$'') of the two particles are always different.''

\eexample
}

\section{Linear transformation}
\marginnote{For proofs and additional information see {\S}32-34 in~\bibentry{halmos-vs}.}
\index{linear transformation}
\label{2019-mm-ch-fdvs-lt}

\subsection{Definition}
A {\em linear transformation}, or, used synonymously, a {\em linear operator},
\index{linear transformation}
\index{linear operator}
$\textsf{\textbf{A}} $ on a vector space ${\frak V}$ is a correspondence that assigns every vector
${\bf x}\in {\frak V}$ a vector $\textsf{\textbf{A}} {\bf x}\in {\frak V}$,
in a linear way; such  that
\begin{equation}
\textsf{\textbf{A}}  (\alpha {\bf x}+ \beta {\bf y}) = \alpha \textsf{\textbf{A}}({\bf x})
+  \beta \textsf{\textbf{A}} ({\bf y}) = \alpha \textsf{\textbf{A}}{\bf x}
+  \beta \textsf{\textbf{A}} {\bf y},
\end{equation}
identically for all vectors ${\bf x},{\bf y}\in {\frak V}$ and all scalars $\alpha , \beta$.

\subsection{Operations}
The {\em sum}
\index{sum of transformations}
$\textsf{\textbf{S}} =\textsf{\textbf{A}} +\textsf{\textbf{B}} $
of two linear transformations $\textsf{\textbf{A}}$ and $\textsf{\textbf{B}} $
is defined by
$\textsf{\textbf{S}} {\bf x}=\textsf{\textbf{A}}{\bf x} +\textsf{\textbf{B}} {\bf x}$
for every ${\bf x}\in {\frak V}$.

The {\em product}
\index{product of transformations}
$\textsf{\textbf{P}} =\textsf{\textbf{A}} \textsf{\textbf{B}} $
of two linear transformations $\textsf{\textbf{A}}$ and $\textsf{\textbf{B}} $
is defined by
$\textsf{\textbf{P}} {\bf x}=\textsf{\textbf{A}}(\textsf{\textbf{B}} {\bf x})$
for every ${\bf x}\in {\frak V}$.

The notation
$\textsf{\textbf{A}}^n\textsf{\textbf{A}}^m=\textsf{\textbf{A}}^{n+m}$
and $(\textsf{\textbf{A}}^n)^m= \textsf{\textbf{A}}^{nm}$,
with $\textsf{\textbf{A}}^1=\textsf{\textbf{A}}$ and
$\textsf{\textbf{A}}^0 =\textsf{\textbf{1}}$ turns out to be useful.

With the exception of commutativity, all formal algebraic properties
of numerical addition and multiplication,
are valid for transformations; that is
$
\textsf{\textbf{A}}\textsf{\textbf{0}}=
\textsf{\textbf{0}}\textsf{\textbf{A}} =\textsf{\textbf{0}}
$,
$
\textsf{\textbf{A}}\textsf{\textbf{1}}=
\textsf{\textbf{1}}\textsf{\textbf{A}} =\textsf{\textbf{A}}
$,
$
\textsf{\textbf{A}} (\textsf{\textbf{B}}+\textsf{\textbf{C}})=
\textsf{\textbf{A}} \textsf{\textbf{B}}
+
\textsf{\textbf{A}} \textsf{\textbf{C}}
$,
$
(\textsf{\textbf{A}}+ \textsf{\textbf{B}})\textsf{\textbf{C}}=
\textsf{\textbf{A}} \textsf{\textbf{C}}
+
\textsf{\textbf{B}} \textsf{\textbf{C}}
$,  and
$
\textsf{\textbf{A}} (\textsf{\textbf{B}}\textsf{\textbf{C}})=
(\textsf{\textbf{A}} \textsf{\textbf{B}})
 \textsf{\textbf{C}}
$.

In {\em matrix notation},  $\textsf{\textbf{1}} \equiv {\Bbb{1}}$,
and the entries of $\textsf{\textbf{0}}$
are $0$ everywhere.

The {\em inverse operator}
\index{inverse operator}
$\textsf{\textbf{A}}^{-1}$
of $\textsf{\textbf{A}}$
is defined by
$\textsf{\textbf{A}}\textsf{\textbf{A}}^{-1}=\textsf{\textbf{A}}^{-1}\textsf{\textbf{A}}=
\textsf{\textbf{1}} \equiv \mathbb{1}$.

The {\em commutator}
\index{commutator}
of two matrices $\textsf{\textbf{A}}$  and $\textsf{\textbf{B}}$ is defined by
\begin{equation}
[\textsf{\textbf{A}}, \textsf{\textbf{B}} ]
=
\textsf{\textbf{A}} \textsf{\textbf{B}}
-
 \textsf{\textbf{B}}      \textsf{\textbf{A}}.
\label{2020-commutator}
\end{equation}
\marginnote{The commutator should not be confused with the bilinear functional
introduced for dual spaces.}

{\color{blue}
\bexample
In terms of this matrix notation, it is quite easy to present an example
for which the commutator
$
[\textsf{\textbf{A}}, \textsf{\textbf{B}} ]
$
does not vanish; that is
$\textsf{\textbf{A}}$  and $\textsf{\textbf{B}}$
do not commute.

Take, for the sake of an example, the
{\em Pauli spin matrices}
\index{Pauli spin matrices}
which are proportional to the angular momentum operators
of spin-$\frac{1}{2}$ particles
along the~$x,y,z$-axis:\marginnote{For
more general angular momentum operators see~\bibentry{schiff-55}.}
\begin{equation}
\begin{split}
\sigma_1=\sigma_x
=
\begin{pmatrix}
0&1\\
1&0
\end{pmatrix}
,   \\
\sigma_2=\sigma_y
=
\begin{pmatrix}
0&-i\\
i&0
\end{pmatrix}
,   \\
\sigma_3=\sigma_z
=
\begin{pmatrix}
1&0\\
0&-1
\end{pmatrix}
.
\end{split}
\label{2019-mm-ch-fdvs-psm}
\end{equation}
Together with the identity, that is, with $\mathbb{1}_2=\textrm{diag}(1,1)$,
they form a complete basis of all $(4\times 4)$ matrices.
Now take, for instance, the commutator
\begin{equation}
\begin{split}
[\sigma_1,\sigma_3]= \sigma_1\sigma_3-\sigma_3\sigma_1\\
\qquad
=
\begin{pmatrix}
0&1\\
1&0
\end{pmatrix}
\begin{pmatrix}
1&0\\
0&-1
\end{pmatrix}
-
\begin{pmatrix}
1&0\\
0&-1
\end{pmatrix}
\begin{pmatrix}
0&1\\
1&0
\end{pmatrix}
\\
\qquad
=  2
\begin{pmatrix}
0&-1\\
1&0
\end{pmatrix}
\neq
\begin{pmatrix}
0&0\\
0&0
\end{pmatrix}
.    \textrm{\eexample}
\end{split}
\end{equation}
}

The {\em polynomial}
\index{polynomial}
can be directly adopted from ordinary arithmetic; that is,
any finite polynomial $p$ of degree $n$
of an operator (transformation) $\textsf{\textbf{A}}$ can be written as
\begin{equation}
p(\textsf{\textbf{A}})= \alpha_0   \textsf{\textbf{1}}
+ \alpha_1   \textsf{\textbf{A}}^1
+ \alpha_2   \textsf{\textbf{A}}^2+
\cdots
+
\alpha_n   \textsf{\textbf{A}}^n
=\sum_{i=0}^n \alpha_i \textsf{\textbf{A}}^i
.
\end{equation}

The Baker-Hausdorff formula
 \begin{equation}
 e^{i\textsf{\textbf{A}}}\textsf{\textbf{B}}e^{-i\textsf{\textbf{A}}}=
B+i[\textsf{\textbf{A}},\textsf{\textbf{B}}]+
{i^2\over 2!}[\textsf{\textbf{A}},[\textsf{\textbf{A}},\textsf{\textbf{B}}]]+\cdots
 \end{equation}
for two arbitrary noncommutative linear operators $\textsf{\textbf{A}}$ and
$\textsf{\textbf{B}}$ is mentioned without proof\cite[-10mm]{messiah-61}).

If $[\textsf{\textbf{A}},\textsf{\textbf{B}}]$ commutes with $\textsf{\textbf{A}}$ and
$\textsf{\textbf{B}}$, then
 \begin{equation}
 e^\textsf{\textbf{A}}e^\textsf{\textbf{B}}=
e^{\textsf{\textbf{A}}+\textsf{\textbf{B}}+{1\over 2}\left[\textsf{\textbf{A}},\textsf{\textbf{B}}\right]}.
 \end{equation}

If  $\textsf{\textbf{A}}$ commutes with $\textsf{\textbf{B}}$, then
 \begin{equation}
 e^\textsf{\textbf{A}}e^\textsf{\textbf{B}}=
e^{\textsf{\textbf{A}}+\textsf{\textbf{B}}}.
 \end{equation}

\subsection{Linear transformations as matrices}

\index{matrix}
\index{transformation matrix}

Let ${\frak V}$ be an $n$-dimensional vector space;
let
${\frak B}=\{\vert {\bf f}_1 \rangle ,\vert {\bf f}_2 \rangle,\ldots ,\vert {\bf f}_n \rangle\}$ be any basis of ${\frak V}$,
and let  $\textsf{\textbf{A}}$ be a linear transformation on ${\frak V}$.

Because every vector is a linear combination of the basis vectors
$\vert {\bf f}_i \rangle$,
every linear transformation can be defined by
``its performance on the basis vectors;'' that is,
by the particular mapping of
all $n$ basis vectors into the transformed vectors, which in turn can be represented as linear combination of the $n$ basis vectors.

Therefore it is possible to define some $n \times n$ matrix with $n^2$ coefficients or coordinates
$\alpha_{ij}$ such that
\begin{equation}
\textsf{\textbf{A}}\vert  {\bf f}_j \rangle = \sum_i \alpha_{ij}\vert {\bf f}_i  \rangle
\label{2015-m-ch-fdlvs-dtlt}
\end{equation}
for all $j=1,\ldots ,n$.
Again, note that this definition of a {\em transformation matrix}
is ``tied to'' a basis.

The ``reverse order'' of indices in (\ref{2015-m-ch-fdlvs-dtlt}) has been chosen
in order for the vector coordinates to transform in the ``right order:''
with~(\ref{2016-m-fdvs-rv0}) on page~\pageref{2016-m-fdvs-rv0}: note that
\begin{equation}
\begin{split}
\textsf{\textbf{A}}\vert  {\bf x} \rangle =
\textsf{\textbf{A}} \sum_j x_j \vert  {\bf f}_j \rangle =
\sum_j \textsf{\textbf{A}} x_j \vert  {\bf f}_j \rangle =
\sum_j x_j \textsf{\textbf{A}} \vert  {\bf f}_j \rangle =
\sum_{i,j} x_j \alpha_{ij} \vert  {\bf f}_i \rangle  \\
= \sum_{i,j} \alpha_{ij}  x_j \vert  {\bf f}_i \rangle =
(i\leftrightarrow j) =
\sum_{j,i} \alpha_{ji}  x_i \vert  {\bf f}_j \rangle .
\end{split}
\label{2016-m-ch-fdlvs-dtltk}
\end{equation}
Because we can formally write
$\textsf{\textbf{A}}\vert  {\bf x} \rangle = \big[\textsf{\textbf{A}}\vert  {\bf x} \rangle\big]_j \vert  {\bf f}_j \rangle$,
the question remains: ``what is $\big[\textsf{\textbf{A}}\vert  {\bf x} \rangle\big]_j$?''
A comparison with~(\ref{2016-m-ch-fdlvs-dtltk}) yields
\begin{equation}
\begin{split}
\sum_j
\left(
\big[\textsf{\textbf{A}}\vert  {\bf x} \rangle\big]_j
-
\sum_{i} \alpha_{ji}  x_i
\right)
\vert  {\bf f}_j \rangle
=0
 .
\end{split}
\label{2016-m-ch-fdlvs-dtltk2}
\end{equation}
Because the basis vectors  in
${\frak B}=\{\vert {\bf f}_1 \rangle ,\vert {\bf f}_2 \rangle,\ldots ,\vert {\bf f}_n \rangle\}$
are linear independent, all  the coefficients
in~(\ref{2016-m-ch-fdlvs-dtltk2})
must vanish; that is, $\big[\textsf{\textbf{A}}\vert  {\bf x} \rangle\big]_j
-
\sum_{i} \alpha_{ji}  x_i    =0$ . Therefore, the $j$th component $x'_j$ of the new, transformed vector $\vert {\bf x}' \rangle$
is
\begin{equation}
\textsf{\textbf{A}}: x_j \mapsto x'_j =
\big[\textsf{\textbf{A}}\vert  {\bf x} \rangle\big]_j  = \sum_{i} \alpha_{ji}  x_i,
\quad
\text{ or }
\quad
{\bf x} \mapsto {\bf x}' =
\textsf{\textbf{A}} {\bf x}
.
\label{2016-m-ch-fdlvs-dtltk1}
\end{equation}

For {\em orthonormal bases}
there is an even closer connection -- representable as scalar product -- between a matrix
defined by an $n$-by-$n$ square array and the representation in terms of the elements of the bases:
 by inserting
two resolutions of the identity
$\mathbb{1}_n = \sum_{i=1}^n
\vert {\bf f}_i\rangle \langle {\bf f}_i \vert$
(see Section~\ref{2016-m-ch-fdvsrotio} on page~\pageref{2016-m-ch-fdvsrotio}) before and after the
linear transformation $\textsf{\textbf{A}}$,
\index{resolution of the identity}
\begin{equation}
\textsf{\textbf{A}}  =
\mathbb{1}_n \textsf{\textbf{A}} \mathbb{1}_n  =
\sum_{i,j=1}^n
\vert {\bf f}_i\rangle \langle {\bf f}_i\vert \textsf{\textbf{A}}\vert {\bf f}_j\rangle \langle {\bf f}_j\vert  =
\sum_{i,j=1}^n \alpha_{ij}
\vert {\bf f}_i\rangle  \langle {\bf f}_j\vert,
\end{equation}
whereby  insertion of (\ref{2015-m-ch-fdlvs-dtlt}) yields
\begin{equation}
\begin{split}
 \langle {\bf f}_i \vert \textsf{\textbf{A}} \vert {\bf f}_j\rangle
= \langle {\bf f}_i \vert \textsf{\textbf{A}}   {\bf f}_j\rangle \\
= \langle {\bf f}_i \vert \left( \sum_l \alpha_{lj}\vert {\bf f}_l \rangle \right)
=  \sum_l \alpha_{lj} \langle {\bf f}_i \vert {\bf f}_l \rangle
=  \sum_l \alpha_{lj} \delta_{il}  = \alpha_{ij} \\
\equiv
\begin{pmatrix}
\alpha_{11}&
\alpha_{12}&
\cdots    &
\alpha_{1n}\\
\alpha_{21}&
\alpha_{22}&
\cdots    &
\alpha_{2n}\\
\vdots&
\vdots&
\cdots    &
\vdots\\
\alpha_{n1}&
\alpha_{n2}&
\cdots   &
\alpha_{nn}
\end{pmatrix}
.
\end{split}
\end{equation}

\section{Change of basis}
\label{2012-m-ch-fdlvs-changeofbasis}
\index{change of basis}
\marginnote{For proofs and additional information see {\S}46 in~\bibentry{halmos-vs}.}
\index{basis change}
\index{change of basis}

Let ${\frak V}$ be an $n$-dimensional vector space and let
${\frak X}
=
\{
{\bf e}_1,
\ldots ,
{\bf e}_n
\}$
and
${\frak Y}
=  \{
{\bf f}_1,
\ldots ,
{\bf f}_n
\}$ be two bases of ${\frak V}$.

Take an arbitrary vector ${\bf z}\in {\frak V}$.
In terms of the two bases
${\frak X}$ and
${\frak Y}$,
${\bf z}$ can be written as
\begin{equation}
{\bf z}=
\sum_{i=1}^n x_i{\bf e}_i
=
\sum_{i=1}^n  y_i{\bf f}_i,
\label{2011-m-btbexy}
\end{equation}
where $x_i$ and $y_i$ stand for the coordinates of the vector  ${\bf z}$
with respect to the bases ${\frak X}$ and
${\frak Y}$,
respectively.

The following questions arise:
\begin{itemize}
\item[(i)]
What is the relation between the ``corresponding'' basis vectors ${\bf e}_i$ and ${\bf f}_j$?
\item[(ii)]
What is the relation between the coordinates $x_i$ (with respect to the basis  ${\frak X}$) and $y_j$ (with respect to the basis  ${\frak Y}$)  of the vector ${\bf z}$ in Equation~(\ref{2011-m-btbexy})?
\item[(iii)]
Suppose one fixes
an $n$-tuple $v = \begin{pmatrix} v_1, v_2, \ldots , v_n \end{pmatrix}$.
What is the relation between
${\bf v} =
\sum_{i=1}^n v_i {\bf e}_i
$
and
${\bf w}=
\sum_{i=1}^n v_i {\bf f}_i
$?
\end{itemize}

\subsection{Settlement of change of basis vectors by definition}

Basis changes can be perceived as linear transformations.
Therefore all earlier considerations \index{linear transformation}
of the previous Section~\ref{2019-mm-ch-fdvs-lt} can also be applied to basis changes.

As an {\it Ansatz} for answering question (i), recall that, just like any other vector in ${\frak V}$,
the new basis vectors ${\bf f}_i$ contained in the new basis ${\frak Y}$
can be (uniquely) written as a {\em linear combination}
(in quantum physics called {\em coherent superposition})
\index{coherent superposition}
\index{superposition}
\index{linear combination}
of the basis vectors
${\bf e}_i$ contained in the old  basis ${\frak X}$.
This can be defined {\it via}
a linear transformation $\textsf{\textbf{A}}$ between the corresponding vectors of the bases
 ${\frak X}$ and
${\frak Y}$ by
\begin{equation}
\begin{pmatrix}{\bf f}_1,\ldots ,{\bf f}_n\end{pmatrix}_i= \left[\begin{pmatrix}{\bf e}_1,\ldots ,{\bf e}_n\end{pmatrix} \cdot \textsf{\textbf{A}}\right]_i
,
\label{2011-m-btbe}
\end{equation}
where $i=1, \ldots , n$ is a column index.
More specifically, let ${a}_{ji}$ be the matrix of the linear transformation $\textsf{\textbf{A}}$
in the basis
${\frak X}
=
\{
{\bf e}_1,
\ldots ,
{\bf e}_n
\}$,
and let us rewrite (\ref{2011-m-btbe}) as a matrix equation
\begin{equation}
{\bf f}_i= \sum_{j=1}^n a_{ji} {\bf e}_j   = \sum_{j=1}^n (a^\intercal )_{ij} {\bf e}_j
.
\label{2011-m-btbe-r}
\end{equation}
If $\textsf{\textbf{A}}$ stands for the matrix whose components (with respect to ${\frak X}$) are  $a_{ji}$,
and $\textsf{\textbf{A}}^\intercal $
stands for the transpose of $\textsf{\textbf{A}}$
whose components (with respect to ${\frak X}$) are  $a_{ij}$,
then
\begin{equation}
\begin{pmatrix}
{\bf f}_1\\
{\bf f}_2\\
\vdots\\
{\bf f}_n
\end{pmatrix}
= \textsf{\textbf{A}}^\intercal
\begin{pmatrix}
{\bf e}_1\\
{\bf e}_2\\
\vdots\\
{\bf e}_n
\end{pmatrix}
.
\label{2011-m-btbe-r1}
\end{equation}

That is, very explicitly,
\begin{equation}
\begin{split}
{\bf f}_1=
\left[\begin{pmatrix}{\bf e}_1,\ldots ,{\bf e}_n\end{pmatrix} \cdot \textsf{\textbf{A}}\right]_1     = \sum_{i=1}^n a_{i1} {\bf e}_i
=
a_{11} {\bf e}_1 + a_{21} {\bf e}_2+\cdots + a_{n1} {\bf e}_n,\\
{\bf f}_2=
\left[\begin{pmatrix}{\bf e}_1,\ldots ,{\bf e}_n\end{pmatrix} \cdot \textsf{\textbf{A}}\right]_2=
 \sum_{i=1}^n a_{i2} {\bf e}_i
=
a_{12} {\bf e}_1 + a_{22} {\bf e}_2+\cdots + a_{n2} {\bf e}_n,\\
 \vdots \qquad \qquad \\
{\bf f}_n=
\left[\begin{pmatrix}{\bf e}_1,\ldots ,{\bf e}_n\end{pmatrix} \cdot \textsf{\textbf{A}}\right]_n      = \sum_{i=1}^n a_{in} {\bf e}_i
= a_{1v} {\bf e}_1 + a_{2n} {\bf e}_2+\cdots + a_{nn} {\bf e}_n .
\end{split}
\label{2011-m-btbe-eigen}
\end{equation}


This {\it Ansatz} includes a convention; namely the {\em order of the indices} of the transformation matrix.
You may have wondered why we have taken the inconvenience of defining
${\bf f}_i$ by $\sum_{j=1}^n a_{ji} {\bf e}_j$ rather than by $\sum_{j=1}^n a_{ij} {\bf e}_j$.
That is, in  Equation~(\ref{2011-m-btbe-r}), why not exchange $a_{ji}$ by $a_{ij}$,
so that the summation index $j$ is ``next to'' ${\bf e}_j$?
This is because we want to transform the coordinates according to this ``more intuitive'' rule, and we cannot have both
at the same time.
More explicitly, suppose that we want to have
\begin{equation}
y_i =\sum_{j=1}^n b_{ij} x_j
,
\label{2016-m-btbe-2-implied}
\end{equation}
or, in operator notation and the coordinates as $n$-tuples,
\begin{equation}
y = \textsf{\textbf{B}} x
.
\label{2016-m-btbe-2-impliedm}
\end{equation}
Then, by insertion of  Eqs.~(\ref{2011-m-btbe-r}) and (\ref{2016-m-btbe-2-implied}) into (\ref{2011-m-btbexy})
we obtain
\marginnote{If, in contrast, we would have started with
\label{2016-m-ch-fdvs-oic}
${\bf f}_i =\sum_{j=1}^n a_{ij} {\bf e}_j$
and still pretended to define
$y_i =\sum_{j=1}^n b_{ij} x_j$,
then we would have ended up with
${\bf z}=
\sum_{i=1}^n x_i{\bf e}_i
= \sum_{i=1}^n \left( \sum_{j=1}^n b_{ij} x_j \right)   \left( \sum_{k=1}^n a_{ik} {\bf e}_k \right)
= \sum_{i,j,k=1}^n  a_{ik} b_{ij}  x_j {\bf e}_k  = \sum_{i=1}^n x_i{\bf e}_i
$
which, in order to represent
$\textsf{\textbf{B}}$   as the inverse of
$\textsf{\textbf{A}}$,  would have forced us to take the transpose of either
$\textsf{\textbf{B}}$ or
$\textsf{\textbf{A}}$ anyway.
}
\begin{equation}
{\bf z}=
\sum_{i=1}^n x_i{\bf e}_i
=
\sum_{i=1}^n  y_i{\bf f}_i
= \sum_{i=1}^n \left( \sum_{j=1}^n b_{ij} x_j \right)   \left( \sum_{k=1}^n a_{ki} {\bf e}_k \right)
= \sum_{i,j,k=1}^n  a_{ki} b_{ij}  x_j {\bf e}_k
,
\label{2016-m-btbexy}
\end{equation}
which, by comparison, can only be satisfied if  $\sum_{i=1}^n  a_{ki} b_{ij} = \delta_{kj}$.
Therefore, $\textsf{\textbf{A}}\textsf{\textbf{B}} = \mathbb{1}_n$ and
$\textsf{\textbf{B}}$   is the inverse of
$\textsf{\textbf{A}}$.
This is quite plausible since any scale basis change needs to be compensated by a reciprocal or inversely proportional
scale change of the coordinates.

\begin{itemize}
\item
Note   that the $n$ equalities (\ref {2011-m-btbe-eigen})
really represent $n^2$ linear equations for the $n^2$
unknowns $a_{ij}$, $1\le i,j\le n$, since every pair of basis vectors
$\{{\bf f}_i,{\bf e}_i\}$, $1\le i\le n$ has $n$ components or coefficients.

\item
If one knows how the basis vectors
$
\{
{\bf e}_1,
\ldots ,
{\bf e}_n
\}$ of ${\frak X}$    transform, then one knows (by linearity) how
all other vectors
${\bf v}=
\sum_{i=1}^n v_i{\bf e}_i
$
(represented in this basis) transform; namely
$\textsf{\textbf{A}}({\bf v})=
\sum_{i=1}^n v_i \left[\begin{pmatrix}{\bf e}_1,\ldots ,{\bf e}_n\end{pmatrix} \cdot \textsf{\textbf{A}}\right]_i
$.

\item
Finally note that, if  ${\frak X}$ is an orthonormal basis,
then the basis transformation has a diagonal form
\begin{equation}
\textsf{\textbf{A}} =   \sum_{i=1}^n  {\bf f}_i {\bf e}_i^\dagger
\equiv
\sum_{i=1}^n \vert {\bf f}_i \rangle \langle {\bf e}_i \vert
\label{2013-m-ch-fdvs-dftm}
\end{equation}
because all the off-diagonal components $a_{ij}$, $i\neq j$ of $\textsf{\textbf{A}}$
explicitly written down in Eqs.(\ref{2011-m-btbe-eigen}) vanish.
This can be easily checked by applying $\textsf{\textbf{A}}$ to the elements ${\bf e}_i $ of the basis ${\frak X}$.
See also Section
\ref{2012-m-ch-citoob} on page \pageref{2012-m-ch-citoob}
for a representation of unitary transformations in terms of basis changes.
In quantum mechanics, the temporal evolution is represented by nothing but a change of orthonormal bases in Hilbert space.
\end{itemize}

\subsection{Scale change of vector components by contra-variation}

Having settled question (i) by the {\it Ansatz}
(\ref{2011-m-btbe}),
we turn to question (ii) next.
Since
\begin{equation}
{\bf z} =
 \sum_{j=1}^n y_j {\bf f}_j=
 \sum_{j=1}^n  y_j \left[\begin{pmatrix}{\bf e}_1,\ldots ,{\bf e}_n\end{pmatrix} \cdot \textsf{\textbf{A}}\right]_j=
 \sum_{j=1}^n  y_j  \sum_{i=1}^n a_{ij} {\bf e}_i=
  \sum_{i=1}^n \left(\sum_{j=1}^n  a_{ij} y^j \right)   {\bf e}_i;
\end{equation}
we obtain by comparison of the coefficients in Equation~(\ref{2011-m-btbexy}),
\begin{equation}
x_i= \sum_{j=1}^n a_{ij} y_j.
\label{2012-m-ch-e-tl1}
\end{equation}
That is, in terms of the ``old'' coordinates $x^i$,
the ``new'' coordinates are
\begin{equation}
\begin{split}
\sum_{i=1}^n (a^{-1})_{ji} x_i= \sum_{i=1}^n (a^{-1})_{ji}  \sum_{k=1}^n  a_{ik} y_k \\
=  \sum_{k=1}^n \left[ \sum_{i=1}^n (a^{-1})_{ji}  a_{ik} \right] y_k
=   \sum_{k=1}^n \delta^j_k y_k
=  y_j
.
\label{2012-m-ch-e-tl2}
\end{split}
\end{equation}

If we prefer to represent the vector coordinates of
${\bf x}$ and ${\bf y}$ as $n$-tuples,
then Eqs.~(\ref{2012-m-ch-e-tl1})  and (\ref{2012-m-ch-e-tl2})
have an interpretation as matrix multiplication; that is,
\begin{equation}
{\bf x} =  \textsf{\textbf{A}}{\bf y}, {\text{ and }}
{\bf y} =  (\textsf{\textbf{A}}^{-1}){\bf x}
.
\label{2012-m-ch-e-tl3}
\end{equation}

Finally, let us answer question (iii)
-- the relation between ${\bf v}=\sum_{i=1}^n v_i {\bf e}_i$ and ${\bf w}=\sum_{i=1}^n v_i {\bf f}_i$ for any $n$-tuple
$v = \begin{pmatrix} v_1, v_2, \ldots , v_n \end{pmatrix}$ --
by substituting the transformation~(\ref{2011-m-btbe-r}) of the basis vectors in ${\bf w}$ and comparing it with ${\bf v}$; that is,
\begin{equation}
{\bf w}= \sum_{j=1}^n v_j {\bf f}_j  = \sum_{j=1}^n v_j \left( \sum_{i=1}^n a_{ij} {\bf e}_i \right)= \sum_{i=1}^n \left( \sum_{j=1}^n a_{ij}  v_j \right)  x_i
\textrm{;  or } {\bf w} = \textsf{\textbf{A}} {\bf v}.
\end{equation}

\begin{marginfigure}%
\begin{center}%
\unitlength 0.3mm 
\linethickness{0.4pt}
\ifx\plotpoint\undefined\newsavebox{\plotpoint}\fi 
\begin{picture}(200,108.5)(30,0)
\put(100,0){\vector(1,0){100}}
\put(100,0){\vector(0,1){100}}
\put(100,0){\color{orange}\vector(1,1){70.25}}
\put(100,0){\color{orange}\vector(-1,1){70.25}}
\put(199.25,6.5){\makebox(0,0)[cc]{${\bf e}_1 =(1,0)^\intercal $}}
\put(100,108.5){\makebox(0,0)[cc]{${\bf e}_2 =(0,1)^\intercal $}}
\put(50,82.5){\makebox(0,0)[cc]{\color{orange}${\bf f}_2 =\frac{1}{\sqrt{2}}(-1,1)^\intercal $}}
\put(160,82.5){\makebox(0,0)[cc]{\color{orange}${\bf f}_1 =\frac{1}{\sqrt{2}}(1,1)^\intercal $}}
{\color{orange}
\qbezier(139.25,1.25)(139.25,20.875)(132.25,29)
\qbezier(99,46)(81.25,45)(70.5,33)
\put(69.569,31.96){\vector(-1,-1){.07}}\multiput(70.609,33.149)(-.0335663,-.0383615){31}{\line(0,-1){.0383615}}
\put(131.407,30.027){\vector(-1,1){.07}}\multiput(132.597,28.69)(-.03303353,.03716272){36}{\line(0,1){.03716272}}
\put(145.232,16.798){\makebox(0,0)[lc]{$\varphi = \frac{\pi}{4}$}}
\put(79.082,55){\makebox(0,0)[cc]{$\varphi = \frac{\pi}{4}$}}
}
\end{picture}
\end{center}
\caption{Basis change by rotation of $\varphi = \frac{\pi}{4}$ around the origin.}
  \label{2012-m-basischange}
\end{marginfigure}

{
\color{blue}
\bexample
\begin{enumerate}

\item
For the sake of an example consider a change of basis in the plane ${\Bbb R}^2$ by rotation of an angle $\varphi = \frac{\pi}{4}$ around the origin,
depicted in Figure~\ref{2012-m-basischange}.
According to Equation~(\ref{2011-m-btbe}),
we have
\begin{equation}
\begin{split}
{\bf f}_1=   a_{11} {\bf e}_1 + a_{21} {\bf e}_2,\\
{\bf f}_2=   a_{12}{\bf e}_1 +  a_{22}{\bf e}_2
,
\end{split}
\end{equation}
which amounts to four linear equations in the four unknowns $a_{11}$, $a_{12}$,
$a_{21}$, and $a_{22}$.

By
inserting the basis vectors
$ {\bf e}_1$, ${\bf e}_2$, ${\bf f}_1$, and ${\bf f}_2$
one obtains for the rotation matrix with respect to the basis ${\frak X}$
\begin{equation}
\begin{split}
\frac{1}{\sqrt{2}}
\begin{pmatrix}
1\\
1
\end{pmatrix}
=
a_{11}
\begin{pmatrix}
1 \\
0
\end{pmatrix}
+
a_{21}
\begin{pmatrix}
0 \\
1
\end{pmatrix} ,
\\
\frac{1}{\sqrt{2}}
\begin{pmatrix}
-1\\
1
\end{pmatrix}
=
a_{12}
\begin{pmatrix}
1 \\
0
\end{pmatrix}
+
a_{22}
\begin{pmatrix}
0 \\
1
\end{pmatrix}
,
\end{split}
\end{equation}
the first pair of equations yielding
$a_{11}=a_{21}=\frac{1}{\sqrt{2}}$,
the second pair of equations yielding
$a_{12}=-\frac{1}{\sqrt{2}}$ and $a_{22}=\frac{1}{\sqrt{2}}$.
Thus,
\begin{equation}
 \textsf{\textbf{A}}=
\begin{pmatrix}
a_{11}&a_{12}\\
a_{12}&a_{22}
\end{pmatrix}
=
\frac{1}{\sqrt{2}}
\begin{pmatrix}
1&-1\\
1&1
\end{pmatrix}
.
\end{equation}

As both coordinate systems ${\frak X}
=
\{
{\bf e}_1,
{\bf e}_2
\}$
and
${\frak Y}
=  \{
{\bf f}_1,
{\bf f}_2
\}$ are orthogonal, we might have just computed
the diagonal form (\ref{2013-m-ch-fdvs-dftm})
\begin{equation}
\begin{split}
 \textsf{\textbf{A}}=
\frac{1}{\sqrt{2}}
\left[
\begin{pmatrix}
1\\
1
\end{pmatrix}
\begin{pmatrix}
1,0
\end{pmatrix}
+
\begin{pmatrix}
-1\\
1
\end{pmatrix}
\begin{pmatrix}
0,1
\end{pmatrix}
\right] \\
=
\frac{1}{\sqrt{2}}
\left[
\begin{pmatrix}
1(1,0)\\
1 (1,0)
\end{pmatrix}
+
\begin{pmatrix}
-1(0,1)\\
1(0,1)
\end{pmatrix}
\right] \\
=
\frac{1}{\sqrt{2}}
\left[
\begin{pmatrix}
1&0\\
1&0
\end{pmatrix}
+
\begin{pmatrix}
0&-1\\
0&1
\end{pmatrix}
\right]
=
\frac{1}{\sqrt{2}}
\begin{pmatrix}
1&-1\\
1&1
\end{pmatrix}
.
\end{split}
\end{equation}
Note, however that coordinates transform contra-variantly with
$\textsf{\textbf{A}}^{-1}$.

Likewise, the rotation matrix with respect to the basis ${\frak Y}$  is
\begin{equation}
\begin{split}
 \textsf{\textbf{A}}'=
\frac{1}{\sqrt{2}}
\left[
\begin{pmatrix}
1\\
0
\end{pmatrix}
\begin{pmatrix}
1,1
\end{pmatrix}
+
\begin{pmatrix}
0\\
1
\end{pmatrix}
\begin{pmatrix}
-1,1
\end{pmatrix}
\right]
=
\frac{1}{\sqrt{2}}
\begin{pmatrix}
1&1\\
-1&1
\end{pmatrix}
.
\end{split}
\end{equation}

\item
By a similar calculation, taking into account the definition for the sine and cosine functions,
one obtains the transformation matrix $\textsf{\textbf{A}}(\varphi )$
associated with an arbitrary angle $\varphi$,
\begin{equation}
 \textsf{\textbf{A}}
=
\begin{pmatrix}
\cos \varphi &-\sin \varphi\\
\sin \varphi &\cos \varphi
\end{pmatrix}
.
\label{2012-m-ch-fdvs-otd2}
\end{equation}
The coordinates transform as
\begin{equation}
 \textsf{\textbf{A}}^{-1}
=
\begin{pmatrix}
\cos \varphi &\sin \varphi\\
-\sin \varphi &\cos \varphi
\end{pmatrix}
.
\end{equation}

\item
Consider the more general rotation depicted in Figure~\ref{2012-m-basischange1}.
\begin{marginfigure}%
\begin{center}%
\unitlength 0.3mm 
\linethickness{0.4pt}
\ifx\plotpoint\undefined\newsavebox{\plotpoint}\fi 
\begin{picture}(110,108.5)(0,0)
\put(10,0){\vector(1,0){100}}
\put(10,0){\vector(0,1){100}}
\put(109.25,6.5){\makebox(0,0)[cc]{${\bf e}_1 =(1,0)^\intercal $}}
\put(10,108.5){\makebox(0,0)[cc]{${\bf e}_2 =(0,1)^\intercal $}}
\put(61,95){\color{orange}\makebox(0,0)[cc]{${\bf f}_2 =\frac{1}{2}(1,\sqrt{3})^\intercal $}}
\put(96.75,56){\color{orange}\makebox(0,0)[cc]{${\bf f}_1 =\frac{1}{2}(\sqrt{3},1)^\intercal $}}
\put(25,55){\color{orange}\makebox(0,0)[cc]{$\varphi = \frac{\pi}{6}$}}
\put(65,11){\color{orange}\makebox(0,0)[cc]{$\varphi = \frac{\pi}{6}$}}
{\color{orange}
\put(96.75,48.375){\vector(2,1){.07}}\multiput(10,0)(.06049486066,.03373416581){1434}{\line(1,0){.06049486066}}
\put(58.375,86.75){\vector(1,2){.07}}\multiput(10,0)(.03373416581,.06049486066){1434}{\line(0,1){.06049486066}}
\put(46.875,20.375){\vector(-1,2){.07}}\qbezier(49.375,.375)(49.375,15.875)(46.875,20.375)
\put(31.375,39.375){\vector(2,-1){.07}}\qbezier(10.5,46.125)(20.562,45.5)(31.375,39.375)
}
\end{picture}
\end{center}
\caption{More general basis change by rotation.}
  \label{2012-m-basischange1}
\end{marginfigure}
Again, by inserting the basis vectors
$ {\bf e}_1,{\bf e}_2, {\bf f}_1$, and ${\bf f}_2$,
one obtains
\begin{equation}
\begin{split}
\frac{1}{{2}}
\begin{pmatrix}
\sqrt{3}\\ 1
\end{pmatrix}
=
a_{11}
\begin{pmatrix}
1\\0
\end{pmatrix}
+
a_{21}
\begin{pmatrix}
0\\1
\end{pmatrix} ,
\\
\frac{1}{{2}}
\begin{pmatrix}
1\\\sqrt{3}
\end{pmatrix}
=
a_{12}
\begin{pmatrix}
1\\0
\end{pmatrix}
+
a_{22}
\begin{pmatrix}
0\\1
\end{pmatrix}
,
\end{split}
\end{equation}
yielding
$a_{11}=a_{22}=\frac{\sqrt{3}}{2}$,
the second pair of equations yielding
$a_{12}= a_{21}=\frac{1}{{2}}$.
Thus,
\begin{equation}
 \textsf{\textbf{A}}=
\begin{pmatrix}
a&b\\
b&a
\end{pmatrix}
=
\frac{1}{{2}}
\begin{pmatrix}
\sqrt{3}&1\\
1&\sqrt{3}
\end{pmatrix}
.
\end{equation}
The coordinates transform according to the inverse transformation, which in this case can be represented by
\begin{equation}
 \textsf{\textbf{A}}^{-1}=
\frac{1}{{a^2-b^2}}
\begin{pmatrix}
a &-b\\
-b&a
\end{pmatrix}
=
\begin{pmatrix}
\sqrt{3}&-1\\
-1&\sqrt{3}
\end{pmatrix}
.
\end{equation}

\end{enumerate}

\eexample
}

\section{Mutually unbiased bases}
\index{mutually unbiased bases}

Two  orthonormal bases
${\frak B} =\{
{\bf e}_1,
\ldots ,
{\bf e}_n
\}$
and
${\frak B}'=\{
{\bf f}_1,
\ldots ,
{\bf f}_n
\}$
are said to be {\em mutually unbiased}
if
their scalar or inner products are
\begin{equation}
\vert \langle {\bf e}_i\vert {\bf f}_j  \rangle \vert^2
=
\frac{1}{n}
\end{equation}
for all $1\le i,j\le n$.
Note without proof -- that is, you do not have to be concerned
that you need to understand  this from what has been said so far --
that ``the elements of two or more mutually unbiased bases are mutually maximally apart.''

{\color{Purple}
In physics, one seeks maximal sets of orthogonal bases who
are maximally apart.\cite{WooFie,durt}
Such maximal sets of bases are used in quantum information theory
to assure the maximal performance of certain protocols
used in quantum cryptography, or for the production of
quantum random sequences by beam splitters.
They are essential for the practical exploitations of quantum complementary properties
and resources.
}

Schwinger presented an algorithm (see Ref.\cite{Schwinger.60} for a proof)
to construct a new mutually unbiased basis ${\frak B}$   from an existing orthogonal one.
The proof idea
is to create a new basis ``inbetween'' the old basis vectors.
by the following construction steps:
\begin{itemize}
\item[(i)]
take the existing orthogonal basis and permute all of its elements by ``shift-permuting'' its elements; that is, by
changing
the basis vectors according to their enumeration $i \rightarrow i+1$ for $i=1,\ldots , n-1$, and $n \rightarrow 1$;
or any other nontrivial (i.e., do not consider identity for any basis element) permutation;
\item[(ii)]
consider the {\em (unitary) transformation} (cf. Sections \ref{2012-m-ch-fdlvs-changeofbasis} and \ref{2012-m-ch-citoob})
corresponding to the basis change from the old basis to the new, ``permutated'' basis;
\item[(iii)]
finally, consider the (orthonormal) {\em eigenvectors} \index{eigenvector}
of this (unitary; cf. page
\pageref{2014-m-ch-fdvs-unitary}) transformation associated with the basis change.
These eigenvectors are the elements of a new basis  ${\frak B}'$.
Together with ${\frak B}$ these two bases
-- that is, ${\frak B}$ and ${\frak B}'$ --  are mutually unbiased.
\end{itemize}

{\color{blue}
\bexample
Consider, for example,
\marginnote{For a {\em Mathematica(R)} program,
see
\url{http://tph.tuwien.ac.at/~svozil/publ/2012-schwinger.m}}
the real plane ${\Bbb R}^2$,
and the basis
$${\frak B}=\{ {\bf e}_1 , {\bf e}_2\} \equiv \{ \vert {\bf e}_1 \rangle , \vert {\bf e}_2\rangle \}
\equiv
\left\{\begin{pmatrix}1\\0\end{pmatrix},\begin{pmatrix}0\\1\end{pmatrix}\right\}.$$
The shift-permutation [step (i)] brings ${\frak B}$ to a new, ``shift-permuted'' basis  ${\frak S}$; that is,
$$\{ {\bf e}_1 , {\bf e}_2\} \mapsto {\frak S}= \{ {\bf f}_1={\bf e}_2,{\bf f}_1={\bf e}_1 \} \equiv \left\{\begin{pmatrix}0\\1\end{pmatrix},\begin{pmatrix}1\\0\end{pmatrix}\right\}.$$
The (unitary) basis transformation [step (ii)] between $ {\frak B} $ and
$  {\frak S}$ can be constructed by a diagonal sum
\begin{equation}
\begin{split}
\textsf{\textbf{U}} ={\bf f}_1  {\bf e}_1^\dagger+ {\bf f}_2  {\bf e}_2^\dagger = {\bf e}_2   {\bf e}_1^\dagger+ {\bf e}_1  {\bf e}_2^\dagger   \\
\equiv \vert {\bf f}_1\rangle \langle {\bf  e}_1\vert + \vert {\bf f}_2\rangle \langle   {\bf e}_2 \vert = \vert {\bf e}_2\rangle \langle    {\bf e}_1\vert + \vert {\bf e}_1\rangle \langle   {\bf e}_2    \vert         \\
\equiv
\begin{pmatrix}
0 \\
1
\end{pmatrix}
(1,0)+
\begin{pmatrix}
1 \\
0
\end{pmatrix}
(0,1)  \\
\qquad \equiv
\begin{pmatrix}
0 (1,0)\\
1 (1,0)
\end{pmatrix}
+
\begin{pmatrix}
1 (0,1)\\
0 (0,1)
\end{pmatrix}
   \\
\equiv
\begin{pmatrix}
0&0\\1&0
\end{pmatrix}
+
\begin{pmatrix}
0&1\\0&0
\end{pmatrix}
=
\begin{pmatrix}
0&1\\1&0
\end{pmatrix}.
\end{split}
\end{equation}
The set of eigenvectors [step (iii)] of this  (unitary) basis transformation $\textsf{\textbf{U}}$ forms a new basis
\begin{equation}
\begin{split}
{\frak B}' =
\{ \frac{1}{\sqrt{2}} ({\bf f}_1 - {\bf e}_1),
 \frac{1}{\sqrt{2}}( {\bf f}_2 + {\bf e}_2 )\}
\\
 \qquad =
\{ \frac{1}{\sqrt{2}} (\vert {\bf f}_1\rangle  - \vert {\bf e}_1\rangle ),
 \frac{1}{\sqrt{2}}(\vert  {\bf f}_2 \rangle + \vert {\bf e}_2\rangle  )\}
\\
 \qquad  =
\{ \frac{1}{\sqrt{2}} (\vert {\bf e}_2\rangle  - \vert {\bf e}_1\rangle ),
 \frac{1}{\sqrt{2}}(\vert  {\bf e}_1 \rangle + \vert {\bf e}_2\rangle  )\}
\\
 \qquad \equiv \left\{ \frac{1}{\sqrt{2}}\left\{\begin{pmatrix}-1\\1\end{pmatrix},\frac{1}{\sqrt{2}}\begin{pmatrix}1\\1\end{pmatrix}\right\}\right\}.
\end{split}
\end{equation}
For a proof of mutually unbiasedness, just form the four inner products of one vector in ${\frak B}$ times one vector in ${\frak B}'$,
respectively.

In three-dimensional complex vector space ${\Bbb C}^3$, a similar construction
from the Cartesian standard basis
${\frak B}=\{ {\bf e}_1 , {\bf e}_2, {\bf e}_3\} \equiv
\{\begin{pmatrix}1,0,0\end{pmatrix}^\intercal ,\begin{pmatrix}0,1,0\end{pmatrix}^\intercal ,\begin{pmatrix}0,0,1\end{pmatrix}^\intercal \}$
yields
\begin{equation}
\begin{split}
 {\frak B}' \equiv   \frac{1}{\sqrt{3}}  \left\{
 \begin{pmatrix}1\\1\\1\end{pmatrix},
 \begin{pmatrix}\frac{1}{2} \left[{\sqrt{3}} i-1\right] \\ \frac{1}{2} \left[-{\sqrt{3}} i-1\right] \\
  1\end{pmatrix},
\begin{pmatrix}
 \frac{1}{ 2} \left[-{\sqrt{3}} i-1\right]\\ \frac{1}{ 2} \left[{\sqrt{3}} i-1\right]\\1  \end{pmatrix}
   \right\} .
\end{split}
\end{equation}
\eexample
}

So far, nobody  has discovered a systematic way to derive and construct a {\em complete} or {\em maximal}
set of mutually unbiased bases in arbitrary dimensions; in particular,
{\em how many} bases are there in such sets.

\section{Completeness or resolution of the identity operator in terms of base vectors}
\label{2016-m-ch-fdvsrotio}
\index{resolution of the identity}
\index{completeness}

The identity $\mathbb{1}_n$ in an $n$-dimensional vector space ${\frak V}$ can be represented in terms of the sum
over all outer (by another naming tensor or dyadic) products
of all vectors of an arbitrary orthonormal basis
\index{outer product}
\index{dyadic product}
\index{tensor product}
${\frak B} =\{
{\bf e}_1,
\ldots ,
{\bf e}_n
\}
\equiv
\{
\vert {\bf e}_1 \rangle ,
\ldots ,
\vert{\bf e}_n \rangle
\}
$; that is,
\begin{equation}
 \mathbb{1}_n = \sum_{i=1}^n \vert {\bf e}_i \rangle \langle {\bf e}_i \vert
\equiv  \sum_{i=1}^n {\bf e}_i  {\bf e}_i^\dagger  .
\label{2016-m-ch-fdlws-roi}
\end{equation}
This is sometimes also referred to as {\em completeness}.
\index{resolution of the identity}
\index{completeness}

{\color{OliveGreen}
\bproof

For a proof,  consider an arbitrary vector $\vert {\bf x} \rangle  \in {\frak V}$.
Then,
\begin{equation}
\begin{split}
 \mathbb{1}_n \vert {\bf x} \rangle
 =
\left(\sum_{i=1}^n \vert {\bf e}_i \rangle \langle {\bf e}_i \vert \right)
\vert {\bf x} \rangle
=
\left(\sum_{i=1}^n \vert {\bf e}_i \rangle \langle {\bf e}_i \vert \right)
\left(\sum_{j=1}^n x_j \vert {\bf e}_j  \rangle \right)\\
=
\sum_{i,j=1}^n x_j  \vert  {\bf e}_i \rangle \langle {\bf e}_i \vert {\bf e}_j  \rangle
=
\sum_{i,j=1}^n x_j  \vert {\bf e}_i \rangle \delta_{ij}
=
\sum_{i=1}^n x_i       \vert {\bf e}_i \rangle
=  \vert {\bf x} \rangle
.
\end{split}
\end{equation}
\eproof
}

{\color{blue}
\bexample
Consider, for example, the basis
${\frak B}=\{ \vert {\bf e}_1 \rangle , \vert {\bf e}_2 \rangle \} \equiv \{(1,0)^\intercal ,(0,1)^\intercal \}$.
Then the two-dimensional resolution of the identity operator $\mathbb{1}_2$
can be written as
\begin{equation}
\begin{split}
\mathbb{1}_2 =   \vert {\bf e}_1 \rangle \langle  {\bf e}_1 \vert   +     \vert {\bf e}_2 \rangle  \langle  {\bf e}_2 \vert \\
=   (1,0)^\intercal  (1,0) +   (0,1)^\intercal  (0,1)
=
\begin{pmatrix}
1 (1,0) \\  0 (1,0)
\end{pmatrix}
 +
\begin{pmatrix}
0 (0,1)\\
1 (0,1)
\end{pmatrix} \\
 =
\begin{pmatrix}
1 & 0 \\
0 & 0
\end{pmatrix}
+
\begin{pmatrix}
0 & 0 \\
0 & 1
\end{pmatrix}
=
\begin{pmatrix}
1 & 0 \\
0 & 1
\end{pmatrix}
.
\end{split}
\end{equation}

Consider, for another example, the basis
${\frak B}' \equiv \{\frac{1}{\sqrt{2}}(-1,1)^\intercal ,\frac{1}{\sqrt{2}}(1,1)^\intercal \}$.
Then the two-dimensional resolution of the identity operator $\mathbb{1}_2$
can be written as
\begin{equation}
\begin{split}
\mathbb{1}_2 =  \frac{1}{\sqrt{2}}   (-1,1)^\intercal   \frac{1}{\sqrt{2}}(-1,1) +   \frac{1}{\sqrt{2}}(1,1)^\intercal  \frac{1}{\sqrt{2}}(1,1)
\\
=
\frac{1}{{2}}
\begin{pmatrix}
-1 (-1,1) \\  1 (-1,1)
\end{pmatrix}
 +
\frac{1}{{2}}
\begin{pmatrix}
1 (1,1)\\
1 (1,1)
\end{pmatrix}
 =
\frac{1}{{2}}
\begin{pmatrix}
1 & -1 \\
-1 & 1
\end{pmatrix}
+
\frac{1}{{2}}
\begin{pmatrix}
1 & 1 \\
1 & 1
\end{pmatrix}
=
\begin{pmatrix}
1 & 0 \\
0 & 1
\end{pmatrix}
.
\end{split}
\end{equation}

\eexample
}

\section{Rank}
\label{2014-m-fdvs-rank}
\index{rank of matrix}
\index{matrix rank}
\index{rank}

The (column or row) {\em rank}, $\rho (  \textsf{\textbf{A}}  )$,
\index{column rank of matrix}
\index{row rank of matrix}
or $\textrm{rk} ( \textsf{\textbf{A}} )$,
of a linear transformation $ \textsf{\textbf{A}} $
in an $n$-dimensional vector space ${\frak V}$
is the maximum number of linearly independent (column or, equivalently,
row) vectors of the associated
$n$-by-$n$ square matrix $ A $, represented by its entries  $a_{ij}$.

This definition can be generalized to arbitrary
$m$-by-$n$ matrices $A$, represented by its entries  $a_{ij}$.
Then, the row and column ranks of $A$ are identical; that is,
\begin{equation}
\textrm{row rk} (A) =
\textrm{column rk} (A)  =
\textrm{rk} (A).
\end{equation}

{\color{OliveGreen}
\bproof

For a proof,  consider Mackiw's argument.\cite{Mackiw-1995}
First we show that $\textrm{row rk} (A)\le \textrm{column rk} (A)$ for any real
(a generalization to complex vector space requires some adjustments)
$m$-by-$n$ matrix $A$. Let the vectors
$\{{\bf e}_1,{\bf e}_2, \ldots ,{\bf e}_r\}$ with ${\bf e}_i \in {\Bbb R}^n$, $1\le i\le r$,
be a basis spanning the
{\em row space} of
\index{row space} $A$; that is, all vectors
that can be obtained by a linear combination of the $m$ row vectors
$$
\begin{pmatrix}
(a_{11},a_{12},\ldots ,a_{1n})\\
(a_{21},a_{22},\ldots ,a_{2n})\\
\vdots                    \\
(a_{m1},a_{n2},\ldots ,a_{mn})
\end{pmatrix}
$$
of $A$ can also be obtained as a linear combination of ${\bf e}_1,{\bf e}_2, \ldots ,{\bf e}_r$.
Note that $r\le m$.

Now form the {\em column vectors} $A{\bf e}_i^\intercal $ for $1\le i\le r$, that is,
$A{\bf e}_1^\intercal ,A{\bf e}_2^\intercal , \ldots ,A{\bf e}_r^\intercal $ {\em via} the usual rules of matrix multiplication.
Let us prove that these resulting column vectors $A{\bf e}_i^\intercal $ are linearly independent.

Suppose they were not (proof by contradiction).
Then, for some scalars
$c_1,c_2, \ldots ,c_r \in {\Bbb R}$,
$$
c_1 A{\bf e}_1^\intercal + c_2 A{\bf e}_2^\intercal + \ldots + c_r A{\bf e}_r^\intercal
=
 A\left(c_1{\bf e}_1^\intercal + c_2  {\bf e}_2^\intercal + \ldots + c_r  {\bf e}_r^\intercal  \right)=0
$$
without all $c_i$'s vanishing.

That is,
$
{\bf v} =
c_1{\bf e}_1^\intercal + c_2  {\bf e}_2^\intercal + \ldots + c_r  {\bf e}_r^\intercal
$, must be in the {\em null space}
\index{null space}
of $A$ defined by all vectors ${\bf x}$ with $A{\bf x}={\bf 0}$,
and $A({\bf v})={\bf 0}$ .
(In this case the inner (Euclidean) product of ${\bf x}$ with all the rows of $A$ must vanish.)
But since the ${\bf e}_i$'s form also a basis of the row vectors,
$
{\bf v}^\intercal
$
is also some vector in the row space of $A$.
The linear independence of the basis elements
${\bf e}_1,{\bf e}_2, \ldots ,{\bf e}_r$   of the row  space of $A$
guarantees that
all the coefficients $c_i$ have to vanish; that is,
$c_1=c_2= \cdots =c_r =0$.

At the same time,
as
for every vector ${\bf x}\in {\Bbb R}^n$,
$A{\bf x}$ is a linear combination of the column vectors
$$
\begin{pmatrix}
\begin{pmatrix}
a_{11}\\
a_{21}\\
\vdots \\
a_{m1}
\end{pmatrix},
&
\begin{pmatrix}
a_{12}\\
a_{22}\\
\vdots \\
a_{m2}
\end{pmatrix},
&
\cdots  ,
&
\begin{pmatrix}
a_{1n}\\
a_{2n}\\
\vdots \\
a_{mn}
\end{pmatrix}
\end{pmatrix}
,
$$
the $r$ linear independent vectors
$A{\bf e}_1^\intercal ,A{\bf e}_2^\intercal , \ldots ,A{\bf e}_r^\intercal $
are all linear combinations of the column vectors of $A$.
Thus, they are in the column space of $A$.
\index{column space}
Hence, $r\le \textrm{column rk}(A)$.
And, as $r= \textrm{row rk}(A)$,
we obtain
$ \textrm{row rk}(A)\le \textrm{column rk}(A)$.

By considering the transposed matrix $A^\intercal $, and by an analogous argument   we obtain
that
$ \textrm{row rk}(A^\intercal )\le \textrm{column rk}(A^\intercal )$.
But
$ \textrm{row rk}(A^\intercal )= \textrm{column rk}(A)$
and
$ \textrm{column rk}(A^\intercal )= \textrm{row rk}(A)$,
and thus
$ \textrm{row rk}(A^\intercal )= \textrm{column rk}(A)\le \textrm{column rk}(A^\intercal )= \textrm{row rk}(A)$.
Finally, by considering both estimates
$ \textrm{row rk}(A)\le \textrm{column rk}(A)$
as well as
$\textrm{column rk}(A)\le \textrm{row rk}(A)$,
we obtain that
$ \textrm{row rk}(A) = \textrm{column rk}(A)$.
\eproof
}

\section{Determinant}
\index{determinant}

\subsection{Definition}

In what follows, the {\em determinant} of a matrix $A$ will be denoted by $\textrm{det} A$ or,
equivalently, by $\vert A \vert$.

Suppose $A=a_{ij}$ is the  $n$-by-$n$ square matrix representation of
a linear transformation $\textsf{\textbf{A}}$
in an $n$-dimensional vector space ${\frak V}$.
We shall define its {\em determinant}
in two equivalent ways.

The
{\em Leibniz formula}
\index{Leibniz formula} defines the determinant of the $n$-by-$n$ square matrix  $A=a_{ij}$ by
\begin{equation}
\textrm{det}A
=\sum_{\sigma \in S_n} \textrm{sgn}(\sigma) \prod_{i=1}^n a_{\sigma(i),j} ,
\end{equation}
where ``sgn'' represents the {\em sign function}
\index{sign function}
of permutations $\sigma$ in the permutation group $S_n$
on $n$ elements $\{1,2, \ldots , n\}$,
which returns $-1$ and $+1$ for odd and even permutations,
respectively.
$\sigma (i)$ stands for the element in position $i$ of $\{1,2, \ldots , n\}$ {\em after} permutation $\sigma$.

An equivalent (no proof is given here) definition
\begin{equation}
\textrm{det}A
=\varepsilon_{i_1 i_2\cdots i_n} a_{1i_1}a_{2i_2} \cdots a_{ni_n},
\end{equation}
makes use of the  totally antisymmetric Levi-Civita symbol  (\ref{2014-m-ch-lcs}) on page \pageref{2014-m-ch-lcs},
\index{Levi-Civita symbol}
\index{antisymmetric tensor} and makes use of the
 Einstein summation convention.
\index{Einstein summation convention}

The second,
{\em Laplace formula}
\index{Laplace formula}
definition of the determinant
is recursive and expands the determinant in cofactors.
It is also called
{\em Laplace expansion},
\index{Laplace expansion}
or
{\em cofactor expansion}
\index{cofactor expansion}.
First,
a {\em minor}
\index{minor}
$M_{ij}$ of an  $n$-by-$n$ square matrix  $A$ is
defined to be the determinant of the
$(n-1)\times (n-1)$ submatrix
that remains after the entire $i$th row and $j$th column have been deleted from $A$.

A {\em cofactor}
\index{cofactor}
$A_{ij}$
of an $n$-by-$n$ square matrix  $A$
is defined in terms of its associated minor by
\begin{equation}
A_{ij}=(-1)^{i+j}M_{ij}.
\end{equation}

The {\em determinant} of a square matrix $A$, denoted by
$\textrm{det} A$ or $\vert A\vert$, is a scalar recursively defined by
\begin{equation}
\textrm{det}A
=\sum_{j=1}^n a_{ij}A_{ij}
=\sum_{i=1}^n a_{ij}A_{ij}
\end{equation}
for any $i$ (row expansion) or $j$ (column expansion), with $i,j=1,\ldots ,n$.
For $1\times 1$ matrices (i.e., scalars), $\textrm{det}A =a_{11}$.

\subsection{Properties}

The following properties of determinants are mentioned (almost) without proof:

\begin{itemize}
\item[(i)]
If $A$ and $B$ are square matrices of the same order, then
$\textrm{det}AB = (\textrm{det}A)  (\textrm{det}B)$.

\item[(ii)]
If either two rows or two columns are exchanged, then the determinant is multiplied
by a factor ``$-1$.''

\item[(iii)]
The determinant of the transposed matrix is equal to the determinant of the original matrix; that is,
$\textrm{det}(A^\intercal ) = \textrm{det}A $ .

\item[(iv)]
The determinant $\textrm{det}A $ of a matrix $A$ is nonzero if and only if $A$ is invertible.
In particular, if $A$ is not invertible, $\textrm{det}A =0$.
If $A$ has an inverse matrix $A^{-1}$, then $\textrm{det}(A^{-1}) = (\textrm{det}A)^{-1} $.

This is a very important property which we shall use in Equation~(\ref{2014-m-eve-ce}) on page \pageref{2014-m-eve-ce}
for the determination of nontrivial
eigenvalues $\lambda$ (including the associated eigenvectors)
\index{eigenvalue}
\index{eigenvector}
\index{eigensystrem}
\index{characteristic equation}
\index{secular determinant}
\index{secular equation}
of a matrix $A$ by solving the secular equation $\textrm{det} (A-\lambda \mathbb{1})=0 $.

\item[(v)]
Multiplication of any row or column with a factor $\alpha$  results in a determinant
which is $\alpha$ times the original determinant.
Consequently,
multiplication of an $n \times n$ matrix with a scalar $\alpha$ results
in a   determinant
which is $\alpha^n$ times the original determinant.

\item[(vi)]
The determinant of an identity matrix is one; that is,
$\textrm{det} \, \mathbb{1}_n =1$.
Likewise, the determinant of a diagonal matrix is just the product of the diagonal entries;
that is,
$\textrm{det} [\textrm{diag}(\lambda_1,\ldots, \lambda_n)] = \lambda_1 \cdots \lambda_n$.

\item[(vii)]
The determinant is not changed if a multiple of an existing row is added to another row.

{\color{OliveGreen}
\bproof
This can be easily demonstrated by considering the Leibniz formula: suppose a multiple $\alpha$
of the $j$'th column is added to the $k$'th column since
\begin{equation}
\begin{split}
\varepsilon_{i_1 i_2\cdots i_j \cdots i_k \cdots i_n} a_{1i_1}a_{2i_2} \cdots a_{j i_j} \cdots (a_{k i_k} + \alpha a_{j i_k}) \cdots a_{ni_n}\\
=
\varepsilon_{i_1 i_2\cdots i_j \cdots i_k \cdots i_n} a_{1i_1}a_{2i_2} \cdots a_{j i_j} \cdots  a_{k i_k} \cdots a_{ni_n} \\
+\alpha \varepsilon_{i_1 i_2\cdots i_j \cdots i_k \cdots i_n} a_{1i_1}a_{2i_2} \cdots a_{j i_j} \cdots   a_{j i_k} \cdots a_{ni_n}.
\end{split}
\end{equation}
The second summation term vanishes, since
$
a_{j i_j} a_{j i_k}
=
a_{j i_k} a_{j i_j}
$
is totally symmetric in the indices $i_j$ and $i_k$,
and the Levi-Civita symbol $\varepsilon_{i_1 i_2\cdots i_j \cdots i_k \cdots i_n}$.
\eproof
}

\item[(viii)]
The absolute value of the determinant of a square matrix $A= \left({\bf e}_1, \ldots {\bf e}_n \right)$ formed by (not necessarily orthogonal)
row (or column) vectors of a basis
$\frak B= \{ {\bf e}_1, \ldots {\bf e}_n\}$
is equal to the {\em volume} of the parallelepiped
\index{volume}
$
\left\{ {\bf x} \mid {\bf x} =\sum_{i=1}^n t_i {\bf e}_i, \; 0 \le t_i \le 1, \; 0\le i \le n \right\}
$
formed by those vectors.

{\color{OliveGreen}
\bproof
This can be demonstrated \marginnote{See, for instance, Section~4.3 of~\bibentry{Strang:2009:ILA}
and \bibentry{Sanderson-3Blue1Brown-LA6}.}
by supposing that
the square matrix $A$ consists of all the $n$ row (column) vectors of an orthogonal basis of dimension $n$.
Then
$AA^\intercal =A^\intercal A$ is a  diagonal matrix  which just contains the square of the
length of all the basis vectors forming a perpendicular parallelepiped which
is just an $n$ dimensional box.
Therefore the volume is just the positive square root of
$\textrm{det} ( AA^\intercal  ) =
(\textrm{det} A) (\textrm{det} A^\intercal  ) = (\textrm{det} A) (\textrm{det} A^\intercal  )=(\textrm{det} A)^2$.

For any nonorthogonal basis, all we need to employ is a Gram-Schmidt process
to obtain a (perpendicular) box of equal volume to the original parallelepiped
formed by the nonorthogonal basis vectors --
any volume that is cut is compensated by adding the same amount to the new volume.
Note that the Gram-Schmidt process operates by adding (subtracting) the projections
of already existing orthogonalized vectors
from the old basis vectors (to render these sums orthogonal to the existing vectors of the new orthogonal basis);
a process which does not change the determinant.
\index{Gram-Schmidt process}
\eproof
}

This result can be used for changing the differential volume element in integrals {\it via} the Jacobian matrix
$J
$
\index{Jacobian matrix}
(\ref{2013-m-t-jm}), as
\begin{equation}
\begin{split}
dx_1'\, dx_2' \cdots dx_n'
= \vert \textrm{det} J \vert dx_1\, dx_2 \cdots dx_n     \\
= \sqrt{\left[\text{det}\left(\frac{dx_i'}{dx_j}\right)\right]^2} dx_1\, dx_2 \cdots dx_n
.
\end{split}
\label{2018-mm-ch-fdvs-jacoinfvol}
\end{equation}
The result applies also for curvilinear coordinates; see Section~\ref{2018-mm-ch-ctensor-volumeclc}
on page~\pageref{2018-mm-ch-ctensor-volumeclc}.

\item[(ix)]
The {\em sign} of a  determinant of a matrix formed by the row (column)
 vectors of a basis indicates the {orientation}
of that basis.
\index{sign}
\index{orientation}

\end{itemize}

\section{Trace}
\label{2013-ch-fdvs-trace}
\index{trace}

\subsection{Definition}
The {\em trace} of an $n$-by-$n$ square matrix $A=a_{ij}$, denoted by
$\textrm{Tr} A$,  is a scalar
defined to be the sum of the elements on the main diagonal
 (the diagonal from the upper left to the lower right) of A; that is  (also in Dirac's bra and ket notation),
\begin{equation}
\textrm{Tr}\,A
= a_{11} +a_{22}+ \cdots +a_{nn}
=\sum_{i=1}^n a_{ii}
=  a_{ii}
\end{equation}

Traces are noninvertible (irreversible) almost by definition: for $n\ge 2$ and for arbitrary values $a_{ii} \in {\Bbb R}, {\Bbb C}$, there are
``many''  ways to obtain the same value of $ \sum_{i=1}^n a_{ii} $.

Traces are linear functionals, because, for two arbitrary matrices $A,B$
and two arbitrary scalars $\alpha, \beta$,
\begin{equation}
\textrm{Tr}\,(\alpha A + \beta B)
=\sum_{i=1}^n (\alpha a_{ii} + \beta b_{ii})
= \alpha \sum_{i=1}^n a_{ii} + \beta \sum_{i=1}^n  b_{ii}
=
\alpha \textrm{Tr}\,(A)+ \beta \textrm{Tr}\,(B)
.
\end{equation}

Traces can be realized {\it via} some arbitrary orthonormal basis ${\frak B} =\{
{\bf e}_1,
\ldots ,
{\bf e}_n
\}$
by ``sandwiching'' an operator $\textsf{\textbf{A}}$ between all basis elements -- thereby effectively taking the diagonal components
of    $\textsf{\textbf{A}}$ with respect to the basis ${\frak B}$ --
and summing over all these scalar components; that is, with definition (\ref{2015-m-ch-fdlvs-dtlt}),\marginnote{Note that antilinearity of the scalar product does not apply for the extraction of $\alpha_{li}$
here, as, strictly speaking, the Euclidean scalar products should be formed {\em after} summation.}
\begin{equation}
\begin{split}
\textrm{Tr}\;\textsf{\textbf{A}}
=\sum_{i=1}^n   \langle {\bf e}_i \vert \textsf{\textbf{A}} \vert {\bf e}_i \rangle
=\sum_{i=1}^n   \langle {\bf e}_i \vert \textsf{\textbf{A}}  {\bf e}_i \rangle  \\
=\sum_{i=1}^n  \sum_{l=1}^n  \langle {\bf e}_i \vert \left(\alpha_{li} \vert  {\bf e}_l \rangle \right)
=\sum_{i=1}^n  \sum_{l=1}^n  \alpha_{li} \langle {\bf e}_i \vert  {\bf e}_l \rangle \\
=\sum_{i=1}^n  \sum_{l=1}^n  \alpha_{li} \delta_{il}
=\sum_{i=1}^n   \alpha_{ii}
.
\end{split}
\end{equation}
This representation is particularly useful in quantum mechanics.

Suppose an operator is defined by the dyadic product
$\textsf{\textbf{A}} =  \vert {\bf u} \rangle \langle {\bf v} \vert
$           of two vectors
$\vert {\bf u} \rangle$
and
$\vert {\bf v} \rangle$.
\marginnote{Cf. example~1.10 of~\bibentry{grau}.}
Then its trace can be rewritten as the scalar product of the two vectors (in exchanged order); that is,
for  some arbitrary orthonormal basis ${\frak B} =\{
 \vert {\bf e}_1\rangle ,
\ldots ,
 \vert {\bf e}_n\rangle
\}$
\begin{equation}
\begin{split}
\textrm{Tr}\;\textsf{\textbf{A}}
=\sum_{i=1}^n   \langle {\bf e}_i \vert \textsf{\textbf{A}} \vert {\bf e}_i \rangle
=\sum_{i=1}^n   \langle {\bf e}_i \vert {\bf u} \rangle \langle {\bf v} \vert   {\bf e}_i \rangle  \\
=\sum_{i=1}^n   \langle {\bf v} \vert   {\bf e}_i \rangle \langle {\bf e}_i \vert {\bf u} \rangle
=   \langle {\bf v} \vert   \mathbb{1}_n \vert {\bf u} \rangle
=   \langle {\bf v} \vert   \mathbb{1}_n  {\bf u} \rangle
=   \langle {\bf v} \vert  {\bf u} \rangle
.
\end{split}
\end{equation}

In general, traces represent noninvertible (irreversible) many-to-one functionals
since the same trace value can be obtained from different inputs.
More explicitly, consider two nonidentical vectors  $\vert {\bf u} \rangle \neq \vert {\bf v} \rangle$ in real Hilbert space.
In this case,
\begin{equation}
\textrm{Tr}\;\textsf{\textbf{A}} =
\textrm{Tr}\;\vert {\bf u} \rangle \langle {\bf v} \rangle =
\langle {\bf v} \vert  {\bf u} \rangle
=
\langle {\bf u} \vert  {\bf v} \rangle =
\textrm{Tr}\;\vert {\bf v} \rangle \langle {\bf u} \rangle =
\textrm{Tr}\;\textsf{\textbf{A}}^\intercal
\end{equation}
This example shows that the traces of two matrices such as $\textrm{Tr}\;\textsf{\textbf{A}}$ and $\textrm{Tr}\;\textsf{\textbf{A}}^\intercal $ can be identical although
the argument matrices  $\textsf{\textbf{A}}=\vert {\bf u} \rangle \langle {\bf v} \rangle
$ and $\textsf{\textbf{A}}^\intercal  = \vert {\bf v} \rangle \langle {\bf u} \rangle$ need not be.

\subsection{Properties}

The following properties of traces are mentioned without proof:

\begin{itemize}
\item[(i)]
$\textrm{Tr}(A+B)=\textrm{Tr}A+\textrm{Tr}B$;
\item[(ii)]
$\textrm{Tr}(\alpha A)= \alpha \textrm{Tr}A$, with $\alpha \in {\Bbb C}$;
\item[(iii)]
$\textrm{Tr}(AB) = \textrm{Tr}(BA)$, hence the trace of the  commutator vanishes; that
is, $\textrm{Tr}([A,B])=0$;
\item[(iv)]
$\textrm{Tr}A = \textrm{Tr}A^\intercal $;
\item[(v)]
$\textrm{Tr}(A\otimes B)= (\textrm{Tr}A) (\textrm{Tr}B)$;
\item[(vi)]
the trace is the sum of the eigenvalues of a {\em normal operator} (cf. page \pageref{2014-m-fdvs-normality});
\index{normal operator}
\index{normal transformation}
\item[(vii)]
$ \textrm{det}(e^A)=e^{\textrm{Tr}A} $;
\item[(viii)]
 the trace is the derivative of the determinant at the identity;
\item[(ix)]
the complex conjugate of the trace of an operator is equal to the trace of its adjoint
(cf. page~\pageref{2014-m-fdvs-adjoint}); that is
$\overline{(  \textrm{Tr} A)}=\textrm{Tr} (A^\dagger)$;
\item[(x)]
the trace is invariant under rotations of the basis as well as
under cyclic permutations.
\item[(xi)]
the trace of an $n \times n$ matrix $A$ for which $AA=\alpha A$ for some $\alpha \in {\Bbb R}$ is
$ \textrm{Tr} A =
\alpha \textrm{rank}(A)$,
where  $\textrm{rank}$ is the rank of $A$ defined on page~\pageref{2014-m-fdvs-rank}.
Consequently, the trace of an idempotent (with $\alpha=1$) operator -- that is, a projection --
\index{idempotence}
is equal to its rank;
and, in particular, the trace of a one-dimensional projection is one.
\index{rank}
\item[(xii)]
Only commutators have trace zero.
\end{itemize}

A {\em trace class} operator is a compact operator for which a trace is finite and independent of the choice of basis.
\index{trace class}

\subsection{Partial trace}
\index{partial trace}
\label{2015-partialtrace}

The quantum mechanics of multi-particle (multipartite) systems allows for configurations -- actually rather processes --
that can be informally described as ``beam dump experiments;'' in which we start out with entangled states
(such as the Bell states on page~\pageref{2014-m-ch-fdvs-bellbasis})  which carry information
about {\em joint properties of the constituent quanta}
and {\em choose to disregard} one quantum state entirely; that is, we pretend
not to care about, and ``look the other way'' with regards to the (possible) outcomes of a measurement on this particle.
In this case, we have to {\em trace out} that particle; and as a result, we obtain a {\em reduced state} without this particle we do
not care about.

Formally the partial trace with respect to the first particle
maps the general density matrix
${ \rho}_{12} = \sum_{i_1 j_1 i_2 j_2} \rho_{i_1 j_1 i_2 j_2}   \vert i_1\rangle \langle j_1\vert \otimes \vert i_2\rangle \langle j_2 \vert$
on a composite Hilbert space
$
{\frak H}_1
\otimes
{\frak H}_2
$
to  a density matrix on the Hilbert space
${\frak H}_2
$ of the second particle by
\begin{equation}
\begin{split}
\textrm{Tr}_1
{ \rho}_{12}
= \textrm{Tr}_1  \left(   \sum_{i_1 j_1 i_2 j_2} \rho_{i_1 j_1 i_2 j_2}   \vert i_1\rangle \langle j_1\vert \otimes \vert i_2 \rangle \langle j_2 \vert \right)
 \\
= \sum_{k_1}   \left\langle {\bf e}_{k_1}  \left\vert \left(   \sum_{i_1 j_1 i_2 j_2} \rho_{i_1 j_1 i_2 j_2}
 \vert i_1\rangle \langle j_1\vert \otimes \vert i_2 \rangle \langle j_2 \vert  \right)  \right\vert {\bf e}_{k_1} \right\rangle
 \\
=  \sum_{i_1 j_1 i_2 j_2} \rho_{i_1 j_1 i_2 j_2}  \left( \sum_{k_1}
 \langle {\bf e}_{k_1}  \vert i_1\rangle \langle j_1\vert    {\bf e}_{k_1}  \rangle  \right)  \vert i_2 \rangle \langle j_2 \vert
 \\
=  \sum_{i_1 j_1 i_2 j_2} \rho_{i_1 j_1 i_2 j_2}  \left( \sum_{k_1}
 \langle j_1\vert    {\bf e}_{k_1}  \rangle  \langle {\bf e}_{k_1}  \vert i_1\rangle \right)  \vert i_2 \rangle \langle j_2 \vert
 \\
=  \sum_{i_1 j_1 i_2 j_2} \rho_{i_1 j_1 i_2 j_2}
 \langle j_1\vert   \; \mathbb{1} \; \vert i_1\rangle   \vert i_2 \rangle \langle j_2 \vert
=  \sum_{i_1 j_1 i_2 j_2} \rho_{i_1 j_1 i_2 j_2}
 \langle j_1 \vert  i_1\rangle    \vert i_2 \rangle \langle j_2 \vert
.
\end{split}
\label{2016-m-ch-fdvs-pt}
\end{equation}
Suppose further that the vectors
$\vert i_1 \rangle$
and
$\vert j_1 \rangle$
associated with the first particle
belong to an orthonormal basis.
Then $\langle j_1 \vert  i_1\rangle =\delta_{i_1 j_1}$ and~(\ref{2016-m-ch-fdvs-pt})
reduces to
\begin{equation}
\begin{split}
\textrm{Tr}_1
{ \rho}_{12}
=  \sum_{i_1 i_2 j_2} \rho_{i_1 i_1 i_2 j_2}
  \vert i_2 \rangle \langle j_2 \vert
.
\end{split}
\label{2016-m-ch-fdvs-pt2}
\end{equation}

The partial trace in general corresponds to a noninvertible map corresponding to an irreversible process; that is,
it is an $m$-to-$n$ with $m>n$, or a many-to-one mapping:
$\rho_{1 1 i_2 j_2} = 1$, $\rho_{1 2 i_2 j_2} =   \rho_{2 1 i_2 j_2} =  \rho_{2 2 i_2 j_2} = 0$
and
$\rho_{2 2 i_2 j_2} = 1$, $\rho_{1 2 i_2 j_2} =   \rho_{2 1 i_2 j_2} =  \rho_{1 1 i_2 j_2} = 0$
are mapped into the same  $ \sum_{i_1} \rho_{i_1 i_1 i_2 j_2}$.
This can be expected, as information about the first particle is ``erased.''

{\color{blue}
\bexample

\label{bellstate}
For an explicit example's sake, consider the Bell state  \index{Bell state}       $\vert \Psi^- \rangle$
defined in Equation~(\ref{2014-m-ch-fdvs-bellbasis}).
\marginnote{The same is true for all elements of the Bell basis.}
Suppose we do not care about the state of the first particle, then we may ask what kind of reduced state results from this
pretension.\marginnote{Be careful here to make the experiment in such a way that in no way you could know the state of the first particle.
You may actually think about this as a measurement of the state of the first particle
by a degenerate observable with only a single, nondiscriminating measurement outcome.}
Then the partial trace is just the trace over the first particle; that is, with subscripts referring to the particle number,
\begin{equation}
\begin{split}
\textrm{Tr}_1\, \vert \Psi^- \rangle \langle  \Psi^-   \vert  \\
=\sum_{i_1=0}^1 \langle i_1 \vert \Psi^- \rangle \langle  \Psi^-  \vert i_1 \rangle \\
=\langle 0_1 \vert \Psi^- \rangle \langle  \Psi^-  \vert 0_1 \rangle
+
\langle 1_1  \vert \Psi^- \rangle \langle  \Psi^-  \vert 1_1 \rangle  \\
=\langle 0_1 \vert  \frac{1}{\sqrt{2}}\left(\vert 0_1   1_2 \rangle - \vert 1_1   0_2 \rangle  \right)  \frac{1}{\sqrt{2}}\left(\langle 0_1   1_2 \vert  - \langle 1_1   0_2 \vert   \right)  \vert 0_1 \rangle\\
\qquad
+
\langle 1_1 \vert  \frac{1}{\sqrt{2}}\left(\vert 0_1   1_2 \rangle - \vert 1_1   0_2 \rangle  \right)  \frac{1}{\sqrt{2}}\left(\langle 0_1   1_2 \vert  - \langle 1_1   0_2 \vert   \right)  \vert 1_1 \rangle  \\
= \frac{1}{2}
\left(
\vert 1_2 \rangle \langle   1_2 \vert
+
\vert 0_2 \rangle \langle   0_2 \vert
\right)
.
\end{split}
\end{equation}

The resulting state is a
{\em mixed state}
\index{mixed state}
defined by the property that its trace is equal to one,
but the trace of its square is smaller than one; in this case the trace is $\frac{1}{2}$, because
\begin{equation}
 \begin{split}
\textrm{Tr}_2\,
\frac{1}{2}
\left(
\vert 1_2 \rangle \langle   1_2 \vert
+
\vert 0_2 \rangle \langle   0_2 \vert
\right)  \\
 = \frac{1}{2}  \langle 0_2 \vert
\left(
\vert 1_2 \rangle \langle   1_2 \vert
+
\vert 0_2 \rangle \langle   0_2 \vert
\right)
\vert 0_2 \rangle
+  \frac{1}{2}
\langle 1_2  \vert
\left(
\vert 1_2 \rangle \langle   1_2 \vert
+
\vert 0_2 \rangle \langle   0_2 \vert
\right)
\vert 1_2 \rangle  \\
=  \frac{1}{2} + \frac{1}{2} =1;
 \end{split}
\end{equation}
but
\begin{eqnarray}
&&\textrm{Tr}_2\,
\left[
\frac{1}{2}
\left(
\vert 1_2 \rangle \langle   1_2 \vert
+
\vert 0_2 \rangle \langle   0_2 \vert
\right)
\frac{1}{2}
\left(
\vert 1_2 \rangle \langle   1_2 \vert
+
\vert 0_2 \rangle \langle   0_2 \vert
\right)
\right]
\nonumber
\\ && =
\textrm{Tr}_2\,
\frac{1}{4}
\left(
\vert 1_2 \rangle \langle   1_2 \vert
+
\vert 0_2 \rangle \langle   0_2 \vert
\right)
= \frac{1}{2}.
\end{eqnarray}
This {\em mixed state} is a 50:50 mixture of the pure particle states  $\vert 0_2 \rangle$ and $\vert 1_2 \rangle$, respectively.
Note that this is different from
a coherent superposition
\index{coherent superposition}
\index{superposition}
$\vert 0_2 \rangle + \vert 1_2 \rangle$
 of the pure particle states  $\vert 0_2 \rangle$ and $\vert 1_2 \rangle$, respectively --
also formalizing a 50:50 mixture with respect to measurements of property $0$ {\it versus} $1$, respectively.

\eexample
}

In quantum mechanics, the ``inverse'' of the partial trace is called
{\em purification}:
\index{purification}
it is the creation of a pure state from a mixed one,
associated with an ``enlargement'' of Hilbert space (more dimensions).
This cannot be done in a unique way (see Section \ref{2015-m-ch-fdvs-purification} below).
\marginnote{For additional information see page~110, Section~2.5 in~\bibentry{nielsen-book10}.}
Some people -- members of the ``church of the larger Hilbert space'' --
believe that mixed states are epistemic (that is, associated with our own personal ignorance rather than with
any ontic, microphysical property), and are always part of an, albeit unknown, pure state in a larger Hilbert space.

\section{Adjoint or dual transformation}
\label{2014-m-fdvs-adjoint}
\index{adjoints}
\index{adjoint operator}
\index{dual operator}

\subsection{Definition}

Let ${\frak V}$ be a vector space and let ${\bf y}$
be any element of its dual space ${\frak V}^\ast$.
For any linear transformation $\textsf{\textbf{A}}$, consider
the bilinear functional
\marginnote{Here $\llbracket \cdot ,\cdot \rrbracket $ is the bilinear functional, not the commutator.}
${\bf y}' ({\bf x}) \equiv \llbracket {\bf x} ,{\bf y}'\rrbracket  =\llbracket \textsf{\textbf{A}}{\bf x},{\bf y}\rrbracket  \equiv {\bf y} (\textsf{\textbf{A}} {\bf x})$.
Let the {\em adjoint} (or {\em dual}) transformation $\textsf{\textbf{A}}^\ast$ be defined by
${\bf y}' ({\bf x}) = \textsf{\textbf{A}}^{\ast}{\bf y} ({\bf x})$ with
\begin{equation}
\textsf{\textbf{A}}^\ast {\bf y} ({\bf x})
\equiv
\llbracket {\bf x},\textsf{\textbf{A}}^\ast{\bf y}\rrbracket
\stackrel{{\tiny \textrm{ def }}}{=}
\llbracket \textsf{\textbf{A}}{\bf x},{\bf y}\rrbracket
\equiv
{\bf y} (\textsf{\textbf{A}}{\bf x}).
\label{2016-m-ch-fdlvs-adjointop}
\end{equation}

\subsection{Adjoint matrix notation}

In matrix notation and in complex vector space with the dot product,
note that there is a correspondence with the inner product
(cf. page \pageref{2011-m-corr-bil-ip})
so that, for all ${\bf z}\in {\frak V}$ and for all ${\bf x}\in {\frak V}$,
there exist a unique ${\bf y}\in {\frak V}$ with
\marginnote{Recall that, for $\alpha, \beta \in {\Bbb C}$,
$\overline{ \left(\alpha \beta \right)}=
\overline{ \alpha }
\overline{  \beta }$, and
$\overline{ \left\llbracket  \overline{ \left(\alpha \right)} \right\rrbracket }  = \alpha $,
and that  the Euclidean scalar product is assumed to be linear in its first argument  and antilinear in its second argument.
}
\begin{equation}
\begin{split}
\llbracket \textsf{\textbf{A}}{\bf x}, {\bf z}\rrbracket
=\langle \textsf{\textbf{A}} {\bf x}\mid {\bf y}\rangle
= \\
=\overline{A_{ij}   x_j}  y_i
=  \overline{x_j A_{ij}} y_i
= \overline{x_j} \overline{(A^\intercal )_{ji}}  y_i
= \overline{ x}  \overline{ \textsf{\textbf{A}}^\intercal}  y ,
\end{split}
\label{2016-m-ch-fdvs-adjdef}
\end{equation}
and another unique vector ${\bf y}'$ obtained from ${\bf y}$ by
some linear operator $\textsf{\textbf{A}}^\ast$
such that ${\bf y}'=\textsf{\textbf{A}}^\ast {\bf y}$ with
\begin{equation}
\begin{split}
\llbracket {\bf x}, \textsf{\textbf{A}}^\ast {\bf z}\rrbracket
= \langle {\bf x}\mid {\bf y}'\rangle
= \langle {\bf x}\mid \textsf{\textbf{A}}^\ast {\bf y}\rangle
 \\
= \overline{ x_i } \left(A_{ij}^\ast y_j \right)
=\\
\llbracket  i \leftrightarrow j\rrbracket
= \overline{  x_j }A_{ji}^\ast    y_i
=   \overline{ x }\textsf{\textbf{A}}^\ast y.
\end{split}
\label{2016-m-ch-fdvs-adjdef2}
\end{equation}
Therefore, by comparing Equations.~(\ref{2016-m-ch-fdvs-adjdef2}) and (\ref{2016-m-ch-fdvs-adjdef}),  we obtain $\textsf{\textbf{A}}^\ast=\overline{ \textsf{\textbf{A}}^\intercal }$, so that
\begin{equation}
\textsf{\textbf{A}}^\ast =\overline{\textsf{\textbf{A}}^\intercal } = \overline{\textsf{\textbf{A}}}^\intercal .
\end{equation}
That is, in matrix notation, the adjoint transformation is just the
transpose of the complex conjugate of the original matrix.

Accordingly, in real inner product spaces,
$\textsf{\textbf{A}}^\ast= \overline{\textsf{\textbf{A}}}^\intercal  =  \textsf{\textbf{A}}^\intercal $ is just the transpose of $\textsf{\textbf{A}}$:
\begin{equation}
\llbracket {\bf x},\textsf{\textbf{A}}^\intercal {\bf y}\rrbracket =
\llbracket \textsf{\textbf{A}}{\bf x},{\bf y}\rrbracket .
\end{equation}
In complex inner product spaces,
define the Hermitian conjugate matrix by
$\textsf{\textbf{\textsf{\textbf{A}}}}^\dagger = \textsf{\textbf{A}}^\ast =\overline{\textsf{\textbf{A}}^\intercal } = \overline{\textsf{\textbf{A}}}^\intercal $, so that
\begin{equation}
\llbracket {\bf x},\textsf{\textbf{A}}^\dagger{\bf y}\rrbracket =
\llbracket \textsf{\textbf{A}}{\bf x},{\bf y}\rrbracket .
\end{equation}

\subsection{Properties}
We mention without proof that the adjoint operator is a linear operator.
Furthermore,
$\textsf{\textbf{0}}^\ast = \textsf{\textbf{0}}$,
$\textsf{\textbf{1}}^\ast = \textsf{\textbf{1}}$,
$(\textsf{\textbf{A}}+\textsf{\textbf{B}})^\ast = \textsf{\textbf{A}}^\ast+\textsf{\textbf{B}}^\ast$,
$(\alpha \textsf{\textbf{A}})^\ast = \alpha \textsf{\textbf{A}}^\ast$,
$( \textsf{\textbf{A}}\textsf{\textbf{B}})^\ast =   \textsf{\textbf{B}}^\ast
 \textsf{\textbf{A}}^\ast$,
and
$( \textsf{\textbf{A}}^{-1})^\ast
=
( \textsf{\textbf{A}}^\ast )^{-1}
$.

{\color{OliveGreen}\bproof
A proof for
$( \textsf{\textbf{A}}\textsf{\textbf{B}})^\ast =   \textsf{\textbf{B}}^\ast  \textsf{\textbf{A}}^\ast
$
is
$
\llbracket {\bf x},(\textsf{\textbf{A}}\textsf{\textbf{B}})^\ast {\bf y}\rrbracket
=
\llbracket \textsf{\textbf{A}}\textsf{\textbf{B}}{\bf x}, {\bf y}\rrbracket
=
\llbracket \textsf{\textbf{B}}{\bf x}, \textsf{\textbf{A}}^\ast{\bf y}\rrbracket
=
\llbracket {\bf x}, \textsf{\textbf{B}}^\ast\textsf{\textbf{A}}^\ast{\bf y}\rrbracket
$.
\eproof
}

Note that, since\marginnote{Recall again that, for $\alpha, \beta \in {\Bbb C}$,
$\overline{ \left(\alpha \beta \right)}=
\overline{ \alpha }
\overline{  \beta }$.
$
( \textsf{\textbf{A}}\textsf{\textbf{B}})^\intercal  = \textsf{\textbf{A}}^\intercal    \textsf{\textbf{B}}^\intercal
$
can be explicitly demonstrated in index notation:
because for any $c_{ij}^\intercal  = c_{ji}$,  and because of linearity of the sum,
$
( \textsf{\textbf{A}}\textsf{\textbf{B}})^\intercal
\equiv
( a_{ik}b_{kj})^\intercal    =  a_{jk}b_{ki}    = b_{ki} a_{jk}  =   b_{ik}^\intercal   a_{kj}^\intercal
\equiv B^\intercal  A^\intercal
$.}
$
( \textsf{\textbf{A}}\textsf{\textbf{B}})^\ast = \textsf{\textbf{A}}^\ast   \textsf{\textbf{B}}^\ast
$, by identifying $\textsf{\textbf{B}}$ with $\textsf{\textbf{A}}$  and by repeating this,
$( \textsf{\textbf{A}}^n)^\ast =  (\textsf{\textbf{A}}^\ast)^n$.
In particular, if
$\textsf{\textbf{E}}$
is a projection,
then
$\textsf{\textbf{E}}^\ast$ is a projection, since
$ (\textsf{\textbf{E}}^\ast)^2 =  ( \textsf{\textbf{E}}^2)^\ast  =\textsf{\textbf{E}}^\ast$.

For finite dimensions,
\begin{equation}
\textsf{\textbf{A}}^{\ast \ast}=
\textsf{\textbf{A}},
\end{equation}

{\color{OliveGreen}\bproof
as, {\it per} definition,
$
\llbracket \textsf{\textbf{A}}{\bf x}, {\bf y}\rrbracket
=
\llbracket {\bf x}, \textsf{\textbf{A}}^\ast{\bf y}\rrbracket
=
\llbracket (\textsf{\textbf{A}}^\ast)^\ast{\bf x}, {\bf y}\rrbracket
$.
\eproof
}

\section{Self-adjoint transformation}
\index{self-adjoint transformation}
\label{2015-m-ch-fdlvs-self-adjoint}

\marginnote{A classical text on this and related subjects is \bibentry{Parlett:1998:SEP:280490}.}

The following definition yields some analogy to real numbers as compared to complex numbers
(``a complex number $z$ is real if $\overline{z}=z$''),
expressed in terms of operators on a complex vector space.

An operator    $\textsf{\textbf{A}}$   on a linear vector space   ${\frak V}$
is called {\em self-adjoint}, if
\begin{equation}
\textsf{\textbf{A}}^{\ast}=
\textsf{\textbf{A}}
\end{equation}
and if the domains of $\textsf{\textbf{A}}$ and $\textsf{\textbf{A}}^{\ast}$
-- that is, the set of vectors on which they are well defined -- coincide.

\marginnote{For infinite dimensions,
a distinction must be made between self-adjoint operators and Hermitian ones; see,
for instance~\bibentry{grau}, \bibentry{Gieres-2000}, \bibentry{2001-Bonneau}.}
In finite dimensional {\em real} inner product spaces,
self-adjoint operators are called {\em symmetric,}
since they are symmetric with respect to transpositions; that is,
\index{symmetric operator}
\begin{equation}
\textsf{\textbf{A}}^{\ast}= \textsf{\textbf{A}}^{T}=
\textsf{\textbf{A}}.
\end{equation}

In finite dimensional
{\em complex} inner product spaces,
self-adjoint operators are called {\em Hermitian,}
since they are identical with respect to Hermitian conjugation (transposition of the matrix and complex conjugation of its
entries); that is,
\index{Hermitian operator}
\begin{equation}
\textsf{\textbf{A}}^{\ast}= \textsf{\textbf{A}}^{\dagger}=
\textsf{\textbf{A}}.
\end{equation}

In what follows, we shall consider only the latter case and identify self-adjoint operators with Hermitian ones.
In terms of matrices, a matrix $A$ corresponding to an operator $\textsf{\textbf{A}}$ in
some fixed basis is self-adjoint
if
\begin{equation}
A^{\dagger}\equiv (\overline{A_{ij}})^\intercal =  \overline{A_{ji}} =A_{ij} \equiv A.
\end{equation}
That is, suppose $A_{ij}$ is the matrix representation
corresponding to a linear transformation $\textsf{\textbf{A}}$  in some basis ${\frak B}$,
then the {\em Hermitian} matrix $\textsf{\textbf{A}}^\ast = \textsf{\textbf{A}}^\dagger$
to the dual basis
${\frak B}^\ast $
is
$\overline{(A_{ij}})^\intercal $.

{\color{blue}
\bexample
For the sake of examples of Hermitian matrices, consider the
{\em Pauli spin matrices} defined earlier in Equation~\ref{2019-mm-ch-fdvs-psm} as well as the unit matrix $\mathbb{1}_2$
\index{Pauli spin matrices}
\begin{equation}
\begin{pmatrix}
0&1\\
1&0
\end{pmatrix}
\text{, }
\begin{pmatrix}
0&-i\\
i&0
\end{pmatrix}
\text{, }
\begin{pmatrix}
1&0\\
0&-1
\end{pmatrix}
\text{, or }
\begin{pmatrix}
1&0\\
0&1
\end{pmatrix}
.
\end{equation}

The following matrices are not self-adjoint:
\begin{equation}
\begin{pmatrix}
0&1\\
0&0
\end{pmatrix}
\text{, }
\begin{pmatrix}
1&1\\
0&0
\end{pmatrix}
\text{, }
\begin{pmatrix}
1&0\\
i&0
\end{pmatrix}
\text{, or }
\begin{pmatrix}
0&i\\
i&0
\end{pmatrix}
.{\textrm{\eexample}}
\end{equation}
}

Note that the coherent real-valued superposition
\index{coherent superposition}
\index{superposition}
of a self-adjoint transformations
(such as the sum or difference of correlations in
the Clauser-Horne-Shimony-Holt expression\cite{filipp-svo-04-qpoly-prl})
is a self-adjoint transformation.

{\color{OliveGreen}\bproof
For a direct proof
suppose that $\alpha_i \in {\Bbb R}$ for all $1\le i \le n$ are $n$ real-valued coefficients and
$\textsf{\textbf{A}}_1, \ldots \textsf{\textbf{A}}_n$ are $n$ self-adjoint operators.
Then
$\textsf{\textbf{B}} = \sum_{i=1}^n \alpha_i \textsf{\textbf{A}}_i$
is self-adjoint, since
\begin{equation}
\textsf{\textbf{B}}^\ast  = \sum_{i=1}^n \overline{\alpha_i} \textsf{\textbf{A}}_i^\ast  = \sum_{i=1}^n  \alpha_i  \textsf{\textbf{A}}_i
=\textsf{\textbf{B}}
.
\end{equation}
\eproof
}

\section{Positive transformation}
\index{positive transformation}
\index{nonnegative transformation}
\label{2015-m-ch-fdlvs-positive}

A linear transformation  $\textsf{\textbf{A}}$ on an inner product space ${\frak V}$ is {\em positive}
(or, used synonymously, {\em nonnegative}),
that is, in symbols $\textsf{\textbf{A}}\ge 0$,
if $\langle \textsf{\textbf{A}}{\bf x}\vert {\bf x}\rangle  \ge 0$
for all ${\bf x}\in {\frak V}$.
If  $\langle \textsf{\textbf{A}}{\bf x}\vert {\bf x}\rangle = 0$ implies
${\bf x}=0$, $\textsf{\textbf{A}}$ is called {\em strictly positive}.
Note that, therefore, $\langle \textsf{\textbf{A}}{\bf x}\vert {\bf x}\rangle$
has to be real-valued.

Positive transformations -- indeed, transformations with real inner products such that
$
\langle \textsf{\textbf{A}} {\bf x}\vert {\bf x}\rangle
= \overline{\langle {\bf x}\vert \textsf{\textbf{A}}{\bf x}\rangle }
=
\langle {\bf x}\vert \textsf{\textbf{A}}{\bf x}\rangle \in \mathbb{R}$
for all vectors  ${\bf x}$
of a Hilbert space ${\frak V}$ --
are self-adjoint.

{\color{OliveGreen}\bproof

In order to prove that positive transformations $\textsf{\textbf{A}}$ are self-adjoint note first that,
from the definition of the adjoint operator (\ref{2016-m-ch-fdlvs-adjointop})
on page~\pageref{2016-m-ch-fdlvs-adjointop},
$
\langle \textsf{\textbf{A}} {\bf x}\vert {\bf x}\rangle
=
\langle {\bf x}\vert \textsf{\textbf{A}}^\ast {\bf x}\rangle \in \mathbb{R}$.
So, for real-valued scalar products involving a single vector ${\bf x}$,
$
\langle \textsf{\textbf{A}} {\bf x}\vert {\bf x}\rangle
= \overline{\langle {\bf x}\vert \textsf{\textbf{A}}{\bf x}\rangle }
=
\langle {\bf x}\vert \textsf{\textbf{A}}{\bf x}\rangle
=
\langle {\bf x}\vert \textsf{\textbf{A}}^\ast {\bf x}\rangle
\in \mathbb{R}
$.

For a direct proof that $\textsf{\textbf{A}}$ is self-adjoint
--
that is,   $\textsf{\textbf{A}}^\ast = \textsf{\textbf{A}}$
--
we need to
consider two arbitrary vectors ${\bf x}, {\bf y} \in {\frak V}$
and prove that positivity (or a real-valued inner product)
implies
$\langle \textsf{\textbf{A}}{\bf x}\vert {\bf y}\rangle
= \langle {\bf x}\vert \textsf{\textbf{A}}^\ast {\bf y}\rangle
= \langle {\bf x}\vert \textsf{\textbf{A}}{\bf y}\rangle  $.
We shall use a different form of the polarization identity
\index{polarization identity}
[which has not been used in~(\ref{2015-m-ch-polidc}) on page~\pageref{2015-m-ch-polidc2}],
so that we can reduce it to the earlier form $\langle \textsf{\textbf{A}} {\bf x}\vert {\bf x}\rangle$
involving the same vectors as arguments,
thereby having in mind
the definition of the adjoint operator (\ref{2016-m-ch-fdlvs-adjointop})
on page~\pageref{2016-m-ch-fdlvs-adjointop},
and write
\begin{equation}
\begin{split}
\langle {\bf x}\vert \textsf{\textbf{A}}^\ast {\bf y}\rangle
= \text{ [definition of adjoint operator] } =
\langle \textsf{\textbf{A}} {\bf x}\vert {\bf y}\rangle \\
=  \text{ [modified polarization identity] } \\
=
\frac{1}{4}\left[
\langle \textsf{\textbf{A}} ({\bf x}+ {\bf y}) \vert {\bf x}+ {\bf y}\rangle
-
\langle \textsf{\textbf{A}}({\bf x}- {\bf y}) \vert {\bf x}- {\bf y}\rangle \right.  \\
\left.
+ i
\langle \textsf{\textbf{A}} ({\bf x}- i{\bf y}) \vert {\bf x}- i{\bf y}\rangle
- i
\langle \textsf{\textbf{A}} ({\bf x}+ i{\bf y}) \vert {\bf x} +i {\bf y}\rangle
\right]
\\
=  \text{ [real-valuedness of $\langle \textsf{\textbf{A}} {\bf x}\vert {\bf x}\rangle
= \overline{\langle {\bf x}\vert \textsf{\textbf{A}}{\bf x}\rangle }
=
\langle {\bf x}\vert \textsf{\textbf{A}}{\bf x}\rangle \in \mathbb{R}$] } \\
=
\frac{1}{4}\left[
\langle {\bf x}+ {\bf y} \vert \textsf{\textbf{A}} ({\bf x}+ {\bf y})\rangle
-
\langle {\bf x}- {\bf y} \vert \textsf{\textbf{A}}({\bf x}- {\bf y})\rangle \right.  \\
\left.
+ i
\langle {\bf x}- i{\bf y} \vert \textsf{\textbf{A}} ({\bf x}- i{\bf y})\rangle
- i
\langle {\bf x}+ i{\bf y} \vert \textsf{\textbf{A}} ({\bf x} +i {\bf y})\rangle
\right]
= \\
\text{ [``inverse'' modified polarization identity] } =
\langle {\bf x}\vert \textsf{\textbf{A}}{\bf y}\rangle
.
\end{split}
\label{2015-m-ch-polidrv}
\end{equation}
\eproof
}

\section{Unitary transformation and isometry}
\index{unitary transformation}
\label{2014-m-ch-fdvs-unitary}
\marginnote[-7mm]{For proofs and additional information see {\S}71-73 in~\bibentry{halmos-vs}.}

\subsection {Definition}
Note that a complex number $z$ has absolute value one if $z\overline{z}=1$, or $\overline{z}=1/z$.
In analogy to this ``modulus one'' behavior,
consider {\em unitary transformations}, or, used synonymously, {\em (one-to-one) isometries}
$\textsf{\textbf{U}}$ for which
\begin{equation}
\textsf{\textbf{U}}^\ast = \textsf{\textbf{U}}^\dagger =\textsf{\textbf{U}}^{-1},
\textrm{ or } \textsf{\textbf{U}}\textsf{\textbf{U}}^\dagger =\textsf{\textbf{U}}^\dagger \textsf{\textbf{U}}=\mathbb{1}.
\end{equation}
The following conditions are equivalent:
\begin{itemize}
\item[(i)]
$\textsf{\textbf{U}}^\ast = \textsf{\textbf{U}}^\dagger =\textsf{\textbf{U}}^{-1}$,
or $\textsf{\textbf{U}}\textsf{\textbf{U}}^\dagger =\textsf{\textbf{U}}^\dagger \textsf{\textbf{U}}=\mathbb{1}$.
\item[(ii)]
$\langle \textsf{\textbf{U}}{\bf x}\mid \textsf{\textbf{U}}{\bf y} \rangle
=
\langle {\bf x}\mid {\bf y} \rangle$ for all ${\bf x} ,{\bf y} \in {\frak V}$;
\item[(iii)]  $\textsf{\textbf{U}}$ is an {\em isometry};
\index{isometry}
that is, preserving the norm
$\| \textsf{\textbf{U}}{\bf x}\|
=
\|{\bf x}\|$ for all ${\bf x}  \in {\frak V}$.
\item[(iv)]
$\textsf{\textbf{U}}$ represents a change of orthonormal basis:\cite{Schwinger.60}
\marginnote{See also \S~74 of~\bibentry{halmos-vs}.}
Let ${\frak B}=\{{\bf f}_1,  {\bf f}_2, \ldots , {\bf f}_n\}$
be an orthonormal basis.
Then
$\textsf{\textbf{U}}{\frak B}={\frak B}'=\{\textsf{\textbf{U}}{\bf f}_1, \textsf{\textbf{U}} {\bf f}_2,
\ldots ,\textsf{\textbf{U}} {\bf f}_n\}$
is also an orthonormal basis of  ${\frak V}$.
Conversely, two arbitrary orthonormal bases
${\frak B}$
and
${\frak B}'$
are connected by a unitary transformation $\textsf{\textbf{U}}$ {\it via} the pairs  ${\bf f}_i$ and $ \textsf{\textbf{U}}{\bf f}_i$
for all $1\le i \le n$, respectively.
More explicitly, denote  $\textsf{\textbf{U}} {\bf f}_i   = {\bf e}_i$; then
(recall ${\bf f}_i$ and ${\bf e}_i$ are elements of the orthonormal bases
$  {\frak B} $  and $ \textsf{\textbf{U}} {\frak B} $, respectively)
$\textsf{\textbf{U}}_{ef}   = \sum_{i=1}^n{\bf e}_i  {\bf f}_i^\dagger
 = \sum_{i=1}^n \vert {\bf e}_i \rangle \langle {\bf f}_i \vert$.
\end{itemize}

{\color{OliveGreen}\bproof
For a direct proof
suppose that (i) holds; that is,
$\textsf{\textbf{U}}^\ast = \textsf{\textbf{U}}^\dagger =\textsf{\textbf{U}}^{-1}$.
then, (ii) follows by
\begin{equation}
\begin{split}
\langle \textsf{\textbf{U}}{\bf x}\mid \textsf{\textbf{U}}{\bf y} \rangle
=
\langle \textsf{\textbf{U}}^\ast\textsf{\textbf{U}}{\bf x}\mid {\bf y} \rangle
=
\langle \textsf{\textbf{U}}^{-1}\textsf{\textbf{U}}{\bf x}\mid {\bf y} \rangle
=
\langle {\bf x}\mid {\bf y} \rangle
\end{split}
\label{2016-m-ch-uniteq}
\end{equation}
for all ${\bf x}, {\bf y}$.

In particular, if ${\bf y} = {\bf x}$, then
\begin{equation}
\begin{split}
\| \textsf{\textbf{U}}{\bf x}\|^2 =
\vert \langle \textsf{\textbf{U}}{\bf x}\mid \textsf{\textbf{U}}{\bf x} \rangle \vert
=
\vert \langle {\bf x}\mid {\bf x} \rangle \vert
=
\|{\bf x}\|^2
\end{split}
\label{2016-m-ch-uniteq2}
\end{equation}
for all ${\bf x}$.

In order to prove (i) from (iii) consider the transformation
$\textsf{\textbf{A}} = \textsf{\textbf{U}}^\ast\textsf{\textbf{U}} - \mathbb{1}$, motivated by~(\ref{2016-m-ch-uniteq2}),
or, by linearity of the inner product in the first argument,
\begin{equation}
\begin{split}
\| \textsf{\textbf{U}}{\bf x}\| - \|{\bf x}\|
= \langle \textsf{\textbf{U}}{\bf x}\mid \textsf{\textbf{U}}{\bf x} \rangle
-
 \langle {\bf x}\mid {\bf x} \rangle  =
\langle \textsf{\textbf{U}}^\ast \textsf{\textbf{U}}{\bf x}\mid {\bf x} \rangle
-
 \langle {\bf x}\mid {\bf x} \rangle=        \\   =
\langle \textsf{\textbf{U}}^\ast \textsf{\textbf{U}}{\bf x}\mid {\bf x} \rangle
-
 \langle \mathbb{1}{\bf x}\mid {\bf x} \rangle  =
 \langle (\textsf{\textbf{U}}^\ast \textsf{\textbf{U}} - \mathbb{1}){\bf x}\mid {\bf x} \rangle
=
0
\end{split}
\end{equation}
for all ${\bf x}$.
$\textsf{\textbf{A}}$ is self-adjoint, since
\begin{equation}
\begin{split}
\textsf{\textbf{A}}^\ast  =\left( \textsf{\textbf{U}}^\ast \textsf{\textbf{U}} \right)^\ast - \mathbb{1}^\ast
=\textsf{\textbf{U}}^\ast \left(\textsf{\textbf{U}}^\ast \right)^\ast - \mathbb{1} =
\textsf{\textbf{U}}^\ast \textsf{\textbf{U}} - \mathbb{1} =\textsf{\textbf{A}}
.
\end{split}
\label{2016-m-ch-uniteq3}
\end{equation}
We need to prove
\marginnote{Cf. page~138, \S~71, Theorem 2 of~\bibentry{halmos-vs}.}
that a necessary and sufficient condition for a self-adjoint linear
transformation $\textsf{\textbf{A}}$
on an inner product space to be $0$ is that $\langle \textsf{\textbf{A}}{\bf x} \mid {\bf x} \rangle = 0$ for
all vectors ${\bf x}$.

Necessity is easy: whenever $\textsf{\textbf{A}} = 0$ the scalar product vanishes.
A proof of sufficiency first notes that, by linearity allowing the expansion of the first summand on the right side,
\begin{equation}
\begin{split}
\langle \textsf{\textbf{A}} {\bf x} \mid {\bf y} \rangle
+
\langle \textsf{\textbf{A}} {\bf y} \mid {\bf x} \rangle
=
\langle \textsf{\textbf{A}} ( {\bf x} +{\bf y} ) \mid {\bf x} + {\bf y} \rangle
-
\langle \textsf{\textbf{A}} {\bf x} \mid {\bf x} \rangle
-
\langle \textsf{\textbf{A}} {\bf y} \mid {\bf y} \rangle
.
\end{split}
\label{2016-m-ch-uniteq4}
\end{equation}

Since $\textsf{\textbf{A}}$ is self-adjoint, the left side is
\begin{equation}
\begin{split}
\langle \textsf{\textbf{A}} {\bf x} \mid {\bf y} \rangle
+
\langle \textsf{\textbf{A}} {\bf y} \mid {\bf x} \rangle
=
\langle \textsf{\textbf{A}} {\bf x} \mid {\bf y} \rangle
+
\langle {\bf y} \mid  \textsf{\textbf{A}}^\ast {\bf x} \rangle
=
\langle \textsf{\textbf{A}} {\bf x} \mid {\bf y} \rangle
+
\langle {\bf y} \mid  \textsf{\textbf{A}} {\bf x} \rangle  \\
=
\langle \textsf{\textbf{A}} {\bf x} \mid {\bf y} \rangle
+
\overline{ \langle \textsf{\textbf{A}} {\bf x} \mid  {\bf y} \rangle }
= 2 \Re \left( \langle \textsf{\textbf{A}} {\bf x} \mid {\bf y} \rangle      \right)
.
\end{split}
\label{2016-m-ch-uniteq5}
\end{equation}
Note that our assumption implied that the right hand side of~(\ref{2016-m-ch-uniteq4}) vanishes.
Thus, \marginnote{$\Re z$ and $\Im z$ stand for the real and imaginary parts of the complex number $z= \Re z + i \Im z$.}
\begin{equation}
\begin{split}
2 \Re  \langle \textsf{\textbf{A}} {\bf x} \mid {\bf y} \rangle
=
0
.
\end{split}
\label{2016-m-ch-uniteq6}
\end{equation}

Since the real part
$\Re  \langle \textsf{\textbf{A}} {\bf x} \mid {\bf y} \rangle$
of $ \langle \textsf{\textbf{A}} {\bf x} \mid {\bf y} \rangle$ vanishes,
what remains is to show that the imaginary part
$\Im  \langle \textsf{\textbf{A}} {\bf x} \mid {\bf y} \rangle$
of  $\langle \textsf{\textbf{A}} {\bf x} \mid {\bf y} \rangle$
vanishes as well.

As long as the Hilbert space is real (and thus the self-adjoint transformation $ \textsf{\textbf{A}}$ is just symmetric)
we are almost finished, as
$ \langle \textsf{\textbf{A}} {\bf x} \mid {\bf y} \rangle$ is real, with vanishing imaginary part.
That is,
$
\Re  \langle \textsf{\textbf{A}} {\bf x} \mid {\bf y} \rangle   =
\langle \textsf{\textbf{A}} {\bf x} \mid {\bf y} \rangle  =0
$.
In this case, we are free to identify ${\bf y} = \textsf{\textbf{A}} {\bf x}$, thus obtaining
$\langle \textsf{\textbf{A}} {\bf x} \mid \textsf{\textbf{A}} {\bf x}  \rangle  =0$
for all vectors ${\bf x}$.
Because of the positive-definiteness
[condition (iii) on page~\pageref{2016-m-ch-fdvs-pd}]
we must have $ \textsf{\textbf{A}} {\bf x}  =0$
for all vectors ${\bf x}$, and thus finally
$\textsf{\textbf{A}}= \textsf{\textbf{U}}^\ast\textsf{\textbf{U}} - \mathbb{1} =0$,
and $\textsf{\textbf{U}}^\ast\textsf{\textbf{U}} = \mathbb{1}$.

In the case of complex Hilbert space, and thus  $\textsf{\textbf{A}}$ being Hermitian,
we can find an unimodular complex number $\theta$ such that $\vert \theta \vert =1$,
and, in particular,
$\theta = \theta ({\bf x} , {\bf y}) = +i$
for $\Im \langle \textsf{\textbf{A}} {\bf x} \mid {\bf y} \rangle < 0$
or
$\theta ({\bf x} , {\bf y}) = -i$
for $\Im \langle \textsf{\textbf{A}} {\bf x} \mid {\bf y} \rangle \ge 0$,
such that
$\theta \langle \textsf{\textbf{A}} {\bf x} \mid {\bf y} \rangle =
\vert \Im \langle \textsf{\textbf{A}} {\bf x} \mid {\bf y} \rangle \vert =
\vert \langle \textsf{\textbf{A}} {\bf x} \mid {\bf y} \rangle \vert$
(recall that the real part of $\langle \textsf{\textbf{A}} {\bf x} \mid {\bf y} \rangle$ vanishes).

Now we are free to substitute  $\theta   {\bf x}$ for ${\bf x}$.
We can again start with our assumption (iii), now with  ${\bf x} \rightarrow \theta   {\bf x}$ and thus
rewritten as $0 =
\langle \textsf{\textbf{A}} (\theta   {\bf x}) \mid {\bf y} \rangle $, which we have already converted into
$0
=\Re  \langle \textsf{\textbf{A}} (\theta    {\bf x}) \mid {\bf y} \rangle$
for self-adjoint (Hermitian) $\textsf{\textbf{A}}$.
By linearity in the first argument of the inner product we obtain
\begin{equation}
\begin{split}
0
=\Re  \langle \textsf{\textbf{A}} (\theta    {\bf x}) \mid {\bf y} \rangle
=\Re  \langle  \theta  \textsf{\textbf{A}}   {\bf x}  \mid {\bf y} \rangle
=\Re   \left( \theta   \langle \textsf{\textbf{A}}   {\bf x} \mid {\bf y} \rangle      \right)  \\
=\Re \left( \vert \langle \textsf{\textbf{A}}   {\bf x} \mid {\bf y} \rangle    \vert  \right)
=  \vert \langle \textsf{\textbf{A}}   {\bf x} \mid {\bf y} \rangle    \vert
=   \langle \textsf{\textbf{A}}   {\bf x} \mid {\bf y} \rangle  .
\end{split}
\label{2016-m-ch-uniteq7}
\end{equation}
Again we can  identify ${\bf y} = \textsf{\textbf{A}} {\bf x}$, thus obtaining
$\langle \textsf{\textbf{A}} {\bf x} \mid \textsf{\textbf{A}} {\bf x}  \rangle  =0$
for all vectors ${\bf x}$.
Because of the positive-definiteness
[condition (iii) on page~\pageref{2016-m-ch-fdvs-pd}]
we must have $ \textsf{\textbf{A}} {\bf x}  =0$
for all vectors ${\bf x}$, and thus finally
$\textsf{\textbf{A}}= \textsf{\textbf{U}}^\ast\textsf{\textbf{U}} - \mathbb{1} =0$,
and $\textsf{\textbf{U}}^\ast\textsf{\textbf{U}} = \mathbb{1}$.

A proof of (iv) from (i) can be given as follows.
Note that every unitary transformation $\textsf{\textbf{U}}$
takes elements of  some ``original'' orthonormal basis
${\frak B}=\{{\bf f}_1,  {\bf f}_2, \ldots , {\bf f}_n\}$
into elements of a ``new'' orthonormal basis defined by
$\textsf{\textbf{U}}{\frak B}={\frak B}'=\{\textsf{\textbf{U}}{\bf f}_1, \textsf{\textbf{U}} {\bf f}_2,
\ldots ,\textsf{\textbf{U}} {\bf f}_n\}$; with  $\textsf{\textbf{U}} {\bf f}_i   = {\bf e}_i$.
Thereby, orthonormality is preserved:
since $\textsf{\textbf{U}}^\ast =\textsf{\textbf{U}}^{-1}$,
\begin{equation}
\begin{split}
\langle  {\bf e}_i \mid   {\bf e}_j  \rangle
= \langle \textsf{\textbf{U}} {\bf f}_i \mid \textsf{\textbf{U}}  {\bf f}_j  \rangle
=
\langle \textsf{\textbf{U}}^\ast \textsf{\textbf{U}} {\bf f}_i \mid  {\bf f}_j  \rangle
=
\langle \textsf{\textbf{U}}^{-1}\textsf{\textbf{U}} {\bf f}_i \mid  {\bf f}_j  \rangle
=
\langle  {\bf f}_i \mid  {\bf f}_j  \rangle = \delta_{ij}.
\end{split}
\label{2016-m-ch-uniteq71}
\end{equation}
$ \textsf{\textbf{U}} {\frak B} $
forms a new basis: both $  {\frak B} $ as well as $ \textsf{\textbf{U}} {\frak B} $
have the same number of mutually orthonormal elements; furthermore, completeness of $ \textsf{\textbf{U}} {\frak B} $
follows from the completeness of $ {\frak B} $:
$
\langle  {\bf x} \mid \textsf{\textbf{U}} {\bf f}_j  \rangle
=
\langle  \textsf{\textbf{U}}^\ast  {\bf x} \mid {\bf f}_j  \rangle
=0
$ for all basis elements ${\bf f}_j$ implies $ \textsf{\textbf{U}}^\ast  {\bf x} =
\textsf{\textbf{U}}^{-1}  {\bf x} =0$ and thus    ${\bf x} = \textsf{\textbf{U}}0=0$.
All that needs to be done is to explicitly identify $\textsf{\textbf{U}}$ with
$\textsf{\textbf{U}}_{ef}   = \sum_{i=1}^n{\bf e}_i  {\bf f}_i^\dagger
 = \sum_{i=1}^n \vert {\bf e}_i \rangle \langle {\bf f}_i \vert$.

Conversely,  since
\begin{equation}
\begin{split}
\textsf{\textbf{U}}_{ef}^\ast
= \sum_{i=1}^n \left(\vert {\bf e}_i \rangle \langle {\bf f}_i \vert \right)^\ast
= \sum_{i=1}^n \left( \langle {\bf f}_i \vert \right)^\ast \left(\vert {\bf e}_i \rangle \right)^\ast
= \sum_{i=1}^n \vert {\bf f}_i \rangle \langle {\bf e}_i \vert = \textsf{\textbf{U}}_{fe},
\end{split}
\label{2016-m-ch-uniteq7134}
\end{equation}   and   therefore
\begin{equation}
\begin{split}
\textsf{\textbf{U}}_{ef}^\ast \textsf{\textbf{U}}_{ef}= \textsf{\textbf{U}}_{fe} \textsf{\textbf{U}}_{ef}
\\
=
\left(\vert {\bf f}_i \rangle \langle {\bf e}_i \vert  \right)
\left(\vert {\bf e}_j \rangle \langle {\bf f}_j \vert \right)
=
 \vert {\bf f}_i \rangle \underbrace{\langle {\bf e}_i \vert   {\bf e}_j \rangle }_{=\delta_{ij}}
\langle {\bf f}_j \vert
=
 \vert {\bf f}_i \rangle   \langle {\bf f}_i \vert
= \mathbb{1},
\end{split}
\label{2016-m-ch-uniteq71345}
\end{equation}
so that
$\textsf{\textbf{U}}_{ef}^{-1} =\textsf{\textbf{U}}_{ef}^\ast$.

An alternative proof of sufficiency makes use of the fact that, if both $\textsf{\textbf{U}} {\bf f}_i$ are  orthonormal bases with ${\bf f}_i \in {\frak B}$
and  $\textsf{\textbf{U}}{\bf f}_i \in \textsf{\textbf{U}}{\frak B}= {\frak B}'$,
so that
$
\langle \textsf{\textbf{U}} {\bf f}_i \mid \textsf{\textbf{U}} {\bf f}_j \rangle
=
\langle {\bf f}_i \mid  {\bf f}_j \rangle$,
then by linearity
$
\langle \textsf{\textbf{U}} {\bf x} \mid \textsf{\textbf{U}} {\bf y} \rangle
=
\langle  {\bf x} \mid  {\bf y} \rangle $
for all ${\bf x},  {\bf y}$,
thus proving (ii) from (iv).

\eproof
}

Note that $\textsf{\textbf{U}}$ preserves length or distances and thus is an {\em isometry}, as for all ${\bf x}, {\bf y}$,
\begin{equation}
\begin{split}
\| \textsf{\textbf{U}}{\bf x} - \textsf{\textbf{U}}{\bf y} \| =
\| \textsf{\textbf{U}}\left( {\bf x} - {\bf y} \right) \| =
\|{\bf x}- {\bf y}\|
.
\end{split}
\label{2016-m-ch-uniteqpl}
\end{equation}

Note also that $\textsf{\textbf{U}}$ preserves the angle $\theta$ between two nonzero vectors ${\bf x}$ and ${\bf y}$  defined by
\begin{equation}
\begin{split}
\cos \theta = \frac{ \langle {\bf x} \mid {\bf y} \rangle } {\|{\bf x}\| \| {\bf y}\| }
\end{split}
\label{2016-m-ch-unitanglepres}
\end{equation}
as it preserves the inner product and the norm.

{\color{blue}
\bexample
Since unitary transformations can also be defined via {\em one-to-one transformations preserving the scalar product,}
functions such as
$f: x \mapsto x' =\alpha x$ with $\alpha \neq e^{i\varphi}$, $\varphi \in {\Bbb R}$,
do not correspond to a  unitary transformation in a one-dimensional Hilbert space, as
the scalar product $f:
\langle x \vert y \rangle
\mapsto
\langle x'\vert y'\rangle = \vert \alpha \vert^2 \langle x\vert y\rangle$
is not preserved; whereas if $\alpha$ is a modulus of one; that is,
with $\alpha = e^{i\varphi}$, $\varphi \in {\Bbb R}$,
$\vert \alpha \vert^2=1$, and the scalar product is preserved.
Thus, $u: x \mapsto x' =e^{i\varphi} x$, $\varphi \in {\Bbb R}$,
represents a unitary transformation.
\eexample
}


\subsection {Characterization in terms of orthonormal basis}
\label{2012-m-ch-citoob}

A complex matrix $\textsf{\textbf{U}}$ is unitary if and only if its row (or column) vectors form
an orthonormal basis.

This can be readily verified\cite{Schwinger.60} by writing $\textsf{\textbf{U}}$
in terms of two orthonormal bases
${\frak B}=\{{\bf e}_1,  {\bf e}_2, \ldots , {\bf e}_n\}\equiv \{\vert {\bf e}_1\rangle , \vert  {\bf e}_2\rangle , \ldots , \vert {\bf e}_n\rangle \}$
${\frak B}'=\{{\bf f}_1,  {\bf f}_2, \ldots , {\bf f}_n\}\equiv \{\vert {\bf f}_1\rangle , \vert  {\bf f}_2\rangle , \ldots , \vert {\bf f}_n\rangle \}$ as
\begin{equation}
\textsf{\textbf{U}}_{ef}= \sum_{i=1}^n  {\bf e}_i {\bf f}_i^\dagger
\equiv \sum_{i=1}^n  \vert {\bf e}_i\rangle \langle {\bf f}_i \vert
.
\label{2019-ch-fdvs-ditoonb}
\end{equation}
Together with $\textsf{\textbf{U}}_{fe}= \sum_{i=1}^n  {\bf f}_i {\bf e}_i^\dagger \equiv  \sum_{i=1}^n  \vert {\bf f}_i\rangle \langle {\bf e}_i \vert $
we form
\begin{equation}
\begin{split}
{\bf e}_k^\dagger \textsf{\textbf{U}}_{ef}  = {\bf e}_k^\dagger\sum_{i=1}^n  {\bf e}_i {\bf f}_i^\dagger
= \sum_{i=1}^n  ({\bf e}_k^\dagger {\bf e}_i) {\bf f}_i^\dagger
= \sum_{i=1}^n  \delta_{ki} {\bf f}_i^\dagger   = {\bf f}_k^\dagger
.
\end{split}
\end{equation}
In a similar way we find that
\begin{equation}
\begin{split}
\textsf{\textbf{U}}_{ef} {\bf f}_k = {\bf e}_k,
{\bf f}_k^\dagger \textsf{\textbf{U}}_{fe}   = {\bf e}_k^\dagger ,
\textsf{\textbf{U}}_{fe} {\bf e}_k  = {\bf f}_k .
\end{split}
\end{equation}
Moreover,
\begin{equation}
\begin{split}
\textsf{\textbf{U}}_{ef}\textsf{\textbf{U}}_{fe}
=
 \sum_{i=1}^n  \sum_{j=1}^n
(\vert {\bf e}_i\rangle \langle {\bf f}_i \vert )
(\vert {\bf f}_j\rangle \langle {\bf e}_j \vert )
=
 \sum_{i=1}^n  \sum_{j=1}^n
\vert {\bf e}_i\rangle \delta_{ij} \langle {\bf e}_j \vert
=
 \sum_{i=1}^n
\vert {\bf e}_i\rangle   \langle {\bf e}_i \vert
=
\mathbb{1}
.
\end{split}
\end{equation}
In a similar way we obtain
$\textsf{\textbf{U}}_{fe}\textsf{\textbf{U}}_{ef}=
\mathbb{1}$.
Since
\begin{equation}
\textsf{\textbf{U}}_{ef}^\dagger = \sum_{i=1}^n  ( {\bf f}_i^\dagger )^\dagger {\bf e}_i^\dagger
= \sum_{i=1}^n  {\bf f}_i {\bf e}_i^\dagger
= \textsf{\textbf{U}}_{fe},
\end{equation}
we obtain that $\textsf{\textbf{U}}_{ef}^\dagger = (\textsf{\textbf{U}}_{ef})^{-1}$
and $\textsf{\textbf{U}}_{fe}^\dagger = (\textsf{\textbf{U}}_{fe})^{-1}$.

Note also that the {\em composition} holds; that is, $\textsf{\textbf{U}}_{ef} \textsf{\textbf{U}}_{fg}=  \textsf{\textbf{U}}_{eg}$.

If we
identify one of the bases  ${\frak B}$ and ${\frak B}'$ by the Cartesian standard basis,
it becomes clear that, for instance,
every unitary operator  $\textsf{\textbf{U}}$  can be written in terms of an orthonormal basis  of the dual space
${\frak B}^\ast=\{ \langle {\bf f}_1  \vert     ,   \langle {\bf f}_2  \vert      \ldots ,  \langle {\bf f}_n  \vert     \}$
by ``stacking'' the conjugate transpose vectors of that orthonormal basis ``on top of each other;''\marginnote{For a quantum mechanical application, see \bibentry{rzbb}}
\index{conjugate transpose}
\index{Hermitian conjugate}
\index{Hermitian adjoint}
that is, by identifying the basis vectors $\vert {\bf e}_i  \rangle $ with elements of the Cartesian standard basis
\index{Cartesian basis}
\index{standard basis}
\marginnote{For proofs and additional information see
 \S 5.11.3, Theorem 5.1.5 and subsequent Corollary in~\bibentry{Joglekar-I}.}
\begin{equation}
\textsf{\textbf{U}}
\equiv
\begin{pmatrix}
1\\
0\\
\vdots\\
0
\end{pmatrix} {\bf f}_1^\dagger
+
\begin{pmatrix}
0\\
1\\
\vdots\\
0
\end{pmatrix} {\bf f}_2^\dagger
+
\cdots +
\begin{pmatrix}
0\\
0\\
\vdots\\
n
\end{pmatrix} {\bf f}_n^\dagger
\equiv
\begin{pmatrix}
{\bf f}_1^\dagger\\
{\bf f}_2^\dagger\\
\vdots\\
{\bf f}_n^\dagger
\end{pmatrix}
\equiv
\begin{pmatrix}
\langle {\bf f}_1\vert \\
\langle {\bf f}_2\vert\\
\vdots\\
\langle {\bf f}_n\vert
\end{pmatrix}
.
\end{equation}
Thereby the conjugate transpose vectors of the orthonormal basis  ${\frak B}$ serve as the
\index{conjugate transpose}
\index{Hermitian conjugate}
\index{Hermitian adjoint}
rows of $\textsf{\textbf{U}}$.

In a similar manner, every unitary operator  $\textsf{\textbf{U}}$  can be written in terms of an orthonormal basis
${\frak B}=\{{\bf e}_1,  {\bf e}_2, \ldots , {\bf e}_n\}$
by ``pasting'' the  vectors of that orthonormal basis ``one after another;''
that is,  by identifying the basis vectors $\langle {\bf f}_i \vert $ of the dual space
with (transposed) elements of the Cartesian standard basis
\begin{equation}
\begin{split}
\textsf{\textbf{U}}
\equiv
{\bf e}_1 \begin{pmatrix} 1,0,\ldots ,0\end{pmatrix} +
{\bf e}_2 \begin{pmatrix} 0,1,\ldots ,0\end{pmatrix} +
\cdots +
{\bf e}_n \begin{pmatrix} 0,0,\ldots ,1\end{pmatrix}   \\
\equiv
\begin{pmatrix}
{\bf e}_1,
{\bf e}_2,
\cdots,
{\bf e}_n
\end{pmatrix}
\equiv
\begin{pmatrix}
\vert {\bf e}_1 \rangle ,
\vert {\bf e}_2 \rangle ,
\cdots,
\vert {\bf e}_n \rangle
\end{pmatrix}
.
\end{split}
\label{2015-m-ch-fdlvs-uniascolv}
\end{equation}
Thereby the vectors of the orthonormal basis  ${\frak B}$ serve as the
columns of $\textsf{\textbf{U}}$.
Note that any permutation of vectors in ${\frak B}$ would also yield unitary matrices.

\subsection{Householder transformation}
\index{Householder transformation}
\label{2021-m-ch-hposholder}

Let $\vert {\bf x} \rangle \in \mathbb{C}^n$ be a nonzero vector. The
Householder transformation\cite{Horn-Johnson-MatrixAnalysis} $\textsf{\textbf{U}}_{\bf x}$ is defined by
\begin{equation}
\textsf{\textbf{U}}_{\bf x}
=
\mathbb{1}- 2 ( \langle {\bf x} \vert   {\bf x} \rangle )^{-1} \vert {\bf x} \rangle \langle  {\bf x}  \vert
\equiv
\mathbb{1}- 2 ( {\bf x}^\dagger   {\bf x} )^{-1}  {\bf x}    {\bf x}^\dagger .
\end{equation}
 If  $\vert {\bf x} \rangle $ is a unit vector, then
$\textsf{\textbf{U}}_{\bf x}
=
\mathbb{1}- 2  \vert {\bf x} \rangle \langle  {\bf x}  \vert
\equiv
\mathbb{1}- 2   {\bf x}    {\bf x}^\dagger $.

The following properties can be asserted by direct proofs:
\begin{itemize}
\item[(i)]
$\textsf{\textbf{U}}_{\bf x}$ is Hermitian; that is,
$
\textsf{\textbf{U}}_{\bf x} = \textsf{\textbf{U}}_{\bf x}^\dagger
$;

\item[(ii)]
$\textsf{\textbf{U}}_{\bf x}$ is unitary; that is,
$
\textsf{\textbf{U}}_{\bf x}  \textsf{\textbf{U}}_{\bf x}^\dagger
=
\textsf{\textbf{U}}_{\bf x}^\dagger  \textsf{\textbf{U}}_{\bf x}
=
\textsf{\textbf{U}}_{\bf x}  \textsf{\textbf{U}}_{\bf x}
=
\left(\mathbb{1}- 2 ( \langle {\bf x} \vert   {\bf x} \rangle )^{-1} \vert {\bf x} \rangle \langle  {\bf x}  \vert\right)
\left(\mathbb{1}- 2 ( \langle {\bf x} \vert   {\bf x} \rangle )^{-1} \vert {\bf x} \rangle \langle  {\bf x}  \vert\right)
= \mathbb{1} - 4 ( \langle {\bf x} \vert   {\bf x} \rangle )^{-1} \vert {\bf x} \rangle \langle  {\bf x}  \vert
+ 4 ( \langle {\bf x} \vert   {\bf x} \rangle )^{-1} \vert {\bf x} \rangle \langle  {\bf x}  \vert =
\mathbb{1}
$.

\item[(iii)]
Hence $\textsf{\textbf{U}}_{\bf x}$ is involutory:
$\textsf{\textbf{U}}_{\bf x}^{-1} =  \textsf{\textbf{U}}_{\bf x}$.
\index{involution}

\item[(iv)]
The eigensystem of $\textsf{\textbf{U}}_{\bf x}$ has two eigenvalues $\pm 1$:
\begin{itemize}
\item[eigenvalue $-1$:]
for the eigenvector $ \vert {\bf x} \rangle$ of    $\textsf{\textbf{U}}_{\bf x}$,
with
$\textsf{\textbf{U}}_{\bf x}\vert {\bf x} \rangle
=
\left(\mathbb{1}- 2 ( \langle {\bf x} \vert   {\bf x} \rangle )^{-1} \vert {\bf x} \rangle \langle  {\bf x}  \vert\right)\vert {\bf x} \rangle
= \vert {\bf x} \rangle- 2 \vert {\bf x} \rangle =- \vert {\bf x} \rangle$
the associated eigenvalue is $-1$.
\item[eigenvalue(s) $+1$:]
The remaining $n-1$ mutually orthogonal eigenvectors span the $n-1$ dimensional subspace orthogonal to $\vert {\bf x} \rangle $.
Every vector in that subspace has eigenvalue $+1$.
(For $n>2$ the spectrum is degenerate.)
\end{itemize}

Stated differently: for all vectors orthogonal to $\vert {\bf x} \rangle $ the
Householder transformation $\textsf{\textbf{U}}_{\bf x}$ acts as identity;
and for $\vert {\bf x} \rangle $ the
Householder transformation $\textsf{\textbf{U}}_{\bf x}$
acts as a reflection on the one-dimensional subspace spanned by $\vert {\bf x} \rangle $.

\item[(v)]
Since the determinant of a matrix is the product of its eigenvalues,
the determinant of a Householder transformation is $-1$.

\item[(vi)]
If
${\frak B}=\{{\bf e}_1,  {\bf e}_2, \ldots , {\bf e}_n\}\equiv \{\vert {\bf e}_1\rangle ,
\vert  {\bf e}_2\rangle , \ldots , \vert {\bf e}_n\rangle \}$
is an orthonormal basis, then the succession
of the respective Householder transformations renders negative unity; that is,
\begin{equation}
\begin{split}
\textsf{\textbf{U}}_{{\bf e}_1} \textsf{\textbf{U}}_{{\bf e}_2} \cdots \textsf{\textbf{U}}_{{\bf e}_n}
=
\left(\mathbb{1}- 2  \vert {\bf e}_1 \rangle \langle  {\bf e}_1  \vert\right)
\left(\mathbb{1}- 2  \vert {\bf e}_2 \rangle \langle  {\bf e}_2  \vert\right)
\cdots
\left(\mathbb{1}- 2  \vert {\bf e}_n \rangle \langle  {\bf e}_n  \vert\right) \\
=
\mathbb{1}- 2 \underbrace{\left( \vert {\bf e}_1 \rangle \langle  {\bf e}_1 \vert +
 \vert {\bf e}_2 \rangle \langle  {\bf e}_2  \vert +
\cdots
+  \vert {\bf e}_n \rangle \langle  {\bf e}_n  \vert\right)}_{\mathbb{1}}=
-\mathbb{1}.
\end{split}
\end{equation}
\end{itemize}

{
\color{blue}
\bexample
For the sake of an example, let
 $\vert {\bf z} \rangle =\begin{pmatrix}1,1\end{pmatrix}^\intercal$,
so that the corresponding Housholder transformation can be written in matrix form as
\[
\textsf{\textbf{U}}_{\bf z}
= \mathbb{1}- 2  ( \langle {\bf z} \vert   {\bf z} \rangle )^{-1} \vert {\bf z} \rangle \langle  {\bf z}  \vert
\equiv \begin{pmatrix}1&0\\0&1\end{pmatrix} -2(2)^{-1} \begin{pmatrix}1&1\\1&1\end{pmatrix}
= -\begin{pmatrix}0&1\\1&0\end{pmatrix}.\]
{\color{black}
\begin{marginfigure}
\begin{center}%
\resizebox{1\textwidth}{!}{
\begin{tikzpicture}  [scale=1]

\tikzstyle{every path}=[line width=1pt]

\newdimen\ms
\ms=0.1cm
\tikzstyle{s1}=[color=red,rectangle,inner sep=3.5]
\tikzstyle{c3}=[circle,inner sep={\ms/8},minimum size=4*\ms]
\tikzstyle{c2}=[circle,inner sep={\ms/8},minimum size=3*\ms]
\tikzstyle{c1}=[circle,inner sep={\ms/8},minimum size=2*\ms]
\tikzstyle{cs1}=[circle,inner sep={\ms/8},minimum size=1*\ms]


\coordinate (zero) at (0,0);
\coordinate (v) at (1,1);
\coordinate (x) at (1,2);
\coordinate (y) at (-2,-1);


  \draw[thin,gray!40] (-2,-2) grid (2,2);
  \draw[<->] (-2,0)--(2,0) node[right]{$x_1$};
  \draw[<->] (0,-2)--(0,2) node[above]{$x_2$};


 \draw[dashed, line width=1pt,blue!40](-2,2)--(2,-2);
 \draw[dotted, line width=1pt,gray!60](-2,-2)--(2,2);

 \draw[dashed, line width=1pt,gray!40](x)--(y);
 \draw[dashed, line width=1pt,gray!40](x)--(1.5,1.5);
 \draw[dashed, line width=1pt,gray!40](y)--(-1.5,-1.5);

 \draw[->,line width=2pt,blue](zero)--(v) node[label=below right:{$\vert {\bf z}\rangle $}] {};

 \draw[->,line width=2pt,green](zero)--(x) node[label=above right:{$\vert {\bf x}\rangle $}] {};

 \draw[->,line width=2pt,red](zero)--(y) node[label=below left:{$\vert {\bf y}\rangle $}] {};

\end{tikzpicture}
}
\end{center}
\caption{\label{2021-mm-fdvs-housholder}
Depiction of the Householder transformation $\textsf{\textbf{U}}_{\bf z}$
with
$\vert {\bf z} \rangle =\begin{pmatrix}1,1\end{pmatrix}^\intercal$
acting on a vector $\vert {\bf x} \rangle =\begin{pmatrix}2,1\end{pmatrix}^\intercal$.
The resulting ``reflected'' vector $\vert {\bf y} \rangle = \textsf{\textbf{U}}_{\bf z} \vert {\bf x} \rangle$
and the original vector $\vert {\bf x} \rangle$
have the same length or norm.
Its component along $\vert {\bf z} \rangle$ is reversed, whereas its component orthogonal to
$\vert {\bf z} \rangle$ remains the same.}
\end{marginfigure}
}

Take
 $\vert {\bf x} \rangle =\begin{pmatrix}2,1\end{pmatrix}^\intercal$,
so that
 $\vert {\bf y} \rangle =-\begin{pmatrix}1,2\end{pmatrix}^\intercal$:
this ``reflected'' vector $\vert {\bf y} \rangle$ and the original vector $\vert {\bf x} \rangle$
have the same length or norm.
The component of $\vert {\bf y} \rangle$  along $\vert {\bf z} \rangle$ is reversed, whereas its component orthogonal to
$\vert {\bf z} \rangle$ remains the same.
This situation is depicted in Figure~\ref{2021-mm-fdvs-housholder}.
\eexample
}

As a consequence of (iii), if $\vert {\bf x} \rangle \neq \vert {\bf y} \rangle$ are two vectors  in $\mathbb{R}^n$
with identical length or norm
$\| {\bf x} \| = \| {\bf y} \|$
then there exists a remarkable ``symmetry delivered by'' a Householder transformation $\textsf{\textbf{U}}_{\bf z}$ such that
$\textsf{\textbf{U}}_{\bf z} \vert {\bf x} \rangle =  \vert {\bf y} \rangle$
and
$\textsf{\textbf{U}}_{\bf z}\textsf{\textbf{U}}_{\bf z} \vert {\bf x} \rangle =  \textsf{\textbf{U}}_{\bf z}\vert {\bf y} \rangle
=  \vert {\bf x} \rangle$.
For this to hold the vector $\vert {\bf z} \rangle$ needs to be a  vector equal to $\vert {\bf x} \rangle - \vert {\bf y} \rangle$:
$  \left( \mathbb{1}- 2  ( \langle {\bf z} \vert   {\bf z} \rangle )^{-1} \vert {\bf z} \rangle \langle  {\bf z}  \vert  \right) \vert {\bf x} \rangle =  \vert {\bf y} \rangle
$
and
$ \vert {\bf x} \rangle  =   \left( \mathbb{1}- 2  ( \langle {\bf z} \vert   {\bf z} \rangle )^{-1}\vert {\bf z} \rangle \langle  {\bf z}  \vert  \right) \vert {\bf y} \rangle
$,
resulting in
$( \langle {\bf z} \vert   {\bf z} \rangle )^{-1} \vert {\bf z} \rangle \langle  {\bf z}  \vert  \left( \vert {\bf x} \rangle  - \vert {\bf y} \rangle \right)
= \vert {\bf x} \rangle  - \vert {\bf y} \rangle
$, and thus $\vert {\bf z} \rangle =  \vert {\bf x} \rangle  - \vert {\bf y} \rangle$.
(For $\vert {\bf x} \rangle = \vert {\bf y} \rangle$ identify with $\vert {\bf z} \rangle$ a vector orthogonal to $\vert {\bf x} \rangle = \vert {\bf y} \rangle$.)
This is not true for $\mathbb{C}^n$, as for instance, there exists no $ \vert {\bf z} \rangle $ which would render
$\textsf{\textbf{U}}_{\bf z} \vert {\bf x} \rangle =  i \vert {\bf x} \rangle$ for nonzero  $ \vert {\bf x} \rangle $,
and an additional unitary transformation is required.

This gives rise to the orthonormalizion of a set of $k$ linear independent nonzero vectors
${\frak S}=\{\vert {\bf s}_1\rangle ,
\vert  {\bf s}_2\rangle , \ldots , \vert {\bf s}_k\rangle \}$   in $\mathbb{R}^n$
by taking some orthonormal basis
${\frak B}=\{{\bf e}_1,  {\bf e}_2, \ldots , {\bf e}_n\}\equiv \{\vert {\bf e}_1\rangle ,
\vert  {\bf e}_2\rangle , \ldots , \vert {\bf e}_n\rangle \}$,
choosing $k$ vectors thereof---say, the first $k$ vectors of the standard Cartesian coordinate system---and identifying
$\vert {\bf s}_i\rangle$ with $\vert {\bf x}_i\rangle$,
and (the extra factor $\| {\bf s}_i \|$ serves to make the vector of equal length or norm)
$\vert {\bf y}_i\rangle$ with $\| {\bf s}_i \| \vert {\bf e}_i\rangle$,
thereby constructing a Housholder transformation followed by normalization (through division by $\| {\bf s}_i \|$)
$\textsf{\textbf{U}}_{{\bf z}_i}$ of $\vert {\bf s}_i\rangle \stackrel{\textsf{\textbf{U}}_{{\bf z}_i}}{\mapsto} \vert {\bf e}_i\rangle$
with respective
$\vert {\bf z}_i\rangle = \vert {\bf s}_i\rangle  - \| {\bf s}_i \| \vert {\bf e}_i\rangle$.
This kind of orthonormalization may yield a span ``outside'' of the  subspace  spanned by the ``original'' vectors.

\section{Orthonormal (orthogonal) transformation}
\index{orthonormal transformation}
\index{orthogonal transformation}
\label{2015-m-ch-fdlvs-orthproj}

Orthonormal (orthogonal) transformations are special cases of unitary transformations restricted to {\em real} Hilbert space.

An {\em orthonormal} or {\em orthogonal transformation} $\textsf{\textbf{R}}$ is a linear transformation
whose corresponding square matrix $R$ has real-valued entries
and mutually orthogonal, normalized row (or, equivalently, column) vectors.
As a consequence (see the equivalence of definitions of unitary definitions and the proofs mentioned earlier),
\begin{equation}
\textsf{\textbf{R}}\textsf{\textbf{R}}^\intercal = \textsf{\textbf{R}}^\intercal \textsf{\textbf{R}}= \mathbb{1},
\textrm{ or } \textsf{\textbf{R}}^{-1}=\textsf{\textbf{R}}^\intercal  .
\end{equation}
As all unitary transformations, orthonormal transformations $\textsf{\textbf{R}}$
preserve a symmetric inner product as well as the norm.

If $\textrm{det} \textsf{\textbf{R}}=1$, $\textsf{\textbf{R}}$ corresponds to a {\em rotation.}
\index{rotation}
If $\textrm{det} \textsf{\textbf{R}}=-1$, $\textsf{\textbf{R}}$ corresponds to a rotation and a {\em reflection.}
\index{reflection}
A reflection is an isometry (a distance preserving map) with a hyperplane as set of fixed points.

As a special case of the decomposition~(\ref{2019-ch-fdvs-ditoonb}) of unitary transformations,
orthogonal transformations ave a decomposition
in terms of two orthonormal bases whose elements have real-valued components
${\frak B}=\{{\bf e}_1,  {\bf e}_2, \ldots , {\bf e}_n\}\equiv \{\vert {\bf e}_1\rangle , \vert  {\bf e}_2\rangle , \ldots , \vert {\bf e}_n\rangle \}$
${\frak B}'=\{{\bf f}_1,  {\bf f}_2, \ldots , {\bf f}_n\}\equiv \{\vert {\bf f}_1\rangle , \vert  {\bf f}_2\rangle , \ldots , \vert {\bf f}_n\rangle \}$,
such that
\begin{equation}
\textsf{\textbf{R}}_{ef}= \sum_{i=1}^n  {\bf e}_i {\bf f}_i^\intercal
\equiv \sum_{i=1}^n  \vert {\bf e}_i\rangle \langle {\bf f}_i \vert
.
\label{2019-ch-fdvs-ditoonbr}
\end{equation}

{\color{blue}
\bexample
For the sake of a two-dimensional  example of rotations in the plane ${\Bbb R}^2$,
take the rotation matrix in Equation~(\ref{2012-m-ch-fdvs-otd2})
representing a rotation of the basis by an angle $\varphi$.

\eexample
}

\section{Permutation}
\index{permutation}
\label{2018-permutation}

Permutations are  ``discrete'' orthogonal transformations
``restricted to binary values''
in the sense that
they merely allow the entries ``$0$'' and ``$1$'' in their respective matrix representations.
With regards to classical and quantum bits\cite{mermin-04,mermin-07}
they serve as a sort of ``reversible classical analog'' for classical reversible computation,
as compared to the more general, continuous unitary transformations of quantum bits introduced earlier.

{\color{blue}
\bexample
Permutation matrices are defined by the requirement that they only contain a single nonvanishing entry ``$1$'' per row and column;
all the other row and column entries vanish; that is, the respective matrix entries are ``$0$.''
For example, the matrices $\mathbb{1}_n=\textrm{diag}(\underbrace{1,\ldots ,1}_{n \textrm{ times}})$,
or
\begin{equation}
\sigma_1=
\begin{pmatrix}
0&1\\
1&0
\end{pmatrix}
\textrm{, or }\;
\begin{pmatrix}
0&1&0\\
1&0&0\\
0&0&1
\end{pmatrix}
\end{equation}
are permutation matrices.
\eexample
}

From the definition and from matrix multiplication follows that,
if $\textsf{\textbf{P}}$ is a permutation represented by its permutation matrix,
then $\textsf{\textbf{P}}\textsf{\textbf{P}}^\intercal =\textsf{\textbf{P}}^\intercal  \textsf{\textbf{P}}=\mathbb{1}_n$.
That is, $\textsf{\textbf{P}}^\intercal $ represents the inverse element of $\textsf{\textbf{P}}$.
As $\textsf{\textbf{P}}$ is real (actually, binary)-valued, it is a {\em normal operator} (cf. page \pageref{2014-m-fdvs-normality}).
\index{normal operator}
\index{normal transformation}

Just as for unitary and orthogonal transformations~(\ref{2019-ch-fdvs-ditoonb})
and~(\ref{2019-ch-fdvs-ditoonbr}),
any permutation matrix can be decomposed as sums of tensor products of row and (dual) column vectors:
The set of all these row and column vectors
with permuted elements:
Suppose
${\frak B}=\{{\bf e}_1,  {\bf e}_2, \ldots , {\bf e}_n\}\equiv \{\vert {\bf e}_1\rangle , \vert  {\bf e}_2\rangle , \ldots , \vert {\bf e}_n\rangle \}$
${\frak B}'=\{{\bf f}_1,  {\bf f}_2, \ldots , {\bf f}_n\}\equiv \{\vert {\bf f}_1\rangle , \vert  {\bf f}_2\rangle , \ldots , \vert {\bf f}_n\rangle \}$,
represent
Cartesian standard basis of $n$-dimensional vector space
and an orthonormal basis whose elements are permutations of elements thereof,
respectively; such that, if $\pi (i)$ stands for the permutation of $i$, ${\bf f}_i = {\bf e}_{\pi (i)}$.
Then
\begin{equation}
\textsf{\textbf{P}}_{ef}= \sum_{i=1}^n  {\bf e}_i {\bf f}_i^\intercal
\equiv \sum_{i=1}^n  \vert {\bf e}_i\rangle \langle {\bf e}_{\pi (i)} \vert
\equiv
\begin{pmatrix}
\langle {\bf e}_{\pi (1)} \vert \\
\langle {\bf e}_{\pi (2)} \vert \\
\vdots \\
\langle {\bf e}_{\pi (n)} \vert \\
\end{pmatrix}
.
\label{2019-ch-fdvs-ditoonbp}
\end{equation}

If $P$ and $Q$ are permutation matrices, so is $PQ$ and $QP$.
The set of all $n!$
permutation $(n\times n)-$matrices corresponding to permutations of $n$ elements of $\{ 1,2,\ldots ,n\}$ form the
{\em symmetric group $S_n$}, with $\mathbb{1}_n$ being the identity element.
\index{symmetric group}

The space spanned the permutation matrices is $\left[(n-1)^2+1\right]$-dimensional;
with $n!>(n-1)^2+1$ for $n>2$.
Therefore,  the bound from above can be improved such that decompositions with $k \le (n-1)^2 +1= n^2-2(n+1)$
exist.\cite[-20mm]{Marcus-Ree-1959}

{\color{blue}
\bexample
For instance, the identity matrix in three dimensions is a permutation
and can be written in terms of the other permutations as
\begin{equation}
\begin{split}
\begin{pmatrix}
1&0&0\\
0&1&0\\
0&0&1
\end{pmatrix}
=
\begin{pmatrix}
1&0&0\\
0&0&1\\
0&1&0
\end{pmatrix}
+
\begin{pmatrix}
0&1&0\\
1&0&0\\
0&0&1
\end{pmatrix}\qquad
\\
-
\begin{pmatrix}
0&1&0\\
0&0&1\\
1&0&0
\end{pmatrix}
-
\begin{pmatrix}
0&0&1\\
1&0&0\\
0&1&0
\end{pmatrix}
+
\begin{pmatrix}
0&0&1\\
0&1&0\\
1&0&0
\end{pmatrix}
.
\end{split}
\end{equation}
\eexample
}

\section{Projection or projection operator}
\label{2011-m-projec}

\index{projection}
\index{projector}
\index{projection operator}

The more I learned about quantum mechanics the more
I realized the importance of projection operators for its conceptualization:\cite[0mm]{v-neumann-49,birkhoff-36}
\begin{itemize}
\item[(i)]
Pure quantum states
\index{pure state}
are represented by a very particular kind of projections;
namely, those that are of the {\em trace class one,} meaning their trace (cf. Section~\ref{2013-ch-fdvs-trace}) is one,
as well as being {\em positive}
(cf. Section~\ref{2015-m-ch-fdlvs-positive}).
Positivity implies
that the projection is {\em self-adjoint} (cf. Section~\ref{2015-m-ch-fdlvs-self-adjoint}),
which is equivalent to the projection being {\em orthogonal}.

{\em Mixed} quantum states
\index{mixed state}
are compositions -- actually, nontrivial convex combinations~\marginnote{For a proof,  see pages 52--53 of~\bibentry{ba-89}.} -- of (pure) quantum states; again they are of the trace class one, self-adjoint, and positive;
yet unlike pure states, they are no projectors (that is, they are not idempotent);
and the trace of their square is not one (indeed, it is less than one).
\item[(ii)]
Mixed states, should they ontologically exist, can be composed of projections by summing over projectors.
\item[(iii)]
Projectors serve as the most elementary observables -- they correspond to yes-no propositions.
\item[(iv)]
In Section~\ref{2012-m-ch-Spectraltheorem} we will learn
that every observable can be decomposed into weighted (spectral) sums of projections.
\item[(v)]
Furthermore, from dimension three onwards, Gleason's theorem (cf. Section~\ref{Gleasontheorem}) allows
quantum probability theory to be based upon maximal (in terms of co-measurability) ``quasi-classical''
blocks of projectors.
\item[(vi)]
Such maximal blocks of projectors can be bundled together to show (cf. Section~\ref{2011-m-KST})
that the corresponding algebraic
structure has no two-valued measure (interpretable as truth assignment), and
therefore cannot be ``embedded'' into a ``larger'' classical (Boolean) algebra.
\end{itemize}

\subsection{Definition}
\marginnote[-15mm]{For proofs and additional information see {\S}41 in~\bibentry{halmos-vs}.}
If ${\frak V}$ is the direct sum of some subspaces
${\frak M}$
and
${\frak N}$
so that every ${\bf z} \in {\frak V}$ can be uniquely written in the form
$
{\bf z}
=
{\bf x}
+
{\bf y}
$, with
${\bf x} \in {\frak M}$
and with
${\bf y} \in {\frak N}$,
then
the {\em projection}, or, used synonymously,
{\em projection operator}
on ${\frak M}$
along ${\frak N}$, is the transformation $\textsf{\textbf{E}}$
defined by $\textsf{\textbf{E}}{\bf z}={\bf x}$.
Conversely,
 $\textsf{\textbf{F}}{\bf z}={\bf y}$  is the projection
on ${\frak N}$
along ${\frak M}$.

A (nonzero) linear transformation $\textsf{\textbf{E}}$ is a {\em projector}
\index{projector}
if and only if one of
the following conditions is satisfied (then all the others are also satisfied):\cite{Trenkler1994}
\begin{itemize}
\item[(i)]
$\textsf{\textbf{E}}$ is idempotent; that is,
\index{idempotence}
$\textsf{\textbf{E}}\textsf{\textbf{E}}=\textsf{\textbf{E}}\neq 0$;

\item[(ii)]
$\textsf{\textbf{E}}^k$ is a projector for all $k \in \mathbb{N}$;

\item[(iii)]
$\textsf{\textbf{1}}-\textsf{\textbf{E}}$ is the {\em complimentary projection} with respect to $\textsf{\textbf{E}}$:
if $\textsf{\textbf{E}}$  is the projection
on ${\frak M}$
along ${\frak N}$,
 $\textsf{\textbf{1}}-\textsf{\textbf{E}}$ is the projection
on ${\frak N}$
along ${\frak M}$; in particular, $\left(\textsf{\textbf{1}}-\textsf{\textbf{E}}\right)\textsf{\textbf{E}}=\textsf{\textbf{E}}-\textsf{\textbf{E}}^2=\textsf{\textbf{E}}-\textsf{\textbf{E}}=0$.

\item[(iv)]
$\textsf{\textbf{E}}^\intercal$ is a projector;

\item[(v)]
$\textsf{\textbf{A}}= 2\textsf{\textbf{E}} - \textsf{\textbf{1}}$ is an involution;
that is, $\textsf{\textbf{A}}^2 = \mathbb{1}=\textsf{\textbf{1}}$;
\index{involution}
see also Section~\ref{2021-m-ch-hposholder} on Householder transformations;
\index{Householder transformation}

\item[(vi)]
$\textsf{\textbf{E}}$ admits the representation
\marginnote{See {\S}~5.8, Corollary~1 in~\bibentry{Lancaster-Tismenetsky}.}
\begin{equation}
\textsf{\textbf{E}} =\sum_{i=1}^k {\bf x}_i {\bf y}_i^\ast,
\label{2018-mm-ch-fdlvs-dop}
\end{equation}
where $k$ is the rank of $\textsf{\textbf{E}}$ and
$\{ {\bf x}_1, \ldots ,{\bf x}_k\}$
and
$\{ {\bf y}_1, \ldots ,{\bf y}_k\}$
are {\em biorthogonal} systems of vectors (not necessarily bases) of the vector space
\index{biorthogonality}
such that ${\bf y}_i^\ast  {\bf x}_j \equiv \langle {\bf y}_i \vert {\bf x}_j \rangle = \delta_{ij}$.
If the systems of vectors are identical; that is, if ${\bf y}_i={\bf x}_i$,
the products $ {\bf x}_i {\bf x}_i^\ast   \equiv \vert {\bf x}_i\rangle \langle {\bf x}_i \vert$ project onto one-dimensional subspaces
spanned by ${\bf x}_i$, and the projection is self-adjoint, and thus orthogonal.

\end{itemize}

{\color{OliveGreen}
\bproof
For a proof of (i) note that, if $\textsf{\textbf{E}}$  is the projection
on ${\frak M}$
along ${\frak N}$,
and if
$
{\bf z}
=
{\bf x}
+
{\bf y}
$, with
${\bf x} \in {\frak M}$
and with
${\bf y} \in {\frak N}$,
the decomposition of ${\bf x}$ yields
${\bf x}+0$, so that
$\textsf{\textbf{E}}^2{\bf z}=\textsf{\textbf{E}}\textsf{\textbf{E}}{\bf z}=\textsf{\textbf{E}}{\bf x}
={\bf x}=\textsf{\textbf{E}}{\bf z}$.
The converse --
idempotence
``$\textsf{\textbf{E}}\textsf{\textbf{E}}=\textsf{\textbf{E}}$''
implies that $\textsf{\textbf{E}}$ is a projection -- is more difficult to prove.

For the necessity of (iii) note that $(\textsf{\textbf{1}}-\textsf{\textbf{E}})^2
=\textsf{\textbf{1}}-\textsf{\textbf{E}}-\textsf{\textbf{E}}+ \textsf{\textbf{E}}^2
=\textsf{\textbf{1}}-\textsf{\textbf{E}}$;
furthermore,
$
\textsf{\textbf{E}}(\textsf{\textbf{1}}-\textsf{\textbf{E}})=
(\textsf{\textbf{1}}-\textsf{\textbf{E}})\textsf{\textbf{E}}=
\textsf{\textbf{E}}- \textsf{\textbf{E}}^2=0
$.

\eproof
}

\marginnote{The vector norm (\ref{2018-mm-ch-vn}) on page~\pageref{2018-mm-ch-vn} induces an operator norm
by $\| \textsf{\textbf{A}} \| = \sup_{\| {\bf x} \| =1} \| \textsf{\textbf{A}} {\bf x} \|$.}
We state without proof\cite{Szyld2006} that, for all projections
which are neither null nor the identity,
\index{norm}
the norm of its complementary projection
is identical with the norm of the projection; that is,
\begin{equation}
\left\| \textsf{\textbf{E}} \right\| = \left\| \textsf{\textbf{1}} - \textsf{\textbf{E}} \right\|
.
\end{equation}


\subsection{Orthogonal (perpendicular) projections}
\index{orthogonal projection}
\index{perpendicular projection}
\marginnote{For proofs and additional information see {\S}42, {\S}75 \& {\S}76 in~\bibentry{halmos-vs}.}

{\em Orthogonal,} or, used synonymously,
{\em perpendicular} projections
are associated with a {\em direct sum decomposition} of the vector space ${\frak V}$;
that is,
\begin{equation}
 {\frak M}\oplus {\frak M}^\perp ={\frak V},
\label{2012-m-ch-fdvs-perp}
\end{equation}
whereby $ {\frak M}= P_{\frak M}({\frak V})$
is the image of some projector $\textsf{\textbf{E}}=P_{\frak M}$
along ${\frak M}^\perp$, and  ${\frak M}^\perp$ is
the kernel of $P_{\frak M}$.
That is, ${\frak M}^\perp = \left\{{\bf x} \in {\frak V} \mid P_{\frak M}({\bf x}) = {\bf 0}\right\}$
\index{kernel}
is the subspace of ${\frak V}$
whose elements are mapped to the zero vector ${\bf 0}$ by $P_{\frak M}$.

Let us, for the sake of concreteness,
\marginnote{\url{http://faculty.uml.edu/dklain/projections.pdf}}
suppose that, in $n$-dimensional complex Hilbert space
${\Bbb C}^n$, we are given a $k$-dimensional subspace
\begin{equation}
{\frak M} = \textrm{span}\left( {\bf x}_1,\ldots ,{\bf x}_k \right)
\equiv \textrm{span}\left(\vert {\bf x}_1\rangle ,\ldots ,\vert {\bf x}_k \right\rangle )
\end{equation}
spanned
\index{linear span}
\index{span}
by  $k \le n$  linear independent base vectors ${\bf x}_1,\ldots ,{\bf x}_k$.
In addition, we are given another (arbitrary) vector ${\bf y} \in {\Bbb C}^n$.

Now consider the following question:
how can we project ${\bf y}$ onto ${\frak M}$ orthogonally (perpendicularly)?
That is, can we find a vector ${\bf y}' \in {\frak M}$ so that ${\bf y}^\perp = {\bf y} - {\bf y}'$
is orthogonal (perpendicular) to all of ${\frak M}$?

The orthogonality of ${\bf y}^\perp$ on the entire ${\frak M}$ can be rephrased in terms
of all the vectors ${\bf x}_1,\ldots ,{\bf x}_k$ spanning ${\frak M}$; that is,
for all ${\bf x}_i \in {\frak M}$, $1\le i\le k$
we must have
$\langle {\bf x}_i \vert {\bf y}^\perp \rangle = 0$.
This can be transformed into matrix algebra by considering the $n \times k$ matrix
[note that ${\bf x}_i$ are column vectors,
and recall the construction in Equation~(\ref{2015-m-ch-fdlvs-uniascolv})]
\begin{equation}
\textsf{\textbf{A}} = \begin{pmatrix}
{\bf x}_1, \ldots ,{\bf x}_k
\end{pmatrix}
\equiv
\begin{pmatrix}
\vert {\bf x}_1\rangle , \ldots , \vert {\bf x}_k \rangle
\end{pmatrix},
\end{equation}
and by requiring
\begin{equation}
\textsf{\textbf{A}}^\dagger \vert {\bf y}^\perp \rangle   \equiv
 \textsf{\textbf{A}}^\dagger  {{\bf y}^\perp} =
 \textsf{\textbf{A}}^\dagger  \left({\bf y} - {\bf y}'\right) =
 \textsf{\textbf{A}}^\dagger {\bf y}  - \textsf{\textbf{A}}^\dagger {{\bf y}'}   =
0,
\end{equation}
yielding
\begin{equation}
 \textsf{\textbf{A}}^\dagger  \vert  {\bf y}  \rangle
\equiv
 \textsf{\textbf{A}}^\dagger   {\bf y}
=
\textsf{\textbf{A}}^\dagger  {{\bf y}'}
\equiv
\textsf{\textbf{A}}^\dagger \vert {\bf y}' \rangle .
\label{2015-m-ch-cfvs-opab1}
\end{equation}

On the other hand, ${\bf y}'$ must be a linear combination of
${\bf x}_1,\ldots ,{\bf x}_k$ with the $k$-tuple of coefficients ${\bf c}$ defined by
\marginnote{Recall that
$(\textsf{\textbf{A}}\textsf{\textbf{B}})^\dagger =\textsf{\textbf{B}}^\dagger \textsf{\textbf{A}}^\dagger $,
and $(\textsf{\textbf{A}}^\dagger )^\dagger =\textsf{\textbf{A}}$.}
\begin{equation}
{\bf y}'
=
c_1 {\bf x}_1 + \cdots + c_k {\bf x}_k
=
\begin{pmatrix}{\bf x}_1, \ldots ,{\bf x}_k
\end{pmatrix}
\begin{pmatrix} c_1\\ \vdots \\ c_k
\end{pmatrix}
=
\textsf{\textbf{A}}  {\bf c}
.
\label{2015-m-ch-cfvs-opab22}
\end{equation}
Insertion into (\ref{2015-m-ch-cfvs-opab1}) yields
\begin{equation}
\textsf{\textbf{A}}^\dagger  {\bf y}
=
\textsf{\textbf{A}}^\dagger  \textsf{\textbf{A}} {\bf c}
.
\label{2015-m-ch-cfvs-opab2}
\end{equation}
Taking the inverse of $\textsf{\textbf{A}}^\dagger  \textsf{\textbf{A}} $
(this is a $k \times k$ diagonal matrix which is invertible, since
the $k$ vectors defining $\textsf{\textbf{A}}$ are linear independent),
and multiplying (\ref{2015-m-ch-cfvs-opab2}) from the left yields
\begin{equation}
{\bf c}
=
\left(\textsf{\textbf{A}}^\dagger  \textsf{\textbf{A}} \right)^{-1} \textsf{\textbf{A}}^\dagger {\bf y}
.
\label{2015-m-ch-cfvs-opab3}
\end{equation}
With (\ref{2015-m-ch-cfvs-opab22}) and (\ref{2015-m-ch-cfvs-opab3})
we find ${\bf y}'$ to be
\begin{equation}
{\bf y}'
=
\textsf{\textbf{A}}  {\bf c} =
\textsf{\textbf{A}}  \left(\textsf{\textbf{A}}^\dagger  \textsf{\textbf{A}} \right)^{-1} \textsf{\textbf{A}}^\dagger {\bf y}
.
\label{2015-m-ch-cfvs-opab4}
\end{equation}
We can define
\begin{equation}
\textsf{\textbf{E}}_{{\frak M}}=
\textsf{\textbf{A}}  \left(\textsf{\textbf{A}}^\dagger  \textsf{\textbf{A}} \right)^{-1} \textsf{\textbf{A}}^\dagger
\label{2015-m-ch-cfvs-opab5}
\end{equation}
to be the {\em projection matrix for the subspace ${\frak M}$}.
Note that
\begin{equation}
\begin{split}
\textsf{\textbf{E}}_{{\frak M}}^\dagger
=
\left[
\textsf{\textbf{A}}  \left(\textsf{\textbf{A}}^\dagger
\textsf{\textbf{A}} \right)^{-1} \textsf{\textbf{A}}^\dagger
\right]^\dagger
=
\textsf{\textbf{A}} \left[  \left(\textsf{\textbf{A}}^\dagger
\textsf{\textbf{A}} \right)^{-1}\right]^\dagger \textsf{\textbf{A}}^\dagger
=
\textsf{\textbf{A}} \left[ \textsf{\textbf{A}}^{-1}
\left(\textsf{\textbf{A}}^\dagger\right)^{-1} \right]^\dagger \textsf{\textbf{A}}^\dagger \\
=
\textsf{\textbf{A}} \textsf{\textbf{A}}^{-1}
\left(\textsf{\textbf{A}}^{-1}\right)^\dagger  \textsf{\textbf{A}}^\dagger
= \textsf{\textbf{A}} \textsf{\textbf{A}}^{-1}
\left(\textsf{\textbf{A}}^\dagger\right)^{-1}  \textsf{\textbf{A}}^\dagger
= \textsf{\textbf{A}} \left(\textsf{\textbf{A}}^\dagger
\textsf{\textbf{A}}\right)^{-1}  \textsf{\textbf{A}}^\dagger
=  \textsf{\textbf{E}}_{{\frak M}},
\end{split}
\label{2015-m-ch-cfvs-opab6}
\end{equation}
that is, $\textsf{\textbf{E}}_{{\frak M}}$ is self-adjoint and thus normal, as well as idempotent:
\begin{equation}
\begin{split}
\textsf{\textbf{E}}_{{\frak M}}^2
=
\left(
\textsf{\textbf{A}}  \left(\textsf{\textbf{A}}^\dagger
\textsf{\textbf{A}} \right)^{-1} \textsf{\textbf{A}}^\dagger
\right)
\left(
\textsf{\textbf{A}}  \left(\textsf{\textbf{A}}^\dagger
\textsf{\textbf{A}} \right)^{-1} \textsf{\textbf{A}}^\dagger
\right) \\
=
\textsf{\textbf{A}}^\dagger  \left(\textsf{\textbf{A}}^\dagger
\textsf{\textbf{A}} \right)^{-1} \left( \textsf{\textbf{A}}^\dagger
\textsf{\textbf{A}}  \right) \left(\textsf{\textbf{A}} ^\dagger
\textsf{\textbf{A}} \right)^{-1} \textsf{\textbf{A}}
=
\textsf{\textbf{A}}^\dagger   \left(\textsf{\textbf{A}} ^\dagger
\textsf{\textbf{A}} \right)^{-1} \textsf{\textbf{A}}
=
\textsf{\textbf{E}}_{{\frak M}}.
\end{split}
\label{2015-m-ch-cfvs-opab7}
\end{equation}

Conversely, every normal projection operator has a ``trivial'' spectral decomposition
(cf. Section~\ref{2012-m-ch-Spectraltheorem} on page \pageref{2012-m-ch-Spectraltheorem})
$\textsf{\textbf{E}}_{{\frak M}}
=
1 \cdot \textsf{\textbf{E}}_{{\frak M}} + 0 \cdot \textsf{\textbf{E}}_{{\frak M}^\perp}
=
1 \cdot \textsf{\textbf{E}}_{{\frak M}} + 0 \cdot \left(\textbf{1} - \textsf{\textbf{E}}_{{\frak M}}\right)$
associated with the two eigenvalues $0$ and $1$, and thus must be orthogonal.

If the basis ${\frak B}= \left\{{\bf x}_1,\ldots ,{\bf x}_k \right\}$ of ${\frak M}$
is orthonormal, then
\begin{equation}
\begin{split}
\textsf{\textbf{A}}^\dagger
\textsf{\textbf{A}}
\equiv
\begin{pmatrix}
\langle {\bf x}_1 \vert \\ \vdots \\ \langle {\bf x}_k \vert
\end{pmatrix}
\begin{pmatrix}
\vert {\bf x}_1\rangle , \ldots , \vert {\bf x}_k \rangle
\end{pmatrix}
=
\begin{pmatrix}
\langle {\bf x}_1 \vert {\bf x}_1\rangle   & \ldots &    \langle {\bf x}_1 \vert {\bf x}_k\rangle \\
\vdots &  \vdots &   \vdots \\
\langle {\bf x}_k \vert {\bf x}_1\rangle   & \ldots &    \langle {\bf x}_k \vert {\bf x}_k\rangle \\
\end{pmatrix}
\equiv
\mathbb{1}_k
\end{split}
\end{equation}
represents a $k$-dimensional resolution of the identity operator.
\index{resolution of the identity}
Thus, $\left(\textsf{\textbf{A}}^\dagger  \textsf{\textbf{A}} \right)^{-1}\equiv \left(\mathbb{1}_k\right)^{-1}$
is also a $k$-dimensional resolution of the identity operator,
and the orthogonal projector $\textsf{\textbf{E}}_{{\frak M}}$
in Equation~(\ref{2015-m-ch-cfvs-opab5})
reduces to
\begin{equation}
\textsf{\textbf{E}}_{{\frak M}}=
\textsf{\textbf{A}}   \textsf{\textbf{A}}^\dagger
\equiv
\sum_{i=1}^k \vert {\bf x}_i \rangle \langle {\bf x}_i \vert
.
\end{equation}

{\color{blue}
\bexample
The simplest example of an orthogonal projection onto a one-dimensional subspace of a Hilbert space
spanned by some unit vector $\vert {\bf x} \rangle$ is the dyadic or outer product
$
\textsf{\textbf{E}}_x = \vert {\bf x} \rangle  \langle {\bf x} \vert$.

If two unit vectors $\vert {\bf x} \rangle$ and $\vert {\bf y} \rangle$ are orthogonal;
that is, if $\langle {\bf x} \vert {\bf y} \rangle = 0$, then
$\textsf{\textbf{E}}_{x,y} = \vert {\bf x} \rangle  \langle {\bf x} \vert  +  \vert {\bf y} \rangle  \langle {\bf y} \vert$
is an orthogonal projector onto a two-dimensional subspace spanned by $\vert {\bf x} \rangle$ and $\vert {\bf y} \rangle$.
\eexample
}

In general, the orthonormal projection corresponding to some arbitrary subspace of some Hilbert space can be (nonuniquely)
constructed by
(i) finding an orthonormal basis spanning that subsystem (this is nonunique),
if necessary by a Gram-Schmidt process;
\index{Gram-Schmidt process}
(ii) forming the projection operators corresponding to the dyadic or outer product
\index{dyadic product}
\index{outer product}
of all these vectors; and
(iii) summing up all these orthogonal operators.

The following propositions are stated mostly without proof.
A  linear transformation $\textsf{\textbf{E}}$ is an orthogonal (perpendicular) projection
if and only if is self-adjoint; that is,
$\textsf{\textbf{E}} = \textsf{\textbf{E}}^2=\textsf{\textbf{E}}^\ast $.

Perpendicular projections are {\em positive} linear transformations,
with
$\left\| \textsf{\textbf{E}}{\bf x} \right\| \le \| {\bf x} \|$
for all
${\bf x} \in {\frak V}$.
Conversely,
if a linear transformation $\textsf{\textbf{E}}$
is idempotent; that is,
$\textsf{\textbf{E}}^2=\textsf{\textbf{E}}$,
and  $\left\| \textsf{\textbf{E}}{\bf x} \right\| \le \| {\bf x} \|$
for all
${\bf x} \in {\frak V}$,
then  is self-adjoint; that is,
$\textsf{\textbf{E}}=\textsf{\textbf{E}}^\ast$.

Recall that
for {\em real} inner product spaces, the self-adjoint operator can be identified with a {\em symmetric} operator
$\textsf{\textbf{E}}=\textsf{\textbf{E}}^\intercal $,
\index{symmetric operator}
whereas
for {\em complex} inner product spaces, the self-adjoint operator can be identified with a {\em Hermitian} operator
$\textsf{\textbf{E}}=\textsf{\textbf{E}}^\dagger$.
\index{Hermitian operator}

If $\textsf{\textbf{E}}_1,\textsf{\textbf{E}}_2, \ldots , \textsf{\textbf{E}}_n$ are (perpendicular)
projections,
then a necessary and sufficient condition that
$\textsf{\textbf{E}} =\textsf{\textbf{E}}_1+\textsf{\textbf{E}}_2+\cdots +\textsf{\textbf{E}}_n$
be a (perpendicular) projection is that
 $\textsf{\textbf{E}}_i \textsf{\textbf{E}}_j =\delta_{ij}\textsf{\textbf{E}}_i =\delta_{ij}\textsf{\textbf{E}}_j$;
and, in particular,
$\textsf{\textbf{E}}_i \textsf{\textbf{E}}_j =0$
whenever $i\neq j$; that is, that all $E_i$ are pairwise orthogonal.

{\color{OliveGreen}\bproof
For a start, consider just two projections
 $\textsf{\textbf{E}}_1$ and $\textsf{\textbf{E}}_2$.
Then we can assert that   $\textsf{\textbf{E}}_1 + \textsf{\textbf{E}}_2$ is a projection if and only if
 $\textsf{\textbf{E}}_1 \textsf{\textbf{E}}_2=\textsf{\textbf{E}}_2 \textsf{\textbf{E}}_1=0$.

Because, for
 $\textsf{\textbf{E}}_1 + \textsf{\textbf{E}}_2$ to be a projection, it must be idempotent; that is,
\index{idempotence}
 \begin{equation}
(\textsf{\textbf{E}}_1 + \textsf{\textbf{E}}_2)^2 =
(\textsf{\textbf{E}}_1 + \textsf{\textbf{E}}_2)(\textsf{\textbf{E}}_1 + \textsf{\textbf{E}}_2)  =
\textsf{\textbf{E}}_1^2 +   \textsf{\textbf{E}}_1\textsf{\textbf{E}}_2 + \textsf{\textbf{E}}_2\textsf{\textbf{E}}_1 + \textsf{\textbf{E}}_2^2
=
 \textsf{\textbf{E}}_1 + \textsf{\textbf{E}}_2 .
\label{2012-m-ch-fdvs-pr3}
\end{equation}
As a consequence, the cross-product terms in (\ref{2012-m-ch-fdvs-pr3}) must vanish; that is,
\begin{equation}
\textsf{\textbf{E}}_1\textsf{\textbf{E}}_2 + \textsf{\textbf{E}}_2\textsf{\textbf{E}}_1 =0.
\label{2012-m-ch-fdvs-pr4}
\end{equation}
Multiplication of (\ref{2012-m-ch-fdvs-pr4}) with $\textsf{\textbf{E}}_1$ from the left and from the right yields
\begin{equation}
\begin{split}
\textsf{\textbf{E}}_1\textsf{\textbf{E}}_1\textsf{\textbf{E}}_2 + \textsf{\textbf{E}}_1\textsf{\textbf{E}}_2\textsf{\textbf{E}}_1 =0, \\
 \textsf{\textbf{E}}_1\textsf{\textbf{E}}_2 + \textsf{\textbf{E}}_1\textsf{\textbf{E}}_2\textsf{\textbf{E}}_1 =0;\textrm{ and} \\
\textsf{\textbf{E}}_1\textsf{\textbf{E}}_2\textsf{\textbf{E}}_1 + \textsf{\textbf{E}}_2\textsf{\textbf{E}}_1\textsf{\textbf{E}}_1 =0, \\
\textsf{\textbf{E}}_1\textsf{\textbf{E}}_2\textsf{\textbf{E}}_1 + \textsf{\textbf{E}}_2\textsf{\textbf{E}}_1  =0.
\end{split}
\label{2012-m-ch-fdvs-pr5}
\end{equation}
Subtraction of the resulting pair of equations yields
\begin{equation}
 \textsf{\textbf{E}}_1\textsf{\textbf{E}}_2 - \textsf{\textbf{E}}_2\textsf{\textbf{E}}_1  =
\left[ \textsf{\textbf{E}}_1,\textsf{\textbf{E}}_2 \right]
=0,
\label{2012-m-ch-fdvs-pr6}
\end{equation}
or
\begin{equation}
 \textsf{\textbf{E}}_1\textsf{\textbf{E}}_2 = \textsf{\textbf{E}}_2\textsf{\textbf{E}}_1 .
\label{2012-m-ch-fdvs-pr67}
\end{equation}
Hence, in order for the cross-product terms in Eqs. (\ref{2012-m-ch-fdvs-pr3} ) and (\ref{2012-m-ch-fdvs-pr4})
to vanish, we must have
\begin{equation}
 \textsf{\textbf{E}}_1\textsf{\textbf{E}}_2 = \textsf{\textbf{E}}_2\textsf{\textbf{E}}_1 =0.
\label{2012-m-ch-fdvs-pr8}
\end{equation}

Proving the reverse statement is straightforward, since (\ref{2012-m-ch-fdvs-pr8}) implies  (\ref{2012-m-ch-fdvs-pr3}).

A generalisation by induction to more than two projections is straightforward,
since, for instance,
$\left(\textsf{\textbf{E}}_1+\textsf{\textbf{E}}_2\right)\textsf{\textbf{E}}_3=0$
implies
$ \textsf{\textbf{E}}_1\textsf{\textbf{E}}_3+\textsf{\textbf{E}}_2\textsf{\textbf{E}}_3=0$.
Multiplication with $\textsf{\textbf{E}}_1$ from the left yields
$
\textsf{\textbf{E}}_1\textsf{\textbf{E}}_1\textsf{\textbf{E}}_3+\textsf{\textbf{E}}_1\textsf{\textbf{E}}_2\textsf{\textbf{E}}_3=
\textsf{\textbf{E}}_1 \textsf{\textbf{E}}_3=
0$.
 \eproof }

\subsection{Construction of orthogonal projections from  single unit vectors}

How can we construct  orthogonal projections from unit vectors or systems of orthogonal projections from some vector in some orthonormal basis
with the standard dot product?

Let ${\bf x}$ be the coordinates of a unit vector;
that is $\|{\bf x} \| =1$.
Transposition is indicated by the superscript ``$\intercal$''
in real vector space.
In complex vector space, the transposition has to be substituted
 for the {\em conjugate transpose} (also denoted as
{\em Hermitian conjugate} or {\em Hermitian adjoint}),
\index{conjugate transpose}
\index{Hermitian conjugate}
\index{Hermitian adjoint}
``$\dagger$,'' standing for transposition and complex conjugation of the coordinates.
More explicitly,
\begin{equation}
\begin{split}
\begin{pmatrix}x_1,\ldots, x_n\end{pmatrix}^\dagger =
\begin{pmatrix}
\overline{x_1}\\ \vdots\\ \overline{x_n}
\end{pmatrix}
,  \textrm{ and }
\begin{pmatrix}
x_1\\ \vdots\\ x_n
\end{pmatrix}^\dagger
= (\overline{x_1},\ldots, \overline{x_n})
.
\end{split}
\end{equation}
Note that, just as for real vector spaces,
$\left({\bf x}^\intercal \right)^\intercal ={\bf x}$,
or, in the bra-ket notation,
$\left(\vert {\bf x}\rangle^\intercal \right)^\intercal =\vert {\bf x}\rangle$,
 so is
$\left({\bf x}^\dagger \right)^\dagger={\bf x}$,
or
$\left(\vert  {\bf x}\rangle ^\dagger \right)^\dagger= \vert{\bf x} \rangle$ for complex vector spaces.

As already mentioned on page~\pageref{2015-m-ch-fdlvs-inter}, Equation~(\ref{2015-m-ch-fdlvs-inter}), for orthonormal bases of complex Hilbert space
we can express the dual vector in terms of the original vector by
taking the conjugate transpose, and {\em vice versa}; that is,
\begin{equation}
\begin{split}
\langle {\bf x} \vert = \left(\vert {\bf x}\rangle \right)^\dagger,
\textrm{ and }
\vert {\bf x}\rangle  = \left(\langle {\bf x}\vert\right)^\dagger.
\end{split}
\end{equation}

In real vector space, the {\em dyadic product}, or {\em tensor product}, or {\em outer product}
\index{outer product}
\index{dyadic product}
\index{tensor product}
\begin{equation}
\begin{split}
\textsf{\textbf{E}}_{\bf x} = {\bf x} \otimes {\bf x}^\intercal  = \vert{\bf x}\rangle \langle {\bf x}\vert
\equiv
\begin{pmatrix}
x_1\\
x_2\\
\vdots \\
x_n
\end{pmatrix}
\begin{pmatrix}x_1,x_2,\ldots ,x_n\end{pmatrix}\\
\\
=
\begin{pmatrix}
x_1\begin{pmatrix}x_1,x_2,\ldots ,x_n\end{pmatrix}\\
x_2\begin{pmatrix}x_1,x_2,\ldots ,x_n\end{pmatrix}\\
\vdots  \\
x_n\begin{pmatrix}x_1,x_2,\ldots ,x_n \end{pmatrix}
\end{pmatrix}
=
\begin{pmatrix}
x_1x_1&x_1x_2& \cdots&x_1x_n\\
x_2x_1&x_2x_2& \cdots&x_2x_n\\
\vdots & \vdots & \vdots &\vdots \\
x_nx_1&x_nx_2& \cdots&x_nx_n
\end{pmatrix}
\end{split}
\end{equation}
is the projection
associated with ${\bf x}$.

If the vector ${\bf x}$ is not normalized,
then the associated projection is
\begin{equation}
\textsf{\textbf{E}}_{\bf x} \equiv \frac{{\bf x} \otimes {\bf x}^\intercal }{\langle {\bf x}\mid {\bf x}\rangle}
\equiv \frac{\vert{\bf x}\rangle \langle {\bf x}\vert}{\langle {\bf x}\mid {\bf x}\rangle}
= \frac{\vert{\bf x}\rangle \langle {\bf x}\vert}{\| {\bf x}\|^2}
\end{equation}
This construction is related to
$P_{\bf x}$ on page \pageref{2011-m-gsp}
by $P_{\bf x}({\bf y})=\textsf{\textbf{E}}_{\bf x}{\bf y}$.

{\color{OliveGreen}
\bproof
For a proof,  consider only normalized vectors ${\bf x}$, and
let $\textsf{\textbf{E}}_{\bf x} = {\bf x}\otimes {\bf x}^\intercal  $,
then
$$
\textsf{\textbf{E}}_{\bf x}\textsf{\textbf{E}}_{\bf x}
=
(\vert{\bf x}\rangle \langle {\bf x}\vert)
(\vert{\bf x}\rangle \langle {\bf x}\vert)
=
\vert{\bf x}\rangle \underbrace{\langle {\bf x}\vert {\bf x}\rangle}_{=1} \langle {\bf x}\vert
=  \textsf{\textbf{E}}_{\bf x}.
$$
More explicitly, by writing out the coordinate tuples, the equivalent proof is
\begin{equation}
\begin{split}
\textsf{\textbf{E}}_{\bf x}\textsf{\textbf{E}}_{\bf x}
\equiv ({\bf x}\otimes {\bf x}^\intercal  ) \cdot ({\bf x}\otimes {\bf x}^\intercal  )
\\
\equiv
\left[
\begin{pmatrix}
x_1\\
x_2\\
\vdots \\
x_n
\end{pmatrix}
(x_1,x_2,\ldots ,x_n)
\right]
\left[
\begin{pmatrix}
x_1\\
x_2\\
\vdots \\
x_n
\end{pmatrix}
(x_1,x_2,\ldots ,x_n)
 \right]
\\
=
\begin{pmatrix}
x_1\\
x_2\\
\vdots \\
x_n
\end{pmatrix}
\underbrace{\left[ (x_1,x_2,\ldots ,x_n)
\begin{pmatrix}
x_1\\
x_2\\
\vdots \\
x_n
\end{pmatrix}
\right]}_{=1}
(x_1,x_2,\ldots ,x_n)
\equiv\textsf{\textbf{E}}_{\bf x}. \textrm{\eproof }
\end{split}
\end{equation}
}

In complex vector space, transposition has to be substituted by the conjugate transposition;
\index{conjugate transpose}
\index{Hermitian conjugate}
\index{Hermitian adjoint}
that is
\begin{equation}
\begin{split}
\textsf{\textbf{E}}_{\bf x} = {\bf x} \otimes {\bf x}^\dagger \equiv \vert{\bf x}\rangle \langle {\bf x}\vert
\end{split}
\end{equation}

{
\color{blue}
\bexample
For two examples, let
${\bf x}=(1,0)^\intercal $
and
${\bf y}=(1,-1)^\intercal $;
then
$$
\textsf{\textbf{E}}_{\bf x}
=
\begin{pmatrix}
1\\
0
\end{pmatrix}
(1,0)
=
\begin{pmatrix}
1(1,0)\\
0(1,0)
\end{pmatrix}
=
\begin{pmatrix}
1&0\\
0&0
\end{pmatrix}
,
$$
and
$$
\textsf{\textbf{E}}_{\bf y}
= \frac{1}{2}
\begin{pmatrix}
1\\
-1
\end{pmatrix}
(1,-1)
= \frac{1}{2}
\begin{pmatrix}
1(1,-1)\\
-1(1,-1)
\end{pmatrix}
= \frac{1}{2}
\begin{pmatrix}
1&-1\\
-1&1
\end{pmatrix}
.
\textrm{\eexample}
$$
}

Note also that
\begin{equation}
\textsf{\textbf{E}}_{\bf x} \vert {\bf y}\rangle
\equiv
\textsf{\textbf{E}}_{\bf x} {\bf y}
=
\langle {\bf x}\vert  {\bf y} \rangle {\bf x},
\equiv
\langle {\bf x}\vert  {\bf y} \rangle \vert {\bf x}\rangle ,
\end{equation}
which can be directly proven by insertion.

\subsection{Examples of oblique projections which are not orthogonal projections}

{\color{blue}
\bexample

Examples for projections which are not orthogonal are
$$\begin{pmatrix}
1&\alpha \\
0&0
\end{pmatrix}
\text{,  or }
\begin{pmatrix}
1&0&\alpha \\
0&1&\beta \\
0&0&0
\end{pmatrix},$$
with $\alpha \neq 0$.
Such projectors are sometimes called
{\em oblique} projections.
\index{oblique projections}

For two-dimensional Hilbert space, the solution
of idempotence
$$
\begin{pmatrix}
a&b \\
c&d
\end{pmatrix}
\begin{pmatrix}
a&b \\
c&d
\end{pmatrix}
=
\begin{pmatrix}
a&b \\
c&d
\end{pmatrix}
$$
yields the three orthogonal projections
$$
\begin{pmatrix}
1&0 \\
0&0
\end{pmatrix},
\begin{pmatrix}
0&0 \\
0&1
\end{pmatrix},\text{ and }
\begin{pmatrix}
1&0 \\
0&1
\end{pmatrix},
$$
as well as a continuum of oblique projections
$$
\begin{pmatrix}
0&0 \\
c&1
\end{pmatrix}
=
\begin{pmatrix}
0 \\
1
\end{pmatrix}
\otimes
\begin{pmatrix}
c,1
\end{pmatrix},
\begin{pmatrix}
1&0 \\
c&0
\end{pmatrix},\text{ and }
\begin{pmatrix}
a&b \\
\frac{a(1-a)}{b}&1-a
\end{pmatrix},
$$
with $a,b,c \neq 0$.

$$
\begin{pmatrix}
0&0 \\
c&1
\end{pmatrix}
=
\begin{pmatrix}
0 \\
1
\end{pmatrix}
\otimes
\begin{pmatrix}
c,1
\end{pmatrix},
\begin{pmatrix}
1&0 \\
c&0
\end{pmatrix}
$$

One can also utilize Equation~(\ref{2018-mm-ch-fdlvs-dop}) and
define two sets of indexed vectors
$\{ {\bf e}_1 , {\bf e}_2 \}$ and
$\{ {\bf f}_1 , {\bf f}_2 \}$
with
$ {\bf e}_1 \equiv \vert {\bf e}_1\rangle =\begin{pmatrix} a,b\end{pmatrix}^\intercal$,
$ {\bf e}_2 \equiv \vert {\bf e}_2\rangle =\begin{pmatrix} c,d\end{pmatrix}^\intercal$,
$ {\bf f}_1 \equiv \vert {\bf f}_1\rangle =\begin{pmatrix} e,f\end{pmatrix}^\intercal$, as well as
$ {\bf f}_2 \equiv \vert {\bf f}_2\rangle =\begin{pmatrix} g,h\end{pmatrix}^\intercal$.
Biorthogonality
\index{biorthogonality}
of this pair of indexed families of vectors is defined by
${\bf f}^\ast_i {\bf e}_j\equiv \langle {\bf f}_i \vert {\bf e}_j\rangle  =\delta_{ij}$.

%
%
%
%
%
%

This results in four families of solutions:
The first solution  requires $ad\neq bc$; with
$e=\frac{d}{ad - bc}$,
$f=-\frac{c}{ad - bc}$,
$g=-\frac{b}{ad - bc}$, and
$h=\frac{a}{ad - bc}$.
It amounts to two  mutually orthogonal (oblique) projections
\begin{equation}
\begin{split}
\textsf{\textbf{G}}_{1,1}=\begin{pmatrix}
a \\
b
\end{pmatrix}
\otimes
\frac{1}{a d - b c}
\begin{pmatrix}
d,-c
\end{pmatrix}
=
\frac{1}{a d - b c}
\begin{pmatrix}
a d & - a c \\
b d & - b c
\end{pmatrix}
      ,
\\
\textsf{\textbf{G}}_{1,2}=
\begin{pmatrix}
c \\
d
\end{pmatrix}
\otimes
\frac{1}{a d - b c}
\begin{pmatrix}
-b,a
\end{pmatrix}
=
\frac{1}{a d - b c}
\begin{pmatrix}
-b c &  a c \\
-b d & a d
\end{pmatrix}
     .
\end{split}
\end{equation}

The second solution  requires $a,c,d\neq 0$; with
$b=g=0$,
$e=\frac{1}{a}$,
$f=-\frac{c}{ad}$,
$h=\frac{1}{d}$.
It amounts to two  mutually orthogonal (oblique) projections
\begin{equation}
\begin{split}
\textsf{\textbf{G}}_{2,1}=\begin{pmatrix}
a \\
0
\end{pmatrix}
\otimes
\begin{pmatrix}
\frac{1}{a},-\frac{c}{ad}
\end{pmatrix}
=
\begin{pmatrix}
1 & - \frac{c}{d} \\
0 & 0
\end{pmatrix}
      ,
\\
\textsf{\textbf{G}}_{2,2}=
\begin{pmatrix}
c \\
d
\end{pmatrix}
\otimes
\begin{pmatrix}
0,\frac{1}{d}
\end{pmatrix}
=
\begin{pmatrix}
0&  \frac{c}{d} \\
0 & 1
\end{pmatrix}
      .
\end{split}
\end{equation}

The third solution  requires $a,d\neq 0$; with
$b=f=g=0$,
$e=\frac{1}{a}$,
$h=\frac{1}{d}$.
It amounts to two mutually orthogonal (orthogonal) projections
\begin{equation}
\begin{split}
\textsf{\textbf{G}}_{3,1}=\begin{pmatrix}
a \\
0
\end{pmatrix}
\otimes
\begin{pmatrix}
0 ,  \frac{1}{a}
\end{pmatrix}
=
\begin{pmatrix}
1 & 0  \\
0 & 0
\end{pmatrix}
      ,
\\
\textsf{\textbf{G}}_{3,2}=
\begin{pmatrix}
0 \\
d
\end{pmatrix}
\otimes
\begin{pmatrix}
0,\frac{1}{d}
\end{pmatrix}
=
\begin{pmatrix}
0& 0 \\
0 & 1
\end{pmatrix}
      .
\end{split}
\end{equation}

The fourth and last solution  requires $a,b,d\neq 0$; with
$c=f=0$,
$e=\frac{1}{a}$,
$g=-\frac{b}{ad}$,
$h=\frac{1}{d}$.
It amounts to two  mutually orthogonal (oblique) projections
\begin{equation}
\begin{split}
\textsf{\textbf{G}}_{4,1}=\begin{pmatrix}
a \\
b
\end{pmatrix}
\otimes
\begin{pmatrix}
\frac{1}{a},0
\end{pmatrix}
=
\begin{pmatrix}
1 &  0\\
\frac{b}{a}& 0
\end{pmatrix}
      ,
\\
\textsf{\textbf{G}}_{4,2}=
\begin{pmatrix}
0 \\
d
\end{pmatrix}
\otimes
\begin{pmatrix}
- \frac{b}{ad},\frac{1}{d}
\end{pmatrix}
=
\begin{pmatrix}
0&  0 \\
-\frac{b}{a} & 1
\end{pmatrix}
      .
\end{split}
\end{equation}

\eexample
}

\section{Proper value or eigenvalue}
\index{proper value}
\marginnote{For proofs and additional information see {\S}54 in~\bibentry{halmos-vs}
and \bibentry{Sanderson-3Blue1Brown-LA14}.}
\index{proper vector}
\index{eigenvalue}
\index{eigenvector}
\index{eigensystem}

\subsection{Definition}

A scalar $\lambda$ is a {\em proper value} or {\em eigenvalue},
and a nonzero vector ${\bf x}$ is a {\em proper vector} or {\em eigenvector}
of a linear transformation $\textsf{\textbf{A}}$
if
\begin{equation}
\textsf{\textbf{A}}{\bf x}=   \lambda {\bf x} =   \lambda \mathbb{1} {\bf x}.
\end{equation}
In an $n$-dimensional
vector space $\frak V$
The set of the set of eigenvalues and the set of the associated eigenvectors
$\{\{\lambda_1,\ldots ,\lambda_k\},\{{\bf x}_1,\ldots ,{\bf x}_n\}\}$
of a linear transformation $\textsf{\textbf{A}}$ form an {\em eigensystem} of $\textsf{\textbf{A}}$.

\subsection{Determination}
\index{characteristic equation}
\index{secular determinant}
\index{secular equation}


Since the eigenvalues and eigenvectors are those scalars $\lambda$  vectors ${\bf x}$ for which $\textsf{\textbf{A}}{\bf x}=   \lambda {\bf x}$,
this equation can be rewritten with a zero vector on the right side of the equation; that is ($\mathbb{1}=\textrm{diag}(1,\ldots ,1)$ stands for the identity matrix),
\begin{equation}
(\textsf{\textbf{A}} - \lambda \mathbb{1}){\bf x}= {\bf 0}.
\label{2011-m-eve}
\end{equation}
Suppose that $\textsf{\textbf{A}} - \lambda \mathbb{1}$ is invertible. Then we could formally write
${\bf x} = (\textsf{\textbf{A}} - \lambda \mathbb{1})^{-1}{\bf 0}$; hence ${\bf x}$ must be the zero vector.

We are not interested in this trivial solution of Equation~(\ref{2011-m-eve}).
Therefore, suppose that, contrary to the previous assumption,
$\textsf{\textbf{A}} - \lambda \mathbb{1}$ is {\em not} invertible.
We have mentioned earlier (without proof~\cite{Sanderson-3Blue1Brown-LA7}) that this implies that its determinant vanishes; that is,
\begin{equation}
\textrm{det} (\textsf{\textbf{A}} - \lambda \mathbb{1}) = \vert \textsf{\textbf{A}} - \lambda \mathbb{1}\vert =0.
\label{2014-m-eve-ce}
\end{equation}
This determinant is often called the {\em secular determinant};
\index{secular determinant}
\index{secular equation}
and the corresponding equation after expansion of the determinant is called the
{\em secular equation}
or {\em characteristic equation}.
Once the eigenvalues, that is, the roots of this polynomial, are determined,
the eigenvectors can be obtained one-by-one by inserting these eigenvalues one-by-one into Equation~(\ref{2011-m-eve}).
\index{root of a polynomial}
\marginnote{The roots of a polynomial $P(x)$ are those values
of the variable $x$ that prompt the polynomial to evaluate to zero.}

{\color{blue}
\bexample
For the sake of an example, consider  the
{matrix}
\begin{equation}
A=
\begin{pmatrix}
1&0&1\\
0&1&0\\
1&0&1
\end{pmatrix}.
\label{2017-m-ch-fdvs-e-eev1}
\end{equation}

The secular equation is
\index{secular equation}
$$
\left|
\begin{matrix}
1-\lambda &0&1\\
0&1-\lambda &0\\
1&0&1-\lambda
\end{matrix}
\right| = 0,
$$
yielding the characteristic equation
$
(1-\lambda )^3 -(1-\lambda ) =(1-\lambda )[(1-\lambda )^2 - 1]=(1-\lambda )[\lambda ^2 - 2\lambda ]= - \lambda (1-\lambda )(2-\lambda ) =0$,
and therefore three  eigenvalues
$\lambda_1=0$,
$\lambda_2=1$, and
$\lambda_3=2$ which are the roots of $\lambda (1-\lambda )(2-\lambda ) =0$.

Next let us determine the eigenvectors of $A$, based on the eigenvalues.
Insertion  $\lambda_1=0$ into Equation~(\ref{2011-m-eve}) yields
\begin{equation}
\left[
\begin{pmatrix}
1&0&1\\
0&1&0\\
1&0&1
\end{pmatrix}  -
\begin{pmatrix}
0&0&0\\
0&0&0\\
0&0&0
\end{pmatrix}
\right]
\begin{pmatrix}
x_1\\
x_2\\
x_3
\end{pmatrix}
=
\begin{pmatrix}
1&0&1\\
0&1&0\\
1&0&1
\end{pmatrix}
\begin{pmatrix}
x_1\\
x_2\\
x_3
\end{pmatrix}
=
\begin{pmatrix}
0\\
0\\
0
\end{pmatrix}
;
\end{equation}
therefore $x_1+x_3=0$ and $x_2=0$.
We are free to choose any (nonzero) $x_1=-x_3$,
but if we are interested in normalized eigenvectors, we obtain
${\bf x}_1 =(1/\sqrt{2})(1,0,-1)^\intercal $.

Insertion  $\lambda_2=1$ into Equation~(\ref{2011-m-eve}) yields
\begin{equation}
\left[
\begin{pmatrix}
1&0&1\\
0&1&0\\
1&0&1
\end{pmatrix}  -
\begin{pmatrix}
1&0&0\\
0&1&0\\
0&0&1
\end{pmatrix}
\right]
\begin{pmatrix}
x_1\\
x_2\\
x_3
\end{pmatrix}
=
\begin{pmatrix}
0&0&1\\
0&0&0\\
1&0&0
\end{pmatrix}
\begin{pmatrix}
x_1\\
x_2\\
x_3
\end{pmatrix}
=
\begin{pmatrix}
0\\
0\\
0
\end{pmatrix}
;
\end{equation}
therefore $x_1=x_3=0$ and $x_2$ is arbitrary.
We are again free to choose any (nonzero) $x_2$,
but if we are interested in normalized eigenvectors, we obtain
${\bf x}_2 = (0,1,0)^\intercal $.

Insertion  $\lambda_3=2$ into Equation~(\ref{2011-m-eve}) yields
\begin{equation}
\left[
\begin{pmatrix}
1&0&1\\
0&1&0\\
1&0&1
\end{pmatrix}  -
\begin{pmatrix}
2&0&0\\
0&2&0\\
0&0&2
\end{pmatrix}
\right]
\begin{pmatrix}
x_1\\
x_2\\
x_3
\end{pmatrix}
=
\begin{pmatrix}
-1&0&1\\
0&-1&0\\
1&0&-1
\end{pmatrix}
\begin{pmatrix}
x_1\\
x_2\\
x_3
\end{pmatrix}
=
\begin{pmatrix}
0\\
0\\
0
\end{pmatrix}
;
\end{equation}
therefore $-x_1+x_3=0$ and $x_2=0$.
We are free to choose any (nonzero) $x_1=x_3$,
but if we are once more interested in normalized eigenvectors, we obtain
${\bf x}_3 =(1/\sqrt{2})(1,0,1)^\intercal $.

Note that the eigenvectors are mutually orthogonal.
We can construct the corresponding orthogonal projections by the outer (dyadic or tensor) product
\index{outer product}
\index{dyadic product}
\index{tensor product}
of the eigenvectors; that is,
\begin{equation}
\begin{split}
\textsf{\textbf{E}}_1 =
{\bf x}_1 \otimes {\bf x}_1^\intercal  =
\frac{1}{2} (1,0,-1)^\intercal (1,0,-1) =
\frac{1}{2}
\begin{pmatrix}
1(1,0,-1)\\
0(1,0,-1)\\
-1(1,0,-1)
\end{pmatrix} =
\frac{1}{2}
\begin{pmatrix}
1&0&-1\\
0&0&0\\
-1&0&1
\end{pmatrix}
\\
\textsf{\textbf{E}}_{2} =
{\bf x}_{2} \otimes {\bf x}_{2}^\intercal  =
 (0,1,0)^\intercal (0,1,0) =
\begin{pmatrix}
0(0,1,0)\\
1(0,1,0)\\
0(0,1,0)
\end{pmatrix} =
\begin{pmatrix}
0&0&0\\
0&1&0\\
0&0&0
\end{pmatrix}
\\
\textsf{\textbf{E}}_{3} =
{\bf x}_{3} \otimes {\bf x}_{3}^\intercal  =
\frac{1}{2} (1,0,1)^\intercal (1,0,1) =
\frac{1}{2}
\begin{pmatrix}
1(1,0,1)\\
0(1,0,1)\\
1(1,0,1)
\end{pmatrix} =
\frac{1}{2}
\begin{pmatrix}
1&0&1\\
0&0&0\\
1&0&1
\end{pmatrix}
\end{split}
\end{equation}
Note also that $A$ can be written as the sum of the products of the
eigenvalues with the associated projections; that is (here, $\textsf{\textbf{E}}$
stands for the corresponding matrix),
$A= 0  \textsf{\textbf{E}}_1 + 1  \textsf{\textbf{E}}_{2} +2\textsf{\textbf{E}}_{3} $.
Also, the projections are mutually orthogonal
-- that is,
$\textsf{\textbf{E}}_1 \textsf{\textbf{E}}_2 = \textsf{\textbf{E}}_1\textsf{\textbf{E}}_3=\textsf{\textbf{E}}_2\textsf{\textbf{E}}_3=0$
--
and add up to the identity; that is,
$\textsf{\textbf{E}}_1+\textsf{\textbf{E}}_2+\textsf{\textbf{E}}_3=\mathbb{1}$.
{\textrm{\eexample}}
}

Henceforth an eigenvalue will be called {\em degenerate} if more
than one linearly independent eigenstates belong to the same eigenvalue.\cite{Praeceptor-1967}
\index{degenerate eigenvalues}
Thus if the some eigenvalues -- the roots of the characteristic
polynomial of a matrix obtained from solving the secular equation
\index{root of a polynomial}
\index{secular equation}
-- are degenerate, then
there exist linearly independent eigenstates whose eigenvalues are not distinct.
In such a case the associated eigenvectors traditionally -- that is, by convention and not by necessity --
are taken to be {\em mutually orthogonormal;} thereby forming an orthonormal basis of the associated subspace
spanned by those associated eigenvectors (with identical eigenvalue):
an explicit construction of this (nonunique) basis
uses a Gram-Schmidt process (cf. Section~\ref{2019-mm-ch-fdvs-GS} on page~\pageref{2019-mm-ch-fdvs-GS})
\index{Gram-Schmidt process}
applied to those linearly independent eigenstates (with identical eigenvalue).

The algebraic multiplicity of an eigenvalue $\lambda$ of a matrix
\index{algebraic multiplicity}
\index{multiplicity}
is the number of times $\lambda$  appears as a root of the characteristic
polynomial of that matrix.
The {\em geometric multiplicity} of an eigenvalue is the number of linearly independent
eigenvectors are associated with it.
\index{geometric multiplicity}
\marginnote{The geometric multiplicity can never exceed the algebraic multiplicity.
For normal operators both multiplicities coincide
because of the spectral theorem (cf. Section~\ref{2012-m-ch-Spectraltheorem} on page~\pageref{2012-m-ch-Spectraltheorem}).}
A more formal motivation will come from the spectral theorem discussed later in Section~\ref{2012-m-ch-Spectraltheorem}
on page~\pageref{2012-m-ch-Spectraltheorem}.

{\color{blue}
\bexample
For the sake of an example, consider  the
{matrix}
\begin{equation}
B=
\begin{pmatrix}
1&0&1\\
0&2&0\\
1&0&1
\end{pmatrix}.
\label{2017-m-ch-fdvs-e-eev2}
\end{equation}

The secular equation yields
\index{secular equation}
$$
\left|
\begin{matrix}
1-\lambda &0&1\\
0&2-\lambda &0\\
1&0&1-\lambda
\end{matrix}
\right| = 0,
$$
which yields the characteristic equation
$
(2-\lambda )(1-\lambda )^2 +[-(2-\lambda )]=
(2-\lambda )[(1-\lambda )^2 -1]=
-\lambda (2-\lambda )^2 =0$,
and therefore just two  eigenvalues
$\lambda_1=0$,  and
$\lambda_2=2$ which are the roots of $\lambda (2-\lambda )^2 =0$.

Let us now determine the eigenvectors of $B$, based on the eigenvalues.
Insertion  $\lambda_1=0$ into Equation~(\ref{2011-m-eve})  yields
\begin{equation}
\left[
\begin{pmatrix}
1&0&1\\
0&2&0\\
1&0&1
\end{pmatrix}  -
\begin{pmatrix}
0&0&0\\
0&0&0\\
0&0&0
\end{pmatrix}
\right]
\begin{pmatrix}
x_1\\
x_2\\
x_3
\end{pmatrix}
=
\begin{pmatrix}
1&0&1\\
0&2&0\\
1&0&1
\end{pmatrix}
\begin{pmatrix}
x_1\\
x_2\\
x_3
\end{pmatrix}
=
\begin{pmatrix}
0\\
0\\
0
\end{pmatrix}
;
\end{equation}
therefore $x_1+x_3=0$ and $x_2=0$.
Again we are free to choose any (nonzero) $x_1=-x_3$,
but if we are interested in normalized eigenvectors, we obtain
${\bf x}_1 =(1/\sqrt{2})(1,0,-1)^\intercal $.

Insertion  $\lambda_2=2$ into Equation~(\ref{2011-m-eve}) yields
\begin{equation}
\left[
\begin{pmatrix}
1&0&1\\
0&2&0\\
1&0&1
\end{pmatrix}  -
\begin{pmatrix}
2&0&0\\
0&2&0\\
0&0&2
\end{pmatrix}
\right]
\begin{pmatrix}
x_1\\
x_2\\
x_3
\end{pmatrix}
=
\begin{pmatrix}
-1&0&1\\
0&0&0\\
1&0&-1
\end{pmatrix}
\begin{pmatrix}
x_1\\
x_2\\
x_3
\end{pmatrix}
=
\begin{pmatrix}
0\\
0\\
0
\end{pmatrix}
;
\end{equation}
therefore $x_1=x_3$; $x_2$ is arbitrary.
We are again free to choose any values of $x_1$, $x_3$ and $x_2$ as long
 $x_1=x_3$ as well as $x_2$ are satisfied.
Take, for the sake of choice, the orthogonal
normalized eigenvectors
${\bf x}_{2,1} = (0,1,0)^\intercal $ and
${\bf x}_{2,2} = (1/\sqrt{2})(1,0,1)^\intercal $,
which are also orthogonal to ${\bf x}_1 =(1/\sqrt{2})(1,0,-1)^\intercal $.

Note again that we can find the corresponding orthogonal projections by the outer (dyadic or tensor) product
\index{outer product}
\index{dyadic product}
\index{tensor product}
of the eigenvectors; that is,  by
\begin{equation}
\begin{split}
\textsf{\textbf{E}}_1 =
{\bf x}_1 \otimes {\bf x}_1^\intercal  =
\frac{1}{2} (1,0,-1)^\intercal (1,0,-1) =
\frac{1}{2}
\begin{pmatrix}
1(1,0,-1)\\
0(1,0,-1)\\
-1(1,0,-1)
\end{pmatrix} =
\frac{1}{2}
\begin{pmatrix}
1&0&-1\\
0&0&0\\
-1&0&1
\end{pmatrix}
\\
\textsf{\textbf{E}}_{2,1} =
{\bf x}_{2,1} \otimes {\bf x}_{2,1}^\intercal  =
(0,1,0)^\intercal (0,1,0) =
\begin{pmatrix}
0(0,1,0)\\
1(0,1,0)\\
0(0,1,0)
\end{pmatrix} =
\begin{pmatrix}
0&0&0\\
0&1&0\\
0&0&0
\end{pmatrix}
\\
\textsf{\textbf{E}}_{2,2} =
{\bf x}_{2,2} \otimes {\bf x}_{2,2}^\intercal  =
\frac{1}{2} (1,0,1)^\intercal (1,0,1) =
\frac{1}{2}
\begin{pmatrix}
1(1,0,1)\\
0(1,0,1)\\
1(1,0,1)
\end{pmatrix} =
\frac{1}{2}
\begin{pmatrix}
1&0&1\\
0&0&0\\
1&0&1
\end{pmatrix}
\end{split}
\end{equation}
Note also that $B$ can be written as the sum of the products of the
eigenvalues with the associated projections; that is (here, $\textsf{\textbf{E}}$
stands for the corresponding matrix),
$B= 0  \textsf{\textbf{E}}_1 + 2 (\textsf{\textbf{E}}_{2,1} + \textsf{\textbf{E}}_{2,2} )$.
Again, the projections are mutually orthogonal
-- that is,
$\textsf{\textbf{E}}_1 \textsf{\textbf{E}}_{2,1} = \textsf{\textbf{E}}_1\textsf{\textbf{E}}_{2,2}=
\textsf{\textbf{E}}_{2,1}\textsf{\textbf{E}}_{2,2}=0$
--
and add up to the identity; that is,
$\textsf{\textbf{E}}_1+\textsf{\textbf{E}}_{2,1}+\textsf{\textbf{E}}_{2,2}=\mathbb{1}$.
This leads us to the much more general spectral theorem.

Another, extreme, example would be the unit matrix in $n$ dimensions; that is,
$\mathbb{1}_n=\textrm{diag}(\underbrace{1,\ldots ,1}_{n \textrm{ times}})$,
which has an $n$-fold degenerate eigenvalue $1$ corresponding to a solution to
$(1-\lambda )^n=0$.
The corresponding projection operator is $\mathbb{1}_n$.  [Note that $(\mathbb{1}_n)^2 =\mathbb{1}_n$
and thus $\mathbb{1}_n$ is a projection.]
If one (somehow arbitrarily but conveniently) chooses a resolution of the identity operator $\mathbb{1}_n$
into projections corresponding to the standard basis (any other orthonormal basis would do as well),
then
\begin{equation}
\begin{split}
\mathbb{1}_n = \textrm{diag}( 1,0,0,\ldots ,0 )
+   \textrm{diag}( 0,1,0,\ldots ,0 )
+ \cdots
+   \textrm{diag}( 0,0,0,\ldots ,1 )\\
\begin{pmatrix}
 1&0&0&\cdots &0\\
 0&1&0&\cdots &0\\
 0&0&1&\cdots &0\\
&&&\vdots&\\
 0&0&0&\cdots &1
\end{pmatrix}  =
\begin{pmatrix}
 1&0&0&\cdots &0\\
 0&0&0&\cdots &0\\
 0&0&0&\cdots &0\\
&&&\vdots&\\
 0&0&0&\cdots &0
\end{pmatrix}  +
\\
\quad
+
\begin{pmatrix}
 0&0&0&\cdots &0\\
 0&1&0&\cdots &0\\
 0&0&0&\cdots &0\\
&&&\vdots&\\
 0&0&0&\cdots &0
\end{pmatrix} +\cdots
+
\begin{pmatrix}
 0&0&0&\cdots &0\\
 0&0&0&\cdots &0\\
 0&0&0&\cdots &0\\
&&&\vdots&\\
 0&0&0&\cdots &1
\end{pmatrix}
,
\end{split}
\end{equation}
where all the matrices in the sum carrying one nonvanishing entry ``$1$''
in their diagonal are  projections.
Note that
\begin{equation}
\begin{split}
{\bf e}_i=\vert {\bf e}_i \rangle  \\
\quad \equiv \begin{pmatrix} \underbrace{0,\ldots , 0}_{i-1 \textrm{ times}}, 1, \underbrace{0,\ldots , 0}_{n-i \textrm{ times}} \end{pmatrix}^\intercal    \\
\quad \equiv  \textrm{diag}(  \underbrace{0,\ldots , 0}_{i-1 \textrm{ times}}, 1, \underbrace{0,\ldots , 0}_{n-i \textrm{ times}}  )
  \\
\quad \equiv  \textsf{\textbf{E}}_{i}.
\end{split}
\end{equation}
{\textrm{\eexample}}
}

The following theorems are enumerated without proofs.

If $\textsf{\textbf{A}}$
is a self-adjoint transformation on an inner product space, then every proper value (eigenvalue)  of $\textsf{\textbf{A}}$
is real.
If $\textsf{\textbf{A}}$ is positive, or strictly positive,
then every proper value of  $\textsf{\textbf{A}}$ is positive, or strictly positive, respectively

Due to their idempotence $\textsf{\textbf{E}}\textsf{\textbf{E}}=\textsf{\textbf{E}}$,
projections have eigenvalues $0$ or $1$.

Every eigenvalue of an isometry has absolute value one.

If  $\textsf{\textbf{A}}$  is either a self-adjoint transformation or an isometry,
then proper vectors of $ \textsf{\textbf{A}}$
belonging to distinct proper values are orthogonal.

\section{Normal transformation}
\index{normal transformation}
\index{normal operator}
\label{2014-m-fdvs-normality}

A transformation $\textsf{\textbf{A}}$ is called {\em normal}
if it commutes with its adjoint; that is,
\begin{equation}
[\textsf{\textbf{A}},\textsf{\textbf{A}}^\ast ]= \textsf{\textbf{A}}\textsf{\textbf{A}}^\ast  -
\textsf{\textbf{A}}^\ast  \textsf{\textbf{A}} =0.
\end{equation}

It follows from their definition that Hermitian and unitary transformations are normal. That is,
$\textsf{\textbf{A}}^\ast =\textsf{\textbf{A}}^\dagger$,
and for Hermitian operators,
$\textsf{\textbf{A}}=\textsf{\textbf{A}}^\dagger$,
and thus
$[\textsf{\textbf{A}},\textsf{\textbf{A}}^\dagger]= \textsf{\textbf{A}}\textsf{\textbf{A}} -
\textsf{\textbf{A}} \textsf{\textbf{A}} =(\textsf{\textbf{A}})^2 -(\textsf{\textbf{A}})^2=0$.
For unitary operators,
$\textsf{\textbf{A}}^\dagger =\textsf{\textbf{A}}^{-1}$,
and thus
$[\textsf{\textbf{A}},\textsf{\textbf{A}}^\dagger]= \textsf{\textbf{A}}\textsf{\textbf{A}}^{-1} -
\textsf{\textbf{A}}^{-1} \textsf{\textbf{A}} =\mathbb{1} -\mathbb{1} =0$.

We mention without proof that
a normal transformation on a finite-dimensional unitary space is
(i) Hermitian,
(ii) positive,
(iii) strictly positive,
(iv) unitary,
(v) invertible,
(vi) idempotent
\index{idempotence}
if and only if all its proper values are
(i) real,
(ii) positive,
(iii) strictly positive,
(iv) of absolute value one,
(v) different from zero,
(vi) equal to zero or one.

\section{Spectrum}
\index{spectrum}
\marginnote{For proofs and additional information see {\S}78 and {\S}80 in~\bibentry{halmos-vs}.}

\subsection{Spectral theorem}
\label{2012-m-ch-Spectraltheorem}

Let
$\frak V$ be
an $n$-dimensional inner (scalar) product space (aka a finite dimensional Hilbert
space\sidenote[][0mm]{\url{https://math.stackexchange.com/questions/168275/proof-that-every-finite-dimensional-normed-vector-space-is-complete}}).
The {\em spectral theorem} states
\index{spectral theorem}
that to every normal transformation $\textsf{\textbf{A}} $ on
$\frak V$
being
\begin{itemize}
\item[(a)] self-adjoint (Hermitian), or
\item[(b)] positive, or
\item[(c)] strictly positive, or
\item[(d)] unitary, or
\item[(e)] invertible, or
\item[(f)] idempotent
\end{itemize}
there exist eigenvalues $
\lambda_1,
\lambda_2, \ldots ,
\lambda_k
$ of   $ \textsf{\textbf{A}}$
which are
\marginnote{Not all matrices are diagonalizable in the way described here,
but a generalization to arbitrary matrices
$\textsf{\textbf{A}}$
resembling the Jordan normal form can be found
at \url{https://terrytao.wordpress.com/2016/10/11/math-246a-notes-4-singularities-of-holomorphic-functions/} by Terence Tao [exercise 29, point (vi)]:
an arbitrary matrix $\textsf{\textbf{A}}$ can be written as
$
\textsf{\textbf{A}}=\sum_{i=1}^k \textsf{\textbf{E}}_i \left(\lambda_i\mathbb{1}+ \textsf{\textbf{N}}_i \right)\textsf{\textbf{E}}_i
$,
where $\textsf{\textbf{N}}_i$ is a nilpotent matrix with $\textsf{\textbf{N}}_i^{d_i}= \textsf{\textbf{0}}$, the matrix with entries zero
(see also \url{https://math.stackexchange.com/questions/3251052/jordan-normal-form-and-spectral-decomposition}). }
\begin{itemize}
\item[(a')] real, or
\item[(b')] positive, or
\item[(c')] strictly positive, or
\item[(d')] of absolute value one, or
\item[(e')] different from zero, or
\item[(f')] equal to zero or one,
\end{itemize}
called the {\em spectrum}
and their associated  orthogonal projections
$
\textsf{\textbf{E}}_1,
\textsf{\textbf{E}}_2, \ldots ,
\textsf{\textbf{E}}_k
$
where $0<k\le n$ is a strictly positive integer so that
\begin{itemize}
\item[(i)]
the $\lambda_i$ are pairwise distinct;
\item[(ii)]
the $\textsf{\textbf{E}}_i$ are pairwise orthogonal and different from $\textsf{\textbf{0}}$;
\item[(iii)]
the set of projectors is complete in the sense that their
sum $\sum_{i=1}^k \textsf{\textbf{E}}_i
= \textsf{\textbf{Z}} \textsf{\textbf{Z}}^\dagger =\mathbb{1}_n$
is a resolution of the identity operator.
\index{resolution of the identity}
\index{completeness}
$\textsf{\textbf{Z}} =
\begin{pmatrix}
{\bf x}_1,\ldots , {\bf x}_n
\end{pmatrix}
$
stands for the matrix assembled by columns of the orthonormalized eigenvectors of  $\textsf{\textbf{A}}$
forming an orthonormal basis.\sidenote[][0mm]{For $k<n$ the  higher-than-one dimensional projections
can be represented by sums of dyadic products of orthonormal bases spanning the associated subspaces of $\frak V$.}
\item[(iv)]
$
\textsf{\textbf{A}}=\sum_{i=1}^k \lambda_i\textsf{\textbf{E}}_i
= \textsf{\textbf{Z}} \boldsymbol{\Lambda}\textsf{\textbf{Z}}^\dagger
$
\index{spectral form}
is the {\em spectral form} of $\textsf{\textbf{A}}$.\sidenote[][-15mm]{For a nondegenerate spectrum $k=n$,
$\mathbb{1}_n =  \sum_{i=1}^n  \vert {\bf x}_i \rangle \langle   {\bf x}_i  \vert$
and
$
\textsf{\textbf{A}}
=\sum_{i=1}^n \lambda_i \vert {\bf x}_i \rangle \langle   {\bf x}_i  \vert
$, where the mutually orthonormal eigenvectors $\vert {\bf x}_i \rangle$ form a basis.}
$
\boldsymbol{\Lambda}=\text{diag}\underbrace{\begin{pmatrix}\lambda_1,\ldots,\lambda_k\end{pmatrix}}_{n\text{ entries}}
$
represents an $n\times n$ diagonal matrix with $k$ mutually distinct
entities.\sidenote[][]{With respect to the orthonormal basis of the vectors
associated with the orthogonal projections  $
\textsf{\textbf{E}}_1,
\textsf{\textbf{E}}_2, \ldots ,
\textsf{\textbf{E}}_k
$  occurring in the spectral form the operator
$\textsf{\textbf{A}}$ can be represented by a {\em diagonal} matrix form $\boldsymbol{\Lambda}$;
see also Fact~1.4 on page~8 of \bibentry{Parlett:1998:SEP:280490}.}

\end{itemize}

{\color{OliveGreen}
\bproof
Rather than proving the spectral theorem in its full generality,
we suppose that the spectrum of a Hermitian (self-adjoint) operator $ \textsf{\textbf{A}}$ is {\em nondegenerate};
that is, all $n$ eigenvalues of $\textsf{\textbf{A}}$ are pairwise distinct:
there do not exist two or more linearly independent eigenstates belonging to the same eigenvalue.
That is, we are assuming a strong form of (i), with $k=n$.

As will be shown this distinctness of the eigenvalues translates into mutual orthogonality of all the eigenvectors of $ \textsf{\textbf{A}}$.
Thereby, the set of $n$ eigenvectors forms some orthogonal (orthonormal) basis of the $n$-dimensional linear vector space $\frak V$.
The respective normalized eigenvectors can then be represented by perpendicular projections which can be summed up to yield the identity
(iii).

More explicitly, suppose wrongly, for the sake of a proof (by contradiction) of the pairwise orthogonality of the eigenvectors (ii),
that two distinct eigenvalues
$\lambda_1$
and
$\lambda_2 \neq \lambda_1$
belong to two respective eigenvectors
$\vert {\bf x}_1\rangle $
and
$\vert {\bf x}_2\rangle $
which are not orthogonal.
Because $\textsf{\textbf{A}}$ is self-adjoint, which implies  real eigenvalues
\marginnote{Self-adjoint operators $\textsf{\textbf{A}}^\ast=\textsf{\textbf{A}}$ have real eigenvalues as
$\overline{\lambda_i}
= $ [unit eigenvectors $\vert {\bf x}_i \rangle $]
$
= \overline{\lambda_i}  \langle {\bf x}_i \vert {\bf x}_i \rangle
= $ [conjugate linearity in the first argument of the scalar product] $
= \langle \lambda_i {\bf x}_i \vert {\bf x}_i \rangle
= \langle \textsf{\textbf{A}}{\bf x}_i \vert {\bf x}_i \rangle
= $ [definition of self-adjoint operator~(\ref{2016-m-ch-fdlvs-adjointop})]
$
= \langle {\bf x}_i \vert \textsf{\textbf{A}}^\ast{\bf x}_i \rangle
= $ [self-adjointness of $\textsf{\textbf{A}}$]
$
=\langle {\bf x}_i \vert \textsf{\textbf{A}}{\bf x}_i \rangle
=\langle {\bf x}_i \vert \lambda_i {\bf x}_i \rangle
= $ [linearity in the second argument of the scalar product]
$
=\lambda_i \langle {\bf x}_i \vert {\bf x}_i \rangle
=\lambda_i
$.},
\begin{equation}
\begin{split}
\lambda_1  \langle {\bf x}_1 \vert {\bf x}_2 \rangle =
\langle \lambda_1  {\bf x}_1 \vert {\bf x}_2 \rangle =
  \langle \textsf{\textbf{A}}{\bf x}_1 \vert {\bf x}_2 \rangle \\=
  \langle {\bf x}_1 \vert \textsf{\textbf{A}}^\ast  {\bf x}_2 \rangle =
  \langle {\bf x}_1 \vert \textsf{\textbf{A}}{\bf x}_2 \rangle =
  \langle {\bf x}_1\vert \left( \lambda_2 \vert {\bf x}_2 \rangle \right)=
  \lambda_2 \langle {\bf x}_1 \vert {\bf x}_2 \rangle,
\end{split}
\end{equation}
which implies that
\begin{equation}
\left(\lambda_1 - \lambda_2\right)
\langle {\bf x}_1 \vert {\bf x}_2 \rangle
=
0
.
\label{2016-m-ch-fdvs-ecfo}
\end{equation}

Equation (\ref{2016-m-ch-fdvs-ecfo}) is satisfied by
either $\lambda_1  = \lambda_2$
--
which is in contradiction to our assumption that $\lambda_1$ and $\lambda_2$ are distinct
--
or by  $\langle {\bf x}_1 \vert {\bf x}_2 \rangle =0$  (thus allowing $\lambda_1  \neq \lambda_2$) --
which is in contradiction to our assumption that $\vert {\bf x}_1\rangle$ and $\vert  {\bf x}_2\rangle$
are nonzero and {\em not} orthogonal.
Hence, if we maintain the distinctness of $\lambda_1$ and $\lambda_2$, the associated eigenvectors need to be orthogonal,
thereby assuring (ii).

Since by our assumption there are $n$ distinct eigenvalues, this implies that, associated with these,
there are $n$ orthonormal eigenvectors.
These $n$ mutually orthonormal eigenvectors
span the entire $n$-dimensional vector space $\frak V$;
and hence their union $\{  {\bf x}_i, \ldots ,   {\bf x}_n\}$ forms an orthonormal basis.
Consequently, the sum of the associated perpendicular projections
$\textsf{\textbf{E}}_i = \frac{\vert {\bf x}_i \rangle \langle {\bf x}_i \vert}{\langle {\bf x}_i \vert {\bf x}_i  \rangle}$
is a resolution of the identity operator $\mathbb{1}_n$
(cf. section~\ref{2016-m-ch-fdvsrotio} on page~\pageref{2016-m-ch-fdvsrotio}); thereby justifying (iii).

In the last step, let us keep in mind the $i$'th projection operator $\textsf{\textbf{E}}_i$
and define the projection of it onto
an arbitrary vector $\vert {\bf z}\rangle \in \frak V$ by
$
\vert {\xi}_i \rangle =  \textsf{\textbf{E}}_i  \vert  {\bf z} \rangle
=
\vert  {\bf x}_i \rangle \langle  {\bf x}_i   \vert  {\bf z} \rangle
= \alpha_i \vert  {\bf x}_i \rangle
$ with $
\alpha_i = \langle  {\bf x}_i   \vert  {\bf z} \rangle
$,
thereby keeping in mind that any such vector $\vert  {\xi}_i \rangle$ (associated with $\textsf{\textbf{E}}_i$)
is an eigenvector of $\textsf{\textbf{A}}$  with the associated eigenvalue $\lambda_i$; that is,
\marginnote{Einstein's summation convention over identical indices does not apply here.}
\begin{equation}
\textsf{\textbf{A}} \vert {\xi}_i  \rangle
 = \textsf{\textbf{A}}\alpha_i \vert  {\bf x}_i \rangle
 = \alpha_i \textsf{\textbf{A}} \vert  {\bf x}_i \rangle
 = \alpha_i \lambda_i \vert  {\bf x}_i  \rangle
 = \lambda_i \alpha_i \vert  {\bf x}_i  \rangle
= \lambda_i \vert {\xi}_i\rangle .
\end{equation}
Then, by the linearity of $\textsf{\textbf{A}}$,
\begin{equation}
\begin{split}
\textsf{\textbf{A}} \vert {\bf z} \rangle =
\textsf{\textbf{A}} \mathbb{1}_n \vert {\bf z} \rangle=
\textsf{\textbf{A}} \left(\sum_{i=1}^n \textsf{\textbf{E}}_i\right) \vert {\bf z}\rangle  =
\textsf{\textbf{A}} \left(\sum_{i=1}^n \textsf{\textbf{E}}_i \vert {\bf z}\rangle \right)      \\
=
\textsf{\textbf{A}} \left(\sum_{i=1}^n \vert {\xi}_i\rangle \right) =
 \sum_{i=1}^n \textsf{\textbf{A}} \vert {\xi}_i\rangle   =
 \sum_{i=1}^n \lambda_i \vert {\xi}_i \rangle   =
 \sum_{i=1}^n \lambda_i \textsf{\textbf{E}}_i   \vert {\bf z}\rangle =
\left( \sum_{i=1}^n \lambda_i \textsf{\textbf{E}}_i \right) \vert {\bf z}\rangle ,
\end{split}
\end{equation}
which is the spectral form of $\textsf{\textbf{A}}$.
\eproof
}

\subsection{Composition of the spectral form by Lagrange polynomial}
\marginnote{A polynomial interpolation is the interpolation of a given data set of points $(x_i,y_i)$ with mutually distinct $x_i$
by the polynomial (of lowest possible degree) that passes through the points of the dataset --
that is, that yields at each input $x_i$ the output $y_i$.}

If the spectrum of a  Hermitian (or, more general, normal) operator $\textsf{\textbf{A}}$ is nondegenerate, that is, $k=n$, then the
$i$th projection
can be written as the outer (dyadic or tensor) product
\index{outer product}
\index{dyadic product}
\index{tensor product}
$
\textsf{\textbf{E}}_i={\bf x}_i \otimes {\bf x}_i^\intercal $
of the $i$th normalized eigenvector ${\bf x}_i $ of $\textsf{\textbf{A}}$.
In this case, the set of all normalized eigenvectors $\{{\bf x}_1, \ldots ,{\bf x}_n\}$ is an orthonormal basis of the vector space $\frak V$.
If the spectrum of $\textsf{\textbf{A}}$ is degenerate, then the projection can be chosen to be the orthogonal sum of projections
corresponding to orthogonal eigenvectors, associated with the same eigenvalues.

Furthermore, for a  Hermitian (or, more general, normal) operator $\textsf{\textbf{A}}$,
if $1\le i \le k$,
then there exist polynomials with real coefficients, such as,  for instance, the
Lagrange basis polynomials
\index{Lagrange polynomial}
\begin{equation}
p_i  (t)
=
\prod_{j\neq i}
\frac{t-\lambda_j}{\lambda_i -\lambda_j}
\label{2011-m-epsf}
\end{equation}
so that
$p_i(\lambda_j) =\delta_{ij}$;
moreover, for every such polynomial,
$p_i(\textsf{\textbf{A}})=\textsf{\textbf{E}}_i$.
\marginnote{For related results see
\url{https://terrytao.wordpress.com/2019/08/13/eigenvectors-from-eigenvalues/}
as well as \bibentry{Mieghem-2014}.
}

{\color{OliveGreen}\bproof

For a proof  it is not too difficult
to show that
$p_i  (\lambda_i)=1$, since in this case in the product of fractions all numerators are equal to denominators.
Furthermore,
$p_i  (\lambda_j)=0$ for $j\neq i $, since some numerator in the product of fractions vanishes; and
therefore,
$p_i  (\lambda_j)=\delta_{ij}$.

Now, substituting for $t$ the spectral form $\textsf{\textbf{A}}=\sum_{i=1}^k \lambda_i\textsf{\textbf{E}}_i$
of $\textsf{\textbf{A}}$, as well as insertion of the
resolution of the identity operator in terms of the projections $\textsf{\textbf{E}}_i$ in the spectral form of
$\textsf{\textbf{A}}$ -- that is, $\mathbb{1}_n=\sum_{i=1}^k \textsf{\textbf{E}}_i$ --
yields
\begin{equation}
p_i  (\textsf{\textbf{A}})
=
\prod_{
j\neq i
}\frac{\textsf{\textbf{A}} - \lambda_j\mathbb{1}_n}{\lambda_i -\lambda_j}
=
\prod_{
j\neq i
}\frac{\sum_{l=1}^k \lambda_l\textsf{\textbf{E}}_l - \lambda_j\sum_{l=1}^k \textsf{\textbf{E}}_l}{\lambda_i -\lambda_j}
.
\end{equation}
Because of the idempotence and pairwise orthogonality of the projections  $\textsf{\textbf{E}}_l$,
\index{idempotence}
\begin{equation}
\begin{split}
p_i  (\textsf{\textbf{A}}) =
\prod_{
j\neq i
}\frac{\sum_{l=1}^k \textsf{\textbf{E}}_l(\lambda_l- \lambda_j)}{\lambda_i -\lambda_j}   \\
= \sum_{l=1}^k \textsf{\textbf{E}}_l
\prod_{
j\neq i
}\frac{\lambda_l- \lambda_j}{\lambda_i -\lambda_j}
= \sum_{l=1}^k \textsf{\textbf{E}}_l
p_i  (\lambda_l)
= \sum_{l=1}^k \textsf{\textbf{E}}_l
\delta_{il} = \textsf{\textbf{E}}_i.
\label{2012-m-ch-fdvs-cc}
\end{split}
\end{equation}
\eproof
}

With the help of the polynomial $p_i(t)$ defined in Equation~(\ref{2011-m-epsf}),
which requires knowledge of the eigenvalues,
the spectral form of a Hermitian (or, more general, normal) operator  $\textsf{\textbf{A}}$ can thus be rewritten as
\begin{equation}
\textsf{\textbf{A}}=\sum_{i=1}^k \lambda_i p_i(\textsf{\textbf{A}})=  \sum_{i=1}^k \lambda_i \prod_{
j\neq i
}\frac{\textsf{\textbf{A}} - \lambda_j\mathbb{1}_n}{\lambda_i -\lambda_j}.
\end{equation}
That is, knowledge of all the eigenvalues entails construction
of all the projections in the spectral decomposition
of a normal transformation.

{\color{blue}
\bexample
For the sake of an example, consider the matrix
\begin{equation}
A=
\begin{pmatrix}
1&0&1\\
0&1&0\\
1&0&1
\end{pmatrix}
\end{equation}
introduced in Equation~(\ref{2017-m-ch-fdvs-e-eev1}).
In particular, the projection $\textsf{\textbf{E}}_1$ associated with the first eigenvalue $\lambda_1=0$
can be obtained from  the set of eigenvalues $\left\{  0,1,2 \right\}$ by
\begin{equation}
\begin{split}
p_1(A)=
\left( \frac{A-\lambda_2\mathbb{1}}{\lambda_1-\lambda_2} \right)
\left( \frac{A-\lambda_3\mathbb{1}}{\lambda_1-\lambda_3} \right) \\
=
\frac{
\left[
\begin{pmatrix}
1&0&1\\
0&1&0\\
1&0&1
\end{pmatrix}
-
1\cdot
\begin{pmatrix}
1&0&0\\
0&1&0\\
0&0&1
\end{pmatrix}
\right]}
{(0-1)}
\cdot
\frac{
\left[
\begin{pmatrix}
1&0&1\\
0&1&0\\
1&0&1
\end{pmatrix}
-
2\cdot
\begin{pmatrix}
1&0&0\\
0&1&0\\
0&0&1
\end{pmatrix}
\right]}
{(0-2)}
\\
=
\frac{1}{2}
\begin{pmatrix}
0&0&1\\
0&0&0\\
1&0&0
\end{pmatrix}
\begin{pmatrix}
-1&0&1\\
0&-1&0\\
1&0&-1
\end{pmatrix}
=
\frac{1}{2}
\begin{pmatrix}
1&0&-1\\
0&0&0\\
-1&0&1
\end{pmatrix}
=\textsf{\textbf{E}}_1.
\end{split}
\end{equation}

For the sake of another, degenerate, example consider again the
{matrix}
\begin{equation}
B=
\begin{pmatrix}
1&0&1\\
0&2&0\\
1&0&1
\end{pmatrix}
\end{equation}
introduced in Equation~(\ref{2017-m-ch-fdvs-e-eev2}).

Again, the projections $\textsf{\textbf{E}}_1, \textsf{\textbf{E}}_2$
can be obtained from  the set of eigenvalues $\left\{  0,2 \right\}$ by
\begin{equation}
\begin{split}
p_1(B)=   \frac{B-\lambda_2\mathbb{1}}{\lambda_1-\lambda_2}
=
\frac{
\begin{pmatrix}
1&0&1\\
0&2&0\\
1&0&1
\end{pmatrix}
-
2\cdot
\begin{pmatrix}
1&0&0\\
0&1&0\\
0&0&1
\end{pmatrix}
}
{(0-2)}
=
\frac{1}{2}
\begin{pmatrix}
1&0&-1\\
0&0&0\\
-1&0&1
\end{pmatrix}
=\textsf{\textbf{E}}_1
,\\
p_2(B)=   \frac{B-\lambda_1\mathbb{1}}{\lambda_2-\lambda_1}
=
\frac{
\begin{pmatrix}
1&0&1\\
0&2&0\\
1&0&1
\end{pmatrix}
-
0\cdot
\begin{pmatrix}
1&0&0\\
0&1&0\\
0&0&1
\end{pmatrix}
}
{(2-0)}
=
\frac{1}{2}
\begin{pmatrix}
1&0&1\\
0&2&0\\
1&0&1
\end{pmatrix}
=\textsf{\textbf{E}}_2.
\end{split}
\end{equation}
Note that, in accordance with the spectral theorem,
$\textsf{\textbf{E}}_1 \textsf{\textbf{E}}_2=0 $,
$\textsf{\textbf{E}}_1+\textsf{\textbf{E}}_2=\mathbb{1}$
and
$0\cdot \textsf{\textbf{E}}_1+2\cdot \textsf{\textbf{E}}_2=B$.

\eexample
}

\section{Functions of normal transformations}
\index{Functions of normal transformation}

Suppose $\textsf{\textbf{A}}=\sum_{i=1}^k \lambda_i  \textsf{\textbf{E}}_i $ is a normal transformation
in its spectral form.
If $f$ is an arbitrary complex-valued function defined at least at the eigenvalues of $\textsf{\textbf{A}}$,
then a linear transformation  $f(\textsf{\textbf{A}})$ can be defined by
\begin{equation}
f(\textsf{\textbf{A}})=
f\left(\sum_{i=1}^k \lambda_i \textsf{\textbf{E}}_i\right)
=\sum_{i=1}^k f(\lambda_i)  \textsf{\textbf{E}}_i
 .
\end{equation}
Note that, if $f$ has a polynomial expansion such as analytic functions, then orthogonality and idempotence
\index{idempotence}
of the projections $\textsf{\textbf{E}}_i $ in the spectral form guarantees this kind of ``linearization.''

{
\color{blue}
\bexample
If the function $f$ is a polynomial of some degree $N$ -- say, if
$f (x) = p(x) = \sum_{l=1}^N \alpha_l x^l$ -- then
\begin{equation}
\begin{split}
p (\textsf{\textbf{A}})
= \sum_{l=1}^N \alpha_l \textsf{\textbf{A}}^l
= \sum_{l=1}^N \alpha_l \left(\sum_{i=1}^k  \lambda_i   \textsf{\textbf{E}}_i\right)^l \\
= \sum_{l=1}^N \alpha_l \underbrace{\left(\sum_{{i_1}=1}^k \lambda_{i_1}  \textsf{\textbf{E}}_{i_1}\right)   \cdots  \left(\sum_{{i_l}=1}^k \lambda_{i_l}  \textsf{\textbf{E}}_{i_l}\right)}_{l \text{ times}}
= \sum_{l=1}^N \alpha_l \left(\sum_{i=1}^k  \lambda_i^l   \textsf{\textbf{E}}_i^l\right)   \\
= \sum_{l=1}^N \alpha_l \left(\sum_{i=1}^k  \lambda_i^l   \textsf{\textbf{E}}_i\right)
= \sum_{i=1}^k    \left( \sum_{l=1}^N \alpha_l \lambda_i^l   \right)  \textsf{\textbf{E}}_i
= \sum_{i=1}^k p ( \lambda_i^l   ) \textsf{\textbf{E}}_i  .
\end{split}
\end{equation}

A very similar argument applies to functional representations as Laurent or Taylor series expansions,
 -- say, $e^\textsf{\textbf{A}} = \sum_{l=0}^\infty   \frac{\textsf{\textbf{A}}^l}{l!} = \sum_{i=1}^k  \left(   \sum_{l=0}^\infty \frac{\lambda_i^l}{l!}\right) \textsf{\textbf{E}}_i
= \sum_{i=1}^k  e^{\lambda_i} \textsf{\textbf{E}}_i $
-- in which case the coefficients $\alpha_l $ have to be identified with the coefficients in the  series expansions.

\marginnote{The denomination ``not'' for  $\textsf{\textbf{not}}$
can be motivated by enumerating its performance at the two ``classical bit states''
$\vert 0 \rangle \equiv (1,0)^\intercal $
and
$\vert 1 \rangle \equiv (0,1)^\intercal $:
$\textsf{\textbf{not}}\vert 0 \rangle = \vert 1 \rangle $
and
$\textsf{\textbf{not}}\vert 1 \rangle = \vert 0 \rangle $.}
For the definition of the ``square root''
for every positive operator $\textsf{\textbf{A}}$, consider
\begin{equation}
\sqrt{\textsf{\textbf{A}}}=\sum_{i=1}^k \sqrt{\lambda_i}  \textsf{\textbf{E}}_i.
\end{equation}
With this definition,
$\left(\sqrt{\textsf{\textbf{A}}}\right)^2=
\sqrt{\textsf{\textbf{A}}}\sqrt{\textsf{\textbf{A}}}= {\textsf{\textbf{A}}}$.

Consider, for instance, the ``square root''  of the $\textsf{\textbf{not}}$ operator
\index{not operator}
\index{square root of not operator}
\begin{equation}
\textsf{\textbf{not}}
=
\begin{pmatrix}
 0&1\\  1&0
\end{pmatrix}.
\end{equation}
To enumerate $\sqrt{\textsf{\textbf{not}}}$  we need to find the {\em spectral form} of $\textsf{\textbf{not}}$ first.
\index{spectral form}
The eigenvalues of  $\textsf{\textbf{not}}$ can be obtained by solving the
secular equation
\index{secular equation}
\begin{equation}
\text{det}
\left(
\textsf{\textbf{not}} - \lambda \mathbb{1}_2
\right)
=
\text{det}
\left(
\begin{pmatrix}
 0&1\\  1&0
\end{pmatrix}
-
\lambda
\begin{pmatrix}
1&0\\  0&1
\end{pmatrix}
\right)
=
\text{det}
\begin{pmatrix}
 -\lambda&1\\
1&-\lambda
\end{pmatrix}
   =\lambda^2-1=0.
\end{equation}
$\lambda^2=1$ yields the two eigenvalues
$\lambda_1=1$
and
$\lambda_2=-1$.
The associated eigenvectors
${\bf x}_1$
and
${\bf x}_2$
can be derived from either the equations
$\textsf{\textbf{not}}\,{\bf x}_1={\bf x}_1$
and
$\textsf{\textbf{not}}\,{\bf x}_2=-{\bf x}_2$,
or by inserting the eigenvalues into the polynomial~(\ref{2011-m-epsf}).

We choose the former method.
Thus, for $\lambda_1=1$,
\begin{equation}
\begin{pmatrix}
 0&1\\  1&0
\end{pmatrix}
\begin{pmatrix}
x_{1,1}\\x_{1,2}
\end{pmatrix}
=\begin{pmatrix}
x_{1,1}\\x_{1,2}
\end{pmatrix}
,
\end{equation}
which yields  $x_{1,1}=x_{1,2}$, and thus, by normalizing the eigenvector,
${\bf x}_1=(1/\sqrt{2})(1,1)^\intercal $.
The associated projection is
\begin{equation}
\textsf{\textbf{E}}_1={\bf x}_1{\bf x}_1^\intercal =\frac{1}{2}
\begin{pmatrix}
 1&1\\  1&1
\end{pmatrix}
.
\end{equation}

Likewise, for $\lambda_2=-1$,
\begin{equation}
\begin{pmatrix}
 0&1\\  1&0
\end{pmatrix}
\begin{pmatrix}
x_{2,1}\\x_{2,2}
\end{pmatrix}
=-\begin{pmatrix}
x_{2,1}\\x_{2,2}
\end{pmatrix}
,
\end{equation}
which yields  $x_{2,1}=-x_{2,2}$, and thus, by normalizing the eigenvector,
${\bf x}_2=(1/\sqrt{2})(1,-1)^\intercal $.
The associated projection is
\begin{equation}
\textsf{\textbf{E}}_2={\bf x}_2{\bf x}_2^\intercal =\frac{1}{2}
\begin{pmatrix}
 1&-1\\  -1&1
\end{pmatrix}
.
\end{equation}

Thus we are finally able to calculate
$\sqrt{\textsf{\textbf{not}}}$
from its spectral form
\begin{equation}
\begin{split}
\sqrt{\textsf{\textbf{not}}}=
\sqrt{\lambda_1}\textsf{\textbf{E}}_1 +
\sqrt{\lambda_2}\textsf{\textbf{E}}_2
\\=  \sqrt{1}
\frac{1}{2} \begin{pmatrix}
 1&1\\  1&1
\end{pmatrix}
+  \sqrt{-1}
\frac{1}{2} \begin{pmatrix}
 1&-1\\  -1&1
\end{pmatrix}
\\
=
\frac{1}{2}
\begin{pmatrix}
 1+i&1-i\\  1-i&1+i
\end{pmatrix}
=
\frac{1}{1-i}
\begin{pmatrix}
 1&-i\\  -i&1
\end{pmatrix}
.
\end{split}
\end{equation}
It can be readily verified that  $\sqrt{\textsf{\textbf{not}}}\sqrt{\textsf{\textbf{not}}}=\textsf{\textbf{not}}$.
Note that this form is not unique:
$\pm_1 \sqrt{\lambda_1}\textsf{\textbf{E}}_1 +
\pm_2 \sqrt{\lambda_2}\textsf{\textbf{E}}_2$, where $\pm_1$ and $\pm_2$ represent separate cases,
yield alternative expressions of $\sqrt{\textsf{\textbf{not}}}$.



\eexample
}

\section{Decomposition of operators}
\index{decomposition}

\subsection{Standard decomposition}
\index{Standard decomposition}

In analogy to the decomposition of every imaginary number $z= \Re z +i \Im z$ with $\Re z,\Im z\in {\Bbb R}$,
every arbitrary transformation $\textsf{\textbf{A}}$ on a finite-dimensional vector space can be decomposed into two Hermitian operators
$\textsf{\textbf{B}}$
and
$\textsf{\textbf{C}}$
such that
\begin{eqnarray}
\textsf{\textbf{A}}&=&\textsf{\textbf{B}} + i \textsf{\textbf{C}}; \textrm{ with }  \nonumber \\
\textsf{\textbf{B}}&=&\frac{1}{2}(\textsf{\textbf{A}} +   \textsf{\textbf{A}}^\dagger ), \\
\textsf{\textbf{C}}&=&\frac{1}{2i}(\textsf{\textbf{A}} -   \textsf{\textbf{A}}^\dagger )\nonumber .
\end{eqnarray}

{\color{OliveGreen}
\bproof
Proof by insertion; that is,
\begin{equation}
\begin{split}
\textsf{\textbf{A}}=\textsf{\textbf{B}} + i \textsf{\textbf{C}}\\
\quad =
\frac{1}{2}(\textsf{\textbf{A}} +   \textsf{\textbf{A}}^\dagger ) + i \left[\frac{1}{2i}(\textsf{\textbf{A}} -   \textsf{\textbf{A}}^\dagger )\right],
\\
\textsf{\textbf{B}}^\dagger=   \left[\frac{1}{2}(\textsf{\textbf{A}} +   \textsf{\textbf{A}}^\dagger )\right]^\dagger
  =    \frac{1}{2}\left[\textsf{\textbf{A}}^\dagger +  ( \textsf{\textbf{A}}^\dagger )^\dagger \right]\\
 =    \frac{1}{2}\left[\textsf{\textbf{A}}^\dagger +    \textsf{\textbf{A}} \right]
 =  \textsf{\textbf{B}} , \\
\textsf{\textbf{C}}^\dagger=   \left[\frac{1}{2i}(\textsf{\textbf{A}} -   \textsf{\textbf{A}}^\dagger )\right]^\dagger
 =   -\frac{1}{2i}\left[\textsf{\textbf{A}}^\dagger -  ( \textsf{\textbf{A}}^\dagger )^\dagger \right]\\
  =   -\frac{1}{2i}\left[\textsf{\textbf{A}}^\dagger -    \textsf{\textbf{A}}\right]
 =  \textsf{\textbf{C}} .
\end{split}
\end{equation}
\eproof
}

\subsection{Polar decomposition}
\index{polar decomposition}
\marginnote{For proofs and additional information see {\S}83 in~\bibentry{halmos-vs}.}

In analogy to the polar representation of every imaginary number $z= R e^{i\varphi}$ with $R,\varphi \in {\Bbb R}$, $R\ge 0$,
$0\le \varphi < 2\pi$,
every arbitrary transformation $\textsf{\textbf{A}}$ on a finite-dimensional inner product space can be decomposed into
a unique positive transform
$\textsf{\textbf{P}}$ and an isometry
$\textsf{\textbf{U}}$, such that $\textsf{\textbf{A}}= \textsf{\textbf{U}} \textsf{\textbf{P}}$.
If $\textsf{\textbf{A}}$ is invertible, then $\textsf{\textbf{U}}$  is uniquely determined by
$\textsf{\textbf{A}}$.
A necessary and sufficient condition that $\textsf{\textbf{A}}$ is normal is that
$\textsf{\textbf{U}} \textsf{\textbf{P}}=\textsf{\textbf{P}} \textsf{\textbf{U}} $.

$\textsf{\textbf{P}}$ can be obtained by taking the square root of $\textsf{\textbf{A}}^\ast \textsf{\textbf{A}}$,
which is self-adjoint as
$\left(\textsf{\textbf{A}}^\ast \textsf{\textbf{A}}\right)^\ast=
\textsf{\textbf{A}}^\ast \left(\textsf{\textbf{A}}^\ast \right)^\ast =\textsf{\textbf{A}}^\ast \textsf{\textbf{A}}$:
multiplication of  $\textsf{\textbf{A}}=\textsf{\textbf{U}} \textsf{\textbf{P}}$ from the left with its adjoint
$\textsf{\textbf{A}}^\ast= \textsf{\textbf{P}}^\ast \textsf{\textbf{U}}^\ast
= \textsf{\textbf{P}}  \textsf{\textbf{U}} ^{-1}$
yields\sidenote[][-10mm]{$\textsf{\textbf{P}}$ is positive and thus self-adjoint; that is,
$\textsf{\textbf{P}}^\ast = \textsf{\textbf{P}}$.}
$
\textsf{\textbf{A}}^\ast \textsf{\textbf{A}}
=
 \textsf{\textbf{P}}   \underbrace{ \textsf{\textbf{U}} ^{-1} \textsf{\textbf{U}}}_{= \textsf{\textbf{I}}} \textsf{\textbf{P}}
=
\textsf{\textbf{P}}^2
$; and therefore,
\begin{equation}
\textsf{\textbf{P}}= \sqrt{\textsf{\textbf{A}}^\ast \textsf{\textbf{A}}}.
\end{equation}
If the inverse $\textsf{\textbf{A}}^{-1} =  \textsf{\textbf{P}}^{-1} \textsf{\textbf{U}}^{-1}$ of $\textsf{\textbf{A}}$
and thus also the inverse  $\textsf{\textbf{P}}^{-1} =  \textsf{\textbf{A}}^{-1} \textsf{\textbf{U}}$
 of $\textsf{\textbf{P}}$
exist, then $\textsf{\textbf{U}} =  \textsf{\textbf{A}}   \textsf{\textbf{P}}^{-1}$ is unique.

\subsection{Decomposition of isometries}

Any unitary or orthogonal transformation   in finite-dimensional inner product space
\index{decomposition}
can be composed of a succession of two-parameter unitary transformations in
two-dimensional subspaces,
and a multiplication of a single diagonal matrix with elements of modulus one
in an algorithmic, constructive and tractable manner.
The method is similar to Gaussian elimination and facilitates the parameterization of elements
of the unitary group in arbitrary dimensions (e.g., Ref.\cite[-40mm]{murnaghan}, Chapter 2).

{\color{Purple}
It has been suggested to implement
these group theoretic results by realizing interferometric analogs
of any discrete unitary and Hermitian operator
in a unified and experimentally feasible way by ``generalized beam splitters.''\cite[-30mm]{rzbb,reck-94}
}

\subsection{Singular value decomposition}

The {\em singular value decomposition}
\index{singular value decomposition}
(SVD)
of an ($m\times n$)  matrix $\textsf{\textbf{A}}$ is a factorization of the form
\begin{equation}
\textsf{\textbf{A}} = \textsf{\textbf{U}} \Sigma \textsf{\textbf{V}} ,
\end{equation}
where
$\textsf{\textbf{U}}$ is a unitary ($m\times m$)  matrix (i.e. an isometry),
$\textsf{\textbf{V}}$ is a unitary ($n\times n$)  matrix,
and
$\Sigma$ is a unique ($m\times n$)   diagonal matrix with nonnegative real numbers on the diagonal;
that is,
\begin{equation}
\Sigma =
\begin{pmatrix}
\sigma_1&&&{|}&&\vdots& \\
  &\ddots &&{|}&\cdots &0&\cdots \\
&&\sigma_r&{|}&&\vdots& \\
-&-&-&&-&-&- \\
&\vdots&&{|}&&\vdots& \\
\cdots &0&\cdots &{|}&\cdots &0&\cdots \\
&\vdots&&{|}&&\vdots& \\
\end{pmatrix}.
\end{equation}
The entries $\sigma_1\ge \sigma_2 \cdots \ge \sigma_r$>0 of $\Sigma$ are called {\em singular values}
of $\textsf{\textbf{A}}$.  No proof is presented here.
\index{singular values}

\subsection{Schmidt decomposition of the tensor product of two vectors}
\index{Schmidt decomposition}
\label{2011-m-Schmidtdecomposition}
\marginnote{For additional information see page~109, Section~2.5 in~\bibentry{nielsen-book10}.}

Let  ${\frak U}$  and   ${\frak V}$ be
two linear vector spaces
of dimension $n\ge m$ and $m$, respectively.
Then, for any vector
${\bf z} \in {\frak U}\otimes {\frak V}$
in the tensor product space,
there exist
orthonormal basis sets of vectors
$\{ {\bf u}_1, \ldots ,{\bf u}_n \}  \subset  {\frak U}$
and
$\{ {\bf v}_1, \ldots ,{\bf v}_m \}  \subset  {\frak V}$
such that
\begin{equation}
\vert {\bf z}\rangle \equiv {\bf z}=\sum_{i=1}^m
\sigma_i  {\bf u}_i \otimes  {\bf v}_i
\equiv
\sum_{i=1}^m \sigma_i   \vert {\bf u}_i  \rangle  \vert {\bf v}_i  \rangle,
\label{2011-e-sd}
\end{equation}
where the $\sigma_i$s are nonnegative scalars and the set of scalars is uniquely determined by
${\bf z}$.
If $  {\bf z}$
is normalized, then the  $\sigma_i$'s are  satisfying
$\sum_i \sigma_i^2=1$;
they are called the
{\em Schmidt coefficients}.
\index{Schmidt coefficients}

{\color{OliveGreen}
\bproof
For a proof by reduction to the singular value decomposition,
let
$\vert i\rangle$
and
$\vert j\rangle$
be any two fixed orthonormal bases of $ {\frak U}$ and $ {\frak V}$, respectively.
Then,
$\vert {\bf z}\rangle $
can be expanded as
$\vert {\bf z}\rangle  = \sum_{ij}a_{ij} \vert i\rangle \vert j\rangle$,
where the $a_{ij}$s can be interpreted as the components of a matrix
$\textsf{\textbf{A}}$.
$\textsf{\textbf{A}}$ can then be subjected to a
singular value decomposition
$\textsf{\textbf{A}} = \textsf{\textbf{U}} \Sigma \textsf{\textbf{V}}$,
or, written in index form [note that $\Sigma=\textrm{diag}(\sigma_1, \ldots, \sigma_n)$ is a diagonal matrix],
$a_{ij}= \sum_l u_{il}\sigma_l v_{lj}$;
and hence  $\vert {\bf z}\rangle  = \sum_{ijl} u_{il}\sigma_l v_{lj}\vert i\rangle \vert j\rangle$.
Finally, by identifying
$\vert {\bf u}_l  \rangle = \sum_i u_{il} \vert i\rangle$
as well as
$\vert {\bf v}_l  \rangle = \sum_l v_{lj} \vert j\rangle$
one obtains the Schmidt decomposition (\ref{2011-e-sd}).
Since $u_{il}$ and $v_{ lj}$ represent unitary matrices,
and because
 $\vert i\rangle$ as well as
 $\vert j\rangle$
are orthonormal,
the newly formed vectors
$\vert {\bf u}_l \rangle$
as well as
$\vert {\bf v}_l  \rangle$
form orthonormal bases as well.
The sum of squares of the $\sigma_i$'s is one if  $\vert {\bf z}\rangle $ is a unit vector,
because  (note that $\sigma_i$s are real-valued)
 $\langle {\bf z}\vert {\bf z}\rangle =1
=   \sum_{lm} \sigma_l \sigma_m   \langle {\bf u}_l  \vert  {\bf u}_m  \rangle   \langle  {\bf v}_l  \vert  {\bf v}_m  \rangle
=   \sum_{lm} \sigma_l \sigma_m  \delta_{lm}
=   \sum_{l} \sigma_l^2
$.
\eproof
}

Note that the Schmidt decomposition cannot, in general, be extended if there are more factors than two.
Note also that the Schmidt decomposition needs not be unique;\cite{ekert:415}
in particular, if some of the Schmidt coefficients $\sigma_i$ are equal.
For the sake of an example of nonuniqueness of the Schmidt decomposition,
take, for instance, the representation of the {\em Bell state} \index{Bell state}
with the two bases
\begin{equation}
\begin{split}
\left\{
\vert {\bf e}_1\rangle \equiv (1,0)^\intercal ,
\vert {\bf e}_2\rangle \equiv (0,1)^\intercal
\right\}
\textrm{ and }\\
\left\{
\vert {\bf f}_1\rangle \equiv \frac{1}{\sqrt{2}}(1,1)^\intercal ,
\vert {\bf f}_2\rangle \equiv \frac{1}{\sqrt{2}}(-1,1)^\intercal
\right\}.
\end{split}
\end{equation}
as follows:
\begin{equation}
\begin{split}
\vert {\Psi^-} \rangle =
\frac{1}{\sqrt{2}}
\left(
\vert {\bf e}_1\rangle
\vert {\bf e}_2\rangle
-
\vert {\bf e}_2\rangle
\vert {\bf e}_1\rangle
\right)\\
\quad \equiv
\frac{1}{\sqrt{2}}
\left[
(1 (0,1),0(0,1))^\intercal - (0 (1,0),1(1,0))^\intercal \right] = \frac{1}{\sqrt{2}} (0,1,-1,0)^\intercal ; \\
\vert {\Psi^-} \rangle =
\frac{1}{\sqrt{2}}
\left(
\vert {\bf f}_1\rangle
\vert {\bf f}_2\rangle
-
\vert {\bf f}_2\rangle
\vert {\bf f}_1\rangle
\right) \\
\quad \equiv
\frac{1}{2\sqrt{2}}
\left[
(1 (-1,1),1(-1,1))^\intercal - ( -1 (1,1),1(1,1))^\intercal \right]  \\
\quad \equiv
\frac{1}{2\sqrt{2}}
\left[
( -1  , 1 ,-1 ,1 )^\intercal  - ( -1,-1 ,1 ,1 )^\intercal \right]
 = \frac{1}{\sqrt{2}} (0,1,-1,0)^\intercal
.
\end{split}
\end{equation}

\section{Purification}
\index{purification}
\label{2015-m-ch-fdvs-purification}

\marginnote{For additional information see page~110, Section~2.5 in~\bibentry{nielsen-book10}.}

In general, quantum states ${\boldsymbol{\rho}}$ satisfy two criteria:\cite{ba-89}  they are
(i) of trace class one:
$\textrm{Tr}({\boldsymbol{\rho}}) =1$;
and
(ii) positive (or, by another term nonnegative):
$
\langle {\bf x} \vert {\boldsymbol{\rho}} \vert {\bf x} \rangle
=\langle {\bf x} \vert {\boldsymbol{\rho}}  {\bf x} \rangle \ge  0
$ for all vectors ${\bf x}$ of the  Hilbert space.

With finite dimension $n$ it follows immediately from (ii)
that $\boldsymbol{\rho}$ is self-adjoint; that is,
${\boldsymbol{\rho}}^\dagger ={\boldsymbol{\rho}}$), and normal, and thus has a spectral decomposition
\begin{equation}
\boldsymbol{\rho} =\sum_{i=1}^n \rho_i \vert \psi_i\rangle \langle \psi_i \vert
\label{2015-sfgs}
\end{equation}
into orthogonal projections $\vert \psi_i\rangle \langle \psi_i \vert$,
with
(i) yielding $\sum_{i=1}^n\rho_i =1$
(hint: take a trace with the orthonormal basis corresponding to all the $\vert \psi_i\rangle$);
(ii) yielding  $\overline{\rho_i}=\rho_i$;
and (iii) implying $\rho_i \ge 0$, and hence [with (i)] $0 \le \rho_i \le 1$
for all $1\le i \le n$.

As has been pointed out earlier, quantum mechanics differentiates between ``two sorts of states,'' namely
pure states and mixed ones:
\begin{itemize}
\item[(i)]
Pure states ${\boldsymbol{\rho}}_p$  are
represented by one-dimensional orthogonal projections; or, equivalently as one-dimensional linear subspaces by some (unit) vector.
They can be written as ${\boldsymbol{\rho}}_p =  \vert \psi \rangle \langle \psi  \vert$ for some unit vector $\vert \psi \rangle$
(discussed in Section~\ref{2011-m-projec}), and
satisfy $({\boldsymbol{\rho}}_p)^2={\boldsymbol{\rho}}_p$.
\item[(ii)]
General mixed states ${\boldsymbol{\rho}}_m$  are ones that are no projections and therefore
satisfy $({\boldsymbol{\rho}}_m)^2 \neq {\boldsymbol{\rho}}_m$.
They can be composed of projections by their spectral form (\ref{2015-sfgs}).
\end{itemize}

The question arises: is it possible to ``purify'' any mixed state by (maybe somewhat superficially) ``enlarging'' its
Hilbert space, such that the resulting state ``living in a larger Hilbert space'' is pure?
This can indeed be achieved by a rather simple procedure:
By considering the spectral form (\ref{2015-sfgs}) of a general mixed state ${\boldsymbol{\rho}}$,
define a new, ``enlarged,'' pure state  $\vert \Psi\rangle \langle \Psi \vert$, with
\begin{equation}
\vert \Psi\rangle = \sum_{i=1}^n \sqrt{\rho_i}  \vert \psi_i\rangle  \vert \psi_i\rangle
.
\label{2015-puran}
\end{equation}

{\color{OliveGreen}\bproof
That $\vert \Psi\rangle \langle \Psi \vert$ is pure can be tediously verified by proving that it is idempotent:
\index{idempotence}
\begin{equation}
\begin{split}
(\vert \Psi\rangle \langle \Psi \vert )^2
=
\left\{
\left[\sum_{i=1}^n \sqrt{\rho_i}  \vert \psi_i\rangle  \vert \psi_i\rangle \right]
\left[\sum_{j=1}^n \sqrt{\rho_j}  \langle  \psi_j\vert \langle \psi_j\vert \right]
\right\}^2
\\=
\left[\sum_{i_1=1}^n \sqrt{\rho_{i_1}}  \vert \psi_{i_1}\rangle  \vert \psi_{i_1}\rangle \right]
\left[\sum_{j_1=1}^n \sqrt{\rho_{j_1}}  \langle  \psi_{j_1}\vert \langle \psi_{j_1}\vert \right]\times \\
\left[\sum_{i_2=1}^n \sqrt{\rho_{i_2}}  \vert \psi_{i_2}\rangle  \vert \psi_{i_2}\rangle \right]
\left[\sum_{j_2=1}^n \sqrt{\rho_{j_2}}  \langle  \psi_{j_2}\vert \langle \psi_{j_2}\vert \right]
\qquad
\\=
\left[\sum_{i_1=1}^n \sqrt{\rho_{i_1}}  \vert \psi_{i_1}\rangle  \vert \psi_{i_1}\rangle \right]
\underbrace{
\left[\sum_{j_1=1}^n \sum_{i_2=1}^n \sqrt{\rho_{j_1}}\sqrt{\rho_{i_2}}  (\delta_{i_2 j_1})^2 \right]
}_
{
 \sum_{j_1=1}^n   \rho_{j_1} = 1
}
\left[\sum_{j_2=1}^n \sqrt{\rho_{j_2}}  \langle  \psi_{j_2}\vert \langle \psi_{j_2}\vert \right]
\\=
\left[\sum_{i_1=1}^n \sqrt{\rho_{i_1}}  \vert \psi_{i_1}\rangle  \vert \psi_{i_1}\rangle \right]
\left[\sum_{j_2=1}^n \sqrt{\rho_{j_2}}  \langle  \psi_{j_2}\vert \langle \psi_{j_2}\vert \right]
 =  \vert \Psi\rangle \langle \Psi \vert
.
\label{2015-puranproof}
\end{split}
\end{equation}
}

Note that this construction is not unique -- any construction
$\vert \Psi' \rangle = \sum_{i=1}^n \sqrt{\rho_i}  \vert \psi_i\rangle  \vert \phi_i\rangle$
involving auxiliary components
$\vert \phi_i\rangle$
representing the elements of some orthonormal basis $\{\vert \phi_1\rangle , \ldots , \vert \phi_n\rangle \}$
would suffice.

The original mixed state ${\boldsymbol{\rho}}$ is obtained from the pure state (\ref{2015-puran})
corresponding to the unit vector $\vert \Psi\rangle = \vert \psi \rangle \vert \psi^a \rangle  = \vert \psi \psi^a \rangle$
-- we might say that ``the superscript $a$ stands for auxiliary'' --
by a partial trace (cf. Section~\ref{2015-partialtrace}) over one of its components, say  $\vert \psi^a\rangle$.
\index{partial trace}

{\color{OliveGreen}\bproof
For the sake of a proof let us ``trace out of the auxiliary components $\vert \psi^a\rangle$,'' that is,
take the trace
\begin{equation}
\textrm{Tr}_{a} (\vert \Psi\rangle \langle \Psi \vert )
=
\sum_{k=1}^n  \langle \psi^a_{k} \vert  (\vert \Psi\rangle \langle \Psi \vert ) \vert \psi^a_{k}\rangle
\end{equation}
of
$\vert \Psi\rangle \langle \Psi \vert$
with respect to one of its components $\vert \psi^a\rangle$:
\begin{equation}
\begin{split}
\textrm{Tr}_{a}\left(
\vert \Psi\rangle \langle \Psi \vert
\right)
\\=
\textrm{Tr}_{a}\left(
\left[\sum_{i=1}^n \sqrt{\rho_{i}}  \vert \psi_{i}\rangle  \vert \psi^a_{i}\rangle \right]
\left[\sum_{j=1}^n \sqrt{\rho_{j}}  \langle  \psi^a_{j}\vert \langle \psi_{j}\vert \right]
\right)
\\=
\sum_{k=1}^n  \left\langle \psi^a_{k} \left\vert
\left[\sum_{i=1}^n \sqrt{\rho_{i}}  \vert \psi_{i}\rangle  \vert \psi^a_{i}\rangle \right]
\left[\sum_{j=1}^n \sqrt{\rho_{j}}  \langle  \psi^a_{j}\vert \langle \psi_{j}\vert \right]
\right\vert \psi^a_{k}\right\rangle
\\=
\sum_{k=1}^n \sum_{i=1}^n \sum_{j=1}^n  \delta_{ki} \delta_{kj}
\sqrt{\rho_{i}} \sqrt{\rho_{j}}
\vert \psi_{i}\rangle
\langle \psi_{j}\vert
\\=
\sum_{k=1}^n
\rho_{k}
\vert \psi_{k}\rangle
\langle \psi_{k}\vert
= {\boldsymbol{\rho}}
.
\label{2015-puranproof1}
\end{split}
\end{equation}
}

\section{Commutativity}
\index{commutativity}
\marginnote{For proofs and additional information see {\S}79 \& {\S}84 in~\bibentry{halmos-vs}.

Recall that, as defined in Eq.~(\ref{2020-commutator}) on page~\pageref{2020-commutator}
the  commutator
\index{commutator}
of two matrices $\textsf{\textbf{A}}$  and $\textsf{\textbf{B}}$ is
$
[\textsf{\textbf{A}}, \textsf{\textbf{B}} ]
=
\textsf{\textbf{A}} \textsf{\textbf{B}}
-
 \textsf{\textbf{B}}      \textsf{\textbf{A}}
$.}

If $\textsf{\textbf{A}}=\sum_{i=1}^k \lambda_i\textsf{\textbf{E}}_i$
is the spectral form of a self-adjoint transformation  $\textsf{\textbf{A}}$
on a finite-dimensional inner product space,
then a necessary and sufficient condition (``if and only if $=$ iff'')
that a linear transformation
 $\textsf{\textbf{B}}$ commutes with
 $\textsf{\textbf{A}}$
is that it commutes with each
$\textsf{\textbf{E}}_i$, $1\le i\le k$.

{\color{OliveGreen}\bproof
Sufficiency is derived easily: whenever   $\textsf{\textbf{B}}$
commutes with all the projectors $\textsf{\textbf{E}}_i$, $1\le i\le k$
in the spectral decomposition
$\textsf{\textbf{A}}
=
\sum_{i=1}^k \lambda_i \textsf{\textbf{E}}_i
$
of   $\textsf{\textbf{A}}$,
then it commutes with $\textsf{\textbf{A}}$; that is,
\begin{equation}
\begin{split}
 \textsf{\textbf{B}}\textsf{\textbf{A}}
=
\textsf{\textbf{B}}\left(\sum_{i=1}^k \lambda_i \textsf{\textbf{E}}_i\right)
=
\sum_{i=1}^k \lambda_i \textsf{\textbf{B}}\textsf{\textbf{E}}_i
\\
=
\sum_{i=1}^k \lambda_i \textsf{\textbf{E}}_i  \textsf{\textbf{B}}
=
\left(\sum_{i=1}^k \lambda_i \textsf{\textbf{E}}_i \right)  \textsf{\textbf{B}}
=\textsf{\textbf{A}} \textsf{\textbf{B}}
.
\end{split}
\end{equation}

Necessity follows from the fact that, if  $\textsf{\textbf{B}}$
commutes with  $\textsf{\textbf{A}}$
then it also commutes with every polynomial of  $\textsf{\textbf{A}}$,
since in this case $\textsf{\textbf{A}}\textsf{\textbf{B}} = \textsf{\textbf{B}} \textsf{\textbf{A}}$, and thus
$
\textsf{\textbf{A}}^m \textsf{\textbf{B}}
=
\textsf{\textbf{A}}^{m-1} \textsf{\textbf{A}} \textsf{\textbf{B}}
=
\textsf{\textbf{A}}^{m-1}  \textsf{\textbf{B}} \textsf{\textbf{A}} =
\ldots = \textsf{\textbf{B}} \textsf{\textbf{A}}^m$.
In particular, it commutes with the polynomial $p_i(\textsf{\textbf{A}})=\textsf{\textbf{E}}_i$
defined by Equation~(\ref{2011-m-epsf}).
\eproof
}

If $\textsf{\textbf{A}}=\sum_{i=1}^k \lambda_i\textsf{\textbf{E}}_i$
and
$\textsf{\textbf{B}}=\sum_{j=1}^l \mu_j\textsf{\textbf{F}}_j$
are the spectral forms of a self-adjoint transformations
$\textsf{\textbf{A}}$ and $\textsf{\textbf{B}}$
on a finite-dimensional inner product space,
then a necessary and sufficient condition (``if and only if $=$ iff'')
that  $\textsf{\textbf{A}}$ and
 $\textsf{\textbf{B}}$ commute
is that the projections
$\textsf{\textbf{E}}_i$, $1\le i\le k$
and
$\textsf{\textbf{F}}_j$, $1\le j\le l$
commute with each other; i.e.,
$\left[\textsf{\textbf{E}}_i,\textsf{\textbf{F}}_j\right] =
\textsf{\textbf{E}}_i \textsf{\textbf{F}}_j-
\textsf{\textbf{F}}_j\textsf{\textbf{E}}_i =0 $.

{\color{OliveGreen}\bproof
Again, sufficiency can be derived as follows: suppose all projection operators $\textsf{\textbf{F}}_j$, $1\le j\le l$
occurring in the spectral decomposition of $\textsf{\textbf{B}}$
commute with all projection operators $\textsf{\textbf{E}}_i$, $1\le i\le k$
in the spectral composition of   $\textsf{\textbf{A}}$,
then
\begin{equation}
\begin{split}
 \textsf{\textbf{B}}\textsf{\textbf{A}}
=
\left(\sum_{j=1}^l \mu_j\textsf{\textbf{F}}_j\right) \left(\sum_{i=1}^k \lambda_i \textsf{\textbf{E}}_i\right)
=
\sum_{i=1}^k \sum_{j=1}^l \lambda_i \mu_j \textsf{\textbf{F}}_j\textsf{\textbf{E}}_i
\\
=
\sum_{j=1}^l \sum_{i=1}^k \mu_j \lambda_i \textsf{\textbf{E}}_i \textsf{\textbf{F}}_j
=
\left(\sum_{i=1}^k \lambda_i \textsf{\textbf{E}}_i \right)  \left(\sum_{j=1}^l \mu_j \textsf{\textbf{F}}_j\right)
=\textsf{\textbf{A}} \textsf{\textbf{B}}
.
\end{split}
\end{equation}

Necessity follows from the fact that, if $\textsf{\textbf{F}}_j$, $1\le j\le l$
commutes with  $\textsf{\textbf{A}}$
then, by the same argument as mentioned earlier,
it also commutes with every polynomial of  $\textsf{\textbf{A}}$;
and
hence also with $p_i(\textsf{\textbf{A}})=\textsf{\textbf{E}}_i$ defined by Equation~(\ref{2011-m-epsf}).
Conversely,
if $\textsf{\textbf{E}}_i$, $1\le i\le k$
commutes with  $\textsf{\textbf{B}}$
then it also commutes with every polynomial of  $\textsf{\textbf{B}}$;
and
hence also with the associated polynomial
$q_j(\textsf{\textbf{B}})=\textsf{\textbf{F}}_j$ defined by Equation~(\ref{2011-m-epsf});
where $q_j(t)$ is a polynomial containing the eigenvalues of $\textsf{\textbf{B}}$.

A more compact proof of necessity uses the two polynomials
$p_i(\textsf{\textbf{A}})=\textsf{\textbf{E}}_i$
and
$q_j(\textsf{\textbf{B}})=\textsf{\textbf{F}}_j$
according to Equation~(\ref{2011-m-epsf})
simultaneously:
If
$[\textsf{\textbf{A}},\textsf{\textbf{B}}] = 0$
then so is
$[p_i(\textsf{\textbf{A}}),q_j(\textsf{\textbf{B}})] = [\textsf{\textbf{E}}_i,\textsf{\textbf{F}}_j]= 0$.
\eproof
}


Suppose, as the simplest case, that $\textsf{\textbf{A}}$ and $\textsf{\textbf{B}}$
both have nondegenerate spectra.
Then all commuting projection operators $\left[
\textsf{\textbf{E}}_{i}
,
\textsf{\textbf{F}}_{j}
\right]=
\textsf{\textbf{E}}_{i}
\textsf{\textbf{F}}_{j}
-
\textsf{\textbf{F}}_{j}
\textsf{\textbf{E}}_{i}
=0$ are of the form
$\textsf{\textbf{E}}_{i} = \vert {\bf e}_i \rangle \langle {\bf e}_i \vert$
and
$\textsf{\textbf{F}}_{j} = \vert {\bf f}_j \rangle \langle {\bf f}_j \vert$
associated with the one-dimensional subspaces of ${\frak V}$
spanned by the normalized vectors ${\bf e}_i$  and ${\bf f}_j$,
respectively.
In this case those projection operators are either identical (that is, the vectors are collinear)
or orthogonal (that is, the vector ${\bf e}_i$ is orthogonal to ${\bf f}_j$).

{\color{OliveGreen}\bproof
For a proof,\marginnote{\color{black}Please note that the Einstein summation convention does not apply here.}
note that if $\textsf{\textbf{E}}_{i}$
and
$\textsf{\textbf{F}}_{j}$ commute, then  multiplying the commutator
$\left[
\textsf{\textbf{E}}_{i}
,
\textsf{\textbf{F}}_{j}
\right]=0$
both with
$\textsf{\textbf{E}}_{i}$ from the right
and  with
$\textsf{\textbf{F}}_{j}$ from the left one obtains
\begin{equation}
\begin{split}
\textsf{\textbf{E}}_{i}
\textsf{\textbf{F}}_{j}
=
\textsf{\textbf{F}}_{j}
\textsf{\textbf{E}}_{i}
,\\
\textsf{\textbf{E}}_{i}
\textsf{\textbf{F}}_{j}\textsf{\textbf{E}}_{i}
=
\textsf{\textbf{F}}_{j}
\textsf{\textbf{E}}_{i}^2
=
\textsf{\textbf{F}}_{j}
\textsf{\textbf{E}}_{i}
,\\
\textsf{\textbf{F}}_{j}\textsf{\textbf{E}}_{i}
\textsf{\textbf{F}}_{j}
=
\textsf{\textbf{F}}_{j}^2
\textsf{\textbf{E}}_{i}
=
\textsf{\textbf{F}}_{j}
\textsf{\textbf{E}}_{i}
,\\
\textsf{\textbf{F}}_{j}\textsf{\textbf{E}}_{i}
\textsf{\textbf{F}}_{j}
=
\textsf{\textbf{E}}_{i}
\textsf{\textbf{F}}_{j}\textsf{\textbf{E}}_{i}
,\\
\vert {\bf f}_j \rangle  \langle {\bf f}_j \vert {\bf e}_i \rangle \langle {\bf e}_i \vert {\bf f}_j \rangle \langle {\bf f}_j \vert
=
 \vert {\bf e}_i \rangle \langle {\bf e}_i \vert {\bf f}_j \rangle \langle {\bf f}_j \vert {\bf e}_i \rangle  \langle {\bf e}_i \vert
,\\
\left|\langle {\bf e}_i \vert {\bf f}_j \rangle\right|^2
\vert {\bf f}_j \rangle \langle {\bf f}_j \vert
=
\left|\langle {\bf e}_i \vert {\bf f}_j \rangle\right|^2
 \vert {\bf e}_i \rangle \langle {\bf e}_i \vert
,                                               \\
\left|\langle {\bf e}_i \vert {\bf f}_j \rangle\right|^2 \left(
\vert {\bf f}_j \rangle \langle {\bf f}_j \vert   -
 \vert {\bf e}_i \rangle \langle {\bf e}_i \vert \right) =    0
,
\end{split}
\end{equation}
which   only holds if either
${\bf e}_i$
and
${\bf f}_j$
are collinear -- in which case
$\textsf{\textbf{E}}_{i} = \textsf{\textbf{F}}_{j}$ --
or orthogonal
 -- in which case
$\textsf{\textbf{E}}_{i} \perp  \textsf{\textbf{F}}_{j}$,
and thus
$\textsf{\textbf{E}}_{i} \textsf{\textbf{F}}_{j} = 0$.
\eproof
}

Therefore, for two or more mutually commuting nondegenerate operators,
the (re)arrangement of the respective orthogonal projection operators (and their associated orthonormal bases)
in the respective spectral forms by
permution and identifying identical projection operators
yields consistent and identical systems of projection operators (and their associated orthonormal bases) --
commuting normal operators share  eigenvectors in their eigensystems, and therefore projection operators in their spectral form;
the only difference being the different eigenvalues.

For two or more mutually commuting operators which may be degenerate this may no longer be the case
because two- or higher dimensional subspaces can be spanned by nonunique bases thereof,
and as a result there may be a mismatch between the such projections.
But it is always possible to co-align the one-dimensional projection operators spanning the subspaces
of commuting operators such that they share a common set of projection operators in their spectral decompositions.


This result can be expressed in the following way:
Consider some set $\textsf{\textbf{M}}
=
\{
\textsf{\textbf{A}}_1,
\textsf{\textbf{A}}_2,
\ldots ,
\textsf{\textbf{A}}_k
\}
$
of  self-adjoint transformations on a finite-dimensional inner product space.
These  transformations  $\textsf{\textbf{A}}_i \in \textsf{\textbf{M}}$, $1\le i \le k$
are mutually commuting -- that is, $[\textsf{\textbf{A}}_i,\textsf{\textbf{A}}_j] =0$ for all $1\le i,j \le k$ -- if and only if there exists
a {\em maximal} (with respect to the set $\textsf{\textbf{M}}$) self-adjoint transformation  $\textsf{\textbf{R}}$ and
a set of real-valued functions
$F
=
\{
f_1,
f_2,
\ldots ,
f_k
\}
$ of a real variable so that
$
\textsf{\textbf{A}}_1=f_1 (\textsf{\textbf{R}})
$,
$
\textsf{\textbf{A}}_2=f_2 (\textsf{\textbf{R}})
$,
$\ldots $,
$\textsf{\textbf{A}}_k=f_k (\textsf{\textbf{R}})$.
If such a {\em maximal operator} $\textsf{\textbf{R}}$ exists, then
it can be written as a function of all transformations in the set $\textsf{\textbf{M}}$; that is,
$\textsf{\textbf{R}}=G(\textsf{\textbf{A}}_1,
\textsf{\textbf{A}}_2,
\ldots ,
\textsf{\textbf{A}}_k)$,
where $G$ is a suitable real-valued function of $n$ variables
(cf. Ref.\cite{v-neumann-31}, Satz 8).
\index{maximal operator}
\index{maximal transformation}

{\color{OliveGreen}\bproof
For a proof involving two operators $\textsf{\textbf{A}}_1$ and $\textsf{\textbf{A}}_2$ we note that
sufficiency  can be derived from   commutativity, which follows from
$
\textsf{\textbf{A}}_1\textsf{\textbf{A}}_2=f_1 (\textsf{\textbf{R}}) f_2 (\textsf{\textbf{R}}) =
f_2 (\textsf{\textbf{R}}) f_1 (\textsf{\textbf{R}})= \textsf{\textbf{A}}_2\textsf{\textbf{A}}_1
$.

Necessity follows by first noticing that, as derived earlier, the projection operators
$\textsf{\textbf{E}}_i$ and $\textsf{\textbf{F}}_j$
in the spectral forms of $\textsf{\textbf{A}}_1=\sum_{i=1}^k \lambda_i\textsf{\textbf{E}}_i$
and
$\textsf{\textbf{A}}_2=\sum_{j=1}^l \mu_j\textsf{\textbf{F}}_j$ mutually commute; that is,
$
\textsf{\textbf{E}}_i \textsf{\textbf{F}}_j
=
\textsf{\textbf{F}}_j \textsf{\textbf{E}}_i
$.

For the sake of construction, design $g(x,y)\in {\Bbb R}$ to be any real-valued function (which can be a polynomial) of two real variables $x,y \in {\Bbb R}$ with the property
that all the coefficients  $c_{ij} = g( \lambda_i,\mu_j )$ are distinct.
Next, define  the maximal operator $\textsf{\textbf{R}}$ by
\begin{equation}
\textsf{\textbf{R}} = g(\textsf{\textbf{A}}_1, \textsf{\textbf{A}}_2) =  \sum_{i=1}^k \sum_{j=1}^l c_{ij}  \textsf{\textbf{E}}_i\textsf{\textbf{F}}_j,
\end{equation}
and the two functions $f_1$ and $f_2$ such that
$f_1(c_{ij}) = \lambda_i$, as well as
$f_2(c_{ij}) = \mu_j$, which result in
\begin{equation}
\begin{split}
f_1(\textsf{\textbf{R}}) = \sum_{i=1}^k \sum_{j=1}^j f_1(c_{ij})  \textsf{\textbf{E}}_i\textsf{\textbf{F}}_j = \sum_{i=1}^k \sum_{j=1}^j \lambda_i  \textsf{\textbf{E}}_i\textsf{\textbf{F}}_j
= \left( \sum_{i=1}^k \lambda_i \textsf{\textbf{E}}_i\right) \underbrace{\left(\sum_{j=1}^l \textsf{\textbf{F}}_j\right)}_\mathbb{1} = \textsf{\textbf{A}}_1
\\
f_2(\textsf{\textbf{R}}) = \sum_{i=1}^k \sum_{j=1}^j f_2(c_{ij})  \textsf{\textbf{E}}_i\textsf{\textbf{F}}_j = \sum_{i=1}^k \sum_{j=1}^j \mu_j  \textsf{\textbf{E}}_i\textsf{\textbf{F}}_j
= \underbrace{\left(\sum_{i=1}^k \textsf{\textbf{E}}_i\right)}_\mathbb{1}\left( \sum_{j=1}^l \mu_j \textsf{\textbf{F}}_j\right)  = \textsf{\textbf{A}}_2
.
\end{split}
\end{equation}
A generalization to arbitrary numbers $n$ of mutually commuting operators follows by induction:
for mutually distinct coefficients $c_{i_1 i_2 \cdots i_n}$ and the polynomials $p,q,\ldots ,r$ referring
to the ones defined in equation~(\ref{2011-m-epsf}),
\begin{equation}
\begin{split}
\textsf{\textbf{R}} = g(\textsf{\textbf{A}}_1, \textsf{\textbf{A}}_2,\dots , \textsf{\textbf{A}}_n)
=  \sum_{i_1=1}^{k_1} \sum_{i_2=1}^{k_2} \cdots  \sum_{i_n=1}^{k_n}
c_{i_1 i_2 \cdots i_n}  p_{i_1}(\textsf{\textbf{A}}_1) q_{i_2}(\textsf{\textbf{A}}_2) \cdots r_{i_n}(\textsf{\textbf{A}}_n)
\\
=  \sum_{i_1=1}^{k_1} \sum_{i_2=1}^{k_2} \cdots  \sum_{i_n=1}^{k_n}
c_{i_1 i_2 \cdots i_n}  \textsf{\textbf{E}}_{i_1}\textsf{\textbf{F}}_{i_2} \cdots \textsf{\textbf{G}}_{i_n}
.
\end{split}
\end{equation}
\eproof
}

The  maximal operator $\textsf{\textbf{R}}$ can be interpreted as
encoding or containing all the information of a collection of commuting operators at once.
Stated pointedly, rather than to enumerate all the $k$ operators in $\textsf{\textbf{M}}$
separately,
a single maximal operator  $\textsf{\textbf{R}}$  represents  $\textsf{\textbf{M}}$;
in this sense, the operators  $\textsf{\textbf{A}}_i \in \textsf{\textbf{M}}$
are all just (most likely incomplete) {\em aspects}  of --
or individual, ``lossy'' (i.e., one-to-many) functional views on -- the  maximal operator $\textsf{\textbf{R}}$.

%
%
%
%
%
%
%
%
%
%
%
%
%

{\color{blue}
\bexample
Let us demonstrate the machinery developed so far by an example.
Consider the normal matrices
$$
\textsf{\textbf{A}} = 
\begin{pmatrix}
0& 1& 0\\ 1& 0& 0\\ 0& 0& 0
\end{pmatrix},\;
\textsf{\textbf{B}} = 
\begin{pmatrix}
2& 3& 0\\ 3& 2& 0\\ 0& 0& 0
\end{pmatrix},\;
\textsf{\textbf{C}} = 
\begin{pmatrix}
5& 7& 0\\ 7& 5& 0\\ 0& 0& 11
\end{pmatrix},
$$
which are mutually commutative; that is,
$
[\textsf{\textbf{A}}, \textsf{\textbf{B}}]=
\textsf{\textbf{A}} \textsf{\textbf{B}}-\textsf{\textbf{B}}\textsf{\textbf{A}}=
[\textsf{\textbf{A}}, \textsf{\textbf{C}}]=
\textsf{\textbf{A}} \textsf{\textbf{C}}-\textsf{\textbf{B}}\textsf{\textbf{C}}=
[\textsf{\textbf{B}}, \textsf{\textbf{C}}]=
\textsf{\textbf{B}} \textsf{\textbf{C}}-\textsf{\textbf{C}}\textsf{\textbf{B}}=0$.

The eigensystems -- that is, the set of the set of eigenvalues and the set of the associated eigenvectors -- of $\textsf{\textbf{A}}$,
$\textsf{\textbf{B}}$
and
$\textsf{\textbf{C}}$
are
\begin{equation}
\begin{split}
\{\{1,-1,  0\}, \{(1, 1, 0)^\intercal , (-1, 1, 0)^\intercal , (0, 0, 1)^\intercal \}\} ,\\
\{\{5, -1, 0\},  \{(1, 1, 0)^\intercal , (-1, 1, 0)^\intercal , (0, 0, 1)^\intercal \}\},\\
\{\{12, -2, 11\},  \{(1, 1, 0)^\intercal , (-1, 1, 0)^\intercal , (0, 0, 1)^\intercal \}\}.
\end{split}
\end{equation}
They share a common orthonormal set of eigenvectors
$$
\left\{
\frac{1}{\sqrt{2}}
\begin{pmatrix}
1\\ 1\\ 0
\end{pmatrix},
\frac{1}{\sqrt{2}}
\begin{pmatrix}
-1\\ 1\\ 0
\end{pmatrix},
\begin{pmatrix}
0\\ 0\\ 1\end{pmatrix}
\right\}
$$
which form an orthonormal basis of ${\Bbb R}^3$ or ${\Bbb C}^3$.
The associated projections are obtained by the outer (dyadic or tensor) products
\index{outer product}
\index{dyadic product}
\index{tensor product}
of these vectors; that is,
\begin{equation}
\begin{split}
\textsf{\textbf{E}}_1= \frac{1}{2}
\begin{pmatrix}
1& 1& 0\\
1& 1& 0\\
0& 0& 0
\end{pmatrix},\\
\textsf{\textbf{E}}_2= \frac{1}{2}
\begin{pmatrix}
1& -1& 0\\
-1& 1& 0\\
0& 0& 0
\end{pmatrix},\\
\textsf{\textbf{E}}_3=
\begin{pmatrix}
0& 0& 0\\
0& 0& 0\\
0& 0& 1
\end{pmatrix}.
\end{split}
\end{equation}
Thus the spectral decompositions of
$\textsf{\textbf{A}}$,
$\textsf{\textbf{B}}$  and
$\textsf{\textbf{C}}$ are
\begin{equation}
\begin{split}
\textsf{\textbf{A}}= \textsf{\textbf{E}}_1  - \textsf{\textbf{E}}_2  + 0  \textsf{\textbf{E}}_3,\\
\textsf{\textbf{B}}= 5\textsf{\textbf{E}}_1  - \textsf{\textbf{E}}_2  + 0 \textsf{\textbf{E}}_3,\\
\textsf{\textbf{C}}= 12\textsf{\textbf{E}}_1  -2 \textsf{\textbf{E}}_2  + 11\textsf{\textbf{E}}_3,
\end{split}
\label{2011-m-empc}
\end{equation}
respectively.

One way to define the  maximal operator  $\textsf{\textbf{R}}$ for this problem
would be
$$
\textsf{\textbf{R}} = \alpha \textsf{\textbf{E}}_1  + \beta \textsf{\textbf{E}}_2  + \gamma  \textsf{\textbf{E}}_3,
$$
with
$\alpha ,  \beta ,   \gamma \in {\Bbb R}-0$ and
$\alpha  \neq \beta  \neq   \gamma \neq \alpha  $.
The functional coordinates
$f_i(\alpha )$, $f_i(\beta)$, and $f_i(\gamma)$,
$i\in \{\textsf{\textbf{A}},\textsf{\textbf{B}},\textsf{\textbf{C}}\}$,  of the three functions
$ f_\textsf{\textbf{A}}(\textsf{\textbf{R}})$,
$ f_\textsf{\textbf{B}}(\textsf{\textbf{R}})$, and
$ f_\textsf{\textbf{C}}(\textsf{\textbf{R}})$
chosen to match the projection coefficients obtained in Equation~(\ref{2011-m-empc});
that is,
\begin{equation}
\begin{split}
\textsf{\textbf{A}}= f_\textsf{\textbf{A}}(\textsf{\textbf{R}})=  \textsf{\textbf{E}}_1  - \textsf{\textbf{E}}_2  + 0  \textsf{\textbf{E}}_3,\\
\textsf{\textbf{B}}=  f_\textsf{\textbf{B}}(\textsf{\textbf{R}})= 5\textsf{\textbf{E}}_1  - \textsf{\textbf{E}}_2  + 0 \textsf{\textbf{E}}_3,\\
\textsf{\textbf{C}}=  f_\textsf{\textbf{C}}(\textsf{\textbf{R}})= 12\textsf{\textbf{E}}_1  -2 \textsf{\textbf{E}}_2  + 11\textsf{\textbf{E}}_3.
\end{split}
\end{equation}
As a consequence, the functions
$\textsf{\textbf{A}}$,
$\textsf{\textbf{B}}$,
$\textsf{\textbf{C}}$ need to satisfy the relations
\begin{equation}
\begin{split}
f_\textsf{\textbf{A}}(\alpha ) =1,\; f_\textsf{\textbf{A}}(\beta ) =-1,\; f_\textsf{\textbf{A}}(\gamma ) =0,\\
f_\textsf{\textbf{B}}(\alpha ) =5,\; f_\textsf{\textbf{B}}(\beta ) =-1,\; f_\textsf{\textbf{B}}(\gamma ) =0,\\
f_\textsf{\textbf{C}}(\alpha ) =12,\; f_\textsf{\textbf{C}}(\beta ) =-2,\; f_\textsf{\textbf{C}}(\gamma ) =11.
\end{split}
\label{2011-m-empc-fsr}
\end{equation}

\eexample
}

It is no coincidence that the projections in the spectral forms of
$\textsf{\textbf{A}}$,
$\textsf{\textbf{B}}$  and
$\textsf{\textbf{C}}$ are identical.
Indeed it can be shown that mutually commuting {\em normal operators} always share the same eigenvectors; and thus also the same projections.
\index{normal operator}
\index{normal transformation}

Let the set $\textsf{\textbf{M}}
=
\{
\textsf{\textbf{A}}_1,
\textsf{\textbf{A}}_2,
\ldots ,
\textsf{\textbf{A}}_k
\}
$
be mutually commuting  normal (or Hermitian, or self-adjoint) transformations on an $n$-dimensional inner product space.
Then there exists an orthonormal basis
${\frak B}= \{
{\bf f}_1,
\ldots ,
{\bf f}_n\}$
such that every ${\bf f}_j \in {\frak B}$  is an eigenvector  of each of the $\textsf{\textbf{A}}_i \in  \textsf{\textbf{M}}$.
Equivalently, there exist $n$ orthogonal projections  (let the vectors ${\bf f}_j$ be represented by the coordinates which are column vectors)
$\textsf{\textbf{E}}_j= {\bf f}_j\otimes {\bf f}_j^\dagger$
such that every $\textsf{\textbf{E}}_j$, $1\le j\le n$ occurs in the spectral form of each of the $\textsf{\textbf{A}}_i \in  \textsf{\textbf{M}}$.

Informally speaking,
a ``generic'' maximal operator $\textsf{\textbf{R}}$ on an $n$-dimensional Hilbert space ${\frak V}$
can be interpreted in terms of a particular orthonormal basis
$\{{\bf f}_1,{\bf f}_2,\ldots ,{\bf f}_n\}$ of ${\frak V}$
-- indeed, the $n$ elements of that basis would have to correspond to the projections occurring
in the spectral decomposition of the self-adjoint operators
generated by $\textsf{\textbf{R}}$.

{\color{Purple}
Likewise, the ``maximal knowledge'' about a quantized physical system -- in terms of empirical operational quantities --
would correspond to such a single maximal operator;
or to the orthonormal basis corresponding to the spectral decomposition of it.
Thus it might not be unreasonable to speculate that a particular (pure) physical state is best characterized by a particular orthonormal basis.
}

\section{Measures on closed subspaces}

In what follows we shall assume that all {\em (probability) measures}
or {\em states}
\index{probability measures}
\index{measures}
\index{states}
behave quasi-classically on sets of mutually commuting self-adjoint operators,
and, in particular, on orthogonal projections. One could call this property {\em subclassicality}.

This can be formalized as follows.
Consider some set
 $
\{
\vert {\bf x}_1 \rangle ,
\vert {\bf x}_2 \rangle ,
\ldots ,
\vert {\bf x}_k \rangle
\}
$
of mutually orthogonal, normalized vectors,
so that $ \langle {\bf x}_i \vert {\bf x}_j \rangle =\delta_{ij}$;
and associated with it,
the set
 $
\{
\textsf{\textbf{E}}_1,
\textsf{\textbf{E}}_2,
\ldots ,
\textsf{\textbf{E}}_k
\}
$
of mutually orthogonal (and thus commuting) one-dimensional projections
$\textsf{\textbf{E}}_i= \vert {\bf x}_i \rangle \langle {\bf x}_i \vert$
on a finite-dimensional inner product space   ${\frak V}$.

We require that probability measures $\mu$ on such mutually commuting sets of observables behave quasi-classically.
Therefore, they
should be {\em additive}; that is,
\begin{equation}
\mu \left(\sum_{i=1}^k \textsf{\textbf{E}}_i \right)= \sum_{i=1}^k \mu \left(\textsf{\textbf{E}}_i \right)
.
\label{2015-m-fdlvs-ff}
\end{equation}
Such a measure is determined by its values on the one-dimensional projections.

Stated differently, we shall assume that,
for any two orthogonal projections $ \textsf{\textbf{E}}$ and $\textsf{\textbf{F}}$
\index{orthogonal projection}
if
 $ \textsf{\textbf{E}}\textsf{\textbf{F}}= \textsf{\textbf{F}}\textsf{\textbf{E}}=0$,
their sum
 $  \textsf{\textbf{G}} =\textsf{\textbf{E}}+\textsf{\textbf{F}}$
has expectation value
\begin{equation}
\mu ( \textsf{\textbf{G}} ) \equiv
\langle \textsf{\textbf{G}}\rangle =
\langle \textsf{\textbf{E}} \rangle +
\langle \textsf{\textbf{F}} \rangle
\equiv
\mu  ( \textsf{\textbf{E}} ) +
\mu  ( \textsf{\textbf{F}} )
.
\end{equation}

Any such measure $\mu $ satisfying (\ref{2015-m-fdlvs-ff})
can be expressed in terms of a (positive) real valued function $f$ on the unit vectors in  ${\frak V}$
by
\begin{equation}
\mu \left( \textsf{\textbf{E}}_x \right)=
f( \vert {\bf x} \rangle )
\equiv f(  {\bf x}  )
,
\label{2015-m-fdlvs-ffmuf}
\end{equation}
(where $\textsf{\textbf{E}}_x= \vert {\bf x}  \rangle \langle {\bf x}  \vert$ for all  unit vectors $\vert {\bf x}  \rangle \in {\frak V}$)
by requiring that,
for every orthonormal basis
${\frak B} =\{
\vert {\bf e}_1 \rangle ,
\vert {\bf e}_2 \rangle ,
\ldots ,
\vert {\bf e}_n \rangle
\}$,  the sum of all basis vectors yields $1$; that is,
\begin{equation}
\sum_{i=1}^n f \left( \vert {\bf e}_i \rangle \right)
\equiv
\sum_{i=1}^n f \left(   {\bf e}_i  \right)
=1
.
\label{2015-m-fdlvs-ff2}
\end{equation}
$f$ is called a (positive) {\em frame function} of weight $1$.
\index{frame function}

\subsection{Gleason's theorem}
\index{Gleason's theorem}
\label{Gleasontheorem}

From now on we shall mostly consider vector spaces of dimension three or greater,
since only in these cases two  orthonormal bases intertwine in a common vector, making possible
some arguments involving multiple intertwining bases  -- in two dimensions,
distinct orthonormal bases contain distinct basis vectors.

{\em Gleason's theorem}\cite{Gleason,r:dvur-93,pitowsky:218,rich-bridge,peres,hamhalter-book}
states that,
for a Hilbert space of dimension three or greater,
every frame function defined in~(\ref{2015-m-fdlvs-ff2})
is of the form  of the inner product
\begin{equation}
f \left(   {\bf x}   \right)
\equiv
f \left( \vert {\bf x} \rangle \right)
=
\langle {\bf x}  \vert \rho {\bf x} \rangle
=
\sum_{i=1}^{k\le n} \rho_i
\langle {\bf x}  \vert \psi_i\rangle \langle \psi_i \vert {\bf x} \rangle
=
\sum_{i=1}^{k\le n} \rho_i
\vert \langle {\bf x}  \vert \psi_i\rangle \vert^2
,
\label{2016-m-fdlvsgleason-f}
\end{equation}
where
(i) $\rho $
is a positive operator (and therefore self-adjoint;
see Section~\ref{2015-m-ch-fdlvs-positive} on page \pageref{2015-m-ch-fdlvs-positive}),
and
(ii) $\rho $ is
of the trace class,
meaning its trace (cf. Section~\ref{2013-ch-fdvs-trace} on page \pageref{2013-ch-fdvs-trace}) is one.
That is, $\rho=
\sum_{i=1}^{k\le n} \rho_i \vert \psi_i \rangle \langle \psi_i \vert$
with $\rho_i \in {\Bbb R}$, $\rho_i \ge 0$, and $\sum_{i=1}^{k\le n} \rho_i =1$.
No proof is given here.

In terms of projections [cf.~Eqs.(\ref{2015-m-ch-fdlvs-bornr}) on page~\pageref{2015-m-ch-fdlvs-bornr}],
(\ref{2016-m-fdlvsgleason-f}) can be written as
\begin{equation}
\mu \left( \textsf{\textbf{E}}_x \right)
=
{\rm Tr}( \rho  \textsf{\textbf{E}}_x )
\label{2016-m-fdlvsgleason-fproj}
\end{equation}

Therefore, for a Hilbert space of dimension three or greater, the  spectral theorem suggests that
the only possible form of the  expectation value
of a self-adjoint operator  $\textsf{\textbf{A}}$
has the form
\begin{equation}
\langle
\textsf{\textbf{A}}
\rangle
=
\textrm{Tr}({  \rho} \textsf{\textbf{A}}).
\label{2016-m-fdlvsgleason-ftc}
\end{equation}
In quantum physical terms, in the formula (\ref{2016-m-fdlvsgleason-ftc}) above
the trace is taken over
the operator product of the  density matrix [which represents a positive (and thus self-adjoint) operator of the trace class]
${  \rho}$
with the observable $\textsf{\textbf{A}}=\sum_{i=1}^k \lambda_i \textsf{\textbf{E}}_i $.

In particular, if $\textsf{\textbf{A}}$ is a projection $\textsf{\textbf{E}}= \vert {\bf e} \rangle \langle {\bf e} \vert$
corresponding to an elementary yes-no proposition
{\it ``the system has property Q,''} then $\langle \textsf{\textbf{E}}\rangle = \textrm{Tr}({  \rho}  \textsf{\textbf{E}})
=
\vert \langle {\bf e}\vert \rho \rangle \vert^2$ corresponds
to the probability of that property $Q$ if the system is in state $\rho= \vert \rho \rangle \langle \rho \vert$
[for a motivation, see again Eqs.~(\ref{2015-m-ch-fdlvs-bornr}) on page~\pageref{2015-m-ch-fdlvs-bornr}].

Indeed, as already observed by Gleason, even for two-dimensional Hilbert spaces, a straightforward {\it Ansatz}
yields a  probability measure satisfying (\ref{2015-m-fdlvs-ff}) as follows.
Suppose some unit vector $\vert \rho \rangle$ corresponding to a pure quantum state (preparation) is selected.
For each one-dimensional closed subspace corresponding
to a one-dimensional orthogonal projection observable (interpretable as an elementary yes-no proposition)
$E=\vert {\bf e}\rangle \langle {\bf e} \vert$
along the unit vector $\vert {\bf e}\rangle$,
define
$w_\rho(\vert {\bf e}\rangle ) =  \vert \langle {\bf e} \vert  \rho \rangle \vert^2$
to be the  square of the length $\vert \langle \rho \vert {\bf e} \rangle \vert$ of the
projection of $\vert \rho \rangle$ onto the subspace spanned by $\vert {\bf e}\rangle$.

The reason for this is that an orthonormal
basis $\{  \vert  {\bf e}_i \rangle   \}$ ``induces''
an {\it ad hoc}  probability measure $w_\rho $
on any such context (and thus basis).
To see this,
consider the length  of
the orthogonal (with respect to the basis vectors)
projections of $\vert  \rho \rangle$
onto all the basis vectors $\vert  {\bf e}_i \rangle$,
that is, the norm of the resulting vector projections of $\vert \rho \rangle$ onto the basis vectors,
respectively.
This amounts to computing the absolute value of the Euclidean scalar products
$\langle  {\bf e}_i \vert  \rho \rangle$
of the state vector with all the basis vectors.

In order that all such  absolute values of the scalar products (or the associated norms)
sum up to one and yield a probability measure as required in Equation~(\ref{2015-m-fdlvs-ff}),
recall that $\vert  \rho \rangle$ is a unit vector
and note that, by the Pythagorean theorem,
these  absolute values of the individual scalar products
-- or the associated norms of the vector projections of $\vert \rho \rangle$ onto the basis vectors --
must be squared.
Thus the value $w_\rho(\vert {\bf e}_i\rangle )$
must be the square of the scalar product of $\vert  \rho \rangle$
with $\vert  {\bf e}_i \rangle$,
corresponding to the square of the length (or norm) of
the respective projection vector of $\vert  \rho \rangle$ onto  $\vert  {\bf e}_i \rangle$.
For complex vector spaces one has to take the absolute square of the scalar product;
that is, $f_\rho (  \vert  {\bf e}_i \rangle   ) = \vert \langle  {\bf e}_i \vert  \rho \rangle \vert ^2$.

\begin{marginfigure}
\begin{center}
\unitlength 0.25mm
\linethickness{0.4pt}
\ifx\plotpoint\undefined\newsavebox{\plotpoint}\fi 
\begin{picture}(178.25,188)(0,0)
\put(71.5,81.25){\color{orange}\vector(1,0){100}}
\put(71.5,81.25){\color{orange}\vector(0,1){100}}
\put(167,109.75){\color{blue}\vector(3,1){.07}}\multiput(71.5,81.25)(.11301775148,.03372781065){845}{\color{blue}\line(1,0){.11301775148}}
\put(142.5,151.25){\vector(1,1){.07}}\multiput(71.5,81.25)(.03421686747,.033734939759){2075}{\line(1,0){.03421686747}}
\put(.5,151.25){\vector(-1,1){.07}}\multiput(71.5,81.25)(-.03421686747,.033734939759){2075}{\line(-1,0){.03421686747}}
\put(141.5,10.25){\vector(1,-1){.07}}\multiput(71.5,81.25)(.033734939759,-.03421686747){2075}{\line(0,-1){.03421686747}}
\put(167.18,109.93) {\color{orange}\line(0,-1){.9828}}
\put(167.18,107.964){\color{orange}\line(0,-1){.9828}}
\put(167.18,105.999){\color{orange}\line(0,-1){.9828}}
\put(167.18,104.033){\color{orange}\line(0,-1){.9828}}
\put(167.18,102.068){\color{orange}\line(0,-1){.9828}}
\put(167.18,100.102){\color{orange}\line(0,-1){.9828}}
\put(167.18,98.137) {\color{orange}\line(0,-1){.9828}}
\put(167.18,96.171) {\color{orange}\line(0,-1){.9828}}
\put(167.18,94.206) {\color{orange}\line(0,-1){.9828}}
\put(167.18,92.24)  {\color{orange}\line(0,-1){.9828}}
\put(167.18,90.275) {\color{orange}\line(0,-1){.9828}}
\put(167.18,88.309) {\color{orange}\line(0,-1){.9828}}
\put(167.18,86.344) {\color{orange}\line(0,-1){.9828}}
\put(167.18,84.378) {\color{orange}\line(0,-1){.9828}}
\put(167.18,82.412) {\color{orange}\line(0,-1){.9828}}
\put(166.93,109.93) {\color{orange}\line(-1,0){.9922}}
\put(164.945,109.93){\color{orange}\line(-1,0){.9922}}
\put(162.961,109.93){\color{orange}\line(-1,0){.9922}}
\put(160.977,109.93){\color{orange}\line(-1,0){.9922}}
\put(158.992,109.93){\color{orange}\line(-1,0){.9922}}
\put(157.008,109.93){\color{orange}\line(-1,0){.9922}}
\put(155.023,109.93){\color{orange}\line(-1,0){.9922}}
\put(153.039,109.93){\color{orange}\line(-1,0){.9922}}
\put(151.055,109.93){\color{orange}\line(-1,0){.9922}}
\put(149.07,109.93) {\color{orange}\line(-1,0){.9922}}
\put(147.086,109.93){\color{orange}\line(-1,0){.9922}}
\put(145.102,109.93){\color{orange}\line(-1,0){.9922}}
\put(143.117,109.93){\color{orange}\line(-1,0){.9922}}
\put(141.133,109.93){\color{orange}\line(-1,0){.9922}}
\put(139.148,109.93){\color{orange}\line(-1,0){.9922}}
\put(137.164,109.93){\color{orange}\line(-1,0){.9922}}
\put(135.18,109.93) {\color{orange}\line(-1,0){.9922}}
\put(133.195,109.93){\color{orange}\line(-1,0){.9922}}
\put(131.211,109.93){\color{orange}\line(-1,0){.9922}}
\put(129.227,109.93){\color{orange}\line(-1,0){.9922}}
\put(127.242,109.93){\color{orange}\line(-1,0){.9922}}
\put(125.258,109.93){\color{orange}\line(-1,0){.9922}}
\put(123.273,109.93){\color{orange}\line(-1,0){.9922}}
\put(121.289,109.93){\color{orange}\line(-1,0){.9922}}
\put(119.305,109.93){\color{orange}\line(-1,0){.9922}}
\put(117.32,109.93) {\color{orange}\line(-1,0){.9922}}
\put(115.336,109.93){\color{orange}\line(-1,0){.9922}}
\put(113.352,109.93){\color{orange}\line(-1,0){.9922}}
\put(111.367,109.93){\color{orange}\line(-1,0){.9922}}
\put(109.383,109.93){\color{orange}\line(-1,0){.9922}}
\put(107.398,109.93){\color{orange}\line(-1,0){.9922}}
\put(105.414,109.93){\color{orange}\line(-1,0){.9922}}
\put(103.43,109.93) {\color{orange}\line(-1,0){.9922}}
\put(101.445,109.93){\color{orange}\line(-1,0){.9922}}
\put(99.461,109.93) {\color{orange}\line(-1,0){.9922}}
\put(97.477,109.93) {\color{orange}\line(-1,0){.9922}}
\put(95.492,109.93) {\color{orange}\line(-1,0){.9922}}
\put(93.508,109.93) {\color{orange}\line(-1,0){.9922}}
\put(91.523,109.93) {\color{orange}\line(-1,0){.9922}}
\put(89.539,109.93) {\color{orange}\line(-1,0){.9922}}
\put(87.555,109.93) {\color{orange}\line(-1,0){.9922}}
\put(85.57,109.93)  {\color{orange}\line(-1,0){.9922}}
\put(83.586,109.93) {\color{orange}\line(-1,0){.9922}}
\put(81.602,109.93) {\color{orange}\line(-1,0){.9922}}
\put(79.617,109.93) {\color{orange}\line(-1,0){.9922}}
\put(77.633,109.93) {\color{orange}\line(-1,0){.9922}}
\put(75.648,109.93) {\color{orange}\line(-1,0){.9922}}
\put(73.664,109.93) {\color{orange}\line(-1,0){.9922}}
\multiput(167.18,109.93)(-.034375,-.0334239){20}{\line(-1,0){.034375}}
\multiput(165.805,108.593)(-.034375,-.0334239){20}{\line(-1,0){.034375}}
\multiput(164.43,107.256)(-.034375,-.0334239){20}{\line(-1,0){.034375}}
\multiput(163.055,105.919)(-.034375,-.0334239){20}{\line(-1,0){.034375}}
\multiput(161.68,104.582)(-.034375,-.0334239){20}{\line(-1,0){.034375}}
\multiput(160.305,103.245)(-.034375,-.0334239){20}{\line(-1,0){.034375}}
\multiput(158.93,101.908)(-.034375,-.0334239){20}{\line(-1,0){.034375}}
\multiput(157.555,100.571)(-.034375,-.0334239){20}{\line(-1,0){.034375}}
\multiput(156.18,99.234)(-.034375,-.0334239){20}{\line(-1,0){.034375}}
\multiput(154.805,97.897)(-.034375,-.0334239){20}{\line(-1,0){.034375}}
\multiput(153.43,96.56)(-.034375,-.0334239){20}{\line(-1,0){.034375}}
\multiput(152.055,95.223)(-.034375,-.0334239){20}{\line(-1,0){.034375}}
\multiput(150.68,93.886)(-.034375,-.0334239){20}{\line(-1,0){.034375}}
\multiput(149.305,92.549)(-.034375,-.0334239){20}{\line(-1,0){.034375}}
\multiput(147.93,91.212)(-.034375,-.0334239){20}{\line(-1,0){.034375}}
\multiput(146.555,89.875)(-.034375,-.0334239){20}{\line(-1,0){.034375}}
\multiput(145.18,88.538)(-.034375,-.0334239){20}{\line(-1,0){.034375}}
\multiput(143.805,87.201)(-.034375,-.0334239){20}{\line(-1,0){.034375}}
\multiput(142.43,85.864)(-.034375,-.0334239){20}{\line(-1,0){.034375}}
\multiput(141.055,84.528)(-.034375,-.0334239){20}{\line(-1,0){.034375}}
\multiput(139.68,83.191)(-.034375,-.0334239){20}{\line(-1,0){.034375}}
\multiput(138.305,81.854)(-.034375,-.0334239){20}{\line(-1,0){.034375}}
\multiput(136.93,80.517)(-.034375,-.0334239){20}{\line(-1,0){.034375}}
\multiput(135.555,79.18)(-.034375,-.0334239){20}{\line(-1,0){.034375}}
\multiput(134.18,77.843)(-.034375,-.0334239){20}{\line(-1,0){.034375}}
\multiput(132.805,76.506)(-.034375,-.0334239){20}{\line(-1,0){.034375}}
\multiput(131.43,75.169)(-.034375,-.0334239){20}{\line(-1,0){.034375}}
\multiput(130.055,73.832)(-.034375,-.0334239){20}{\line(-1,0){.034375}}
\multiput(128.68,72.495)(-.034375,-.0334239){20}{\line(-1,0){.034375}}
\multiput(127.305,71.158)(-.034375,-.0334239){20}{\line(-1,0){.034375}}
\multiput(125.93,69.821)(-.034375,-.0334239){20}{\line(-1,0){.034375}}
\multiput(124.555,68.484)(-.034375,-.0334239){20}{\line(-1,0){.034375}}
\multiput(123.18,67.147)(-.034375,-.0334239){20}{\line(-1,0){.034375}}
\multiput(121.805,65.81)(-.034375,-.0334239){20}{\line(-1,0){.034375}}
\multiput(120.43,64.473)(-.034375,-.0334239){20}{\line(-1,0){.034375}}
\multiput(119.055,63.136)(-.034375,-.0334239){20}{\line(-1,0){.034375}}
\multiput(117.68,61.799)(-.034375,-.0334239){20}{\line(-1,0){.034375}}
\multiput(116.305,60.462)(-.034375,-.0334239){20}{\line(-1,0){.034375}}
\multiput(114.93,59.125)(-.034375,-.0334239){20}{\line(-1,0){.034375}}
\multiput(113.555,57.788)(-.034375,-.0334239){20}{\line(-1,0){.034375}}
\multiput(112.18,56.451)(-.034375,-.0334239){20}{\line(-1,0){.034375}}
\multiput(110.805,55.114)(-.034375,-.0334239){20}{\line(-1,0){.034375}}
\multiput(109.43,53.778)(-.034375,-.0334239){20}{\line(-1,0){.034375}}
\multiput(108.055,52.441)(-.034375,-.0334239){20}{\line(-1,0){.034375}}
\multiput(106.68,51.104)(-.034375,-.0334239){20}{\line(-1,0){.034375}}
\multiput(105.305,49.767)(-.034375,-.0334239){20}{\line(-1,0){.034375}}
\multiput(167.43,109.93)(-.0344388,.0329082){20}{\line(-1,0){.0344388}}
\multiput(166.052,111.246)(-.0344388,.0329082){20}{\line(-1,0){.0344388}}
\multiput(164.675,112.562)(-.0344388,.0329082){20}{\line(-1,0){.0344388}}
\multiput(163.297,113.879)(-.0344388,.0329082){20}{\line(-1,0){.0344388}}
\multiput(161.92,115.195)(-.0344388,.0329082){20}{\line(-1,0){.0344388}}
\multiput(160.542,116.511)(-.0344388,.0329082){20}{\line(-1,0){.0344388}}
\multiput(159.164,117.828)(-.0344388,.0329082){20}{\line(-1,0){.0344388}}
\multiput(157.787,119.144)(-.0344388,.0329082){20}{\line(-1,0){.0344388}}
\multiput(156.409,120.46)(-.0344388,.0329082){20}{\line(-1,0){.0344388}}
\multiput(155.032,121.777)(-.0344388,.0329082){20}{\line(-1,0){.0344388}}
\multiput(153.654,123.093)(-.0344388,.0329082){20}{\line(-1,0){.0344388}}
\multiput(152.277,124.409)(-.0344388,.0329082){20}{\line(-1,0){.0344388}}
\multiput(150.899,125.726)(-.0344388,.0329082){20}{\line(-1,0){.0344388}}
\multiput(149.522,127.042)(-.0344388,.0329082){20}{\line(-1,0){.0344388}}
\multiput(148.144,128.358)(-.0344388,.0329082){20}{\line(-1,0){.0344388}}
\multiput(146.766,129.675)(-.0344388,.0329082){20}{\line(-1,0){.0344388}}
\multiput(145.389,130.991)(-.0344388,.0329082){20}{\line(-1,0){.0344388}}
\multiput(144.011,132.307)(-.0344388,.0329082){20}{\line(-1,0){.0344388}}
\multiput(142.634,133.624)(-.0344388,.0329082){20}{\line(-1,0){.0344388}}
\multiput(141.256,134.94)(-.0344388,.0329082){20}{\line(-1,0){.0344388}}
\multiput(139.879,136.256)(-.0344388,.0329082){20}{\line(-1,0){.0344388}}
\multiput(138.501,137.573)(-.0344388,.0329082){20}{\line(-1,0){.0344388}}
\multiput(137.124,138.889)(-.0344388,.0329082){20}{\line(-1,0){.0344388}}
\multiput(135.746,140.205)(-.0344388,.0329082){20}{\line(-1,0){.0344388}}
\multiput(134.368,141.522)(-.0344388,.0329082){20}{\line(-1,0){.0344388}}
\put(175,111.25){\makebox(0,0)[cc]{\color{blue}\tiny $\vert \rho \rangle$}}
\put(185,81.5){\makebox(0,0)[cc]{\color{orange}\tiny $\vert {\bf e}_1 \rangle$}}
\put(147.75,157){\makebox(0,0)[cc]{\tiny $\vert {\bf f}_1 \rangle$}}
\put(71.5,188){\makebox(0,0)[cc]{\color{orange}\tiny $\vert {\bf e}_2 \rangle$}}
\put(0,160){\makebox(0,0)[cc]{\tiny $\vert {\bf f}_2 \rangle$}}
\put(145.5,7.25){\makebox(0,0)[cc]{\tiny $- \vert {\bf f}_2 \rangle$}}
\put(135,120){\makebox(0,0)[cc]
{\color{orange}\tiny $\vert \langle \rho \vert {\bf e}_1 \rangle \vert$}}
\put(185,97){\makebox(0,0)[cc]
{\color{orange}\tiny $\vert \langle \rho \vert {\bf e}_2 \rangle \vert$}}
\put(145,64.5){\makebox(0,0)[cc]
{\tiny $\vert \langle \rho \vert {\bf f}_1 \rangle \vert$}}
\put(170,135){\makebox(0,0)[cc]
{\tiny $\vert \langle \rho \vert {\bf f}_2 \rangle \vert$}}
\end{picture}
\end{center}
\caption{Different orthonormal bases
{\color{orange}
$\{
\vert {\bf e}_1 \rangle ,
\vert {\bf e}_2 \rangle
\}$}
and
$\{
\vert {\bf f}_1 \rangle ,
\vert {\bf f}_2 \rangle
\}$
offer different ``views''
on the pure state {\color{blue} $\vert \rho \rangle$}.
As {\color{blue} $\vert \rho \rangle$} is a unit vector
it follows  from the Pythagorean theorem that
${\color{orange}
\vert \langle \rho \vert {\bf e}_1 \rangle \vert^2
+
\vert \langle \rho \vert {\bf e}_2 \rangle \vert^2}=
\vert \langle \rho \vert {\bf f}_1 \rangle \vert^2
+
\vert \langle \rho \vert {\bf f}_2 \rangle \vert^2
=1
$, thereby
motivating the use of the aboslute value (modulus) squared of the amplitude for
quantum probabilities on pure states.}
  \label{2015-m-fdlvs-vv}
\end{marginfigure}

Pointedly stated, from this point of view the probabilities $w_\rho (  \vert  {\bf e}_i \rangle   )$
are just the (absolute) squares of the coordinates
of a unit vector  $\vert \rho \rangle$ with respect to some orthonormal basis $\{  \vert  {\bf e}_i \rangle   \}$,
representable by the square $\vert \langle  {\bf e}_i \vert  \rho \rangle \vert ^2$ of the length of the vector projections of
  $\vert \rho \rangle$ onto the basis vectors   $\vert {\bf e}_i \rangle$
--
one might also say that each orthonormal basis allows ``a view'' on the pure state $\vert  \rho \rangle$.
In two dimensions this is illustrated for two bases in Figure~\ref{2015-m-fdlvs-vv}.
The squares come in because the absolute values of the individual components do not add up to one, but their squares do.
These considerations apply to Hilbert spaces of any, including two, finite dimensions.
In this nongeneral, {\it ad hoc} sense the Born rule for a system in a pure state and an elementary proposition observable
(quantum encodable by a one-dimensional projection operator) can be motivated by the requirement of additivity
for arbitrary finite-dimensional Hilbert space.

\subsection{Kochen-Specker theorem}
\index{Kochen-Specker theorem}
\label{2011-m-KST}

In what follows the overall strategy is to identify (finite) configurations of quantum observables which are then
interpreted ``as if'' they were classical observables; thereby deriving some conditions (of classical experience) which are either broken
by the quantum predictions (i.e., quantum probabilities and expectations), or yield complete contradictions.
The arguably strongest form of such a statement is the fact that, for   Hilbert spaces of dimension three or greater,
there does not exist any
two-valued probability measures interpretable as classical and consistent, overall truth assignment.\cite[-15mm]{specker-60,kochen1}
Consequently, the classical strategy
to construct probabilities by a convex combination of all two-valued states fails entirely.

{\em Greechie (orthogonality) diagrams},\cite{greechie:71} are hypergraphs whose
points  represent basis vectors.
\index{Greechie diagram}
If they belong to the same basis -- in this context also called {\em context} -- they are connected by  smooth curves.
\index{context}

{\color{OliveGreen}\bproof
A parity proof by contradiction
exploits the particular subset of real four-dimensional Hilbert space with a ``parity property,'' as depicted in Figure~\ref{2018-m-ch-fdlvs-ksc}.
\index{parity property}
It represents the most compact way of deriving the Kochen-Specker theorem in four dimensions.
The configuration consists of 18 biconnected (two contexts intertwine per atom)
atoms $a_1, \ldots , a_{18}$ in 9 contexts.
It has a (quantum) realization in $\mathbb{R}^4$
consisting of the 18 projections associated with the one dimensional subspaces spanned by
the vectors from the origin $(0,0,0,0)^\intercal$ to
$a_1=\left(   0,0,1,-1     \right)^\intercal    $,
$a_2=\left(   1,-1,0,0     \right)^\intercal    $,
$a_3=\left(   1,1,-1,-1    \right)^\intercal   $,
$a_4=\left(   1,1,1,1      \right)^\intercal     $,
$a_5=\left(   1,-1,1,-1    \right)^\intercal  $,
$a_6=\left(   1,0,-1,0     \right)^\intercal   $,
$a_7=\left(   0,1,0,-1   \right)^\intercal   $,
$a_8=\left(   1,0,1,0    \right)^\intercal    $,
$a_9=\left(   1,1,-1,1   \right)^\intercal   $,
$a_{10}=\left(-1,1,1,1   \right)^\intercal    $,
$a_{11}=\left(1,1,1,-1   \right)^\intercal    $,
$a_{12}=\left(1,0,0,1    \right)^\intercal     $,
$a_{13}=\left(0,1,-1,0   \right)^\intercal    $,
$a_{14}=\left(0,1,1,0    \right)^\intercal    $,
$a_{15}=\left(0,0,0,1    \right)^\intercal    $,
$a_{16}=\left(1,0,0,0    \right)^\intercal    $,
$a_{17}=\left(0,1,0,0    \right)^\intercal    $,
$a_{18}=\left(0,0,1,1    \right)^\intercal    $,
 respectively.\cite[-15mm]{cabello:210401}

\begin{figure}
\begin{center}
\begin{tikzpicture}  [scale=0.6]

        \tikzstyle{every path}=[line width=1pt]
        \tikzstyle{c2}=[circle,inner sep=3pt,minimum size=15pt]
        \tikzstyle{c1}=[circle,inner sep=3pt,minimum size=0pt]
        \tikzstyle{l1}=[draw=none,circle,minimum size=35]
        \tikzstyle{l2}=[draw=none,circle,minimum size=12]

        \path
              (240:5) coordinate(1)
              (-0.833,-4.33) coordinate(2)
              (0.833,-4.33) coordinate(3)
              (300:5) coordinate(4)
              (3.33,-2.88) coordinate(5)
              (4.167,-1.44) coordinate(6)
     (0:5) coordinate(7)
              (4.167,1.44) coordinate(8)
              (3.33,2.88) coordinate(9)
              (60:5) coordinate(10)
              (0.833,4.33) coordinate(11)
              (-0.833,4.33) coordinate(12)
              (120:5) coordinate(13)
              (-3.33,2.88) coordinate(14)
              (-4.167,1.44) coordinate(15)
              (180:5) coordinate(16)
              (-4.167,-1.44) coordinate(17)
              (-3.33,-2.88) coordinate(18);

\node[draw=none,color=blue] at (0,-6) {$a$};
\node[draw=none,color=red] at (6,-3) {$b$};
\node[draw=none,color=green] at (6,3) {$c$};
\node[draw=none,color=violet] at (0,6) {$d$};
\node[draw=none,color=gray] at (-6,3) {$e$};
\node[draw=none,color=magenta] at (-6,-3) {$f$};
\node[draw=none,color=cyan] at (-0.8,0.7) {$i$};
\node[draw=none,color=orange] at (0.8,0.7) {$h$};
\node[draw=none,color=lime] at (0,-0.9) {$g$};

        \draw [color=green] (7) -- (8) -- (9)-- (10);
\draw [color=violet] (10) -- (11) -- (12) -- (13);
\draw [color=gray] (13) -- (14) -- (15) -- (16);
\draw [color=magenta] (16) -- (17) -- (18) -- (1);
\draw [color=blue] (1) -- (2) -- (3) -- (4);
\draw [color=red] (4) -- (5) -- (6) -- (7);

        \draw [color=lime] (8) -- (15);
        \draw [color=lime](17) -- (6);
        \draw [color=lime] (8) arc (450:270:2 and 1.44);
        \draw [color=lime] (15) arc (90:270:2 and 1.44);

        \draw [color=cyan] (9) -- (2);
        \draw [color=cyan] (11) -- (18);
        \draw [rotate=240,color=cyan] (9) arc (90:270:2 and 1.44);
        \draw[rotate=60,color=cyan] (18) arc (90:270:2 and 1.44);

        \draw [color=orange] (12) -- (5);
        \draw [color=orange] (14) -- (3);
        \draw[rotate=300,color=orange] (12) arc (90:270:2 and 1.44);
        \draw[rotate=120,color=orange] (3) arc (90:270:2 and 1.44);

        \draw (1) coordinate[c2,fill=blue,label=85:$a_1$];
        \draw (1) coordinate[c1,fill=magenta];
        \draw (2) coordinate[c2,fill=cyan,label=270:$a_2$];
        \draw (2) coordinate[c1,fill=blue];
\draw (3) coordinate[c2,fill=orange,label=270:$a_3$];
        \draw (3) coordinate[c1,fill=blue];
        \draw (4) coordinate[c2,fill=red,label=95:$a_4$];
        \draw (4) coordinate[c1,fill=blue];
\draw (5) coordinate[c2,fill=orange,label=0:$a_5$];
        \draw (5) coordinate[c1,fill=red];
\draw (6) coordinate[c2,fill=lime,label=290:$a_6$];
        \draw (6) coordinate[c1,fill=red];
        \draw (7) coordinate[c2,fill=green,label=180:$a_7$];
        \draw (7) coordinate[c1,fill=red];
\draw (8) coordinate[c2,fill=lime,label=30:$a_8$];
        \draw (8) coordinate[c1,fill=green];
\draw (9) coordinate[c2,fill=cyan,label=0:$a_9$];
        \draw (9) coordinate[c1,fill=green];
        \draw (10) coordinate[c2,fill=violet,label=265:$a_{10}$];
        \draw (10) coordinate[c1,fill=green];
\draw (11) coordinate[c2,fill=cyan,label=91:$a_{11}$];
        \draw (11) coordinate[c1,fill=violet];
\draw (12) coordinate[c2,fill=orange,label=90:$a_{12}$];
        \draw (12) coordinate[c1,fill=violet];
        \draw (13) coordinate[c2,fill=gray,label=285:$a_{13}$];
        \draw (13) coordinate[c1,fill=violet];
\draw (14) coordinate[c2,fill=orange,label=180:$a_{14}$];
        \draw (14) coordinate[c1,fill=gray];
\draw (15) coordinate[c2,fill=lime,label=160:$a_{15}$];
        \draw (15) coordinate[c1,fill=gray];
        \draw (16) coordinate[c2,fill=magenta,label=0:$a_{16}$];
        \draw (16) coordinate[c1,fill=gray];
\draw (17) coordinate[c2,fill=lime,label=215:$a_{17}$];
        \draw (17) coordinate[c1,fill=magenta];
\draw (18) coordinate[c2,fill=cyan,label=180:$a_{18}$];
        \draw (18) coordinate[c1,fill=magenta];

    \end{tikzpicture}
\end{center}
\label{2018-m-ch-fdlvs-ksc}
\caption{Orthogonality diagram (hypergraph) of a configuration of observables without any two-valued state,
used in a parity proof of the Kochen-Specker theorem
presented in~\protect\bibentry{cabello-96}.}
\end{figure}

Note that, on the one hand,
each atom/point/vector/projector belongs
to exactly two -- that is, an {\em even} number of -- contexts; that is, it is biconnected.
Therefore,
any enumeration of  all the contexts occurring in the graph
would contain an {\em even} number of $1$s assigned.
Because due to noncontextuality and biconnectivity,
any atom $a$ with $v(a)=1$ along one context must have the same value 1 along the second context
which is intertwined with the first one -- to the values 1 appear in pairs.

Alas, on the other hand, in such an enumeration
there are nine  -- that is, an {\em odd} number of -- contexts.
Hence,
in order to obey the quantum predictions,
any  two-valued state (interpretable as truth assignment)
would need to have an {\em odd} number of $1$s -- exactly one for each context.
Therefore, there cannot exist any two-valued state on Kochen-Specker type graphs with the  ``parity property.''

More concretely,  note that, within each one of those 9 contexts,
the sum of any state on the atoms of that context must add up to 1.
That is, 
one obtains a system of 9 equations
\begin{equation}
\begin{split}
\color{blue}        v(a)= v( a_1 ) + v( a_2 ) + v( a_3 ) + v( a_4 ) = 1 ,                  \\
\color{red}         v(b)= v( a_4 ) + v( a_5 ) + v( a_6 ) + v( a_7 ) = 1 ,                  \\
\color{green}       v(c)= v( a_7 ) + v( a_8 ) + v( a_9 ) + v( a_{10} ) = 1 ,               \\
\color{violet}      v(d)= v( a_{10} ) + v( a_{11} ) + v( a_{12} ) + v( a_{13} ) = 1 ,      \\
\color{gray}        v(e)= v( a_{13} ) + v( a_{14} ) + v( a_{15} ) + v( a_{16} ) = 1 ,      \\
\color{magenta}     v(f)= v( a_{16} ) + v( a_{17} ) + v( a_{18} ) + v( a_1 ) = 1 ,         \\
\color{lime}        v(g)= v( a_6 ) + v( a_8 ) + v( a_{15} ) + v( a_{17} ) = 1 ,            \\
\color{orange}      v(h)= v( a_3 ) + v( a_5 ) + v( a_{12} ) + v( a_{14} ) = 1 ,            \\
\color{cyan}        v(i)= v( a_2 ) + v( a_9 ) + v( a_{11} ) + v( a_{18} ) = 1 .
\label{2017-b-e-cabp-conf}
\end{split}
\end{equation}
By summing up the left hand side and the right hand sides of the equations, and since all atoms are biconnected,
one obtains
\begin{equation}
2 \left[\sum_{i=1}^{18} v(a_i)\right] = 9.
\label{2017-b-e-cabp}
\end{equation}
Because $v(a_i)\in \{0,1\}$ the sum in (\ref{2017-b-e-cabp}) must add up to some natural number $M$.
Therefore, Equation~(\ref{2017-b-e-cabp}) is impossible to solve in the domain of natural numbers,
as on the left and right-hand sides, there appear even ($2M$) and odd ($9$) numbers, respectively.

Of course, one could also prove the nonexistence of any  two-valued state (interpretable as truth assignment)
by exhaustive attempts
(possibly exploiting symmetries) to assign values $0$s and $1$s to the atoms/points/vectors/projectors occurring in the graph
in such a way that both the quantum predictions as well as context independence are satisfied.
This latter method needs to be applied in cases with Kochen-Specker type diagrams (hypergraphs) without the  ``parity property;''
such as in the original Kochen-Specker proof.\cite[-50mm]{kochen1}

\bproof
}

Note also that in this original paper Kochen and Specker pointed out (in Theorem~0 on page~67)
that a much smaller set of quantum propositions  in intertwining contexts (orthonormal basis) suffices
to prove nonclassicality: all it needs is a configuration with a {\em nonseparating set}
of two-valued states; that is, there exist at least two observables with the same truth assignments for all
such truth assignments -- pointedly stated, the classical truth assignments
are unable to separate between those two observables.

Any such construction is usually based on a succession of  auxiliary  {\em gadget graphs}\cite[-0mm]{tutte_1954,SZABO2009436,Ramanathan-18}
\index{gadget graph}
stitched together to yield the desired property.
Thereby, gadgets are formed from gadgets of ever-increasing size and functional performance
(see also Chapter~12 of Ref.\cite[-0mm]{svozil-2016-pu-book}):
\begin{enumerate}

\item 0th order gadget:  a single context (aka clique/block/Boolean (sub)algebra/maximal observable/orthonormal basis);

\item 1st order ``firefly'' gadget:] two contexts connected in a single intertwining atom;

\item 2nd order gadget:  two 1st order  firefly  gadgets connected in a single intertwining atom;

\item 3rd order house/pentagon/pentagram gadget:  one firefly and one 2nd order gadget connected in two intertwining atoms to form a cyclic orthogonality diagram (hypergraph);

\item 4rth order true-implies-false (TIFS)/01-(maybe better 10)-gadget:  e.g., a Specker bug consisting of two pentagon gadgets connected by an entire context;
as well as extensions thereof to arbitrary angles for terminal (``extreme'') points;

\item 5th order  true-implies-true (TITS)/11-gadget:  e.g.,  Kochen and Specker's $\Gamma_1$, consisting of one 10-gadget and one firefly gadget,
connected at the respective terminal points;

\item 6th order gadget:  e.g.,  Kochen and Specker's $\Gamma_3$,
consisting of a combo of two 11-gadgets, connected  by their common firefly gadgets;

\item 7th order construction:  consisting of one 10- and one 11-gadget, with identical terminal points serving as constructions of Pitowsky's
principle of indeterminacy;~\cite{pitowsky:218,2015-AnalyticKS,svozil-2018-whycontexts}

\item 8th order construction:  concatenation of (10- and) 11-gadgets pasted/stitched together to form a graph used for proofs of the Kochen-Specker theorem;
e.g.,  Kochen and Specker's $\Gamma_2$.
\end{enumerate}


\begin{center}
{\color{olive}   \Huge
 \floweroneleft
}
\end{center}

\chapter{Multilinear algebra and tensors}
\label{ch:t}

In the following chapter multilinear extensions of linear functionals will be discussed.
Tensors will be introduced as multilinear forms,
and their transformation properties will be derived.

For many physicists, the following derivations might appear confusing and overly formalistic
as they might have difficulties to ``see the forest for the trees.''
For those, a brief overview sketching the most important aspects of tensors might serve as a first orientation.

Let us start by defining, or rather {\em declaring} or supposing the following:  basis vectors of some given (base) vector space are said to ``(co-)vary.''
This is just a ``fixation,'' a designation of notation;
important insofar as it implies that the respective coordinates, as well as the dual basis vectors ``contra-vary;''
and the coordinates of dual space vectors ``co-vary.''

Based on this declaration or rather convention
--
that is,
relative to the behavior with respect to variations of scales of the reference axes (the basis vectors) in the base vector space
--
there exist two important categories: entities which co-vary, and entities which vary inversely, that is, contra-vary, with such changes.
\begin{itemize}

\item  Contravariant entities such as vectors in the base vector space:
These vectors of the base vector space are called contravariant because
their {\em components} contra-vary (that is, vary inversely) with respect to variations of the basis vectors.
By identification, the components of contravariant vectors (or tensors) are also contravariant.
In general, a multilinear form on a vector space is called contravariant if its components
(coordinates) are contravariant; that is, they contra-vary with respect to variations of the basis vectors.
\index{contravariant vector}

\item
Covariant entities such as vectors in the dual space:
\marginnote{The dual space is spanned by all linear functionals on that vector space (cf. Section~\ref{2011-m-dvs} on page \pageref{2011-m-dvs}).
}
The vectors of the dual space are called covariant because
their {\em components} contra-vary with respect to variations of the basis vectors of the dual space,
which in turn contra-vary with respect to variations of the basis vectors of the base space.
Thereby the double contra-variations (inversions) cancel out,
so that effectively the vectors of the dual space co-vary with the vectors of the basis of the base vector space.
By identification, the components of covariant vectors (or tensors) are also covariant.
In general, a multilinear form on a vector space is called covariant if its components
(coordinates) are covariant; that is, they co-vary with respect to variations of the basis vectors of the base vector space.
\index{covariant vector}

\item
Covariant and contravariant indices will be denoted by subscripts (lower indices) and superscripts (upper indices), respectively.

\item
Covariant and contravariant entities transform inversely.
Informally, this is due to the fact that their changes must compensate each other,
as covariant and contravariant entities are ``tied together'' by some invariant
(id)entities such as vector encoding and dual basis formation.

\item
Covariant entities can be transformed into contravariant ones
by the application of metric tensors,
and, {\it vice versa,} by the inverse of metric tensors.
\end{itemize}

\section{Notation}

In what follows, vectors and tensors will be encoded in terms of indexed coordinates or components
(with respect to a specific basis).
The biggest advantage is that such coordinates or components are scalars which can be
exchanged and rearranged according to commutativity, associativity, and distributivity,
as well as differentiated.

Let us consider
\marginnote{For a more systematic treatment, see for instance, the introductions~\bibentry{Klingbeil} and \bibentry{Dirschmid}.}
the vector space ${\frak V} = \mathbb{R}^n$ of dimension $n$.
A covariant basis
\marginnote{For a detailed explanation of covariance and contravariance, see  Section~\ref{2016-m-tensor-cob} on page~\pageref{2016-m-tensor-cob}.
}
${\mathfrak B}=\{{\bf e}_1,{\bf e}_2,\ldots ,{\bf e}_n\}$ of ${\frak V}$
consists of
$n$ covariant basis vectors ${\bf e}_i$.
A contravariant basis
${\mathfrak B}^\ast =\{{\bf e}_1^\ast,{\bf e}_2^\ast,\ldots ,{\bf e}_n^\ast\}
= \{{\bf e}^1,{\bf e}^2,\ldots ,{\bf e}^n\}$ of the dual space ${\frak V}^\ast$
(cf. Section~\ref{2011-m-Dualbasis} on page \pageref{2011-m-Dualbasis})
consists of
$n$ basis vectors ${\bf e}_i^\ast$, where ${\bf e}_i^\ast = {\bf e}^i$ is just a different notation.

Every contravariant vector ${\bf x} \in {\frak V}$ can be coded by,  or expressed in terms of, its contravariant vector components
$x^1, x^2, \ldots , x^n \in \mathbb{R}$
by
${\bf x} =\sum_{i=1}^n x^{i} {\bf e}_{i}$.
Likewise, every covariant vector ${\bf x} \in {\frak V}^\ast$ can be coded by, or expressed in terms of, its covariant vector components
$x_1, x_2, \ldots , x_n \in \mathbb{R}$
by
${\bf x} =\sum_{i=1}^n x_{i} {\bf e}_{i}^\ast =\sum_{i=1}^n x_{i} {\bf e}^{i}$.
\marginnote{Note that in both covariant and contravariant cases the upper-lower pairings ``${~\cdot_i}~\cdot^i$'' and  ``${~\cdot^i}~\cdot_i$''of the indices match.
}

Suppose that there are $k$ arbitrary contravariant vectors ${\bf x}_1,{\bf x}_2,\ldots ,{\bf x}_k$ in ${\frak V}$ which are indexed by a subscript
(lower index). This lower index should not be confused with a covariant lower index.
Every such vector ${\bf x}_j$, $1 \le j \le k$ has contravariant vector components
$x^{1_j}_j, x^{2_j}_j, \ldots , x^{n_j}_j \in \mathbb{R}$
with respect to a particular basis ${\mathfrak B}$
such that
\marginnote{This notation ``$x^{i_j}_j$'' for the $i$th component of the $j$th vector is redundant as it requires two indices $j$; we could have just denoted it by
``$x^{i_j}$.''
The lower index $j$ does {\em not} correspond to any covariant entity but just indexes the $j$th vector ${\bf x}_j$.
}
\begin{equation}
{\bf x}_j =\sum_{{i_j}=1}^n x^{i_j}_j{\bf e}_{{i_j}}.
\label{2016-m-tensor-contrvvs}
\end{equation}

Likewise, suppose that there are $k$ arbitrary covariant vectors ${\bf x}^1,{\bf x}^2,\ldots ,{\bf x}^k$ in the dual space ${\frak V}^\ast$ which are indexed by a superscript
(upper index). This upper index should not be confused with a contravariant upper index.
Every such vector ${\bf x}^j$, $1 \le j \le k$ has covariant vector components
$x_{1_j}^j, x_{2_j}^j, \ldots , x_{n_j}^j \in \mathbb{R}$
with respect to a particular basis ${\mathfrak B}^\ast$
such that
\marginnote{Again, this notation ``$x_{i_j}^j$'' for the $i$th component of the $j$th vector is redundant as it requires two indices $j$; we could have just denoted it by
``$x_{i_j}$.''
The upper index $j$ does {\em not} correspond to any contravariant entity but just indexes the $j$th vector ${\bf x}^j$.
}
\begin{equation}
{\bf x}^j =\sum_{{i_j}=1}^n x_{i_j}^j{\bf e}^{{i_j}}.
\label{2016-m-tensor-covvs}
\end{equation}

Tensors are constant with respect to variations of points of $\mathbb{R}^n$.
In contradistinction, {\em tensor fields} depend on points of $\mathbb{R}^n$ in a nontrivial (nonconstant) way.
Thus, the components of a tensor field
depend on the coordinates.
{
\color{blue}
\bexample
For example, the contravariant vector defined by the coordinates
$
\begin{pmatrix}
5.5,
3.7,
\ldots ,
10.9
\end{pmatrix}^\intercal
$
with respect to a particular basis ${\mathfrak B}$
is a tensor; while, again with respect to a particular basis ${\mathfrak B}$,
$
\begin{pmatrix}
\sin x_1,
\cos x_2,
\ldots ,
e^{x_n}
\end{pmatrix}^\intercal
$ or
$
\begin{pmatrix}
x_1,
x_2,
\ldots ,
x_n
\end{pmatrix}^\intercal
$,
which depend on the coordinates $x_1, x_2, \ldots , x_n \in \mathbb{R}$,
are tensor fields.
\eexample
}

We adopt Einstein's summation convention to sum over equal indices.
If not explained otherwise (that is, for orthonormal bases) those pairs have exactly one lower and one upper index.

In what follows, the notations
``$x\cdot y$'',
``$(x,y)$'' and
``$\langle x\mid y\rangle $'' will be used synonymously for the {\em
scalar product}
or
{\em inner product}.
\index{scalar product}
\index{inner product}
Note, however, that the ``dot notation $x\cdot y$''
may be a little bit misleading; for example, in the case of the ``pseudo-Euclidean'' metric
represented by the matrix
 ${\rm diag}(+,+,+,\cdots ,+,-)$, it is no more the standard Euclidean dot product
${\rm diag}(+,+,+,\cdots ,+,+)$.


\section{Change of basis}
\label{2016-m-tensor-cob}

\subsection{Transformation of the covariant basis}

Let
${\mathfrak B}$
and
${\mathfrak B'}$
be two arbitrary bases of
$\mathbb{R}^n$.
Then every vector ${\bf f}_i$ of
${\mathfrak B'}$
can be represented as linear combination of basis vectors of
${\mathfrak B}$ [see also Eqs.~(\ref{2011-m-btbe}) and (\ref{2011-m-btbe-r})]:
\begin{equation}
{\bf f}_i=\sum_{j=1}^n {a^j}_i {\bf e}_j, \qquad i=1,\ldots , n
.
\label{2001-mu-tensors}
\end{equation}

The matrix
\begin{equation}
\textsf{\textbf{A}}
\equiv {a^j}_i \equiv
\begin{pmatrix}
{a^1}_1&{a^1}_2& \cdots & {a^1}_n\\
{a^2}_1&{a^2}_2& \cdots & {a^2}_n\\
\vdots &\vdots & \ddots & \vdots \\
{a^n}_1&{a^n}_2& \cdots & {a^n}_n
\end{pmatrix}
.
\label{2016-m-ch-tensor-tm}
\end{equation}
is called the {\em transformation matrix}.
As defined in (\ref{2016-m-ch-fdvs-matrixind}) on page \pageref{2016-m-ch-fdvs-matrixind},
the second (from the left to the right), rightmost (in this case lower) index $i$ varying in row vectors is the column index;
and, the first, leftmost (in this case upper) index $j$  varying in columns is the row index, respectively.


Note that, as discussed earlier, it is necessary to fix a convention for the transformation of the covariant basis vectors discussed on page~\pageref{2016-m-ch-fdvs-oic}.
This then specifies the exact form of the (inverse, contravariant) transformation of the components or coordinates of vectors.

Perhaps not very surprisingly, compared to the transformation (\ref{2001-mu-tensors}) yielding the ``new'' basis
${\mathfrak B'}$  in terms of elements of the ``old'' basis ${\mathfrak B}$,
a transformation  yielding the ``old'' basis
${\mathfrak B}$  in terms of elements of the ``new'' basis ${\mathfrak B'}$ turns out to be just the inverse ``back'' transformation of the former:
substitution of (\ref{2001-mu-tensors}) yields
\begin{equation}
{\bf e}_i=\sum_{j=1}^n {{a'}^j}_i {\bf f}_j
=\sum_{j=1}^n {{a'}^j}_i  \sum_{k=1}^n {a^k}_j {\bf e}_k
=\sum_{k=1}^n \left(\sum_{j=1}^n  {{a'}^j}_i  {a^k}_j \right) {\bf e}_k,
\end{equation}
which, due to the linear independence of the basis vectors ${\bf e}_i$ of ${\mathfrak B}$,
can only be satisfied if
\begin{equation}
{a^k}_j  {{a'}^j}_i =\delta_i^k
\qquad
{\rm or}
\qquad
\textsf{\textbf{A}}\textsf{\textbf{A}}'=\mathbb{1}.
\end{equation}

Thus $\textsf{\textbf{A}}'$ is the {\em inverse matrix}
$\textsf{\textbf{A}}^{-1}$  of $\textsf{\textbf{A}}$. In index notation,
\begin{equation}
{{a'}^j}_i ={{(a^{-1})}^j}_i
,
\label{2001-mu-tensor-tl2}
\end{equation}
and
\begin{equation}
{\bf e}_i=\sum_{j=1}^n {{(a^{-1})}^j}_i  {\bf f}_j
.
\label{2012-m-ch-tlcbv}
\end{equation}

\subsection{Transformation of the contravariant coordinates}

Consider an arbitrary contravariant vector ${\bf x} \in \mathbb{R}^n$ in two basis representations:
(i)
with contravariant components $x^i$ with respect to the basis
${\mathfrak B}$,
and  (ii) with ${ y }^i$  with respect to the basis
${\mathfrak B'}$.
Then, because both coordinates with respect to the two different bases
have to encode the same vector, there has to be a ``compensation-of-scaling'' such that
\begin{equation}
{\bf x}
=\sum_{i=1}^n x^i {\bf e}_i
=\sum_{i=1}^n { y }^i {\bf f}_i
.
\end{equation}
Insertion of the basis transformation (\ref{2001-mu-tensors}) and
relabelling of the indices $i \leftrightarrow j$ yields
\begin{equation}
\begin{split}
{\bf x}
=\sum_{i=1}^n x^i {\bf e}_i
=\sum_{i=1}^n { y }^i {\bf f}_i
=\sum_{i=1}^n { y }^i \sum_{j=1}^n {a^j}_i {\bf e}_j\\
=
\sum_{i=1}^n\sum_{j=1}^n {a^j}_i { y }^i  {\bf e}_j
=
\sum_{j=1}^n\left[ \sum_{i=1}^n {a^j}_i  { y }^i \right] {\bf e}_j
=
\sum_{i=1}^n\left[ \sum_{j=1}^n {a^i}_j  { y }^j \right] {\bf e}_i
.
\end{split}
\label{2016-m-ch-tensor-tcvc}
\end{equation}
A comparison of coefficients
yields the transformation laws of vector components
[see also Equation~(\ref{2012-m-ch-e-tl1})]
\begin{equation}
x^i   = \sum_{j=1}^n {a^i}_j  { y }^j .
\label{2012-m-ch-di-choic}
\end{equation}

In the matrix notation introduced in Equation~(\ref{2016-m-fdvs-rv}) on page~\pageref{2016-m-fdvs-rv},
(\ref{2012-m-ch-di-choic}) can  be written as
\begin{equation}
X   =  \textsf{\textbf{A}}  Y .
\label{2016-m-ch-di-choic-mn}
\end{equation}

A similar ``compensation-of-scaling'' argument using
(\ref{2012-m-ch-tlcbv})
yields the transformation laws for
\begin{equation}
{ y }^j   = \sum_{i=1}^n {(a^{-1})^j}_i {x}^i
\label{2015-m-ch-tensor-tlcc}
\end{equation}
with respect to the covariant basis vectors.
In the matrix notation introduced in Equation~(\ref{2016-m-fdvs-rv}) on page~\pageref{2016-m-fdvs-rv},
(\ref{2015-m-ch-tensor-tlcc}) can simply be written as
\begin{equation}
 Y    =  \left(\textsf{\textbf{A}}^{-1}\right) X.
\label{2016-m-ch-di-tlcc-mn}
\end{equation}

If the basis transformations involve nonlinear coordinate changes -- such as from the
Cartesian to the polar or spherical coordinates discussed later -- we have to employ differentials
\begin{equation}
dx^j   = \sum_{i=1}^n {a^j}_i \,d{ y }^i  ,
\label{2012-m-ch-di-choic11}
\end{equation}
so that, by partial differentiation,
\begin{equation}
{a^j}_i ={\partial x^j \over \partial  y ^i}   .
\label{2001-mu-tensor-tl11}
\end{equation}

By assuming that the coordinate transformations are linear, ${a_i}^j$ can be expressed in terms of the coordinates $x^j$
\begin{equation}
{a^j}_i =\frac{  x^j }{   y ^i}  .
\label{2001-mu-tensor-tl1}
\end{equation}

Likewise,
\begin{equation}
d{ y }^j   =
\sum_{i=1}^n {{({a^{-1}})}^j}_i\, d{x}^i,
\end{equation}
so that, by partial differentiation,
\begin{equation}
{{(a^{-1})}^j}_i =
{\partial { y }^j \over \partial x^i}   =J_{ji},
\label{2001-mu-tensor-tl2nl}
\end{equation}
where $J_{ji} = {\partial { y }^j\over \partial {x}^i}$ stands for
the $j$th row and $i$th column component of the {\em Jacobian matrix}
\index{Jacobian matrix}
\begin{equation}
J(x^1,x^2,\ldots ,x^n)
\stackrel{{\text{\tiny def}}}{=}
\begin{pmatrix}
\frac{\partial }{\partial { x }^1}&\cdots&\frac{\partial }{\partial {x}^n}
\end{pmatrix}
\times
\begin{pmatrix}
y^1\\
\vdots\\
y^n
\end{pmatrix}
\equiv
\begin{pmatrix}
{\partial { y }^1\over \partial {x}^1}&\cdots&{\partial { y }^1\over \partial {x}^n}\\
\vdots&\ddots &\vdots\\
{\partial { y }^n\over \partial {x}^1}&\cdots&{\partial { y }^n\over \partial {x}^n}
\end{pmatrix}
.
\label{2013-m-t-jm}
\end{equation}
Potential confusingly,
its determinant
\begin{equation}
J \stackrel{{\text{\tiny def}}}{=}
\frac{
\partial \begin{pmatrix}y^1, \ldots , y^n \end{pmatrix}
}
{
\partial \begin{pmatrix} x^1, \ldots , x^n \end{pmatrix}
}
=
\text{det}
\begin{pmatrix}
{\partial { y }^1\over \partial {x}^1}&\cdots&{\partial { y }^1\over \partial {x}^n}\\
\vdots&\ddots &\vdots\\
{\partial { y }^n\over \partial {x}^1}&\cdots&{\partial { y }^n\over \partial {x}^n}
\end{pmatrix}
\label{2018-mm-t-jd}
\end{equation}
is also often referred to as ``the Jacobian.''
\index{Jacobian determinant}
\index{Jacobian}

\subsection{Transformation of the contravariant (dual) basis}
\index{dual basis}
\index{reciprocal basis}
\index{contravariant basis}

Consider again, as a starting point, a covariant basis
${\mathfrak B}=\{{\bf e}_1,{\bf e}_2,\ldots ,{\bf e}_n\}$ consisting of
$n$ basis vectors ${\bf e}_i$.
A {\em contravariant basis} can be defined by identifying it with the dual basis
introduced earlier in Section~\ref{2011-m-Dualbasis} on page~\pageref{2011-m-Dualbasis},
in particular, Equation~(\ref{2011-m-Dualbasis-e1}).
Thus a {\em contravariant} basis
${\mathfrak B^\ast}=\{{\bf e}^1,{\bf e}^2,\ldots ,{\bf e}^n\}$ is a set of $n$ contravariant
basis vectors ${\bf e}^i$
which satisfy Eqs.~(\ref{2011-m-Dualbasis-e1})-(\ref{2011-m-Dualbasis-e3})
\begin{equation}
{\bf e}^j \left( {\bf e}_i \right) = \left\llbracket {\bf e}_i,{\bf e}^j \right\rrbracket  =  \left\llbracket {\bf e}_i,  {\bf e}_j^\ast \right\rrbracket
= \delta^j_i = \delta_{ij}
.
\label{2001-mu-tensors0}
\end{equation}

In terms of the bra-ket notation, (\ref{2001-mu-tensors0}) somewhat superficially transforms into
(a formal justification for this identification is the Riesz representation theorem)
\index{Riesz representation theorem}
\index{Fr\'echet-Riesz representation theorem}
\begin{equation}
\left\llbracket  \vert {\bf e}_i \rangle , \langle  {\bf e}^j \vert \right\rrbracket =
\langle  {\bf e}^j \vert {\bf e}_i \rangle =
  \delta_{ij}
.
\label{2001-mu-tensors0bk}
\end{equation}
Furthermore, the resolution of identity~(\ref{2016-m-ch-fdlws-roi}) can be rewritten as
\begin{equation}
 \mathbb{1}_n = \sum_{i=1}^n \vert {\bf e}^i \rangle \langle {\bf e}_i \vert
.
\label{2016-m-ch-tensor-roi}
\end{equation}

As demonstrated earlier in Equation~(\ref{2016-m-fdlvs-recoverc}) the vectors ${\bf e}^\ast_i = {\bf e}^i$ of the dual basis can be used to ``retrieve'' the components of arbitrary vectors
${\bf x} = \sum_j x^j {\bf e}_j$  through
\begin{equation}
{\bf e}^i ( {\bf x} ) =
{\bf e}^i \left( \sum_i x^j {\bf e}_j \right) =
\sum_i  x^j {\bf e}^i \left({\bf e}_j \right) =
\sum_i  x^j \delta^i_j =
 x^i.
\label{2016-m-tensor-recoverc}
\end{equation}
Likewise, the basis vectors ${\bf e}_i$ of the ``base space'' can be used to obtain the coordinates of any dual vector ${\bf x} = \sum_j x_j {\bf e}^j$  through
\begin{equation}
{\bf e}_i ( {\bf x} ) =
{\bf e}_i \left( \sum_i x_j {\bf e}^j \right) =
\sum_i  x_j {\bf e}_i \left({\bf e}^j \right) =
\sum_i  x_j \delta_i^j =
 x_i.
\label{2016-m-tensor-recoverc2}
\end{equation}

As also noted earlier,  for orthonormal bases and Euclidean scalar (dot) products (the coordinates of) the dual basis vectors of an orthonormal basis can be coded identically
as  (the coordinates of) the original basis vectors; that is,
in this case,
(the coordinates of) the dual basis vectors are just rearranged as the transposed form of the original basis vectors.

In the same way as argued for changes of covariant bases (\ref{2001-mu-tensors}),
that is, because every vector in the new basis of the dual space can be represented as a linear combination of the vectors of the original dual basis -- we can make the formal {\it Ansatz}:
\begin{equation}
{\bf f}^j=\sum_i{b^j}_i{\bf e}^i,
\label{2016-m-ch-tensor-tocontravb}
\end{equation}
where $ \textsf{\textbf{B}} \equiv {b^j}_i$ is
the transformation matrix associated with the contravariant basis.
How is $b$, the transformation of the contravariant basis, related to $a$,
the transformation of the covariant basis?

Before answering this question, note that, again -- and just as the necessity to fix a convention for the transformation of the covariant basis vectors discussed on page~\pageref{2016-m-ch-fdvs-oic} --
we have to choose by {\em convention} the way transformations are represented.
In particular,
if in (\ref{2016-m-ch-tensor-tocontravb}) we would have reversed the indices ${b^j}_i \leftrightarrow {b_i}^j$, thereby effectively transposing the transformation matrix  $ \textsf{\textbf{B}} $,
this would have resulted in a changed (transposed) form of the transformation laws,
as compared to both the transformation $a$ of the covariant basis, and of the transformation of covariant vector components.

By exploiting (\ref{2001-mu-tensors0}) twice we can find the connection between
the transformation of covariant and contravariant basis elements and thus
tensor components; that is (by assuming Einstein's summation convention we are omitting to write sums explicitly),
\begin{equation}
\begin{split}
\delta_i^j= \delta_{ij}=
{\bf f}^j({\bf f}_i) =
\left\llbracket {\bf f}_i , {\bf f}^j \right\rrbracket =
\left\llbracket  {a^k}_i {\bf e}_k ,  {b^j}_l {\bf e}^l \right\rrbracket  =
\\
=
{a^k}_i {b^j}_l  \left\llbracket  {\bf e}_k , {\bf e}^l \right\rrbracket
={a^k}_i {b^j}_l   \delta_k^l
={a^k}_i {b^j}_l   \delta_{kl}
={b^j}_k {a^k}_i  .
\end{split}
\end{equation}
Therefore,
\begin{equation}
\textsf{\textbf{B}}
= \textsf{\textbf{A}}^{-1}
\textrm{, or }
{b^j}_i =  {\left( a^{-1} \right)^j}_i,
\label{2012-m-ch-tensor-tocontrav}
\end{equation}
and
\begin{equation}
{\bf f}^j=
\sum_i{\left( a^{-1}\right)^j}_i{\bf e}^i
.
\label{2012-m-ch-tensor-tocontrav1}
\end{equation}
In short, by comparing (\ref{2012-m-ch-tensor-tocontrav1})  with (\ref{2015-m-ch-tensor-tlcc}), we find that the vectors of the contravariant dual basis transform just like the components of contravariant vectors.

\subsection{Transformation of the covariant coordinates}

For the same, compensatory, reasons yielding the ``contra-varying'' transformation of the contravariant coordinates
with respect to variations of the covariant bases [reflected in Eqs.~(\ref{2001-mu-tensors}), (\ref{2015-m-ch-tensor-tlcc}), and (\ref{2001-mu-tensor-tl2nl})]
the coordinates with respect to the dual, contravariant, basis vectors, transform {\em covariantly.}
\index{contravariance}
We may therefore say that
``basis vectors ${\bf e}_i$, as well as dual components (coordinates) $x_i$ vary covariantly.''
Likewise,
``vector components (coordinates) $x^i$, as well as dual basis vectors ${\bf e}^\ast_i= {\bf e}^i$ vary contra-variantly.''

A similar calculation as for the contravariant components (\ref{2016-m-ch-tensor-tcvc}) yields a transformation for the covariant components:
\begin{equation}
\begin{split}
{\bf x}
=\sum_{i=1}^n x_j {\bf e}^j
=\sum_{i=1}^n { y }_i {\bf f}^i
=\sum_{i=1}^n { y }_i \sum_{j=1}^n {b^i}_j {\bf e}^j
=
\sum_{j=1}^n \left( \sum_{i=1}^n {b^i}_j  { y }_i \right) {\bf e}^j
.
\end{split}
\label{2016-m-ch-tensor-tcontravc}
\end{equation}
Thus, by comparison we obtain
\begin{equation}
\begin{split}
 x_i = \sum_{j=1}^n {b^j}_i  { y }_j =  \sum_{j=1}^n  {\left( a^{-1} \right)^j}_i  { y }_j    \textrm{, and }
\\
 y_i = \sum_{j=1}^n {\left( b^{-1}\right)^j}_i  { x }_j
    = \sum_{j=1}^n {a^j}_i   { x }_j.
\end{split}
\label{2016-m-ch-tensor-tcontravcc}
\end{equation}
In short, by comparing (\ref{2016-m-ch-tensor-tcontravcc})  with (\ref{2001-mu-tensors}), we find that the components of covariant vectors transform just like the vectors of the covariant basis vectors of ``base space.''

\subsection{Orthonormal bases}
For orthonormal bases of $n$-dimensional Hilbert space,
\begin{equation}
\delta_i^j = {\bf e}_i\cdot {\bf e}^j
\textrm { if and only if }
{\bf e}_i= {\bf e}^i  \textrm { for all } 1\le i,j \le n.
\end{equation}
Therefore, the vector space and its dual vector space are ``identical''
in the sense that the coordinate tuples representing their bases are identical
(though relatively transposed).
That is, besides transposition, the two bases are identical
\begin{equation}
{\mathfrak B}\equiv {\mathfrak B}^\ast
\end{equation}
and  formally any distinction between covariant and contravariant vectors becomes
irrelevant. Conceptually, such a distinction persists, though.
In this sense, we might ``forget about the difference between
covariant and contravariant orders.''


\section{Tensor as multilinear form}

A {\em multilinear form}
$\alpha :{ \frak V }^k \mapsto \mathbb{R}$  or  $\mathbb{C}$
is a map from (multiple) arguments ${\bf x}_i$ which are elements of some vector space  $\frak V$
into some scalars in $\mathbb{R}$ or $\mathbb{C}$,  satisfying
\begin{equation}
\begin{split}
\alpha ( {\bf x}_1,{\bf x}_2,\ldots , A {\bf y}+ B {\bf z}, \ldots ,{\bf x}_k)
=
A\alpha ( {\bf x}_1,{\bf x}_2,\ldots , {\bf y}, \ldots ,{\bf x}_k)   \\
   \qquad +
B\alpha ( {\bf x}_1,{\bf x}_2,\ldots , {\bf z}, \ldots ,{\bf x}_k)
\end{split}
\end{equation}
for every one of its (multi-)arguments.

{
\color{blue}
Note that linear functionals on ${\frak V}$, which constitute the elements of the dual space ${\frak V}^\ast$
(cf. Section~\ref{2011-m-dvs} on page \pageref{2011-m-dvs}) is just a particular example of a  multilinear form
-- indeed rather a  linear form -- with just one argument, a vector in ${\frak V}$.
\eexample
}

In what follows we shall concentrate on {\em real-valued} multilinear forms which map
$k$ vectors in
$\mathbb{R}^n$
into
$\mathbb{R}$.

\section{Covariant tensors}
\index{covariance}

Mind the notation introduced earlier; in particular in Eqs.~(\ref{2016-m-tensor-contrvvs}) and (\ref{2016-m-tensor-covvs}).
A covariant tensor of rank $k$
\index{rank of tensor}
\index{tensor rank}
\index{tensor type}
\begin{equation}
\alpha:{ \frak V }^k \mapsto \mathbb{R}
\end{equation}
is a multilinear form
\begin{equation}
\alpha ( {\bf x}_1,{\bf x}_2,\ldots ,{\bf x}_k)=
\sum_{i_1=1}^n
\sum_{i_2=1}^n
\cdots
\sum_{i_k=1}^n
x^{i_1}_1 x^{i_2}_2\ldots x^{i_k}_k
\alpha ( {\bf e}_{i_1},{\bf e}_{i_2},\ldots ,{\bf e}_{i_k}).
\end{equation}
The
\begin{equation}
A_{{i_1}{i_2}\cdots {i_k}}
\stackrel{{\tiny \textrm{ def }}}{=}
\alpha ( {\bf e}_{i_1},{\bf e}_{i_2},\ldots ,{\bf e}_{i_k})
\end{equation}
 are the
{\em covariant components} or
{\em covariant coordinates}
of the tensor $\alpha $ with respect to the basis
${\mathfrak B}$.

Note that, as each of the $k$ arguments of  a tensor of type (or rank) $k$ has to be evaluated
at each of the $n$ basis vectors ${\bf e}_1,{\bf e}_2,\ldots ,{\bf e}_n$ in an $n$-dimensional vector space,
$A_{{i_1}{i_2}\cdots {i_k}} $ has $n^k$  coordinates.

{\color{OliveGreen}
\bproof
To prove that tensors are multilinear forms, insert
\begin{equation}
\begin{split}
 \alpha ( {\bf x}_1,{\bf x}_2,\ldots , A{\bf x}^1_j+B{\bf x}_j^2,\ldots ,{\bf x}_k)
\\
=
\sum_{i_1=1}^n
\sum_{i_2=1}^n
\cdots
\sum_{i_k=1}^n
x^{i_i}_1x^{i_2}_2\ldots  [A(x^1)^{i_j}_j+B(x^2)^{i_j}_j]\ldots x^{i_k}_k
\alpha ( {\bf e}_{i_1},{\bf e}_{i_2},\ldots,{\bf e}_{i_j},\ldots ,{\bf e}_{i_k})
\\
= A
\sum_{i_1=1}^n
\sum_{i_2=1}^n
\cdots
\sum_{i_k=1}^n
x^{i_i}_1x^{i_2}_2\ldots  (x^1)^{i_j}_j\ldots x^{i_k}_k
\alpha ( {\bf e}_{i_1},{\bf e}_{i_2},\ldots,{\bf e}_{i_j},\ldots ,{\bf e}_{i_k})
\\
\quad +
B
\sum_{i_1=1}^n
\sum_{i_2=1}^n
\cdots
\sum_{i_k=1}^n
x^{i_i}_1x^{i_2}_2\ldots  (x^2)^{i_j}_j\ldots x^{i_k}_k
\alpha ( {\bf e}_{i_1},{\bf e}_{i_2},\ldots,{\bf e}_{i_j},\ldots ,{\bf e}_{i_k})
\\
=
A \alpha ( {\bf x}_1,{\bf x}_2,\ldots , {\bf x}^1_j,\ldots ,{\bf x}_k)+
B \alpha ( {\bf x}_1,{\bf x}_2,\ldots , {\bf x}_j^2,\ldots ,{\bf x}_k)\nonumber
\end{split}
\end{equation}
\eproof
}

\subsection{Transformation of covariant tensor components}

Because of multilinearity  and by insertion into
(\ref{2001-mu-tensors}),
\begin{eqnarray}
&&\alpha ( {\bf f}_{j_1},{\bf f}_{j_2},\ldots ,{\bf f}_{j_k})=
\alpha \left(
\sum_{i_1=1}^n {a^{i_1}}_{j_1} {\bf e}_{i_1},
\sum_{i_2=1}^n {a^{i_2}}_{j_2} {\bf e}_{i_2},
\ldots ,
\sum_{i_k=1}^n {a^{i_k}}_{j_k} {\bf e}_{i_k}
\right)
\nonumber \\&& \quad
=
\sum_{i_1=1}^n\sum_{i_2=1}^n\cdots \sum_{i_k=1}^n
{a^{i_1}}_{j_1}{a^{i_2}}_{j_2}\cdots {a^{i_k}}_{j_k} \alpha ( {\bf e}_{i_1},{\bf e}_{i_2},\ldots ,{\bf e}_{i_k})
\label{2012-m-ch-tensor-etotc1}
\end{eqnarray}
or
\begin{equation}
A'_{{j_1}{j_2}\cdots {j_k}}=
\sum_{i_1=1}^n\sum_{i_2=1}^n\cdots \sum_{i_k=1}^n
{a^{i_1}}_{j_1}{a^{i_2}}_{j_2}\cdots {a^{i_k}}_{j_k} A_{i_1 i_2\ldots i_k}.
\label{2011-m-tvtcov}
\end{equation}

In effect, this yields a transformation factor ``${a^i}_j$'' for every ``old index $i$'' and ``new index $j$.''

\section{Contravariant tensors}
\index{contravariance}

Recall the inverse scaling of contravariant vector coordinates with respect to covariantly varying basis vectors.
Recall further that the dual base vectors are defined in terms of the base vectors by
a kind of ``inversion'' of the latter, as expressed by $[{\bf e}_i,  {\bf e}_j^*]=\delta_{ij}$
in Equation~(\ref{2011-m-Dualbasis-e1}).
Thus, by analogy, it can be expected that similar considerations apply to
the scaling of dual base vectors with respect to the scaling of covariant base vectors:
in order to compensate those scale changes, dual basis vectors should contra-vary,
and, again analogously,  their respective dual coordinates, as well as the dual vectors,
should vary covariantly.
Thus, both vectors in the dual space, as well as their components or coordinates, will be called covariant vectors,
as well as covariant coordinates, respectively.
\index{covariance}
\index{covariant vector}
\index{contravariance}

\subsection{Definition of contravariant tensors}

The entire tensor formalism developed so far can be transferred and applied to define {\em contravariant} tensors
as multilinear forms with contravariant components
\begin{equation}
\beta:{ \frak V^\ast }^k \mapsto \mathbb{R}
\end{equation}
by
\begin{equation}
\beta ( {\bf x}^1,{\bf x}^2,\ldots ,{\bf x}^k)=
\sum_{i_1=1}^n
\sum_{i_2=1}^n
\cdots
\sum_{i_k=1}^n
x_{i_1}^1 x_{i_2}^2 \ldots x_{i_k}^k
\beta ( {\bf e}^{i_1},{\bf e}^{i_2},\ldots ,{\bf e}^{i_k}).
\end{equation}
By definition
\begin{equation}
B^{{i_1}{i_2}\cdots {i_k}}=\beta ( {\bf e}^{i_1},{\bf e}^{i_2},\ldots ,{\bf e}^{i_k})
\label{2011-m-tvtcontrav}
\end{equation}
 are the contravariant
{\em components} of the contravariant tensor $\beta $ with respect to the basis
${\mathfrak B}^\ast$.

\subsection{Transformation of contravariant tensor components}

The argument concerning transformations of covariant tensors and components
can be carried through to the contravariant case.
Hence, the contravariant components transform as
\begin{eqnarray}
&&\beta ( {{\bf f}}^{j_1},{{\bf f}}^{j_2},\ldots ,{{\bf f}}^{j_k})=
\beta \left(
\sum_{i_1=1}^n {b^{j_1}}_{i_1} {\bf e}^{i_1},
\sum_{i_2=1}^n {b^{j_2}}_{i_2} {\bf e}^{i_2},
\ldots ,
\sum_{i_k=1}^n {b^{j_k}}_{i_k} {\bf e}^{i_k}
\right)
\nonumber \\&& \quad
=
\sum_{i_1=1}^n\sum_{i_2=1}^n\cdots \sum_{i_k=1}^n
{b^{j_1}}_{i_1}{b^{j_2}}_{i_2}\cdots {b^{j_k}}_{i_k} \beta ( {\bf e}^{i_1},{\bf e}^{i_2},\ldots ,{\bf e}^{i_k})
 \label{2012-m-ch-tensor-etotccon1}
\end{eqnarray}
or
\begin{equation}
B'^{{j_1}{j_2}\cdots {j_k}}=
\sum_{i_1=1}^n\sum_{i_2=1}^n\cdots \sum_{i_k=1}^n
{b^{j_1}}_{i_1}{b^{j_2}}_{i_2}\cdots {b^{j_k}}_{i_k} B^{i_1 i_2\ldots i_k}.
 \label{2012-m-ch-tensor-etotccon2}
\end{equation}

Note that, by Equation~(\ref{2012-m-ch-tensor-tocontrav}),
$ {b^j}_i =  {\left( a^{-1} \right)^j}_i$.
In effect, this yields a transformation factor ``${\left( a^{-1} \right)^j}_i$'' for every ``old index $i$'' and ``new index $j$.''

\section{General tensor}

A (general) Tensor $T$ can be defined as a multilinear form  on the
$r$-fold product of a vector space ${\frak V}$, times the
$s$-fold product of the dual vector space ${\frak V}^\ast$.
If all $r$ components appear on the left and all $s$ components right side
-- in general covariant and contravariant components can have mixed orders --
one can denote this by
\begin{equation}
T: \left( {\frak V} \right)^r \times \left( {\frak V}^\ast \right)^s
=
\underbrace{{\frak V}\times \cdots \times {\frak V}}_{r\textrm{ \scriptsize copies}}
\times
\underbrace{{\frak V}^\ast \times \cdots \times {\frak V}^\ast}_{s\textrm{ \scriptsize copies}}
\mapsto \mathbb{F}
.
 \label{2012-m-ch-tensor-gdt}
\end{equation}
Most commonly, the scalar field
$\mathbb{F}$
will be identified with the set $\mathbb{R}$ of reals,
or with the set $\mathbb{C}$ of complex numbers.
Thereby,
$r$ is called the
{\em covariant order}, and
\index{covariance}
$s$ is called the
{\em contravariant order}
\index{contravariance}
of $T$.
A tensor of covariant order $r$ and contravariant order $s$
is then pronounced a tensor of
{\em type} (or {\em rank})
\index{tensor rank}
\index{tensor type}
$(r,s)$.
By convention, covariant indices are denoted by {\em subscripts},
whereas the contravariant indices are denoted by {\em superscripts}.

With the standard, ``inherited'' addition and scalar multiplication,
the set ${\frak T}_r^s$ of all tensors of type $(r,s)$
forms a linear vector space.

Note that a tensor of type $(1,0)$ is called  a
{\em covariant vector}
\index{covariance},
or just a
{\em vector}.
\index{vector}
A tensor of type $(0,1)$ is called a
{\em contravariant vector}.
\index{contravariance}

Tensors can change their type by the invocation of the {\em metric tensor}.
That is, a covariant tensor (index) $i$ can be made into a contravariant tensor (index) $j$
by summing over the index $i$ in a product involving the tensor and $g^{ij}$.
Likewise,  a contravariant tensor (index) $i$ can be made into a covariant tensor (index) $j$
by summing over the index $i$ in a product involving the tensor and $g_{ij}$.

Under basis or other linear transformations,
covariant tensors with index $i$ transform by summing over this index with (the transformation matrix) ${a_i}^j$.
Contravariant tensors with index $i$ transform by summing over this index with the inverse (transformation matrix)  ${(a^{-1})_i}^j$.

\section{Metric}

A {\em metric} or {\em metric tensor} $g$ is a measure of {\em distance} between two points in a vector space.

\subsection{Definition}
\index{metric tensor}
\index{metric}
\label{2011-m-metrict}

Formally, a metric, or metric tensor,  can be defined as a functional $g: \mathbb{R}^n\times\mathbb{R}^n\mapsto \mathbb{R}$
which maps two vectors (directing from the origin to the two points)
into a scalar
with the following properties:
\begin{itemize}
\item
$g$ is symmetric; that is, $g({\bf x},{\bf y})=g({\bf y},{\bf x})$;
\item
$g$ is bilinear; that is,
$g(
\alpha {\bf x} + \beta {\bf y}, {\bf z})
= \alpha g( {\bf x},{\bf z}) + \beta g({\bf y}, {\bf z})
$ (due to symmetry $g$ is also bilinear in the second argument);
\item
$g$ is nondegenerate; that is,
for every ${\bf x}\in {\frak V}$, ${\bf x}\neq 0$, there exists a
${\bf y}\in {\frak V}$ such that $g({\bf x},{\bf y})\neq 0$.
\end{itemize}

\subsection{Construction from a scalar product}

In real Hilbert spaces the {\em metric} tensor can be defined {\it via}  the scalar product  by
\begin{equation}
g_{ij}= \langle {\bf e}_i \mid {\bf e}_j\rangle .
\label{2016-m-ch-tensor-gij}
\end{equation}
and
\begin{equation}
g^{ij}= \langle {\bf e}^i \mid {\bf e}^j\rangle .
\label{2016-m-ch-tensor-gij2}
\end{equation}

For orthonormal bases, the metric tensor can be
represented as a Kronecker delta function, and thus remains form invariant.
Moreover, its covariant and contravariant components are identical; that is,
$g_{ij}=\delta_{ij}=\delta^i_j=\delta_i^j=\delta^{ij}=g^{ij}$.

\subsection{What can the metric tensor do for you?}

We shall see that with the help of the metric tensor we can ``raise and lower indices;''
that is, we can transform lower (covariant) indices into upper (contravariant) indices, and {\it vice versa}.
This can be seen as follows.
Because of linearity, any contravariant basis vector ${\bf e}^i$
can be written as a linear sum of covariant (transposed, but we do not mark transposition here) basis vectors:
\begin{equation}
{\bf e}^i=A^{ij}{\bf e}_j.
\end{equation}
Then,
\begin{equation}
g^{ik} = \langle {\bf e}^i \vert {\bf e}^k \rangle  =\langle A^{ij}{\bf e}_j\vert  {\bf e}^k  \rangle
=A^{ij}\langle {\bf e}_j \vert  {\bf e}^k\rangle =A^{ij}\delta_j^k=A^{ik}
\end{equation}
and thus
\begin{equation}
{\bf e}^i=g^{ij}{\bf e}_j
\end{equation}
and, by a similar argument,
\begin{equation}
{\bf e}_i=g_{ij}{\bf e}^j.
\end{equation}

This property can also be used to raise or lower the indices not only of basis vectors but also of tensor components; that is,
to change from contravariant to covariant and conversely from covariant
to contravariant.
For example,
\begin{equation}
{\bf x} =
x^i {\bf e}_i = x^i g_{ij} {\bf e}^j   = x_j {\bf e}^j,
\end{equation}
and hence $x_j = x^i g_{ij}$.

What is  ${g^i}_{j}$?
A straightforward calculation yields, through insertion of Eqs.~(\ref{2016-m-ch-tensor-gij}) and (\ref{2016-m-ch-tensor-gij2}),
as well as the resolution of unity (in a modified form involving upper and lower indices;
cf. Section~\ref{2016-m-ch-fdvsrotio} on page \pageref{2016-m-ch-fdvsrotio}),
\begin{equation}
{g^i}_{j} = g^{ik}g_{kj}  =
\langle {\bf e}^i \underbrace{\mid {\bf e}^k \rangle \langle {\bf e}_k \mid}_{\mathbb{1}} {\bf e}_j\rangle
= \langle {\bf e}^i \mid {\bf e}_j\rangle
= \delta^i_j = \delta_{ij}.
\label{2014-m-ch-tensor-deltag}
\end{equation}
A similar calculation yields ${g_i}^{j} = \delta_{ij}$.

The metric tensor has been defined in terms of the scalar product.
The converse can be true as well.
(Note, however, that the metric need not be positive.)
In Euclidean space with the dot (scalar, inner) product
the metric tensor represents the scalar product between vectors: let
${\bf x}=x^i{\bf e}_i \in \mathbb{R}^n$ and ${\bf y}=y^j{\bf e}_j \in \mathbb{R}^n$ be two vectors.
Then (``$\intercal$'' stands for the transpose),
\begin{equation}
{\bf x}\cdot {\bf y}\equiv ({\bf x},{\bf y})\equiv \langle {\bf x}\mid {\bf y}\rangle
= x^i {\bf e}_i\cdot y^j {\bf e}_j
= x^iy^j {\bf e}_i\cdot  {\bf e}_j
=x^iy^j g_{ij}= x^\intercal  g y.
\end{equation}

It also characterizes the length of a vector: in the above
equation, set ${\bf y}={\bf x}$. Then,
\begin{equation}
{\bf x}\cdot {\bf x}\equiv ({\bf x},{\bf x})\equiv \langle {\bf x}\mid {\bf x}\rangle
=x^ix^j g_{ij}\equiv x^\intercal  g x,
\end{equation}
and thus, if the metric is positive definite,
\begin{equation}
\|  x\|  =\sqrt{x^ix^j g_{ij}}= \sqrt{x^\intercal  g x}.
\end{equation}

The square of the {\em line element} or {\em length element}
\index{line element}
\index{length element}
$ds =\| d{\bf x} \| $ of an infinitesimal vector $d{\bf x} $ is
\begin{equation}
d s^2  = g_{ij}dx^i dx^j= d{\bf x}^\intercal  g d{\bf x}.
\end{equation}

In (special) relativity with indefinite (Minkowski) metric, $ds^2$, or its finite difference form $\Delta s^2$, is used to define timelike, lightlike and spacelike distances:
\index{spacelike distance}
\index{timelike distance}
\index{lightlike distance}
with $g_{ij}=\eta_{ij}\equiv \text{diag}(1,1,1,-1)$,
$\Delta s^2 >0$ indicates spacelike distances,
$\Delta s^2 <0$ indicates timelike distances,
and $\Delta s^2 =0$ indicates lightlike distances.

\subsection{Transformation of the metric tensor}

Insertion into the definitions and coordinate transformations
 (\ref{2001-mu-tensor-tl2}) and (\ref{2012-m-ch-tlcbv})
yields
\begin{equation}
\begin{split}
g_{ij}={\bf e}_i\cdot {\bf e}_j
={a'^l}_i{\bf e'}_l\cdot {{a'}^m}_j{\bf e'}_m
={a'^l}_i{{a'}^m}_j {\bf e'}_l\cdot {\bf e'}_m \\
= {a'^l}_i{{a'}^m}_j  {g'}_{lm}
= {\partial { y }^l\over \partial x^i}{\partial { y }^m\over \partial x^j} {g'}_{lm}
.
\label{2011-m-emtdc}
\end{split}
\end{equation}

Conversely,  (\ref{2001-mu-tensors}) as well as    (\ref{2001-mu-tensor-tl1})
yields
\begin{equation}
\begin{split}
g'_{ij}={\bf f}_i\cdot {\bf f}_j
={a^l}_i{\bf e}_l\cdot {a^m}_j{\bf e}_m
={a^l}_i {a^m}_j {\bf e}_l\cdot {\bf e}_m  \\
= {a^l}_i {a^m}_j  {g}_{lm}
= {\partial {x}^l\over \partial { y }^i}{\partial {x}^m\over \partial { y }^j} {g}_{lm}
.
\end{split}
\label{2018-mm-ch-tensor-gpijgeneral}
\end{equation}

If the geometry (i.e., the basis) is locally orthonormal, ${g}_{lm}=\delta_{lm}$,
then
$g'_{ij}={\partial {x}^l\over \partial { y }^i}{\partial {x}_l\over \partial { y }^j}$.

Just to check consistency with Equation~(\ref{2014-m-ch-tensor-deltag}) we can compute,
for suitable differentiable coordinates $X$ and $ Y $,
\begin{equation}
\begin{split}
{g'_i}^j={\bf f}_i\cdot {{\bf f}}^j
={a^l}_i{\bf e}_l\cdot {(a^{-1})_m}^j{\bf e}^m
={a^l}_i {(a^{-1})_m}^j {\bf e}_l\cdot {\bf e}^m  \\
= {a^l}_i {(a^{-1})_m}^j \delta_l^m
= {a^l}_i {(a^{-1})_l}^j  \\
= {\partial {x}^l\over \partial { y }^i}{\partial { y }_l\over \partial {x}_j}
= {\partial {x}^l\over \partial {x}^j}{\partial { y }_l\over \partial { y }_i}
= \delta^{lj}\delta_{li}
= \delta^l_i
.
\end{split}
\end{equation}

In terms of the
{\em  Jacobian matrix} defined in Equation~(\ref{2013-m-t-jm})
\index{Jacobian matrix}
the metric tensor in Equation~(\ref{2011-m-emtdc})
can be rewritten as
\begin{equation}
g = J^\intercal  g' J
\equiv g_{ij}= J_{li}J_{mj}g'_{lm}
.
\label{2011-m-emtdcJ}
\end{equation}
The metric tensor and the Jacobian (determinant)
are thus related by
\begin{equation}
\textrm{det }g = (\textrm{det }J^\intercal ) (\textrm{det } g')(\textrm{det } J)
.
\label{2011-m-emtdcJd}
\end{equation}
If the manifold is embedded into an Euclidean space,
then $g'_{lm}=\delta_{lm}$
and  $g = J^\intercal   J $.

\subsection{Examples}

In what follows a few metrics are enumerated and briefly commented.
For a more systematic treatment, see, for instance, Snapper and Troyer's {\em Metric Affine geometry}.\cite{snapper-troyer}

Note also that due to the properties of the metric tensor, its coordinate representation has to be a {\em symmetric matrix}
with {\em nonzero diagonals}.
For the symmetry $g( {\bf x}  ,{\bf y} )=g({\bf y} ,{\bf x} )$ implies that $g_{ij}x_iy^j= g_{ij}y^ix_j=  g_{ij}x_jy^i= g_{ji}x_iy^j$ for all coordinate tuples
$x^i$ and $y^j$. And for any zero diagonal entry (say, in the $k$'th position of the diagonal we can choose a nonzero vector  ${\bf z}$
whose coordinates are all zero except the $k$'th coordinate. Then $g ({\bf z},{\bf x})=0$ for all ${\bf x}$ in the vector space.

\subsection*{$n$-dimensional Euclidean space}

\begin{equation}
g\equiv \{g_{ij}\}={\rm diag} (\underbrace{1,1,\ldots ,1}_{n\; {\rm times}})
\end{equation}

One application in physics is quantum mechanics,
where $n$ stands for the dimension of a complex Hilbert space.
Some definitions can be easily adapted to accommodate the complex numbers.
E.g., axiom 5 of the scalar product becomes
$(x,y)=\overline{(x,y)}$, where ``$\overline{(x,y)}$'' stands for complex conjugation of $(x,y)$.
Axiom 4 of the scalar product becomes
$(x,\alpha y)=\overline{\alpha} (x,y)$.

\subsection*{Lorentz plane}

\begin{equation}
g\equiv \{g_{ij}\}={\rm diag} (1,-1)
\end{equation}

\subsection*{Minkowski space of dimension $n$}

In this case the metric tensor is called the
{\em Minkowski metric}
\index{Minkowski metric}
and is often denoted by  ``$\eta$'':
\begin{equation}
\eta \equiv \{\eta_{ij}\}={\rm diag} (\underbrace{1,1,\ldots ,1}_{n-1\; {\rm times}},-1)
\label{2012-m-ch-tensor-minspn}
\end{equation}

One application in physics is the theory of special relativity,
where $D=4$.
Alexandrov's theorem states that the mere requirement of the preservation of
zero distance (i.e., lightcones), combined with bijectivity (one-to-oneness) of the transformation law
yields the Lorentz transformations.\cite[-70mm]{alex1,alex2,alex3,alex-col,borchers-heger,benz,lester,svozil-2001-convention}

\subsection*{Negative Euclidean space of dimension $n$}

\begin{equation}
g\equiv \{g_{ij}\}={\rm diag} (\underbrace{-1,-1,\ldots ,-1}_{n\; {\rm times}})
\end{equation}

\subsection*{Artinian four-space}

\begin{equation}
g\equiv \{g_{ij}\}={\rm diag} (+1,+1,-1 ,-1)
\end{equation}

\subsection*{General relativity}

In general relativity, the metric tensor $g$ is linked to the energy-mass distribution.
There, it appears as the primary concept when compared to the scalar product.
In the case of zero gravity, $g$ is just the  Minkowski metric (often denoted by  ``$\eta$'')
${\rm diag} (1,1,1,-1) $ corresponding to ``flat'' space-time.

The best known non-flat metric is the Schwarzschild metric
\begin{equation}
g
\equiv
\begin{pmatrix}
(1-2m/r)^{-1}&0&0&0\\
0&r^2&0&0\\
0&0&r^2\sin^2 \theta &0\\
0&0&0&- \left( 1-{2m/r}\right)
\end{pmatrix}
\end{equation}
with respect to the spherical space-time coordinates $r,\theta ,\phi ,t$.

{
\color{blue}
\bexample

\subsection*{Computation of the metric tensor of the circle of radius $r$}
Consider the transformation from the standard orthonormal
threedimensional ``Cartesian'' coordinates
$x_1=x$,
$x_2=y$,
into polar coordinates
\index{spherical coordinates}
$x_1'=r$,
$x_2'=\varphi$.
In terms of  $r$ and $\varphi$, the Cartesian coordinates can be written as
\begin{equation}
\begin{split}
 x_1=r \cos \varphi \equiv x_1' \cos x_2'  , \\
 x_2=r \sin \varphi \equiv x_1'\sin x_2'  .
\end{split}
\end{equation}
Furthermore,  since the basis we start with is the Cartesian orthonormal basis,
$g_{ij}=\delta_{ij}$; therefore,
\begin{equation}
g'_{ij}= {\partial {x}^l\over \partial { y }^i}{\partial {x}_k\over \partial { y }^j} g_{lk}
= {\partial {x}^l\over \partial { y }^i}{\partial {x}_k\over \partial { y }^j} \delta_{lk}
= {\partial {x}^l\over \partial { y }^i}{\partial {x}_l\over \partial { y }^j}.
\end{equation}
More explicitely, we obtain for the coordinates of the transformed metric tensor $g'$
\begin{equation}
\begin{split}
g'_{11}
= {\partial {x}^l\over \partial { y }^1}{\partial {x}_l\over \partial { y }^1} \\
=
{\partial (r \cos \varphi) \over \partial {r}}{\partial (r \cos \varphi) \over \partial {r}}
+
{\partial (r \sin \varphi) \over \partial {r}}{\partial (r \sin \varphi) \over \partial {r}}       \\
=
( \cos \varphi )^2
+
(\sin \varphi )^2 =1,\\
g'_{12}
= {\partial {x}^l\over \partial { y }^1}{\partial {x}_l\over \partial { y }^2} \\
=
{\partial (r \cos \varphi) \over \partial {r}}{\partial (r \cos \varphi) \over \partial {\varphi }}
+
{\partial (r \sin \varphi) \over \partial {r}}{\partial (r \sin \varphi) \over \partial {\varphi }}  \\
=
(\cos \varphi ) (- r \sin \varphi )
+
(\sin \varphi )( r \cos \varphi )  =0,\\
g'_{21}
= {\partial {x}^l\over \partial { y }^2}{\partial {x}_l\over \partial { y }^1} \\
=
{\partial (r \cos \varphi) \over \partial {\varphi }}{\partial (r \cos \varphi) \over \partial {r}}
+
{\partial (r \sin \varphi) \over \partial {\varphi }}{\partial (r \sin \varphi) \over \partial {r}}   \\
=
(- r \sin \varphi ) ( \cos \varphi )
+
( r\cos  \varphi )( \sin \varphi )  =0,\\
g'_{22}
= {\partial {x}^l\over \partial { y }^2}{\partial {x}_l\over \partial { y }^2} \\
=
{\partial (r \cos \varphi) \over \partial {\varphi}}{\partial (r \cos \varphi) \over \partial {\varphi }}
+
{\partial (r \sin \varphi) \over \partial {\varphi}}{\partial (r \sin \varphi) \over \partial {\varphi }}      \\
=
(- r \sin \varphi )^2
+
( r \cos \varphi )^2  =r^2;
\end{split}
\label{2018-mm-ch-tensor-gijpola}
\end{equation}
that is, in matrix notation,
\begin{equation}
g'
=
\begin{pmatrix}
1&0\\
0&r^2
\end{pmatrix}
,
\end{equation}
and thus
\begin{equation}
(d s')^2  = g_{ij}'d{\bf x'}^i d{\bf x'}^j=   (dr)^2 + r^2 (d\varphi )^2.
\label{2014-gppolarc}
\end{equation}


\subsection*{Computation of the metric tensor of the ball}
Consider the transformation from the standard orthonormal
threedimensional ``Cartesian'' coordinates
$x_1=x$,
$x_2=y$,
$x_3=z$,
into spherical coordinates
\index{spherical coordinates}
$x_1'=r$,
$x_2'=\theta$,
$x_3'=\varphi$.
In terms of  $r,\theta , \varphi$, the Cartesian coordinates can be written as
\begin{equation}
\begin{split}
 x_1=r \sin \theta \cos \varphi \equiv x_1' \sin x_2' \cos x_3'  , \\
 x_2=r \sin \theta \sin \varphi \equiv x_1'\sin x_2' \sin x_3'  ,    \\
 x_3=r \cos \theta  \equiv x_1'\cos x_2'  .
\end{split}
\end{equation}
Furthermore,  since the basis we start with is the Cartesian orthonormal basis,
$g_{ij}=\delta_{ij}$; hence finally
\begin{equation}
g'_{ij}= {\partial {x}^l\over \partial { y }^i}{\partial {x}_l\over \partial { y }^j}
\equiv {\rm diag}(1,r^2,r^2\sin^2 \theta ),
\end{equation}
and
\begin{equation}
(ds')^2 =(dr)^2+r^2(d\theta )^2+r^2\sin^2 \theta (d\varphi )^2.
\end{equation}

The expression $ds^2 =(dr)^2+r^2(d\varphi )^2$
for polar coordinates in two dimensions (i.e., $n=2$) of Equation~(\ref{2014-gppolarc})  is recovered by setting $\theta = \pi/2 $ and $d\theta =0$.

\subsection*{Computation of the metric tensor of the Moebius strip}
The parameter representation of the Moebius strip is
\begin{equation}
\Phi (u,v) =\left(
\begin{array}{c}
(1+v\cos \frac{u}{2})\sin u \\
(1+v\cos \frac{u}{2})\cos u \\
v\sin \frac{u}{2}
\end{array}
\right),
\end{equation}
where
$u\in [0,2\pi ]$ represents the position of the point on the circle,  and where $2a>0$ is the ``width'' of the Moebius strip,
and where $v\in [-a,a]$.

\begin{equation}
\begin{split}
\Phi _{v}=\frac{\partial \Phi }{\partial v}=\allowbreak
\begin{pmatrix}
\cos \frac{u}{2}\sin u \\
\cos \frac{u}{2}\cos u \\
\sin \frac{u}{2}
\end{pmatrix}
 \\
\Phi _{u}=\frac{\partial \Phi }{\partial u}=\allowbreak
\begin{pmatrix}
-\frac{1}{2}v\sin \frac{u}{2}\sin u+\left( 1+v\cos \frac{u}{2}\right) \cos u \\
-\frac{1}{2}v\sin \frac{u}{2}\cos u-\left( 1+v\cos \frac{u}{2}\right) \sin u \\
\frac{1}{2}v\cos \frac{u}{2}
\end{pmatrix}
\end{split}
\end{equation}

\begin{equation}
\begin{split}
\left(\frac{\partial \Phi }{\partial v}\right)^\intercal\frac{\partial \Phi }{\partial u}
=  \allowbreak
\begin{pmatrix}
\cos \frac{u}{2}\sin u \\
\cos \frac{u}{2}\cos u \\
\sin \frac{u}{2}
\end{pmatrix}^\intercal
\begin{pmatrix}
-\frac{1}{2}v\sin \frac{u}{2}\sin u+\left( 1+v\cos \frac{u}{2}\right) \cos u \\
-\frac{1}{2}v\sin \frac{u}{2}\cos u-\left( 1+v\cos \frac{u}{2}\right) \sin u \\
\frac{1}{2}v\cos \frac{u}{2}
\end{pmatrix}
\\
=
-\frac{1}{2}\left( \cos \frac{u}{2}\sin ^{2}u\right) v\sin \frac{u}{2}-%
\frac{1}{2}\left( \cos \frac{u}{2}\cos ^{2}u\right) v\sin \frac{u}{2}
\\
+%
\frac{1}{2}\sin \frac{u}{2} v\cos \frac{u}{2}=\allowbreak 0
\end{split}
\end{equation}

\begin{equation}
\begin{split}
\left(\frac{\partial \Phi }{\partial v}\right)^\intercal \frac{\partial \Phi }{\partial v}
=\allowbreak
\begin{pmatrix}
\cos \frac{u}{2}\sin u \\
\cos \frac{u}{2}\cos u \\
\sin \frac{u}{2}
\end{pmatrix}^\intercal
\begin{pmatrix}
\cos \frac{u}{2}\sin u \\
\cos \frac{u}{2}\cos u \\
\sin \frac{u}{2}
\end{pmatrix}
 \\
=
\cos ^{2}\frac{u}{2}\sin ^{2}u+\cos ^{2}\frac{u}{2}\cos ^{2}u+\sin ^{2}%
\frac{u}{2}=\allowbreak 1
\end{split}
\end{equation}

\begin{equation}
\begin{split}
\left(\frac{\partial \Phi }{\partial u}\right)^\intercal \frac{\partial \Phi }{\partial u}
\\
\begin{pmatrix}
= -\frac{1}{2}v\sin \frac{u}{2}\sin u+\left( 1+v\cos \frac{u}{2}\right) \cos
u \\
-\frac{1}{2}v\sin \frac{u}{2}\cos u-\left( 1+v\cos \frac{u}{2}\right) \sin
u \\
\frac{1}{2}v\cos \frac{u}{2}
\end{pmatrix}^\intercal     \cdot
\\
\cdot
\begin{pmatrix}
-\frac{1}{2}v\sin \frac{u}{2}\sin u+\left( 1+v\cos \frac{u}{2}\right) \cos
u \\
-\frac{1}{2}v\sin \frac{u}{2}\cos u-\left( 1+v\cos \frac{u}{2}\right) \sin
u \\
\frac{1}{2}v\cos \frac{u}{2}
\end{pmatrix}
 \\
=
\frac{1}{4}v^{2}\sin ^{2}\frac{u}{2}\sin ^{2}u+\cos
^{2}u+2 v \cos ^{2}u \cos \frac{u}{2}+
v^{2}\cos ^{2}u  \cos ^{2}\frac{u}{2}
\\
+\frac{1}{4}v^{2}\sin ^{2}\frac{u}{2}\cos^{2}u
+\sin ^{2}u+2  v\sin ^{2}u\cos \frac{u}{2}+ v^{2} \sin^{2}u\cos ^{2}\frac{u}{2}
 \\
+\frac{1}{4}v^{2}\cos ^{2}\frac{1}{2}%
u =\allowbreak \frac{1}{4}v^{2}+v^{2}\cos ^{2}\frac{u}{2}+1+2v\cos \frac{%
1}{2}u
 \\
=\left(1+v\cos \frac{u}{2}\right)^{2}+\frac{1}{4}v^{2}
\end{split}
\end{equation}

Thus the metric tensor is given by
\begin{equation}
\begin{split}
g'_{ij}
= {\partial {x}^s\over \partial { y }^i}{\partial {x}^t\over \partial { y }^j}g_{st}
= {\partial {x}^s\over \partial { y }^i}{\partial {x}^t\over \partial { y }^j}\delta_{st}\\
\quad \equiv
\left(
\begin{array}{cc}
\Phi _{u}\cdot \Phi _{u} & \Phi _{v}\cdot \Phi _{u} \\
\Phi _{v}\cdot \Phi _{u} & \Phi _{v}\cdot \Phi _{v}
\end{array}
\right) ={\rm diag}\left(
\left(1+v\cos \frac{u}{2}\right)^{2}+\frac{1}{4}v^{2} , 1\right).
\end{split}
\end{equation}

\eexample
}

\section{Decomposition of tensors}

Although a tensor of type (or rank) $n$ transforms like the tensor product of $n$ tensors of type 1,
not all type-$n$ tensors can be decomposed into a single
tensor product of $n$ tensors of type (or rank) 1.

Nevertheless,
by a generalized Schmidt decomposition (cf. page \pageref{2011-m-Schmidtdecomposition}),
any type-$2$ tensor  can be decomposed into
the sum of
tensor products of two tensors of type 1.

\section{Form invariance of tensors}

A tensor (field) is
form-invariant  with respect to some basis change
\index{form invariance}
if its representation in the new basis has the same form as in the old basis.
For instance, if the ``12122--component'' $T_{12122} (x)$ of the tensor $T$
with respect to the old basis and old coordinates $x$   equals some function $f(x)$ (say, $f(x)=x^2$),
then, a necessary condition for $T$ to be form invariant is that, in terms of the new basis,
that component  $T'_{12122} (x')$  equals the same function $f(x')$ as before, but in the new coordinates $x'$
[say, $f(x')=(x')^2$].
A sufficient condition for form invariance of $T$ is that {\em all}
coordinates or components of $T$ are form-invariant in that way.

Although form invariance is a gratifying feature for the reasons explained shortly,
a tensor (field) needs not necessarily
be form invariant with respect to all or even any (symmetry) transformation(s).

A physical motivation for the use of form-invariant tensors can be given as follows.
What makes some tuples (or matrix, or tensor components in general)  of
numbers or scalar functions a tensor? It is the
interpretation of the scalars as tensor components {\em with respect to
a particular basis}. In another basis, if we were talking about the same
tensor, the tensor components; that is, the numbers or scalar functions,
would be different.
Pointedly stated, the tensor coordinates represent some
encoding of a multilinear function with respect to a particular basis.

Formally, the tensor coordinates are numbers; that is, scalars,
which are grouped together in vector tuples or matrices or whatever form we consider useful.
As the tensor coordinates are scalars, they can be treated as scalars.
For instance, due to commutativity and associativity, one can exchange
their order. (Notice, though, that this is generally not the case for
differential operators such as $\partial_i=\partial / \partial {\bf x}^i$.)

A {\em form invariant} tensor with respect to  certain transformations
is a tensor which retains
the same functional form if the transformations are performed; that is,
if the basis changes accordingly.
That is, in this case,
the functional form of mapping numbers or coordinates or other entities remains unchanged, regardless of the coordinate change.
Functions remain the same but with the new parameter components as
argument. For instance; $4\mapsto 4$ and $f(x_1,x_2,x_3)\mapsto
f( y_1, y_2, y_3)$.

Furthermore, if a tensor is invariant with respect to one transformation, it need not
be invariant with respect to another transformation, or with respect to
changes of the scalar product; that is, the metric.

Nevertheless, totally symmetric (antisymmetric) tensors remain totally
symmetric (antisymmetric) in all cases:
\begin{equation}
A_{i_1i_2 \ldots i_si_t\ldots i_k}
=
\pm A_{i_1i_2 \ldots i_ti_s\ldots i_k}
\end{equation}
implies
\begin{equation}
\begin{split}
A'_{j_1i_2 \ldots j_s j_t\ldots j_k}
=
{a^{i_1}}_{j_1}{a^{i_2}}_{j_2}
\cdots
{a_{j_s}}^{i_s}{a_{j_t}}^{i_t}
\cdots
{a^{i_k}}_{j_k} A_{i_1 i_2\ldots i_s i_t\ldots  i_k}
 \\
=
\pm {a^{i_1}}_{j_1}{a^{i_2}}_{j_2}\cdots
{a_{j_s}}^{i_s}{a_{j_t}}^{i_t}\cdots
{a^{i_k}}_{j_k} A_{i_1 i_2\ldots i_t i_s\ldots  i_k}
  \\
=
\pm {a^{i_1}}_{j_1}{a^{i_2}}_{j_2}
\cdots
{a_{j_t}}^{i_t}{a_{j_s}}^{i_s}
\cdots
{a^{i_k}}_{j_k} A_{i_1 i_2\ldots i_t i_s\ldots  i_k}
  \\
=
\pm A'_{j_1i_2 \ldots j_t j_s\ldots j_k}    .
\end{split}
\end{equation}

In physics, it would be nice if the natural laws could be written into a
form which does not depend on the particular reference frame or  basis
used.
Form invariance thus is a gratifying physical feature, reflecting the
{\em symmetry} against changes of coordinates and bases.

After all, physicists want the formalization of their fundamental laws not to artificially depend on,
say, spacial directions, or on some particular basis, if there is no physical reason why this should be so.
Therefore, physicists tend to be crazy to write down everything in a
form-invariant manner.

One strategy to accomplish  form invariance  is to start out with form-invariant
tensors and compose -- by tensor products and index reduction -- everything from them. This method guarantees form
invariance.

{
\color{blue}
\bexample

The ``simplest'' form-invariant tensor under all transformations is the constant tensor of rank $0$.

Another constant form invariant tensor  under all transformations is represented by the Kronecker symbol $\delta^i_j$,
because
\begin{equation}
{(\delta ')}^i_j= {(a^{-1})^i}_{k} {a^{l}}_j\delta^{k}_{l}={(a^{-1})^i}_{k} {a^{k}}_j= \delta^i_j
.
\end{equation}

A simple form invariant tensor field is a vector ${\bf x}$,
because if $T({\bf x})= x^i t_i= x^i {\bf e}_i={\bf x}$, then
the ``inner transformation''
${\bf x} \mapsto  {\bf x}'$
and the ``outer transformation''
$T \mapsto  T'= \textsf{\textbf{A}}T$
just compensate each other; that is, in coordinate representation, Eqs.(\ref{2012-m-ch-di-choic}) and (\ref{2011-m-tvtcov}) yield
\begin{equation}
T'({\bf x}')= {x'}^i t'_i= {(a^{-1})^i}_l x^l   {a^j}_i t_j = {a^j}_i {(a^{-1})^i}_l  {\bf e}_j  x^l
= \delta_l^j x^l {\bf e}_j  = {\bf x} = T({\bf x}).
\end{equation}

For the sake of another demonstration of form invariance, consider the following two factorizable tensor fields:
while
\begin{equation}
{S}(x)=
\begin{pmatrix}
  {  x}_2  \\
- {  x}_1
\end{pmatrix}
\otimes
\begin{pmatrix}
   {  x}_2  \\
 - {  x}_1
\end{pmatrix}^\intercal
=
\begin{pmatrix}    {  x}_2 ,- {  x}_1  \end{pmatrix}^\intercal
\otimes
\begin{pmatrix}   {  x}_2 ,- {  x}_1  \end{pmatrix}
\equiv
\begin{pmatrix}
   {  x}_2^2          & -{ x}_1{x}_2  \\
 - {  x}_1{  x}_2     & { x}_1^2
\end{pmatrix}
\label{2012-m-ch-tensor-etotccon1factorized}
\end{equation}
is a form invariant tensor field with respect to the basis $\{(0,1),(1,0)\}$
and orthogonal transformations (rotations around the origin)
\begin{equation}
\begin{pmatrix}
  \cos \varphi & \sin \varphi  \\
 -\sin \varphi & \cos \varphi
\end{pmatrix}
,
\end{equation}
\begin{equation}
{ T}(x)=
\begin{pmatrix}
{  x}_2  \\
{  x}_1
\end{pmatrix}
\otimes
\begin{pmatrix}
{  x}_2  \\
{  x}_1 \end{pmatrix}^\intercal
=
\begin{pmatrix}   {  x}_2 ,  {  x}_1  \end{pmatrix}^\intercal
\otimes
\begin{pmatrix}    {  x}_2 ,  {  x}_1 \end{pmatrix}
\equiv
\begin{pmatrix}
{  x}_2^2 & { x}_1{  x}_2  \\
{  x}_1{  x}_2          & { x}_1^2
\end{pmatrix}
\end{equation}
is not.

This can be proven by considering the single factors from which $S$ and $T$ are composed.
Eqs. (\ref{2012-m-ch-tensor-etotc1})-(\ref{2011-m-tvtcov})
and
(\ref{2012-m-ch-tensor-etotccon1})-(\ref{2012-m-ch-tensor-etotccon2})
show that the form
invariance of the factors implies the form invariance of the tensor products.

For instance, in our example, the factors $\begin{pmatrix}  {  x}_2 ,- {  x}_1  \end{pmatrix}^\intercal $
of $S$ are invariant, as they transform as
$$
\begin{pmatrix} \cos \varphi & \sin \varphi  \\
                         -\sin \varphi & \cos \varphi
\end{pmatrix}
\begin{pmatrix}
{  x}_2  \\
 - {  x}_1
\end{pmatrix}
=
\begin{pmatrix}
 {  x}_2 \cos \varphi  - x_1 \sin \varphi  \\
            - x_2 \sin \varphi         - {  x}_1 \cos \varphi
\end{pmatrix}
=
\begin{pmatrix}
{  x}_2'  \\
 - {  x}_1'
\end{pmatrix},
$$
where the transformation of the coordinates
$$
\begin{pmatrix}
{  x}_1'  \\
  {  x}_2'
\end{pmatrix}
=
\begin{pmatrix}
 \cos \varphi & \sin \varphi  \\
   -\sin \varphi & \cos \varphi
\end{pmatrix}
\begin{pmatrix}
 {  x}_1  \\
 {  x}_2
\end{pmatrix}
=
\begin{pmatrix}
{  x}_1 \cos \varphi  + x_2 \sin \varphi  \\
- x_1 \sin \varphi         + {  x}_2 \cos \varphi
\end{pmatrix}
$$
has been used.

Note that  the notation identifying tensors of type (or rank) two with matrices,
creates an ``artefact'' insofar as the transformation of the ``second index'' must then be represented by
the exchanged multiplication order, together with the transposed transformation matrix;
that is,
\begin{equation}
a_{ik}a_{jl}A_{kl}
=  a_{ik}A_{kl}a_{jl}
=  a_{ik}A_{kl}\left(a^\intercal  \right)_{lj}
\equiv a\cdot A\cdot a^\intercal  .
\end{equation}
Thus for a
transformation  of
the transposed tuple  $\begin{pmatrix}   {  x}_2 ,- {  x}_1  \end{pmatrix}$
we must consider the {\em transposed} transformation matrix arranged {\em after} the factor; that is,
\begin{equation}
\begin{split}
\begin{pmatrix}  {  x}_2 , - {  x}_1 \end{pmatrix}
\begin{pmatrix}  \cos \varphi & -\sin \varphi  \\
  \sin \varphi & \cos \varphi
\end{pmatrix}
\\
=
\begin{pmatrix}
{  x}_2 \cos \varphi  - x_1 \sin \varphi ,
 - x_2 \sin \varphi         - {  x}_1 \cos \varphi
\end{pmatrix}
=
\left(
{  x}_2'  ,
 - {  x}_1'
\right).
\end{split}
\end{equation}

In contrast, a similar calculation shows that the factors
$\begin{pmatrix}  {  x}_2 ,  {  x}_1  \end{pmatrix}^\intercal $
of $T$ do not transform invariantly.
However, noninvariance with respect to certain transformations does not imply that
$T$ is not a valid, ``respectable'' tensor field; it is just not form-invariant under rotations.
\eexample
}

Nevertheless, note again that, while the tensor product of form-invariant tensors is again a form-invariant tensor,  not every form
invariant tensor might be decomposed into products of form-invariant tensors.

{
\color{blue}
\bexample
Let
$\vert + \rangle  \equiv   (1,0)^\intercal$
and
$\vert - \rangle  \equiv   (0,1)^\intercal$.
For a nondecomposable tensor, consider the sum of two-partite tensor products (associated with two ``entangled'' particles)
\index{entanglement}
Bell state (cf. Equation~(\ref{2014-m-ch-fdvs-bellbasis}) on page \pageref{2014-m-ch-fdvs-bellbasis}) in the standard basis     \index{Bell state}
\begin{equation}
\begin{split}
\vert \Psi^-\rangle = \frac{1}{\sqrt{2}}\left(\vert +-\rangle   - \vert -+\rangle  \right)
 \equiv  \left( 0,\frac{1}{\sqrt{2}},- \frac{1}{\sqrt{2}} ,  0 \right)^\intercal , \\
\vert \Psi^-\rangle \langle  \Psi^- \vert \equiv  \frac{1}{2}
\begin{pmatrix}
0&0&0&0\\
0&1&-1&0\\
0&-1&1&0\\
0&0&0&0
\end{pmatrix}
.
\end{split}
\label{2011-m-bellstatenondec}
\end{equation}
\marginnote{$\vert \Psi^-\rangle $,   together with the other three Bell states
$\vert \Psi^+\rangle = \frac{1}{\sqrt{2}}\left(\vert +-\rangle   + \vert -+\rangle  \right) $,
$\vert \Phi^+\rangle = \frac{1}{\sqrt{2}}\left(\vert --\rangle   + \vert ++\rangle  \right) $,
and
$\vert \Phi^-\rangle = \frac{1}{\sqrt{2}}\left(\vert --\rangle   - \vert ++\rangle  \right) $,
forms an orthonormal basis of $\mathbb{C}^4$.
}

Why is $\vert \Psi^-\rangle$ not decomposable into a product form of two vectors?
In order to be able to answer this question
(see also Section \ref{2012-m-c-fdvs-entanglement} on page \pageref{2012-m-c-fdvs-entanglement}), consider
the most general two-partite state
\begin{equation}
\vert \psi \rangle
=
\psi_{--}\vert -- \rangle
+
\psi_{-+}\vert -+ \rangle
+
\psi_{+-}\vert +- \rangle
+
\psi_{++}\vert ++ \rangle
,
\end{equation}
with $\psi_{ij}\in \mathbb{C}$,
and compare it to the most general state obtainable through products of single-partite states
$\vert \phi_1\rangle  = \alpha_-  \vert - \rangle    + \alpha_+  \vert + \rangle$,
and
$\vert \phi_2\rangle  = \beta_-  \vert - \rangle    + \beta_+  \vert + \rangle$
with $\alpha_{i}, \beta_i \in \mathbb{C}$;
that is,
\begin{equation}
\begin{split}
\vert \phi \rangle  =\vert \phi_1\rangle    \vert \phi_2\rangle
 =
(\alpha_-  \vert - \rangle    + \alpha_+  \vert + \rangle )
(\beta_-  \vert - \rangle    + \beta_+  \vert + \rangle )   \\
 =\alpha_- \beta_- \vert -- \rangle    + \alpha_-\beta_+  \vert -+ \rangle +
\alpha_+ \beta_- \vert +- \rangle    + \alpha_+\beta_+  \vert ++ \rangle
.
\end{split}
\end{equation}
$
\vert -- \rangle  \equiv (1,0,0,0)^\intercal
$,
$
\vert -+ \rangle    \equiv (0,1,0,0)^\intercal
$,
$
\vert +- \rangle     \equiv (0,0,1,0)^\intercal
$, and
$
\vert ++ \rangle      \equiv (0,0,0,1)^\intercal
$
are linear independent (indeed, orthonormal),
a comparison of $\vert \psi \rangle  $ with  $\vert \phi \rangle$ yields
$
\psi_{--}=  \alpha_- \beta_-
$, $
\psi_{-+}=   \alpha_-\beta_+
$, $
\psi_{+-}=  \alpha_+ \beta_-
$, and$
\psi_{++}= \alpha_+\beta_+
$.
The divisions
$\psi_{--}/ \psi_{-+} =   \beta_- / \beta_+ =   \psi_{+-} / \psi_{++}$
yield
 a necessary and sufficient condition for a two-partite quantum state to be decomposable
into a product of single-particle quantum states:  its amplitudes must obey
 \begin{equation}
\psi_{--}\psi_{++}  =  \psi_{-+}   \psi_{+-} .
\end{equation}
This is not satisfied for the Bell state $\vert \Psi^-\rangle$ in Equation~(\ref{2011-m-bellstatenondec}),
because in this case $\psi_{--}=\psi_{++} =0$
and  $ \psi_{-+} = - \psi_{+-} =1/\sqrt{2}$.
In physics this is referred to as {\em entanglement}.\cite[-80mm]{CambridgeJournals:1737068,CambridgeJournals:2027212,schrodinger}
\index{entanglement}

Note also that $\vert \Psi^-\rangle$ is a {\em singlet state},
as it is form invariant under the following generalized rotations in two-dimensional complex Hilbert subspace; that is,
(if you do not believe this please check yourself)
\begin{equation}
\begin{split}
\vert + \rangle =
e^{ i{\frac{\varphi}{2}} }
\left(
\cos \frac{\theta}{2} \vert +'  \rangle
-
\sin \frac{\theta}{2} \vert -'   \rangle
\right),
\\
 \vert - \rangle =
e^{ -i{\frac{\varphi}{2}} }
\left(
\sin \frac{\theta}{2} \vert +'   \rangle
+
\cos \frac{\theta}{2} \vert -'  \rangle
\right)
\end{split}
\end{equation}
in the spherical coordinates $\theta , \varphi$,
but it cannot be composed or written as a product of a {\em single} (let alone form invariant) two-partite tensor product.

\eexample
}

%
%

In order to prove form invariance of a constant tensor,
one has to transform the tensor according to the standard transformation laws
(\ref{2011-m-tvtcov}) and (\ref{2011-m-tvtcontrav}), and compare the result with the input;
that is, with the untransformed, original, tensor.
This is sometimes referred to as the ``outer transformation.''

In order to prove form invariance of a tensor field,
one has to additionally transform the spatial coordinates on which the field depends;
that is, the arguments of that field; and then compare.
This is sometimes referred to as the ``inner transformation.''
This will become clearer with the following example.

{
\color{blue}
\bexample

Consider again the tensor field defined
earlier in Equation
(\ref{2012-m-ch-tensor-etotccon1factorized}),
but let us not choose the ``elegant''
ways of proving form invariance by factoring; rather we explicitly
consider the transformation of all the  components
$$S_{ij}(x_1,x_2)
=
\begin{pmatrix}
 -x_1x_2 & - x_2^2  \\
 x_1^2 & x_1x_2
\end{pmatrix}
$$
with respect to the standard basis  $\{(1,0), (0,1)\}$.

Is $S$ form-invariant with respect to rotations around the origin?
That is, $S$ should be form invariant with respect to transformations
$x_i' = a_{ij} x_j$
with
$$
a_{ij}=\begin{pmatrix}
 \cos \varphi & \sin \varphi  \\
  -\sin \varphi & \cos \varphi
\end{pmatrix}.
$$

Consider the ``outer'' transformation first.
As has been pointed out earlier,
the term on the right hand side in $
S_{ij}'= a_{ik}a_{jl}S_{kl}
$
can be rewritten as a product of three matrices; that is,
$$
a_{ik}a_{jl}S_{kl}\left(x_n\right)
=  a_{ik}S_{kl}a_{jl}
=  a_{ik}S_{kl}\left(a^\intercal  \right)_{lj}
\equiv a\cdot S\cdot a^\intercal  .
$$
$a^\intercal $ stands for the transposed matrix; that is,
$(a^\intercal )_{ij}=a_{ji}$.
\begin{equation}
\begin{split}
  \left(
    \begin{array}{cc}
      \cos \varphi  & \sin \varphi \\
      -\sin \varphi & \cos \varphi
    \end{array}
  \right)
  \left(
    \begin{array}{cc}
      -x_1x_2 & -x_2^2 \\
      x_1^2   & x_1x_2
    \end{array}
  \right)
  \left(
    \begin{array}{cc}
      \cos \varphi  & -\sin \varphi \\
      \sin \varphi & \cos \varphi
    \end{array}
  \right)
\\
  =\left(
    \begin{array}{cc}
      -x_1 x_2 \cos \varphi + x_1^2 \sin \varphi &
        -x_2^2 \cos \varphi + x_1 x_2 \sin \varphi \\
      x_1 x_2 \sin \varphi + x_1^2 \cos \varphi &
        x_2^2 \sin \varphi + x_1 x_2 \cos \varphi
    \end{array}
  \right)
  \left(
    \begin{array}{cc}
      \cos \varphi  & -\sin \varphi \\
      \sin \varphi & \cos \varphi
    \end{array}
  \right)
\\
  =\left(\!\!\!
    \begin{array}{cc}
      \cos \varphi
        \left(-x_1 x_2 \cos \varphi + x_1^2 \sin \varphi\right)+\!&\!
      -\sin \varphi
        \left(-x_1 x_2 \cos \varphi + x_1^2 \sin \varphi\right)+\\
      \quad +\sin \varphi
        \left(-x_2^2 \cos \varphi + x_1 x_2 \sin \varphi\right)\!&\!
      \quad +\cos \varphi
        \left(-x_2^2 \cos \varphi + x_1 x_2 \sin \varphi\right)\\
      \\[1ex]
      \cos \varphi
        \left(x_1 x_2 \sin \varphi + x_1^2 \cos \varphi\right)+\!&\!
      -\sin \varphi
        \left(x_1 x_2 \sin \varphi + x_1^2 \cos \varphi\right)+\\
      \quad +\sin \varphi
        \left(x_2^2 \sin \varphi + x_1 x_2 \cos \varphi\right)\!&\!
      \quad +\cos \varphi
        \left(x_2^2 \sin \varphi + x_1 x_2 \cos \varphi\right)
    \end{array}
  \!\right)
\\
  =\left(
    \begin{array}{cc}
      x_1 x_2 \left(\sin^2 \varphi - \cos^2 \varphi \right) + &
      2x_1 x_2 \sin \varphi \cos \varphi \\
      \qquad + \left( x_1^2-x_2^2 \right) \sin \varphi \cos \varphi &
      \qquad - x_1^2 \sin^2 \varphi - x_2^2 \cos^2 \varphi \\
      \\[1ex]
      2x_1 x_2 \sin \varphi \cos \varphi + &
      -x_1 x_2 \left(\sin^2 \varphi - \cos^2 \varphi \right) - \\
      \qquad + x_1^2 \cos^2 \varphi + x_2^2 \sin^2 \varphi &
      \quad -\left(x_1^2-x_2^2\right) \sin \varphi \cos \varphi
    \end{array}
  \right)
\end{split}
\nonumber
\end{equation}

Let us now perform the ``inner'' transform
$$
  x'_i =a_{ij}x_j \Longrightarrow
  \begin{array}{rcl}
    x'_1 & = & x_1 \cos \varphi + x_2 \sin \varphi \\
    x'_2 & = & -x_1 \sin \varphi + x_2 \cos \varphi .
  \end{array}
$$

Thereby we assume (to be corroborated) that the functional form in the new coordinates are identical
to the functional form of the old coordinates.
A comparison yields
\begin{eqnarray*}
  -x'_1 \,x'_2 & = &
  -\left(x_1 \cos \varphi + x_2 \sin \varphi \right)
  \left(-x_1 \sin \varphi + x_2 \cos \varphi \right) = \\
  & = &
  -\left(
    -x_1^2 \sin \varphi \cos \varphi +
      x_2^2 \sin \varphi \cos \varphi -
      x_1 x_2 \sin^2 \varphi + x_1 x_2 \cos^2 \varphi
  \right) = \\
  & = &
  x_1 x_2 \left(\sin^2 \varphi - \cos^2 \varphi \right) +
    \left(x_1^2 - x_2^2\right) \sin \varphi \cos \varphi \\
  (x'_1)^2  & = &
    \left(x_1 \cos \varphi + x_2 \sin \varphi \right)
    \left(x_1 \cos \varphi + x_2 \sin \varphi \right) = \\
  & = & x_1^2 \cos^2 \varphi + x_2^2 \sin^2 \varphi +
    2 x_1 x_2 \sin \varphi \cos \varphi \\
  (x'_2)^2  & = &
    \left(-x_1 \sin \varphi + x_2 \cos \varphi \right)
    \left(-x_1 \sin \varphi + x_2 \cos \varphi \right) = \\
  & = & x_1^2 \sin^2 \varphi + x_2^2 \cos^2 \varphi -
    2 x_1 x_2 \sin \varphi \cos \varphi
\end{eqnarray*}
and hence
$$S' (x'_1,x'_2)=\left(
    \begin{array}{cc}
      -x'_1x'_2 & -(x'_2)^2 \\
      (x'_1)^2   & x'_1x'_2
    \end{array}
\right)
$$ is invariant with respect to rotations by angles $\varphi$, yielding the new basis
$\{ (\cos \varphi ,-\sin \varphi
),(\sin
\varphi ,\cos \varphi )\}$.

Incidentally,
as has been stated earlier, $S(x)$ can be written as the product of two invariant tensors $b_i(x)$ and $c_j(x)$:
$$S_{ij}(x)=b_i(x)c_j(x),$$
with
$
b(x_1,x_2)=(-x_2,x_1),
$ and
$
c(x_1,x_2)=(x_1,x_2)
$.
This can be easily checked by comparing the components:
\begin{eqnarray*}
b_1c_1&=&-x_1x_2 = S_{11},\\
b_1c_2&=&-x_2^2 = S_{12},\\
b_2c_1&=&x_1^2 = S_{21},\\
b_2c_2&=&x_1x_2 = S_{22}.\\
\end{eqnarray*}

Under rotations, $b$ and $c$  transform into
\begin{eqnarray*}
a_{ij}b_j&=&
  \left(
    \begin{array}{cc}
      \cos \varphi  & \sin \varphi \\
      -\sin \varphi & \cos \varphi
    \end{array}
  \right)
  \left(
    \begin{array}{c}
      -x_2\\
       x_1
    \end{array}
  \right)
=
  \left(
    \begin{array}{c}
      -x_2\cos \varphi +x_1 \sin \varphi \\
      x_2\sin \varphi +x_1 \cos \varphi
    \end{array}
  \right)
=
  \left(
    \begin{array}{c}
      -x'_2 \\
      x'_1
    \end{array}
  \right)    \\
 a_{ij}c_j&=&
  \left(
    \begin{array}{cc}
      \cos \varphi  & \sin \varphi \\
      -\sin \varphi & \cos \varphi
    \end{array}
  \right)
  \left(
    \begin{array}{c}
      x_1\\
       x_2
    \end{array}
  \right)
=
  \left(
    \begin{array}{c}
      x_1\cos \varphi +x_2 \sin \varphi \\
      -x_1\sin \varphi +x_2 \cos \varphi
    \end{array}
  \right)
=
  \left(
    \begin{array}{c}
      x'_1    \\
      x'_2
    \end{array}
  \right) .
\end{eqnarray*}

This factorization of $S$ is nonunique, since
Equation
(\ref{2012-m-ch-tensor-etotccon1factorized})
uses a different factorization; also, $S$ is decomposable into, for example,
$$S(x_1,x_2)=
  \left(
    \begin{array}{cc}
      -x_1x_2 & -x_2^2 \\
      x_1^2   & x_1x_2
    \end{array}
  \right)     =
  \left(
    \begin{array}{cc}
      -x_2^2 \\
      x_1 x_2
    \end{array}
  \right)
\otimes \left({x_1\over x_2},1\right).
$$

\eexample
}

\section{The Kronecker symbol $\delta$}
\index{delta tensor}
For vector spaces of dimension $n$ the totally symmetric Kronecker symbol $\delta$,
sometimes referred to
as the delta symbol $\delta$--tensor, can be defined by
\begin{equation}
\delta_{i_1 i_2\cdots i_k}
=
\left\{
\begin{array}{rl}
+1&\textrm{ if }  i_1 = i_2 = \cdots = i_k \\
0&\textrm{ otherwise (that is, some indices are not identical).}
\end{array}
\right.
\end{equation}

Note that, with the Einstein summation convention,
\index{Einstein summation convention}
\begin{equation}
\begin{split}
\delta_{ij} a_j    = a_j  \delta_{ij} =
\delta_{i1} a_1
+
\delta_{i2} a_2
+
\cdots
+
\delta_{in} a_n  =a_i
,  \\
\delta_{ji} a_j = a_j  \delta_{ji} =
\delta_{1i} a_1
+
\delta_{2i} a_2
+
\cdots
+
\delta_{ni} a_n  =a_i
.
\end{split}
\end{equation}

\section{The Levi-Civita symbol $\varepsilon$}
\index{Levi-Civita symbol}
\index{antisymmetric tensor}
For vector spaces of dimension $n$ the totally antisymmetric Levi-Civita symbol $\varepsilon$, sometimes referred to
as the Levi-Civita symbol $\varepsilon$--tensor, can be defined by the number of permutations of its indices; that is,
\begin{equation}
\varepsilon_{i_1 i_2\cdots i_k}
=
\left\{
\begin{array}{rl}
+1&\textrm{ if } (i_1 i_2\ldots i_k) \textrm{ is an {\em even} permutation of } (1,2,\ldots k)\\
-1&\textrm{ if } (i_1 i_2\ldots i_k) \textrm{ is an {\em odd} permutation of } (1,2,\ldots k)\\
0&\textrm{ otherwise (that is, some indices are identical).}
\end{array}
\right.
\label{2014-m-ch-lcs}
\end{equation}
Hence, $\varepsilon_{i_1 i_2\cdots i_k}$ stands for the sign of the permutation in the case of a permutation, and zero otherwise.

{
\color{blue}
\bexample

In two dimensions,
$$\varepsilon_{ij}\equiv
\left(
\begin{array}{rrrr}
\varepsilon_{11}&\varepsilon_{12}\\
\varepsilon_{21}&\varepsilon_{22}
\end{array}
\right)
=
\left(
\begin{array}{rrrr}
0&1\\
-1&0
\end{array}
\right)
.
$$
\eexample
}

In threedimensional Euclidean space,
the cross product, or vector product
\index{cross product}
\index{vector product}
of two vectors
${\bf x}\equiv x_i$
and
${\bf y}\equiv y_i$
can be written as
${\bf x} \times {\bf y}\equiv \varepsilon_{ijk}x_jy_k$.

{\color{OliveGreen}
\bproof
For a direct proof, consider, for arbitrary threedimensional vectors ${\bf x}$ and ${\bf y}$,
and by enumerating all nonvanishing terms; that is, all permutations,
\begin{equation}
\begin{split}
{\bf x} \times {\bf y}\equiv \varepsilon_{ijk}x_jy_k
\equiv
\begin{pmatrix}
\varepsilon_{123}x_2y_3 + \varepsilon_{132}x_3y_2 \\
\varepsilon_{213}x_1y_3 + \varepsilon_{231}x_3y_1 \\
\varepsilon_{312}x_2y_3 + \varepsilon_{321}x_3y_2
\end{pmatrix}   \\
=
\begin{pmatrix}
\varepsilon_{123}x_2y_3 - \varepsilon_{123}x_3y_2 \\
-\varepsilon_{123}x_1y_3 + \varepsilon_{123}x_3y_1 \\
\varepsilon_{123}x_2y_3 - \varepsilon_{123}x_3y_2
\end{pmatrix}
=
\begin{pmatrix}
x_2y_3 - x_3y_2 \\
-x_1y_3 + x_3y_1 \\
x_2y_3 - x_3y_2
\end{pmatrix}
.
\end{split}
\end{equation}
\eproof
}

\section{Nabla, Laplace, and D'Alembert operators}
\index{nabla operator}
\index{Laplace operator}
\index{D'Alembert operator}

The {\em nabla operator}
\begin{equation}
\nabla_i \equiv \left(
\frac{\partial }{\partial x^1},
\frac{\partial }{\partial x^2},
\ldots ,
\frac{\partial }{\partial x^n}
\right).
\end{equation}
is a vector differential operator in an $n$-dimensional vector space $\frak V$.
In index notation, $\nabla_i$ is also written as
\begin{equation}
\nabla_i  =\partial_i =\partial_{x^i}
= \frac{\partial }{\partial x^i}
.
\end{equation}

Why is the lower index indicating covariance used when differentiation with respect to upper indexed, contravariant coordinates?
The nabla operator transforms in the following manners:
$\nabla_i  =\partial_i =\partial_{x^i}$ transforms like a {\em covariant} basis vector
[cf. Eqs.~(\ref{2012-m-ch-tlcbv}) and (\ref{2001-mu-tensor-tl2nl})], since
\begin{equation}
\partial_i =
\frac{\partial }{\partial x^i}
=
\frac{\partial { y }^j}{\partial x^i}
\;
\frac{\partial }{\partial { y }^j}
=
\frac{\partial { y }^j}{\partial x^i}
\;
\partial'_j
=
{{\left(a^{-1}\right)}^j}_i
\partial'_j
=
J_{ji}
\partial'_j,
\end{equation}
where $J_{ij}$ stands for the {\em  Jacobian matrix} defined in Equation~(\ref{2013-m-t-jm}).
\index{Jacobian matrix}

As very similar calculation demonstrates that $\partial^i=\frac{\partial }{\partial x_i}$ transforms like a {\em contravariant} vector.

In three dimensions and in the standard Cartesian basis with the Euclidean metric,
covariant and contravariant entities coincide,
and
\begin{equation}
\begin{split}
\nabla
=
\begin{pmatrix}
\frac{\partial }{\partial x^1},
\frac{\partial }{\partial x^2},
\frac{\partial }{\partial x^3}
\end{pmatrix}
={\bf e}^1\frac{\partial }{\partial x^1}
+{\bf e}^2\frac{\partial }{\partial x^2}
+{\bf e}^3\frac{\partial }{\partial x^3}
=
\\
= \begin{pmatrix}
\frac{\partial }{\partial x_1},
\frac{\partial }{\partial x_2},
\frac{\partial }{\partial x_3}
\end{pmatrix}^\intercal
={\bf e}_1\frac{\partial }{\partial x_1}
+{\bf e}_2\frac{\partial }{\partial x_2}
+{\bf e}_3\frac{\partial }{\partial x_3}
.
\end{split}
\end{equation}

It is often used to define basic differential operations;
in particular, (i) to denote the {\em gradient} of a scalar field $f(x_1,x_2,x_3)$ (rendering a vector field with respect to a particular basis),
(ii) the {\em divergence} of a vector field ${\bf v}(x_1,x_2,x_3)$
(rendering a scalar field with respect to a particular basis), and
(iii) the {\em curl} (rotation) of a vector field  ${\bf v}(x_1,x_2,x_3)$ (rendering a vector field with respect to a particular basis)
as follows:
\index{gradient}
\index{divergence}
\index{curl}
\begin{eqnarray}
\textrm{grad } f &=& \nabla f =\begin{pmatrix}
\frac{\partial f}{\partial x_1},
\frac{\partial f}{\partial x_2},
\frac{\partial f}{\partial x_3}
\end{pmatrix}^\intercal   ,\\
\textrm{div }  {\bf v} &=& \nabla \cdot {\bf v} =
\frac{\partial v_1}{\partial x_1}+
\frac{\partial v_2}{\partial x_2}+
\frac{\partial v_3}{\partial x_3}
  ,\\
\textrm{rot } {\bf v} &=& \nabla \times {\bf v} = \begin{pmatrix}
\frac{\partial v_3}{\partial x_2}-
\frac{\partial v_2}{\partial x_3}
,
\frac{\partial v_1}{\partial x_3}-
\frac{\partial v_3}{\partial x_1}
,
\frac{\partial v_2}{\partial x_1}-
\frac{\partial v_1}{\partial x_2}
\end{pmatrix}^\intercal           \\
&\equiv& \varepsilon_{ijk} \partial_j v_k.
\end{eqnarray}

The {\em Laplace operator}
\index{Laplace operator}
is defined by
\begin{equation}
\Delta = \nabla^2= \nabla \cdot \nabla =
\frac{\partial^2 }{\partial  x_1 ^2}+
\frac{\partial^2 }{\partial  x_2 ^2}+
\frac{\partial^2 }{\partial  x_3 ^2}
.
\end{equation}

In special relativity and electrodynamics,  as well as in  wave theory and quantized field theory, with the Minkowski space-time
of dimension four
(referring to the metric tensor with the signature ``$\pm ,\pm ,\pm ,\mp$''),
the {\em D'Alembert operator}
\index{D'Alembert operator}
is defined by the Minkowski metric $\eta = {\rm diag} (1,1,1,-1)$
\begin{equation}
\begin{split}
\Box  = \partial_i \partial^i
=
\eta_{ij}  \partial^i \partial^j=
\nabla^2- \frac{\partial^2 }{\partial t^2}=
\nabla \cdot \nabla - \frac{\partial^2 }{\partial t^2}\\
=
\frac{\partial^2 }{\partial  x_1^2}+
\frac{\partial^2 }{\partial  x_2^2}+
\frac{\partial^2 }{\partial  x_3^2}- \frac{\partial^2 }{\partial t^2}
.
\end{split}
\end{equation}


\section{Tensor analysis in orthogonal curvilinear coordinates}
\index{curvilinear coordinates}

\subsection{Curvilinear coordinates}
In terms of (orthonormal) Cartesian coordinates $\begin{pmatrix}x_1, x_2,\ldots , x_n\end{pmatrix}^\intercal$
of the Cartesian standard basis ${\frak B}=\left\{ {\bf e}_1, {\bf e}_2,\ldots ,{\bf e}_n \right\}$,
curvilinear coordinates
\begin{equation}
\begin{pmatrix}  u_1(x_1, x_2,\ldots ,x_n)\\
 u_2(x_1, x_2,\ldots ,x_n)\\
\vdots \\
 u_n(x_1, x_2,\ldots ,x_n)\\
\end{pmatrix}
\text{, and }
\begin{pmatrix}
x_1(u_1, u_2,\ldots ,u_n)\\
x_2(u_1, u_2,\ldots ,u_n)\\
\vdots \\
 x_n(u_1, u_2,\ldots ,u_n)
\end{pmatrix}
\label{2018-mm-ch-ten-cc}
\end{equation}
are coordinates, defined relative to the local curvilinear basis
${\frak B}'=\left\{ {\bf e}_{u_1}, {\bf e}_{u_2},\ldots ,{\bf e}_{u_n} \right\}$
(defined later)
in which the coordinate lines (defined later) may be curved.
\marginnote{Coordinates with straight coordinate lines, like Cartesian coordinates, are special cases of curvilinear coordinates.}
Therefore, curvilinear coordinates should be ``almost everywhere'' (but not always are)
locally invertible (surjective, one-to-one) maps whose differentiable functions
$u_i(x_1, x_2,\ldots ,x_n)$ and $x_i(u_1, u_2,\ldots ,u_n)$
are continuous (better smooth, that is, infinitely often differentiable).
Points
\marginnote{The origin in polar or spherical coordinates is a singular point because,
for zero radius all angular parameters yield this same point.
For the same reason for the cylinder coordinates the line of zero radius
at the center of the cylinder consists of  singular points.}
in which this is not the case are called
{\em singular points}.
\index{singular points}
This translates into the requirement that the
Jacobian matrix
\index{Jacobian matrix}
$
J (u_1, u_2,\ldots ,u_n)
$
with components in the $i$th row and $j$ column
$
\frac{\partial x_i}{\partial u_j}
$
defined in (\ref{2013-m-t-jm})
is invertible; that is,
its Jacobian determinant
$
\frac{
\partial
(
x_1, x_2,\ldots ,x_n
)
}{
\partial
(
u_1, u_2,\ldots ,u_n
)
}
$
defined in
(\ref{2018-mm-t-jd})
must not vanish.
$
\frac{
\partial
(
x_1, x_2,\ldots ,x_n
)
}{
\partial
(
u_1, u_2,\ldots ,u_n
)
}
=0
$
indicates singular point(s).

Some $i$th {\em coordinate line}
\index{coordinate lines}
is a curve  (a one-dimensional subset of $\mathbb{R}^n$)
\begin{equation}
\left\{ {\bf x}(c_1, \ldots ,u_i,\ldots , c_n)
\middle|
{\bf x} = x^k(c_1, \ldots ,u_i,\ldots , c_n) \, {\bf e}_k ,\;
u_i\in \mathbb{R}, \;
u_{j\neq i} = c_j \in \mathbb{R}
\right\}
\label{2018-mm-ch-ten-cl}
\end{equation}
where $u_i$ varies and all other coordinates $u_j{\neq i} = c_j$, $1\le j\neq i\le n$
remain constant with fixed  $c_j \in \mathbb{R}$.

Another way of perceiving this is to consider coordinate hypersurfaces of constant $u_i$.
The coordinate lines are just intersections on $n-1$ of these coordinate hypersurfaces.

In three dimensions, there are  three coordinate surfaces (planes) corresponding to constant
$u_1=c_1$,
$u_2=c_2$, and
$u_3=c_3$ for fixed $c_1,c_2,c_3 \in \mathbb{R}$, respectively.
Any of the three intersections of two of these three planes
fixes two parameters out of three, leaving the third one to freely vary;
thereby forming the respective coordinate lines.

Orthogonal curvilinear coordinates are coordinates for which all coordinate
lines are mutually orthogonal ``almost everywhere''
(that is, with the possible exception of singular points).

{
\color{blue}
\bexample

Examples of orthogonal curvilinear coordinates
are polar coordinates in $\mathbb{R}^2$,
as well as cylindrical and spherical coordinates in $\mathbb{R}^3$.

\begin{itemize}
\item[(i)]
Polar coordinates\marginnote{The Jacobian $J(r,\theta )=
\begin{pmatrix}
\frac{\partial r}{\partial x} &
\frac{\partial r}{\partial y}  \\
\frac{\partial \theta }{\partial x} &
\frac{\partial \theta }{\partial y}
\end{pmatrix}
=
\frac{1}{r}
\begin{pmatrix}
r\cos \theta &
r\sin \theta \\
- \sin \theta  &
 \cos \theta
\end{pmatrix}
$
is not invertible at $r=0$.
Therefore, points with $r=0$  are singular points of the transformation
which is not invertible there.}
$\begin{pmatrix}
u_1=r, u_2=\theta
\end{pmatrix}^\intercal$
can be written in terms of Cartesian coordinates
$\begin{pmatrix}
x,y \end{pmatrix}^\intercal$ as
\begin{equation}
\begin{split}
x = r \cos \theta ,\quad
y = r \sin \theta   ;\text{ and}\quad   \\
r=\sqrt{x^2+y^2}\text{, }
\theta = \arctan \left(\frac{y}{x}\right)
,
\end{split}
\label{2018-mm-ch-pc}
\end{equation}
with $r \ge 0$
and $-\pi < \theta \le \pi$.
The first coordinate lines
are straight lines going through the origin at some fixed angle $\theta $.
The second coordinate lines
form concentric circles of some fixed radius $r=R$ around the origin.

\item[(ii)]
Cylindrical coordinates
\index{cylindrical coordinates}
$\begin{pmatrix}
u_1=r, u_2=\theta, u_3=z
\end{pmatrix}^\intercal$
are just extensions of polar coordinates into three-dimensional vector space, such that the additional coordinate
$u_3$ coincides with the additional Cartesian coordinate $z$.

\item[(iii)]
Spherical coordinates
\index{spherical coordinates}
$\begin{pmatrix}
u_1=r, u_2=\theta, u_3=\varphi
\end{pmatrix}^\intercal$
can be written in terms of Cartesian coordinates as
\begin{equation}
\begin{split}
x  = r\sin \theta \cos \varphi , \quad
y = r\sin \theta \sin \varphi ,  \quad
z = r\cos \theta  \text{; and}\\
r=\sqrt{x^2+y^2+z^2}\text{, }
\theta = \arccos \left(\frac{z}{r}\right), \quad
\varphi=  \arctan \left(\frac{y}{x}\right),
\end{split}
\label{2018-mm-ch-sc}
\end{equation}
whereby  $\theta$ is the polar angle in the $x$--$z$-plane measured
from the $z$-axis, with $0 \le \theta \le \pi$,
and $\varphi $ is  the azimuthal angle in the $x$--$y$-plane, measured
from the $x$-axis with $0 \le \varphi < 2 \pi$.
\end{itemize}

The Jacobian $J(r,\theta ,\varphi )$ in terms of Cartesian coordinates $(x,y,z)$
can be obtained from a rather tedious calculation:
\begin{equation}
\begin{split}
J(r,\theta ,\varphi ) =\begin{pmatrix}
\frac{\partial r}{\partial x} &
\frac{\partial r}{\partial y} &
\frac{\partial r}{\partial z} \\
\frac{\partial \theta }{\partial x} &
\frac{\partial \theta }{\partial y} &
\frac{\partial \theta }{\partial z} \\
\frac{\partial \varphi }{\partial x} &
\frac{\partial \varphi }{\partial y} &
\frac{\partial \varphi }{\partial z}
\end{pmatrix}
\\
=
\begin{pmatrix}
\frac{x}{\sqrt{x^2+y^2+z^2}}                           & \frac{y}{\sqrt{x^2+y^2+z^2}}             & \frac{z}{\sqrt{x^2+y^2+z^2}}   \\
\frac{xz}{(x^2+y^2+z^2)\sqrt{x^2+y^2}} &
\frac{yz}{(x^2+y^2+z^2)\sqrt{x^2+y^2}} &
-\frac{\sqrt{x^2+y^2}}{x^2+y^2+z^2} \\
- \frac{y}{ x^2+y^2 } &
 \frac{x}{ x^2+y^2 } &
0 &
\end{pmatrix}
\\
=  \frac{1}{r}
\begin{pmatrix}
  r \sin \theta \cos\varphi  &
 r \sin \theta \sin \varphi  &
 r \cos \theta   \\
\cos \theta \cos \varphi &
 \cos \theta \sin \varphi  &
-  \sin \theta    \\
-\frac{ \sin \varphi }{   \sin \theta}&
\frac{ \cos \varphi }{   \sin \theta}&
0
\end{pmatrix}
.
\end{split}
\label{2018-mm-ch-sf-ijbsc}
\end{equation}
\marginnote{Points with $r=0$ are singular points; the transformation is not invertible there.}

The inverse Jacobian matrix
$J(x,y,z)$ in terms of spherical coordinates $(r, \theta , \varphi )$ is
\marginnote{Note that
$
\left|\frac{\partial x }{\partial u }\frac{\partial u }{\partial x }\right|
\le
\frac{\partial x }{\partial u }\frac{\partial u }{\partial x} +
\frac{\partial x }{\partial v }\frac{\partial v }{\partial x} +
\frac{\partial x }{\partial w }\frac{\partial w }{\partial x}
= 1
$.
}
\begin{equation}
\begin{split}
J(x,y,z) = \left[J(r,\theta ,\varphi )\right]^{-1}
=
\begin{pmatrix}
\frac{\partial x}{\partial r} &
\frac{\partial x}{\partial \theta } &
\frac{\partial x}{\partial \varphi } \\
\frac{\partial y}{\partial r} &
\frac{\partial y}{\partial \theta } &
\frac{\partial y}{\partial \varphi } \\
\frac{\partial z}{\partial r} &
\frac{\partial z}{\partial \theta } &
\frac{\partial z}{\partial \varphi }
\end{pmatrix}
\\
=
\begin{pmatrix}
\sin \theta \cos \varphi &  r\cos \theta \cos \varphi &  -r\sin \theta \sin \varphi \\
\sin \theta \sin \varphi &  r\cos \theta \sin \varphi &  -r\sin \theta \cos \varphi \\
\cos \theta & - \sin \theta & 0
\end{pmatrix}
.
\end{split}
\label{2018-mm-ch-sf-ijbscinv}
\end{equation}

\newcommand{\comm}[1]{}
\comm{
=============================================================================

r[x_,y_,z_] := Sqrt[x^2 + y^2 + z^2];
\[Theta][x_,y_,z_] := ArcCos[z/Sqrt[x^2 + y^2 + z^2]];
\[Phi][x_,y_,z_] := ArcTan[ y / x ];

x[r_,\[Theta]_,\[Phi]_] := r*Sin[\[Theta]]*Cos[\[Phi]];
y[r_,\[Theta]_,\[Phi]_] :=  r*Sin[\[Theta]]*Sin[\[Phi]];
z[r_,\[Theta]_,\[Phi]_] :=  r*Cos[\[Theta]];

JacoSp= FullSimplify[ {
 { D[Sqrt[x^2 + y^2 + z^2], x] /. x -> r*Sin[\[Theta]]*Cos[\[Phi]] /. y -> r*Sin[\[Theta]]*Sin[\[Phi]] /. z -> r*Cos[\[Theta]]             ,
  D[Sqrt[x^2 + y^2 + z^2], y] /. x -> r*Sin[\[Theta]]*Cos[\[Phi]] /. y -> r*Sin[\[Theta]]*Sin[\[Phi]] /. z -> r*Cos[\[Theta]]             ,
 D[Sqrt[x^2 + y^2 + z^2], z] /. x -> r*Sin[\[Theta]]*Cos[\[Phi]] /. y -> r*Sin[\[Theta]]*Sin[\[Phi]] /. z -> r*Cos[\[Theta]]              },
{  D[ArcCos[z/Sqrt[x^2 + y^2 + z^2]], x] /. x -> r*Sin[\[Theta]]*Cos[\[Phi]] /. y -> r*Sin[\[Theta]]*Sin[\[Phi]] /. z -> r*Cos[\[Theta]]      ,
  D[ArcCos[z/Sqrt[x^2 + y^2 + z^2]], y] /. x -> r*Sin[\[Theta]]*Cos[\[Phi]] /. y -> r*Sin[\[Theta]]*Sin[\[Phi]] /. z -> r*Cos[\[Theta]]      ,
 D[ArcCos[z/Sqrt[x^2 + y^2 + z^2]], z] /. x -> r*Sin[\[Theta]]*Cos[\[Phi]] /. y -> r*Sin[\[Theta]]*Sin[\[Phi]] /. z -> r*Cos[\[Theta]]     },
{  D[ ArcTan[ y / x ], x] /. x -> r*Sin[\[Theta]]*Cos[\[Phi]] /. y -> r*Sin[\[Theta]]*Sin[\[Phi]] /. z -> r*Cos[\[Theta]]                  ,
  D[ ArcTan[ y / x ], y] /. x -> r*Sin[\[Theta]]*Cos[\[Phi]] /. y -> r*Sin[\[Theta]]*Sin[\[Phi]] /. z -> r*Cos[\[Theta]]                  ,
 D[ ArcTan[ y / x ], z] /. x -> r*Sin[\[Theta]]*Cos[\[Phi]] /. y -> r*Sin[\[Theta]]*Sin[\[Phi]] /. z -> r*Cos[\[Theta]]                 }
       },Element[r > 0 | 0 <= \[Theta] <= Pi | 0 <= \[Phi] <= 2 Pi  , Reals] ];

MatrixForm[JacoSp]

JacoSp2= FullSimplify[ {
{
D[ x[r,\[Theta],\[Phi]], r],
D[ x[r,\[Theta],\[Phi]], \[Theta]],
D[ x[r,\[Theta],\[Phi]], \[Phi]]
},
{
D[ y[r,\[Theta],\[Phi]], r],
D[ y[r,\[Theta],\[Phi]], \[Theta]],
D[ y[r,\[Theta],\[Phi]], \[Phi]]
},
{
D[ z[r,\[Theta],\[Phi]], r],
D[ z[r,\[Theta],\[Phi]], \[Theta]],
D[ z[r,\[Theta],\[Phi]], \[Phi]]
}
}
]

MatrixForm[JacoSp2]

FullSimplify[JacoSp.JacoSp2]
=============================================================================
}

\eexample
}

\subsection{Curvilinear bases}

Let us henceforth concentrate on three dimensions. In terms of Cartesian coordinates
${\bf r}= \begin{pmatrix}x,y,z\end{pmatrix}^\intercal$
a curvilinear basis
\index{curvilinear basis}
can be defined by noting that
$
\frac{\partial  {\bf r}}{\partial u}$,
$\frac{\partial  {\bf r}}{\partial v}$, and
$\frac{\partial  {\bf r}}{\partial w}$
are {\em tangent vectors} ``along''  the coordinate curves of varying
$u$, $v$, and $w$, with all other coordinates $\{ v,w \}$, $\{ u,w \}$, and $\{ u,v \}$ constant, respectively.
They are mutually orthogonal for orthogonal curvilinear coordinates.
Their lengths, traditionally denoted by $h_u$, $h_v$, and $h_w$, are obtained from their Euclidean norm
and identified with the square root of the diagonal elements of the metric tensor~(\ref{2018-mm-ch-tensor-gpijgeneral}):
\begin{equation}
\begin{split}
h_u \stackrel{{\text{\tiny def}}}{=}   \left\| \frac{\partial  {\bf r}}{\partial u} \right\| =
\sqrt{ \left(\frac{\partial  x}{\partial u}\right)^2+
  \left(\frac{\partial  y}{\partial u}\right)^2+
  \left(\frac{\partial  z}{\partial u}\right)^2}  = \sqrt{g_{uu}}
,        \\
h_v \stackrel{{\text{\tiny def}}}{=}    \left\| \frac{\partial  {\bf r}}{\partial v} \right\| =
\sqrt{ \left(\frac{\partial  x}{\partial v}\right)^2+
  \left(\frac{\partial  y}{\partial v}\right)^2+
  \left(\frac{\partial  z}{\partial v}\right)^2} =\sqrt{g_{vv}}
,        \\
h_w \stackrel{{\text{\tiny def}}}{=}    \left\| \frac{\partial  {\bf r}}{\partial w} \right\| =
\sqrt{ \left(\frac{\partial  x}{\partial w}\right)^2+
  \left(\frac{\partial  y}{\partial w}\right)^2+
  \left(\frac{\partial  z}{\partial w}\right)^2}  =\sqrt{g_{ww}}
.
\end{split}
\label{2018-mm-ch-tensor-l}
\end{equation}

The associated unit vectors ``along''  the coordinate curves of varying
$u$, $v$, and $w$ are defined by
\begin{equation}
\begin{split}
{\bf e}_u = \frac{1}{ h_u } \frac{\partial  {\bf r}}{\partial u},\quad
{\bf e}_v = \frac{1}{ h_v } \frac{\partial  {\bf r}}{\partial v},\quad
{\bf e}_w = \frac{1}{ h_w } \frac{\partial  {\bf r}}{\partial w},
\\
\text{or }\quad
\frac{\partial  {\bf r}}{\partial u} = h_u {\bf e}_u,\quad
\frac{\partial  {\bf r}}{\partial v} = h_v {\bf e}_v,\quad
\frac{\partial  {\bf r}}{\partial w} = h_w {\bf e}_w
.
\end{split}
\label{2018-mm-ch-tensor-uv}
\end{equation}

In case of orthogonal curvilinear coordinates these unit vectors form an orthonormal basis
\begin{equation}
{\frak B}'=\left\{ {\bf e}_{u}(u,v,w), {\bf e}_{v}(u,v,w),{\bf e}_{w} (u,v,w)\right\}
\label{2018-mm-ch-tensor-ob}
\end{equation}
at the point $\begin{pmatrix}u,v,w\end{pmatrix}^\intercal$ so that
\begin{equation}
{\bf e}_{u_i} \cdot {\bf e}_{u_j} = \delta_{ij}
\text{, with }
u_i,u_j \in \{u,v,w\}.
\label{2018-mm-ch-tensor-uvob}
\end{equation}
Unlike the Cartesian standard basis which remains the same in all points,
the curvilinear basis is {\em locally} defined because the orientation of the curvilinear basis vectors
could (continuously or smoothly, according to the assumptions for curvilinear coordinates) vary for different points.

\subsection{Infinitesimal increment, line element, and volume}
\label{2018-mm-ch-ctensor-volumeclc}

The {\em infinitesimal increment}
\index{infinitesimal increment}
of the Cartesian coordinates~(\ref{2018-mm-ch-ten-cc}) in three dimensions
$
\begin{pmatrix}
x(u,v,w), y(u,v,w), z(u,v,w)
\end{pmatrix}^\intercal
$
can be expanded
in
the orthogonal curvilinear coordinates
$\begin{pmatrix}
u, v, w
\end{pmatrix}^\intercal$ as
\begin{equation}
\begin{split}
d {\bf r}
=
\frac{\partial  {\bf r}}{\partial u} du
+ \frac{\partial  {\bf r}}{\partial v}dv
+ \frac{\partial  {\bf r}}{\partial w}dw
\\
=
{ h_u }{\bf e}_u   du + { h_v }{\bf e}_v   dv + { h_w } {\bf e}_w dw
,
\end{split}
\label{2018-mm-ch-tensor-infincr}
\end{equation}
where~(\ref{2018-mm-ch-tensor-uv}) has been used.
Therefore, for orthogonal curvilinear coordinates,
\begin{equation}
\begin{split}
{\bf e}_u \cdot d {\bf r} =
{\bf e}_u \cdot \left({ h_u }{\bf e}_u   du + { h_v }{\bf e}_v   dv + { h_w } {\bf e}_w dw \right) \\
=
{ h_u } \underbrace{{\bf e}_u \cdot {\bf e}_u}_{=1} du
+ { h_v } \underbrace{{\bf e}_u {\bf e}_v}_{=0}   dv
+ { h_w } \underbrace{{\bf e}_u {\bf e}_w}_{=0} dw
=  { h_u } du \text{, }
\\
{\bf e}_v \cdot d {\bf r} =   { h_v } dv \text{,  }
{\bf e}_w \cdot d {\bf r} =   { h_w } dw
.
\end{split}
\label{2018-mm-ch-tensor-etdr}
\end{equation}

In a similar derivation using the orthonormality of the curvilineas basis~(\ref{2018-mm-ch-tensor-ob})
the
(Euclidean) {\em line element }
for orthogonal curvilinear coordinates
\index{line element}
can be defined and evaluated as
\begin{equation}
\begin{split}
ds \stackrel{{\text{\tiny def}}}{=}
\sqrt{d {\bf r} \cdot d {\bf r}}
\\ =
\sqrt{
\left(
{ h_u }{\bf e}_u   du + { h_v }{\bf e}_v   dv + { h_w } {\bf e}_w dw
\right)
\cdot
\left(
{ h_u }{\bf e}_u   du + { h_v }{\bf e}_v   dv + { h_w } {\bf e}_w dw
\right)
}
\\ =
\sqrt{
(h_u du)^2 + (h_v dv)^2 + (h_w dw)^2
}
=
\sqrt{
h_u^2 du^2 + h_v^2 dv^2 + h_w^2 dw^2
}
.
\end{split}
\label{2018-mm-ch-tensor-le}
\end{equation}
That is, effectively, for the line element $ds$ the infinitesimal Cartesian coordinate increments
$
d {\bf r}=
\begin{pmatrix}
dx, dy, dz
\end{pmatrix}^\intercal
$
can be rewritten in terms of the ``normalized'' (by $h_u$, $h_v$, and $h_w$)
orthogonal curvilinear coordinate increments
$
d {\bf r}=
\begin{pmatrix}
h_u  du , h_v dv, h_w dw
\end{pmatrix}^\intercal
$
by substituting
$dx$ with $h_u  du$,
$dy$ with $h_v  dv$, and
$dz$ with $h_w  dw$, respectively.

The infinitesimal three-dimensional volume $dV$ of the parallelepiped ``spanned'' by the
unit vectors
${\bf e}_{u}$, ${\bf e}_{v}$, and ${\bf e}_{w}$
of the
curvilinear basis~(\ref{2018-mm-ch-tensor-ob})
is given by
\begin{equation}
\begin{split}
dV =
\left|
(h_u {\bf e}_u du) \cdot
(h_v {\bf e}_v dv) \times
(h_w {\bf e}_w dw)
\right|
\\ =
\underbrace{\left| {\bf e}_u \cdot  {\underbrace{{\bf e}_v \times {\bf e}_w}_{{\bf e}_u}} \right|}_{=1}
h_u h_v h_w  du dv dw
.
\end{split}
\label{2018-mm-ch-tensor-veoclc}
\end{equation}

This result can be generalized to arbitrary dimensions:
according to Equation~(\ref{2018-mm-ch-fdvs-jacoinfvol})
on page~\pageref{2018-mm-ch-fdvs-jacoinfvol}
the volume of the infinitesimal parallelepiped
can be written in terms  of the Jacobian determinant~(\ref{2018-mm-t-jd}) on page~\pageref{2018-mm-t-jd}
as
\begin{equation}
dV =
\left| J \right|
d_{u_1} d_{u_2} \cdots d_{u_n}
 =
\left|
\frac{
\partial \begin{pmatrix}x_1, \ldots , x_n \end{pmatrix}
}
{
\partial \begin{pmatrix} u_1, \ldots , u_n \end{pmatrix}
}
\right|
d_{u_1} d_{u_2} \cdots d_{u_n}
.
\label{2018-mm-ch-tensor-veoclcjd}
\end{equation}

{
\color{blue}
\bexample
For the sake of examples, let us again consider polar, cylindrical and spherical coordinates.
\begin{itemize}
\item[(i)]
For polar coordinates [cf. the metric~(\ref{2018-mm-ch-tensor-gijpola}) on page~\pageref{2018-mm-ch-tensor-gijpola}],
\index{polar coordinates}
\begin{equation}
\begin{split}
h_r=\sqrt{g_{rr}}=\sqrt{ \cos^2\theta + \sin^2 \theta }=1,\\
h_\theta=\sqrt{g_{\theta \theta}}=\sqrt{ r^2 \sin^2\theta + r^2 \cos^2 \theta }=r, \\
{\bf e}_r =
\begin{pmatrix}
\cos \theta \\ \sin \theta
\end{pmatrix}
,\quad
{\bf e}_\theta = \frac{1}{r}
\begin{pmatrix}
- r \sin \theta \\ r \cos \theta
\end{pmatrix}
=
\begin{pmatrix}
- \sin \theta \\ \cos \theta
\end{pmatrix}
,\quad
{\bf e}_r \cdot {\bf e}_\theta = 0
,
\\
d{\bf r}= {\bf e}_r dr + r {\bf e}_\theta d \theta
=  \begin{pmatrix}
\cos \theta \\ \sin \theta
\end{pmatrix}  dr
+
\begin{pmatrix}
- r\sin \theta \\ r\cos \theta
\end{pmatrix} d \theta ,
\\
ds= \sqrt{ (dr)^2+ r^2 (d \theta )^2},
\\
dV
=\text{ det }
\begin{pmatrix}
\frac{\partial x}{\partial r} & \frac{\partial x}{\partial \theta} \\
\frac{\partial y}{\partial r} & \frac{\partial y}{\partial \theta} \\
\end{pmatrix}
dr d \theta
=\text{ det }
\begin{pmatrix}
\cos \theta &  -r \sin \theta  \\
\sin \theta & r \cos \theta  \\
\end{pmatrix}
dr d \theta
\\
=
r(\cos^2 \theta +\sin^2 \theta) = r
dr d \theta
.
\end{split}
\label{2018-mm-ch-exhhpc}
\end{equation}
In the Babylonian spirit it is always prudent to check the validity of the expressions  for some  known instances,
say the circumference
$C=\int_{r=R,0\le \theta <2\pi} ds
=\int_{r=R,0\le \theta <2\pi} \sqrt{ \underbrace{(dr)^2}_{=0}+ r^2 (d \theta )^2}
=  \int_0^{2\pi}R d \theta
= 2\pi R$ of a circle of radius $R$.
The volume of this circle is
$V
=\int_{0\le r\le R,0\le \theta <2\pi} dV
=\int_{0\le r\le R,0\le \theta <2\pi} r dr d\theta =
\left( \left.\frac{r^2}{2}\right|_{r=0}^{r=R}\right)
\left( \left.\theta \right|_{\theta=0}^{\theta=2\pi}\right)
=\frac{R^2}{2} 2\pi = R^2 \pi$.

\item[(ii)]
For cylindrical coordinates,
\begin{equation}
\begin{split}
h_r=\sqrt{ \cos^2\theta + \sin^2 \theta }=1,\quad
h_\theta=\sqrt{ r^2 \sin^2\theta + r^2 \cos^2 \theta }=r, \quad
h_z=1,\\
ds= \sqrt{ (dr)^2+ r^2 (d \theta )^2 + (dz)^2},\\
dV= r dr d\theta dz
.
\end{split}
\label{2018-mm-ch-exhhcyc}
\end{equation}
Therefore, a cylinder of radius $R$ and height $H$ has the volume
$V
=\int_{0\le r\le R,0\le \theta <2\pi,0\le z \le H} dV
=\int_{0\le r\le R,0\le \theta <2\pi} rdr d\theta dz =
\left( \left.\frac{r^2}{2}\right|_{r=0}^{r=R}\right)
\left( \left.\theta \right|_{\theta=0}^{\theta=2\pi}\right)
\left( \left.z\right|_{z=0}^{z=H}\right)
=\frac{R^2}{2} 2\pi H = R^2 H \pi$.

\item[(iii)]
For spherical coordinates,
\begin{equation}
\begin{split}
h_r=\sqrt{ \sin^2\theta \cos^2\varphi + \sin^2 \theta \sin^2\varphi +\cos^2 \theta}=1,\\
h_\theta=\sqrt{ r^2 \cos^2\theta \cos^2 \varphi + r^2 \cos^2 \theta \sin^2 \varphi + r^2 \sin^2 \theta }=r,\\
h_\varphi=\sqrt{ r^2 \sin^2\theta \sin^2 \varphi + r^2 \sin^2 \theta \cos^2 \varphi  }=r\sin \theta,\\
ds= \sqrt{ (dr)^2+ r^2 (d \theta )^2 + (r \sin \theta)^2 (d\varphi )^2},
dV= r^2 \sin \theta dr d\theta d\varphi
.
\end{split}
\label{2018-mm-ch-exhhspher}
\end{equation}
Therefore, a sphere of radius $R$  has the volume
$V
=\int_{0\le r\le R,0\le \theta \le \pi,0\le \varphi \le 2\pi} dV
=\int_{0\le r\le R,0\le \theta \le \pi,0\le \varphi \le 2\pi} r^2 \sin \theta dr d\theta d\varphi
=
\left( \left.\frac{r^3}{3}\right|_{r=0}^{r=R}\right)
\left( \left.-\cos \theta \right|_{\theta=0}^{\theta=\pi}\right)
\left( \left.\varphi \right|_{\varphi =0}^{\varphi =2\pi}\right)
=\frac{R^3}{3} 2 (2\pi) = \frac{4 \pi}{3}  R^3$.
\end{itemize}
\eexample
}

\subsection{Vector differential operator and gradient}
\index{gradient}

The {\em gradient} $\nabla f$ of a scalar field $f(u,v,w)$ in orthogonal curvilinear coordinates can,
by insertion of $1=\frac{h_u}{h_u}=\frac{h_v}{h_v}=\frac{h_w}{h_w}$ and
with Eqs.~(\ref{2018-mm-ch-tensor-etdr}), be defined by
the infinitesimal change of $f$  as the coordinates vary infinitesimally:
\begin{equation}
\begin{split}
df  =
\frac{\partial f}{\partial u} du +
\frac{\partial f}{\partial v} dv +
\frac{\partial f}{\partial w} dw
\\
=
\frac{1}{h_u} \left(\frac{\partial f}{\partial u}\right) \underbrace{h_u du}_{={\bf e}_u \cdot {\bf dr}} +
\frac{1}{h_v} \left(\frac{\partial f}{\partial v}\right) \underbrace{h_v dv}_{={\bf e}_v \cdot {\bf dr}}  +
\frac{1}{h_w} \left(\frac{\partial f}{\partial w}\right) \underbrace{h_w dw}_{={\bf e}_w \cdot {\bf dr}}
\\
=
\left[
\frac{1}{h_u}  {\bf e}_u \left(\frac{\partial f}{\partial u}\right)     +
\frac{1}{h_v}  {\bf e}_v \left(\frac{\partial f}{\partial v}\right)      +
\frac{1}{h_w}  {\bf e}_w \left(\frac{\partial f}{\partial w}\right)   \right]
\cdot {\bf dr}
\\
=
\left[
\frac{{\bf e}_u}{h_u}  \left(\frac{\partial  }{\partial u} f \right)  +
\frac{{\bf e}_v}{h_v}  \left(\frac{\partial  }{\partial v} f \right)    +
\frac{{\bf e}_w}{h_w}  \left(\frac{\partial  }{\partial w} f \right)
\right]
\cdot {\bf dr}
=\nabla f \cdot d {\bf r}
,
\end{split}
\label{2018-mm-ch-del}
\end{equation}
such that the vector differential operator $\nabla$, when applied to a scalar field $f(u,v,w)$,
can be identified with
\begin{equation}
 \nabla  f
=
\frac{{\bf e}_u}{h_u}  \frac{\partial  f}{\partial u}    +
\frac{{\bf e}_v}{h_v}  \frac{\partial  f}{\partial v}      +
\frac{{\bf e}_w}{h_w} \frac{\partial  f}{\partial w}
=
\left(
\frac{{\bf e}_u}{h_u}  \frac{\partial   }{\partial u}    +
\frac{{\bf e}_v}{h_v}  \frac{\partial   }{\partial v}      +
\frac{{\bf e}_w}{h_w} \frac{\partial   }{\partial w}
\right) f,
\end{equation}
and
\begin{equation}
 \nabla
=
\frac{{\bf e}_u}{h_u}  \frac{\partial   }{\partial u}    +
\frac{{\bf e}_v}{h_v}  \frac{\partial   }{\partial v}      +
\frac{{\bf e}_w}{h_w} \frac{\partial   }{\partial w}
.
\label{2018-mm-ch-veo}
\end{equation}
Note that\cite{chow_2000}
\begin{equation}
\begin{split}
 \nabla  u =
\left(\frac{{\bf e}_u}{h_u}  \frac{\partial  }{\partial u}    +
\frac{{\bf e}_v}{h_v}  \frac{\partial  }{\partial v}      +
\frac{{\bf e}_w}{h_w} \frac{\partial  }{\partial w}\right) u =
 \frac{{\bf e}_u}{h_u} \text{, }\quad
 \nabla  v =  \frac{{\bf e}_v}{h_v}  ,\quad
 \nabla  w =  \frac{{\bf e}_w}{h_w}  \text{,}  \\
 \text{or } h_u \nabla  u =  {\bf e}_u,\quad
h_v \nabla  v =  {\bf e}_v ,\quad
h_w \nabla  w =  {\bf e}_w
.
\end{split}
\label{2018-mm-ch-veouvw}
\end{equation}
Because
${\bf e}_u$, ${\bf e}_v$, and ${\bf e}_w$ are unit vectors, taking the norms (lengths)
of~(\ref{2018-mm-ch-veouvw}) yields
\begin{equation}
\begin{split}
\frac{1}{h_u}  = | \nabla  u | ,\quad
\frac{1}{h_v}  = | \nabla  v | ,\quad
\frac{1}{h_w}  = | \nabla  w |
.
\end{split}
\label{2018-mm-ch-veouvwtn}
\end{equation}
Using~(\ref{2018-mm-ch-veouvw}) we obtain for
(both left-- and right--handed)
orthogonal curvilinear coordinates
\begin{equation}
\begin{split}
 h_v h_w ( \nabla  v \times \nabla  w ) =
{\bf e}_v \times {\bf e}_w  = {\bf e}_u ,\\
 h_u h_w ( \nabla  u \times \nabla  w ) =
{\bf e}_u \times {\bf e}_w  =
- {\bf e}_w \times {\bf e}_u  =
- {\bf e}_v ,                                 \\
 h_u h_v ( \nabla  u \times \nabla  v ) =
{\bf e}_u \times {\bf e}_v  =
{\bf e}_w
.
\end{split}
\label{2018-mm-ch-veouvwtnex}
\end{equation}

It is important to keep in mind that,
for both left-- and right--handed orthonormal bases
${\frak B}'=\{
{\bf e}_u,
{\bf e}_v,
{\bf e}_w
\}$, the following  relations for the cross products hold:
\begin{equation}
\begin{split}
{\bf e}_u \times {\bf e}_v  =
- {\bf e}_v \times {\bf e}_u
= {\bf e}_w
,\\
{\bf e}_u \times {\bf e}_w   =
- {\bf e}_w \times {\bf e}_u
= - {\bf e}_v
, \\
{\bf e}_v \times {\bf e}_w    =
- {\bf e}_w \times {\bf e}_v
= {\bf e}_u
.
\end{split}
\end{equation}

{
\color{blue}
\bexample
For the sake of examples, let us again consider polar, cylindrical and spherical coordinates.
\begin{itemize}
\item[(i)]
For polar coordinates recall that $h_r=1$, $h_\theta=r$
and $
{\bf e}_r =
\begin{pmatrix}
\cos \theta , \sin \theta
\end{pmatrix}^\intercal
$ as well as $
{\bf e}_\theta
=
\begin{pmatrix}
- \sin \theta , \cos \theta
\end{pmatrix}^\intercal  $.
Therefore,
\begin{equation}
\nabla =
\frac{ {\bf e}_r }{ h_r } \frac{\partial}{\partial r}
+
\frac{ {\bf e}_\theta }{ h_\theta } \frac{\partial}{\partial \theta}
=
\begin{pmatrix}
\cos \theta \\ \sin \theta
\end{pmatrix}
\frac{\partial}{\partial r}
+
\frac{ 1 }{ r }
\begin{pmatrix}
- \sin \theta \\ \cos \theta
\end{pmatrix}
\frac{\partial}{\partial \theta}
.
\end{equation}

\item[(ii)]
For cylindrical coordinates,
$h_r=1$, $h_\theta=r$, $h_z=1$, and
$
{\bf e}_r =
\begin{pmatrix}
\cos \theta , \sin \theta ,0
\end{pmatrix}^\intercal
$, $
{\bf e}_\theta
=
\begin{pmatrix}
- \sin \theta , \cos \theta ,0
\end{pmatrix}$ as well as $
{\bf e}_z
=
\begin{pmatrix}
0,0,1
\end{pmatrix}^\intercal  $.
Therefore,
\begin{equation}
\nabla =
\begin{pmatrix}
\cos \theta \\ \sin \theta \\0
\end{pmatrix}
\frac{\partial}{\partial r}
+
\frac{ 1 }{ r }
\begin{pmatrix}
- \sin \theta \\ \cos \theta  \\0
\end{pmatrix}
\frac{\partial}{\partial \theta}
+
\begin{pmatrix}
0 \\ 0  \\1
\end{pmatrix}
\frac{\partial}{\partial z}
.
\end{equation}

\item[(iii)]
For spherical coordinates,
$
h_r=1$,
$
h_\theta=r$,
$
h_\varphi=r\sin \theta$, and
$
{\bf e}_r =
\begin{pmatrix}
\sin \theta \cos \varphi , \sin \theta \sin \varphi,\cos \theta
\end{pmatrix}^\intercal
$, $
{\bf e}_\theta
=
\begin{pmatrix}
\cos \theta \cos \varphi , \cos \theta \sin \varphi,-\sin \theta
\end{pmatrix}$ as well as $
{\bf e}_\varphi
=
\begin{pmatrix}
-\sin \varphi , \cos \varphi ,0
\end{pmatrix}^\intercal  $.
Therefore,
\begin{equation}
\nabla =
\begin{pmatrix}
\sin \theta \cos \varphi \\ \sin \theta \sin \varphi\\ \cos \theta
\end{pmatrix}
\frac{\partial}{\partial r}
+
\frac{ 1 }{ r }
\begin{pmatrix}
\cos \theta \cos \varphi \\ \cos \theta \sin \varphi\\ -\sin \theta
\end{pmatrix}
\frac{\partial}{\partial \theta}
+
\frac{ 1 }{ r \sin \theta}
\begin{pmatrix}
-\sin \varphi \\ \cos \varphi \\0
\end{pmatrix}
\frac{\partial}{\partial \varphi }
.
\end{equation}
\end{itemize}
\eexample
}

\subsection{Divergence in three dimensional orthogonal curvilinear coordinates}

Equations
(\ref{2018-mm-ch-veouvw})
 and
(\ref{2018-mm-ch-veouvwtnex})
are instrumental for a derivation of other vector differential operators.
The divergence $\text{div}\;{\bf a}(u,v,w) = \nabla \cdot {\bf a}(u,v,w)$  of a vector field
${\bf a}(u,v,w) =
a_1(u,v,w) {\bf e}_u +
a_2(u,v,w) {\bf e}_v +
a_3(u,v,w) {\bf e}_w$ can,
in orthogonal curvilinear coordinates,\marginnote{Note that,
because of the product rule for differentiation,
$
\nabla f {\bf a} = (\nabla f)\cdot {\bf a} + f \nabla \cdot {\bf a}
$.}
be written as
\begin{equation}
\begin{split}
\nabla \cdot {\bf a}
=
\nabla \cdot \left(
a_1  {\bf e}_u +
a_2  {\bf e}_v +
a_3  {\bf e}_w
\right)
\\ =
\nabla \cdot \Big[
a_1   h_v h_w ( \nabla  v \times \nabla  w ) -
a_2   h_u h_w ( \nabla  u \times \nabla  w ) +
a_3   h_u h_v ( \nabla  u \times \nabla  v )
\Big]
\\ =
\nabla \cdot \left(a_1   h_v h_w   \nabla  v \times \nabla  w \right) -
\nabla \cdot \left(a_2   h_u h_w   \nabla  u \times \nabla  w \right) +
\nabla \cdot \left(a_3   h_u h_v   \nabla  u \times \nabla  v \right)
\\ =
\left(\nabla a_1 h_v h_w\right) \cdot    \left(\nabla  v \times \nabla  w \right)  +
a_1 h_v h_w   \underbrace{\nabla \cdot  ( \nabla  v) \times \nabla  w ) }_{\begin{array}{c}
\epsilon_{ijk} \nabla_i [(\nabla_j v) (\nabla_k w)]\\=
\epsilon_{ijk} (\nabla_i \nabla_j v) (\nabla_k w)\\+
\epsilon_{ijk} (\nabla_j v) (\nabla_i \nabla_k w)=0
\end{array}
}
\qquad
\\
- \left(\nabla a_2 h_u h_w\right) \cdot    \left(\nabla  u \times \nabla  w \right)  + 0
+ \left(\nabla a_3 h_u h_v\right) \cdot    \left(\nabla  u \times \nabla  v \right)  + 0
\\ =
  \left(\nabla a_1 h_v h_w\right) \cdot    \frac{ {\bf e}_u}{h_v h_w}
+ \left(\nabla a_2 h_u h_w\right) \cdot    \frac{ {\bf e}_v}{h_u h_w}
+ \left(\nabla a_3 h_u h_v\right) \cdot    \frac{ {\bf e}_w}{h_u h_v}
\\ =
  \left[\left(
\frac{{\bf e}_u}{h_u}  \frac{\partial   }{\partial u}    +
\frac{{\bf e}_v}{h_v}  \frac{\partial   }{\partial v}      +
\frac{{\bf e}_w}{h_w} \frac{\partial   }{\partial w}
\right) a_1 h_v h_w\right] \cdot    \frac{ {\bf e}_u}{h_v h_w}
\qquad
 \\
+ \left[\left(
\frac{{\bf e}_u}{h_u}  \frac{\partial   }{\partial u}    +
\frac{{\bf e}_v}{h_v}  \frac{\partial   }{\partial v}      +
\frac{{\bf e}_w}{h_w} \frac{\partial   }{\partial w}
\right) a_2 h_u h_w\right] \cdot    \frac{ {\bf e}_v}{h_u h_w} \qquad \\
+ \left[\left(
\frac{{\bf e}_u}{h_u}  \frac{\partial   }{\partial u}    +
\frac{{\bf e}_v}{h_v}  \frac{\partial   }{\partial v}      +
\frac{{\bf e}_w}{h_w} \frac{\partial   }{\partial w}
\right) a_3 h_u h_v\right] \cdot    \frac{ {\bf e}_w}{h_u h_v} \qquad \\
= \frac{ 1 }{h_u h_v h_w}
\left(
\frac{\partial   }{\partial u} a_1 h_v h_w+
\frac{\partial   }{\partial v} h_u a_2 h_w+
\frac{\partial   }{\partial w} h_u h_v a_3
\right) \\
= \frac{ 1 }{\sqrt{g_{uu} g_{vv} g_{ww}}}
\left(
\frac{\partial   }{\partial u} a_1 \sqrt{g_{vv} g_{ww}}+
\frac{\partial   }{\partial v} a_2 \sqrt{g_{uu} g_{ww}}+
\frac{\partial   }{\partial w} a_3 \sqrt{g_{uu} g_{vv}}
\right),
\end{split}
\label{2018-mm-ch-divergenceoclc}
\end{equation}
where, in the final phase of the proof, the formula~(\ref{2018-mm-ch-veo}) for the gradient,
as well as the mutual ortogonality of the unit basis vectors
${\bf e}_u$,
${\bf e}_v$, and
${\bf e}_w$
have been used.



{
\color{blue}
\bexample
Take, for example, spherical coordinates with
$
h_r=1$,
$
h_\theta=r$, and
$
h_\varphi=r\sin \theta$.
Equation~(\ref{2018-mm-ch-divergenceoclc}) yields
\begin{equation}
\text{div}\;{\bf a} = \nabla \cdot {\bf a} =
\frac{1}{r^2}
\frac{\partial}{\partial r}
\left( r^2 a_1\right)
+
\frac{1}{r \sin \theta}
\frac{\partial}{\partial \theta}
\left( \sin \theta a_2\right)
+
\frac{1}{r \sin \theta}
\frac{\partial}{\partial \varphi}
 a_3
.
\end{equation}
\eexample
}

\subsection{Curl in three dimensional orthogonal curvilinear coordinates}
\index{curl}

Using~(\ref{2018-mm-ch-veouvw}) and~(\ref{2018-mm-ch-veo})
the curl differential operator
$\text{curl} \; {\bf a}(u,v,w) = \nabla \times {\bf a}(u,v,w)$  of a vector field
${\bf a}(u,v,w) =
a_1(u,v,w) {\bf e}_u +
a_2(u,v,w) {\bf e}_v +
a_3(u,v,w) {\bf e}_w$ can,
in  (both left-- and right--handed)
orthogonal curvilinear coordinates,
be written as

\begin{equation}
\begin{split}
\nabla \times {\bf a}
=
\nabla \times \left(
a_1  {\bf e}_u +
a_2  {\bf e}_v +
a_3  {\bf e}_w
\right)
\\ =
\nabla \times \left(
a_1   h_u  \nabla  u +
a_2   h_v  \nabla  v +
a_3   h_w  \nabla  w
\right)
\\ =
\nabla \times a_1   h_u  \nabla  u +
\nabla \times a_2   h_v  \nabla  v +
\nabla \times a_3   h_w  \nabla  w
\\ =
\left(\nabla a_1 h_u \right) \times \nabla  u  +
a_1 h_u    \underbrace{\nabla \times  \nabla  u  }_{=0}
\\
+\left(\nabla a_2 h_v \right) \times \nabla  v  +  0
+\left(\nabla a_3 h_w \right) \times \nabla  w  +  0
\qquad
\\ =
\left[\left(
\frac{{\bf e}_u}{h_u}  \frac{\partial   }{\partial u}    +
\frac{{\bf e}_v}{h_v}  \frac{\partial   }{\partial v}      +
\frac{{\bf e}_w}{h_w} \frac{\partial   }{\partial w}
\right) a_1 h_u \right] \times \frac{{\bf e}_u}{h_u}  \qquad \\
+ \left[\left(
\frac{{\bf e}_u}{h_u}  \frac{\partial   }{\partial u}    +
\frac{{\bf e}_v}{h_v}  \frac{\partial   }{\partial v}      +
\frac{{\bf e}_w}{h_w} \frac{\partial   }{\partial w}
\right) a_2 h_v \right] \times \frac{{\bf e}_v}{h_v}   \qquad \\
+\left[\left(
\frac{{\bf e}_u}{h_u}  \frac{\partial   }{\partial u}    +
\frac{{\bf e}_v}{h_v}  \frac{\partial   }{\partial v}      +
\frac{{\bf e}_w}{h_w} \frac{\partial   }{\partial w}
\right) a_3 h_w \right] \times \frac{{\bf e}_w}{h_w}   \qquad
\\ =
\frac{{\bf e}_v}{h_uh_w} \frac{\partial   }{\partial w}
\left( a_1 h_u \right)
-
\frac{{\bf e}_w}{h_uh_v} \frac{\partial   }{\partial v}
\left( a_1 h_u \right)
\\
-\frac{{\bf e}_u}{h_vh_w} \frac{\partial   }{\partial w}
\left( a_2 h_v \right)
+
\frac{{\bf e}_w}{h_uh_v} \frac{\partial   }{\partial u}
\left( a_2 h_v \right)
\qquad \\
+ \frac{{\bf e}_u}{h_vh_w} \frac{\partial   }{\partial v}
\left( a_3 h_w \right)
-
\frac{{\bf e}_v}{h_uh_w} \frac{\partial   }{\partial u}
\left( a_3 h_w \right)
 \qquad
\\ =
\frac{{\bf e}_u}{h_vh_w}
\left[
\frac{\partial}{\partial v} (a_3 h_w) -
\frac{\partial}{\partial w} (a_2 h_v)
\right]
\\ +
\frac{{\bf e}_v}{h_uh_w}
\left[
\frac{\partial}{\partial w} (a_1 h_u) -
\frac{\partial}{\partial u} (a_3 h_w)
\right]
\qquad \\ +
\frac{{\bf e}_w}{h_uh_v}
\left[
\frac{\partial}{\partial u} (a_2 h_v) -
\frac{\partial}{\partial v} (a_1 h_u)
\right]
\qquad
\\
=
\frac{1}{h_u h_v h_w}
\text{det}
\begin{pmatrix}
h_u {\bf e}_u &
h_v {\bf e}_v &
h_w {\bf e}_w \\
\frac{\partial }{\partial u} &
\frac{\partial }{\partial v} &
\frac{\partial }{\partial w} \\
a_1 h_u &
a_2 h_v &
a_3 h_w
\end{pmatrix}
\\
=
\frac{1}{\sqrt{g_{uu} g_{vv} g_{ww}}}
\text{det}
\begin{pmatrix}
\sqrt{ g_{uu} } {\bf e}_u &
\sqrt{ g_{vv} } {\bf e}_v &
\sqrt{ g_{ww} } {\bf e}_w \\
\frac{\partial }{\partial u} &
\frac{\partial }{\partial v} &
\frac{\partial }{\partial w} \\
a_1 \sqrt{ g_{uu} } &
a_2 \sqrt{ g_{vv} } &
a_3 \sqrt{ g_{ww} }
\end{pmatrix}
.
\end{split}
\label{2018-mm-ch-curloclc}
\end{equation}


{
\color{blue}
\bexample
Take, for example, spherical coordinates with
$
h_r=1$,
$
h_\theta=r$,
$
h_\varphi=r\sin \theta$, and
$
{\bf e}_r =
\begin{pmatrix}
\sin \theta \cos \varphi , \sin \theta \sin \varphi,\cos \theta
\end{pmatrix}^\intercal
$, $
{\bf e}_\theta
=
\begin{pmatrix}
\cos \theta \cos \varphi , \cos \theta \sin \varphi,-\sin \theta
\end{pmatrix}$ as well as $
{\bf e}_\varphi
=
\begin{pmatrix}
-\sin \varphi , \cos \varphi ,0
\end{pmatrix}^\intercal  $.
Equation~(\ref{2018-mm-ch-curloclc}) yields
\begin{equation}
\begin{split}
\text{rot}\;{\bf a} = \nabla \times {\bf a} =
\frac{1}{r \sin \theta}
\begin{pmatrix}
\sin \theta \cos \varphi \\ \sin \theta \sin \varphi\\ \cos \theta
\end{pmatrix}
\left(
\frac{\partial}{\partial \theta} (a_3 \sin \theta )
-
\frac{\partial}{\partial \varphi} a_2
\right)
\\
+
\frac{1}{r}
\begin{pmatrix}
\cos \theta \cos \varphi \\ \cos \theta \sin \varphi\\ -\sin \theta
\end{pmatrix}
\left(
\frac{1}{\sin \theta}
\frac{\partial}{\partial \varphi} a_1
-
\frac{\partial}{\partial r}  (r a_3)
\right)
+
\frac{1}{r}
\begin{pmatrix}
-\sin \varphi \\ \cos \varphi \\ 0
\end{pmatrix}
\left(
\frac{\partial}{\partial r} (r a_2)
-
\frac{\partial}{\partial \theta } a_1
\right)
.
\end{split}
\end{equation}
\eexample
}

\subsection{Laplacian in three dimensional orthogonal curvilinear coordinates}
\index{Laplacian}
\index{Laplace operator}

Using~(\ref{2018-mm-ch-veo}) and~(\ref{2018-mm-ch-divergenceoclc})
the second order Laplacian differential operator
$\Delta a(u,v,w) = \nabla \cdot [\nabla a(u,v,w)]$  of a  field
$a(u,v,w)$ can,
in   orthogonal curvilinear coordinates,
be written as
\begin{equation}
\begin{split}
\Delta a(u,v,w) = \nabla \cdot [\nabla a(u,v,w)]=
\nabla \cdot
\left(
\frac{{\bf e}_u}{h_u}  \frac{\partial   }{\partial u}    +
\frac{{\bf e}_v}{h_v}  \frac{\partial   }{\partial v}      +
\frac{{\bf e}_w}{h_w} \frac{\partial   }{\partial w}
\right) a \\
=
\frac{ 1 }{h_u h_v h_w}
\left[
\frac{\partial   }{\partial u} \frac{h_v h_w}{h_u}\frac{\partial   }{\partial u}  +
\frac{\partial   }{\partial v} \frac{h_u h_w}{h_v}\frac{\partial   }{\partial v}  +
\frac{\partial   }{\partial w} \frac{h_u h_v}{h_w}\frac{\partial   }{\partial w}
\right] a
,
\end{split}
\label{2018-mm-ch-laplaceocc}
\end{equation}
so that the Lapace operator
in   orthogonal curvilinear coordinates can be identified with
\begin{equation}
\begin{split}
\Delta
=
\frac{ 1 }{h_u h_v h_w}
\left[
\frac{\partial   }{\partial u} \frac{h_v h_w}{h_u}\frac{\partial   }{\partial u}  +
\frac{\partial   }{\partial v} \frac{h_u h_w}{h_v}\frac{\partial   }{\partial v}  +
\frac{\partial   }{\partial w} \frac{h_u h_v}{h_w}\frac{\partial   }{\partial w}
\right]
\\
=
\frac{ 1 }{\sqrt{ g_{uu}  g_{vv} g_{ww} }}
\left[
\frac{\partial   }{\partial u} \sqrt{  \frac{g_{vv} g_{ww}  }{  g_{uu}  }}\frac{\partial   }{\partial u}  +
\right.  \qquad \qquad  \\  + \left.
\frac{\partial   }{\partial v} \sqrt{ \frac{g_{uu}   g_{ww}  }{  g_{vv} }}\frac{\partial   }{\partial v}  +
\frac{\partial   }{\partial w} \sqrt{ \frac{g_{uu}  g_{vv}   }{   g_{ww} }}\frac{\partial   }{\partial w}
\right] \\
=
\frac{ 1 }{\sqrt{ g_{uu}  g_{vv} g_{ww} }}
\sum_{t=u,v,w}
\frac{\partial   }{\partial t}  \frac{\sqrt{ g_{uu}  g_{vv} g_{ww} }}{g_{tt}}\frac{\partial   }{\partial t}
.
\end{split}
\label{2018-mm-ch-laplaceoc}
\end{equation}

{
\color{blue}
\bexample
For the sake of examples, let us again consider cylindrical and spherical coordinates.

\begin{itemize}
\item[(i)]
The Laplace operator in cylindrical coordinates
\index{cylindrical coordinates}
can be computed by insertion of~(\ref{2018-mm-ch-exhhcyc})
$h_u=h_r= 1$, $h_v=h_\theta= r$, and $h_w=h_z= 1$:
\begin{equation}
\Delta
=
\frac{ 1 }{r}
\frac{\partial   }{\partial r} r  \sin \theta\frac{\partial   }{\partial r}  +
\frac{ 1 }{r^2} \frac{\partial^2   }{\partial \theta^2}  +
\frac{\partial^2   }{\partial \varphi^2}
.
\label{2018-mm-ch-laplaceocylin}
\end{equation}

\item[(ii)]
The Laplace operator in spherical coordinates
\index{spherical coordinates}
can be computed by insertion of~(\ref{2018-mm-ch-exhhspher})
$h_u=h_r= 1$, $h_v=h_\theta= r$, and $h_w=h_\varphi= r \sin \theta$:
\begin{equation}
\Delta
=
\frac{1}{r^2} \left[ \frac{\partial}{\partial r}\left( r^2\frac{\partial}{\partial r}\right)
+
\frac{1}{\sin \theta}   \frac{\partial}{\partial \theta }
\sin \theta \frac{\partial}{\partial \theta }
+
\frac{1}{\sin^2 \theta} \frac{\partial^2}{\partial \varphi^2 }
\right]
.
\label{2018-mm-ch-laplaceocsc}
\end{equation}
\end{itemize}

\eexample
}


\section{Index trickery and examples}

The biggest ``trick'' or advantage in using indexed entities is the consequence
that, instead of ``bulk'' entities ``packaged'' in ``lumps'' we are actually {\em dealing with scalars}.
That means that we can exploit the usual laws associated with operations among scalars, such as addition or multiplication.
In particular, if no differential operators acting on fields are involved we can
commute indexed terms, or use associativity and distributivity.

We have already mentioned {\em Einstein's summation convention}
\index{Einstein summation convention}
requiring that, when an index variable appears twice in a single term, one has to
sum over all of the possible index values. For instance, $a_{ij}b_{jk}$ stands for $\sum_j a_{ij}b_{jk}$.

There are other tricks which are commonly used.
Here, some of them are enumerated:

\begin{itemize}
\item[(i)]
Indices which appear as internal sums can be renamed arbitrarily
(provided their name is not already taken by some other index).
That is, $a_ib^i=a_jb^j$ for arbitrary $a,b,i,j$.
\item[(ii)]
With the Euclidean metric, $\delta_{ii}=n$.
\item[(iii)]
$\frac{\partial x^i }{\partial x^j}=\delta^i_j=\delta^{ij}$ and
$\frac{\partial x_i }{\partial x_j}=\delta_i^j=\delta^{ij}$.
\item[(iv)]
With the Euclidean metric, $\frac{\partial x^i }{ \partial x^i}=n$.
\item[(v)]
$\varepsilon_{ij}\delta_{ij}=-\varepsilon_{ji}\delta_{ij}=-\varepsilon_{ji}\delta_{ji}=(i \leftrightarrow j)=-\varepsilon_{ij}\delta_{ij}=0$,
since $a=-a$ implies $a=0$;
likewise, $\varepsilon_{ij}x_i x_j=0$.
In general, the Einstein summations $s_{ij\ldots }a_{ij\ldots}$ over objects $s_{ij\ldots }$ which are {\em symmetric} with respect to index exchanges
over objects $a_{ij\ldots}$ which are {\em antisymmetric}  with respect to index exchanges yields zero.
\item[(vi)]
For threedimensional vector spaces ($n=3$)  and the Euclidean metric,
the {\em Grassmann identity} holds:
\index{Grassmann identity}
\begin{equation}
 \varepsilon_{ijk}\varepsilon_{klm}
=  \delta_{il}\delta_{jm}-\delta_{im}\delta_{jl}.
\label{2011-m-egi}
\end{equation}
{\color{OliveGreen}
\bproof
For the sake of a proof, consider
\begin{equation}
\begin{split}
{\bf x} \times ({\bf y} \times {\bf z}) \equiv \\
\textrm{in index notation}\\
x_j  \varepsilon_{ijk} y_l z_m\varepsilon_{klm}   =
x_j y_l z_m  \varepsilon_{ijk} \varepsilon_{klm}   \equiv \\
\textrm{in coordinate notation}\\
\begin{pmatrix}
x_1 \\ x_2\\ x_3
\end{pmatrix}
\times
\left[
\begin{pmatrix}
y_1 \\y_2\\y_3
\end{pmatrix}
\times
\begin{pmatrix}
z_1 \\z_2\\z_3
\end{pmatrix}
\right] =
\begin{pmatrix}
x_1 \\ x_2\\ x_3
\end{pmatrix}
\times
\begin{pmatrix}
y_2z_3 -y_3z_2 \\
y_3z_1 -y_1z_3 \\
y_1z_2 -y_2z_1 \\
\end{pmatrix}
\\ =
\begin{pmatrix}
x_2 (y_1z_2 -y_2z_1) - x_3(y_3z_1 -y_1z_3) \\
x_3 (y_2z_3 -y_3z_2) - x_1(y_1z_2 -y_2z_1) \\
x_1 (y_3z_1 -y_1z_3) - x_2(y_2z_3 -y_3z_2)
\end{pmatrix}
\\ =
\begin{pmatrix}
x_2 y_1z_2 -x_2y_2z_1 - x_3y_3z_1 +x_3y_1z_3 \\
x_3 y_2z_3 -x_3y_3z_2 - x_1y_1z_2 +x_1y_2z_1 \\
x_1 y_3z_1 -x_1y_1z_3 - x_2y_2z_3 +x_2y_3z_2
\end{pmatrix}
\\  =
\begin{pmatrix}
y_1 (x_2z_2 + x_3z_3) - z_1(x_2y_2 + x_3y_3) \\
y_2 (x_3z_3 + x_1z_1) - z_2(x_1y_1 + x_3y_3) \\
y_3 (x_1z_1 + x_2z_2) - z_3(x_1y_1 + x_2y_2)
\end{pmatrix}
\end{split}
\end{equation}
The ``incomplete''  dot products can be completed through addition and subtraction of the same term, respectively; that is,
\begin{equation}
\begin{split}
\begin{pmatrix}
y_1(x_1z_1 + x_2z_2 + x_3z_3) - z_1(x_1y_1 + x_2y_2 + x_3y_3)   \\
y_2(x_1z_1 + x_2z_2 + x_3z_3) - z_2(x_1y_1 + x_2y_2 + x_3y_3)     \\
y_3(x_1z_1 + x_2z_2 + x_3z_3) - z_3(x_1y_1 + x_2y_2 + x_3y_3)       \\
\end{pmatrix}
\\
\equiv   \textrm{in vector notation}\\
{\bf y} \left( {\bf x} \cdot {\bf z}\right)
-
{\bf z} \left( {\bf x} \cdot {\bf y}\right)
\\
\equiv   \textrm{in index notation}\\
x_j y_l z_m \left(\delta_{il}\delta_{jm}-\delta_{im}\delta_{jl}\right).
\end{split}
\end{equation}
\eproof
}

\item[(vii)]
For threedimensional vector spaces ($n=3$) and the Euclidean metric the Grassmann identity~(\ref{2011-m-egi})
\index{Grassmann identity}  implies\\
\begin{equation}
\begin{split}
\| a\times b \| =
\sqrt{\varepsilon_{ijk}\varepsilon_{ist}a_j a_s b_k b_t} =
\sqrt{\| a\|^2
\| b\|^2
-(a\cdot b)^2}\\=
\sqrt{{\rm det}
\left(
\begin{array}{cc}
a\cdot a&a\cdot b\\
a\cdot b&b\cdot b
\end{array}\right)} =
\sqrt{\| a\|^2
\| b\|^2 \left (1- \cos^2 \angle_{ab} \right)}
 =
\| a\|
\| b\|
\sin \angle_{ab}.
\end{split}
\end{equation}
\item[(viii)]
Let $u,v\equiv x_1',x_2'$ be two parameters associated with an
orthonormal Cartesian basis $\{(0,1),(1,0)\}$, and let
$\Phi :(u,v)\mapsto \mathbb{R}^3$
be a mapping from some area of $\mathbb{R}^2$ into a twodimensional
surface of $\mathbb{R}^3$. Then the metric tensor is given by
$g_{ij}=
{\partial \Phi^k \over \partial  y ^i}
{\partial \Phi^m \over \partial  y^j} \delta_{km}.$

\end{itemize}

{
\color{blue}
\bexample

Consider the following examples in three-dimensional vector space.
Let $r^2 = \sum_{i=1 }^3 x_i^2$.

\begin{enumerate}
\item
\begin{equation}
\begin{split}
  \partial_jr =  \partial_j \sqrt{\sum_ix_i^2} =
  \frac{1}{2}\frac{1}{\sqrt{\sum_ix_i^2}}\,2x_j =
  \frac{x_j}{r}
\end{split}
\end{equation}
By using the chain  rule one obtains
\begin{equation}
\begin{split}
  \partial_jr^\alpha =
  \alpha r^{\alpha-1}\left(\partial_jr\right) =
  \alpha r^{\alpha-1}\left(\frac{x_j}{r}\right)=
  \alpha r^{\alpha-2}x_j
\end{split}
\label{2011-m-eet1}
\end{equation}
and thus $\nabla r^\alpha = \alpha r^{\alpha-2}{\bf x}$.

\item
\begin{equation}
\begin{split}
  \partial_j \log r=\frac{1}{r}\left(\partial_jr\right)
\end{split}
\end{equation}
With $
  \partial_jr = \frac{x_j}{r}
$  derived earlier in Equation~(\ref{2011-m-eet1}) one obtains
$
 \partial_j \log r= \frac{1}{r}\frac{x_j}{r}=
  \frac{x_j}{r^2}
$,
and thus $\nabla  \log r =\frac{{\bf x}}{r^2}$.

\item
\begin{equation}
\begin{split}
  \partial_j
  \left[
    \left(
      \sum_i\left(x_i-a_i\right)^2
    \right)^{-\frac{1}{2}}+
    \left(
      \sum_i\left(x_i+a_i\right)^2
    \right)^{-\frac{1}{2}}
  \right]=
\\
  = -\frac{1}{2}\left[\frac{1}{\left(\sum_i\left(x_i-a_i\right)^2\right)^
    \frac{3}{2}}\,2\left(x_j-a_j\right)
  +\frac{1}{\left(\sum_i\left(x_i+a_i\right)^2\right)^\frac{3}{2}}\,
    2\left(x_j+a_j\right)\right]\\
 = -\left(\sum_i\left(x_i-a_i\right)^2\right)^{-\frac{3}{2}}\left(x_j-a_j\right)-
    \left(\sum_i\left(x_i+a_i\right)^2\right)^{-\frac{3}{2}}\left(x_j+a_j\right)   .
\end{split}
\end{equation}

\item
For three dimensions and for $r \neq 0$,
\begin{equation}
\nabla \bigl({{\bf r} \over r^3} \bigr)\equiv
  \partial_i\left(\frac{r_i}{r^3}\right)=
  \frac{1}{r^3}\underbrace{\partial_i r_i}_{=3}+
  r_i\left(-3\frac{1}{r^4}\right)\left(\frac{1}{2r}\right)2r_i=
  3\frac{1}{r^3}-3\frac{1}{r^3}=0 .
\label{2011-m-eet2}
\end{equation}

\item  With this solution (\ref{2011-m-eet2}) one obtains, for three dimensions and $r \neq 0$,
\begin{equation}
\Delta \bigl({1 \over r} \bigr)\equiv
  \partial_i\partial_i\frac{1}{r}=\partial_i\left(-\frac{1}{r^2}\right)
  \left(\frac{1}{2r}\right)2r_i=-\partial_i\frac{r_i}{r^3}=0   .
\end{equation}

\item  With the earlier solution (\ref{2011-m-eet2}) one obtains
\begin{equation}
\begin{split}
\Delta \bigl({{\bf r} {\bf p} \over r^3} \bigr)\equiv
  \partial_i\partial_i\frac{r_jp_j}{r^3}=
    \partial_i
    \left[
      \frac{p_i}{r^3}+r_jp_j\left(-3\frac{1}{r^5}\right)r_i
    \right] \\
  = p_i\left(-3\frac{1}{r^5}\right)r_i+
    p_i\left(-3\frac{1}{r^5}\right)r_i \\
   +r_jp_j
    \left[
      \left(15\frac{1}{r^6}\right)
      \left(\frac{1}{2r}\right)2r_i
    \right]r_i+
    r_jp_j\left(-3\frac{1}{r^5}\right)
    \underbrace{\partial_i r_i}_{=3} \\
  = r_i p_i \frac{1}{r^5}(-3-3+15-9)=0
\end{split}
\end{equation}

\item    With $r\neq 0$ and constant $\bf p$ one obtains
\marginnote[-10mm]{Note that, in three dimensions, the Grassmann identity~(\ref{2011-m-egi})
\index{Grassmann identity}
 $\varepsilon_{ijk}\varepsilon_{klm}       =          \delta_{il}\delta_{jm}-\delta_{im}\delta_{jl}$
holds.}
\begin{equation}
\begin{split}
  \nabla \times ({\bf p} \times \frac{{\bf r}}{r^3})
 \equiv
  \varepsilon_{ijk}\partial_j\varepsilon_{klm}p_l\frac{r_m}{r^3}=
  p_l \varepsilon_{ijk} \varepsilon_{klm}
  \left[\partial_j\frac{r_m}{r^3}\right]  \\
 = p_l
    \varepsilon_{ijk}\varepsilon_{klm}
  \left[
    \frac{1}{r^3}\partial_j r_m + r_m
    \left(-3\frac{1}{r^4}\right)\left(\frac{1}{2r}\right)2r_j
  \right]  \\
  = p_l\varepsilon_{ijk}\varepsilon_{klm}
  \left[
    \frac{1}{r^3}\delta_{jm} - 3\frac{r_j r_m}{r^5}
  \right]  \\
  = p_l(\delta_{il}\delta_{jm}-\delta_{im}\delta_{jl})
  \left[
    \frac{1}{r^3}\delta_{jm} - 3\frac{r_j r_m}{r^5}
  \right]  \\
  = p_i \underbrace{\left(3\frac{1}{r^3}-3\frac{1}{r^3}\right)}_{=0}-
  p_j
  \Biggl({1\over {r^3}}
    \underbrace{{\partial_j r_i}}_{=\delta_{ij}}-
    3\frac{r_j r_i}{r^5}
  \Biggr)  \\
  = -\frac{{\bf p}}{r^3}+3\frac{\left({\bf r} {\bf p}\right){\bf r}}{r^5}
.
\end{split}
\end{equation}

\item
\begin{equation}
\begin{split}
{\nabla} \times({\nabla }\Phi )\\
\equiv
\varepsilon_{ijk} \partial_j \partial_k \Phi \\ =
\varepsilon_{ikj} \partial_k \partial_j \Phi  \\=
\varepsilon_{ikj} \partial_j \partial_k \Phi  \\=
-\varepsilon_{ijk} \partial_j \partial_k \Phi =0.
\end{split}
\end{equation}
This is due to the fact that $\partial_j \partial_k$ is  symmetric, whereas
$\varepsilon_{ijk}$ is totally antisymmetric.

\item        For a proof of
$({{\bf x} } \times {{\bf y}})\times {{\bf z}}
\neq
{{\bf x} } \times ({{\bf y}}\times {{\bf z}})$ consider
\begin{equation}
\begin{split}
({{\bf x} } \times {{\bf y}})\times {{\bf z}}\\
\equiv
\underbrace{\varepsilon_{ijm}}_{\text{ second } \times}
\underbrace{\varepsilon_{jkl}}_{\text{ first } \times}
x_k y_l z_m\\=
-\varepsilon_{imj}
\varepsilon_{jkl}
x_k y_l z_m\\ =
-(\delta_{ik}\delta_{ml}-
\delta_{im}\delta_{lk})
x_k y_l z_m\\ =
-x_i {{\bf y}}\cdot {{\bf z}}+
y_i {{\bf x}}\cdot {{\bf z}}.
\end{split}
\end{equation}
{\it versus}
\begin{equation}
\begin{split}
{{\bf x} } \times ({{\bf y}}\times {{\bf z}})\\
\equiv
\underbrace{\varepsilon_{ikj}}_{\text{ first } \times}
\underbrace{\varepsilon_{jlm}}_{\text{ second } \times}
x_k y_l z_m\\ =
(\delta_{il}\delta_{km}-
\delta_{im}\delta_{kl})
x_k y_l z_m\\ =
y_i {{\bf x}}\cdot {{\bf z}}-
z_i {{\bf x}}\cdot {{\bf y}}.
\end{split}
\end{equation}

\item
Let ${\bf w} = { {{\bf p}} \over r} $ with  $p_i=p_i\left(t  - {r \over c}\right)$,
whereby   $t$ and $c$  are constants. Then,
\begin{eqnarray*}
  \mbox{div}{\bf w}& = &
\nabla \cdot {\bf w}\\
\equiv \partial_i w_i & = & \partial_i
  \left[
    \frac{1}{r} p_i \left(t-\frac{r}{c}\right)
  \right]= \\
  & = & \left(-\frac{1}{r^2}\right)\left(\frac{1}{2r}\right)
    2r_i p_i+
    \frac{1}{r}p_i'\left(-\frac{1}{c}\right)
    \left(\frac{1}{2r}\right)2 r_i  \\
  & = & -\frac{r_i p_i}{r^3}-\frac{1}{c r^2}p_i' r_i
.
\end{eqnarray*}
Hence,
$
  \mbox{div}{\bf w}=\nabla \cdot {\bf w}=
  -\left(\frac{{\bf r} {\bf p}}{r^3}+\frac{{\bf r} {\bf p}'}{c r^2}\right)
$.

\begin{eqnarray*}
\mbox{rot}{\bf w}& =& \nabla \times {\bf w}  \\
  \varepsilon_{ijk}\partial_j w_k & = &
     \varepsilon_{ijk}
    \left[
      \left(-\frac{1}{r^2}\right)\left(\frac{1}{2r}\right)
      2 r_j p_k +
      \frac{1}{r}p_k'
      \left(-\frac{1}{c}\right)\left(\frac{1}{2r}\right)2r_j
    \right] \\
  & = & -\frac{1}{r^3}\varepsilon_{ijk}r_j p_k -\frac{1}{cr^2}
    \varepsilon_{ijk}r_j p_k' = \\
  & \equiv & -\frac{1}{r^3}\left({\bf r} \times {\bf p}\right)-
    \frac{1}{cr^2}\left({\bf r} \times {\bf p}'\right)   .
\end{eqnarray*}

\item
Let us verify  some specific examples of Gauss' (divergence) theorem,
\index{Gauss' theorem}
stating that the outward flux of a vector field through a closed surface
is equal to the volume integral of the divergence of the region inside the surface.
That is, the sum of all sources subtracted by the sum of all sinks represents the net flow out of a region or volume of threedimensional space:
\begin{equation}
\int \limits_V \nabla \cdot {\bf w} \, dv   =\int \limits_{F_V} {\bf w} \cdot d{\bf f}
.   \label{2011-m-gauss}
\end{equation}

Consider the vector field ${\bf w} = \begin{pmatrix} 4x , -2y^2 , z^2\end{pmatrix}^\intercal $
and the (cylindric) volume bounded by the planes  $z=0$ und $z=3$,
as well as by the surface
$x^2 + y^2 = 4$.

Let us first look at the left hand side $\int \limits_V \nabla \cdot {\bf w} \, dv $
of Equation~(\ref{2011-m-gauss}):
$$
  \nabla {\bf w}=\textrm{div } {\bf w}=4-4y+2z
$$
\begin{eqnarray*}
  \Longrightarrow \int \limits_V \! \textrm{div } {\bf w} dv& = &
  \int \limits_{z=0}^3 \! dz \int \limits_{x=-2}^2 \!\! dx
  \int \limits_{y=-\sqrt{4-x^2}}^{\sqrt{4-x^2}} \!\!\!dy\,
    \left(4-4y+2z\right) \\
  & & \mbox{cylindric coordinates: }
    \begin{pmatrix}
      x   =   r \cos \varphi ,
      y   =   r \sin \varphi ,
      z   =   z
    \end{pmatrix}^\intercal
   \\
  & = & \int \limits_{z=0}^3 \!\! dz \int \limits_0^2 r\, dr
  \int \limits_0^{2\pi} \!\! d\varphi \left(4-4r \sin \varphi+2z\right)  \\
  & = & \int \limits_{z=0}^3 \!\! dz \int \limits_0^2 r \, dr
  \left(4 \varphi +4r\cos\varphi+2\varphi z\right)
  \Biggl|_{\varphi=0}^{2 \pi}  \\
  & = & \int \limits_{z=0}^3 \!\! dz \int \limits_0^2 r\, dr
  \left(8 \pi +4r+4 \pi z -4r\right)  \\
  & = & \int \limits_{z=0}^3 \!\! dz \int \limits_0^2 r\, dr
  \left(8 \pi +4 \pi z\right) \\
  & = & 2 \left( 8 \pi z +4 \pi \frac{z^2}{2}\right)\Biggl|_{z=0}^{z=3}=
    2 (24+18) \pi = 84 \pi
\end{eqnarray*}

Now consider the right hand side $\int \limits_F {\bf w} \cdot d{\bf f}$
of Equation~(\ref{2011-m-gauss}).
The surface consists of three  parts:
the lower plane $F_1$ of the cylinder is characterized by $z=0$;
the upper plane $F_2$  of the cylinder is characterized by  $z=3$;
the surface on the side of the cylinder $F_3$
 is characterized by   $x^2+y^2=4$.
$d {\bf f}$ must be normal to these surfaces, pointing outwards; hence
(since the area of a circle of radius $r=2$ is $\pi r^2 = 4\pi$),
 \begin{eqnarray*}
  F_1 : \int \limits_{{\cal F}_1} {\bf w} \cdot d{\bf f}_1  & = &
    \int \limits_{{\cal F}_1}
    \left(
      \begin{array}{c}
        4x \\
        -2y^2 \\
        z^2=0
      \end{array}
    \right)
    \left(
      \begin{array}{c}
        0 \\
        0 \\
        -1
      \end{array}
    \right)
    \, d\, x d\, y = 0 \\
  F_2 : \int \limits_{{\cal F}_2} {\bf w} \cdot d{\bf f}_2 & = &
    \int \limits_{{\cal F}_2}
    \left(
      \begin{array}{c}
        4x \\
        -2y^2 \\
        z^2=9
      \end{array}
    \right)
    \left(
      \begin{array}{c}
        0 \\
        0 \\
        1
      \end{array}
    \right)
    \, d\, x d\, y   \\
  & = & 9 \int \limits_{K_{r=2}} \!\! d\, f=9 \cdot 4 \pi=36 \pi \\
  F_3 : \int \limits_{{\cal F}_3} {\bf w} \cdot d{\bf f}_3 & = &
    \int \limits_{{\cal F}_3}
    \left(
      \begin{array}{c}
        4x \\
        -2y^2 \\
        z^2
      \end{array}
    \right)
    \left(
      \frac{\partial {\bf x}}{\partial \varphi} \times
      \frac{\partial {\bf x}}{\partial z}
    \right)
    \, d\varphi \, dz \quad (r=2)
\end{eqnarray*}
$$
  \frac{\partial {\bf x}}{\partial \varphi} =
  \left(
    \begin{array}{c}
      -r \sin \varphi \\
       r \cos \varphi \\
       0
    \end{array}
  \right)=
  \left(
    \begin{array}{c}
      -2 \sin \varphi \\
       2 \cos \varphi \\
       0
    \end{array}
  \right); \enspace
  \frac{\partial {\bf x}}{\partial z} =
  \left(
    \begin{array}{c}
      0 \\
      0 \\
      1
    \end{array}
  \right) ,
$$
and therefore
$$
    \frac{\partial {\bf x}}{\partial \varphi} \times
    \frac{\partial {\bf x}}{\partial z}
 =
  \left(
    \begin{array}{c}
      2 \cos \varphi \\
      2 \sin \varphi \\
      0
    \end{array}
  \right),
$$
and
\begin{eqnarray*}
  F_3 & = & \int \limits_{\varphi =0}^{2 \pi} \!\! d\varphi
    \int \limits_{z=0}^3 \!\! dz
    \left(
      \begin{array}{c}
        4 \cdot 2 \cos \varphi \\
        -2(2 \sin \varphi)^2 \\
        z^2
      \end{array}
    \right)
    \left(
      \begin{array}{c}
        2 \cos \varphi \\
        2 \sin \varphi \\
        0
      \end{array}
    \right)  \\
    &= & \int \limits_{\varphi =0}^{2 \pi} \!\! d\varphi
    \int \limits_{z=0}^3 \!\! dz
    \left(16 \cos^2 \varphi -16 \sin^3 \varphi\right)  \\
  & = & 3 \cdot 16 \int \limits_{\varphi =0}^{2 \pi} \!\! d\varphi
    \left( \cos^2\varphi - \sin^3 \varphi \right)  \\
  & = &
    \Biggl[
      \begin{array}{rcl}
        \int \cos^2 \varphi \, d\varphi & = & \frac{\varphi}{2}+
          \frac{1}{4} \sin 2 \varphi \\
        \int \sin^3 \varphi \, d\varphi & = & -\cos \varphi+
          \frac{1}{3} \cos^3 \varphi
      \end{array}
    \Biggr]  \\
    & = & 3 \cdot 16
    \left\{
      \frac{2 \pi}{2}-
      \underbrace
        {\left[
          \left(1+\frac{1}{3}\right)-\left(1+\frac{1}{3}\right)
        \right]}
      _{=0}
    \right\}=48 \pi
\end{eqnarray*}
For the flux through the surfaces one thus obtains
$$ \oint \limits_F {\bf w} \cdot d{\bf f}=F_1+F_2+F_3=84 \pi .$$

\item
Let us verify  some specific examples of Stokes' theorem in three dimensions,
\index{Stokes' theorem}
stating that
\begin{equation}
\int \limits_{\cal F} \textrm{rot } {\bf b} \cdot d{\bf f}   =\oint \limits_{{\cal C}_{\cal F}} {\bf b} \cdot d{\bf s}
.   \label{2011-m-stokes}
\end{equation}

Consider the vector field ${\bf b} = \begin{pmatrix} yz , -xz , 0 \end{pmatrix}^\intercal $
and the volume bounded by spherical cap
formed by the plane at $z = a / \sqrt{2}$ of a sphere of radius $a$ centered around the origin.

Let us first look at the left hand side $\int \limits_{\cal F} \textrm{rot } {\bf b} \cdot d{\bf f} $
of Equation~(\ref{2011-m-stokes}):
$$
  {\bf b}=
  \left(
    \begin{array}{c}
      yz \\
      -xz \\
      0
    \end{array}
  \right)
  \Longrightarrow
  \textrm{rot } {\bf b} = \nabla \times {\bf b} =
  \left(
    \begin{array}{c}
      x \\
      y \\
      -2z
    \end{array}
  \right)
$$
Let us transform this into spherical coordinates:
$$
  {\bf x}=
  \left(
    \begin{array}{c}
      r\sin \theta \cos \varphi \\
      r\sin \theta \sin \varphi \\
      r\cos \theta
    \end{array}
  \right)
$$
$$
  \Rightarrow
  \frac{\partial {\bf x}}{\partial \theta}=
  r
  \left(
    \begin{array}{c}
      \cos \theta \cos \varphi \\
      \cos \theta \sin \varphi \\
      -\sin \theta
    \end{array}
  \right);\quad
  \frac{\partial {\bf x}}{\partial \varphi}=
  r
  \left(
    \begin{array}{c}
      -\sin \theta \sin \varphi \\
      \sin \theta \cos \varphi \\
      0
    \end{array}
  \right)
$$
$$
  d{\bf f}=
  \left(
    \frac{\partial {\bf x}}{\partial \theta} \times
    \frac{\partial {\bf x}}{\partial \varphi}
  \right)
  d\theta \, d\varphi=
  r^2
  \left(
    \begin{array}{c}
      \sin^2 \theta \cos \varphi \\
      \sin^2 \theta \sin \varphi \\
      \sin \theta \cos \theta
    \end{array}
  \right)
  d\theta \, d\varphi \label{eqn:1.14.1}
$$
$$
  \nabla \times {\bf b}=
  r
  \left(
    \begin{array}{c}
      \sin \theta \cos \varphi \\
      \sin \theta \sin \varphi \\
      -2\cos \theta
    \end{array}
  \right)
$$

\begin{eqnarray*}
  \int \limits_{\cal F} \textrm{rot } {\bf b} \cdot d{\bf f} & = &
  \int \limits_{\theta=0}^{\pi/4} \!\! d\theta
  \int \limits_{\varphi=0}^{2 \pi} \!\! d\varphi \, a^3
  \left(
    \begin{array}{c}
      \sin \theta \cos \varphi \\
      \sin \theta \sin \varphi \\
      -2\cos \theta
    \end{array}
  \right)
  \left(
    \begin{array}{c}
      \sin^2 \theta \cos \varphi \\
      \sin^2 \theta \sin \varphi \\
      \sin \theta \cos \theta
    \end{array}
  \right)  \\
  & = & a^3
  \int \limits_{\theta=0}^{\pi/4} \!\! d\theta
  \int \limits_{\varphi=0}^{2 \pi} \!\! d\varphi
  \Biggl[
    \sin^3\theta
    \underbrace{\left(\cos^2\varphi+\sin^2\varphi\right)}_{=1}-
      2\sin\theta\cos^2\theta
  \Biggr]   \\
  & = & 2 \pi a^3
  \left[
    \int \limits_{\theta=0}^{\pi/4} \!\! d\theta
    \left(1-\cos^2\theta\right)\sin\theta-
    2\int \limits_{\theta=0}^{\pi/4} \!\! d\theta
    \sin \theta \cos^2 \theta
  \right]   \\
  & = & 2 \pi a^3
  \int \limits_{\theta=0}^{\pi/4}d\theta
  \sin \theta \left(1-3\cos^2\theta\right)   \\
  &&   \left[
    \begin{array}{l}
      \mbox{transformation of variables: } \\
      \cos \theta = u \Rightarrow du=-\sin \theta d\theta
      \Rightarrow d\theta=-\frac{du}{\sin \theta}
    \end{array}
  \right] \\
  & = & 2 \pi a^3
  \int \limits_{\theta=0}^{\pi/4}(-du)\left(1-3u^2\right)=
    2 \pi a^3
    \left( \frac{3u^3}{3}-u \right)\Biggr|_{\theta=0}^{\pi/4}  \\
  & = & 2 \pi a^3
    \left(\cos^3\theta-\cos\theta \right)\Biggr|_{\theta=0}^{\pi/4}=
    2 \pi a^3
    \left(\frac{2\sqrt{2}}{8}-\frac{\sqrt{2}}{2} \right)   \\
  & = & \frac{2 \pi a^3}{8}\left(-2\sqrt{2}\right)=
    -\frac{\pi a^3 \sqrt{2}}{2}
\end{eqnarray*}

Now consider the right hand side $\oint \limits_{{\cal C}_{\cal F}} {\bf b} \cdot d{\bf s}$
of Equation~(\ref{2011-m-stokes}).
The radius $r'$ of the circle  surface
$\{(x,y,z) \mid x,y \in \mathbb{R} ,z=a/\sqrt{2}\}$ bounded by the sphere with radius $a$
is determined by
$ a^2 =(r')^2 +(a/ \sqrt{2})^2$; hence, $r' =a/\sqrt{2}$.
The curve of integration ${\cal C}_{\cal F}$ can be parameterized by
$$\{(x,y,z) \mid
x={a\over \sqrt{2}} \cos \varphi,
y={a\over \sqrt{2}} \sin \varphi,
z={a\over \sqrt{2}}\}.$$
Therefore,
$$
  {\bf x} = a
  \left(
    \begin{array}{c}
      \frac{1}{\sqrt2}\cos\varphi \\[1ex]
      \frac{1}{\sqrt2}\sin\varphi \\[1ex]
      \frac{1}{\sqrt2}
    \end{array}
  \right)=
  \frac{a}{\sqrt{2}}
  \left(
    \begin{array}{c}
      \cos\varphi \\
      \sin\varphi \\
      1
    \end{array}
  \right)
\in {\cal C}_{\cal F}
$$
Let us transform this into polar coordinates:
$$
  d{\bf s}=\frac{d{\bf x}}{d\varphi} \,d\varphi =
  \frac{a}{\sqrt{2}}
  \left(
    \begin{array}{c}
      -\sin\varphi \\
      \cos\varphi \\
      0
    \end{array}
  \right) d\varphi
$$
$$
  {\bf b}=
  \left(
    \begin{array}{c}
      \frac{a}{\sqrt{2}}\sin\varphi\cdot\frac{a}{\sqrt{2}} \\
      -\frac{a}{\sqrt{2}}\cos\varphi\cdot\frac{a}{\sqrt{2}} \\
      0
    \end{array}
  \right)=
  \frac{a^2}{2}
  \left(
    \begin{array}{c}
      \sin\varphi \\
      -\cos\varphi \\
      0
    \end{array}
  \right)
$$
Hence the circular integral is given by
$$
  \oint \limits_{{\cal C}_F} {\bf b} \cdot d{\bf s}=
  \frac{a^2}{2}\frac{a}{\sqrt{2}}
  \int \limits_{\varphi=0}^{2 \pi}
  \underbrace
    {\left(-\sin^2\varphi-\cos^2\varphi\right)}
  _{=-1}
  \, d\varphi =
  -\frac{a^3}{2\sqrt{2}}2 \pi=-\frac{a^3 \pi}{\sqrt{2}}
.
$$

\item
In machine learning, a linear regression {\it Ansatz}\cite{Goodfellow-et-al-2016-Book} is to find a linear model for the prediction of some unknown
\index{linear regression}
observable, given some anecdotal instances of its performance.
More formally, let
$y$ be an arbitrary real-valued observable which depends
on $n$ real-valued parameters $x_1, \ldots , x_n$  by linear means; that is, by
\begin{equation}
y = \sum_{i=1}^n x_i r_i = \langle  {\bf x}\vert {\bf r} \rangle,
\label{2016-ml-ansatz-lr}
\end{equation}
where $\langle {\bf x} \vert = (\vert {\bf x} \rangle )^\intercal $ is the transpose
of the vector $\vert {\bf x} \rangle$.
The tuple
\begin{equation}
\vert {\bf r} \rangle = \begin{pmatrix} r_1, \ldots , r_n \end{pmatrix}^\intercal
\label{2016-ml-ansatz-vectorweights}
\end{equation}
contains the unknown weights of the approximation --
the ``theory,'' if you like --
and $\langle  {\bf a}\vert {\bf b} \rangle = \sum_i a_ib_i$ stands for the Euclidean scalar product of the tuples interpreted
as (dual) vectors in $n$-dimensional (dual) vector space $\mathbb{R}^n$.

Given are $m$ known instances of (\ref{2016-ml-ansatz-lr}); that is, suppose $m$ real-valued pairs
$\begin{pmatrix}z_j, \vert {\bf x}_j \rangle \end{pmatrix}$ are known.
These data can be bundled into an $m$-tuple
\begin{equation}
\vert {\bf z} \rangle \equiv \begin{pmatrix} z_{j_1}, \ldots , z_{j_m} \end{pmatrix}^\intercal ,
\end{equation}
and an $(m \times n)$-matrix
\begin{equation}
\textsf{\textbf{X}} \equiv
\begin{pmatrix}
x_{{j_1}{i_1}} & \ldots & x_{{j_1}{i_n}}\\
\vdots & \vdots & \vdots \\
x_{{j_m}{i_1}} & \ldots & x_{{j_m}{i_n}}
\end{pmatrix}
\end{equation}
where $j_1,\ldots , j_m$ are arbitrary permutations of $1,\ldots ,m$,
and the matrix rows are just the vectors
$\vert {\bf x}_{j_k} \rangle \equiv \begin{pmatrix} x_{{j_k}{i_1}} & \ldots , x_{{j_k}{i_n}} \end{pmatrix}^\intercal $.

The task is to compute a ``good'' estimate of $\vert {\bf r} \rangle$;
that is, an estimate of $\vert {\bf r} \rangle$
which allows an ``optimal'' computation of the prediction $y$.

Suppose that a good way to measure the performance
of the prediction from some particular definite but unknown $\vert {\bf r} \rangle $
with respect to the $m$ given data
$\begin{pmatrix}z_j, \vert {\bf x}_j \rangle \end{pmatrix}$
is by the mean squared error (MSE)
\marginnote{Note that $\langle {\bf z} \vert  \textsf{\textbf{X}} \vert {\bf r} \rangle
=\langle {\bf z} \vert \left( \langle  {\bf r} \vert \textsf{\textbf{X}}^\intercal\right)^\intercal
=\left(\langle {\bf z} \vert \left( \langle  {\bf r} \vert \textsf{\textbf{X}}^\intercal\right)^\intercal \right)^\intercal
=\left[  \left( \langle  {\bf r} \vert \textsf{\textbf{X}}^\intercal \right)^\intercal  \right]^\intercal \vert {\bf z} \rangle   $.}
\begin{equation}
\begin{split}
\text{MSE}
=
\frac{1}{m}
\left\|
\vert {\bf y} \rangle - \vert {\bf z} \rangle
\right\|^2
=
\frac{1}{m}
\left\|
\textsf{\textbf{X}} \vert {\bf r} \rangle
 - \vert {\bf z}   \rangle
\right\|^2
\\
=
\frac{1}{m}
\left(
\textsf{\textbf{X}} \vert {\bf r} \rangle
 - \vert {\bf z}   \rangle
\right)^\intercal
\left(
\textsf{\textbf{X}} \vert {\bf r} \rangle
 - \vert {\bf z}   \rangle
\right)
\\
=
\frac{1}{m}
\left(
\langle {\bf r} \vert \textsf{\textbf{X}}^\intercal
- \langle {\bf z} \vert
\right)
\left(
\textsf{\textbf{X}} \vert {\bf r} \rangle
- \vert {\bf z}   \rangle
\right)
\\
=
\frac{1}{m} \left(
\langle {\bf r} \vert \textsf{\textbf{X}}^\intercal  \textsf{\textbf{X}} \vert {\bf r} \rangle
- \langle {\bf z} \vert  \textsf{\textbf{X}} \vert {\bf r} \rangle
- \langle {\bf r} \vert \textsf{\textbf{X}}^\intercal    \vert {\bf z}   \rangle
+ \langle {\bf z} \vert    {\bf z}   \rangle
\right)
\\
=
\frac{1}{m} \left[
\langle {\bf r} \vert \textsf{\textbf{X}}^\intercal  \textsf{\textbf{X}} \vert {\bf r} \rangle
- \langle {\bf z} \vert \left( \langle  {\bf r} \vert \textsf{\textbf{X}}^\intercal \right)^\intercal
- \langle {\bf r} \vert \textsf{\textbf{X}}^\intercal    \vert {\bf z}   \rangle
+ \langle {\bf z} \vert    {\bf z}   \rangle
\right]
\\
=
\frac{1}{m} \left\{
\langle {\bf r} \vert \textsf{\textbf{X}}^\intercal  \textsf{\textbf{X}} \vert {\bf r} \rangle
- \left[  \left( \langle  {\bf r} \vert \textsf{\textbf{X}}^\intercal  \right)^\intercal   \right]^\intercal  \vert {\bf z} \rangle
- \langle {\bf r} \vert \textsf{\textbf{X}}^\intercal    \vert {\bf z}   \rangle
+ \langle {\bf z} \vert    {\bf z}   \rangle
\right\}
\\
=
\frac{1}{m} \left(
\langle {\bf r} \vert \textsf{\textbf{X}}^\intercal  \textsf{\textbf{X}} \vert {\bf r} \rangle
- 2 \langle {\bf r} \vert \textsf{\textbf{X}}^\intercal    \vert {\bf z}   \rangle
+ \langle {\bf z} \vert    {\bf z}   \rangle
\right)
.
\end{split}
\label{2016-ml-MSE}
\end{equation}

In order to minimize the mean squared error (\ref{2016-ml-MSE}) with respect to variations of $\vert {\bf r} \rangle$
one obtains a condition for ``the linear theory'' $\vert {\bf y} \rangle$
by setting its derivatives (its gradient) to zero; that is
\begin{equation}
\partial_{ \vert {\bf r} \rangle } \text{MSE}
=
{\bf 0}.
\label{2016-ml-MSE-min}
\end{equation}
A lengthy but straightforward computation yields
\begin{equation}
\begin{split}
\frac{\partial }{\partial r_i}
\left(
r_j  \textsf{\textbf{X}}^\intercal _{jk} \textsf{\textbf{X}}_{kl} r_l -2 r_j \textsf{\textbf{X}}^\intercal _{jk}z_k + z_jz_j
\right)
\\
=
\delta_{ij}  \textsf{\textbf{X}}^\intercal _{jk} \textsf{\textbf{X}}_{kl} r_l
+
r_j \textsf{\textbf{X}}^\intercal _{jk} \textsf{\textbf{X}}_{kl}  \delta_{il}
-2 \delta_{ij} \textsf{\textbf{X}}^\intercal _{jk}z_k
\\
=
\textsf{\textbf{X}}^\intercal _{ik} \textsf{\textbf{X}}_{kl} r_l
+
r_j \textsf{\textbf{X}}^\intercal _{jk} \textsf{\textbf{X}}_{ki}
-2 \textsf{\textbf{X}}^\intercal _{ik}z_k
\\
=
\textsf{\textbf{X}}^\intercal _{ik} \textsf{\textbf{X}}_{kl} r_l
+
\textsf{\textbf{X}}^\intercal _{ik} \textsf{\textbf{X}}_{kj} r_j
-2 \textsf{\textbf{X}}^\intercal _{ik}z_k
\\
=
2 \textsf{\textbf{X}}^\intercal _{ik} \textsf{\textbf{X}}_{kj} r_j
-2 \textsf{\textbf{X}}^\intercal _{ik} z_k
\\
\equiv
2\left( \textsf{\textbf{X}}^\intercal  \textsf{\textbf{X}} \vert {\bf r} \rangle - \textsf{\textbf{X}}^\intercal    \vert {\bf z} \rangle
\right)
=0
\end{split}
\label{2016-ml-MSE-min-res-der}
\end{equation}
and finally, upon multiplication with
$ \left( \textsf{\textbf{X}}^\intercal  \textsf{\textbf{X}}  \right)^{-1}$ from the left,
\begin{equation}
\vert {\bf r} \rangle
=  \left( \textsf{\textbf{X}}^\intercal  \textsf{\textbf{X}} \right)^{-1}
\textsf{\textbf{X}}^\intercal  \vert {\bf z} \rangle
.
\label{2016-ml-MSE-min-res}
\end{equation}
A short plausibility check for $n=m=1$ yields the linear dependency
$\vert {\bf z} \rangle  =  \textsf{\textbf{X}} \vert {\bf r} \rangle$.

\end{enumerate}

\eexample
}

\section{Some common misconceptions}

\subsection{Confusion between component representation and ``the real thing''}

Given a particular basis, a tensor is uniquely characterized by its components.
However, without at least implicit reference to a particular basis, the enumeration of components of tuples are just blurbs,
and such ``tensors'' remain undefined.

{
\color{blue}
\bexample
Example (wrong!): a type-1 tensor (i.e., a vector) is given by
$(1,2)^\intercal $.

Correct: with respect (relative) to the basis $\{(0,1)^\intercal,(1,0)^\intercal\}$,  a (rank, degree, order) type-1 tensor (a vector) is given by
$(1,2)^\intercal$.
\eexample
}

\subsection{Matrix as a representation of a tensor of type (order, degree, rank) two}

A matrix ``is'' not a tensor; but a tensor of  type (order, degree, rank) 2 can be represented or encoded as a matrix
 with respect (relative) to some basis.
{
\color{blue}
\bexample
Example (wrong!): A matrix is  a  tensor of type (or order, degree, rank) 2.
Correct: with respect to the basis $\{(0,1)^\intercal,(1,0)^\intercal\}$,  a matrix represents a type-2 tensor.
The matrix components are the tensor components.
\eexample
}

Also, for non-orthogonal bases, covariant, contravariant, and mixed tensors correspond to different matrices.

\begin{center}
{\color{olive}   \Huge
\floweroneright
}
\end{center}

\chapter{Groups as permutations}
\label{2012-m-ch-gt}
\index{group theory}

\newthought{Group theory} is about {\em transformations, actions,} and the {\em symmetries} presenting themselves in terms of {\em invariants} with respect to those transformations and actions.
\index{symmetry}
\index{invariant}
One of the central axioms is the {\em reversibility} -- in mathematical terms, the {\em invertibility} -- of all operations: every transformation has a unique inverse transformation.
Another one is associativity; that is, the property that the order of the transformations is irrelevant.
These properties have far-reaching implications: from a functional perspective, group theory amounts to the study of {\em permutations} among the sets involved; nothing more and nothing less.
\index{permutation}

Rather than citing standard texts on group theory\cite{rotman} the reader is encouraged to consult
two internet resources: Dimitri Vvedensky's group theory course notes,\cite{vvedensky-grouptheory}
as well as
John Eliott's youtube presentation\cite{eliott-gt-yt} for an online course on group theory.
Hall's introductions to Lie groups\cite{hall-2000,hall-2015} contain fine presentations thereof.

\section{Basic definition and properties}

\subsection{Group axioms}

A {\em group} is a set of objects ${\frak G}$ which satisfy the following conditions (or, stated differently, axioms):

\begin{itemize}
\item[(i)] {\bf closure}:
There exists a map, or {\em composition rule}  $\circ : {\frak G} \times {\frak G} \rightarrow {\frak G}$, from  ${\frak G} \times {\frak G}$ into ${\frak G}$
which is {\em closed} under any composition
of elements; that is, the combination $a\circ b$ of any two elements
$a,b \in {\frak G}$ results in an element of the group ${\frak G}$.
That is,
the composition never yields anything ``outside'' of the group;
\item[(ii)]
{\bf associativity}:
for all $a$, $b$, and $c$ in ${\frak G}$,
the following equality holds: $a \circ (b \circ c) = (a \circ b) \circ c$.
Associativity amounts to the requirement that the order of the operations is irrelevant,
thereby restricting group operations to permutations;
\item[(iii)]
{\bf identity (element)}:
there exists an element of ${\frak G}$,
called the  {\em identity} (element) and denoted by $I$, such that for all $a$ in ${\frak G}$,
$a \circ I = I \circ  a=  a$.
\item[(iv)]
{\bf inverse (element)}:
for every $a$ in ${\frak G}$, there exists an element $a^{-1} \in {\frak G}$, such that $a^{-1} \circ  a  = a \circ  a^{-1}   =I$.
\item[(v)]
{\bf (optional) commutativity}:
if, for all $a$ and $b$ in ${\frak G}$, the following equalities hold: $a \circ b = b \circ a$,
then the group ${\frak G}$ is called {\em Abelian (group)}; otherwise it is called {\em nonabelian (group)}.
\index{Abelian group}
\index{nonabelian group}
\end{itemize}

A {\em subgroup} of a group is a subset which also satisfies the above axioms.
\index{subgroup}

In discussing groups one should keep in mind that there are two abstract spaces involved:

\begin{itemize}
\item[(i)] {\em Representation space} is the space of elements on which the group elements -- that is, the group transformations -- act.

\item[(ii)]  {\em Group space} is the space of elements of the group transformations.
\end{itemize}

{
\color{blue}
\bexample
Examples of groups operations and their respective representation spaces are:
\begin{itemize}
\item
addition of vectors in real or complex vector space;

\item
multiplications in ${\Bbb R} - {0}$ and ${\Bbb C} - {0}$, respectively;

\item
permutations (cf. Section~\ref{2018-permutation})
\index{permutation}
acting on products of the two $2$-tuples
$(0,1)^\intercal $ and
$(1,0)^\intercal $
(identifiable as the two classical bit states\cite{mermin-07});

\item
orthogonal transformations
\index{orthogonal transformation}
 (cf. Section~\ref{2015-m-ch-fdlvs-orthproj})
in real  vector space;

\item
unitary transformations
\index{unitary transformation}
 (cf. Section~\ref{2014-m-ch-fdvs-unitary})
in  complex vector space;

\item
real or complex nonsingular (invertible; that is, their determinant does not vanish)
matrices $\textrm{GL}(n,{\Bbb R})$ or $\textrm{GL}(n,{\Bbb C})$ on real or complex vector spaces, respectively.

\item
the {\em free group}
\index{free group}
of words (or terms) generated by two symbols $a$ and $b$ and their
inverses $a^{-1}$ and $b^{-1}$,
respectively. Examples of such words are $aab$, $ba^{-1}b^{-1}$, and so on.
\marginnote{In this example the group composition symbol ``$\circ$'' is omitted.
All  words or terms should be understood in their ``reduced form'',
in which all instances of $a^{-1}a=aa^{-1}=b^{-1}b=bb^{-1}=\emptyset$
are already eliminated.}

Let $F$ denote the (infinite) set of such words (or terms);
and let $F_a$,
$F_{a^{-1}}$,
$F_b$,
$F_{b^{-1}}\subset F$
denote the four sets starting with the symbols
$a$, $a^{-1}$,
$b$, and $b^{-1}$,
respectively.
By construction
$F_a$,
$F_{a^{-1}}$,
$F_b$, and
$F_{b^{-1}}$
are pairwise disjoint,
and, by symmetry, contain the same number of elements.
Therefore we may say that each one of these four sets
$F_a$,
$F_{a^{-1}}$,
$F_b$,  and
$F_{b^{-1}}$
represents ``one quarter of the entire set $F$.''

Furthermore,
an arbitrary element of $F_{a^{-1}}$ must be of the form
$a^{-1}w$, with $w \in F_{a^{-1}} \cup F_b \cup F_{b^{-1}}\subset F$.
Stated differently,
$w$ cannot be in $F_a$, since by definition all words in $F_a$ start with the
symbol $a$, and the latter
would immediately ``get annihilated'' by $a^{-1}$ from the left,
the starting symbol of $F_{a^{-1}}$ (that is, $a^{-1}a=\emptyset$).
Therefore the ``concatenation''
of $F_{a^{-1}}$ by $a$ from the left yields
``three quarters of the entire set $F$,''
since
$aF_{a^{-1}} = F_{a^{-1}}\cup F_b \cup F_{b^{-1}}$.
Likewise,
$bF_{b^{-1}} = F_a \cup F_{a^{-1}}\cup  F_{b^{-1}}$.
These constructions yield two
compositions or resolutions of $F$; namely
$F= F_a \cup a F_{a^{-1}}$
as well as
$F= F_b \cup b F_{b^{-1}}$.
This might be considered ``paradoxical''
\marginnote{These constructions are rooted in ``paradoxes of infinity,''
\index{Hilbert's hotel}
such as {\em Hilbert's hotel}.}
because, at the same time,
$F= F_a \cup  F_{a^{-1}} \cup F_b \cup F_{b^{-1}}$;
with pairwise disjoint  $F_a$,
$F_{a^{-1}}$,
$F_b$, and
$F_{b^{-1}}$.

We may identify the two words with different rotations (of a certain notrivial, independent, kind\cite{Hausdorff1914})
of points on the sphere.
This can be applied to the parametrization of
a sphere giving rise to the Banach-Tarski paradox.\cite{wagon1}
\index{Banach-Tarski paradox}

\end{itemize}

\eexample
}

\subsection{Discrete and continuous groups}

The {\em order} $\vert {\frak G} \vert$ of a group ${\frak G}$ is the number of distinct elements of that group.
\index{order}
If the order is finite or denumerable, the group is called {\em discrete}.
\index{discrete group}
If the group contains a continuity of elements, the group is called {\em continuous}.
\index{continuous group}

A {\em continuous group} can geometrically be imagined as a linear  space  (e.g., a linear vector or matrix space)
in which every point in this linear space is an element of that group.

\subsection{Generators and relations in finite groups}

The following notation will be used: $a^n = \underbrace{a \circ \cdots \circ a}_{n \text{ times}}$.

Elements of finite groups eventually ``cycle back;'' that is, multiple (but finite) operations of the same arbitrary element $a \in {\frak G}$ will eventually yield the identity:
$\underbrace{a \circ \cdots \circ a}_{k \text{ times}}= a^k=I$.
The {\em period} of $a \in {\frak G}$ is defined by
$\{e,a^1,a^2,\ldots , a^{k-1}\}$.

A {\em generating set}
of a group is a {\em minimal subset} -- a ``basis'' of sorts -- of that group
such that every element of the group can be expressed as the composition of elements of this subset and their inverses.
Elements of the generating set are called {\em generators}.
\index{generator}
These independent elements form a basis for all group elements.
\index{basis of group}
The {\em dimension} of a group is the {\em number of independent transformations} of that group,
which is the number of elements in a generating set.
\index{dimension}
The {\em coordinates} are defined relative to (in terms of) the basis elements.

{\em Relations}
\index{relations} are equations in those generators which hold for the group so that all other equations which hold for the group can be derived from those relations.

\subsection{Uniqueness of identity and inverses}
One important consequence of the axioms is the {\em uniqueness} of the identity and the inverse elements.
{\color{OliveGreen}
\bproof
In a proof by contradiction of the uniqueness of the identity, suppose that $I$ is not unique; that is,
there would exist (at least) two identity elements $I,I' \in {\frak G}$  with $I\neq I'$
such that $I\circ a= I' \circ a =a$.
This assumption yields a complete contradiction, since right composition with the inverse $a^{-1}$ of $a$, together with associativity, results in
\begin{equation}
\begin{split}
(I\circ a) \circ   a^{-1}= (I' \circ a) \circ a^{-1}\\
I\circ (a \circ   a^{-1})= I' \circ (a \circ a^{-1})\\
I\circ I= I' \circ I \\
I= I'
.
\end{split}
\label{2017-m-ch-gt-pouide}
\end{equation}

Likewise, in a proof by contradiction of the uniqueness of the inverse, suppose that the inverse is not unique; that is,
given some element $a\in {\frak G}$, then there would exist (at least) two inverse elements $g,g' \in {\frak G}$ with $g\neq g'$
such that $g\circ a= g' \circ a =I$.
This assumption yields a complete contradiction, since right composition with the inverse $a^{-1}$ of $a$, together with associativity, results in
\begin{equation}
\begin{split}
(g\circ a) \circ   a^{-1}= (g' \circ a)  \circ   a^{-1}\\
g\circ (a \circ   a^{-1})= g' \circ (a  \circ   a^{-1})\\
g\circ I= g' \circ I\\
g = g'
.
\end{split}
\label{2017-m-ch-gt-pouie}
\end{equation}
\eproof
}

\subsection{Cayley or group composition table}
For finite groups (containing finite sets of objects $\vert G\vert <\infty$) the composition rule can be nicely represented in matrix form by a
{\em Cayley table,}
or {\em composition table},
\index{Cayley table}
\index{composition table}
as enumerated in Table~\ref{2017-m-ch-gt-t-gct}.

\begin{table}
\begin{center}
\begin{tabular}{c|ccccccccccccccccccccccccccccccc}
$\circ$&$a$&$b$&$c$&$\cdots$\\
\hline
$a$&$a \circ a$&$a \circ b$&$a \circ c$&$\cdots$\\
$b$&$b \circ a$&$b \circ b$&$b \circ c$&$\cdots$  \\
$c$&$c \circ a$&$c \circ b$&$c \circ c$&$\cdots$    \\
$\vdots$&$\vdots$&$\vdots$&$\vdots$&$\ddots$
\end{tabular}
\caption{Group composition table\label{2017-m-ch-gt-t-gct}}
\end{center}
\end{table}

\subsection{Rearrangement theorem}
\label{2019-mm-ch-gt-rt}

Note that every row and every column of this table (matrix) enumerates the entire set ${\frak G}$ of the group;
more precisely, (i)  every row and every column
contains {\em each} element of the group ${\frak G}$;
(ii) but only  {\em once}.
This amounts to the {\em rearrangement theorem}
\index{rearrangement theorem}
stating that, for all $a \in  {\frak G}$, composition with $a$ permutes the elements of ${\frak G}$ such that
$a \circ {\frak G} =  {\frak G} \circ a = {\frak G}$.
That is, $a \circ {\frak G}$ contains each group element once and only once.

{\color{OliveGreen}
\bproof
Let us first prove (i): every row and every column is an enumeration of the set of objects of ${\frak G}$.

In a direct proof for rows, suppose that, given some $a\in {\frak G}$,  we want to know the ``source'' element $g$
which is send  into an arbitrary ``target'' element $b\in {\frak G}$ {\it via} $a \circ g = b$.
For a determination of this $g$ it suffices to explicitly form
\begin{equation}
g = I \circ  g =  (a^{-1} \circ a) \circ g = a^{-1} \circ (a \circ g) =  a^{-1}  \circ b,
\end{equation}
which is the element ``sending $a$, if multiplied from the right hand side (with respect to $a$), into $b$.''

Likewise, in a direct proof for columns, suppose that, given some $a\in {\frak G}$,  we want to know the ``source'' element $g$
which is send  into an arbitrary ``target'' element $b\in {\frak G}$ {\it via} $g \circ a = b$.
For a determination of this $g$ it suffices to explicitly form
\begin{equation}
g = g \circ I  =  g \circ ( a\circ a^{-1}  ) = ( g \circ a )\circ a^{-1}   = b  \circ  a^{-1},
\end{equation}
which is the element ``sending $a$, if multiplied from the left hand side (with respect to $a$), into $b$.''

Uniqueness (ii) can be proven by complete contradiction:
suppose there exists a row with two identical entries $a$ at different places, ``coming (via a single $c$ depending on the row)
from different sources $b$ and $b'$;''
that is, $c \circ b =   c \circ b' = a$, with $b \neq b'$.
But then, left composition with $c^{-1}$, together with associativity, yields
\begin{equation}
\begin{split}
c^{-1} \circ (c \circ b) = c^{-1} \circ  (c \circ b') \\
(c^{-1} \circ c) \circ b = (c^{-1} \circ  c) \circ b' \\
I \circ b = I \circ b' \\
b = b' .
\end{split}
\label{2017-m-ch-gt-pouer}
\end{equation}

Likewise,
suppose there exists a column with two identical entries $a$ at different places, ``coming (via a single $c$ depending on the column)
from different sources $b$ and $b'$;''
that is, $b \circ  c =  b' \circ  c = a$, with $b \neq b'$.
But then, right composition with $c^{-1}$, together with associativity, yields
\begin{equation}
\begin{split}
(b \circ  c) \circ c^{-1}=  (b' \circ  c) \circ c^{-1}\\
b \circ  (c \circ c^{-1})=  b' \circ  (c \circ c^{-1})\\
b \circ  I=  b' \circ  I\\
b  =  b'
.
\end{split}
\label{2017-m-ch-gt-pouec}
\end{equation}
\eproof
}

Exhaustion (i) and uniqueness (ii) impose rather stringent conditions on the composition rules,
which essentially have to permute elements of the set of the group ${\frak G}$.
Syntactically,  simultaneously {\em every row and every column} of a matrix representation of some group composition table
must contain the entire set ${\frak G}$.

Note also that Abelian groups have composition tables which are {\em symmetric along its diagonal axis}; that is, they are identical to their transpose. This is a direct consequence
of the Abelian property $a \circ b = b \circ a$.

\section{Zoology of finite groups up to order 6}
To give a taste of group zoology there is only one group of order 2, 3 and 5;
all three are Abelian. One (out of two groups) of order 6 is nonabelian.
\marginpar{\url{http://www.math.niu.edu/~beachy/aaol/grouptables1.html}, accessed on March 14th, 2018.}

\subsection {Group of order 2}

Table~\ref{2017-m-ch-gt-t-gc1t}
enumerates all $2^4$ binary functions of two bits;
only the two mappings represented by Tables~\ref{2017-m-ch-gt-t-gc1t}(7) and \ref{2017-m-ch-gt-t-gc1t}(10) represent groups,
with the identity elements 0 and 1, respectively. Once the identity element is identified, and subject to the substitution $0 \leftrightarrow 1$
the two groups are identical; they are the cyclic group $C_2$ of order 2.

\begin{table}
\begin{center}
\begin{tabular}{cccccccccccccccccccccccccccccccc}
\begin{tabular}{c|ccccccccccccccccccccccccccccccc}

$\circ$&0&1\\
\hline
0&0&0 \\  1&0&0
\end{tabular}
&
\begin{tabular}{c|ccccccccccccccccccccccccccccccc}
$\circ$&0&1\\
\hline
0&0&0 \\  1&0&1
\end{tabular}
&
\begin{tabular}{c|ccccccccccccccccccccccccccccccc}
$\circ$&0&1\\
\hline
0&0&0 \\  1&1&0
\end{tabular}
&
\begin{tabular}{c|ccccccccccccccccccccccccccccccc}
$\circ$&0&1\\
\hline
0&0&0 \\  1&1&1
\end{tabular}
\\
(1)&(2)&(3)&(4)\\
\\
\begin{tabular}{c|ccccccccccccccccccccccccccccccc}
$\circ$&0&1\\
\hline
0&0&1 \\  1&0&0
\end{tabular}
&
\begin{tabular}{c|ccccccccccccccccccccccccccccccc}
$\circ$&0&1\\
\hline
0&0&1 \\  1&0&1
\end{tabular}
&
{\color{blue}
\begin{tabular}{c|ccccccccccccccccccccccccccccccc}
$\circ$&0&1\\
\hline
0&0&1 \\  1&1&0
\end{tabular}
}
&
\begin{tabular}{c|ccccccccccccccccccccccccccccccc}
$\circ$&0&1\\
\hline
0&0&1 \\  1&1&1
\end{tabular}
\\
(5)&(6)&(7)&(8)\\
\begin{tabular}{c|ccccccccccccccccccccccccccccccc}
$\circ$&0&1\\
\hline
0&1&0 \\  1&0&0
\end{tabular}
&
{\color{blue}
\begin{tabular}{c|ccccccccccccccccccccccccccccccc}
$\circ$&0&1\\
\hline
0&1&0 \\  1&0&1
\end{tabular}
}
&
\begin{tabular}{c|ccccccccccccccccccccccccccccccc}
$\circ$&0&1\\
\hline
0&1&0 \\  1&1&0
\end{tabular}
&
\begin{tabular}{c|ccccccccccccccccccccccccccccccc}
$\circ$&0&1\\
\hline
0&1&0 \\  1&1&1
\end{tabular}
\\
(9)&(10)&(11)&(12)\\
\begin{tabular}{c|ccccccccccccccccccccccccccccccc}
$\circ$&0&1\\
\hline
0&1&1 \\  1&0&0
\end{tabular}
&
\begin{tabular}{c|ccccccccccccccccccccccccccccccc}
$\circ$&0&1\\
\hline
0&1&1 \\  1&0&1
\end{tabular}
&
\begin{tabular}{c|ccccccccccccccccccccccccccccccc}
$\circ$&0&1\\
\hline
0&1&1 \\  1&1&0
\end{tabular}
&
\begin{tabular}{c|ccccccccccccccccccccccccccccccc}
$\circ$&0&1\\
\hline
0&1&1 \\  1&1&1
\end{tabular}
\\
(13)&(14)&(15)&(16)\\
\end{tabular}
\caption{Different mappings; only (7) and (10) satisfy exhaustion (i) and uniqueness (ii);
they represent  permutations which induce associativity. Therefore only (7) and (10) represent group composition tables, with identity elements 0 and 1, respectively.\label{2017-m-ch-gt-t-gc1t}}
\end{center}
\end{table}

\subsection {Group of order 3, 4 and 5}
For a systematic enumeration of groups, it appears better to start with the identity element, and then use all properties (and equivalences) of composition tables to construct a valid one.
From the $3^{3^2}=3^9$ possible trivalent functions of a ``trit'' there exists only a single group with three elements ${\frak G}=\{I,a,b\}$; and its construction is enumerated in
Table~\ref{2017-m-ch-gt-t-gc1t3}.

\begin{table}
\begin{center}
\begin{tabular}{cccccccccccccccccccccccccccccccc}
\begin{tabular}{c|ccccccccccccccccccccccccccccccc}
$\circ$&$I$&$a$&$b$\\
\hline
$I$    &$I$&$a$&$b$\\
$a$    &$a$ & $t_{22}$&$t_{23}$\\
$b$    &$b$&  $t_{32}$&$t_{33}$ \\
\end{tabular}
&  &
\begin{tabular}{c|ccccccccccccccccccccccccccccccc}
$\circ$&$I$&$a$&$b$\\
\hline
$I$    &$I$&$a$&$b$\\
$a$    &$a$&$b$&$I$\\
$b$    &$b$&$I$&$a$\\
\end{tabular}
&  &
\begin{tabular}{c|ccccccccccccccccccccccccccccccc}
$\circ$&$I$&$a$&$a^2$\\
\hline
$I$    &$I$&$a$&$a^2$\\
$a$    &$a$&$a^2$&$I$\\
$a^2$    &$a^2$&$I$&$a$\\
\end{tabular}
\\(1)& &(2)& &(3)\\
\end{tabular}
\caption{Construction of the only group with three elements, the cyclic group $C_3$ of order 3}
\label{2017-m-ch-gt-t-gc1t3}
\end{center}
\end{table}
During the construction of the only group with three elements, the cyclic group $C_3$ of order 3,
note that
$t_{22}$ cannot be $a$ because this value already occurs in the second row and column, so it has to be either
$I$ or $b$. Yet $t_{22}$ cannot be $I$ because this would require $t_{23}=t_{32}=b$, but $b$ is already in the third row and column.
Therefore, $t_{22} = b$, implying  $t_{23}=t_{32}=I$, and in the next step, $t_{33} = a$.
The third Table~\ref{2017-m-ch-gt-t-gc1t3}(3) represents the composition table in terms of multiples of the generator $a$ with the relations $b=a^2$ and $a^3=I$.

There exist two groups with four elements, the cyclic group $C_4$ as well as the Klein four group. Both are enumerated in
Table~\ref{2017-m-ch-gt-t-gc1t4}.

\begin{table}
\begin{center}
\begin{tabular}{cccccccccccccccccccccccccccccccc}
\begin{tabular}{c|ccccccccccccccccccccccccccccccc}
$ \circ $&$1      $&$a      $&$a^2     $&$a^3                                            $   \\
\hline
$1      $&$1      $&$a      $&$a^2     $&$a^3                                            $   \\
$a      $&$a      $&$a^2     $&$a^3     $&$1                                             $   \\
$a^2     $&$a^2     $&$a^3     $&$1      $&$a                                            $   \\
$a^3     $&$a^3     $&$1      $&$a      $&$a^2                                           $   \\
\end{tabular}
&  &
\begin{tabular}{c|ccccccccccccccccccccccccccccccc}
$ \circ $&$1      $&$a      $&$b      $&$ab                                              $   \\
\hline
$1      $&$1      $&$a      $&$b      $&$ab                                              $   \\
$a      $&$a      $&$1      $&$ab     $&$b                                               $   \\
$b      $&$b      $&$ab     $&$1      $&$a                                               $   \\
$ab     $&$ab     $&$b      $&$a      $&$1                                               $   \\
\end{tabular}
\\(1)& &(2)&\\
\end{tabular}
\caption{Composition tables of the two groups of order 4 in terms of their generators.
The first table (1) represents the cyclic group $C_4$  of order 4 with the generator $a$ relation $a^4=1$.
The second table (2) represents the Klein four group with the generators $a$ and $b$ and the relations $a^2=b^2=1$ and $ab=ba$.
\label{2017-m-ch-gt-t-gc1t4}}
\end{center}
\end{table}


There exist only a single group with five elements ${\frak G}=\{I,a,b,c,d\}$  enumerated in  Table~\ref{2017-m-ch-gt-t-gc1t5}.
\begin{table}
\begin{center}
\begin{tabular}{cccccccccccccccccccccccccccccccc}
\begin{tabular}{c|ccccccccccccccccccccccccccccccc}
$\circ$&$I$&$a$&$a^2$&$a^3$&$a^4$\\
\hline
$I$    &$I$&$a$&$a^2$&$a^3$&$a^4$\\
$a$    &$a$&$a^2$&$a^3$&$d$&$I$\\
$a^2$    &$a^2$&$c$&$a^4$&$I$&$a$\\
$a^3$    &$a^3$&$a^4$&$I$&$a$&$a^2$\\
$a^4$    &$a^4$&$I$&$a$&$a^2$&$a^3$\\
\end{tabular}
\end{tabular}
\caption{The only group with 5 elements is the cyclic group $C_5$ of order 5, written in terms of multiples of the generator $a$ with $a^5=I$.
\label{2017-m-ch-gt-t-gc1t5}}
\end{center}
\end{table}

\subsection {Group of order 6}

There exist two groups with six elements ${\frak G}=\{I,a,b,c,d,e\}$, as enumerated in
Table~\ref{2017-m-ch-gt-t-gc1t6}. The second group is nonabelian; that is, the group  composition is not equal its transpose.
\begin{table}
\begin{center}
\begin{tabular}{cccccccccccccccccccccccccccccccc}
\begin{tabular}{c|ccccccccccccccccccccccccccccccc}
$ \circ $&$1      $&$a      $&$a^2     $&$a^3     $&$a^4     $&$a^5                      $   \\
\hline
$1      $&$1      $&$a      $&$a^2     $&$a^3     $&$a^4     $&$a^5                      $   \\
$a      $&$a      $&$a^2     $&$a^3     $&$a^4     $&$a^5     $&$1                       $   \\
$a^2     $&$a^2     $&$a^3     $&$a^4     $&$a^5     $&$1      $&$a                      $   \\
$a^3     $&$a^3     $&$a^4     $&$a^5     $&$1      $&$a      $&$a^2                     $   \\
$a^4     $&$a^4     $&$a^5     $&$1      $&$a      $&$a^2     $&$a^3                     $   \\
$a^5     $&$a^5     $&$1      $&$a      $&$a^2     $&$a^3     $&$a^4                     $   \\
\end{tabular}
\\ (1)\\
\begin{tabular}{c|ccccccccccccccccccccccccccccccc}
$    \circ      $&$1      $&$a      $&$a^2             $&$b      $&$ab     $&$a^2b       $   \\
\hline
$1              $&$1      $&$a      $&$a^2             $&$b      $&$ab     $&$a^2b       $   \\
$a              $&$a      $&$a^2     $&$1              $&$ab     $&$a^2b    $&$b         $   \\
$a^2             $&$a^2     $&$1      $&$a              $&$a^2b    $&$b      $&$ab       $   \\
$b              $&$b      $&$a^2b    $&$ab             $&$1      $&$a^2     $&$a         $   \\
$ab             $&$ab     $&$b      $&$a^2b            $&$a      $&$1      $&$a^2        $   \\
$a^2b            $&$a^2b    $&$ab     $&$b              $&$a^2     $&$a      $&$1        $   \\
\end{tabular}
\\(2)\\
\end{tabular}
\caption{The two groups with 6 elements; the latter one being nonabelian.
The generator of the cyclic group of order 6 is $a$ with the relation $a^6=I$.
The generators of the second group are $a,b$
with the relations $a^3 = 1$, $b^2 = 1$, $ba = a^{-1}b$.
\label{2017-m-ch-gt-t-gc1t6}}
\end{center}
\end{table}

\subsection{Cayley's theorem}

Properties (i) and (ii) -- exhaustion and uniqueness -- is a translation into the equivalent properties of bijectivity; together with the coinciding
\hbox{(co-)}domains this is just saying
that every element $a \in {\frak G}$ ``induces'' a {\em permutation};
\index{permutation}
that is, a map identified as $a(g) = a \circ g$ onto its domain ${\frak G}$.

Indeed,  {\em Cayley's (group representation) theorem}
\index{Cayley's theorem}
states that every group ${\frak G}$ is isomorphic to a subgroup
of the symmetric group; that is, it is isomorphic to some permutation group.
In particular, every finite group ${\frak G}$ of order $n$  can be imbedded as
a subgroup
of the symmetric group $\textrm{S}(n)$.

Stated pointedly: permutations exhaust the possible structures of (finite) groups.
The study of subgroups of the symmetric groups is no less general than the study of all groups.

{\color{OliveGreen}
\bproof
For a proof, consider the rearrangement theorem mentioned earlier, and identify ${\frak G}=\{a_1,a_2,\ldots\}$
with the ``index set'' $\{1,2,\ldots\}$ of the same number of elements as ${\frak G}$ through a bijective map $f(a_i)=i$,
$i= 1,2, \ldots$.
\eproof
}

\section{Representations by homomorphisms}

How can abstract groups be concretely represented in terms of matrices or operators?
Suppose we can find a structure- and distinction-preserving mapping $\varphi$ -- that is, an injective mapping preserving the group operation $\circ$  --
between elements of a group ${\frak G}$
and the groups of general either real or complex nonsingular  matrices $\textrm{GL}(n,{\Bbb R})$ or $\textrm{GL}(n,{\Bbb C})$, respectively.
Then this mapping is called
a {\em representation}
\index{representation} of the group ${\frak G}$.
In particular,
for this $\varphi  : {\frak G} \mapsto  \textrm{GL}(n,{\Bbb R})$ or $\varphi : {\frak G} \mapsto \textrm{GL}(n,{\Bbb C})$,
\begin{equation}
\varphi (a\circ b)   = \varphi (a)\cdot \varphi (b),
\end{equation}
for all
$a,b, a\circ b \in {\frak G}$.

{\color{blue}
\bexample
Consider, for the sake of an example, the
{\em Pauli spin matrices}
\index{Pauli spin matrices}
which are proportional to the angular momentum operators along the $x,y,z$-axis:\cite{schiff-55}
\begin{equation}
\begin{split}
\sigma_1=\sigma_x
=
\begin{pmatrix}
0&1\\
1&0
\end{pmatrix}
,
\;
\sigma_2=\sigma_y
=
\begin{pmatrix}
0&-i\\
i&0
\end{pmatrix}
,
\;
\sigma_3=\sigma_z
=
\begin{pmatrix}
1&0\\
0&-1
\end{pmatrix}
.
\end{split}
\end{equation}

Suppose these matrices $\sigma_1,\sigma_2,\sigma_3$
serve as generators of a group.
With respect to this basis system of matrices $\{ \sigma_1,\sigma_2,\sigma_3\}$
a general point in group in group space might be labelled by a three-dimensional
vector with the coordinates $(x_1,x_2,x_3)$
(relative to the basis $\{ \sigma_1,\sigma_2,\sigma_3\}$);
that is,
\begin{equation}
{\bf x} =   x_1\sigma_1 + x_2\sigma_2 +x_3 \sigma_3.
\end{equation}
If we form the exponential $  A  ({\bf x})= e^{\frac{i}{2} {\bf x}}$,
we can show (no proof is given here)
that $  A  ({\bf x})$ is a two-dimensional matrix representation of the group $\textrm{SU}(2)$,
the special unitary group of degree $2$ of $2\times 2$ unitary matrices with determinant $1$.
\eexample
}

\section{Partitioning of finite groups by cosets}
\index{coset}

There exists a straightforward method in which subgroups can be used for the generation of partitions of
a  finite  group:
\begin{enumerate}
\item
Start with an arbitrary subgroup ${\frak H} \subset {\frak G}$ of a group ${\frak G}$;
\item
Take some arbitrary element $g \in {\frak G}$, and either form the
{\em left coset} $g \circ {\frak H}$
\index{left coset}
of ${\frak H}$ in ${\frak G}$ with respect to $g$;
or the
{\em right coset} $ {\frak H} \circ g$
\index{right coset}
of ${\frak H}$ in ${\frak G}$ with respect to $g$.
\item
Do this for all $g \in {\frak G}$, and form the union of all these cosets.
\end{enumerate}
The resulting union set is a partition of ${\frak G}$.

{\color{OliveGreen}
\bproof
A proof for left cosets needs to show that
these cosets are mutually disjoint,
and that their union yields the entire group.
More explicitly, suppose that the two sets formed by
$g_1 \circ {\frak H}$ and $g_2 \circ {\frak H}$
are {\em not} disjoint.
By this assumption there exist some $u_1, u_2 \in {\frak H}$
with $g_1 \circ u_1 = g_2 \circ u_2$.
Now take some arbitrary $u_3 \in {\frak H}$ and form
\begin{equation}
g_2 \circ u_3 =
\underbrace{g_2 \circ u_2}_{ =g_1 \circ u_1}  \circ u_2^{-1} \circ u_3 =
g_1 \circ
\underbrace{u_1 \circ
\underbrace{u_2^{-1} \circ u_3}_{\in {\frak H}}
}_{\in {\frak H}} \in g_1{\frak H}
,
\end{equation}
and thus we obtain
$g_2 \circ {\frak H} \subset g_1 \circ {\frak H}$.
A similar, symmetric argument yields
$g_1 \circ {\frak H} \subset g_2 \circ {\frak H}$;
therefore,
$g_2 \circ {\frak H} = g_1 \circ {\frak H}$.
That is, stated pointedly, if
the two sets
$g_1 \circ {\frak H}$ and $g_2 \circ {\frak H}$
are not disjoint they must be
identical.
In the first case of identical sets
$g_1 \circ {\frak H}=g_2 \circ {\frak H}$,
${\frak H}=(g_1)^{-1} \circ g_2 \circ {\frak H}$,
and thus, by the rearrangement theorem
(cf. Section~\ref{2019-mm-ch-gt-rt}, page~\pageref{2019-mm-ch-gt-rt}),
\index{rearrangement theorem}
$(g_1)^{-1} \circ g_2 \in {\frak H}$.
At the same time, if one considers all
$g \in {\frak G}$,
and forms $g \circ {\frak H}$, already the elements $g \circ I = g$
recovers the entire group ${\frak G}$. (Note that $I \in {\frak H}$.)
\eproof
}

For any finite group ${\frak G}$ and any subgroup ${\frak H} \subset {\frak G}$,
the relation
$x \sim y:
x\circ {\frak H} = y\circ {\frak H}
\Leftrightarrow x^{-1}y \in {\frak H}$
defines an equivalence relation\marginnote{This result is part of {\em Lagrange's theorem}  in the mathematics of group theory.\index{Lagrange's theorem}}
\index{equivalence relation}
on ${\frak G}$.
Thereby, the set $x\circ {\frak H}$ with $x\in {\frak G}$ is a
left coset of ${\frak H}$ in ${\frak G}$ with respect to $g$.
A similar statement applies to right cosets.

{\color{blue}
\bexample
In the following example we shall consider the symmetric group $\textrm{S}(3)$
\index{symmetric group}
on a set of $3$ elements, say, the set of three numbers $\{1,2,3\}$.
In cycle notation
\marginnote{The cycle notation is a compact representation of permutations,
suppressing constant elements not changed,
and writing the changed elements (numbers) without commas,
starting with a left (unclosed) bracket sign ``$($''
and from an arbitrary element $i$
(mostly the first if an order exists),
and writing consecutive permutations
$\sigma(i)$,
$\sigma( \sigma(i))$,
$\sigma( \sigma(\sigma(i)))$,~$\ldots $
 of this element until the original ``seed'' $i$
is reached again;
at this point the initial, unclosed bracket is closed by a right bracket sign
 ``$)$''; e.g.,
$(1 \sigma(  1 )
\ldots \sigma( \sigma(i))\sigma( \sigma(\sigma(\ldots (1) \ldots )))$.
}
the group can be written as
\begin{equation}
\textrm{S}(3)
=
\{
()\equiv I,
(12),
(13),
(23),
(123),
(132)
\}
.
\label{2019-ch-gt-S3}
\end{equation}

The respective subgroups of $\textrm{S}(3)$
are
\begin{equation}
\begin{split}
{\frak H}_1
= \{()\},\quad
{\frak H}_2
=
\{
() ,
(12)
\},\quad
{\frak H}_3
=
\{
() ,
(13)
\},\\
{\frak H}_4
=
\{
() ,
(23)
\},\quad
{\frak H}_5
=
\{
() ,
(123),(132)
\}
.
 \end{split}
\end{equation}

Take, for the sake of an example, as a starting point the
subgroup ${\frak H}_2 = \{ () , (12) \}$ of $\textrm{S}(3)$,
and generate the associated partition of $\textrm{S}(3)$
by forming the {left cosets} $g \circ {\frak H}$
for all group elements  $g \in {\frak G}$;
that is,
\begin{equation}
\begin{split}
()\circ {\frak H}_2 = \{ ()\circ () , ()\circ (12) \}
 = \{ () , (12) \} = {\frak H}_2
,\\
(12)\circ {\frak H}_2 = \{ (12)\circ () , (12)\circ (12) \}
 = \{ (12) , () \} = {\frak H}_2
,\\
(13)\circ {\frak H}_2 = \{ (13)\circ () , (13)\circ (12) \}
 = \{ (13) , (123) \}
,\\
(23)\circ {\frak H}_2 = \{ (23)\circ () , (23)\circ (12) \}
 = \{ (23) , (132) \}
,\\
(123)\circ {\frak H}_2 = \{ (123)\circ () , (123)\circ (12) \}
 = \{ (123) , (13) \}
,\\
(132)\circ {\frak H}_2 = \{ (132)\circ () , (132)\circ (12) \}
 = \{ (132) , (23) \}
;
 \end{split}
\end{equation}
thereby effectively rendering the following partitioning of $\textrm{S}(3)$
enumerated in~(\ref{2019-ch-gt-S3}):
\begin{equation}
P_{{\frak H}_2}\left[ \textrm{S}(3) \right] =
\{
\underbrace{\{(), (12)\}}_{{\frak H}_2},
\{ (13) , (123) \},
\{ (23) , (132) \}
\}
.
\end{equation}

Similar calculations yield the partitions associated with different subgroups:
\begin{equation}
\begin{split}
P_{{\frak H}_1}\left[ \textrm{S}(3) \right] =
\{
\{()   \}  ,
\{(12) \}  ,
\{(13) \}  ,
\{(23) \}  ,
\{(123)\}  ,
\{(132)\}
\}  ,\\
P_{{\frak H}_3}\left[ \textrm{S}(3) \right] =
\{
{\frak H}_3,
\{ (12) , (132) \},
\{ (23) , (123) \}
\}  ,\\
P_{{\frak H}_4}\left[ \textrm{S}(3) \right] =
\{
{\frak H}_4,
\{ (12) , (123) \},
\{ (13) , (132) \}
\}  ,\\
P_{{\frak H}_5}\left[ \textrm{S}(3) \right] =
\{
{\frak H}_5,
\{ (12) , (13) ,(23) \}
\}  ,\;
P_{\textrm{S}(3)}\left[ \textrm{S}(3) \right]  =\{ \textrm{S}(3)\}
.
\end{split}
\end{equation}

\eexample
}

\marginnote{For quantum computation links to the hidden subgroup problem see Section~5.4.3 of \bibentry{nielsen-book10}.}
In quantum information theory
the {\em hidden subgroup problem}
\index{hidden subgroup problem}
is the problem to find (the generators of) some unknown
subgroup  ${\frak H}$
which is
``hidden'' by a
function $f({\frak G}) = X$
which maps elements of a group  ${\frak G}$
onto some set $X$;
while at the same time being constant on the cosets of ${\frak G}$;
more precisely, $f(g_1)=f(g_2)$ if and only if $g_1$ and $g_2$
belong to the same coset $g_1{\frak H}=g_2{\frak H}$ of ${\frak G}$
--
the function $f$ represents or ``encodes''
the cosets of ${\frak G}$ by being constant on any single coset
while being different between the different cosets of ${\frak G}$.

\section{Lie theory}
\index{Lie group}

Lie groups\cite[-0mm]{hall-2000,hall-2015} are continuous groups described by several real parameters.

\subsection{Generators}
We can generalize this example by defining
the {\em generators}
\index{generator}
of a continuous group as the first coefficient of a Taylor expansion
around unity; that is if the dimension of the group is $n$, and the Taylor expansion is
\begin{equation}
  G  ({\bf X}) =   \sum_{i=1}^n X_i   T  _i + \ldots ,
\end{equation}
then the matrix generator $T_i$ is defined by
\begin{equation}
  T  _i = \left. \frac{\partial   G  ({\bf X})}{\partial X_i} \right|_{ {\bf X}=0}.
\end{equation}

\subsection{Exponential map}
There is an exponential connection
$\exp : {\frak X} \mapsto {\frak G}$
between a matrix Lie group
and the Lie algebra ${\frak X}$ generated by the generators
$ T_i $.

\subsection{Lie algebra}
\index{Lie algebra}

A Lie algebra is a vector space ${\frak X}$,
together with a binary
{\em Lie bracket}
\index{Lie bracket}
operation $[\cdot,\cdot ]: {\frak X} \times {\frak X}  \mapsto {\frak X} $
satisfying
\begin{itemize}
\item[(i)]
bilinearity;
\item[(ii)]
antisymmetry: $[X,Y]=-[Y,X]$, in particular $[X,X]=0$;
\item[(iii)]
the Jacobi identity:
$[X,[Y,Z]] +  [Z,[X,Y]] + [Y,[Z,X]] =0$
\end{itemize}
for all $X,Y,Z \in {\frak X}$.

\section{Zoology of some important continuous groups}

\subsection{General linear group $\textrm{GL}(n,{\Bbb C})$}

The {\em general linear group} $\textrm{GL}(n,{\Bbb C})$
\index{general linear group}
contains all  nonsingular (i.e., invertible; there exist an inverse)
$n\times n$ matrices with complex entries.
The composition rule ``$\circ$''
is identified with matrix multiplication (which is associative); the neutral element is the unit
matrix ${\Bbb I}_n=\textrm{diag}(\underbrace{1,\ldots ,1}_{n \textrm{ times}})$.

\subsection{Orthogonal group over the reals $\textrm{O}(n,{\Bbb R})=\textrm{O}(n)$}

The {\em orthogonal group}\cite{murnaghan} $\textrm{O}(n)$ over the reals ${\Bbb R}$
\index{orthogonal group}
\index{orthogonal matrix}
can be represented by real-valued  orthogonal [i.e., $  A  ^{-1}=    A   ^\intercal $]
$n\times n$ matrices.
The composition rule ``$\circ$''
is identified with matrix multiplication (which is associative); the neutral element is the unit
matrix ${\Bbb I}_n=\textrm{diag}(\underbrace{1,\ldots ,1}_{n \textrm{ times}})$.


Because of orthogonality, only half of the off-diagonal entries are independent of one another,
resulting in $n(n-1)/2$ independent real parameters; the dimension of $ \textrm{O}(n)$.

{\color{OliveGreen}
\bproof
This can be demonstrated by writing any matrix $A \in \textrm{O}(n)$ in terms of its column vectors:
Let $a_{ij}$ be the $i$th row and $j$th column component of $A$.
Then $A$ can be written in terms of its column vectors as
$A=\begin{pmatrix}
{\bf a}_1,
{\bf a}_2,
\cdots ,
{\bf a}_n
\end{pmatrix}$,
where the $n$ tuples of scalars ${\bf a}_j = \begin{pmatrix}
{a}_{1j},
{a}_{2j},
\cdots ,
{a}_{nj}
\end{pmatrix}^\intercal $
contain the components $a_{ij}$, $ 1 \le i,j, \le n$ of the original matrix  $A$.

Orthogonality implies the following $n^2$ equalities: as
\begin{equation}
A^\intercal  =\begin{pmatrix}
{\bf a}_1^\intercal \\
{\bf a}_2^\intercal \\
\vdots \\
{\bf a}_n^\intercal
\end{pmatrix},
\text { and }
AA^\intercal =A^\intercal  A=
\begin{pmatrix}
{\bf a}_1^\intercal {\bf a}_1 & {\bf a}_1^\intercal {\bf a}_2  &\cdots & {\bf a}_1^\intercal {\bf a}_n  \\
{\bf a}_2^\intercal {\bf a}_1 & {\bf a}_2^\intercal {\bf a}_2  &\cdots & {\bf a}_2^\intercal {\bf a}_n  \\
\vdots   & \vdots &\ddots & \vdots  \\
{\bf a}_n^\intercal  {\bf a}_1& {\bf a}_n^\intercal {\bf a}_2  &\cdots & {\bf a}_n^\intercal {\bf a}_n  \\
\end{pmatrix}
={\Bbb I}_n,
\end{equation}
Because
\begin{equation}
\begin{split}
{\bf a}_i^\intercal {\bf a}_j =
\begin{pmatrix}
{a}_{1i},
{a}_{2i},
\cdots ,
{a}_{ni}
\end{pmatrix}
\cdot
\begin{pmatrix}
{a}_{1j},
{a}_{2j},
\cdots ,
{a}_{nj}
\end{pmatrix}^\intercal
\\
=
{a}_{1i}  {a}_{1j} +
\cdots +
{a}_{ni}  {a}_{nj}
=
{a}_{1j} {a}_{1i}  +
\cdots +
 {a}_{nj} {a}_{ni} =
{\bf a}_j^\intercal {\bf a}_i,
\end{split}
\end{equation}
this yields, for the first, second, and so on, until the $n$'th row,
$n + (n-1) +  \cdots +1= \sum_{i=1}^{n} i= n(n+1)/2$ nonredundand equations, which reduce the original number of $n^2$ free real parameters to
$n^2 - n(n+1)/2 = n(n-1)/2$.
\eproof
}

\subsection{Rotation group $\textrm{SO}(n)$}

The {\em special orthogonal group} or, by another name, the {\em rotation group} $\textrm{SO}(n)$
\index{special orthogonal group}
\index{rotation group}
\index{rotation matrix}
contains all  orthogonal
$n\times n$ matrices with unit determinant.
$\textrm{SO}(n)$ containing orthogonal matrices with determinants $1$ is a subgroup of $\textrm{O}(n)$,
the other component being orthogonal matrices with determinants $-1$.

The rotation group in two-dimensional configuration space  $\textrm{SO}(2)$
corresponds to planar rotations around the origin. It has dimension 1 corresponding to one parameter $\theta$.
Its elements can be written as
\begin{equation}
R(\theta )  =
\begin{pmatrix}
\cos \theta & \sin \theta\\
- \sin \theta  & \cos \theta
\end{pmatrix}
.
\end{equation}

\subsection{Unitary group  $\textrm{U}(n,{\Bbb C}) = \textrm{U}(n)$}

The {\em unitary group}\cite{murnaghan} $\textrm{U}(n)$
\index{unitary group}
\index{unitary matrix}
contains all  unitary [i.e., $  A  ^{-1}=  A  ^\dagger =(\overline{  A  })^\intercal $]
$n\times n$ matrices.
The composition rule ``$\circ$''
is identified with matrix multiplication (which is associative); the neutral element is the unit
matrix ${\Bbb I}_n=\textrm{diag}(\underbrace{1,\ldots ,1}_{n \textrm{ times}})$.

For similar reasons as mentioned earlier only half of the off-diagonal entries
-- in total $(n-1) + (n-2)+  \cdots +1= \sum_{i=1}^{n-1} i= n(n-1)/2$ --
are independent of one another, yielding twice as much -- that is, $n(n-1)$ --
conditions for the real parameters.
Furthermore
the diagonal elements of
$  A  A  ^\dagger = {\mathbb {I}}_n$
must be real and one, yielding $n$ conditions.
The resulting number of independent real parameters is $2 n^2 - n(n-1) - n =n^2$.

Not that, for instance,
$\textrm{U}(1)$ is the set of complex numbers $z=e^{i\theta}$ of unit modulus  $|z|^2=1$. It forms an Abelian group.

\subsection{Special unitary group $\textrm{SU}(n)$}

The {\em special unitary group} $\textrm{SU}(n)$
\index{special unitary group}
contains all  unitary
$n\times n$ matrices with unit determinant.
$\textrm{SU}(n)$ is a subgroup of $\textrm{U}(n)$.

Since there is one extra condition  $\textrm{det} A =1$
(with respect to unitary matrices)
the number of independent parameters for $\textrm{SU}(n)$ is  $n^2-1$.

We mention without proof that $\textrm{U}(2)$, which generates all normalized vectors -- identified with pure quantum states--
in two-dimensional
Hilbert space from some given arbitrary vector, is $2:1$ isomorphic to the rotation group $\textrm{SO}(3)$;
that is, more precisely $SU(2)/\{ \pm \mathbb{I}\} = SU(2)/\mathbb{Z}_2\cong SO(3)$.
This is the basis of the Bloch sphere representation of pure states in two-dimensional Hilbert space.
\index{Bloch sphere}

\subsection{Symmetric group $\textrm{S}(n)$}

The {\em symmetric group}
\marginnote{The symmetric group should not be confused with a symmetry group.}
\index{symmetric group}  $\textrm{S}(n)$ on a finite set of $n$ elements (or symbols)
is the group whose elements are all the permutations of the $n$ elements,
and whose group operation is the composition of such permutations.
The identity is the identity permutation.
The {\em permutations} are bijective functions from the set of elements onto itself.
\index{permutation}
The order (number of elements) of $\textrm{S}(n)$ is $n!$.

\subsection{Poincar\'e group}
\index{Poincar\'e group}

The {Poincar\'e group} is the group of {\em isometries}
--
that is,
bijective maps preserving distances
--
in space-time modelled by ${\Bbb R}^4$
endowed with a scalar product and thus
of a norm induced by the
{\em Minkowski metric}
\index{Minkowski metric}
$
\eta \equiv \{\eta_{ij}\}={\rm diag} (1,1,1,-1)
$
introduced in (\ref{2012-m-ch-tensor-minspn}).

It has dimension ten ($4+3+3=10$), associated with
the ten fundamental (distance preserving) operations
from which general isometries can be composed:
(i) translation through time and any of the three dimensions of space ($1+3=4$),
(ii) rotation (by a fixed angle) around any of the three spatial axes ($3$),
and a (Lorentz) boost, increasing the velocity
in any of the three spatial directions
of two uniformly moving bodies ($3$).

The rotations and Lorentz boosts form the
{\em  Lorentz group}.
\index{Lorentz group}

\begin{center}
{\color{olive}   \Huge
 \decosix
}
\end{center}

\chapter{Projective and incidence geometry}
\label{2012-m-ch-projgeom}
\index{projective geometry}
\index{incidence geometry}

\newthought{Projective geometry} is about the {\em geometric properties} that are invariant under
{\em projective transformations.}
\index{projective transformations}
{\em Incidence geometry} is about which points lie on which line.
\index{incidence geometry}

\section{Notation}
In what follows, for the sake of being able to formally represent {\em geometric transformations as
``quasi-linear'' transformations} and matrices,
the coordinates of $n$-dimensional Euclidean space will be augmented with one additional coordinate
which is set to one.
The following presentation will use two dimensions, but a generalization to arbitrary finite dimensions should be straightforward.
For instance, in the plane $\mathbb{R}^2$, we define new ``three-component'' coordinates
(with respect to some basis) by
\begin{equation}
{\bf x} =
\begin{pmatrix}
x_1\\
x_2
\end{pmatrix}
\equiv
\begin{pmatrix}
x_1\\
x_2 \\
1
\end{pmatrix}
 =  {\bf X}
.
\end{equation}
In order to differentiate these new coordinates ${\bf X}$
from the usual ones ${\bf x}$, they will be written in capital letters.

\section{Affine transformations map lines into lines as well as parallel lines to parallel lines}
\index{affine transformations}

In what follows we shall consider transformations which map lines into lines; and, likewise,
parallel lines to parallel lines.
A theorem of affine geometry,\cite[-40mm]{Stothers-ag,Gruenberg-77,Artstein-Avidan-2016}
essentially states that these are the {\em affine transformations}
\begin{equation}
f({\bf x})
=   \textsf{\textbf{A}}{\bf x} + {\bf t}
\end{equation}
with the {\em translation}
\index{translation}  ${\bf t}$, encoded by a tuple $(t_1,t_2)^\intercal $,
and an arbitrary linear transformation
$\textsf{\textbf{A}}$ represented by its associated matrix.
Examples of $\textsf{\textbf{A}}$ are
 {\em rotations,} as well as {\em dilatations}  and {\em skewing transformations}.
\index{rotation}
\index{dilatation}
\index{skewing}

Those two operations -- the linear transformation $\textsf{\textbf{A}}$ combined with a ``standalone'' translation by the vector ${\bf t}$ --
can be ``wrapped together'' to form the ``enlarged'' transformation matrix  (with respect to some basis;
``${\bf 0}^\intercal $'' indicates a row matrix with entries zero)
\begin{equation}
\textsf{\textbf{f}}=
\begin{pmatrix}
\textsf{\textbf{A}}&{\bf t}\\
{\bf 0}^\intercal &1
\end{pmatrix}
\equiv
\begin{pmatrix}
a_{11}&a_{12}&{t}_1\\
a_{21}&a_{22}&{t}_2\\
0&0&1
\end{pmatrix}
.
\label{2018-mm-ch-projgeom-def}
\end{equation}
Therefore, the affine transformation $f$ can be represented in the ``quasi-linear'' form
\begin{equation}
\textsf{\textbf{f}}({\bf X})
=
\textsf{\textbf{f}}{\bf X}
=
\begin{pmatrix}
\textsf{\textbf{A}}&{\bf t}\\
{\bf 0}^\intercal &1
\end{pmatrix}
{\bf X}
.
\label{2018-mm-ch-projgeom-def1}
\end{equation}

{\color{OliveGreen}
\bproof
Let us prove sufficiency of the aforementioned theorem of affine geometry
by explicitly showing that an arbitrary affine transformation of the form~(\ref{2018-mm-ch-projgeom-def}),
when applied to the parameter form  of the line
\begin{equation}
{\bf L}= \left\{ \begin{pmatrix}
y_1\\ y_2\\1
\end{pmatrix}
\middle|
\begin{pmatrix}
y_1\\ y_2 \\1
\end{pmatrix}
=
\begin{pmatrix}
x_1\\ x_2  \\0
\end{pmatrix}
s
+
\begin{pmatrix}
a_1\\ a_2  \\1
\end{pmatrix}
=
\begin{pmatrix}
x_1 s + a_1\\
x_2 s + a_2 \\1
\end{pmatrix} \text{, }\;
s\in \mathbb{R}
\right\}
,
\label{2018-mm-ch-projgeom-linep}
\end{equation}
again yields a line of the form~(\ref{2018-mm-ch-projgeom-linep}).
Indeed, applying~(\ref{2018-mm-ch-projgeom-def}) to~(\ref{2018-mm-ch-projgeom-linep})
yields
\begin{equation}
\begin{split}
\textsf{\textbf{f}} {\bf L}
=
\begin{pmatrix}
a_{11}&a_{12}&{t}_1\\
a_{21}&a_{22}&{t}_2\\
0&0&1
\end{pmatrix}
\begin{pmatrix}
x_1 s + a_1\\
x_2 s + a_2 \\
1
\end{pmatrix}
=
\begin{pmatrix}
a_{11}(x_1s +a_1) +a_{12}(x_2s+a_2) +{t}_1\\
a_{21}(x_1s +a_1) +a_{22}(x_2s+a_2) +{t}_2\\
 1
\end{pmatrix}
\\
=
\begin{pmatrix}
\underbrace{(a_{11}x_1 +a_{12}x_2)s}_{=x_1's} + \underbrace{a_{11}a_1 +a_{12}a_2 +{t}_1}_{=a_1'}\\
\underbrace{(a_{11}x_1 +a_{22}x_2)s}_{=x_2's} + \underbrace{a_{21}a_1 +a_{22}a_2 +{t}_2}_{=a_2'}\\
 1
\end{pmatrix}  = {\bf L}'
.
\end{split}
\label{2018-mm-ch-projgeom-linepap}
\end{equation}

Another, more elegant, way of demonstrating this property of affine maps
in a standard notation\cite{Stothers-ag} is
by representing a line with direction vector ${\bf x}$
through the point ${\bf a}$ by ${\bf l}= s {\bf x} + {\bf a}$,
with ${\bf x}= \begin{pmatrix}
x_1,  x_2
\end{pmatrix}^\intercal$
and ${\bf a}= \begin{pmatrix}
a_1,  a_2
\end{pmatrix}^\intercal$, and arbitrary $s$.
Applying an affine transformation
$\textsf{\textbf{f}}= \textsf{\textbf{A}}+ {\bf t}$
with ${\bf t}= \begin{pmatrix}
t_1,  t_2
\end{pmatrix}^\intercal$,
because of linearity of the matrix $\textsf{\textbf{A}}$,
yields
\begin{equation}
\begin{split}
\textsf{\textbf{f}}({\bf l})
=
\textsf{\textbf{A}}\left(s {\bf x} + {\bf a}\right) + {\bf t}
=
 s \textsf{\textbf{A}}{\bf x}
+ \textsf{\textbf{A}}{\bf a} + {\bf t} = {\bf l}',
\end{split}
\label{2018-mm-ch-projgeom-linepap1}
\end{equation}
which is again a line; but one with direction vector $\textsf{\textbf{A}}{\bf x}$
through the point $\textsf{\textbf{A}}{\bf a} + {\bf t}$.

The preservation of the ``parallel line'' property can be proven
by considering a second line
${\bf m}$ supposedly parallel to the first line ${\bf l}$, which means that
${\bf m}$ has an identical direction vector ${\bf x}$ as ${\bf l}$.
Because  the affine transformation $\textsf{\textbf{f}}({\bf m})$
yields an identical direction vector $\textsf{\textbf{A}}{\bf x}$
for ${\bf m}$ as for ${\bf l}$, both transformed lines remain parallel.

\eproof
}

It is not too difficult to prove [by the compound of two transformations
of the affine form~(\ref{2018-mm-ch-projgeom-def})]
that two or more successive affine transformations again render an affine transformation.

A {\em proper} affine transformation is invertible, reversible and one-to-one.
We state without proof that this is equivalent to the
invertibility of $\textsf{\textbf{A}}$  and thus  $\left| \textsf{\textbf{A}} \right| \neq 0$.
If $\textsf{\textbf{A}}^{-1}$ exists then
the inverse transformation with respect to~(\ref{2018-mm-ch-projgeom-def1})    is
\begin{equation}
\begin{split}
\textsf{\textbf{f}}^{-1}     =
\begin{pmatrix}
\textsf{\textbf{A}}&{\bf t}\\
{\bf 0}^\intercal &1
\end{pmatrix}^{-1}
=
\begin{pmatrix}
\textsf{\textbf{A}}^{-1} &-\textsf{\textbf{A}}^{-1} {\bf t}\\
{\bf 0}^\intercal &1
\end{pmatrix}
 \qquad    \qquad
\\
=
\frac{1}{a_{11}a_{22}-a_{12}a_{21}}
\begin{pmatrix}
a_{22}&-a_{12} & (- a_{22} t_1 + a_{12}t_2)\\
-a_{21}&a_{11} &   (a_{21} t_1 - a_{11}t_2)\\
0&0 &(a_{11}a_{22}-a_{12}a_{21})
\end{pmatrix}
.
\end{split}
\label{2018-mm-ch-projgeom-def1inv}
\end{equation}
This can be
directly
checked by concatenation of
$\textsf{\textbf{f}}$ and $\textsf{\textbf{f}}^{-1}$; that is, by
$\textsf{\textbf{f}} \textsf{\textbf{f}}^{-1}=\textsf{\textbf{f}}^{-1} \textsf{\textbf{f}}=\mathbb{I}_3$:
with
$\textsf{\textbf{A}}^{-1} = \frac{1}{a_{11}a_{22}-a_{12}a_{21}}\begin{pmatrix}
a_{22}&-a_{12}\\
-a_{21}&a_{22}
\end{pmatrix}$.
Consequently the proper affine transformations form a group (with the unit element represented by a diagonal matrix with entries $1$), the {\em affine group}.
\index{group}
\index{affine group}

As mentioned earlier affine transformations preserve
the ``parallel line'' property. But what about non-collinear lines?
The fundamental theorem of affine geometry\cite{Stothers-ag}
\index{fundamental theorem of affine geometry}
states that,
given two lists
$L=\{{\bf a},{\bf b},{\bf c}\}$
and
$L'=\{{\bf a}',{\bf b}',{\bf c}'\}$
of non-collinear
\marginnote{A set of points are non-collinear if
they dont lie on the same line; that is, their associated vectors from the origin are linear independent.}
points of $\mathbb{R}^2$;
then there is a {\em unique}  proper affine transformation  mapping $L$ to $L'$ (and {\it vice versa}).

{\color{OliveGreen}
\bproof
For the sake of convenience we shall first prove  the ``$\{{\bf 0},{\bf e}_1,{\bf e}_2\}$ theorem'' stating that
if $L=\{{\bf p},{\bf q},{\bf r}\}$
is a list of non-collinear points of $\mathbb{R}^2$, then
there is a {\em unique}  proper affine transformation  mapping
$\{{\bf 0},{\bf e}_1,{\bf e}_2\}$ to $L=\{{\bf p},{\bf q},{\bf r}\}$;
whereby
${\bf 0}= \begin{pmatrix} 0,0\end{pmatrix}^\intercal$,
${\bf e}_1= \begin{pmatrix} 1,0\end{pmatrix}^\intercal$,
and
${\bf e}_2= \begin{pmatrix} 0,1\end{pmatrix}^\intercal$:
First note that because ${\bf p}$, ${\bf q}$, and ${\bf r}$ are non-collinear by assumption,
$({\bf q}-{\bf p})$ and $({\bf r}-{\bf p})$ are non-parallel.
Therefore, $({\bf q}-{\bf p})$ and $({\bf r}-{\bf p})$ are linear independent.

Next define $\textsf{\textbf{f}}$ to be some affine transformation
which maps $\{{\bf 0},{\bf e}_1,{\bf e}_2\}$ to $L=\{{\bf p},{\bf q},{\bf r}\}$;
such that
\begin{equation}
\begin{split}
\textsf{\textbf{f}}({\bf 0})= \textsf{\textbf{A}}{\bf 0} + {\bf b}= {\bf b} = {\bf p} ,\quad
\textsf{\textbf{f}}({\bf e}_1)= \textsf{\textbf{A}}{\bf e}_1 + {\bf b} = {\bf q},\quad
\textsf{\textbf{f}}({\bf e}_2)= \textsf{\textbf{A}}{\bf e}_2 + {\bf b} = {\bf r}
.
\end{split}
\label{2018-mm-ch-projgeom-0-pqr}
\end{equation}

Now consider a column vector representation of $\textsf{\textbf{A}} = \begin{pmatrix} {\bf a}_1,{\bf a}_2\end{pmatrix}$
with
${\bf a}_1 = \begin{pmatrix}  a_{11},a_{21}\end{pmatrix}^\intercal$ and
${\bf a}_2 = \begin{pmatrix}  a_{12},a_{22}\end{pmatrix}^\intercal$,
respectively.
Because of the special form of
${\bf e}_1= \begin{pmatrix} 1,0\end{pmatrix}^\intercal$
and
${\bf e}_2= \begin{pmatrix} 0,1\end{pmatrix}^\intercal$,
\begin{equation}
\begin{split}
 \textsf{\textbf{f}}({\bf e}_1)= \textsf{\textbf{A}}{\bf e}_1 + {\bf b}
=  {\bf a}_1 + {\bf b}= {\bf q},\quad
 \textsf{\textbf{f}}({\bf e}_2)= \textsf{\textbf{A}}{\bf e}_2 + {\bf b}
=  {\bf a}_2 + {\bf b}= {\bf r}.
\end{split}
\label{2018-mm-ch-projgeom-0-pqr2}
\end{equation}
Therefore,
\begin{equation}
\begin{split}
{\bf a}_1=  {\bf q} - {\bf b} =  {\bf q} - {\bf p} , \quad
{\bf a}_2=  {\bf r} - {\bf b} =  {\bf r} - {\bf p} .
\end{split}
\label{2018-mm-ch-projgeom-0-pqr3}
\end{equation}
Since by assumption $({\bf q}-{\bf p})$ and $({\bf r}-{\bf p})$ are linear independent,
so are ${\bf a}_1$ and ${\bf a}_2$.
Therefore, $\textsf{\textbf{A}} = \begin{pmatrix} {\bf a}_1,{\bf a}_2\end{pmatrix}$ is invertible;
and together with the translation vector ${\bf b} = {\bf p}$, forms a unique
affine transformation $\textsf{\textbf{f}}$
which maps $\{{\bf 0},{\bf e}_1,{\bf e}_2\}$ to $L=\{{\bf p},{\bf q},{\bf r}\}$.

The fundamental theorem of affine geometry can be obtained by  a conatenation of (inverse) affine transformations of
$\{{\bf 0},{\bf e}_1,{\bf e}_2\}$:
as by the ``$\{{\bf 0},{\bf e}_1,{\bf e}_2\}$ theorem''
there exists
a unique (invertible) affine transformation $\textsf{\textbf{f}}$ connecting $\{{\bf 0},{\bf e}_1,{\bf e}_2\}$ to $L$,
as well as
a unique affine transformation $\textsf{\textbf{g}}$ connecting $\{{\bf 0},{\bf e}_1,{\bf e}_2\}$ to $L'$,
the concatenation of  $\textsf{\textbf{f}}^{-1}$ with $\textsf{\textbf{g}}$
forms a compound affine transformation $\textsf{\textbf{g}}\textsf{\textbf{f}}^{-1}$
mapping $L$ to $L'$.

\eproof
}

\subsection{One-dimensional case}
In {one dimension}, that is,  for ${\bf z}\in {\Bbb C}$, among the five basic operations
\begin{itemize}
\item[(i)] scaling:  $\textsf{\textbf{f}}({\bf z}) = r  {\bf z}  \textrm{ for } r\in {\Bbb R}$,
\item[(ii)] translation:  $\textsf{\textbf{f}}({\bf z}) = {\bf z} + {\bf w}  \textrm{ for } w\in {\Bbb C}$,
\item[(iii)] rotation: $ \textsf{\textbf{f}}({\bf z}) = e^{i\varphi}{\bf z}    \textrm{ for } \varphi\in {\Bbb R}$,
\item[(iv)] complex conjugation: $\textsf{\textbf{f}}({\bf z}) = \overline{{\bf z}}$,
\item[(v)] inversion: $\textsf{\textbf{f}}({\bf z}) = {\bf z}^{-1}$,
\end{itemize}
there are three types of
affine transformations (i)--(iii)  which can be combined.


{
\color{blue}
\bexample

An example of a one-dimensional case is the ``conversion'' of probabilities to expectation values in a dichotonic system; say, with observables
in $\{-1,+1\}$.
Suppose $p_{+1} = 1-p_{-1}$ is the probability of the occurrence of the observable ``$+1$''.
Then the expectation value is given by $E =  (+1)p_{+1} + (-1)p_{-1} =  p_{+1} - (1-p_{+1}) = 2 p_{+1}-1$; that is, a scaling of $p_{+1}$ by a factor of $2$, and a translation
by $-1$.
Its inverse is $p_{+1} = (E+1)/2= E/2 + 1/2$.
The respective matrix representation are
$\begin{pmatrix}
2&-1\\
0&1
\end{pmatrix}$
and
$\frac{1}{2}\begin{pmatrix}
1&1\\
0&2
\end{pmatrix}$.

For more general dichotomic observables
in $\{a,b\}$,
 $E =  a p_{a} + bp_{b} =  a p_{a} + b(1-p_{a}) = (a-b) p_{a}+b$, so that the matrices representing these affine transformations are
$\begin{pmatrix}
(a-b)&b\\
0&1
\end{pmatrix}$
and
$\frac{1}{a-b}\begin{pmatrix}
1&-b\\
0&a-b
\end{pmatrix}$.

\eexample
}

\section{Similarity transformations}
\index{similarity transformations}

{\em Similarity transformations}  involve translations ${\bf t}$, rotations $\textsf{\textbf{R}}$ and a dilatation $r$
and can be represented by the matrix
\begin{equation}
\begin{pmatrix}
r \textsf{\textbf{R}}&{\bf t}\\
{\bf 0}^\intercal &1
\end{pmatrix}
\equiv
\begin{pmatrix}
m\cos \varphi &-m\sin \varphi &{t}_1\\
m\sin \varphi &m\cos \varphi &{t}_2\\
0&0&1
\end{pmatrix}
.
\end{equation}




\section{Fundamental theorem of affine geometry revised}
\index{fundamental theorem of affine geometry}
\marginnote{For a proof and further references, see \bibentry{lester}.}

Any bijection from ${\Bbb R}^n$, $n\ge 2$,  onto itself
which maps all lines onto lines is an affine transformation.

\section{Alexandrov's theorem}
\index{Alexandrov's theorem}
\marginnote{For a proof and further references, see \bibentry{lester}.}

Consider the Minkowski space-time
${\Bbb M}^n$; that is,  ${\Bbb R}^n$, $n\ge 3$, and the Minkowski metric
[cf. (\ref{2012-m-ch-tensor-minspn}) on page \pageref{2012-m-ch-tensor-minspn}]
$\eta \equiv \{\eta_{ij}\}={\rm diag} (\underbrace{1,1,\ldots ,1}_{n-1\; {\rm times}},-1)$.
Consider further  bijections $\textsf{\textbf{f}}$ from  ${\Bbb M}^n$
onto itself preserving light cones; that is
for all ${\bf x}, {\bf y} \in {\Bbb M}^n$,
$$ \eta_{ij} (x^i-y^i)(x^j-y^j)=0 \textrm{ if and only if }
\eta_{ij} (\textsf{\textbf{f}}^i(x)-\textsf{\textbf{f}}^i(y))
(\textsf{\textbf{f}}^j(x)-\textsf{\textbf{f}}^j(y))=0.$$
Then $\textsf{\textbf{f}}(x)$ is the product of a Lorentz transformation
and a positive scale factor.

\begin{center}
{\color{olive}   \Huge
 \floweroneleft
}
\end{center}

\chapter*{\color{BurntOrange} \thispagestyle{empty} {\fontsize{40}{168} \selectfont Part II} \\ {\fontsize{30}{68} \selectfont Functional analysis}
\addcontentsline{toc}{part}{Part II:  Functional analysis}
{\newpage \thispagestyle{empty}   $\;$ \vskip 9 true cm \begin{center}\includegraphics[width=0.7\textwidth, angle=-20]{2019-mm-swimmer.png}\end{center}}}
\chapter{Brief review of complex analysis}
\label{2011-m-ch-ca}

Is it not amazing that complex numbers\cite[-20mm]{Hlawka-zz} can be used for physics?
Robert Musil (an Austrian novelist and  mathematician), in
{\it ``Verwirrungen des Z\"ogling T\"orle\ss''}\sidenote[][-15mm]{German original \url{http://www.gutenberg.org/ebooks/34717}:
{\it
``In solch einer Rechnung sind am Anfang ganz solide Zahlen,
die Meter oder Gewichte, oder irgend etwas anderes Greifbares darstellen
k\"onnen und wenigstens wirkliche Zahlen sind.
Am Ende der Rechnung stehen ebensolche.
Aber diese beiden h\"angen miteinander durch etwas zusammen, das es gar nicht gibt.
Ist das nicht wie eine Br\"ucke,
von der nur Anfangs- und Endpfeiler vorhanden sind
und die man dennoch so sicher \"uberschreitet,
als ob sie ganz dast\"unde?
F\"ur mich hat so eine Rechnung etwas Schwindliges;
als ob es ein St\"uck des Weges wei\ss{} Gott wohin ginge.
Das eigentlich Unheimliche ist mir aber die Kraft,
die in solch einer Rechnung steckt und einen so festh\"alt,
da\ss{} man doch wieder richtig landet.''}},
has expressed the amazement of a youngster confronted with the applicability of imaginaries,
by stating that, at the beginning of any computation involving imaginary numbers
are ``solid'' numbers which could represent something measurable, like lengths or weights,
or something else tangible; or are at least real numbers.
At the end of the computation, there are also such ``solid'' entities.
But the beginning and the end of the computation are connected by something
seemingly nonexisting.
Does this not appear, Musil's {\it  Z\"ogling T\"orle\ss} wonders,
like a bridge crossing an abyss
with only a bridge pier at the very beginning and one at the very end,
which could nevertheless be crossed with certainty and securely,
as if this bridge would exist entirely?

In what follows, a very brief review of
{\em complex analysis},
\index{complex analysis}
or, by another term,
{\em theory of complex functions},
\index{theory of complex functions}
will be presented.
For much more detailed introductions to complex analysis,
including proofs,
take, for instance, a ``classical'' introduction,\cite[-7mm]{Remmert-1991-tocf}
among a zillion of other very good ones.\cite{freitag-busam-en,whittaker:1927:cma,Greene,Hille62,ahlfors:1966:ca}
We shall study complex analysis not only for its beauty but also because it yields
very important analytical methods and tools;
for instance for the solution of (differential) equations and the
computation of definite integrals.
These methods will then be required for the computation of distributions and Green's functions,
as well for the solution of differential equations of mathematical physics -- such as the Schr\"odinger equation.

One motivation for introducing
{\em imaginary numbers}
\index{imaginary numbers}
is the (if you perceive it that way)
``malady''
that not every polynomial such as $P(x)=x^2+1$ has a root $x$
-- and thus not every (polynomial) equation $P(x)=x^2+1=0$ has a
solution $x$  --
which is a real number.
Indeed, you need the imaginary unit $i^2 =-1$ for a factorization $P(x)=(x+i)(x-i)$ yielding the two roots $\pm i$
to achieve this.
In that way, the introduction of imaginary numbers is a further step towards omni-solvability.
No wonder that the fundamental theorem of algebra, stating that every
non-constant polynomial with complex coefficients has at least one complex root
\index{root of a polynomial}
--
and thus total factorizability of polynomials into linear factors follows!

If not mentioned otherwise, it is assumed that the
{\em Riemann surface,}
\index{Riemann surface}
representing a ``deformed version'' of the complex plane for functional purposes,
is simply connected.
Simple connectedness means that the Riemann surface
is path-connected so that every path between two points can be continuously transformed, staying within the domain,
into any other path while preserving the two endpoints between the paths.
In particular, suppose that
there are no ``holes'' in the Riemann surface; it is not ``punctured.''

Furthermore, let $i$ be the
{\em imaginary unit}
\index{imaginary unit} with the property that
 $i^2=-1$ is the solution of the equation $x^2+1=0$.
The introduction of imaginary numbers  guarantees that all quadratic
equations have two roots (i.e., solutions).

By combining imaginary and real numbers,
any {\em complex number} can be defined to be some linear combination of the real unit number ``$1$''
\index{complex numbers}
with the imaginary unit number $i$
that is,
$
z= 1 \times (\Re z)  +  i \times (\Im z)
$,
with the real valued factors $(\Re z) $
and $(\Im z)$, respectively.
By this definition,
a complex number $z$ can be decomposed into real numbers $x$, $y$, $r$ and $\varphi$ such that
\begin{equation}
z \stackrel{{\tiny \textrm{ def }}}{=}  \Re z + i \Im  z  =x +iy  =r e^{i\varphi}= r e^{i \textrm{arg}(z)},
\label{2017-m-ca-cn}
\end{equation}
with $x = r\cos \varphi$
and $y = r\sin \varphi$,
where {Euler's formula}
\index{Euler's formula}
\begin{equation}
e^{i\varphi} = \cos \varphi +i \sin \varphi
\label{2012-m-ch-ca-ef}
\end{equation}
has been used.
If  $z = \Re z $ we call $z$ a real number.
If  $z = i\Im z $ we call $z$ a purely imaginary number.
The {\em argument}
\index{argument}
or {\em phase}
\index{phase}
$\textrm{arg}(z)$ of the complex number $z$ is the angle $\varphi$ (usually in radians)
measured counterclockwise from the positive real axis to the vector representing $z$ in the complex plane.
The principal value $\textrm{Arg} (z)$ is usually defined to lie in   the interval $( - \pi, \pi ]$; that is,
\index{principal value}
\begin{equation}
-\pi < \textrm{Arg} (z) \le +\pi
.
\label{2018-m-ch-ca-ef}
\end{equation}

Note that the function $\varphi \mapsto e^{i\varphi}$ in (\ref{2017-m-ca-cn}) is not injective. In particular,
$\exp ({i\varphi})= \exp ( {i(\varphi} + 2\pi  k)$ for arbitrary $k \in \mathbb{Z}$.
This has no immediate consequence on $z$;
but it yields differences for {\em functions} thereof, like the square root or the logarithm.
A remedy is the introduction of Riemann surfaces which are ``extended'' and ``deformed'' versions of the complex plane.

The {\em modulus} or {\em absolute value}
\index{modulus}
\index{absolute value}
of a complex number $z$ is defined by
\begin{equation}
| z |
\stackrel{{\tiny \textrm{ def }}}{=} +
\sqrt{ (\Re z)^2 + (\Im z)^2 }.
\label{2012-m-ch-ca-defmodulus}
\end{equation}

Many rules of classical arithmetic can be carried over to complex arithmetic.\cite{apostol,freitag-busam}
Note, however,
that, because of noninjectivity of $\exp ( i\varphi)$ for arbitrary values of $\varphi$,
for instance, $\sqrt{a}\sqrt{b}= \sqrt{ab}$
is only valid if at least one factor $a$ or $b$ is positive;
otherwise one could construct wrong deductions $-1=i^2 \stackrel{{\tiny \textrm{ ? }}}{=} \sqrt{i^2}\sqrt{i^2} \stackrel{{\tiny \textrm{ ? }}}{=} \sqrt{-1}\sqrt{-1} \stackrel{{\tiny \textrm{ ? }}}{=} \sqrt{(-1)^2}=1$.
More generally, for two arbitrary numbers, $u$ and $v$,
$\sqrt{u}\sqrt{v}$ is not always equal to $\sqrt{u v}$.
\marginnote{Nevertheless,
$\sqrt{|u|}\sqrt{|v|}=\sqrt{|u v|}$.
}
The $n$'th root of a complex number $z$ parameterized by many (indeed, an infinity of) angles $\varphi$ is no unique function any longer,
as $\sqrt[\leftroot{0}\uproot{3}n]{z} = \sqrt[\leftroot{0}\uproot{3}n]{\vert z\vert } \exp \left(i\varphi / n + 2\pi i k/n\right)$
with $k \in \mathbb{Z}$.
Thus, in particular, for the square root with $n=2$,
$\sqrt{u}\sqrt{v}= \sqrt{|u|\,|v|}
\exp \left[ (i/2)(\varphi_u   +\varphi_v )
\right]
\underbrace{\exp \left[i \pi (k_u + k_v)\right]}_{\pm 1}
$.
Therefore, with $u={-1} = \exp [i\pi (1 +2 k)]$ and
$v={-1} = \exp [i\pi (1 +2 k')]$ and $k,k' \in \mathbb{Z}$, one obtains
$
\sqrt{-1}\sqrt{-1}=
\underbrace{\exp \left[ (i/2)(\pi   +\pi )\right]}_{1}
\underbrace{\exp \left[i \pi (k  + k')\right]}_{\pm 1}
=\mp 1
$,
for even and odd $k + k'$, respectively.

For many mathematicians
{\em Euler's identity}
\index{Euler identity}
\begin{equation}
e^{i\pi}=-1 \textrm{, or } e^{i\pi}+1=0,
\end{equation}
is the ``most beautiful'' theorem.\cite{springerlink:10.1007/BF03023741}

Euler's formula (\ref{2012-m-ch-ca-ef}) can be used to derive {\em de Moivre's formula}
\index{Moivre's formula} for integer $n$ (for non-integer $n$ the formula is multi-valued for different arguments $\varphi$):
\begin{equation}
e^{in\varphi} = (\cos \varphi +i \sin \varphi )^n      = \cos (n\varphi) +i \sin (n\varphi).
\end{equation}

\section{Geometric representations of complex numbers and functions thereof}

\subsection{The complex plane}

It is quite suggestive to consider the complex numbers $z$, which are linear combinations
of the real and the imaginary unit,
in the {\em complex plane} ${\Bbb C} = {\Bbb R} \times {\Bbb R}$
\index{complex plane}
as a geometric representation of complex numbers.
Thereby,  the real and the imaginary unit are identified with the (orthonormal) basis vectors
of the
{\em standard (Cartesian) basis};
that is, with the tuples
\index{Cartesian basis}
\index{standard basis}
\begin{equation}
1 \equiv \begin{pmatrix}1,0\end{pmatrix},\text{ and }
i \equiv \begin{pmatrix}0,1\end{pmatrix}.
\end{equation}
Figure~\ref{2018-m-f-cp} depicts this schema,
including the location of the points corresponding to the real and imaginary units $1$ and $i$,
respectively.
\begin{marginfigure}
\begin{center}
\begin{tikzpicture}  [scale=0.5]

\tikzstyle{every path}=[line width=2pt]


\draw[line width=0.5pt,orange] (-3,-3) grid[xstep=1, ystep=1]  (3,3);


\draw[draw=gray!80,->] (0,-3) + (0,-0.5cm)  -- (0,3) -- +(0,0.5cm) node[above right] {$\Im z$};
\draw[draw=gray!80,->] (-3,0) +(-0.5cm,0) -- (3,0) -- +(0.5cm,0) node[below right] {$\Re z$};


\draw [line width=0.5pt,color=blue,dashed] (-1cm,0cm) arc [start angle=-180,end angle=180,x radius=1cm, y radius=1cm];


\filldraw  (0,1) circle[radius=4pt] node[above right] {$i$};
\filldraw  (1,0) circle[radius=4pt] node[above right] {$1$};
\filldraw  (3,1) circle[radius=4pt] node[above right] {$3+i$};
\filldraw  (1,-2) circle[radius=4pt] node[below right] {$1-2i$};
\filldraw  (-2,2) circle[radius=4pt] node[above left] {$-2+2i$};
\filldraw  (-2.5,-3) circle[radius=4pt] node[below left] {$-\frac{5}{2}-3i$};
\filldraw  (0,0) circle[radius=4pt] node[below left,yshift=0.0ex,xshift=0.0ex] {$0$};

\end{tikzpicture}
\end{center}
\caption{Complex plane with dashed unit circle around origin and some points}
\label{2018-m-f-cp}
\end{marginfigure}

The addition and multiplication of two complex numbers represented by $\begin{pmatrix}x,y\end{pmatrix}$ and $\begin{pmatrix}u,v\end{pmatrix}$ with $x,y,u,v \in {\Bbb R}$
are then defined by
\begin{equation}
\begin{split}
\begin{pmatrix}x,y\end{pmatrix} + \begin{pmatrix}u,v\end{pmatrix} = \begin{pmatrix}x+u,y+v\end{pmatrix},\\
\begin{pmatrix}x,y\end{pmatrix} \cdot \begin{pmatrix}u,v\end{pmatrix} = \begin{pmatrix}xu-yv,xv+yu\end{pmatrix},
\end{split}
\end{equation}
and the neutral elements for addition and multiplication are $\begin{pmatrix}0,0\end{pmatrix}$ and $\begin{pmatrix}1,0\end{pmatrix}$, respectively.

We shall also consider the {\em extended plane}  ${\overline{\Bbb C}} = {\Bbb C} \cup \{\infty \}$
\index{extended plane}
consisting of the entire complex plane ${\Bbb C}$ {\em together} with the point ``$\infty$''
representing infinity.
Thereby, $\infty$
is introduced as an ideal element,
completing the one-to-one (bijective) mapping $w=\frac{1}{z}$,
which otherwise would have no image at $z=0$, and no pre-image (argument)
at $w=0$.

\subsection{Multi-valued relationships, branch points, and branch cuts}

Earlier we encountered problems with the square root function on complex numbers.
We shall use this function as a sort of ``Rosetta stone'' for an understanding of conceivable ways of coping with nonunique functional values.
Note that even in the real case there  are issues:
for positive real numbers we can uniquely define the square root function $y=\sqrt{x}$, $x\in \mathbb{R}$, $x\ge 0$ by its inverse -- that is,
the square function -- such that $y^2=y\cdot y =x$.
However, this latter way of defining the square root function is no longer uniquely possible if we allow negative arguments $x\in \mathbb{R}$, $x < 0$,
as this would render the value assignment nonunique:  $(-y)^2=(-y)\cdot (-y) =x$: we would essentially end up with two ``branches'' of $\sqrt : x \mapsto \{y,-y\}$
meeting at the origin,
as depicted in Figure~\ref{2018-m-ca-sqrb}.
\begin{marginfigure}
{\color{black}
\begin{center}
\begin{tikzpicture}[ scale=0.6,
 declare function={
    func1(\x)= (sqrt(\x ));
    func2(\x)= (-sqrt(\x ));
                  } ]

\tikzstyle{every path}=[line width=3pt]

\begin{axis}[axis lines=middle, draw=gray!80,axis equal, enlargelimits=true ,
xtick={-1,0,1,2},
ytick={-1,0,1,2},
ticklabel style = {font=\Large },
every axis x label/.style={
    at={(ticklabel* cs:1)},
    anchor=west,
    font=\huge ,
},
every axis y label/.style={
    at={(ticklabel* cs:1)},
    anchor=south,
    font=\huge ,
},
xlabel={$x$},
ylabel={$y^2=x$}
]

\addplot [
orange,
domain=0:2,
line width=2pt
]  {func1(x)};

\addplot [
blue,dashed,
domain=0:2,
line width=2pt
]  {func2(x)};

\end{axis}
\end{tikzpicture}
\end{center}
}
\caption{The two branches of a nonunique value assignment $y(x)$ with $x =\left[y(x)\right]^2$.
}
\label{2018-m-ca-sqrb}
\end{marginfigure}

It has been mentioned earlier that the Riemann surface of a function is an extended complex plane which makes the function a function; in particular,
it guarantees that a function is uniquely defined; that is, it renders a unique complex value on that the Riemann surface (but not necessarily on the complex plane).

To give an example mentioned earlier: the square root function $\sqrt{z} =
\sqrt{\vert z\vert } \exp \left(i\varphi / 2 +  i \pi  k \right)$, with $k\in \mathbb{Z}$, or in this case rather $k\in \{0,1\}$,
of a complex number cannot be uniquely defined on the complex plane {\it via} its inverse function.
Because the
inverse (square) function of square root function is not injective, as it
maps {\em different} complex numbers,
represented by different arguments
$z=r \exp(i\varphi )$
and
$z'= r \exp[i(\pi + \varphi )]$
to {\em the same} value
$z^2=r^2 \exp(2i\varphi )= r^2 \exp[2i(\pi + \varphi )]=(z')^2$
on the complex plane.
So, the ``inverse'' of the square function $z^2=(z')^2$ is nonunique: it could be either one of the two different numbers $z$ and $z'$.

In order to establish uniqueness for complex extensions of the square root function
one assumes that its domain is an intertwine
of two different ``branches;'' each branch being a copy of the complex plane:
the first branch ``covers'' the complex half-space with
$-\frac{\pi}{2} < \textrm{arg}(z) \le \frac{\pi}{2}$, whereas the second one ``covers'' the complex half-space with
$ \frac{\pi}{2} < \textrm{arg}(z) \le -\frac{\pi}{2}$.
They are intertwined in the branch cut starting from the origin, spanned along the negative real axis.

Functions like the square root functions are called {\em multi-valued functions} (or {\em multifunctions}).
\index{multifunction}
\index{multi-valued function}
They require  Riemann surfaces which are {\em not} simply connected.
An argument
$z$ of the function $f$ is called
{\em branch point}
\index{branch point}
if there is a closed curve $C_z$ around $z$ whose image $f(C_z)$ is an open curve.
That is, the multifunction $f$ is discontinuous in $z$.
Intuitively speaking, branch points are the points where the various sheets of a multifunction come together.

A {\em branch cut} is a curve (with ends possibly open, closed, or half-open)
in the complex plane across which an analytic multifunction is discontinuous.
Branch cuts are often taken as lines.

 \section{Riemann surface}
Suppose $f(z)$ is a multi-valued function.
Then the various $z$-surfaces on which $f(z)$ is uniquely defined,
together with their connections through branch points and branch cuts,
constitute the Riemann surface of $f$.
The required leaves are called {\em Riemann sheet}.
\index{sheet}

A point $z$ of the function $f(z)$ is called a {\em branch point of order $n$} if through it and through the associated cut(s)
$n+1$ Riemann sheets are connected.

{
\color{blue}
\bexample
A good strategy for finding the Riemann surface of a function is to figure out what the {\em inverse} function does: if the inverse function is not injective on the complex plane, then the function is nonunique.
For example, in the case of the square root function, the inverse function is the square
$w : z \mapsto z^2= r^2 \exp [2 i \textrm{arg}(z)]$ which covers the complex plane twice during the variation of the principal value
$-\pi < \textrm{arg} (z)\le +\pi$.
Thus the Riemann surface of an inverse function of the square function, that is, the square function has to have two sheets to be able to cover
the original complex plane of the argument. otherwise, with the exception of the origin, the square root would be nonunique,
and the same point on the complex $w$ plane would correspond to two distinct points
in the original $z$-plane.
This is depicted in Figure~\ref{f-2018-m-ca-csrt}, where $c\neq d$ yet $c^2=d^2$.

{\color{black}
\begin{figure}
\begin{center}
\begin{tabular}{ccccccc}
\begin{tikzpicture}  [scale=0.25]

\tikzstyle{every path}=[line width=2pt]


\draw[draw=gray!80,->] (0,-5) + (0,-0.5cm)  -- (0,5) -- +(0,0.5cm) node[above right] {$\Im z$};
\draw[draw=gray!80,->] (-5,0) +(-0.5cm,0) -- (5,0) -- +(0.5cm,0) node[below right] {$\Re z$};


\draw [color=orange,->] (0cm,-2cm) arc [start angle=-90,end angle=87,x radius=2cm, y radius=2cm];
\draw [color=blue,dashed,->] (0cm,2cm) arc [start angle=90,end angle=267,x radius=2cm, y radius=2cm];


\filldraw  (0,-2) circle[radius=4pt] node[below right] {$a$};
\filldraw  (0,2) circle[radius=4pt] node[above right] {$b$};
\filldraw [color=orange] (2,0) circle[radius=4pt] node[above right] {$c$};
\filldraw [color=blue] (-2,0) circle[radius=4pt] node[above left] {$d$};
\end{tikzpicture}
&
\begin{tikzpicture}  [scale=0.25]

\tikzstyle{every path}=[line width=2pt]


\draw[draw=gray!80,->] (0,-5) + (0,-0.5cm)  -- (0,5) -- +(0,0.5cm) node[above right] {$\Im w$};
\draw[draw=gray!80,->] (-5,0) +(-0.5cm,0) -- (5.3,0) -- +(0.5cm,0) node[below right] {$\Re w$};


\draw [color=orange,->] (-4cm,0cm) arc [start angle=-180,end angle=177,x radius=4cm, y radius=4cm];


\filldraw  (-4,0) circle[radius=6pt] node[above right] {$b^2$};
\filldraw [draw=gray!90] (-4,0) circle[radius=3pt] node[below right] {$a^2$};
\filldraw [color=orange] (4,0) circle[radius=4pt] node[above right] {$c^2$};
\end{tikzpicture}
&
\begin{tikzpicture}  [scale=0.25]

\tikzstyle{every path}=[line width=2pt]


\draw[draw=gray!80,->] (0,-5) + (0,-0.5cm)  -- (0,5) -- +(0,0.5cm) node[above right] {$\Im w$};
\draw[draw=gray!80,->] (-5,0) +(-0.5cm,0) -- (5.3,0) -- +(0.5cm,0) node[below right] {$\Re w$};


\draw [color=blue,dashed,->] (-4cm,0cm) arc [start angle=-177,end angle=180,x radius=4cm, y radius=4cm];


\filldraw  (-4,0) circle[radius=6pt] node[above right] {$a^2$};
\filldraw [draw=gray!90] (-4,0) circle[radius=3pt] node[below right] {$b^2$};
\filldraw [color=blue] (4,0) circle[radius=4pt] node[above right] {$d^2$};
\end{tikzpicture}
\\
$z= r e^{  i \textrm{arg}(z)}$
&
$w(z)= z^2= r^2 e^{ 2 i \textrm{arg}(z)}$
&
$w(z)= z^2= r^2 e^{ 2 i \textrm{arg}(z)}$
\\
$-\pi < \textrm{arg}(z) \le +\pi$
&
$-\frac{\pi}{2} < \textrm{arg}(z) \le +\frac{\pi}{2}$
&
$+\frac{\pi}{2} < \textrm{arg}(z) \le -\frac{\pi}{2}$
\\
\end{tabular}
\end{center}
\label{f-2018-m-ca-csrt}
\caption{Sketch of the Riemann surface of the square root function, requiring two sheets with the origin as branch point,
as argued from the inverse (in this case square) function.}
\end{figure}
}

\eexample
}

\section{Differentiable, holomorphic (analytic) function}

Consider the function $f(z)$ on the domain $G\subset {\rm Domain}(f)$.

$f$ is called {\em differentiable} at the point $z_0$ if the  differential quotient
\index{differentiable function}
\begin{equation}
\left. {df\over dz}\right|_{z_0}=
\left. f'(z)\right|_{z_0} =
\left. {\partial f\over \partial  x}\right|_{z_0} =
\left. {1\over i}{\partial f\over \partial y}\right|_{z_0}
\end{equation}
 exists.
\index{differentiable}

If $f$ is (arbitrarily often) differentiable in the  domain $G$ it is called {\em holomorphic.}
\index{holomorphic function}
We shall state without proof that, if a holomorphic function is differentiable, it is also arbitrarily often differentiable.

If a function  can be expanded as a convergent power series, like $f (z) = \sum_{n=0}^\infty a_n z^n$, in the  domain $G$ then it is called
{\em analytic} in the domain $G$.
\index{analytic function}
We state without proof that holomorphic functions are analytic, and {\it vice versa};
that is the terms ``holomorphic'' and ``analytic'' will be used synonymously.

 \section{Cauchy-Riemann equations}
\index{Cauchy-Riemann equations}
 The function $f(z)=u(z)+iv(z)$ (where $u$ and $v$ are real valued functions) is
 {analytic or holomorphic} if and only if
 ($a_b=\partial a/\partial b$)
 \begin{equation}
u_x=v_y, \qquad u_y=-v_x .
\end{equation}
{\color{OliveGreen}
\bproof
For a proof, differentiate along the real, and then along the imaginary axis,
taking
\begin{equation}
\begin{split}
f'(z) =\lim_{x\rightarrow 0}\frac{f(z+x)-f(z)}{x}=\frac{\partial f}{\partial x}=   \frac{\partial u}{\partial x}+i\frac{\partial v}{\partial x},\\
\textrm { and } f'(z) =\lim_{y\rightarrow 0}\frac{f(z+iy)-f(z)}{iy}=\frac{\partial f}{\partial iy}= -i\frac{\partial f}{\partial y}=   -i\frac{\partial u}{\partial y}+ \frac{\partial v}{\partial y}.
\end{split}
\end{equation}
For $f$ to be analytic, both partial derivatives have to be identical, and thus $\frac{\partial f}{\partial x}=\frac{\partial f}{\partial iy}$, or
\begin{equation}
\frac{\partial u}{\partial x}+i\frac{\partial v}{\partial x}=   -i\frac{\partial u}{\partial y}+ \frac{\partial v}{\partial y}.
\end{equation}
By comparing the real and imaginary parts of this equation, one obtains the two real Cauchy-Riemann equations
\begin{equation}
\begin{split}
\frac{\partial u}{\partial x}=   \frac{\partial v}{\partial y},\\
\frac{\partial v}{\partial x}=   -\frac{\partial u}{\partial y}.
\end{split}
\end{equation}
\eproof
}

  \section{Definition analytical function}
{\color{OliveGreen}
\bproof
If $f$ is analytic in $G$, all derivatives of $f$ exist, and all mixed derivatives are independent on the order of differentiations.
Then the  Cauchy-Riemann equations  imply that
\begin{equation}
\begin{split}
\frac{\partial }{\partial x}\left(\frac{\partial u}{\partial x}\right)=
\frac{\partial }{\partial x}\left(\frac{\partial v}{\partial y}\right)=
\frac{\partial }{\partial y}\left(\frac{\partial v}{\partial x}\right)=
 -\frac{\partial }{\partial y}\left(\frac{\partial u}{\partial y}\right),     \\
\textrm{ and }\frac{\partial }{\partial y}\left(\frac{\partial v}{\partial y}\right)=
\frac{\partial }{\partial y}\left(\frac{\partial u}{\partial x}\right)=
\frac{\partial }{\partial x}\left(\frac{\partial u}{\partial y}\right)=
-\frac{\partial }{\partial x}\left(\frac{\partial v}{\partial x}\right)
,
\end{split}
\end{equation}
and thus
\eproof
}
\begin{equation}
 \left({\partial^2\over \partial x^2}
 + {\partial^2\over \partial y^2}\right)u=0      \textrm{, and }
 \left({\partial^2\over \partial x^2}
 + {\partial^2\over \partial y^2}\right)v=0 .
 \end{equation}

 If $f=u+iv$ is analytic in $G$, then the lines of constant $u$ and $v$ are orthogonal.

 {\color{OliveGreen}
\bproof
 The tangential vectors of the lines of constant $u$ and $v$ in the two-dimensional complex plane are defined by the two-dimensional nabla operator
\index{nabla operator}
$\nabla u(x,y)$ and $\nabla v(x,y)$.
Since, by the  Cauchy-Riemann equations $u_x=v_y$ and $u_y=-v_x$
\begin{equation}
\nabla u(x,y)\cdot \nabla v(x,y)
=
\left(
\begin{array}{c}
u_x\\
u_y
\end{array}
\right)
\cdot
\left(
\begin{array}{c}
v_x\\
v_y
\end{array}
\right)
=  u_x  v_x + u_y v_y   =   u_x  v_x  + (-v_x) u_x =0
\end{equation}
these tangential vectors are normal.
\eproof
}

$f$
is
{\em angle (shape)  preserving}
\index{conformal map}
{\em conformal} if and only if it is holomorphic and its derivative is everywhere non-zero.

 {\color{OliveGreen}
\bproof

Consider an analytic function $f$ and an arbitrary path $C$ in the complex plane of the arguments parameterized
by $z(t)$, $t\in {\Bbb R}$.
The image of $C$ associated with $f$ is  $f(C) = C': f(z(t))$, $t\in {\Bbb R}$.

The tangent vector of $C'$ in $t=0$ and $z_0=z(0)$ is
\begin{equation}
\begin{split}
\left. \frac{d }{dt} f(z(t))\right|_{t=0}
=
\left. \frac{d }{dz} f(z)\right|_{z_0}
\left. \frac{d }{dt} z(t)\right|_{t=0}
=
\lambda_0
e^{i\varphi_0}
\left. \frac{d }{dt} z(t)\right|_{t=0} .
\end{split}
\end{equation}
Note that the first term $\left. \frac{d }{dz} f(z)\right|_{z_0}$
is independent of the curve $C$ and only depends on $z_0$.
Therefore, it can be written as a product of a  squeeze (stretch) $\lambda_0 $
and a rotation $e^{i\varphi_0}$.
This is independent of the curve; hence
two curves $C_1$ and $C_2$ passing through $z_0$ yield the same
transformation of the image $\lambda_0
e^{i\varphi_0}$.
\eproof
}

 \section{Cauchy's integral theorem}
\index{Cauchy's integral theorem}
 If $f$ is analytic on $G$ and on its borders $\partial G$, then any closed line integral of $f$ vanishes
 \begin{equation}
\oint_{\partial G}f(z)dz=0
.
\label{2018-m-ch-ca-cit}
\end{equation}

No proof is given here.

In particular,
 $\oint_{C\subset \partial G}f(z)dz
$ is independent of the particular curve and only depends on the initial and the endpoints.

 {\color{OliveGreen}
\bproof
 For a proof, subtract two line integral which follow arbitrary paths  $C_1$ and $C_2$ to a common initial and end point,
and which have the same integral kernel.
Then reverse the integration direction of one of the line integrals.
According to Cauchy's integral theorem, the resulting integral over the closed loop has to vanish.
\eproof
}

Often it is useful to parameterize a contour integral by some form of
 \begin{equation}
\int_{C}f(z)dz= \int_{a}^b f(z(t))\frac{dz(t)}{dt} dt.
\end{equation}

{
\color{blue}
\bexample
Let $f(z) = 1/z$ and $C: z(\varphi )=R e^{i\varphi}$, with $R>0$ and $-\pi < \varphi \le \pi$. Then
\begin{equation}
\begin{split}
\oint_{\vert z\vert =R}
f (z) dz
 =
\int_{-\pi}^\pi
f (z(\varphi ))\frac{dz(\varphi )}{d\varphi } d\varphi   \\
=
\int_{-\pi}^\pi
\frac{1}{R e^{i\varphi}}R \, i\, e^{i\varphi} d\varphi   \\
 =
\int_{-\pi}^\pi
i\varphi   \\
=    2\pi i
\end{split}
\end{equation}
is independent of $R$.
\eexample
}

 \section{Cauchy's integral formula}
\index{Cauchy's integral formula}

If $f$ is analytic on $G$ and on its borders $\partial G$, then
\begin{equation}
f(z_0)={1\over 2\pi i}\oint_{\partial G}{f(z)\over z-z_0}dz
 .
\label{2012-m-ch-ca-cif}
\end{equation}

No proof is given here.

Note that because of Cauchy's integral formula, analytic
functions have an integral representation.
This  has far-reaching consequences:
because analytic functions have integral
representations, their higher derivatives also have integral representations.
And, as a result,
if a function has one complex derivative, then it has infinitely many complex derivatives.
This statement can be formally expressed by
the {\em generalized Cauchy  integral formula} or, by another term,
by {\em Cauchy's differentiation formula}
 \index{Cauchy's differentiation formula}
 \index{generalized Cauchy integral formula}
states that if $f$ is analytic on $G$ and on its borders $\partial G$, then
\begin{equation}
f^{(n)}(z_0)={n!\over 2\pi i}\oint_{\partial G}{f(z)\over
 (z-z_0)^{n+1}}dz  .
\label{2012-m-ch-cagcif}
\end{equation}

No proof is given here.

{
\color{blue}
\bexample
Cauchy's integral formula presents a powerful method to compute integrals.
Consider the following examples.

\begin{enumerate}

\item First,
let us calculate  $$\oint_{\vert z\vert =3} \frac{3z+2}{z(z+1)^3} dz.$$
The kernel has two poles at $z=0$ and $z=-1$ which are both inside the domain of the contour defined by $\vert z\vert =3$.
By using Cauchy's integral formula we obtain for ``small'' $\epsilon$
\begin{equation}
\begin{split}
\oint_{\vert z\vert =3} \frac{3z+2}{z(z+1)^3} dz  \\
 =\oint_{\vert z\vert =\epsilon} \frac{3z+2}{z(z+1)^3} dz    + \oint_{\vert z+1\vert =\epsilon} \frac{3z+2}{z(z+1)^3} dz \\
  =\oint_{\vert z\vert =\epsilon} \frac{3z+2}{(z+1)^3} \frac{1}{z} dz    + \oint_{\vert z+1\vert =\epsilon} \frac{3z+2}{z}\frac{1}{(z+1)^3} dz  \\
  =
\left.
\frac{2\pi i}{0!}
\underbrace{\frac{d^0}{dz^0}}_{1}
\frac{3z+2}{(z+1)^3}
\right|_{z=0}
+
\left.
\frac{2\pi i}{2!}
\frac{d^2}{dz^2}
\frac{3z+2}{z}
\right|_{z=-1} \\
  =
\left.
\frac{2\pi i}{0!}
\frac{3z+2}{(z+1)^3}
\right|_{z=0}
+
\left.
\frac{2\pi i}{2!}
\underbrace{\frac{d^2}{dz^2}
\frac{3z+2}{z}}_{4(-1)^2 z^{-3}}
\right|_{z=-1} \\
 = 4\pi i - 4 \pi i  =0.
\end{split}
\end{equation}

\item
Consider
\begin{equation}
\begin{split}
\oint_{\vert z\vert =3}
\frac{e^{2z}}{(z+1)^4 }dz\\
  =
\frac{2\pi i}{3!}
\frac{3!}{2\pi i}
\oint_{\vert z\vert =3}
\frac{e^{2z}}{(z- (-1))^{3+1} }dz  \\
  =
\frac{2\pi i}{3!}
\frac{d^3}{dz^3}
\left| e^{2z} \right|_{z=-1}  \\
  =
\frac{2\pi i}{3!}
2^3  \left| e^{2z} \right|_{z=-1}    \\
 =
\frac{8 \pi i e^{-2}}{3}.
\end{split}
\end{equation}

\end{enumerate}
\eexample
}

Suppose $g(z)$ is a function with a pole of order $n$ at the point
 $z_0$; that is
 \begin{equation}
g(z)= {f(z)\over (z-z_0)^n},
\end{equation}
 where $f(z)$ is an analytic function. Then,
 \begin{equation}
\oint_{\partial G}g(z)dz={2\pi i\over (n-1)!}f^{(n-1)}(z_0) .
\end{equation}

 \section{Series representation of complex differentiable functions}

As a consequence of Cauchy's (generalized) integral formula,
analytic functions have power series representations.

{\color{OliveGreen}
\bproof

For the sake of a proof,
we shall recast the denominator   $z-z_0$
in Cauchy's integral formula
(\ref{2012-m-ch-ca-cif})
as a geometric series as follows  (we shall assume that $|z_0 - a| < |z - a|$)
\begin{equation}
\begin{split}
\frac{1}{z - z_0}=
\frac{1}{(z - a) - (z_0 - a)}\\=
\frac{1}{ (z - a)} \left[\frac  {1}{1 - \frac{ z_0 - a }{ z - a }}\right]\\
=
\frac{1}{(z - a)} \left[ \sum_{n=0}^\infty \frac{ (z_0 - a)^n }{ (z - a)^n }\right]\\
=
 \sum_{n=0}^\infty \frac{ (z_0 - a)^n }{ (z - a)^{n+1} } .
\label{2012-m-ch-rgs}
\end{split}
\end{equation}
By substituting this in
Cauchy's integral formula
(\ref{2012-m-ch-ca-cif})
and using
Cauchy's generalized integral formula
(\ref{2012-m-ch-cagcif})
yields    an expansion of the analytical function $f$ around $z_0$ by a power series
\begin{equation}
\begin{split}
f(z_0)={1\over 2\pi i}\oint_{\partial G}{f(z)\over z-z_0}dz \\
={1\over 2\pi i}\oint_{\partial G} f(z) \sum_{n=0}^\infty \frac{ (z_0 - a)^n }{ (z - a)^{n+1}} dz \\
=  \sum_{n=0}^\infty (z_0 - a)^n  {1\over 2\pi i}\oint_{\partial G}  \frac{ f(z) }{ (z - a)^{n+1}} dz \\
=  \sum_{n=0}^\infty  \frac{ f^{n}(z_0) }{ n! }   (z_0 - a)^n  .
\end{split}
\label{2012-m-ch-ca-ssolut}
\end{equation}

\eproof
}

 \section{Laurent and Taylor series}
 \index{Laurent series}
 \index{Taylor series}


Every function $f$ which is analytic in a concentric region
$R_1< \vert z-z_0\vert <R_2$ can in this region be uniquely
 written as a {\em Laurent series}
\begin{equation}
\begin{split}
f(z)=\sum_{k=-\infty}^\infty (z-z_0)^k a_k \text{, with coefficients}
\\
a_k={1\over 2\pi i}\oint_C (\chi -z_0)^{-k-1}f(\chi ) d\chi .
\label{011-m-ch-ca-else1}
\end{split}
\end{equation}
The closed contour $C$ must be in the concentric region.

The coefficient $a_{-1}$ is called the {\em residue} and denoted by ``$\text{Res}$:''
\index{residue}
\begin{equation}
\text{Res}(f(z_0)) \stackrel{{\tiny \textrm{ def }}}{=} a_{-1}={1\over 2\pi i}\oint_C f(\chi )d\chi
.
\label{011-m-ch-ca-else2}
\end{equation}

{\color{OliveGreen}
\bproof

For  a proof,   as in Eqs. (\ref{2012-m-ch-rgs})
we shall recast $(a-b)^{-1}$   for $|a|>|b|$
as a geometric series
\begin{equation}
\begin{split}
\frac{1}{a - b}=
\frac{1}{a} \left(\frac  {1}{1 - \frac{ b }{ a }}\right)
=
\frac{1}{a} \left( \sum_{n=0}^\infty \frac{ b^n }{  a^n }\right)
=
\sum_{n=0}^\infty \frac{ b^n }{  a^{n+1} } \\
[\textrm{substitution }  n+1 \rightarrow -k, \, n \rightarrow -k-1\, k \rightarrow -n-1]
=
\sum_{k=-1}^{-\infty} \frac{ a^k }{  b^{k+1} },
\end{split}
\end{equation}
and, for   $|a|<|b|$,
\begin{equation}
\begin{split}
\frac{1}{a - b}= - \frac{1}{b - a}=
-
\sum_{n=0}^\infty \frac{ a^n }{  b^{n+1} } \\
[\textrm{substitution }  n+1 \rightarrow -k, \, n \rightarrow -k-1\, k \rightarrow -n-1]
=
-
\sum_{k=-1}^{-\infty} \frac{ b^k }{  a^{k+1} }.
\end{split}
\end{equation}
Furthermore since $a+b = a-(-b)$,
we obtain, for $|a|>|b|$,
\begin{equation}
\frac{1}{a + b}
=
\sum_{n=0}^\infty (-1)^n \frac{ b^n }{  a^{n+1} }
=
\sum_{k=-1}^{-\infty}(-1)^{-k-1} \frac{ a^k }{  b^{k+1} }
=
-
\sum_{k=-1}^{-\infty}(-1)^{ k } \frac{ a^k }{  b^{k+1} }
,
\end{equation}
and,  for $|a|<|b|$,
\begin{equation}
\begin{split}
\frac{1}{a + b}
=
-
\sum_{n=0}^\infty (-1)^{n+1} \frac{ a^n }{  b^{n+1} }
=
\sum_{n=0}^\infty (-1)^{n} \frac{ a^n }{  b^{n+1} }  \\
=
\sum_{k=-1}^{-\infty}(-1)^{-k-1} \frac{ b^k }{  a^{k+1} }
=
-
\sum_{k=-1}^{-\infty}(-1)^{ k } \frac{ b^k }{  a^{k+1} }
.
\end{split}
\end{equation}

Suppose that some function $f(z)$ is analytic in an annulus bounded by the radius $r_1$ and $r_2 > r_1$.
By substituting this in
Cauchy's integral formula
(\ref{2012-m-ch-ca-cif})
for an annulus bounded by the radius $r_1$ and $r_2 > r_1$
(note that the orientations of the boundaries with respect to the annulus are opposite,
rendering a relative factor ``$-1$'')
and using
Cauchy's generalized integral formula
(\ref{2012-m-ch-cagcif})
yields an expansion of the analytical function $f$ around $z_0$ by the Laurent series
for a point $a$ on the annulus; that is,
for a path  containing the point $z$ around a circle with radius
$r_1$, $|z - a| < |z_0 - a|$;
likewise,
for a path  containing the point $z$ around a circle with radius
$r_2 > a > r_1$, $|z - a| > |z_0 - a|$,
\begin{equation}
\begin{split}
f(z_0)=
{1\over 2\pi i}\oint_{r_1}{f(z)\over z-z_0}dz
-
{1\over 2\pi i}\oint_{r_2}{f(z)\over z-z_0}dz
\\
=
{1\over 2\pi i}\left[
\oint_{r_1} f(z) \sum_{n=0}^\infty \frac{ (z_0 - a)^n }{ (z - a)^{n+1}} dz
+
\oint_{r_2} f(z) \sum_{n=-1}^{-\infty} \frac{ (z_0 - a)^n }{ (z - a)^{n+1}} dz
\right]
\\
=
{1\over 2\pi i}\left[
\sum_{n=0}^\infty (z_0 - a)^n  \oint_{r_1}   \frac{  f(z) }{ (z - a)^{n+1}} dz
+
\sum_{n=-1}^{-\infty} (z_0 - a)^n \oint_{r_2} \frac{ f(z) }{ (z - a)^{n+1}} dz
\right]
\\
=
\sum_{n = -\infty}^\infty (z_0 - a)^n  \left[ {1\over 2\pi i} \oint_{r_1\le r \le r_2} \frac{ f(z) }{ (z - a)^{n+1}}  dz \right] .
\end{split}
\label{2012-m-ch-ca-ssolutlarent}
\end{equation}

\eproof
}

Suppose that  $g(z)$ is a function with a pole of order $n$ at the point
 $z_0$; that is
 $g(z)= {h(z)/ (z-z_0)^n}$ ,
where $h(z)$ is an analytic function. Then the terms  $k\le -(n+1)$
vanish in the Laurent series.
This follows from  Cauchy's integral formula
\begin{equation}
a_k ={1\over 2\pi i}\oint_C(\chi -z_0)^{-k-n-1}h(\chi )d\chi =0
\end{equation}
 for $-k-n-1\ge 0$.

Note that, if $f$ has a simple pole (pole of order 1) at $z_0$,
then it can be rewritten into $f(z)=g(z)/(z-z_0)$ for some analytic function $g(z)=(z-z_0) f(z)$
that remains after the singularity has been ``split'' from $f$.
Cauchy's integral formula (\ref{2012-m-ch-ca-cif}),
and the residue can be rewritten as
\begin{equation}
a_{-1}={1\over 2\pi i}
\oint_{\partial G}{g(z)\over z-z_0}dz =
g(z_0)
 .
\label{2012-m-ch-ca-cif1}
\end{equation}
For poles of higher order, the generalized Cauchy integral formula
(\ref{2012-m-ch-cagcif}) can be used.

Suppose that $f(z)$ is analytic at and in a region $G$ ``around'' $z_0$.
Then the Laurent series (\ref{011-m-ch-ca-else1})
``turns into'' a {\em Taylor series}
\index{Taylor series}
expansion of $f(z)$:
\begin{equation}
f(z) =
\sum_{k=0}^\infty \frac{f^{(k)}(z_0)}{k!} (z - z_0)^k
\text{, with } z\in G
.
\label{2019-mm-ca-ts}
\end{equation}

{\color{OliveGreen}
\bproof
For a proof relative to the validity of the Laurent series (\ref{011-m-ch-ca-else1}),
suppose that $f(z)$ is analytic at and ``in a region around'' $z_0$,
and note the following:
\begin{itemize}

\item[(i)]
Because of Cauchy's integral theorem~(\ref{2018-m-ch-ca-cit}),
\index{Cauchy's integral theorem}
\begin{equation}
a_{k<0} =
{1\over 2\pi i}\oint_C(\chi -z_0)^{|k|-1}f(\chi )d\chi =0 ,
\end{equation}
since, for
$k<0$, $-k-1 = |k|-1\ge 0$, and, therefore,
$(\chi -z_0)^{|k|-1}f(\chi )$ is analytic, too.

\item[(ii)]
Because of Cauchy's integral formula~(\ref{2012-m-ch-ca-cif}),
\index{Cauchy's integral formula}
\begin{equation}
a_{k=0} =
{1\over 2\pi i}\oint_C \frac{f(\chi )}{\chi -z_0}d\chi = f(z_0).
\end{equation}

\item[(iii)]
Because of the generalized Cauchy  integral formula (aka
Cauchy's differentiation formula)~(\ref{2012-m-ch-cagcif}),
 \index{Cauchy's differentiation formula}
 \index{generalized Cauchy integral formula}
\begin{equation}
a_{k>0} =
{1\over 2\pi i}\oint_C \frac{f(\chi )}{(\chi -z_0)^{k+1}} d\chi
= \frac{f^{(k)}(z_0)}{k!}.
\end{equation}

\end{itemize}

\eproof
}

 \section{Residue theorem}
 \index{residue theorem}
\label{2021-mm-ch-ca-rt}

Suppose $f$ is analytic on a  simply connected open subset $G$
with the exception of finitely many (or denumerably many) points  $z_i$.
Then,
\begin{equation}
\oint_{\partial G} f(z)dz=2\pi i \sum_{z_i} {\rm Res}f(z_i) .
\end{equation}

No proof is given here.
\marginnote[0mm]{For proofs and additional information see  Chapter~6 of~\bibentry{Brown-Churchill}.}

The residue theorem presents a powerful tool for calculating integrals, both real and complex.
Let us first mention a rather general case of a situation often used.
Suppose we are interested in the integral
  $$I=\int_{-\infty}^{\infty} R(x) dx$$
with rational kernel $R$; that is, $R(x)= P(x)/Q(x)$,
where $P(x)$ and $Q(x)$
are polynomials (or can at least be bounded by a polynomial) with no common root (and therefore factor).
Suppose further that the degrees of the polynomial are
$$
\textrm{deg } P(x) \le
\textrm{deg } Q(x) -2.
$$
This condition is needed to assure that the additional upper or lower path we want to add when completing the contour
does not contribute; that is, vanishes.

Now first let us analytically continue $R(x)$ to the complex plane $R(z)$; that is,
$$I=\int_{-\infty}^{\infty} R(x) dx =\int_{-\infty}^{\infty} R(z) dz.$$
Next let us close the contour by adding a (vanishing) curve integral
$$\int_{\curvearrowleft} R(z) dz =
0$$
in the upper (lower) complex plane
$$I=\int_{-\infty}^{\infty} R(z) dz +\int_{\curvearrowleft} R(z) dz=\oint_{\rightarrow \& \curvearrowleft} R(z) dz.$$
The added integral vanishes because
it can be approximated by
$$\left| \int_{\curvearrowleft} R(z)  dz\right| \le \lim_{r\rightarrow \infty} \left(\frac{\textrm{const.}}{r^2} \pi r \right) =0.$$

With the contour closed the residue theorem can be applied  for an evaluation of $I$; that is,
$$I= 2\pi i \sum_{z_i} {\rm Res}R(z_i)$$
for all singularities $z_i$ in the region enclosed by ``$\rightarrow \& \curvearrowleft$. ''

Let us consider some examples.

{
\color{blue}
\bexample

\renewcommand{\labelenumi}{(\roman{enumi})}
\begin{enumerate}

\item  Consider   $$I=\int_{-\infty}^{\infty}\frac{dx}{x^2+1} .$$
The analytic continuation of the kernel and the addition with vanishing a semicircle ``far away'' closing the integration path
in the {\em upper} complex half-plane of $z$ yields
\begin{equation}
\begin{split}
I=\int_{-\infty}^{\infty}\frac{dx}{x^2+1} \\
     =  \int_{-\infty}^{\infty}\frac{dz}{z^2+1}  \\
     = \int_{-\infty}^{\infty}\frac{dz}{z^2+1} + \int_{\curvearrowleft} \frac{dz}{z^2+1} \\
     = \int_{-\infty}^{\infty}\frac{dz}{(z+i)(z-i)} +  \int_{\curvearrowleft} \frac{dz}{(z+i)(z-i)} \\
     = \oint\frac{1}{(z-i)} f(z) dz \textrm{ with } f(z)=\frac{1}{(z+i)} \\
     = 2\pi i \textrm{Res}\left.\left(\frac{1}{(z+i)(z-i)} \right)\right|_{z=+i} \\
     = 2\pi i f(+i)  \\
     = 2\pi i \frac{1}{(2i)}     = \pi.   \\
\end{split}
\end{equation}
Here, Equation (\ref{2012-m-ch-ca-cif1}) has been used.
Closing the integration path
in the {\em lower} complex half-plane of $z$ yields (note that in this case the contour integral is negative because of the path orientation)
\begin{equation}
\begin{split}
I=\int_{-\infty}^{\infty}\frac{dx}{x^2+1} \\
     =  \int_{-\infty}^{\infty}\frac{dz}{z^2+1}  \\
     = \int_{-\infty}^{\infty}\frac{dz}{z^2+1}  + \int_{\textrm{lower path}} \frac{dz}{z^2+1} \\
     = \int_{-\infty}^{\infty}\frac{dz}{(z+i)(z-i)} + \int_{\textrm{lower path}} \frac{dz}{(z+i)(z-i)} \\
     = \oint\frac{1}{(z+i)} f(z) dz \textrm{ with } f(z)=\frac{1}{(z-i)} \\
     = -2\pi i \textrm{Res}\left.\left(\frac{1}{(z+i)(z-i)} \right)\right|_{z=-i} \\
     = -2\pi i f(-i)  \\
     = 2\pi i \frac{1}{(2i)}       = \pi.   \\
\end{split}
\end{equation}

\item  Consider   $$F(p)=\int_{-\infty}^{\infty}\frac{ e^{ipx}}{x^2+a^2} dx$$ with $a\neq 0$.

The analytic continuation of the kernel yields
$$F(p)=\int_{-\infty}^{\infty}\frac{ e^{ipz}}{z^2+a^2} dz
=  \int_{-\infty}^{\infty}\frac{ e^{ipz}}{(z-ia)(z+ia)} dz
.$$

Suppose first that $p>0$. Then, if $z=x+iy$,  $e^{ipz}=e^{ipx}e^{-py}\rightarrow 0$
for $z \rightarrow \infty $
in the {\em upper} half plane.
Hence,  we can close the contour in the upper half plane and obtain  $F(p)$
with the help of the residue theorem.

If $a>0$ only the pole at $z=+ia$ is enclosed in the contour; thus we obtain
\begin{equation}
\begin{split}
F(p) =\left.  2\pi i  {\rm Res} \frac{ e^{ipz}}{(z+ia)} \right|_{z=+ia} \\
       =  2\pi i   \frac{e^{i^2 pa}}{2ia} \\
      =  \frac{\pi}{a}    e^{- pa}.
\end{split}
\end{equation}

If $a<0$ only the pole at $z=-ia$ is enclosed in the contour; thus we obtain
\begin{equation}
\begin{split}
F(p) =\left.  2\pi i  {\rm Res} \frac{ e^{ipz}}{(z-ia)} \right|_{z=-ia} \\
       =  2\pi i   \frac{e^{-i^2 pa}}{-2ia} \\
       =  \frac{\pi}{-a}    e^{-i^2 pa} \\
       =  \frac{\pi}{-a}    e^{ pa}
.
\end{split}
\end{equation}
Hence, for $a\neq 0$,
\begin{equation}
F(p) =  \frac{\pi}{\vert a\vert }    e^{ -\vert p a\vert}.
\end{equation}

For $p<0$ a very similar consideration, taking the {\em lower} path for continuation --
and thus acquiring a minus sign because of the ``clockwork''
orientation of the path as compared to its interior --
yields
\begin{equation}
F(p) =  \frac{\pi}{\vert a\vert }    e^{  - \vert pa\vert }.
\end{equation}

\item
If some function $f(z)$ can be expanded into a Taylor series or Laurent series,
 \index{Taylor series}
 \index{Laurent series}
the residue can be directly obtained by the coefficient of the $\frac{1}{z}$ term.
For instance, let $f(z) = e^{1\over z}$ and $C: z(\varphi )=R e^{i\varphi}$, with $R=1$ and $-\pi < \varphi \le \pi$.
This function is singular only in the origin $z=0$,
but this is an {\em essential singularity} near which the function exhibits extreme behavior.
Nevertheless, $f(z) = e^{1\over z}$
can be expanded into a Laurent series
$$
f(z) = e^{1\over z} =\sum_{l=0}^\infty \frac{1}{l!}   \left(\frac{1}{z}\right)^l
$$
around this singularity.
The residue can be found  by using the series expansion of $f(z)$;
that is, by {\em comparing} its coefficient of the $1/z$ term.
Hence,  $\left.\textrm{ Res }\left( e^{1\over z} \right) \right|_{z=0}$
is the coefficient $1$ of the $1/z$ term.
Thus,
\begin{equation}
\oint_{\vert z\vert =1}
e^{1\over z} dz
=
2\pi i
\textrm{ Res }\left. \left( e^{1\over z} \right)\right|_{z=0} = 2\pi i.
\end{equation}

For $f(z) = e^{-{1\over z}}$, a similar argument yields  $\left.\textrm{ Res }\left( e^{-{1\over z}} \right) \right|_{z=0} = -1$
and thus  $
\oint_{\vert z\vert =1}
e^{-{1\over z}} dz =  -2\pi i$.

An alternative attempt to compute the residue, with $z=e^{i\varphi}$, yields
\begin{equation}
\begin{split}
a_{-1}= \left.\textrm{ Res }\left( e^{\pm {1\over z}} \right) \right|_{z=0} =
 {1\over 2\pi i}\oint_C e^{\pm{ 1\over z}}dz \\
 = {1\over 2\pi i}\int_{-\pi}^\pi  e^{\pm {1\over e^{i\varphi}}}\frac{dz(\varphi )}{d\varphi } d\varphi  \\
= \pm {1\over 2\pi i}\int_{-\pi}^\pi  e^{\pm {1\over e^{i\varphi}}}\, i\, e^{\pm i\varphi} d\varphi    \\
 = \pm {1\over 2\pi }\int_{-\pi}^\pi  e^{\pm e^{\mp i\varphi}}\,  e^{\pm i\varphi} d\varphi    \\
  = \pm {1\over 2\pi }\int_{-\pi}^\pi  e^{\pm e^{\mp i\varphi} \pm i\varphi} d\varphi    .
\end{split}
\end{equation}

\end{enumerate}
\eexample
}

\section{Some special functional classes}

\subsection{Criterion for coincidence}

The requirement that a function is holomorphic (analytic, differentiable)
puts some stringent conditions on its type, form, and on its behavior.
For instance,
let
$z_0\in G$ the limit of a sequence
$\{z_n\}\in G$, $z_n\ne z_0$.
Then it can be shown that,
if two analytic functions
$f$ and $g$
on the domain $G$  coincide in the points $z_n$,
then they coincide on the entire domain $G$.

\subsection{Entire function}

An function is said to be an {\em entire function}
if it is defined and differentiable (holomorphic, analytic)
in the entire {\em finite complex plane} ${\Bbb C}$.
\index{entire function}

An entire function may be
either a
{\em rational function}
\index{rational function}
$f(z)=P(z)/Q(z)$
which can be written as the ratio of two polynomial functions
$P(z)$ and $Q(z)$,
or it may be a
{\em transcendental function}
\index{transcendental function}
such as $e^z$ or $\sin z$.

The
{\em  Weierstrass factorization theorem}
\index{Weierstrass factorization theorem}
\index{power series}
states that an entire function can be represented by a
(possibly infinite\cite{Gamelin-ca})
product involving its zeroes [i.e., the points $z_k$
at which the function vanishes $f(z_k)=0$].
For example (for a proof, see Equation (6.2) of,\cite{conway-focvI})
\begin{equation}
\sin z = z \prod_{k=1}^\infty \left[ 1-  \left(\frac{z}{\pi k}\right)^2\right].
\end{equation}

\subsection{Liouville's theorem for bounded entire function}
\label{2012-m-ch-ca-lt}
{\em Liouville's theorem}
\index{Liouville theorem}
states that
a bounded [that is, the (principal, positive)
square root of its absolute square is finite everywhere in ${\Bbb C}$]
entire function which is defined at infinity
is a constant.
Conversely, a nonconstant entire function cannot be bounded.
\marginnote{It may (wrongly)
appear that $\sin z$ is nonconstant and bounded. However, it is only bounded on the real axis;
indeed,  $\sin iy = (1/2i)(e^{-y}-e^y) = i \sinh y$.
Likewise, $\cos iy =   \cosh y$.}

{\color{OliveGreen}
\bproof
For a proof, consider the integral representation of the derivative $f'(z)$
of some bounded entire function $\vert f(z)\vert <C <\infty$ with bound $C$,
obtained through Cauchy's integral formula (\ref{2012-m-ch-cagcif}),
taken along a circular path with arbitrarily but ``large'' radius $r \gg 1$ of length $2\pi r$ in the limit of infinite radius;
that is,
 \begin{equation}
\begin{split}
\left| f'(z_0) \right| = \left|
{1\over 2\pi i}\oint_{\partial G}{f(z)\over
 (z-z_0)^{2}}dz
\right|   \\
=
\left|
{1\over 2\pi i}
\right|
\left|
\oint_{\partial G}{  f(z)  \over
 (z-z_0)^{2}}dz \right|
<
{1\over 2\pi }
\oint_{\partial G}{ \left| f(z)\right|  \over
 (z-z_0)^{2}}dz   \\
<  \frac{1}{2\pi } 2\pi r  \frac{C}{r^2}
=  \frac{C}{r} \stackrel{r\rightarrow \infty}{\longrightarrow} 0   .
\end{split}
 \end{equation}
\marginnote[-20mm]{Note that, as
 $\overline{(uv)}= (\overline{u}) {\;} (\overline{v}) $, so is
 $\vert uv \vert^2 = uv \overline{(uv)}= u(\overline{u}) {\;} v(\overline{v}) =\vert u \vert^2 {\;} \vert v \vert^2$.}
As a result, $f(z_0)=0$ and thus $f = A \in {\Bbb C}$.
\eproof
}

A {\em generalized Liouville theorem}
\index{generalized Liouville theorem}
states that if $f: {\Bbb C} \rightarrow {\Bbb C}$ is an entire function,
and if,
for some real number $C$ and some positive integer $k$, $f(z)$ is bounded by
$\vert f(z)\vert \le C \vert z \vert^k$ for all $z$ with $\vert z \vert \ge 0$,
then $f(z)$ is a polynomial in $z$ of degree at most $k$.

{\color{OliveGreen}
\bproof
For a proof of the generalized Liouville theorem
we exploit the fact that $f$ is analytic on the entire complex plane.
Thus it can be expanded into a Taylor series~(\ref{2019-mm-ca-ts}) about $z_0$:
\index{Taylor series}
 \begin{equation}
f(z) = \sum_{l=0}^\infty a_l (z-z_0)^l \text{, with } a_l=\frac{f^{(l)}(z_0)}{l!}  .
 \end{equation}

Now consider the integral representation of the $l$th derivative $f^{(l)}(z)$
of some bounded entire function $\vert f(z)\vert < C \vert z \vert^k $ with bound $C <\infty$,
obtained through Cauchy's integral formula (\ref{2012-m-ch-cagcif}),  and
taken along a circular path with arbitrarily but ``large'' radius $r \gg 1$ of length $2\pi r$ in the limit of infinite radius;
that is,
 \begin{equation}
\begin{split}
\left| f^{(l)}(z_0) \right| = \left|
{1\over 2\pi i}\oint_{\partial G}{f(z)\over
 (z-z_0)^{l+1}}dz
\right|   \\
=
\left|
{1\over 2\pi i}
\right|
\left|
\oint_{\partial G}{  f(z)  \over
 (z-z_0)^{l+1}}dz \right|
<
{1\over 2\pi }
\oint_{\partial G}{ \left| f(z)\right|  \over
 (z-z_0)^{l+1}}dz   \\
<  \frac{1}{2\pi } 2\pi r   C r^{k-l-1}
=  C r^{k-l}  \stackrel{\overset{r\rightarrow \infty}{l>k}}{\longrightarrow} 0   .
\end{split}
 \end{equation}
As a result, $f(z)= \sum_{l=0}^k a_l (z-z_0)^l$, with $a_l\in {\Bbb C}$.
\eproof
}

Liouville's theorem is important for an investigation into the general form of
the Fuchsian differential equation
on page
\pageref{2014-m-ch-sf-glt}.
\index{Fuchsian equation}

\subsection{Picard's theorem}
\label{2012-m-ch-ca-pt}
{\em Picard's theorem}
\index{Picard theorem}
states that any entire function that misses two or more points
$f: {\Bbb C} \mapsto  {\Bbb C} - \{z_1,z_2,\ldots \}$
is constant.
Conversely, any nonconstant entire function covers the entire
complex plane ${\Bbb C}$ except a single point.

An example of a nonconstant entire function is $e^z$ which never reaches the point $0$.

\subsection{Meromorphic function}

If $f$ has no singularities other than poles in the domain
$G$ it is called {\em meromorphic} in the domain $G$.
\index{meromorphic function}

We state without proof (e.g., Theorem 8.5.1 of Ref.~\cite[-10mm]{Hille62})
that a function $f$ which is meromorphic in the extended plane
is a rational function $f(z)=P(z)/Q(z)$
which can be written as the ratio of two polynomial functions
$P(z)$ and $Q(z)$.

\section{Fundamental theorem of algebra}
\index{fundamental theorem of algebra}
\marginnote[-20mm]{For a discussion and proofs, see, for instance,
Chapter~19 of \bibentry{ziegler-aigner},
or Chapter~4 (by Remmert) of \bibentry{Numbers-Ebbinghaus}.}

The {\em factor theorem} states that a polynomial $P(z)$  in $z$  of degree $k$
\index{factor theorem}
has a factor $z-z_0$ if and only if $P (z_0)=0$, and can thus be written as $P (z)= (z-z_0)Q(z)$,
where $Q(z)$ is a polynomial  in $z$  of degree $k-1$.
Hence, by iteration,
\begin{equation}
P(z)= \alpha \prod_{i=1}^k \left(z-z_i\right),
\end{equation}
where $\alpha \in {\Bbb C}$.

No proof is presented here.

The {\em fundamental theorem of algebra} states that
every polynomial (with arbitrary complex coefficients) has a root [i.e. solution of $f(z)=0$] in the complex plane.
Therefore, by the factor theorem, the number of roots of a polynomial, up to multiplicity, equals its degree.
\index{root of a polynomial}

Again, no proof
is presented here.

\section{Asymptotic series}
\index{Asymptotic power series}
\label{2019-mm-ch-ca-Ritt}

Asymtotic series
\index{asymptotic series}
occur in physics in the context of ``perturbative'' or series solutions of ordinary differential equations.
they will be studied in the last Chapter~\ref{2011-m-ch-ds}.
In what follows we shall closely follow Remmert's exposition.\cite[-0mm]{Remmert-1991-tocf}

In what follows a formal (power) series $s_n(z) =\sum_{j=0}^n a_j z^j$
is called an {\em asymptotic development}
\index{asymptotic development}
or, equivalently, an
{\em asymptotic representation}
\index{asymptotic representation} or
{\em asymptotic expansion}
\index{asymptotic expansion}
of some holomorphic function $f$
in a domain $G$ at the border point $0 \in \partial G \in G$ if  the asymptotic series ``approximates'' $f$ at $0$; that is, if
 \begin{equation}
\lim_{z\rightarrow 0} \frac{1}{z^n}\left[ f(z) - \sum_{j=0}^n a_j z^j  \right] =0
\text{ for every } n\in \mathbb{N}.
 \end{equation}
Alternatively and equivalently,
asymptoticity
\index{asymptoticity} can be defined as follows:\cite[-0mm]{Olver-1997,Bender-Orszag,Boyd99thedevil}
a (power) series $\sum_{j=0}^n a_j z^j$ is asymptotic to a function $f(z)$
if, for every $n\in \mathbb{N}$ and sufficiently small $r$,
\begin{equation}
\left| f(z) - \sum_{j=0}^n a_j z^j  \right| = O\left( r^{N+1} \right)
;
 \end{equation}
where $O$ represents the
big $O$ notation,
or, used synonymously,
the Bachmann-Landau notation or asymptotic notation\sidenote{The symbol ``$O$'' stands for ``of the order of'' or ``absolutely bound by'' in the following way:
if $g(x)$ is a positive function,  then $f(x)=O\left(g(x)\right)$
implies that there exist a positive real number $m$
such that $\vert f(x) \vert < m g(x)$. \label{2018-mm-ch-di-otof}}.
\index{order of}
\index{big $O$ notation}
\index{Bachmann-Landau notation}
\index{asymptotic notation}

In this case we introduce the following ``$\sim_G$''notation:
 \begin{equation}
f(z) \sim_G  \sum_{j=0}^\infty a_j z^j.
\label{2019-mm-ch-ca-ritt}
 \end{equation}
Note that the asymptotic expansion of any holomorphic function $f$ in a domain $G$ at the border point $0 \in \partial G \in G$
is {\em unique}; the coefficients $a_j$ can be found iteratively by
\begin{align*}
a_0 &= \lim_{z \rightarrow 0} f(z)\text{, and} \\
a_n &= \lim_{z\rightarrow 0} \frac{1}{z^n}\left[ f(z) - \sum_{j=0}^{n-1} a_j z^j  \right]
\text{ for } n>0.
\end{align*}

{
\color{blue}
\bexample
To obtain a feeling for this type of asymptotic expansion, consider the following holomorphic functions:
\begin{enumerate}

\item
the constant function $f(z)=c$.
In this case, $a_0=c$, and all other $a_n=0$ for $n>0$;


\item
the function $f(z)=c e^z$.
In this case, by using the Taylor expansion for $e^z$, one recovers that same Taylor series:
 \index{Taylor series}
\begin{align*}
a_0 &= \lim_{z\rightarrow 0}  c e^z = c e^0 =c,\\
a_1 &= \lim_{z\rightarrow 0} \frac{1}{z} \left[c e^z - c \right] = \lim_{z\rightarrow 0} \left[c z^0 +O(z^1)\right] =c,\\
a_2 &= \lim_{z\rightarrow 0} \frac{1}{z^2} \left[c e^z - c - c z \right] = \lim_{z\rightarrow 0} \left[c \frac{1}{2!} z^0 +O(z^1)\right] =\frac{c}{2!},\\
&\vdots                                                                                                                                                  \\
a_k &= \lim_{z\rightarrow 0} \frac{1}{z^k} \left[c e^z - c\sum_{j=1}^{n-1} \frac{z^j}{j!}  \right] = \lim_{z\rightarrow 0} \left[c \frac{1}{n!} z^0 +O(z^1)\right] =\frac{c}{n!}.
\end{align*}

\end{enumerate}
\eexample
}

Is the converse also true? That is, given an arbitrary asymptotic sequence on a domain;
does there exist an associated holomorphic function such that the former sequence yields
an asymptotic expansion of the latter function?

A similar question can be asked for Taylor expansions: let $(a_j)_{j=0}^\infty$ be an infinite sequence of numbers whose Taylor series
 \index{Taylor series}
$\sum_{j=0}^\infty \frac{1}{j!}a_jz^j$ at $z=0$ converges (with positive radius of convergence $r$).
This Taylor series then defines a unique {\em analytic} function $f(z) = \sum_{j=0}^\infty \frac{1}{j!}a_jz^j$
which is uniquely defined in a circular domain with radius $r$ and center $z=0$.

However, if we allow also functions $g(z)$ which are not necessarily analytic, then
the Taylor series  $g(z) = \sum_{j=0}^\infty \frac{1}{j!}a_jz^j$ is not unique,
because $g(z) = f(z) + d(z)$ would also be represented by one and the same Taylor series if only
$d(0)=0$ as well as all of the derivatives vanish at $z=0$; that is, if
$d^{(n)}(0)=0$ for $n=0,1,2,\ldots$~.
Take, for example,
$d(z) = \exp\left({-\frac 1{z^2}}\right)$ for $x \neq 0$ and $d(0)=0$,
which is a variant of the class-I test function with compact support
[cf. Equation~(\ref{2018-m-ch-di-tf1}) on page~\pageref{2018-m-ch-di-tf1}]:
$d$ is smooth but not analytic.

Let us now come back to the general case of not necessarily converging power series with arbitrary
coefficients $a_i$.
The following theorem of Ritt
\index{Ritt's theorem}
\index{theorem of Ritt}
gives a positive answer but does not guarantee uniqueness of the function:
Associated with every infinite power series
$\sum_{j=0}^\infty a_j z^j$
with arbitrary complex coefficients $a_j$
corresponds a holomorphic function $f$ in a proper circular sector $G$ at z=0
such that (\ref{2019-mm-ch-ca-ritt}) holds; that is,\sidenote[][-0mm]{For proper definitions, proofs and further details
see~\bibentry{Pittnauer-73}, \bibentry{Remmert-1991-tocf} and \bibentry{Costin-2009}.}
$ \sum_{j=0}^\infty a_j z^j  \sim_G  f(z) $.

The idea of Ritt's theorem is elegant and not too difficult to comprehend:
define a series
\begin{equation}
f(z)
\stackrel{\text{def}}{=}
 \sum_{j=0}^\infty a_j z^j f_j(z)  \equiv    \sum_{j=0}^\infty a_j z^j f(j,z)
 \end{equation}
with additional ``convergence factors'' $f_j(z) \equiv f(j,z)$
which should perform according to two criteria:
\begin{enumerate}
\item
$f_j(z)\equiv f(j,z)$ should become ``very small'' as a function of $j$; that is, as $j$ grows; so much so that it ``compensates'' for the term $z^j$; and
\item
at the same time, for every fixed $j$, $ \lim_{z\rightarrow 0} f_j(z)\equiv \lim_{z\rightarrow 0}  f(j,z) = 1$; that is, the
convergence factors should all converge ``sufficiently rapidly'' so as to obtain (\ref{2019-mm-ch-ca-ritt});
that is,  $ \sum_{j=0}^\infty a_j z^j  \sim_G  f(z) $.
\end{enumerate}

There may be many functional forms of ``convergence factors''
satisfying the above criteria; therefore the construction cannot yield uniqueness.
One candidate for the  ``convergence factors'' is
\begin{equation}
f_j(z) \equiv f(j,z)
= 1 - e^{-\frac{b_j}{\sqrt{z}}},
 \end{equation}
with
$\sqrt{z} = e^\frac{\log z}{2}$ and
real positive coefficients $b_j>0$ properly chosen such that,
for all $j\in \mathbb{N} $,
\begin{equation}
\left| 1 - e^{-\frac{b_j}{\sqrt{z}}}  \right|
\le
\frac{b_j}{   \left| \sqrt{z} \right|}\text{,
as  well as
}
\lim_{z\rightarrow 0} \frac{e^{-\frac{b_j}{\sqrt{z}}}}{z^j} .
\end{equation}

Other (uniform with respect to the summation index $j$)
convergence or cutoff factors discussed by Tao\cite{Tao-2013}  in \S~3.7 are
compactly supported bounded functions that equals $1$ at $z=0$;
see~(\ref{2019-m-ch-di-tf1-mod1}) on page~\pageref{2019-m-ch-di-tf1-mod1}
for an example.

\section{Jordan's Lemma}
\label{2021-mm-ch-ca-jl}
\index{Jordan's lemma}
\marginnote[0mm]{Here we closely follow Section~88 of~\bibentry{Brown-Churchill}.}
Jordan's lemma is often invoked in the calculation of countour integrals, to ``get rid'' of the
simple arc, or a Jordan arc---a line integral that
\index{Jordan arc}
``closes the  contour'' in either the upper or the lower complex plane.

We shall first derive Jordan's inequality.
\index{Jordan's inequality}
Suppose $R>0$.
A  direction $\theta$   reflected about a line with direction  $\frac{\pi}{2}$
yields $ \sin  \theta  =  \sin \left( \pi - \theta \right)$.
Therefore,
\begin{equation}
\begin{split}
\label{2021-mm-ch-ca-ji}
\int_0^\pi e^{-R \sin \theta} d\theta
=
\int_0^\frac{\pi}{2} e^{-R \sin \theta} d\theta
+
\int_\frac{\pi}{2}^\pi e^{-R \sin \theta} d\theta
\\
\text{[substitution } u= \pi -\theta,\; d\theta = -du \text{ in the second integral]}
\\
= \int_0^\frac{\pi}{2} e^{-R \sin \theta} d\theta
+
\int_\frac{\pi}{2}^0  e^{-R \sin \left( \pi -u\right) } \left(-du\right)
\\
= \int_0^\frac{\pi}{2} e^{-R \sin \theta} d\theta
+
\int_0^\frac{\pi}{2}  e^{-R \sin u } du
= 2 \int_0^\frac{\pi}{2} e^{-R \sin \theta} d\theta
.
\end{split}
\end{equation}

In the domain $0\le \theta \le \frac{\pi}{2}$ of this integration
$\sin \theta \ge \frac{2}{\pi} \theta$ holds.\sidenote{This can be made plausible by drawing the graphs of
$\sin \theta$ and $\frac{2}{\pi} \theta$ in the interval $\left[0,\frac{\pi}{2}\right]$; see also
formula~4.3.79, page~75 of~\bibentry{abramowitz:1964:hmf}}
Therefore, and because $R>0$,
$e^{-R \sin \theta} \le e^{- \frac{2}{\pi}R \theta}$.
Insertion of this estimate into~(\ref{2021-mm-ch-ca-ji}) yields   Jordan's inequality
\begin{equation}
\label{2021-mm-ch-ca-jii}
\begin{split}
\int_0^\pi e^{-R \sin \theta} d\theta=
2 \int_0^\frac{\pi}{2} e^{-R \sin \theta} d\theta
\le
2 \int_0^\frac{\pi}{2}e^{- \frac{2}{\pi}R \theta} d\theta
\\ \qquad
=
2 \int_0^{-R} e^u \left( -\frac{\pi}{2R} du \right)
=
\frac{\pi}{R} \int_{-R}^0 e^u  du =  \frac{\pi}{R} \left( e^0 -e^{-R}\right)
\le  \frac{\pi}{R}
.
\end{split}
\end{equation}

{
\color{blue}
\bexample
{\color{black}
\begin{marginfigure}
\begin{center}
\begin{tikzpicture}  [scale=0.4]

\tikzstyle{every path}=[line width=2pt]


\draw[draw=gray!80,->] (0,-5) + (0,-0.5cm)  -- (0,5) -- +(0,1cm) node[above right] {$\Im z$};
\draw[draw=gray!80,->] (-5,0) +(-0.5cm,0) -- (5,0) -- +(1cm,0) node[below right] {$\Re z$};

\draw[orange,->] (-5,0) -- (5,0);
\draw[blue,dashed,->] (-5,0) -- (4.8,0);


\draw [color=orange,->] (5,0) arc [start angle=0,end angle=180,x radius=5cm, y radius=5cm];
\node[orange] at (5.3,4.5) {$C_\infty \text{ for } k> 0$};
\draw [color=blue,dashed,->] ( 5,0) arc [start angle=360,end angle=180,x radius=5cm, y radius=5cm];
\node[blue] at (5.3,-4.5) {$C_\infty \text{ for } k < 0$};


\end{tikzpicture}
\caption{Extension of the integral over the real $z$-axis by including the Jordan arcs $C_\infty$
in the upper ($k > 0$) and lower ($k < 0$)  imaginary half planes of $z$, respectively.}
\label{2021-mm-ch-ca-ja}
\end{center}
\end{marginfigure}
}
Let us, for the sake of an example,
compute the integral along the real line
$\int_{-\infty}^\infty e^{i k x} dx$.
We extend this integral by the Jordan arc  $C_\infty = \lim_{R\rightarrow \infty} C_R$
after
analytic continuation of the integral kernel $ e^{i k x} \rightarrow  e^{i k z}$
to the complex $z$-plane.

This contour integral has an analytic interior.
Therefore, according to the residue theorem,
 \index{residue theorem}
it vanishes:
\begin{equation}
\int_{-\infty}^\infty e^{i k x} dx + \int_{C_\infty} e^{i k z} dz =0
.
\end{equation}
We can use Jordan's inequality to argue that the second summand,
the line integral along the Jordan arc $C_\infty$, vanishes.
Note that, say, for the Jordan arc in the upper half $z$-plane,
\begin{equation}
\int_{C_R} e^{i k z} dz
=
\int_0^\pi e^{i k R(\cos \theta + i \sin \theta} d\theta
=
\int_0^\pi e^{i k R\cos \theta}  e^{-k R\sin \theta} d\theta.
\end{equation}
The first factor $e^{i k R\cos \theta}$ is an oscillating term of modulus one.
Its absolute value is bounded from above by one.
Therefore the absolute value of the integral can be estimated
with the help of Jordan's inequality~(\ref{2021-mm-ch-ca-jii}):
\begin{equation}
\left|
\int_{C_\infty} e^{i k z} dz
\right|
\le
\int_0^\pi e^{-k R\sin \theta} d\theta
< \frac{\pi}{k R}.
\end{equation}
It vanishes as $R$ approaches infinity.
Therefore we obtain
$\int_{C_\infty} e^{i k z} dz =0$
and $\int_{-\infty}^\infty e^{i k x} dx =0$.

\eexample
}

We can proceed to derive Jordan's lemma.
\index{Jordan's Lemma}
Suppose that a function $f(z)$
\begin{itemize}
\item[(i)]
is analytic in the upper half plane ``outside of'' (exterior to)
some circle of finite value $R_0=\vert z\vert \ge 0$.

\item[(ii)]
for all arguments $z$ of $f(z)$ on some semicircle---a Jordan arc
$C_R=\left\{ z = Re^{i\theta} \middle| R>R_0 \text{ and } 0\le \theta \le \pi\right\}$---in this outside region
there is a positive $M_R\ge \vert f(z) \vert$ which vanishes as $R$ tends to infinity; that is,
$\lim_{R\rightarrow \infty} M_R=0$.
\end{itemize}
Relative to these assumptions Jordan's lemma states
that, for every positive constant $k$,

\begin{equation}
\label{2021-mm-ch-ca-jle}
\lim_{R\rightarrow \infty}
\int_{C_R} f(z) e^{ikz} dz
=0
.
\end{equation}

{\color{OliveGreen}
\bproof
For a proof we rewrite~(\ref{2021-mm-ch-ca-jle})
in terms of the polar representation of  $z=R e^{i \theta}$ and $dz = i R e^{i \theta} d\theta$ as
\begin{equation}
\label{2021-mm-ch-ca-jlp1}
\int_{C_R} f(z) e^{ikz} dz
=
\int_0^\pi f \left( R e^{i \theta} \right) e^{ik R e^{i \theta}} \left(i R e^{i \theta} d\theta \right)
.
\end{equation}

Since, by assumption, $C_R$ is located in the exterior region $R>R_0$,
$ \vert f \left( R e^{i \theta} \right) \vert \le M_R$,
and since the modulus
$\left|i R e^{i \theta}  \right|= R$
and the modulus
$
\left| e^{ik R e^{i \theta}}\right|
=
\left| e^{ik R \left(\cos \theta + i \sin \theta \right) }\right|
=
\left| e^{ik R \cos \theta}\right| \cdot
\left| e^{-k R  \sin \theta }\right|
=
 e^{-k R  \sin \theta }
$,
Jordan's inequality~(\ref{2021-mm-ch-ca-jii}) can be used to estimate the absolute value of
\begin{equation}
\label{2021-mm-ch-ca-jlp2}
\left|\int_{C_R} f(z) e^{ikz} dz \right|
\le
R M_R
\int_0^\pi e^{- (k R) \sin \theta} d \theta < R M_R \frac{\pi}{k R}   = \frac{\pi}{k} M_R
\underset{M_R\rightarrow 0}{\xrightarrow{R\rightarrow \infty}}
0.
\end{equation}

\eproof
}

\begin{center}
{\color{olive}   \Huge
 \floweroneright
}
\end{center}

\chapter{Brief review of Fourier transforms}
\index{Fourier transform}

\subsection{Functional spaces}

That complex continuous waveforms or functions are comprised
of a number of harmonics seems to be an idea at least as old as the Pythagoreans.
In physical terms, Fourier analysis\cite{koerner,Howell,herman-fa}
attempts to decompose a function into its constituent harmonics, known as a frequency spectrum.
Thereby the goal is the expansion of periodic and aperiodic functions into sine and cosine functions.
Fourier's observation or conjecture is, informally speaking, that any   ``suitable''
function $f(x)$
can be expressed as a  possibly infinite  sum (i.e., linear combination), of sines
and cosines of the form
\begin{equation}
\begin{split}
f(x)
=
\sum_{k=-\infty}^\infty
\left[ A_k \cos (C k x) + B_k \sin (C k x)\right]\\
=
\left\{
\sum_{k=-\infty}^{-1} + \sum_{k=0}^\infty
\right\}
\left[ A_k \cos (C k x) + B_k \sin (C k x)\right]\\
=
\sum_{k=1}^{\infty}
\left[ A_{-k} \cos (-C k x) + B_{-k} \sin (-C k x)\right]  \qquad \qquad \\
+
\sum_{k=0}^\infty
\left[ A_k \cos (C k x) + B_k \sin (C k x)\right]\\
=
 A_0 +
\sum_{k=1}^{\infty}
\left[ \left( A_k+A_{-k} \right) \cos (C k x) + \left( B_k-B_{-k} \right) \sin (C k x)\right]
\\
=  \frac{a_0}{2} +
\sum_{k=1}^\infty
\left[ a_k  \cos (C k x) + b_k  \sin (C k x)\right]
,
\label{2012-m-ch-fourier_conjecture}
\end{split}
\end{equation}
with   $a_0=2 A_0$,
$a_k = A_k+ A_{-k}$,
and
$b_k = B_k- B_{-k}$.

Moreover, it is conjectured that any ``suitable''
function $f(x)$
can be expressed as a  possibly infinite  sum (i.e. linear combination), of exponentials; that is,
\begin{equation}
f(x)= \sum_{k=-\infty}^\infty
D_k e^{ikx}.
\label{2012-m-ch-fourier_conjecture1}
\end{equation}

More generally, it is conjectured that any ``suitable''
function $f(x)$
can be expressed as a  possibly infinite  sum (i.e. linear combination), of other (possibly orthonormal) functions $g_k( x)$; that is,
\begin{equation}
f(x)= \sum_{k=-\infty}^\infty
\gamma_k g_k(x).
\label{2012-m-ch-fourier_conjecturegen}
\end{equation}

The bigger picture can then be viewed in terms of
{\em functional (vector) spaces}: these are spanned by the elementary functions $g_k$, which serve as elements of a
{\em functional basis} of a possibly infinite-dimensional vector space.
\index{functional spaces}
Suppose, in further analogy to the set of all such functions ${\frak G}=\bigcup_k g_k(x)$
to the (Cartesian) standard basis, we can consider these elementary functions $g_k$ to be
{\em orthonormal} in the sense of a {\em generalized functional scalar product}
 [cf. also Section \ref{2012-m-sf-fs} on page \pageref{2012-m-sf-fs}; in particular
Equation (\ref{2011-m-ch-sfesp})]
\index{inner product}
\index{scalar product}
\begin{equation}
\langle   g_k \mid g_l\rangle
=
\int_{a}^b g_k(x)g_l(x)  \rho(x) dx =\delta_{kl}.
\label{2012-m-ch-sfesp1}
\end{equation}
For most of our purposes, $\rho (x)=1$.
One could arrange the coefficients $\gamma_k$ into a tuple (an ordered list of elements)
$(\gamma_1,\gamma_2, \ldots)$
and consider them as components or coordinates of a vector with respect to the
linear orthonormal functional basis ${\frak G}$.

\subsection{Fourier series}

Suppose that a function $f(x)$ is periodic -- that is, it repeats  its values in the interval $[-\frac{L}{2},\frac{L}{2}]$ --  with period $L$.
(Alternatively, the function may be only defined in this interval.)
A function $f(x)$
is {\em periodic}
\index{periodic function}
if there exist a period $L\in {\Bbb R}$ such that, for all $x$ in the domain of $f$,
\begin{equation}
f(L+x)=f(x).
\end{equation}

With certain ``mild'' conditions
-- that is, $f$ must be piecewise continuous, periodic with period $L$,  and (Riemann) integrable --
$f$ can be decomposed into a {\em Fourier series}
\index{Fourier series}
\begin{equation}
\begin{split}
f(x)={a_0\over2}+\sum_{k=1}^\infty
\left[a_k\cos\left(\frac{2\pi }{L}kx\right)+b_k\sin\left(\frac{2\pi }{L}kx\right)\right] \textrm{, with }\\
   a_k={2\over L}\int\limits_{-\frac{L}{2}}^\frac{L}{2}  f(x)\cos\left(\frac{2\pi }{L}kx\right) dx \textrm{ for } k \ge 0\\
   b_k={2\over L}\int\limits_{-\frac{L}{2}}^\frac{L}{2} f(x)\sin\left(\frac{2\pi }{L}kx\right) dx \textrm{ for } k > 0  .
\end{split}
\label{2012-m-ch-fs}
\end{equation}

{\color{OliveGreen}
\bproof

\marginnote{For proofs and additional information see \S~8.1 in \bibentry{Howell}.}
For a (heuristic) proof, consider the Fourier conjecture (\ref{2012-m-ch-fourier_conjecture}),
and compute the coefficients $A_k$, $B_k$, and $C$.

First, observe that we have assumed that $f$ is periodic with period $L$.
This should be reflected in the sine and cosine terms of   (\ref{2012-m-ch-fourier_conjecture}),
which themselves are periodic functions, repeating their values in the interval $[- \pi , \pi ]$; with period $2\pi$.
Thus in order to map the functional period of $f$ into the sines and cosines, we can ``stretch/shrink''
$L$ into $2\pi$; that is,
$C$ in Equation  (\ref{2012-m-ch-fourier_conjecture}) is identified with
\begin{equation}
C=\frac{2\pi}{L}.
\label{2012-m-ch-pfc1}
\end{equation}
Thus we obtain
\begin{equation}
f(x)= \sum_{k=-\infty}^\infty
\left[ A_k \cos \left(\frac{2\pi}{L} k x\right) + B_k \sin \left(\frac{2\pi}{L} k x\right)\right].
\label{2012-m-ch-pfc2}
\end{equation}

Now use the following properties:
(i)
for $k=0$, $\cos (0)=1$ and $\sin (0)=0$.
Thus, by comparing the coefficient $a_0$ in
(\ref{2012-m-ch-fs}) with $A_0$ in (\ref{2012-m-ch-fourier_conjecture})
we obtain  $A_0=\frac{a_0}{2}$.

(ii) Since $\cos (x)= \cos (-x)$ is an {\em even function} of $x$, we can rearrange the summation
by combining identical functions  $\cos (-\frac{2\pi}{L} k x) = \cos (\frac{2\pi}{L} k x) $,
thus obtaining $a_k = A_{-k}+A_k$ for $k>0$.

(iii) Since $\sin (x)= -\sin (-x)$ is an {\em odd function} of $x$, we can rearrange the summation
by combining identical functions  $\sin (-\frac{2\pi}{L} k x) =  -\sin (\frac{2\pi}{L} k x)$,
thus obtaining $b_k = -B_{-k}+B_k$ for $k>0$.

Having obtained the same form of the Fourier series of $f(x)$ as exposed in (\ref{2012-m-ch-fs}),
we now turn to the derivation of the coefficients $a_k$ and $b_k$.
$a_0$ can be derived by just considering the functional scalar product in
Equation (\ref{2012-m-ch-sfesp1})
of $f(x)$ with the constant identity function $g(x)=1$; that is,
\begin{equation}
\begin{split}
\langle   g  \mid f \rangle
=
\int_{-\frac{L}{2}}^\frac{L}{2} f(x)  dx \\
 =
\int_{-\frac{L}{2}}^\frac{L}{2} \left\{ \frac{a_0}{2}+\sum_{n=1}^\infty
\left[a_n\cos\left(\frac{2\pi}{L} n x\right)+b_n\sin\left(\frac{2\pi}{L} n x\right)\right]\right\}  dx  =
a_0\frac{L}{2}
,
\end{split}
\label{2012-m-ch-sfespnn}
\end{equation}
and hence
\begin{equation}
a_0 = \frac{2}{L} \int_{-\frac{L}{2}}^\frac{L}{2} f(x)  dx
\label{2012-m-ch-sfespnn2}
\end{equation}

In a similar manner, the other coefficients can be computed by considering
$\left\langle   \cos \left(\frac{2\pi }{L}kx\right)  \mid f(x) \right\rangle$
$\left\langle   \sin \left(\frac{2\pi }{L}kx\right)  \mid f(x) \right\rangle$
and exploiting the
{\em orthogonality relations for sines and cosines}
\index{orthogonality relations for sines and cosines}
\begin{equation}
\begin{split}
\int_{-\frac{L}{2}}^\frac{L}{2}
\sin\left(\frac{2\pi }{L}kx\right)
\cos\left(\frac{2\pi }{L}lx\right)
dx
=0,\\
\int_{-\frac{L}{2}}^\frac{L}{2}
\cos\left(\frac{2\pi }{L}kx\right)
\cos\left(\frac{2\pi }{L}lx\right)
dx
\\
=
\int_{-\frac{L}{2}}^\frac{L}{2}
\sin\left(\frac{2\pi }{L}kx\right)
\sin\left(\frac{2\pi }{L}lx\right)
dx
=\frac{L}{2}\delta_{kl}.
\end{split}
\label{2012-m-ch-orsc}
\end{equation}

\eproof
}

{
\color{blue}
\bexample
For the sake of an example, let us compute the  Fourier series of
$$
f(x)=\vert x\vert
=
\begin{cases}
 -x, & \textrm{ for }  -\pi \le x<0 ,\\
 +x, & \textrm{ for }  0\le x\le \pi  .
\end{cases}
$$

First observe that $L=2\pi$, and that
$f(x)=f(-x)$; that is, $f$ is an {\em even} function of $x$;
hence
$b_n=0$, and the coefficients $a_n$ can be obtained by considering only the integration
between $0$ and $\pi$.

For $n=0$,
$$a_0={1\over\pi}\int\limits_{-\pi}^\pi dxf(x)={2\over\pi}
\int\limits_0^\pi xdx=\pi.$$

For $n>0$,
\begin{eqnarray*}
   a_n&=&{1\over\pi}\int\limits_{-\pi}^\pi f(x)\cos(nx)dx=
         {2\over\pi}\int\limits_0^\pi x\cos(nx)dx=\\
      &=&{2\over\pi}\left[\left.{\sin(nx)\over n}x\right|_0^\pi-
         \int\limits_0^\pi{\sin(nx)\over n}dx\right]={2\over\pi}\left.
         {\cos(nx)\over n^2}\right|_0^\pi=\\
      &=&{2\over\pi}{\cos(n\pi)-1\over n^2}=-{4\over\pi n^2}\sin^2{n\pi\over2}=
         \left\{\begin{array}{cl}
              0&\mbox{for even $n$}\\
              \displaystyle-{4\over\pi n^2}&\mbox{for odd $n$}
              \end{array}\right.
\end{eqnarray*}
Thus,
\begin{eqnarray*}
    f(x)&=&{\pi\over2}-{4\over\pi}\left(\cos x+
               {\cos 3x\over9}+{\cos 5x\over 25}+\cdots\right)=\\
            &=&{\pi\over2}-{4\over\pi}\sum_{n=0}^\infty
               {\cos[(2n+1)x]\over(2n+1)^2}.
\end{eqnarray*}

\eexample
}

One could arrange the coefficients
$(a_0,a_1,b_1,a_2,b_2, \ldots)$
into a tuple (an ordered list of elements)
and consider them as components or coordinates of a vector spanned by the
linear independent sine and cosine functions which serve as a basis of an infinite
dimensional vector space.

\subsection{Exponential Fourier series}
\index{exponential Fourier series}

Suppose again that a function is periodic
with period $L$.
Then, under certain ``mild'' conditions
-- that is, $f$ must be piecewise continuous, periodic with period $L$,  and (Riemann) integrable --
$f$ can be decomposed  into an {\em exponential Fourier series}
\begin{equation}
\begin{split}
f(x)= \sum _{k=-\infty}^\infty c_k e^{ikx} \textrm{, with } \\
c_k=\frac{1}{L}\int_{-\frac{L}{2}}^{\frac{L}{2}} f(x') e^{-ikx'} dx'.
\end{split}
\label{2011-m-fa-e1fc}
\end{equation}

{\color{OliveGreen}
\bproof

The exponential form of the Fourier series
can be derived from the Fourier series (\ref{2012-m-ch-fs}) by
Euler's formula  (\ref{2012-m-ch-ca-ef}), in particular,
\index{Euler's formula}
$e^{ik\varphi} = \cos (k\varphi )+i \sin (k\varphi )$, and thus
$$
\cos (k\varphi ) =\frac{1}{2}\left(e^{ik\varphi}+e^{-ik\varphi}\right)
,\textrm{ as well as }
\sin (k\varphi ) =\frac{1}{2i}\left(e^{ik\varphi}-e^{-ik\varphi} \right)
.
$$
By comparing the coefficients of (\ref{2012-m-ch-fs}) with the coefficients of (\ref{2011-m-fa-e1fc}),
we obtain
\begin{equation}
\begin{split}
a_k= c_k+c_{-k} \textrm{ for } k \ge 0,\\
b_k= i(c_k-c_{-k})\textrm{ for } k > 0  ,\\
\end{split}
\label{2011-m-fa-e1fccc1}
\end{equation}
or
\begin{equation}
c_k=
\left\{
\begin{split}
\frac{1}{2}(a_k-ib_k) \textrm{ for } k > 0,\\
\frac{a_0}{2} \textrm{ for } k = 0,\\
\frac{1}{2}(a_{-k}+ib_{-k}) \textrm{ for } k < 0.
\end{split}
\right.
\label{2011-m-fa-e1fccc2}
\end{equation}

Eqs. (\ref{2011-m-fa-e1fc}) can be combined into
\begin{equation}
f(x)= \frac{1}{L}\sum _{\check{k}=-\infty}^\infty  \int_{-\frac{L}{2}}^\frac{L}{2} f(x') e^{-i{\check{k}(x'-x)}} dx'
.
\label{2011-m-eft1}
\end{equation}

\eproof
}

\subsection{Fourier transformation}

Suppose we define
$\Delta {k} = 2\pi /L$, or $1/L = \Delta {k} /2\pi$.
Then Equation (\ref{2011-m-eft1}) can be rewritten  as
\begin{equation}
f(x)= \frac{1}{2\pi}
\sum _{k=-\infty}^\infty  \int_{-\frac{L}{2}}^\frac{L}{2} f(x') e^{-i{{k}(x'-x)}} dx' \Delta {k}
.
\end{equation}
Now,
in the ``aperiodic'' limit $L\rightarrow \infty$ we obtain  the {\em Fourier transformation}
and the {\em Fourier inversion}
$
{\cal F}^{-1}[{\cal F} [f(x)]]=
{\cal F} [{\cal F}^{-1}[f(x)]]= f(x)
$
by
\index{Fourier transformation}
\index{Fourier inversion}
\begin{equation}
\begin{split}
f(x)= \frac{1}{2\pi}
 \int_{-\infty}^\infty   \int_{-\infty}^\infty f(x') e^{-i{{k}(x'-x)}} dx' d{k} \textrm{, whereby} \\
  {\cal F}^{-1}[\tilde{f}](x)=f(x)=  \alpha \int_{-\infty}^\infty \tilde{f}(k) e^{\pm i{kx}} dk \textrm{, and} \\
 {\cal F}[f](k)=\tilde{f}(k)=  \beta \int_{-\infty}^\infty  f(x') e^{\mp i{kx'}} dx'
.
\end{split}
\label{2011-m-efta}
\end{equation}
$ {\cal F}[f(x)]= \tilde{f}(k)$
is called the {\em Fourier transform}
\index{Fourier transform}
of $f(x)$.
{\em Per} convention, either one of the two sign pairs $+-$ or $-+$ must be chosen.
The factors $\alpha$ and $\beta$ must be chosen such that
\begin{equation}
 \alpha  \beta    = \frac{1}{2\pi};
\label{2011-m-efta1}
\end{equation}
that is, the factorization can be ``spread evenly among $\alpha$ and $\beta$,''
such that  $\alpha=\beta=1/\sqrt{2\pi}$, or ``unevenly,''
such as, for instance,
 $\alpha=1$ and $\beta=1/ 2\pi $,
or    $\alpha=1/ 2\pi $ and $\beta=1$.

Most generally, the Fourier transformations can be rewritten (change of integration constant),
 with arbitrary $A,B\in {\Bbb R}$,
as
\begin{equation}
\begin{split}
 {\cal F}^{-1}[\tilde{f}](x)=f(x)= B \int_{-\infty}^\infty \tilde{f}(k) e^{  iA{kx}} dk \textrm{, and} \\
 {\cal F}[f](k)=\tilde{f}(k)=  \frac{A}{2\pi B} \int_{-\infty}^\infty  f(x') e^{- iA{kx'}} dx'
.
\end{split}
\label{2011-m-efta1mg}
\end{equation}

The choice $A=2\pi $ and $B=1$ renders a  symmetric form of (\ref{2011-m-efta1mg}); more precisely,
\begin{equation}
\begin{split}
 {\cal F}^{-1}[\tilde{f}](x)=f(x)=   \int_{-\infty}^\infty \tilde{f}(k) e^{  2\pi i{kx}} dk \textrm{, and} \\
 {\cal F}[f](k)=\tilde{f}(k)=    \int_{-\infty}^\infty  f(x') e^{-  2\pi  i {kx'}} dx'
.
\end{split}
\label{2011-m-efta1mgAB1}
\end{equation}

{
\color{blue}
\bexample

For the sake of an example,
assume $A=2\pi $ and $B=1$  in Equation (\ref{2011-m-efta1mg}), therefore starting with (\ref{2011-m-efta1mgAB1}),
and consider the  Fourier transform of the  {\em Gaussian function}
\index{Gaussian function}
\begin{equation}
\varphi (x)= e^{-\pi x^2}   .
\label{2012-m-ch-fa-gaussian}
\end{equation}

As a hint, notice that
the analytic continuation of $e^{-t^2}$ is analytic in the region $0\le \vert {\rm Im}\,t\vert \le \sqrt{\pi }\vert k \vert$.
Furthermore, as will be shown in Eqs. (\ref{2012-m-ch-di-gi2}), the {\em Gaussian integral}
\index{Gaussian integral} is
\begin{equation}
\int_{-\infty}^\infty
 e^{-t^2} dt =\sqrt{\pi }  .
\end{equation}
With $A=2\pi $ and $B=1$  in Equation (\ref{2011-m-efta1mg}), the Fourier transform of the  Gaussian function is
\begin{equation}
\begin{split}
    {\cal F}[\varphi  ](k)=\widetilde{ \varphi} (k) =  \int\limits_{-\infty}^\infty
                 e^{-\pi  x^2}  e^{- 2\pi ikx} dx
\\
\textrm{[completing the exponent]}
              =
\int\limits_{-\infty}^\infty
                 e^{-{\pi  k^2}}e^{-\pi \left(x+{i}k\right)^2}   dx
\end{split}
\end{equation}
The variable transformation  $t=\sqrt{\pi} (x + {i}k) $
yields
$dt/ dx=\sqrt{\pi} $; thus $dx = dt/ \sqrt{\pi}$, and
\begin{equation}
   {\cal F}[\varphi  ](k)=\widetilde{ \varphi} (k)=\frac{ e^{-{\pi k^2}}}{\sqrt{\pi} }
                 \int\limits_{-\infty+i\sqrt{\pi} {k}}^{+\infty+i\sqrt{\pi} {k}}
                 e^{-t^2}  dt
\label{2012-m-ch-dergauin1}
\end{equation}

\begin{marginfigure}
\begin{center}
\begin{tikzpicture}  [scale=0.3]

\tikzstyle{every path}=[line width=2pt]


\draw[draw=gray!80,->] (0,-5) + (0,-0.5cm)  -- (0,5) -- +(0,0.5cm) node[above right] {$\Im t$};
\draw[draw=gray!80,->] (-7,0) +(-0.5cm,0) -- (7,0) -- +(0.5cm,0) node[below right] {$\Re t$};

\draw[orange,->] (-3,4) -- (3,4)  node[above right] {$k\ge 0$};
\draw[orange,->] (3,0) -- (-3.4,0) ;
\draw[color=blue,dashed,->] (-3,-4) -- (3,-4)  node[below right] {$k\le 0$};
\draw[color=blue,dashed,->] (3,0) -- (-3,0);


\draw [color=orange,->] (-5,0) arc [start angle=270,end angle=90,x radius=2cm, y radius=2cm];
\draw [color=orange,->] (5,4) arc [start angle=90,end angle=-60,x radius=2cm, y radius=2cm];
\draw [color=blue,dashed,->] (-5,0) arc [start angle=90,end angle=270,x radius=2cm, y radius=2cm];
\draw [color=blue,dashed,->] ( 5,-4) arc [start angle=-90,end angle=60,x radius=2cm, y radius=2cm];

\node[orange] at (-4,4) {$\cdots$};
\node[orange] at (4,4) {$\cdots$};
\node[blue] at (-4,-4) {$\cdots$};
\node[blue] at (4,-4) {$\cdots$};

\end{tikzpicture}
\end{center}
\caption{Integration paths to compute the  Fourier transform of the  Gaussian.}
\label{2011-m-ftgauss}
\end{marginfigure}
Let us rewrite the integration (\ref{2012-m-ch-dergauin1}) into the Gaussian integral by considering the closed paths
(depending on whether $k$ is positive or negative) depicted in Fig.~\ref{2011-m-ftgauss}.
whose ``left and right pieces vanish'' strongly as the real part goes to (minus) infinity.
Moreover,
by the Cauchy's integral theorem, Equation~(\ref{2018-m-ch-ca-cit}) on page~\pageref{2018-m-ch-ca-cit},
\index{Cauchy's integral theorem}
\begin{equation}
   \oint\limits_{\cal C} dt e^{-t^2}=\int\limits_{+\infty}^{-\infty}
  e^{-t^2} dt +\int\limits_{-\infty+{i\sqrt{\pi} }k}^{+\infty+{i\sqrt{\pi} }k}
   e^{-t^2} dt =0,
\end{equation}
because $e^{-t^2}$ is analytic in the region $0\le \vert {\rm Im}\,t\vert \le \sqrt{\pi }\vert k \vert$.
Thus, by substituting
\begin{equation}
 \int\limits_{-\infty+{i}\sqrt{\pi} k}^{+\infty+{i}\sqrt{\pi} k}
   e^{-t^2} dt =\int\limits_{-\infty}^{+\infty}e^{-t^2}  dt   ,
\end{equation}
in (\ref{2012-m-ch-dergauin1})
and  by insertion of the value
$\sqrt\pi$ for the {\em Gaussian integral},
as shown in  Equation (\ref{2012-m-ch-di-gi2}), we finally obtain
\begin{equation}
     {\cal F}[\varphi  ](k)=\widetilde{ \varphi} (k)=\frac{e^{-{\pi k^2 }}}{ \sqrt{\pi} }\underbrace{\int
   \limits_{-\infty}^{+\infty}e^{-t^2}dt}_{\mbox{$\sqrt\pi$}}=
   e^{-{\pi k^2}} .
\label{2012-m-ch-fa-gift}
\end{equation}

A similar calculation yields
\begin{equation}
     {\cal F}^{-1}[\widetilde{\varphi}  ](x)= \varphi (x)= e^{-{\pi x^2}} .
\label{2012-m-ch-fa-giftinverse}
\end{equation}
\eexample
}

Eqs.
(\ref{2012-m-ch-fa-gift})
and
(\ref{2012-m-ch-fa-giftinverse})
establish the fact that
the Gaussian function
$\varphi (x) = e^{-{\pi x^2}}$ defined in
(\ref{2012-m-ch-fa-gaussian})
is an {\em eigenfunction}
\index{eigenfunction}
\index{eigenvector}
of the Fourier transformations
${\cal F}$
and
${\cal F}^{-1}$ with associated eigenvalue $1$.
\marginnote{See Section~6.3 in \bibentry{strichartz}.}

With a slightly different definition the Gaussian function $f(x) = e^{-{x^2/2}}$ is also an eigenfunction of the operator
\begin{equation}
{\cal H} = - \frac{d^2}{d x^2} + x ^2
\label{2012-m-ch-fa-hphoe}
\end{equation}
corresponding to a harmonic oscillator.
The resulting eigenvalue equation is
\begin{equation}
\begin{split}
{\cal H} f(x) = \left(- \frac{d^2}{d x^2} +  x^2 \right) e^{-\frac{x^2}{2}}
=  -\frac{d }{d x }\left(-  x e^{-\frac{x^2}{2}} \right) +  x^2 e^{-\frac{x^2}{2}}  \\
=  e^{-\frac{x^2}{2}} -  x^2 e^{-\frac{x^2}{2}}  +  x^2 e^{-\frac{x^2}{2}}
= e^{-\frac{x^2}{2}}
=  f(x);
\label{2012-m-ch-fa-hphoeee}
\end{split}
\end{equation}
with eigenvalue $1$.
\if01
Let us introduce the
{\em creation}
and
{\em annihilation}
opertors
\index{creation opertors}
\index{annihilation opertors}
${\cal A}^\dagger =  \frac{d}{d x} -  x$
and
${\cal A}  =  - \frac{d}{d x} -  x$,
respectively.
Then,
${\cal A}^\dagger {\cal A}= {\cal H} -1$ and
${\cal A} {\cal A}^\dagger = {\cal H} +1$,
or
${\cal A}{\cal A}^\dagger {\cal A}= {\cal A}{\cal H} -{\cal A}$ and
${\cal A} {\cal A}^\dagger {\cal A}= {\cal H}{\cal A} +{\cal A}$,
and thus
$ {\cal A}{\cal H} -{\cal A} - {\cal H}{\cal A} - {\cal A} = 0$,
hence
$ [{\cal A},{\cal H}] = 2{\cal A}$.
\fi

Instead of going too much into the details here, it may suffice to say
that the
{\em Hermite functions}
\index{Hermite functions}
\begin{equation}
h_n(x) =\pi^{-1/4}(2^n n!)^{-1/2}\left(  \frac{d}{dx} -x\right)^n e^{-{x^2/2}}
= \pi^{-1/4}(2^n n!)^{-1/2} H_n(x) e^{-{x^2/2}}
\end{equation}
are all eigenfunctions of the Fourier transform with the eigenvalue $i^n \sqrt{2\pi }$.
The polynomial $H_n(x)$ of degree $n$ is called {\em Hermite polynomial}. \index{Hermite polynomial}
Hermite functions form a complete system, so that any function $g$ (with $\int \vert g (x) \vert^2 dx <\infty$) has a
{\em Hermite expansion}
\index{Hermite expansion}
\begin{equation}
g(x) = \sum_{k=0}^\infty \langle g , h_n\rangle h_n(x)
.
\end{equation}
This is an example of an
{\em eigenfunction expansion}.
\index{eigenfunction expansion}

\begin{center}
{\color{lightgray}   \Huge
\aldine
}
\end{center}

\chapter{Distributions as generalized functions}
\label{2011-m-ch:gf}

\section{Coping with discontinuities and singularities}

What follows are ``recipes'' and a ``cooking course'' for some ``dishes'' Heaviside, Dirac and others
have enjoyed ``eating,'' alas without being able to ``explain their digestion''
(cf. the citation by Heaviside on page~\pageref{2013-m-ch-intro-cooking}).

Insofar theoretical physics is natural philosophy,
the question arises if ``measurable'' physical entities need to be ``smooth'' and ``continuous'',\cite{trench}
as ``Nature abhors sudden discontinuities,''
or if we are willing to allow and conceptualize singularities of different sorts.
Other, entirely different,
scenarios are discrete,
computer-generated universes.
This little course is no place for preference and judgments regarding these matters.
Let me just point out that contemporary mathematical physics is not only leaning toward,
but appears to be deeply committed to discontinuities;
both in classical and quantized field theories dealing with
 ``point charges,''
as well as in general relativity,  the (nonquantized field theoretical)
geometrodynamics of gravitation,
dealing with singularities such as ``black holes'' or ``initial singularities'' of various sorts.

Discontinuities were introduced quite naturally as electromagnetic pulses,
which can, for instance, be described with the {\em Heaviside function}
$H(t)$ representing vanishing, zero field strength until time $t=0$, when suddenly a constant electrical field is
``switched on eternally.''
It is quite natural to ask what the derivative of the unit step function $H(t)$ might be.
---
At this point, the reader is kindly asked to stop reading for a moment and contemplate
on what kind of function that might be.

Heuristically, if we call this derivative the {\em (Dirac) delta function} $\delta$ defined by
$\delta (t)= \frac{d H(t)}{dt}$,
we can assure ourselves of two of its properties
(i) ``$\delta (t) =0$ for $t\neq 0$,''
as well as the antiderivative of the Heaviside function, yielding
(ii) ``$\int_{-\infty}^\infty \delta (t) dt  = \int_{-\infty}^\infty \frac{d H(t)}{dt} dt  =
H(\infty ) - H(-\infty ) = 1-0=1$.''

\marginnote{This heuristic definition of the Dirac delta function $\delta_y (x)= \delta (x,y)= \delta (x-y)$
with a discontinuity at $y$
is not unlike the discrete Kronecker symbol $\delta_{ij}$.
We may even define the Kronecker symbol $\delta_{ij}$
as the difference quotient of some ``discrete Heaviside function''
$H_{ij} =1$ for $i\ge j$, and $H_{i,j} =0$ else:
$\delta_{ij} = H_{ij} -H_{(i-1)j} = 1$ only for $i=j$; else it vanishes.}

Indeed, we could follow a pattern of ``growing discontinuity,''
reachable by ever higher and higher derivatives of the
absolute value (or modulus); that is, we shall pursue the path sketched by
$$
\vert x\vert
\stackrel{\frac{d}{dx}  }{  \longrightarrow}
\textrm{sgn}(x) ,\,
H(x)
\stackrel{\frac{d}{dx} }{ \longrightarrow}
\delta (x)
\stackrel{\frac{d^n}{dx^n} }{ \longrightarrow}
\delta^{(n)} (x)
.
$$

Objects like $\vert x\vert$,  $H(x)=\frac{1}{2}\left[ 1+\textrm{sgn}(x)\right]$ or $\delta (x)$ may be heuristically understandable
as ``functions'' not unlike
the regular analytic functions; alas
their $n$th derivatives cannot be straightforwardly defined.
In order to cope with a formally precise definition
and derivation of (infinite) pulse functions and to achieve this goal,
a theory of  {\em  generalized functions,}
or, used synonymously,
{\em distributions}
has been developed.
In what follows we shall
develop the theory of distributions;
always keeping in mind the assumptions
regarding (dis)continuities
that make necessary this part of the calculus.

The {\it Ansatz} pursued\cite{richards_youn_1990}
will be to ``pair'' (that is, to multiply) these generalized functions $F$ with suitable ``good''
test functions $\varphi$,
and integrate over these functional pairs $F \varphi$.
Thereby we obtain a linear continuous functional
$F[\varphi]$,
also denoted by
$\langle F, \varphi \rangle $.
This strategy allows for the
``transference'' or ``shift'' of operations on, and transformations of, $F$
-- such as differentiations or Fourier transformations, but also multiplications with polynomials or other smooth functions --
to the test function $\varphi$ according to
{\em adjoint identities}
\index{adjoint identities}
\marginnote{See Sect. 2.3 in \bibentry{strichartz}.}
\begin{equation}
\langle \textsf{\textbf{T}} F, \varphi \rangle
=
\langle F, \textsf{\textbf{S}} \varphi \rangle.
\end{equation}
For example,
for the $n$'th derivative,
\begin{equation}
\textsf{\textbf{S}} = (-1)^n \textsf{\textbf{T}} = (-1)^n \frac{d^{n}}{dx^{n}};
\end{equation}
and for the Fourier transformation,
\begin{equation}
\textsf{\textbf{S}} =   \textsf{\textbf{T}} = {\cal F} .
\end{equation}
For some (smooth) functional multiplier $g(x)\in C^\infty$ ,
\begin{equation}
\textsf{\textbf{S}} =   \textsf{\textbf{T}} = g(x) .
\label{2013-m-ch-di-multipl}
\end{equation}

One more issue is the problem of the meaning and existence of
{\em weak solutions} (also called   generalized solutions)
\index{weak solution}
of differential equations  for which, if interpreted in terms of regular functions,
the derivatives may not all exist.

{
\color{blue}
\bexample
Take, for example, the wave equation in one spatial dimension
$
\frac{\partial^2 }{\partial t^2} u(x,t)
=c^2
\frac{\partial^2 }{\partial x^2} u(x,t)
.
$
It has a  solution of the form\cite{Barut1990349} $ u(x,t)= f(x-ct) + g(x+ct)$,
where $f$ and $g$ characterize  a travelling ``shape'' of inert, unchanged form.
There is no obvious physical reason why the pulse shape function $f$ or $g$ should be differentiable,
alas if it is not, then $u$ is not differentiable either.
What if we, for instance, set $g=0$, and identify $f(x-ct)$ with the Heaviside infinite pulse function $H(x-ct)$?

\eexample
}

\section{General distribution}

\marginnote{A nice video on ``Setting Up the Fourier Transform of a Distribution''
by Professor Dr. Brad G. Osgood @ Stanford University is available {\it via}
URL
\url{https://youtu.be/47yUeygfj3g}}
Suppose we have some ``function'' $F(x)$; that is, $F(x)$ could be either
a regular analytical function, such as $F(x)=x$,
or some other, ``weirder, singular, function,'' such as the Dirac delta function,
or the derivative of the Heaviside (unit step) function, which might be ``highly discontinuous.''
As an {\it Ansatz},  we may associate with this ``function'' $F(x)$
a
{\em distribution,}
\index{distribution}
or, used synonymously,
a
{\em generalized function}
\index{generalized function}
$F[\varphi ]$
or $\langle F , \varphi \rangle $
which
in the ``weak sense'' is  defined as a {\em continuous linear functional}
by integrating $F(x)$ together with some ``good'' {\em test function} $\varphi$
as follows:\cite[-10mm]{schwartz}
\begin{equation}
 F(x) \longleftrightarrow \langle F , \varphi \rangle \equiv F[\varphi] =\int_{-\infty}^{\infty} F(x) \varphi (x) dx.
\end{equation}
We say that $F[\varphi ]$ or $\langle F , \varphi \rangle $ is the distribution {\em associated with}
or {\em induced by}
$F(x)$.
We can distinguish between a
{\em regular}
and a
{\em singular}
distribution:
\index{regular distribution}
\index{singular functional}
a regular distribution can be defined by a continuous function $F$; otherwise it is called singular.

One interpretation of
$F[\varphi ]\equiv \langle F , \varphi \rangle $
is that  $\varphi$ stands for a sort of ``measurement device'' probing
$F$, the ``system to be measured.''
In this interpretation,
$F[\varphi ]\equiv \langle F , \varphi \rangle $
is the ``outcome'' or ``measurement result.''
Thereby, it completely suffices to say what $F$ ``does to'' some test function $\varphi$; there is nothing more to it.

{
\color{blue}
\bexample
For example, the Dirac Delta function $\delta(x)$, as defined later in Equation~(\ref{2018-m-ch-di-delta}),
is completely characterised by
$$\delta(x)  \longleftrightarrow \delta [\varphi ]\equiv \langle \delta , \varphi \rangle =\varphi (0);$$
likewise,
the shifted Dirac Delta function $\delta_y(x)\equiv \delta (x-y)$ is completely characterised by
$$\delta_y (x) \equiv \delta(x-y) \longleftrightarrow \delta_y
[\varphi ]\equiv \langle \delta_y , \varphi \rangle =\varphi (y).$$
\eexample
}

Many other generalized ``functions'' which are usually not integrable in the interval
$( -\infty , +\infty )$ will, through the pairing with a
``suitable'' or ``good'' test function $\varphi$,
induce a distribution.

{
\color{blue}
\bexample
For example, take
$$1 \longleftrightarrow 1 [\varphi ]\equiv \langle 1 , \varphi \rangle
=\int_{-\infty}^\infty
\varphi (x)
dx  , $$
or
$$x \longleftrightarrow x [\varphi ]\equiv \langle x , \varphi \rangle =\int_{-\infty}^\infty
x\varphi (x)
dx,$$
or
$$e^{2\pi i ax} \longleftrightarrow e^{2\pi i ax} [\varphi ]\equiv \langle e^{2\pi i ax}  ,
\varphi \rangle
=\int_{-\infty}^\infty
e^{2\pi i ax} \varphi (x)
dx  .$$
\eexample
}

\subsection{Duality}

Sometimes, $F[\varphi ]\equiv \langle F , \varphi \rangle $   is also written in a scalar product notation; that is,
$F[\varphi] =\langle F \mid \varphi \rangle$.
This emphasizes the pairing aspect of $F[\varphi ]\equiv \langle F , \varphi \rangle $.
In this view, the set of all distributions $F$ is the {\em dual space} of the set of test functions $\varphi$.
\index{dual space}

\subsection{Linearity}
\index{linearity of distributions}

Recall that a {\em linear} functional is some mathematical entity which maps a function or another mathematical object
into scalars in a linear manner; that is, as the integral is linear, we obtain
\begin{equation}
F[c_1\varphi_1+c_2\varphi_2 ]=
c_1F[\varphi_1]  +
c_2F[\varphi_2];
\end{equation}
or, in the bracket notation,
\begin{equation}
\langle F ,   c_1\varphi_1+c_2\varphi_2 \rangle  =
c_1 \langle F ,  \varphi_1   \rangle  +
c_2 \langle F ,   \varphi_2   \rangle .
\end{equation}
This linearity is guaranteed by integration.

\subsection{Continuity}

One way of expressing {\em continuity} is
\index{continuity of distributions} the following:
\begin{equation}
\textrm{if }
\varphi_n \stackrel{n\rightarrow \infty}{\longrightarrow} \varphi
\textrm{, then }
F[\varphi_n ] \stackrel{n\rightarrow \infty}{\longrightarrow} F[\varphi  ],
\end{equation}
or, in the bracket notation,
\begin{equation}
\textrm{if }
\varphi_n \stackrel{n\rightarrow \infty}{\longrightarrow} \varphi
\textrm{, then }
\langle F ,    \varphi_n  \rangle \stackrel{n\rightarrow \infty}{\longrightarrow} \langle F ,    \varphi   \rangle .
\end{equation}

\section{Test functions}

Test functions are useful for a consistent definition of generalized functions.
Nevertheless, the results obtained should be independent of their particular form.

\subsection{{\it Desiderata} on test functions}

By invoking test functions, we would like to be able to differentiate distributions very much like ordinary functions.
We would also like to transfer differentiations to the functional context.
How can this be implemented in terms of possible ``good'' properties we require from the behavior of test functions, in accord with our wishes?

Consider the partial integration
obtained from $(uv)' = u'v+uv'$; thus
$\int (uv)' = \int u'v+\int uv'$,
and finally   $\int u'v = \int (uv)'  -\int uv'$,
thereby effectively allowing us to ``shift'' or ``transfer''
the differentiation of the original function to the test function.
By identifying $u$ with the generalized function $g$ (such as, for instance  $\delta$),
and $v$ with the test function $\varphi$, respectively, we obtain
\begin{equation}
\begin{split}
\langle g' ,    \varphi  \rangle
\equiv
g'[\varphi] =
\int_{-\infty}^\infty
g'(x)\varphi(x)
dx      \\
=
\left.
g(x)\varphi(x)\right|_{-\infty}^\infty
-  \int_{-\infty}^\infty
g(x)\varphi'(x)
dx           \\
=
\underbrace{g(\infty)\varphi(\infty)}_{\textrm{should vanish}} - \underbrace{g(-\infty)\varphi(-\infty)}_{\textrm{should vanish}}
-  \int_{-\infty}^\infty
g(x)\varphi'(x)
dx  \\
= - g[\varphi'] \equiv   -   \langle g ,    \varphi'  \rangle
.
\end{split}
\label{2012-m-ch-di-desiderata}
\end{equation}
We can justify the two main requirements of ``good'' test functions, at least for a wide variety of purposes:
\begin{enumerate}
\item
that they ``sufficiently'' vanish at infinity -- this can, for instance, be achieved by requiring that their support
(the set of arguments $x$ where $g(x)\neq 0$) is finite; and
\item
that they are continuously differentiable -- indeed, by induction, that they are arbitrarily often differentiable.
\end{enumerate}

In what follows we shall enumerate three types of suitable test functions satisfying these {\it desiderata}.
One should, however, bear in mind that the class of ``good'' test functions depends on the distribution.
Take, for example, the Dirac delta function $\delta (x)$. It is so ``concentrated'' that any (infinitely often)
differentiable -- even constant -- function $f(x)$ defined ``around $x=0$''
can serve as a ``good'' test function (with respect to $\delta$),
as $f(x)$ is only evaluated at $x=0$; that is, $\delta[f]=f(0)$.
This is again an indication of the {\em duality} between distributions on the one hand,
and their test functions on the other hand.

Note that if $\varphi (x)$ is a ``good'' test function, then
\begin{equation}
x^\alpha P_n (x)\varphi (x), \alpha \in \mathbb{R} n \in \mathbb{N}
\end{equation}
with any Polynomial $P_n (x)$, and, in particular, $x^n\varphi (x)  $, is also
 a ``good'' test function.

\subsection{Test function class I}

Recall that we require\cite{schwartz} our test functions $\varphi$
to be infinitely often differentiable. Furthermore, in order to get rid of terms at infinity ``in a straightforward, simple way,''
suppose that their support is compact.
Compact support means that $\varphi (x)$ does not vanish only at a finite, bounded region of $x$.
Such a ``good'' test function is, for instance,
\begin{equation}
\varphi_{\sigma ,a}(x)
=
\begin{cases}
\exp \left\{ -\left[ 1 - \left( \frac{x-a}{\sigma }\right)^2 \right]^{-1} \right\} & \textrm{ for } \left\vert  {x-a\over \sigma }\right\vert <1, \\
                                0 & \textrm{ else.}
\end{cases}
\label{2018-m-ch-di-tf1}
\end{equation}

{\color{OliveGreen}
\bproof
In order to show that $\varphi_{\sigma ,a}$ is a suitable test function,
we have to prove its infinite differentiability, as well as the compactness of its support
$M_{\varphi_{\sigma, a}}$.
Let
$$
   \varphi_{\sigma,a}(x):=\varphi\left({ x- a
   \over\sigma}\right)
$$
and thus
\begin{equation}
   \varphi(x)=
\begin{cases}
\exp \left( \frac{1}{x^2-1}\right) &\textrm{ for } |x|<1\\
                 0 &\textrm{ for }|x|\geq 1
.
\end{cases}
\label{2019-m-ch-di-tf1-mod}
\end{equation}
This function is drawn in Figure~\ref{2011-m-fd1}.
{\color{black}
\begin{marginfigure}
\begin{center}
\begin{tikzpicture}[ scale=0.6,
 declare function={
    func(\x)= (\x < 1) * (0)   +
              and(\x >= -1,\x <= 1) * (exp(1/( \x*\x-1)))     +
              (\x > 1) * (0)
   ;
                  } ]


\begin{axis}[ ticklabel style = {font=\Large },
]
\addplot [
orange,
domain=-2:2,
samples=201,
line width=3pt
]  {func(x)};
\end{axis}
\end{tikzpicture}
\end{center}
\caption{Plot of a test function $\varphi(x)$. }
\label{2011-m-fd1}
\end{marginfigure}
}

First, note, by definition, the support  $M_\varphi=(-1,1)$,
because $\varphi (x)$   vanishes outside  $(-1,1)$).

Second, consider the differentiability of $\varphi (x)$;
that is $\varphi\in C^\infty({\Bbb R})$?
Note that
$\varphi^{(0)}=\varphi$ is continuous;
and that $\varphi^{(n)}$ is of the form
$$
   \varphi^{(n)}(x)=\left\{\begin{array}{cl}
                         {P_n(x)\over(x^2-1)^{2n}}    e^{1\over x^2-1}&\mbox{for $|x|<1$}\\
                         0&\mbox{for $|x|\geq1$,}
                    \end{array}\right.
$$
where $P_n(x)$ is a finite polynomial in $x$
 ($\varphi(u)=e^u\Longrightarrow
\varphi'(u)={d\varphi\over du}{du\over dx^2}{dx^2\over dx}=\varphi(u)
\left(-{1\over(x^2-1)^2}\right)2x$ etc.) and $[x=1-\varepsilon]\Longrightarrow
x^2=1-2\varepsilon+\varepsilon^2\Longrightarrow x^2-1=
\varepsilon(\varepsilon-2)$
\begin{eqnarray*}
   \lim_{x\uparrow1}\varphi^{(n)}(x)&=&\lim_{\varepsilon\downarrow0}
      {P_n(1-\varepsilon)\over\varepsilon^{2n}(\varepsilon-2)^{2n}}
      e^{1\over\varepsilon(\varepsilon-2)}=\\
   &=&\lim_{\varepsilon\downarrow0}{P_n(1)\over\varepsilon^{2n}2^{2n}}
      e^{-{1\over2\varepsilon}}=\left[\varepsilon={1\over R}\right]=
      \lim_{R\to\infty}{P_n(1)\over2^{2n}}R^{2n}e^{-{R\over2}}=0,
\end{eqnarray*}
because the power $e^{-x}$ of $e$
decreases stronger
than any polynomial  $x^n$.

Note that the complex continuation
$\varphi (z)$ is not an analytic function and cannot be expanded as a Taylor series on the
entire complex plane ${\Bbb C}$ although it is infinitely often
differentiable on the real axis; that is, although
$\varphi\in C^\infty({\Bbb R})$.
This can be seen from a uniqueness theorem of complex analysis.
Let
 $B\subseteq{\Bbb C}$ be a domain, and
let
$z_0\in B$ the limit of a sequence
$\{z_n\}\in B$, $z_n\ne z_0$.
Then it can be shown that, if two
analytic functions
$f$ and $g$ on $B$  coincide in the points $z_n$,
then they coincide on the entire domain $B$.

Now, take  $B={\Bbb R}$ and
the  vanishing analytic function $f$; that is,
$f(x)=0$.
$f(x)$ coincides with $\varphi (x)$ only in
 ${\Bbb R} - M_\varphi$.
As a result, $\varphi$ cannot be analytic.

Indeed, suppose one does not consider  the piecewise definition~(\ref{2018-m-ch-di-tf1}) of $\varphi_{\sigma, a}(x)$
(which ``gets rid'' of the ``pathologies'')
but just concentrates on its ``exponential part'' as a standalone function on the entire real continuum,
then $\exp \left\{ -\left[ 1 - \left( \frac{x-a}{\sigma }\right)^2 \right]^{-1} \right\}$ diverges at $x=a\pm\sigma$  when
computed from the ``outer regions''  $\left\vert  (x-a)/ \sigma \right\vert \ge 1$.
Therefore this function cannot be Taylor expanded around these two singular points;
and hence smoothness (that is, being in  $C^\infty$)
not necessarily implies that its continuation into the complex plain results in an analytic function. (The converse is true though: analyticity implies smoothness.)
\eproof
}

Another possible test function\cite{sommer-tao-glaettung} is a variant of $\varphi(x)$ defined in~(\ref{2019-m-ch-di-tf1-mod}), namely
\begin{equation}
   \eta (x)=
\begin{cases}
\exp \left( \frac{x^2}{x^2-1}\right) &\textrm{ for } |x|<1\\
                 0 &\textrm{ for }|x|\geq 1
.
\end{cases}
\label{2019-m-ch-di-tf1-mod1}
\end{equation}
$\eta $ has the same compact  support  $M_\varphi=(-1,1)$ as $\varphi(x)$; and it is also in $C^\infty({\Bbb R})$.
Furthermore, $\eta (0) =1$, a property required for smoothing functions used in the summation of divergent series
reviewed in Section~\ref{2019-mm-ds-zetafr}.

\subsection{Test function class II}

Other ``good'' test functions are\cite{schwartz}
\begin{equation}
\left\{\phi_{c,d}(x)\right\}^\frac{1}{n}
\end{equation}
obtained by choosing $n\in {\Bbb N} - 0$
and $-\infty \le c<d\le \infty$ and by defining
\begin{equation}
\phi_{c,d}(x)
=
\begin{cases}
e^{-\left( \frac{1}{x-c} + \frac{1}{d-x} \right)} & \textrm{ for }  c<x<d ,   \\
                                                0 & \textrm{ else.}
\end{cases}
\end{equation}

\subsection{Test function class III: Tempered distributions and Fourier transforms}
\label{2019-mm-ft-td}

A particular class of ``good'' test functions -- having the property that they vanish
``sufficiently fast'' for large arguments, but are nonzero at any finite argument --
are capable of rendering Fourier transforms of generalized functions. Such generalized functions are called
{\em tempered distributions}.
\index{tempered distributions}

One example of a test function yielding tempered distribution is the {\em Gaussian function}
\index{Gaussian function}
\begin{equation}
\varphi (x)= e^{-\pi x^2}.
\label{2012-m-ch-di-td}
\end{equation}
We can multiply the Gaussian function with polynomials (or take its derivatives) and thereby obtain a particular class of test functions
inducing tempered distributions.

The Gaussian function is normalized such that
\begin{equation}
\begin{split}
\int_{-\infty}^{ \infty} \varphi (x)dx =
\int_{-\infty}^{ \infty} e^{-\pi x^2}dx \\
  [\textrm{variable substitution }\; x = \frac{t}{\sqrt{\pi}}, \, dx = \frac{dt}{\sqrt{\pi}} ]\\
= \int_{-\infty}^{ \infty}  e^{-\pi \left(\frac{t}{\sqrt{\pi}}\right)^2}d\left(\frac{t}{\sqrt{\pi}}\right)  \\
=\frac{1}{ \sqrt{\pi} } \int_{-\infty}^{ \infty}  e^{- t^2}dt  \\
=\frac{1}{\sqrt{\pi} } \sqrt{\pi} = 1.
\end{split}
\end{equation}
In this evaluation, we have used the {\em Gaussian integral}
\index{Gaussian integral}
\begin{equation}
I= \int_{-\infty}^{ \infty}  e^{-x^2}dx=   \sqrt{\pi},
\label{2018-m-ch-di-gi}
\end{equation}
which can be obtained by considering its square and transforming into polar coordinates $r,\theta$; that  is,
\begin{equation}
\begin{split}
I^2 =
\left(\int_{-\infty}^{ \infty}  e^{-x^2}dx\right)\left(\int_{-\infty}^{ \infty}  e^{-y^2}dy\right)  \\
  =
 \int_{-\infty}^{ \infty} \int_{-\infty}^{ \infty}   e^{-\left(x^2+y^2\right)}dx \,dy   \\
  =
 \int_{0}^{2\pi } \int_{0}^{ \infty}   e^{-r^2}r \, d\theta \,dr   \\
  =
 \int_{0}^{2\pi }  d\theta \int_{0}^{ \infty}   e^{-r^2}r  \,dr   \\
  =
 2\pi   \int_{0}^{ \infty}   e^{-r^2}r  \,dr   \\
\left[
u=r^2, \frac{du}{dr} =2r, dr =  \frac{du}{2r}
\right]   \\
  =
  \pi   \int_{0}^{ \infty}   e^{-u}  \,du  \\
  =
  \pi     \left( \left. - e^{-u} \right|_{0}^{ \infty} \right)  \\
  =
  \pi     \left( - e^{-\infty} + e^{0} \right)  \\
  =
  \pi .
\end{split}
\label{2012-m-ch-di-gi2}
\end{equation}

The  Gaussian  test function (\ref{2012-m-ch-di-td})
has the advantage that, as has been shown in
(\ref{2012-m-ch-fa-gift}),
with a particular kind of definition for the Fourier transform,
namely  $A=2\pi $ and $B=1$  in Equation~(\ref{2011-m-efta1mg}),\marginnote{$A$ and $B$ refer to Equation~(\ref{2011-m-efta1mg}), page~\pageref{2011-m-efta1mg}.}
its functional form does not change under Fourier transforms.
More explicitly, as derived in Equations~(\ref{2012-m-ch-fa-gift})
and
(\ref{2012-m-ch-fa-giftinverse}),
\begin{equation}
    {\cal F}[\varphi (x)](k)=\widetilde{\varphi} (k) =  \int_{-\infty}^\infty
                e^{-\pi  x^2}  e^{- 2\pi ikx} dx
 = e^{-{\pi k^2}} .
\end{equation}

Just as for differentiation discussed later it is possible to ``shift'' or ``transfer'' the
Fourier transformation from the distribution to the test function
as follows.
\index{Fourier transformation}
Suppose we are interested in the Fourier transform ${\cal F}[F]$ of some distribution $F$.
Then, with the convention
 $A=2\pi $ and $B=1$  adopted in Equation~(\ref{2011-m-efta1mg}), we must consider
\begin{equation}
\begin{split}
\langle  {\cal F}[F ], \varphi \rangle \equiv {\cal F}[F ][\varphi ]
=
\int_{-\infty}^\infty {\cal F}[F](x) \varphi (x) dx
\\ \qquad =
\int_{-\infty}^\infty \left[ \int_{-\infty}^\infty F(y) e^{- 2\pi ixy} dy \right] \varphi (x) dx
\\ \qquad =
\int_{-\infty}^\infty  F(y)  \left[ \int_{-\infty}^\infty \varphi (x) e^{- 2\pi ixy}  dx \right] dy
\\ \qquad =
\int_{-\infty}^\infty  F(y)  {\cal F}[ \varphi ](y) dy
\\ \qquad =
\langle  F , {\cal F}[\varphi ]\rangle \equiv F [{\cal F}[\varphi ]]
.
\end{split}
\label{2019-mm-ft-1}
\end{equation}
in the same way we obtain the
{\em Fourier inversion}
\index{Fourier inversion}
for distributions
\begin{equation}
\langle   {\cal F}^{-1}[{\cal F}[F ]], \varphi \rangle
=
\langle   {\cal F}[{\cal F}^{-1}[F ]], \varphi \rangle
=
\langle    F  , \varphi \rangle
.
\end{equation}

Note that, in the case of test functions with compact support -- say, $\widehat{\varphi} (x) = 0$ for $\vert x \vert > a > 0$ and finite $a$
--  if the order of integrations is exchanged, the ``new test function''
\begin{equation}
{\cal F}[ \widehat{\varphi}] (y)=
\int_{-\infty}^\infty  \widehat{\varphi} (x) e^{- 2\pi ixy}  dx
=
\int_{-a}^a  \widehat{\varphi} (x) e^{- 2\pi ixy}  dx
\label{2019-mm-ft-2}
\end{equation}
obtained through a Fourier transform of  $\widehat{\varphi} (x)$,
does not necessarily inherit a compact support  from $\widehat{\varphi} (x)$;
in particular,
${\cal F}[ \widehat{\varphi}] (y)$
may not necessarily vanish [i.e.  ${\cal F}[ \widehat{\varphi}] (y) = 0$] for $\vert y \vert > a > 0$.

{
\color{blue}
\bexample
Let us, with these conventions, compute the Fourier transform of the tempered Dirac delta distribution.
Note that, by the very definition of the  Dirac delta distribution,
\begin{equation}
\begin{split}
\langle    {\cal F}[\delta  ] , \varphi \rangle
=
\langle   \delta , {\cal F}[\varphi ]\rangle
\\ =
{\cal F}[\varphi](0)=  \int_{-\infty}^\infty  e^{- 2\pi ix 0} \varphi (x) dx
=  \int_{-\infty}^\infty  1 \varphi (x) dx
=  \langle   1 , \varphi \rangle
.
\end{split}
\end{equation}
Thus we may identify  ${\cal F}[\delta  ]$ with $1$; that is,
\begin{equation}
{\cal F}[\delta  ] = 1
.
\end{equation}
This is an extreme example of an {\em infinitely concentrated} object whose Fourier transform is
{\em infinitely spread out}.

A very similar calculation renders the tempered distribution associated with the Fourier transform of the shifted Dirac delta distribution
\begin{equation}
{\cal F}[\delta_y  ] = e^{- 2\pi ix y}
.
\end{equation}
\eexample
}

Alas, we shall pursue a different, more conventional, approach, sketched in Section \ref{2012-m-ch-di-ftgeneraldefcon}.

\subsection{Test function class $C^\infty$}

If the generalized functions are ``sufficiently concentrated'' so that they themselves guarantee that the terms
$g(\infty)\varphi(\infty)$ as well as $g(-\infty)\varphi(-\infty)$
in Equation~(\ref{2012-m-ch-di-desiderata}) to vanish,
we may just require the test functions to be infinitely differentiable -- and thus in $C^\infty$ --
for the sake of making possible a transfer of differentiation.
(Indeed, if we are willing to sacrifice even infinite differentiability, we can widen this class of test functions even more.)
We may, for instance, employ constant functions such as $\varphi (x)=1$ as test functions,
thus giving meaning to, for instance,
$\langle \delta , 1\rangle= \int_{-\infty}^\infty \delta (x) dx$,
or
$\langle f(x)\delta , 1\rangle= \langle f(0)\delta , 1\rangle= f(0)\int_{-\infty}^\infty  \delta (x) dx$.

However, one should keep in mind that constant functions, or arbitrary smooth functions, do not comply with the generally accepted notion of a test function.
Test functions are usually assumed to have either a compact support or at least decrease sufficiently fast to allow,
say,  vanishing nonintegral surface terms in integrations by parts.

\section{Derivative of distributions}

Equipped with ``good'' test functions
which have a finite support and are
infinitely often (or at least sufficiently often) differentiable,
we can now give meaning to the transferral  of differential quotients from
the objects entering the integral towards the test function by {\em partial integration}.
First note again that $(uv)' = u'v+uv'$
and thus
$\int (uv)' = \int u'v+\int uv'$
and finally   $\int u'v = \int (uv)'  -\int uv'$.
Hence,     by identifying $u$ with $g$, and $v$ with the test function $\varphi$, we obtain
\begin{equation}
\begin{split}
\langle {F}' , \varphi \rangle \equiv {F}'\left[\varphi\right] =
\int_{-\infty}^\infty
\left( \frac{d}{dx} F(x)\right) \varphi (x) dx
\\
\qquad =
\underbrace{\left. F(x) \varphi (x) \right|_{x=-\infty}^\infty}_{=0}
- \int_{-\infty}^\infty
F(x)\left( \frac{d}{dx} \varphi (x) \right) dx \\
\qquad =
- \int_{-\infty}^\infty
F(x)\left( \frac{d}{dx} \varphi (x) \right) dx \\
\qquad =-F\left[\varphi  '\right] \equiv - \langle {F} , \varphi '\rangle .
\end{split}
\end{equation}
By induction
\begin{equation}
\left\langle \frac{d^{n}}{dx^{n}}{F} , \varphi \right\rangle
\equiv
\langle {F}^{(n)} , \varphi \rangle \equiv F^{(n)}\left[\varphi\right]
 = (-1)^n F\left[\varphi  ^{(n)}\right]
 = (-1)^n   \langle {F} , \varphi^{(n)}\rangle.
\end{equation}

{
\color{blue}
\bexample

In anticipation of the definition (\ref{2018-m-ch-di-delta}) of the delta function by $\delta [\varphi ]=\varphi (0)$
we immediately obtain its derivative by $\delta' [\varphi ]= - \delta[\varphi '  ]=-\varphi' (0)$.

For the sake of a further example using   adjoint identities
\index{adjoint identities},
to swapping products and differentiations forth and back
through the $F$--$\varphi$ pairing, let us compute
$g(x)\delta' (x)$ where $g \in C^\infty$; that is
\begin{equation}
\begin{split}
g \delta' [\varphi ] \equiv
\langle g \delta'   , \varphi \rangle
=
\langle \delta'   , g  \varphi \rangle
 =
- \langle \delta   , (g  \varphi )'\rangle
 =  \\
- \langle \delta   ,  g  \varphi  '+ g'  \varphi  \rangle
 =
-  g(0)  \varphi ' (0) - g'(0)  \varphi(0)
 =
  \langle g(0) \delta '   -  g'(0)\delta , \varphi   \rangle
\\ \equiv
\left(g(0)\delta ' -g'(0)\delta \right) [\varphi ]
= g(0)\delta '[\varphi] -g'(0)\delta [\varphi ]
.
\end{split}
\end{equation}
\eexample
}
Therefore,  in the functional sense,
\begin{equation}
g(x)\delta' (x)=g(0) \delta '(x)   -  g'(0)\delta (x) .
\label{2012-m-ch-di-sederi}
\end{equation}

\section{Fourier transform  of distributions}
\label{2012-m-ch-di-ftgeneraldefcon}

We mention without proof that, if $\{ f_n(x)\}$ is a sequence of functions converging, for $n\rightarrow \infty$,
toward a function $f$ in the functional sense (i.e. {\it via}
integration of $f_n$ and $f$ with ``good'' test functions),
then the Fourier transform $\widetilde f$ of $f$ can be defined by\cite{Lighthill,Howell,doi:10.1080/0020739900210418}
\begin{equation}
\begin{split}
 {\cal F}[f]=\widetilde{f}(k)= \lim_{n\rightarrow \infty}
 \int_{-\infty}^\infty  f_n(x) e^{-i{kx}} dx
.
\end{split}
\end{equation}

While this represents a method to calculate  Fourier transforms of distributions, there are other, more direct ways of
obtaining them.
These were mentioned earlier.

\section{Dirac  delta function}
\index{Dirac delta function}
\index{delta function}

The theory of distributions has been stimulated by physics.
Historically, the Heaviside step function, which will be discussed later --
was used for the description of electrostatic pulses.

In the days when Dirac developed quantum mechanics
(cf. \S 15 of Ref.~\cite[-10mm]{dirac})
there was a need to define
``singular scalar products'' such as ``$\langle x \mid y \rangle = \delta (x-y)$,''
with some generalization of the Kronecker delta function $\delta_{ij}$, depicted in Figure~\ref{2011-m-fdeltaplot},
which is zero whenever $x\neq y$;
and yet at the same time ``large enough'' and
``needle shaped'' as depicted in Figure~\ref{2011-m-fdeltaplot} to yield unity when
integrated over the entire reals; that is, ``$\int_{-\infty}^\infty \langle x \mid y \rangle dy =\int_{-\infty}^\infty \delta (x-y) dy =1$.''
\begin{marginfigure}%
\begin{center}
\begin{tikzpicture}[ scale=0.6,
 declare function={
    func(\x)= 0;
                  } ]

\begin{axis}[
ticklabel style = {font=\Large },
ymin=-0.5,
ymax=5.0,
xmin=-2.5,
xmax=2.5,
height=5.5cm,width=5cm,
scale only axis
]
\addplot [
orange,
domain=-2:2,
line width=3pt
]  {func(x)};

\draw[orange,line width=3pt,->] (axis cs:0,0) --   (axis cs:0,4.2) node[above]{{\Large $\delta$}};

\end{axis}
\end{tikzpicture}
\end{center}
\caption{Dirac's $\delta$-function as a ``needle shaped'' generalized function.}
  \label{2011-m-fdeltaplot}
\end{marginfigure}

Naturally, such ``needle shaped functions'' were viewed suspiciously by many mathematicians
at first, but later they embraced these types of functions\cite[20mm]{gelfand:1964:gf} by developing a theory of
{\em functional analysis}
\index{functional analysis},
{\em generalized functions}
\index{generalized functions}
or, by another naming,
{\em distributions}.
\index{distributions}

In what follows we shall first define the Dirac delta function by delta sequences; that is, by sequences of functions which render the
delta function in the limit.
Then the delta function will be formally defined in~(\ref{2018-m-ch-di-delta}) by $\delta [\varphi ]=\varphi (0)$.

\subsection{Delta sequence}
One of the first attempts to formalize these objects with ``large discontinuities''
was in terms
of functional limits.
Take, for instance, the {\em delta sequence}
\index{delta sequence}
of ``strongly peaked'' pulse functions
depicted in Figure~\ref{2011-m-fdeltaplotnseq};
defined by
\begin{equation}
\delta_n(x-y) =
\left\{
\begin{array}{rl}
n & \textrm{ for } y - \frac{1}{2n}  < x < y+ \frac{1}{2n} \\
0& \textrm{ else. }
\end{array}
\right.
\label{2011-m-deltseq}
\end{equation}
In  the functional sense the ``large $n$ limit''   of
the sequences $\{f_n(x-y)\}$
 becomes the delta function $\delta (x-y)$:
\begin{equation}
\lim_{n\rightarrow \infty} \delta_n(x-y)= \delta (x-y) ;
\end{equation}
that is,
\begin{equation}
\lim_{n\rightarrow \infty} \int \delta_n(x-y) \varphi (x) dx = \delta_y [\varphi ]=\varphi (y).
\end{equation}
\begin{marginfigure}%
\begin{center}
\begin{tikzpicture}[ scale=0.6,
 declare function={
    func(\x)= 0;
    func1(\x)= (\x <= 1/2) * (0)   +
              and(\x >= -1/2,\x <= 1/2) * (1)     +
              (\x >= 1/2) * (0) ;
    func2(\x)= (\x < 1/4) * (0)   +
              and(\x >= -1/4,\x <= 1/4) * (2)     +
              (\x > 1/4) * (0) ;
    func3(\x)= (\x < 1/8) * (0)   +
              and(\x >= -1/8,\x <= 1/8) * (4)     +
              (\x > 1/8) * (0) ;
                  } ]

\begin{axis}[
ticklabel style = {font=\Large },
ymin=-0.5,
ymax=5.0,
xmin=-1.5,
xmax=1.5,
height=5.5cm,width=5cm,
scale only axis
]

\addplot [
orange,
domain=-1.25:1.25,
line width=3pt
]  {func(x)};

\draw[orange,line width=3pt,->] (axis cs:0,0) -- (axis cs:0,4.7) node[right]{{\Large $\;\delta$}};

\addplot [
blue!70,
domain=-1/8:1/8,
line width=3pt
]  {func3(x)} node[right]{{\color{blue}$\delta_3$}};

\draw[blue!70,line width=0.2pt] (axis cs:-1/8,0) -- (axis cs:-1/8,4);
\draw[blue!70,line width=0.2pt] (axis cs:1/8,0) -- (axis cs:1/8,4);

\draw[blue!70,line width=3pt] (axis cs:-1.25,0) -- (axis cs:-1/8,0);
\draw[blue!70,line width=3pt] (axis cs:1/8,0) -- (axis cs:1.25,0);

\addplot [
blue!50,
domain=-0.25:0.25,
line width=3pt
]  {func2(x)} node[right]{{\color{blue}$\delta_2$}};

\draw[blue!50,line width=0.2pt] (axis cs:-1/4,0) -- (axis cs:-1/4,2);
\draw[blue!50,line width=0.2pt] (axis cs:1/4,0) -- (axis cs:1/4,2);

\draw[blue!50,line width=3pt] (axis cs:-1.25,0) -- (axis cs:-1/4,0);
\draw[blue!50,line width=3pt] (axis cs:1/4,0) -- (axis cs:1.25,0);

\addplot [
blue!30,
domain=-0.5:0.5,
line width=3pt
]  {func1(x)} node[right]{{\color{blue}$\delta_1$}};

\draw[blue!30,line width=0.2pt] (axis cs:-0.5,0) -- (axis cs:-0.5,1);
\draw[blue!30,line width=0.2pt] (axis cs:0.5,0) -- (axis cs:0.5,1);

\draw[blue!30,line width=3pt] (axis cs:-1.25,0) -- (axis cs:-0.5,0);
\draw[blue!30,line width=3pt] (axis cs:0.5,0) -- (axis cs:1.25,0);

\end{axis}
\end{tikzpicture}
\end{center}
\caption{\label{2011-m-fdeltaplotnseq}Delta sequence approximating Dirac's $\delta$-function as a more and more ``needle shaped'' generalized function.}
\end{marginfigure}

Note that, for all $n\in \mathbb{N}$ the area of $\delta_n(x-y)$ above the $x$-axes
is $1$ and independent of $n$, since the width is $1/n$ and the height is $n$, and the of width and height is $1$.

{
\color{blue}
\bexample
Let us proof that the sequence $\{ \delta_n\}$ with
$$\delta_n(x-y) =
\left\{
\begin{array}{rl}
n & \textrm{ for } y - \frac{1}{2n}  < x < y+ \frac{1}{2n} \\
0& \textrm{ else }
\end{array}
\right.$$
defined in Equation~(\ref{2011-m-deltseq}) and depicted
in Figure~
\ref{2011-m-fdeltaplotnseq}
is a delta sequence;
that is, if, for large $n$, it converges to $\delta$ in a functional sense.
In order to verify this claim, we have to integrate $\delta_n(x)$
with ``good'' test functions $ \varphi (x)$ and take the limit $n\rightarrow \infty$;
if the result is $ \varphi (0)$, then we can identify $\delta_n(x)$ in this limit with $\delta (x)$
(in the functional sense).
Since $\delta_n(x)$ is uniform convergent, we can exchange the limit with the integration; thus
\begin{equation}
\begin{split}
\lim_{n\rightarrow \infty} \int_{-\infty}^\infty \delta_n(x-y) \varphi (x) dx \\
 \textrm{{\Large[}variable transformation:}  \\
  x'=x-y, x= x'+y, dx' =  dx, -\infty\le x' \le \infty \textrm{\Large]} \\
 =
\lim_{n\rightarrow \infty} \int_{-\infty}^\infty \delta_n(x') \varphi (x'+y) dx'
 =
\lim_{n\rightarrow \infty} \int_{- \frac{1}{2n}}^\frac{1}{2n} n \varphi (x'+y) dx'    \\
 \textrm{{\Large[}variable transformation:}  \\
  u=2nx', x'=\frac{u}{2n},
   du = 2n dx', -1\le u \le 1\textrm{\Large]} \\
 = \lim_{n\rightarrow \infty} \int_{- 1}^1 n \varphi \left(\frac{ u}{2n}+y\right) \frac{du}{2n}
 =
\lim_{n\rightarrow \infty} \frac{1}{2} \int_{- 1}^1 \varphi \left(\frac{ u}{2n}+y\right) du     \\
 =
\frac{1}{2} \int_{- 1}^1 \lim_{n\rightarrow \infty} \varphi \left(\frac{ u}{2n}+y\right) du
  =
\frac{1}{2} \varphi (y) \int_{- 1}^1  du
  =
\varphi (y)
.
\end{split}
\end{equation}
Hence, in the functional sense,
this limit yields the shifted $\delta$-function $\delta_y$.
Thus we obtain
$
\lim_{n\rightarrow \infty} \delta_n[\varphi] = \delta_y [\varphi] = \varphi (y)
$.
\eexample
}

Other delta sequences can be {\it ad hoc}  enumerated as follows.
They all converge towards the delta function in the sense of linear functionals (i.e. when integrated over a test function).
\begin{eqnarray}
\delta_n(x)
&=& \frac{n}{\sqrt{\pi}} e^{-n^2 x^2},
\label{2012-m-ch-di-g1}\\
&=&
\frac{1}{\pi}\frac{n}{1+ n^2x^2},    \label{2012-m-ch-di-l1} \\
&=&
\frac{1}{\pi}   \frac{\sin (n x)}{ x}, \label{2012-m-ch-di-d1}\\
&=&
= (1\mp i)\left( {n\over 2\pi }\right)^{1\over 2} e^{\pm inx^2}  \\
&=&
\frac{1}{\pi x}  \frac{e^{inx}-e^{-inx}}{2i} ,\\
&=&
\frac{1}{\pi}  \frac{n  e^{-x^2}}{1+n^2x^2} ,\\
&=&
\frac{1}{2\pi } \int_{-n}^n e^{ixt} dt  = \frac{1}{2\pi i x} \left. e^{ixt}\right|_{-n}^n    ,\\
&=&
\frac{1}{2\pi} \frac{\sin \left[\left( n+\frac{1}{2}\right) x \right]  }{\sin \left( \frac{1}{2}x \right)   },\\
&=&
\frac{n}{ \pi}\left(\frac{\sin (nx)}{nx}\right)^2.
\end{eqnarray}
Other commonly used limit forms of the $\delta $-function are the Gaussian, Lorentzian, and Dirichlet forms
\begin{eqnarray}
\delta_\epsilon (x) &=&   \frac{1}{\sqrt{\pi } \epsilon } e^{-\frac{x^2}{\epsilon^2}} ,
\label{2012-m-ch-di-g2} \\
&=&  \frac{1}{\pi} \frac{\epsilon }{x^2+\epsilon^2}
=   {1\over 2\pi i}
\left(
{1 \over x-i\epsilon }
-
{1 \over x+i\epsilon }
 \right)
 , \label{2012-m-ch-di-l2}  \\
&=&  \frac{1}{\pi } \frac{\sin \left(\frac{x}{\epsilon }\right)}{x} ,  \label{2012-m-ch-di-d2}
\end{eqnarray}
respectively.
Note that
(\ref{2012-m-ch-di-g2}) corresponds to (\ref{2012-m-ch-di-g1}),
(\ref{2012-m-ch-di-l2}) corresponds to (\ref{2012-m-ch-di-l1}) with $\epsilon=n^{-1}$,
and
(\ref{2012-m-ch-di-d2}) corresponds to (\ref{2012-m-ch-di-d1}).
Again, the limit
$
\delta (x)= \lim_{\epsilon \rightarrow 0} \delta_\epsilon (x)
$
has to be understood in the functional sense; that is, by integration over a test function,
so that
\begin{equation}
\lim_{\epsilon \rightarrow 0}\, \delta_\epsilon [\varphi]=
\lim_{\epsilon \rightarrow 0}\, \int_{-\infty}^\infty \delta_\epsilon (x) \varphi (x) dx  =    \delta [\varphi]=\varphi (0).
\end{equation}

\subsection{$\delta \left[ \varphi \right]$ distribution}
\index{delta function}

The distribution (linear functional) associated with the $\delta$ function
can be defined by mapping any test function into a scalar as follows:
\begin{equation}
\delta_y [\varphi ]\stackrel{{\tiny \textrm{ def }}}{=}\varphi (y);
\label{2018-m-ch-di-delta}
\end{equation}
or, as it is often expressed,
\begin{equation}
\int_{-\infty}^\infty
\delta(x-y) \varphi (x) dx = \varphi(y).
\end{equation}
Other common ways of expressing this {\em delta function distribution} is by writing
\begin{equation}
\delta(x-y) \longleftrightarrow
\langle \delta_y , \varphi \rangle \equiv
\langle \delta_y \vert \varphi \rangle \equiv
\delta_y [\varphi] = \varphi (y).
\end{equation}
For $y=0$, we just obtain
\begin{equation}
\delta(x) \longleftrightarrow
\langle \delta , \varphi \rangle \equiv
\langle \delta \vert \varphi \rangle \equiv
\delta [\varphi] \stackrel{{\tiny \textrm{ def }}}{=} \delta_0 [\varphi] =  \varphi (0).
\end{equation}
Note that $\delta_y[\varphi]$ is a singular distribution,
as no regular function is capable of such a performance.
\index{singular distribution}

\subsection{Useful formul\ae{} involving $\delta$}

The following formul\ae{} are sometimes enumerated without proofs.

 \begin{equation}
 f (x)\delta (x-x_0)
 =
f (x_0)\delta (x-x_0)
\label{2018-m-ch-di-totenk}
 \end{equation}
{\color{OliveGreen}
\bproof
This results from a direct application of Equation~(\ref{2013-m-ch-di-multipl}); that is,
 \begin{equation}
f(x) \delta
[\varphi ]
=
\delta
\left[  f  \varphi \right]
=
f(0)\varphi(0) = f(0)   \delta
[\varphi ]
,
 \end{equation}
and
 \begin{equation}
f(x) \delta_{x_0}
[\varphi ]
=
\delta_{x_0}
\left[  f  \varphi \right]
=
f({x_0})\varphi({x_0}) = f({x_0})   \delta_{x_0}
[\varphi ]
.
 \end{equation}
For a more explicit direct proof, note that formally
 \begin{equation}
 \begin{split}
\int _{-\infty}^\infty f (x)\delta (x-x_0)  \varphi(x) dx
=
\int _{-\infty}^\infty\delta (x-x_0)  ( f (x)\varphi(x) ) dx = f(x_0) \varphi(x_0)
,
 \end{split}
 \end{equation}
and hence $f(x) \delta_{x_0}[ \varphi ] =   f(x_0)\delta_{x_0}[ \varphi ]$.
\eproof
}

 \begin{equation}
 \delta (-x)=\delta (x)
 \end{equation}
{\color{OliveGreen}
\bproof
For a proof, note that $\varphi (x)\delta (-x) = \varphi (0)\delta (-x)$, and that, in particular,
with the substitution $x \rightarrow -x$ and a redefined test function $\psi (x) =\varphi(-x)$:
 \begin{equation}
 \begin{split}
\int _{-\infty}^\infty \delta (-x)  \varphi(x) dx   =
 \int_\infty ^{-\infty}\delta (-(-x)) \underbrace{\varphi(-x)}_{=\psi(x)} d(-x) \\
   =
-\underbrace{\psi(0)}_{\varphi(0)} \int _\infty ^{-\infty}\delta (x)  d x     =
\varphi(0) \int _{-\infty}^\infty \delta (x)  d x =\varphi(0)=\delta[\varphi].
 \end{split}
 \end{equation}
\eproof
}

For the $\delta$ distribution with its ``extreme concentration'' at the origin, a ``nonconcentrated test function'' suffices; in particular, a constant ``test'' function
-- even without compact support and sufficiently strong damping at infinity --
such as $\varphi (x) = 1$ is fine.
This is the reason why test functions need not show up explicitly in expressions, and, in particular, integrals, containing $\delta$.
Because, say, for suitable functions $g(x)$ ``well behaved'' at the origin, formally by invoking~(\ref{2018-m-ch-di-totenk})
 \begin{equation}
 \begin{split}
\int_{-\infty}^{\infty} g(x)\delta(x-y) \,dx =
\int_{-\infty}^{\infty} g(y)\delta(x-y) \,dx \\=
g(y)\int_{-\infty}^{\infty} \delta(x-y)  \,dx =g(y).
 \end{split}
\end{equation}

 \begin{equation}
 x\delta (x)=0
 \end{equation}
{\color{OliveGreen}
\bproof
For a proof invoke~(\ref{2018-m-ch-di-totenk}), or explicitly consider
 \begin{equation}
x \delta
[\varphi ]
=
\delta
\left[  x  \varphi \right]
=
0\varphi(0) = 0.
 \end{equation}
\eproof
}
For $a\neq 0$,
 \begin{equation}
 \delta (ax)={1\over \vert a\vert }\delta (x),
\label{2011-m-distdp}
 \end{equation}
and, more generally,
 \begin{equation}
 \delta (a(x-x_0))={1\over \vert a\vert }\delta (x-x_0)
\label{2011-m-distdpmg}
 \end{equation}
{\color{OliveGreen}
\bproof
For the sake of a proof,  consider the case $a>0$ as well as $x_0=0$ first:
 \begin{equation}
 \begin{split}
\int _{-\infty}^\infty \delta (ax)  \varphi (x)  dx
\\
  [\textrm{variable substitution }\; y = ax, x=\frac{y}{a}, dx=\frac{1}{a} dy]\\
  =
\frac{1}{a}\int _{-\infty}^\infty \delta (y)  \varphi\left (\frac{y}{a}\right) dy  \\
  =    \frac{1}{a}  \varphi (0)=    \frac{1}{\vert a\vert}  \varphi (0);
 \end{split}
 \end{equation}
and, second, the case $a<0$:
 \begin{equation}
 \begin{split}
\int _{-\infty}^\infty \delta (ax)  \varphi (x)  dx     \\
 [\textrm{variable substitution }\; y = ax, x=\frac{y}{a}, dx=\frac{1}{a} dy]\\
 =
\frac{1}{a}\int _\infty^{-\infty} \delta (y)  \varphi\left (\frac{y}{a}\right) dy \\
 =
- \frac{1}{a}\int _{-\infty}^\infty \delta (y)  \varphi\left (\frac{y}{a}\right) dy       \\
 =   - \frac{1}{a}  \varphi (0)=    \frac{1}{\vert a\vert}  \varphi (0).
 \end{split}
 \end{equation}
In the case of $x_0\neq 0$ and $\pm a>0$, we obtain
\begin{equation}
 \begin{split}
\int _{-\infty}^\infty \delta (a(x-x_0))  \varphi (x)  dx
\\
 [\textrm{variable substitution }\; y = a(x-x_0), x=\frac{y}{a}+x_0, dx=\frac{1}{a} dy]\\
 =
\pm \frac{1}{a}\int _{-\infty}^\infty \delta (y)  \varphi \left(\frac{y}{a}+x_0\right) dy  \\
 =    \pm \frac{1}{a}  \varphi (0)=\frac{1}{|a|}  \varphi (x_0).
 \end{split}
 \end{equation}
\eproof
}

If there exists a simple singularity $x_0$ of $f(x)$ in the
integration interval, then
\begin{equation}
 \delta (f(x))={1\over \vert f'(x_0)\vert }\delta(x-x_0)
.
 \end{equation}
More generally,  if $f$ has only simple roots and $f'$ is nonzero there,
\begin{equation}
 \delta (f(x))=\sum_{x_i}{\delta(x-x_i)\over \vert f'(x_i)\vert }
\label{2011-m-distdp1}
 \end{equation}
where the sum extends over all simple roots $x_i$ in
the integration interval.
In particular,
 \begin{equation}
 \delta (x^2-x_0^2)={1\over 2\vert x_0\vert }[\delta (x-x_0)+\delta
 (x+x_0)] \end{equation}
{\color{OliveGreen}
\bproof
For a sloppy proof, note that since $f$ has only simple roots,
\marginnote{An example is  a polynomial of degree $k$ of the form
$f= A \prod_{i=1}^k (x-x_i)$;
with mutually distinct $x_i$, $1\le i \le k$.}
it can be expanded around these roots as
\index{order of}
\index{big $O$ notation}
\index{Bachmann-Landau notation}
\index{asymptotic notation}
\marginnote{Again the symbol ``$O$'' stands for ``of the order of'' or ``absolutely bound by'' in the following way:
if $g(x)$ is a positive function,  then $f(x)=O(g(x))$ implies that there exist a positive real number $m$
such that $\vert f(x) \vert < m g(x)$.}
\[
\begin{split}
f(x) =\underbrace{f(x_0)}_{=0} +(x-x_0) f'(x_0) + O\left((x-x_0)^2\right) \\
= (x-x_0) \left[ f'(x_0) +  O\left(\vert x-x_0 \vert \right) \right] \approx  (x-x_0) f'(x_0)
,
\end{split}
\]
with nonzero $f'(x_0) \in {\Bbb R}$.
\marginnote{The simplest nontrivial case is $f(x) = a + bx  = b \left(\frac{a}{b}+x\right)$,
for which $x_0 = -\frac{a}{b}$ and $f' \left(x_0=\frac{a}{b}\right) = b$.}
By identifying  $f'(x_0)$ with $a$ in
Equation~(\ref{2011-m-distdp})
we obtain
Equation~(\ref{2011-m-distdp1}).

For a proof\cite{Cortizo-95} the integration which originally extend over the set of real numbers $\mathbb{R}$
can be reduced to intervals $[x_i-r_i,x_i+r_i]$, containing the roots $x_i$ of $f(x)$.
so that  the ``radii'' $r_i$  are ``small enough'' for these intervals to be pairwise disjoint,
and $f(x) \neq 0$ for any $x$ outside of the union set of these intervals.
Therefore the integration over the entire reals can be reduced to the sum of the integrations
over the intervals; that is,
\begin{equation}
\int_{-\infty}^{+\infty} \delta (f(x)) \varphi(x) dx
=
\sum_i
\int_{x_i-r_i}^{x_i+r_i} \delta (f(x)) \varphi(x) dx
.
\label{2017-m-ch-di-sum}
\end{equation}
The terms in the sum can be evaluated separately; so let us concentrate on the $i$'th term
$\int_{x_i-r_i}^{x_i+r_i} \delta (f(x)) \varphi(x) dx    $ in~(\ref{2017-m-ch-di-sum}).
Restriction to a sufficiently small single region $[x_i-r_i,x_i+r_i]$, and the assumption of simple roots
guarantees that $f(x)$ is invertible within that region; with the inverse $f_i^{-1}$; that is,
\begin{equation}
f_i^{-1}(f(x))=x \text{ for } x \in  [x_i-r_i,x_i+r_i]
;
\end{equation}
and, in particular, $f(x_i)=0$ and $f_i^{-1}(0) =f_i^{-1}(f(x_i))=x_i$.
Furthermore, this inverse $f_i^{-1}$ is monotonic, differentiable and its derivative is nonzero within $[f(x_i-r_i),f(x_i+r_i)]$.
Define
\begin{equation}
\begin{split}
y = f(x),\\
x = f_i^{-1} (y), \text{ and}\\
dy = f'(x) dx , \text{ or } dx = \frac{dy}{f'(x)}
,
\end{split}
\end{equation}
 so that, for  $f'(x_i)>0$,
\begin{equation}
\begin{split}
\int_{x_i-r_i}^{x_i+r_i} \delta (f(x)) \varphi(x) dx
=
\int_{x_i-r_i}^{x_i+r_i} \delta (f(x)) \frac{\varphi(x)}{f'(x)} f'(x)  dx   \\
=
\int_{f(x_i-r_i)}^{f(x_i+r_i)} \delta (y) \frac{\varphi(f_i^{-1} (y))}{f'(f_i^{-1} (y))}   dy  \\
=  \frac{\varphi(f_i^{-1} (0))}{f'(f_i^{-1} (0))}
=  \frac{\varphi(f_i^{-1} (f(x_i)))}{f'(f_i^{-1} (f(x_i)))}
=  \frac{\varphi( x_i)}{f'( x_i )}.
\end{split}
\end{equation}
Likewise, for  $f'(x_i)<0$,
\begin{equation}
\begin{split}
\int_{x_i-r_i}^{x_i+r_i} \delta (f(x)) \varphi(x) dx
=
\int_{x_i-r_i}^{x_i+r_i} \delta (f(x)) \frac{\varphi(x)}{f'(x)} f'(x)  dx   \\
=
\int_{f(x_i-r_i)}^{f(x_i+r_i)} \delta (y) \frac{\varphi(f_i^{-1} (y))}{f'(f_i^{-1} (y))}   dy
=
- \int_{f(x_i+r_i)}^{f(x_i-r_i)} \delta (y) \frac{\varphi(f_i^{-1} (y))}{f'(f_i^{-1} (y))}   dy  \\
=  - \frac{\varphi(f_i^{-1} (0))}{f'(f_i^{-1} (0))}
=  - \frac{\varphi(f_i^{-1} (f(x_i)))}{f'(f_i^{-1} (f(x_i)))}
=  - \frac{\varphi( x_i)}{f'( x_i )} .
\end{split}
\end{equation}
\eproof
}

 \begin{equation}
 \vert x\vert \delta (x^2)=\delta (x)
 \end{equation}
{\color{OliveGreen}
\bproof
For a proof consider
 \begin{equation}
 \begin{split}
\vert x \vert \delta (x^2)[\varphi ]
=
\int_{-\infty}^{\infty}\vert x \vert \delta (x^2)\varphi (x) dx\\
=
\lim_{a\rightarrow 0^+}
\int_{-\infty}^{\infty}\vert x \vert \delta (x^2-a^2)\varphi (x) dx    \\
=
\lim_{a\rightarrow 0^+}
\int_{-\infty}^{\infty}\frac{\vert x \vert}{2a} \left[ \delta (x -a ) + \delta (x + a )\right] \varphi (x) dx \\
=
\lim_{a\rightarrow 0^+} \left[
\int_{-\infty}^{\infty}\frac{\vert x \vert}{2a}   \delta (x -a ) \varphi (x) dx  +
\int_{-\infty}^{\infty}\frac{\vert x \vert}{2a}  \delta (x + a )\varphi (x) dx \right]  \\
=
\lim_{a\rightarrow 0^+} \left[
 \frac{\vert a \vert}{2a}   \varphi   (a)   +
 \frac{\vert -a \vert}{2a}  \varphi (-a)   \right]  \\
=
\lim_{a\rightarrow 0^+} \left[
 \frac{1}{2 }     \varphi (a)    +
 \frac{1}{2 }  \varphi (-a)   \right]  \\
=
\frac{1}{2 }     \varphi (0)   +
 \frac{1}{2 }  \varphi (0)
= \varphi (0)
=
\delta[\varphi ]
.
 \end{split}
 \end{equation}
\eproof
}

 \begin{equation}
 -x\delta '(x)=\delta (x),
 \end{equation}
which is a direct consequence of Equation~(\ref{2012-m-ch-di-sederi}).
{\color{OliveGreen}
\bproof
More explicitly, we can use partial integration and obtain
 \begin{equation}
 \begin{split}
-\int _{-\infty}^\infty x \delta' (x)  \varphi (x)  dx  \\
 =  - \left. x \delta (x)\right|_{-\infty}^\infty  + \int _{-\infty}^\infty \delta (x)
\frac{d}{dx}\left(x \varphi (x)\right)dx\\
 =
\int _{-\infty}^\infty \delta (x)
x \varphi' (x)  dx
+
\int _{-\infty}^\infty \delta (x)
\varphi (x) dx\\
 =
0 \varphi' (0)  +
\varphi (0)= \varphi (0).
 \end{split}
 \end{equation}
\eproof
}

 \begin{equation}
 \delta^{(n)}(-x) =(-1)^n\delta^{(n)}(x)
,
 \end{equation}
where the index $^{(n)}$ denotes $n$-fold differentiation,
can be proven by
[recall that, by the chain rule of differentiation, $\frac{d}{dx} \varphi (-x)  = - \varphi' (-x)$]
{\color{OliveGreen}
\bproof
 \begin{equation}
 \begin{split}
\int _{-\infty}^\infty \delta^{(n)} (-x)  \varphi (x)  dx  \\
[\textrm{variable substitution }\; x \rightarrow -x]\\
=
-\int _\infty^{-\infty} \delta^{(n)} (x)  \varphi (-x) dx
=
\int _{-\infty}^\infty \delta^{(n)} (x)  \varphi (-x) dx \\
 =
(-1)^n \int _{-\infty}^\infty \delta (x) \left[ \frac{d^n}{dx^n}  \varphi  (-x)\right] dx \\
 =
(-1)^n \int _{-\infty}^\infty \delta (x) \left[(-1)^n  \varphi^{(n)} (-x)\right] dx \\
 =
\int _\infty^{-\infty} \delta (x) \varphi^{(n)} (-x) dx \\
[\textrm{variable substitution }\; x \rightarrow -x]\\
=
-\int _\infty^{-\infty} \delta (-x) \varphi^{(n)} (x) dx
=
\int  _{-\infty}^\infty \delta (x) \varphi^{(n)} (x) dx \\
=
(-1)^n\int _\infty^{-\infty} \delta^{(n)} (x) \varphi (x) dx   .
 \end{split}
 \end{equation}
\eproof
}

Because of an additional factor $(-1)^n$ from the chain rule,
in particular, from the $n$-fold ``inner'' differentiation of $-x$, follows that
 \begin{equation}
 \frac{d^n}{dx^n}\delta(-x) = (-1)^n \delta^{(n)}(-x)  = \delta^{(n)}(x).
 \end{equation}

 \begin{equation}
 x^{m+1}\delta^{(m)}(x)=0
,
 \end{equation}
 where the index $^{(m)}$ denotes $m$-fold differentiation;
 \begin{equation}
 x^2\delta '(x)=0,
 \end{equation}
which is a  consequence of Equation~(\ref{2012-m-ch-di-sederi}).
More generally,
formally, $
x^n\delta^{(m)}(x) =  (-1)^n n! \delta_{nm}\delta (x)
$, or
 \begin{equation}
x^n\delta^{(m)} [\varphi ] =  (-1)^n n! \delta_{nm}\delta [\varphi ]
.
 \end{equation}
{\color{OliveGreen} \bproof
This can be demonstrated by considering
 \begin{equation}
 \begin{split}
x^n\delta^{(m)}[\varphi ]
=
\int_{-\infty}^\infty   x^n \delta^{(m)}(x)\varphi (x) dx
=
\int_{-\infty}^\infty  \delta^{(m)}(x) x^n \varphi (x) dx
\\
=
(-1)^m
\int_{-\infty}^\infty  \delta(x)\frac{d^m}{dx^m} \left[x^n\varphi (x)\right] dx
=
(-1)^m
\left.  \frac{d^m}{dx^m} \left[x^n\varphi (x)\right] \right|_{x=0}\\
\textrm{[after $n$ derivations
the only remaining nonvanishing term}\\
\textrm{is of degree $n=m$, with $x^m\varphi (x)$ resulting in $m! \varphi (0)$]}\\
=
(-1)^m  m! \delta_{nm}  \varphi (0)
=
(-1)^n  n! \delta_{nm} \delta [\varphi ].
\end{split}
\end{equation}

A shorter proof employing the polynomial $x^n$ as a ``test'' function may also be enumerated by
\begin{equation}
 \begin{split}
 \langle x^n \delta^{(m)} \vert 1 \rangle
=
 \langle \delta^{(m)} \vert  x^n \rangle
=
(-1)^n
 \langle \delta \vert \frac{d^m}{d x^m}  x^n \rangle    \\
=
(-1)^n n! \delta_{nm}
 \underbrace{\langle \delta \vert 1 \rangle}_{1}
  .
\end{split}
 \end{equation}
\eproof
}

Suppose $H$ is the
{Heaviside step function}
\index{Heaviside step function}
\index{unit step function}
as defined later in Equation~(\ref{2011-m-di-edhf}), then
 \begin{equation}
H' [\varphi] =\delta[\varphi]
.
\label{2019-mm-ch-di-ad1}
\end{equation}
{\color{OliveGreen}
\bproof
For a proof, note that
 \begin{equation}
 \begin{split}
H'[\varphi] =\frac{d}{dx}H[\varphi ] = -H[\varphi']=
-\int_{-\infty}^\infty H (x)  \varphi'(x) dx
\\
=
-\int_{0}^\infty   \varphi'(x) dx
=
-\left. \varphi (x) \right|_{x=0}^{x=\infty} =   - \underbrace{\varphi (\infty )}_{=0} +\varphi (0)= \varphi (0) =\delta[\varphi ]
.
 \end{split}
\label{2019-mm-ch-di-ad2}
 \end{equation}
\eproof
}

 \begin{equation}
 {d^2\over dx^2}[x H (x)] = {d \over dx }[H(x) + \underbrace{x \delta  (x)}_{0}]={d \over dx } H(x)  = \delta (x)
 \end{equation}

If $ \delta^{(3)} ({\bf r})=
\delta (x)
\delta (y)
\delta (r)$ with ${\bf r}=(x,y,z)$ and $\vert {\bf r}\vert=r$, then
 \begin{equation}
 \delta^{(3)} ({\bf r})=\delta (x)\delta (y)\delta (z)=-{1\over 4\pi }\Delta {1\over  r }
 \end{equation}
 \begin{equation}
 \delta^{(3)}  ({\bf r})=-{1\over 4\pi }(\Delta +k^2){e^{ikr}\over r} = -{1\over 4\pi }(\Delta +k^2){\cos kr\over r},
 \end{equation}
and therefore
\begin{equation}
 (\Delta +k^2){\sin kr\over r} = 0.
 \end{equation}

In quantum field theory,  phase space integrals of the form
 \begin{equation}
 {1\over 2E}=\int dp^0 \, H (p^0)\delta (p^2-m^2)
 \end{equation}
 with $E=({\vec p}^2+m^2)^{(1/2)}$
 are exploited.

{\color{OliveGreen}
\bproof
For a proof consider
 \begin{equation}
 \begin{split}
\int_{-\infty}^\infty     H (p^0)\delta (p^2-m^2)  dp^0
=
\int_{-\infty}^\infty    H (p^0)\delta \left((p_0)^2-{\bf p}^2-m^2\right) dp^0\\
=
\int_{-\infty}^\infty   H (p^0)\delta \left((p_0)^2-E^2\right)  dp^0  \\
=
\int_{-\infty}^\infty   H (p^0)\frac{1}{2E} \left[\delta \left( p_0 - E \right) + \delta \left( p_0 + E \right) \right]  dp^0 \\
=
\frac{1}{2E} \int_{-\infty}^\infty   \Big[\underbrace{H (p^0) \delta \left( p_0 - E \right)}_{=\delta \left( p_0 - E \right)}
 +
\underbrace{H (p^0) \delta \left( p_0 + E \right)}_{=0}
\Big] dp^0\\
=  \frac{1}{2E} \underbrace{\int_{-\infty}^\infty   \delta \left( p_0 - E \right)  dp^0}_{=1} =\frac{1}{2E}
.
\end{split}
 \end{equation}
\eproof
}

\subsection{Fourier transform of~~$\delta$}
\index{delta function}
The Fourier transform of the $\delta$-function can be obtained straightforwardly
by insertion into Equation~(\ref{2011-m-efta1mg});\sidenote[][-10mm]{The convention $A=B=1$ differs from the convention
$A=2\pi$ and $B=1$  used earlier in Section~\ref{2019-mm-ft-td}, page~\pageref{2019-mm-ft-1}.
$A$ and $B$ refer to Equation~(\ref{2011-m-efta1mg}), page~\pageref{2011-m-efta1mg}.}   that is,
with $A=B=1$
\begin{equation}
\begin{split}
  {\cal F}[\delta (x)]=\widetilde{\delta}(k)=   \int_{-\infty}^\infty  \delta(x) e^{-i{kx}} dx   \\
  =    e^{-i{0k}}  \int_{-\infty}^\infty  \delta(x)  dx   \\
  =    1, \textrm{ and thus}
\\
 {\cal F}^{-1}[\widetilde{\delta} (k)]=
 {\cal F}^{-1}[1]=\delta (x)  \\
  = \frac{1}{2\pi}  \int_{-\infty}^\infty    e^{i{kx}} dk\\
  =
\frac{1}{2\pi}  \int_{-\infty}^\infty  \left[  \cos(kx)+ i \sin(kx) \right] dk\\
  =
\frac{1}{ \pi}  \int_{0}^\infty    \cos(kx) dk   +
\frac{ i }{2\pi}  \int_{-\infty}^\infty   \sin(kx)   dk\\
  =
\frac{1}{ \pi}  \int_{0}^\infty    \cos(kx) dk
.
\end{split}
\label{2011-m-eftdelta}
\end{equation}
That is, the Fourier transform of the $\delta$-function is just a constant.
$\delta$-spiked signals carry all frequencies in them.
Note also that  ${\cal F}[\delta_y ]={\cal F}[\delta  (y-x)]=e^{i{ky}}{\cal F}[\delta (x)]=e^{i{ky}}{\cal F}[\delta ]$.

From Equation~(\ref{2011-m-eftdelta} ) we can compute
\begin{equation}
\begin{split}
{\cal F}[1]=\widetilde{1}(k)=   \int_{-\infty}^\infty    e^{-i{kx}} dx   \\
 [\textrm{variable substitution}\; x \rightarrow -x]\\
 =   \int_{+\infty}^{-\infty}    e^{-i{k(-x)}} d(-x)   \\
 =  - \int_{+\infty}^{-\infty}    e^{i{kx}} dx   \\
 =    \int_{-\infty}^{+\infty}    e^{i{kx}} dx   \\
 =    2\pi \delta (k)
.
\end{split}
\label{2011-m-eftdelta1}
\end{equation}

\subsection{Eigenfunction expansion of~~$\delta$}
\label{2012-m-efed1}

The  $\delta$-function can be expressed  in terms of, or ``decomposed'' into, various
{\em eigenfunction expansions}.
\index{eigenfunction expansion}
We mention without proof\cite{duffy2001} that, for $0< x,x_0 <L$,
two such expansions in terms of trigonometric functions are
\begin{equation}
\begin{split}
\delta (x-x_0) =
\frac{2}{L}
\sum_{k=1}^\infty
\sin \left( \frac{\pi k x_0}{L}\right)
\sin \left( \frac{\pi k x}{L}\right)\\
  =
\frac{1}{L}
+
\frac{2}{L}
\sum_{k=1}^\infty
\cos \left( \frac{\pi k x_0}{L}\right)
\cos \left( \frac{\pi k x}{L}\right).
\end{split}
\label{2012-m-efed}
\end{equation}

This ``decomposition of unity'' is analogous to the expansion of the identity in terms of orthogonal projectors
$\textsf{\textbf{E}}_i$  (for one-dimensional projectors, $\textsf{\textbf{E}}_i=\vert i\rangle \langle i \vert $)
encountered in the spectral theorem \ref{2012-m-ch-Spectraltheorem}.

Other decomposions are in terms of orthonormal (Legendre) polynomials (cf. Sect. \ref{2013-m-sf-lp} on page \pageref{2013-m-sf-lp}),
or other functions of mathematical physics discussed later.

\subsection{Delta function expansion}
\label{2012-m-dfex}

Just like ``slowly varying'' functions can be expanded into a Taylor series in terms of the power functions $x^n$,
highly localized functions can be expanded in terms of derivatives of the $\delta$-function in the form\cite{lindell:438}
\begin{equation}
\begin{split}
f(x) \sim
f_0 \delta (x) +
f_1 \delta' (x) +
f_2 \delta'' (x) + \cdots +f_n \delta^{(n)}(x) + \cdots =\sum_{k=1}^\infty f_k \delta^{(k)}(x),\\
\textrm{with } f_k= \frac{(-1)^k}{k!} \int_{-\infty}^\infty f(y) y^k \, dy
.
\end{split}
\label{2012-m-edfd}
\end{equation}
The sign ``$\sim$'' denotes the functional character of this ``equation''~(\ref{2012-m-edfd}).

{\color{OliveGreen}
\bproof
The delta expansion (\ref{2012-m-edfd}) can be proven by considering a smooth function $g(x)$, and integrating over its expansion; that is,
\begin{equation}
\begin{split}
 \int_{-\infty}^\infty  f(x) \varphi (x) dx  \\
  =
 \int_{-\infty}^\infty   \left[
f_0 \delta (x) +
f_1 \delta' (x) +
f_2 \delta'' (x)
+ \cdots  +
f_n \delta^{(n)}(x)
+ \cdots \right]\varphi (x)  dx \\
  =
f_0  \varphi (0) - f_1  \varphi' (0) + f_2  \varphi'' (0) +\cdots  + (-1)^n  f_n  \varphi^{(n)}(0)
+ \cdots
,
\end{split}
\label{2012-m-edfd1}
\end{equation}
and comparing the coefficients in (\ref{2012-m-edfd1})
with the coefficients  of  the Taylor series expansion of $\varphi$ at $x=0$
\begin{equation}
\begin{split}
 \int_{-\infty}^\infty  \varphi (x) f(x)  =
 \int_{-\infty}^\infty  \left[
 \varphi (0) +x  \varphi' (0) + \cdots + \frac{x^n}{n!} \varphi^{(n)} (0)  + \cdots
 \right] f(x) dx \\
 =
 \varphi (0) \int_{-\infty}^\infty  f(x) dx  +
\varphi' (0) \int_{-\infty}^\infty x f(x) dx   + \cdots +  \varphi^{(n)} (0)\underbrace{\int_{-\infty}^\infty \frac{x^n}{n!} f(x) dx}_{(-1)^n f_n}  + \cdots
,
\end{split}
\label{2012-m-edfd2tse1}
\end{equation}
so that
$f_n =  (-1)^n  \int_{-\infty}^\infty \frac{x^n}{n!} f(x) dx$.
\eproof
}


\section{Cauchy principal value}
\index{Cauchy principal value}
\index{principal value}

\subsection{Definition}

The {\em  (Cauchy) principal value} ${\cal P}$ (sometimes also denoted by $\textrm{p.v.}$)
is a value associated with an integral as follows:
suppose $f(x)$ is not locally integrable around $c$; then
\begin{equation}
\begin{split}
{\cal P}
\int_{a}^b f(x) dx
= \lim_{\varepsilon \rightarrow 0^+}
\left[
\int_{a}^{c-\varepsilon} f(x) dx
+
\int_{c+\varepsilon}^{b} f(x) dx
\right]
\\
=  \lim_{\varepsilon \rightarrow 0^+}
\int_{[a,c-\varepsilon] \cup [c+\varepsilon,b]}   f(x) dx
.
\end{split}
\end{equation}

{
\color{blue}
\bexample
For example, the integral
$ \int_{-1}^1 \frac{dx}{x}$ diverges, but
\begin{equation}
\begin{split}
{\cal P}
\int_{-1}^1 \frac{dx}{x}
= \lim_{\varepsilon \rightarrow 0^+}
\left[
\int_{-1}^{ -\varepsilon}\frac{dx}{x}
+
\int_{ +\varepsilon}^{1} \frac{dx}{x}
\right]
\\
[\textrm{variable substitution }\; x\rightarrow -x \textrm{ in the first integral}]
\\
=
\lim_{\varepsilon \rightarrow 0^+}
\left[
\int_{+1}^{ +\varepsilon}\frac{dx}{x}
+
\int_{ +\varepsilon}^{1} \frac{dx}{x}
\right]
\\
=
\lim_{\varepsilon \rightarrow 0^+}
\left[
\log \varepsilon - \log 1  + \log 1  - \log \varepsilon
\right]
=0.
\end{split}
\end{equation}
\eexample
}

\subsection{Principle value and pole function $\frac{1}{x}$ distribution}

The ``standalone function'' $\frac{1}{x}$
does not define a distribution  since it is not integrable in
the vicinity of $x=0$.
This issue can be ``alleviated'' or ``circumvented''  by considering the principle value  ${\cal P}\frac{1}{x}$.
In this way the
principle value can be transferred to the context of distributions
by defining
a {\em principal value distribution} in a functional sense:
\index{principal value distribution}
\begin{equation}
\begin{split}
{\cal P} \left(\frac{1}{x}\right) \left[\varphi\right]
=
\lim_{\varepsilon \rightarrow 0^+}
\int_{ \vert x \vert > \varepsilon }   \frac{1}{x}\varphi(x) dx
\\
= \lim_{\varepsilon \rightarrow 0^+}
\left[
\int_{-\infty}^{ -\varepsilon} \frac{1}{x}\varphi(x) dx
+
\int_{ +\varepsilon}^{\infty} \frac{1}{x}\varphi(x) dx
\right]
\\
[\textrm{variable substitution } \; x\rightarrow -x \textrm{ in the first integral}]
\\
= \lim_{\varepsilon \rightarrow 0^+}
\left[
\int_{+\infty}^{ +\varepsilon} \frac{1}{x}\varphi(-x) dx
+
\int_{ +\varepsilon}^{\infty} \frac{1}{x}\varphi(x) dx
\right]
\\
= \lim_{\varepsilon \rightarrow 0^+}
\left[
-\int_{ +\varepsilon}^{\infty} \frac{1}{x}\varphi(-x) dx
+
\int_{ +\varepsilon}^{\infty} \frac{1}{x}\varphi(x) dx
\right]
\\
=
\lim_{\varepsilon \rightarrow 0^+}
\int_{\varepsilon}^{+\infty}   \frac{\varphi(x)-\varphi(-x)}{x} dx
\\
=
\int_{0}^{+\infty}   \frac{\varphi(x)-\varphi(-x)}{x} dx
.
\end{split}
\label{2012-m-ch-di-prinvaldi}
\end{equation}

%

\section{Absolute value distribution}

The distribution associated with the absolute value $\left|x\right|$ is defined by
 \begin{equation}
\left|x\right| \left[ \varphi \right] =
\int_{-\infty}^\infty  \left|x\right|  \varphi(x) dx.
 \end{equation}
$\left|x\right| \left[ \varphi \right]$
can be evaluated and represented as follows:
\begin{equation}
\begin{split}
\left|x\right| \left[ \varphi \right]
=
\int_{-\infty}^\infty  \left|x\right|  \varphi(x) dx
\\
=
\int_{-\infty}^0 (-x)  \varphi(x) dx
+
\int_{0}^\infty   x   \varphi(x) dx
=
-\int_{-\infty}^0  x   \varphi(x) dx
+
\int_{0}^\infty   x   \varphi(x) dx
\\
[\textrm{variable substitution }\; x\rightarrow -x, dx \rightarrow -dx \textrm{ in the first integral}]
\\
=
-\int_{+\infty}^0  x   \varphi(-x) dx
+
\int_{0}^\infty   x   \varphi(x) dx
=
\int_0^{\infty}  x   \varphi(-x) dx
+
\int_{0}^\infty   x   \varphi(x) dx
\\
=
\int_0^{\infty}  x   \left[ \varphi(x) + \varphi(-x) \right] dx .
\end{split}
\end{equation}

An alternative derivation uses the reflection symmetry at zero:
\begin{equation}
\begin{split}
\left|x\right| \left[ \varphi \right]
=
\int_{-\infty}^\infty  \left|x\right|  \varphi(x) dx
=
\int_{-\infty}^\infty    \frac{\left|x\right|}{2}\left[\varphi(x) + \varphi (-x)\right] dx   \\
=
\int_{0}^\infty  x  \left[\varphi(x) + \varphi (-x)\right] dx
.
\end{split}
\end{equation}

\section{Logarithm distribution}

\subsection{Definition}
Let, for $x\neq 0$,
\begin{equation}
\begin{split}
\log \left|x\right| \left[ \varphi \right]
=
\int_{-\infty}^\infty  \log \left|x\right|  \varphi(x) dx
\\
=
\int_{-\infty}^0 \log (-x)   \varphi(x) dx
+
\int_{0}^\infty   \log x    \varphi(x) dx
\\
[\textrm{variable substitution }\; x\rightarrow -x, dx \rightarrow -dx \textrm{ in the first integral}]
\\
=
\int_{+\infty}^0  \log  (-(-x))     \varphi(-x) d(-x)
+
\int_{0}^\infty   \log x    \varphi(x) dx
\\
=
-\int_{+\infty}^0  \log x    \varphi(-x) d x
+
\int_{0}^\infty   \log x    \varphi(x) dx
\\
=
\int_0^{\infty}  \log x    \varphi(-x) d x
+
\int_{0}^\infty   \log x    \varphi(x) dx
\\
=
\int_0^{\infty}  \log x  \left[  \varphi(x)
+
   \varphi(-x) \right] dx
 .
\end{split}
\label{2012-m-ch-di-logdidef}
\end{equation}

\subsection{Connection with pole function}
Note that
\begin{equation}
 {\cal P} \left(\frac{1}{x}\right) \left[\varphi\right]  = \frac{d}{dx} \log \vert x \vert \left[\varphi\right]
,
\label{2014-m-ch-di-cwtpf}
\end{equation}
and thus,
for the principal value of a pole of degree $n$,
\begin{equation}
  {\cal P} \left(\frac{1}{x^n}\right) \left[\varphi\right]  =  \frac{(-1)^{n-1}}{(n-1)!}
\frac{d^n}{dx^n} \log \vert x \vert \left[\varphi\right]
.
\label{2014-m-ch-di-cwtpf2}
\end{equation}


{\color{OliveGreen}
\bproof

For a proof of Equation~(\ref{2014-m-ch-di-cwtpf}) consider the functional derivative
$\log' \vert x\vert [\varphi ]$
by insertion into Equation~(\ref{2012-m-ch-di-logdidef}); as well as by using the symmetry of the resulting integral kernel
\marginnote{Note that
every function $f(x)$, by addition of the neutral term zero $0 = \frac{1}{2}\left[ f(-x) - f(-x) \right]$,
can be decomposed into its symmetric $s(x)$ and antisymmetric part $a(x)$ (with respect to the origin $x=0$) as follows:\\
$f(x)=   \frac{1}{2}\left[ f(x) + f(x) \right] + 0= \frac{1}{2}\left[ f(x) + f(x) \right]+  \frac{1}{2}\left[ f(-x) - f(-x) \right]
= \underbrace{\frac{1}{2}\left[ f(x) + f(-x) \right]}_{s(x)} +  \underbrace{\frac{1}{2}\left[ f(x) - f(-x) \right]}_{a(x)}$. \\
By identifying $f=\varphi '$ and observing that $\log \vert x\vert$ is symmetric (with respect to the origin $x=0$)
only the symmetric part  $\frac{1}{2}\left[ \varphi ' (x) + \varphi ' (-x) \right] \log \vert x\vert$ of $\varphi ' (x) \log \vert x\vert$ ``survives''. \\
Furthermore,
for $x < 0$,
 $\frac{d}{dx} \log  \vert x\vert = \frac{d}{dx}\log (- x) = \left[\frac{d}{dy}\log (y) \right]_{y=-x} \frac{d}{dx}(-x) = \frac{-1}{-x} = \frac{1}{x}$.}
 at zero:
\begin{equation}
\begin{split}
\log' \vert x\vert [\varphi ]
=
- \log \vert x\vert [\varphi ' ]
=
- \int_{-\infty}^\infty  \log \vert x\vert  \varphi '(x) dx
\\
=
- \frac{1}{2} \int_{-\infty}^\infty  \log \vert x\vert  \left[ \varphi '(x)   +\varphi '(- x) \right] dx  \\
=
- \frac{1}{2} \int_{-\infty}^\infty  \log \vert x\vert  \frac{d}{dx}  \left[ \varphi  (x)   -\varphi  (- x) \right] dx
\\
=
  \frac{1}{2} \int_{-\infty}^\infty   \left( \frac{d}{dx} \log \vert x\vert \right) \left[ \varphi  (x)   -\varphi  (- x) \right] dx    \\
=
  \frac{1}{2} \left\{ \int_{-\infty}^0   \left[ \frac{d}{dx} \log (- x) \right] \left[ \varphi  (x)   -\varphi  (- x) \right] dx
+ \right. \qquad \qquad \\ \left.
+   \int_{0}^\infty   \left( \frac{d}{dx} \log   x \right) \left[ \varphi  (x)   -\varphi  (- x) \right] dx \right\}
 \\
=
  \frac{1}{2} \left\{ \int_{\infty}^0   \left( \frac{d}{dx} \log  x \right) \left[ \varphi  (-x)   -\varphi  ( x) \right] dx
+ \right. \qquad \qquad \\ \left.
+  \int_{0}^\infty   \left( \frac{d}{dx} \log   x  \right) \left[ \varphi  (x)   -\varphi  (- x) \right] dx  \right\}
 \\
=
  \int_{0}^\infty   \frac{1}{x}   \left[ \varphi  (x)   -\varphi  (- x) \right] dx
=
{\cal P} \left(\frac{1}{x}\right) \left[\varphi\right]
,
\end{split}
\end{equation}

The more general Equation~(\ref{2014-m-ch-di-cwtpf2}) follows by direct differentiation.

\eproof
}

\section{Pole function $\frac{1}{x^n}$ distribution}

For $n\ge 2$, the integral over $\frac{1}{x^n}$ is undefined even if we take the principal value.
Hence the direct route to an evaluation is blocked, and we have to take an indirect approach {\it via}
derivatives of\cite{sommer-di} $\frac{1}{x}$.
Thus, let
\begin{equation}
\begin{split}
\frac{1}{x^2} \left[ \varphi \right]
=
-\frac{d}{dx} \frac{1}{x} \left[ \varphi \right] \\
=\frac{1}{x} \left[ \varphi ' \right] =
\int_0^{\infty}  \frac{1}{x}  \left[  \varphi'(x)
-
   \varphi'(-x) \right] dx
\\
 =
{\cal P} \left(\frac{1}{x}\right) \left[\varphi ' \right]
 .
\end{split}
\label{2013-m-v-di-pfd}
\end{equation}

Also,
\begin{equation}
\begin{split}
\frac{1}{x^3} \left[ \varphi \right]
=
-\frac{1}{2}\frac{d}{dx} \frac{1}{x^2} \left[ \varphi \right]
=\frac{1}{2}\frac{1}{x^2} \left[ \varphi ' \right] =  \frac{1}{2x} \left[ \varphi '' \right]\\
=
\frac{1}{2} \int_0^{\infty}  \frac{1}{x}  \left[  \varphi''(x)
-
   \varphi''(-x) \right] dx
\\
 =
\frac{1}{2} {\cal P} \left(\frac{1}{x}\right) \left[\varphi '' \right]
 .
\label{2013-m-v-di-pfd3}
\end{split}
\end{equation}

More generally, for $n>1$, by induction, using (\ref{2013-m-v-di-pfd}) as induction basis,
\begin{equation}
\begin{split}
\frac{1}{x^n} \left[ \varphi \right]  \\
=
-\frac{1}{n-1}\frac{d}{dx} \frac{1}{x^{n-1}} \left[ \varphi \right]
=
 \frac{1}{n-1}  \frac{1}{x^{n-1}} \left[ \varphi' \right]  \\
=
-\left( \frac{1}{n-1} \right) \left(\frac{1}{n-2}\right) \frac{d}{dx} \frac{1}{x^{n-2}} \left[ \varphi' \right]
=
 \frac{1}{(n-1)(n-2)}   \frac{1}{x^{n-2}} \left[ \varphi'' \right]  \\
=   \cdots =
\frac{1}{(n-1)!} \frac{1}{x} \left[ \varphi^{({n-1})} \right]  \\
=\frac{1}{(n-1)!}
\int_0^{\infty}  \frac{1}{x}  \left[  \varphi^{(n-1)}(x)
-
   \varphi^{(n-1)}(-x) \right] dx
\\
 =
\frac{1}{(n-1)!} {\cal P} \left(\frac{1}{x}\right) \left[\varphi^{(n-1)} \right]
 .
\end{split}
\end{equation}

\section{Pole function $\frac{1}{x\pm i\alpha}$ distribution}

We are interested in the limit $\alpha  \rightarrow 0$ of $\frac{1}{x+i\alpha}$.
Let  $\alpha >0$. Then,
\begin{equation}
\begin{split}
\frac{1}{x+i\alpha} \left[ \varphi \right]
=
\int_{-\infty}^\infty  \frac{1}{x+i\alpha}  \varphi(x) dx
\\
=
\int_{-\infty}^\infty   \frac{x-i\alpha}{ (x+i\alpha)(x-i\alpha) }   \varphi(x) dx
\\
=
\int_{-\infty}^\infty   \frac{x-i\alpha}{x^2+ \alpha^2}   \varphi(x) dx
\\
=
\int_{-\infty}^\infty   \frac{x}{x^2+ \alpha^2}   \varphi(x) dx
-i\alpha \int_{-\infty}^\infty   \frac{1}{x^2+ \alpha^2}   \varphi(x) dx
.
\end{split}
\label{2012-m-ch-di-mf}
\end{equation}

Let us treat the two summands of (\ref{2012-m-ch-di-mf}) separately.
(i) Upon variable substitution  $x = \alpha y$, $dx =\alpha dy$ in the second integral in (\ref{2012-m-ch-di-mf}) we obtain
\begin{equation}
\begin{split}
\alpha \int_{-\infty}^\infty   \frac{1}{x^2+ \alpha^2}   \varphi(x) dx
=
\alpha \int_{-\infty}^\infty   \frac{1}{\alpha^2y^2+ \alpha^2}   \varphi(\alpha y) \alpha dy
\\
=
\alpha^2 \int_{-\infty}^\infty   \frac{1}{\alpha^2(y^2+ 1)}   \varphi(\alpha y)   dy
\\
=
  \int_{-\infty}^\infty   \frac{1}{  y^2+ 1 }   \varphi(\alpha y)   dy
\end{split}
\end{equation}
In the limit $\alpha  \rightarrow 0$, this is
\begin{equation}
\begin{split}
\lim_{\alpha  \rightarrow 0} \int_{-\infty}^\infty   \frac{1}{  y^2+ 1 }   \varphi(\alpha y)   dy
=
\varphi(0) \int_{-\infty}^\infty   \frac{1}{  y^2+ 1 }       dy
\\
 =
\varphi(0) \left. \left( \arctan y \right) \right|_{y=-\infty}^{y=\infty}
\\
  =
\pi \varphi(0) =
\pi \delta [\varphi ]
.
\end{split}
\end{equation}

(ii)
The first integral in (\ref{2012-m-ch-di-mf}) is
\begin{equation}
\begin{split}
\int_{-\infty}^\infty   \frac{x}{x^2+ \alpha^2}   \varphi(x) dx
\\
=
\int_{-\infty}^0   \frac{x}{x^2+ \alpha^2}   \varphi(x) dx
+
\int_{0}^\infty   \frac{x}{x^2+ \alpha^2}   \varphi(x) dx
\\
=
\int_{+\infty}^0   \frac{-x}{(-x)^2+ \alpha^2}   \varphi(-x) d(-x)
+
\int_{0}^\infty   \frac{x}{x^2+ \alpha^2}   \varphi(x) dx
\\
=
-\int_{0}^\infty   \frac{ x}{x^2+ \alpha^2}   \varphi(-x) dx
+
\int_{0}^\infty   \frac{x}{x^2+ \alpha^2}   \varphi(x) dx
\\
=
 \int_{0}^\infty   \frac{ x}{x^2+ \alpha^2} \left[  \varphi(x) - \varphi(-x) \right] dx
.
\end{split}
\end{equation}
In the limit $\alpha  \rightarrow 0$, this becomes
\begin{equation}
\begin{split}
\lim_{\alpha  \rightarrow 0} \int_{0}^\infty   \frac{ x}{x^2+ \alpha^2} \left[  \varphi(x) - \varphi(-x) \right] dx
=
\int_{0}^\infty   \frac{ \varphi(x) - \varphi(-x) }{x} dx
\\
{\cal P} \left(\frac{1}{x}\right) \left[\varphi\right]
,
\end{split}
\end{equation}
where in the last step the principle value distribution (\ref{2012-m-ch-di-prinvaldi})
has been used.

Putting all parts together, we obtain
\begin{equation}
\frac{1}{x+i0^+} \left[ \varphi \right]
= \lim_{\alpha  \rightarrow 0^+}\frac{1}{x+i\alpha} \left[ \varphi \right]
=  {\cal P} \left(\frac{1}{x}\right) \left[\varphi\right]
-i \pi \delta [\varphi ] = \left\{
{\cal P} \left(\frac{1}{x}\right) -i \pi \delta
\right\}  [\varphi ].
\label{2012-m-ch-di-Sokhotskyformula1}
\end{equation}
A very similar calculation yields
\begin{equation}
\frac{1}{x-i0^+} \left[ \varphi \right]
=
\lim_{\alpha  \rightarrow 0^+}\frac{1}{x-i\alpha} \left[ \varphi \right]
=  {\cal P} \left(\frac{1}{x}\right) \left[\varphi\right]
+i \pi \delta [\varphi ] = \left\{
{\cal P} \left(\frac{1}{x}\right) +i \pi \delta
\right\}  [\varphi ].
\label{2012-m-ch-di-Sokhotskyformula2}
\end{equation}
These equations
(\ref{2012-m-ch-di-Sokhotskyformula1})
and
(\ref{2012-m-ch-di-Sokhotskyformula2})
are often called the
{\em Sokhotsky  formula}, also known as the {\em Plemelj  formula}, or the {\em Plemelj-Sokhotsky formula}.\cite[-20mm]{Sokhotski-1873,Plemelj1908}
\index{Sokhotsky formula}
\index{Plemelj formula}
\index{Plemelj-Sokhotsky formula}


\section{Heaviside or unit step function}
\index{Heaviside step function}
\index{unit step function}

\subsection{Ambiguities in definition}
Let us now turn
to Heaviside's electromagnetic {\em pulse function}, often referred to as Heaviside's unit step function.
\index{pulse function}
One of the possible definitions of the Heaviside step function $H(x)$, and maybe the most common one --
they differ by the difference of the value(s) of $H(0)$ at the origin $x=0$, a difference which is irrelevant measure theoretically for ``good'' functions
since it is only about an isolated point --
is
\begin{equation}
H(x-x_0)
=
\left\{
\begin{array}{rl}
1&\textrm{ for } x\ge x_0\\
0&\textrm{ for } x < x_0
\end{array}
\right.
\label{2011-m-di-edhf}
\end{equation}
Alternatively one may define $H(0)=\frac{1}{2}$, as plotted in Figure~\ref{2011-m-fhsf}.
\begin{equation}
H(x-x_0)
=   \frac{1}{2} + \frac{1}{\pi } \lim_{\varepsilon \rightarrow 0^+} \textrm{arctan}\left( \frac{x-x_0}{\varepsilon} \right)
=
\left\{
\begin{array}{rl}
1&\textrm{ for } x > x_0\\
\frac{1}{2}&\textrm{ for } x = x_0\\
0&\textrm{ for } x < x_0
\end{array}
\right.
\label{2012-m-di-hsfoh}
\end{equation}
and, since this affects only an isolated point at $x=x_0$, we may happily do so if we prefer.
\begin{marginfigure}
\begin{center}
\begin{tikzpicture}[ scale=0.6,inner sep=0.8em ]

\tikzstyle{every path}=[line width=2pt]

\begin{axis}[axis lines=middle, draw=gray!80,
    ticklabel style = {font=\Large },
    xmin=-2.5, xmax=2.5,
    ymin=-0.25, ymax=1.25,
     ytick={0.5,1},
     extra y ticks={0},
     extra y tick style={
       tick label style={anchor=south west, xshift=6pt},
     },
]
\addplot[orange,smooth,domain=0:2.5] {1};
\addplot[orange,smooth,domain=-2.5:0] {0};
\draw[black,fill=white] (axis cs:0,0) circle(1mm)  (axis cs:0,1) circle(1mm);
\draw[fill,orange] (axis cs:0,1/2) circle(1mm);
\end{axis}
\end{tikzpicture}
\end{center}
\caption{Plot of the Heaviside step function  $H(x)$.
Its value at $x=0$ depends on its definition.}
\label{2011-m-fhsf}
\end{marginfigure}

It is also very common to define the  unit step function as the
{\em antiderivative  of the $\delta$ function};
\index{antiderivative}
likewise the delta function is the derivative of the Heaviside step function; that is,
\begin{equation}
\begin{split}
H'[\varphi ]=\delta [\varphi ]\text{, or formally,}\\
H(x-x_0)
=
\int_{-\infty}^{x-x_0} \delta (t) dt,\text{ and }
\frac{d}{dx} H(x-x_0)=\delta (x-x_0).
\end{split}
\end{equation}

{\color{OliveGreen}
\bproof
The latter equation can, in the functional sense -- that is, by integration over a test function --
be proven by
\begin{equation}
\begin{split}
H'[\varphi ]= \langle   H' , \varphi   \rangle = - \langle  H ,  \varphi'   \rangle
 =
-\int_{-\infty}^\infty H(x) \varphi' (x)  dx \\
 =
-\int_{0}^\infty  \varphi' (x)  dx
  =
-\left.   \varphi  (x) \right|_{x=0}^{x= \infty} \\
  =
 -   \underbrace{\varphi  (\infty)}_{=0} +  \varphi  (0)
  =    \langle   \delta , \varphi  \rangle  =\delta [\varphi ]
\end{split}
\end{equation}
for all test functions $\varphi (x)$. Hence we can -- in the functional sense  -- identify $\delta$ with $H'$.
More explicitly, through integration by parts, we obtain
\begin{equation}
\begin{split}
\int _{-\infty}^\infty \left[\frac{d}{dx} H(x-x_0)\right] \varphi (x) dx       \\
  =
\left. H(x-x_0) \varphi (x)\right| _{-\infty}^\infty - \int _{-\infty}^\infty H(x-x_0) \left[\frac{d}{dx} \varphi (x)\right] dx \\
  =
\underbrace{H(\infty)}_{=1}\underbrace{\varphi(\infty)}_{=0} - \underbrace{H(-\infty)}_{=0}\underbrace{\varphi(-\infty)}_{=0}
   - \int _{x_0}^\infty \left[\frac{d}{dx} \varphi (x)\right] dx  \\
 =  - \int _{x_0}^\infty \left[\frac{d}{dx} \varphi (x)\right] dx \\
 =    -  \left.  \varphi (x)   \right| _{x=x_0}^{x= \infty}
 =    - [  \underbrace{\varphi  (\infty)}_{=0}  - \varphi (x_0)]
 =     \varphi (x_0).
\end{split}
\end{equation}

\eproof
}

\subsection{Unit step function sequence}

\marginnote{For a great variety of unit step function sequences see
\url{http://mathworld.wolfram.com/HeavisideStepFunction.html}.}
As mentioned earlier, a commonly used limit form  of the Heaviside step function is
\begin{equation}
H(x)= \lim_{\epsilon \rightarrow 0} H_\epsilon (x)
=\lim_{\epsilon \rightarrow 0}   \left[ \frac{1 }{2} + \frac{1 }{\pi} \text{arctan}\left(  \frac{x}{\epsilon }\right) \right] .
\end{equation}
respectively.

Another limit representation of the Heaviside function is in terms of the
{\em Dirichlet's discontinuity factor} as follows:
\index{Dirichlet's discontinuity factor}
\begin{equation}
\begin{split}
H(x)= \lim_{t \rightarrow \infty} H_t (x)
\\
= \frac{1}{2}+\frac{1}{\pi }\lim_{t \rightarrow \infty}\int_0^t \frac{\sin (kx)}{k} dk
\\
=
\frac{1}{2}+\frac{1}{\pi }\int_0^\infty \frac{\sin (kx)}{k} dk
.
\end{split}
\label{2012-m-ch-di-dcf}
\end{equation}

{\color{OliveGreen}
\bproof
A proof\cite{maor1998} uses a variant
of the {\em sine integral function}
\index{sine integral}
\begin{equation}
\textrm{Si}(y) = \int_0^y \frac{\sin t}{t} \,dt
\end{equation}
which in the limit of large argument $y$ converges towards
the {\em Dirichlet integral}   (no proof is given here)
\index{Dirichlet integral}
\begin{equation}
\lim_{y \rightarrow  \infty}\textrm{Si}(y) = \int_0^\infty  \frac{\sin t}{t} \,dt= \frac{\pi}{2}.
\label{2012-m-ch-di-diint}
\end{equation}

Suppose we
replace $t$ with $t=kx$
in the Dirichlet integral (\ref{2012-m-ch-di-diint}),
whereby $x\neq 0$ is a nonzero constant; that is,
\begin{equation}
\int_0^\infty  \frac{\sin (kx)}{kx} \,d(kx)
= H(x)\int_0^{\infty}  \frac{\sin (kx)}{k} \,dk
+H(-x)\int_0^{-\infty}  \frac{\sin (kx)}{k} \,dk
.
\label{2012-m-ch-di-diint2}
\end{equation}
Note that the integration border $\pm \infty$ changes, depending on whether $x$ is positive or negative,
respectively.

If $x$ is positive, we leave the integral
(\ref{2012-m-ch-di-diint2})
as is, and we recover the original Dirichlet integral (\ref{2012-m-ch-di-diint}),
which is $\frac{\pi}{2}$.
If $x$ is negative,
in order to recover the original Dirichlet integral form with the upper limit $\infty$,
we have to perform yet another substitution $k \rightarrow -k$
on (\ref{2012-m-ch-di-diint2}), resulting in
\begin{equation}
= \int_0^{-\infty}  \frac{\sin (-kx)}{-k} \,d(-k)
= -\int_0^{\infty}  \frac{\sin (kx)}{k} \,dk
= - \textrm{Si}(\infty ) =  -\frac{\pi}{2}
,
\label{2015-m-ch-di-diint3}
\end{equation}
since the sine function is an odd function; that is, $\sin(-\varphi )=-\sin \varphi $.

The Dirichlet's discontinuity factor (\ref{2012-m-ch-di-dcf})
is obtained by normalizing the absolute value of
(\ref{2012-m-ch-di-diint2}) [and thus also
(\ref{2015-m-ch-di-diint3})]
to $\frac{1}{2}$
by multiplying  it with $1/\pi$, and by adding $\frac{1}{2}$.
\eproof
}

\subsection{Useful formul\ae{} involving $H$}

Some other formul\ae{}  involving the unit step function
are
 \begin{eqnarray}
H (-x)&=& 1-H(x)  \text{, or }   H(x) = 1 - H(-x),
\\
H (\alpha x)&=&
\begin{cases}
H (x) & \text{ for real } \alpha > 0,\\
1-H (x) & \text{ for real } \alpha < 0,
\end{cases}
\\
H (x)
&=&{1\over 2}
+
\sum_{l=0}^\infty (-1)^l {(2l)!(4l+3)\over 2^{2l+2}l!(l+1)!}
P_{2l+1} (x),
 \end{eqnarray}
where $P_{2l+1} (x)$ is a Legendre polynomial.
Furthermore,
\begin{equation}
\delta(x)=
\lim_{\varepsilon \rightarrow 0^+}
\frac{1}{\varepsilon } H\left( \frac{\varepsilon }{2} -\vert x\vert\right) .
\end{equation}
{\color{OliveGreen}
\bproof
The latter equation can
be proven by
\begin{equation}
\begin{split}
\lim_{\varepsilon \rightarrow 0^+}  \int_{-\infty}^\infty
\frac{1}{\varepsilon } H\left( \frac{\varepsilon }{2} -\vert x\vert\right) \varphi(x) dx
=
\lim_{\varepsilon \rightarrow 0^+} \frac{1}{\varepsilon } \int_{-\frac{\varepsilon}{2}}^{\frac{\varepsilon}{2}}
\varphi(x) dx
\\
\textrm{[mean value theorem: $\exists$  $y$ with $-\frac{\varepsilon}{2}\le y \le \frac{\varepsilon}{2}$ such that]} \\
=\lim_{\varepsilon \rightarrow 0^+} \frac{1}{\varepsilon }\varphi(y) \underbrace{\int_{-\frac{\varepsilon}{2}}^{\frac{\varepsilon}{2}} dx}_{=\varepsilon }
=\lim_{\varepsilon \rightarrow 0^+}  \varphi(y)  =   \varphi(0) =\delta [\varphi ].
\end{split}
\end{equation}
\eproof
}

A Fourier integral representation~(\ref{2019-mm-ch-di-fthusf}) of $H(x)$ derived later
is\sidenote[][-5mm]{The second integral is the complex conjugate of the first integral,
$
\overline{ab}=
\overline{a}
\overline{b}
$,
and
$
\overline{\left[\frac{1}{k-i\epsilon }\right]}
=
\overline{\left[\frac{k+i\epsilon }{(k+i\epsilon)(k-i\epsilon ) }\right]}
=
\frac{k-i\epsilon }{k+\epsilon^2}
=
\frac{k-i\epsilon }{(k+i\epsilon)(k-i\epsilon ) }
=
\frac{1}{ k+i\epsilon }
$.}
 \begin{equation}
H (x)
=
\lim_{\epsilon \rightarrow 0^+} \frac {1}{2\pi i}
\int_{-\infty}^\infty
 \frac{1}{t - i\epsilon}e^{ ixt} dt
=
\lim_{\epsilon \rightarrow 0^+} \frac{- 1}{2\pi i}
\int_{-\infty}^\infty
 \frac{1}{t + i\epsilon}e^{- ixt} dt
.
 \end{equation}

\subsection{$H  \left[ \varphi \right]$ distribution}
The distribution associated with the Heaviside function $H(x)$ is defined by
\begin{equation}
H \left[ \varphi \right] =
\int_{-\infty}^\infty  H(x)  \varphi(x) dx.
 \end{equation}
$H  \left[ \varphi \right]$
can be evaluated and represented as follows:
\begin{equation}
\begin{split}
H  \left[ \varphi \right]
=
\int_{-\infty}^0  \underbrace{H(x)}_{=0} \varphi(x) dx
+
\int_{0}^\infty  \underbrace{H(x)}_{=1} \varphi(x) dx
=
\int_{0}^\infty      \varphi(x) dx
.
\end{split}
\end{equation}

Recall that, as has been pointed out in Equations~(\ref{2019-mm-ch-di-ad1}) and~(\ref{2019-mm-ch-di-ad2}),
$H\left[ \varphi \right]$ is the antiderivative of the delta function; that is, $H'\left[ \varphi \right]=\delta\left[ \varphi \right]$.

\subsection{Regularized unit step function}
In order to be able to define the Fourier transformation
associated with the Heaviside function  we sometimes
consider the distribution of the {\em regularized Heaviside function}
\index{regularized Heaviside function}
\begin{equation}
H_\varepsilon (x) =H(x)e^{-\varepsilon x},
\label{2012-m-ch-di-rhfun}
\end{equation}
 with $\varepsilon >0$,
such that $\lim_{\varepsilon \rightarrow 0^+}  H_\varepsilon (x) =H (x)$.

\subsection{Fourier transform  of  the unit step function}
\index{Heaviside function}
\index{unit step function}

The Fourier transform\sidenote[][]{The convention $A=B=1$ is used.
$A$ and $B$ refer to Equation~(\ref{2011-m-efta1mg}), page~\pageref{2011-m-efta1mg}.}  of the Heaviside (unit step) function
cannot be directly obtained by insertion into Equation~(\ref{2011-m-efta1mg})
because the associated integrals do not exist.
We shall thus use the  regularized Heaviside function (\ref{2012-m-ch-di-rhfun}), and arrive at
{\em Sokhotsky's  formula} (also known as the {\em Plemelj's  formula}, or the {\em Plemelj-Sokhotsky formula})
\index{Sokhotsky formula}
\index{Plemelj formula}
\index{Plemelj-Sokhotsky formula}
\begin{equation}
\begin{split}
 {\cal F}[H(x)]=\widetilde{H}(k)=   \int_{-\infty}^\infty  H(x) e^{-i{kx}} dx   \\
=    \pi \delta(k) -  i {\cal P}  \frac{1}{k} \\
=   -i \left(i\pi \delta(k) + {\cal P} \frac{1}{k}\right)\\
=   \lim_{\varepsilon \rightarrow 0^+} -\frac{i}{k-i\varepsilon }
\end{split}
\label{2019-mm-ch-di-fthusf}
\end{equation}

{\color{OliveGreen}
\bproof
We shall compute the Fourier transform
 of the regularized Heaviside function
$H_\varepsilon (x) =H(x)e^{-\varepsilon x}$, with $\varepsilon >0$, of Equation~(\ref{2012-m-ch-di-rhfun}); that is,\cite[-10mm]{sommer-di}
\begin{equation}
\begin{split}
 {\cal F}[H_\varepsilon (x)] =
 {\cal F}[H(x)e^{-\varepsilon x}]
=\widetilde{H_\varepsilon }(k)
\\
=   \int_{-\infty}^\infty  H_\varepsilon (x) e^{-i{kx}} dx
\\
=   \int_{-\infty}^\infty  H(x)e^{-\varepsilon x}  e^{-i{kx}} dx
\\
=   \int_{-\infty}^\infty  H(x) e^{-i{kx}+(-i)^2\varepsilon x}  dx
\\
=   \int_{-\infty}^\infty  H(x) e^{-i(k - i \varepsilon ) x}  dx
\\
=   \int_{0}^\infty  e^{-i(k - i \varepsilon ) x}  dx
\\
=  \left. \left[ -\frac{ e^{-i(k - i \varepsilon ) x}}{i(k - i \varepsilon ) } \right] \right|_{x=0}^{x=\infty}
=  \left. \left[ -\frac{ e^{-ik} e^{- \varepsilon   x}}{i(k - i \varepsilon ) } \right] \right|_{x=0}^{x=\infty}
\\
=  \left[ -\frac{ e^{-ik\infty} e^{- \varepsilon  \infty}}{i(k - i \varepsilon ) } \right]
-  \left[ -\frac{ e^{-ik0} e^{- \varepsilon  0}}{i(k - i \varepsilon ) } \right]
\\
=    0 - \frac{ (-1)}{i(k - i \varepsilon ) }
=     -\frac{ i}{k - i \varepsilon  }
.
\end{split}
\end{equation}
Taking the limit and using
Sokhotsky's  formula~(\ref{2012-m-ch-di-Sokhotskyformula2})
we therefore conclude that
\begin{equation}
 {\cal F}[H(x)] =
 {\cal F}[H_{0^+} (x)] =
 \lim_{\varepsilon \rightarrow 0^+} {\cal F}[H_\varepsilon (x)] =
  \pi \delta (k)   -i {\cal P} \left(\frac{1}{k}\right)
.
\label{2012-m-ch-di-fthfu}
\end{equation}

\eproof
}


\section{The sign function}

\subsection{Definition}
The
{\em sign function}
\index{sign function}
is defined by
\begin{equation}
\begin{split}
\textrm{sgn}(x-x_0)
=
\lim_{\varepsilon \rightarrow 0^+}
\frac{2}{\pi}
\textrm{arctan}\left( \frac{x-x_0}{\varepsilon } \right)
=
\left\{
\begin{array}{rl}
-1&\textrm{ for } x < x_0\\
0&\textrm{ for } x = x_0 \\
+1&\textrm{ for } x > x_0
\end{array}
\right.
.
\end{split}
\end{equation}
It is plotted in Figure~\ref{2011-m-fsf}.
\begin{marginfigure}
\begin{center}
\begin{tikzpicture}[ scale=0.6,inner sep=0.8em]

\tikzstyle{every path}=[line width=2pt]

\begin{axis}[
axis lines=middle,axis equal,
    ticklabel style = {font=\Large },
    draw=gray!80,
    xmin=-2.5, xmax=2.5,
    ymin=-1.5, ymax=1.5,
     extra x ticks={0},
     extra x tick style={tick label style={anchor=south west, xshift=4pt}},
     ytick={0,1},
     extra y ticks={-1},
     extra y tick style={tick label style={anchor=west, xshift=6pt}}
]

\addplot[orange,smooth,domain=0:2.5] {1};
\addplot[orange,smooth,domain=-2.5:0] {-1};
\draw[black,fill=white] (axis cs:0,-1) circle(1mm)  (axis cs:0,1) circle(1mm);
\draw[fill,orange] (axis cs:0,0) circle(1mm);

\end{axis}
\end{tikzpicture}
\end{center}
\caption{Plot of the sign function.}
\label{2011-m-fsf}
\end{marginfigure}

\subsection{Connection to the Heaviside function}

In terms of  the  Heaviside step function, in particular, with
$H(0)=\frac{1}{2}$ as in Equation~(\ref{2012-m-di-hsfoh}),
the sign function can be written by ``stretching'' the former (the Heaviside step function) by a factor of two,
and shifting it by one negative unit---an affine map---as follows
\begin{equation}
\begin{split}
\textrm{sgn}(x-x_0) = 2H(x-x_0) -1,\\
H(x-x_0) = \frac{1}{2} \left[ \textrm{sgn}(x-x_0)+1\right];  \\
\textrm{and also}\\
\textrm{sgn}(x-x_0) = H(x-x_0) - H(x_0-x).
\end{split}
\label{2011-m-cbhsf}
\end{equation}
Therefore, the derivative of the sign function is
\begin{equation}
\frac{d}{dx}\textrm{sgn}(x-x_0) = \frac{d}{dx} \left[2H(x-x_0) -1\right] =2 \delta(x-x_0).
\end{equation}

Note also that  $\textrm{sgn}(x-x_0) =  - \textrm{sgn}(x_0-x)$.

\subsection{Sign sequence}

The sequence of functions
\begin{equation}
\begin{split}
\textrm{sgn}_n(x-x_0)
=
\left\{
\begin{array}{rl}
- e^{-\frac{x}{n}}&\textrm{ for } x < x_0\\
+ e^{\frac{-x}{n}}&\textrm{ for } x > x_0
\end{array}
\right.
\end{split}
\label{2012-m-ch-di-lsegn}
\end{equation}
is a limiting sequence of
$
\textrm{sgn}(x-x_0)\stackrel{x\neq x_0}{=} \lim_{n\rightarrow \infty} \textrm{sgn}_n(x-x_0)
$.

We can also use the Dirichlet integral
\index{Dirichlet integral}
to express a limiting sequence for the sign function,
in a similar way as the derivation of Eqs.~(\ref{2012-m-ch-di-dcf}); that is,
\begin{equation}
\begin{split}
\textrm{sgn}(x)= \lim_{t \rightarrow \infty} \textrm{sgn}_t (x)
\\
= \frac{2}{\pi }\lim_{t \rightarrow \infty}\int_0^t \frac{\sin (kx)}{k} dk
\\
=
\frac{2}{\pi }\int_0^\infty \frac{\sin (kx)}{k} dk
.
\end{split}
\label{2015-m-ch-di-sign}
\end{equation}

Note (without proof) that
\begin{eqnarray}
\mbox{sgn}(x)
&=&{4\over \pi }\sum_{n=0}^\infty {\sin [
(2n+1)x]\over
(2n+1)}\\
&=&{4\over \pi }\sum_{n=0}^\infty (-1)^n{\cos [
(2n+1)(x-\pi /2)]\over
(2n+1)}\;,\; -\pi <x<\pi  .
 \end{eqnarray}

\subsection{Fourier transform of $\textrm{sgn}$}

Since the Fourier transform is linear,
we may use the connection between the sign and the Heaviside functions $\textrm{sgn}(x) = 2H(x) -1$, Equation~(\ref{2011-m-cbhsf}),
together with the
Fourier transform of the Heaviside function
${\cal F}[H(x)] =  \pi \delta (k)   -i {\cal P} \left(\frac{1}{k}\right)$,
Equation~(\ref{2012-m-ch-di-fthfu}) and the Dirac delta function
${\cal F}[1] = 2\pi \delta (k)$, Equation~(\ref{2011-m-eftdelta1}),
to compose and compute the Fourier transform  of $\textrm{sgn}$:
\begin{equation}
\begin{split}
{\cal F}[\textrm{sgn}(x)] = {\cal F}[2H(x) -1] = 2{\cal F}[H(x)] - {\cal F}[1]
\\   =
2 \left[\pi \delta (k)   -i {\cal P} \left(\frac{1}{k}\right)\right] -  2\pi \delta (k)
\\   =
-2i {\cal P} \left(\frac{1}{k}\right)
.
\end{split}
\label{2011-m-ch-di-fdsgnfu}
\end{equation}

\if01
We shall consider the Fourier transform of the limit of the sequence (\ref{2012-m-ch-di-lsegn})
and assume (without proof) that, in the functional sense,
this yields the   Fourier transform of $\textrm{sgn}$; that is,
$
\lim_{n\rightarrow \infty}
{\cal F}[ \textrm{sgn}_n(x) ]
=
{\cal F}[ \textrm{sgn}(x) ]
$.

So, consider
\begin{equation}
\begin{split}
  {\cal F}[\textrm{sgn}_n (x)]= \int_{-\infty}^\infty  \textrm{sgn}_n (x) e^{-i{kx}} dx   \\
  =
\int_{0}^\infty  e^{-\frac{x}{n}} e^{-i{kx}} dx
-
\int_{-\infty}^0  e^{ \frac{x}{n}} e^{-i{kx}} dx
\\
  =
\int_{0}^\infty  e^{-\left(\frac{1}{n}+i{k}\right) x} dx
-
\int_{-\infty}^0  e^{ \left(\frac{1}{n}-i{k}\right) x} dx
\\
[\textrm{substitution }x \rightarrow -x \textrm{ in second integral}]\\
  =
\int_{0}^\infty  e^{-\left(\frac{1}{n}+i{k}\right) x} dx
-
\int_{0}^\infty  e^{-\left(\frac{1}{n}-i{k}\right) x} dx
\\
  =
\left.  \frac{ e^{-\left(\frac{1}{n}+i{k}\right) x}}{{-\left(\frac{1}{n}+i{k}\right)  }} \right|_{0}^\infty
-
\left.  \frac{ e^{-\left(\frac{1}{n}-i{k}\right) x}}{-\left(\frac{1}{n}-i{k}\right)} \right|_{0}^\infty
\\
  =
  \frac{ 1}{  \frac{1}{n}+i{k} }
-
 \frac{ 1 }{ \frac{1}{n}-i{k} }
.
\end{split}
\label{2011-m-ftsgn1a}
\end{equation}
Thus,
\begin{equation}
\begin{split}
{\cal F}[\textrm{sgn}  (x)] =\lim_{n\rightarrow \infty}  {\cal F}[\textrm{sgn}_n (x)]
\\
  =  \lim_{n\rightarrow \infty}  \left(
  \frac{ 1}{ \frac{1}{n}+i{k}   }
-
 \frac{ 1 }{  \frac{1}{n}-i{k}   }          \right)
\\
  =
  \frac{ 1}{   i{k}  }
+
 \frac{ 1 }{  i{k} }
 =
  -\frac{2i}{ k }
.
\end{split}
\label{2011-m-ftsgn1b}
\end{equation}
\fi


\begin{marginfigure}
\begin{center}
\begin{tikzpicture}[ scale=0.6,
 declare function={
    func(\x)= (\x <= 0) * (-\x)   +
              (\x > 0) * (\x)
   ;
                  } ]

\tikzstyle{every path}=[line width=2pt]

\begin{axis}[axis lines=middle,
draw=gray!80,
enlargelimits=true,axis equal,
ticklabel style = {font=\Large },
]
\addplot [
orange,
domain=-2:2,
line width=3pt
]  {func(x)};
\end{axis}
\end{tikzpicture}
\caption{Plot of the absolute value function $f(x)=\left|x\right|$.}
\label{2011-m-avm}
\end{center}
\end{marginfigure}

\section{Absolute value function (or modulus)}
\index{absolute value}
\index{modulus}

\subsection{Definition}
The {\em absolute value} (or {\em modulus})
of $x$ is defined by
\begin{equation}
\left|
x-x_0\right|
=
\left\{
\begin{array}{ll}
x-x_0&\textrm{ for } x > x_0\\
0&\textrm{ for } x = x_0\\
x_0-x&\textrm{ for } x < x_0
\end{array}
\right.
\label{2011-m-di-avm}
\end{equation}
It is plotted in Figure~\ref{2011-m-avm}.

\subsection{Connection of absolute value with the sign and Heaviside functions}

Its relationship to the sign function is twofold:
on the one hand, there is
 \begin{equation}
 \left|x\right| = x \,\textrm{sgn} (x),
 \end{equation}
and thus, for $x\neq 0$,
 \begin{equation}
\textrm{sgn} (x)  = \frac{\left|x\right|}{x} = \frac{x}{\left|x\right|}.
 \end{equation}

On the other hand, the derivative of the absolute value function is the sign function, at least up to a singular point at $x=0$,
and thus the absolute value function can be interpreted as the integral of the sign function (in the distributional sense);
that is,
\begin{equation}
\begin{split}
\frac{d}{dx} \left|x\right| \left[ \varphi \right]
=
\textrm{sgn}\left[ \varphi \right] \text{, or, formally,}
\\
\frac{d }{dx} \left|x\right|
=
\textrm{sgn} (x)
=
\left\{
\begin{array}{rl}
1&\textrm{ for } x > 0\\
0&\textrm{ for } x = 0\\
-1&\textrm{ for } x < 0
\end{array}
\right.
,\\
\text{and } \left|x\right| =  \int \textrm{sgn} (x) dx.
\end{split}
 \end{equation}

{\color{OliveGreen}
\bproof
This can  be formally proven by inserting
$\left|x\right|  =  {x}\, \textrm{sgn} (x)$; that is,
\begin{equation}
\begin{split}
\frac{d }{dx}  \left|x\right|
=
\frac{d }{dx}x\,\textrm{sgn} (x)
=
\textrm{sgn} (x)  + x\frac{d}{dx} \textrm{sgn} (x)
\\
=
\textrm{sgn} (x)  + x\frac{d }{dx}\left[2H(x)-1\right]
=
\textrm{sgn} (x) - 2\underbrace{x\delta(x)}_{=0}.
\end{split}
 \end{equation}

Another proof is via linear functionals:
\begin{equation}
\begin{split}
\frac{d }{dx}  \left|x\right|   \left[ \varphi \right]
= -  \left|x\right|   \left[ \varphi' \right]
= -\int_{-\infty}^\infty  \left|x\right|   \varphi'  (x) dx
\\
= -\int_{-\infty}^0 \underbrace{\left|x\right|}_{=-x}   \varphi'  (x) dx
- \int_{0}^\infty  \underbrace{\left|x\right|}_{=x}    \varphi'  (x) dx
\\
= \int_{-\infty}^0  x    \varphi'  (x) dx - \int_{0}^\infty   x    \varphi'  (x) dx
\\
= \underbrace{\left. x \varphi   (x)\right|_{-\infty}^0}_{=0}  -  \int_{-\infty}^0    \varphi   (x) dx  -  \underbrace{\left. x \varphi   (x)\right|_0^{\infty}}_{=0}  + \int_{0}^\infty        \varphi'  (x) dx
\\
=   \int_{-\infty}^0    (-1) \varphi   (x) dx      + \int_{0}^\infty  (+1)    \varphi'  (x) dx  \\
=   \int_{-\infty}^\infty     \textrm{sgn}(x) \varphi   (x) dx
= \textrm{sgn}\left[ \varphi \right]
.
\end{split}
 \end{equation}

\eproof
}

\section{Some examples}

{
\color{blue}
\bexample

Let us compute some concrete examples related to distributions.
\begin{enumerate}
\item
For a start, let us prove that
\begin{equation}
\lim_{\epsilon \rightarrow 0}{\epsilon \sin^2 \frac{x}{\epsilon}\over \pi
 x^2}= \delta (x).\end{equation}
As a hint, take  $\int_{-\infty}^{+\infty} {\sin^2x\over x^2}dx =\pi $.

Let us prove this conjecture by integrating over a good test function $\varphi$
\begin{equation}
\begin{split}
   {1\over\pi}\lim_{\epsilon \rightarrow 0}\int\limits_{-\infty}^{+\infty}
   {\varepsilon\sin^2\left({x\over\varepsilon}\right)\over x^2} \varphi(x) dx
\\
 \textrm{[variable substitution}\;   y={x\over\varepsilon} , {dy\over dx}={1\over\varepsilon}, dx=\varepsilon dy
\textrm{]}
\\  =
  {1\over\pi}\lim_{\epsilon \rightarrow 0}
   \int\limits_{-\infty}^{+\infty}\varphi(\varepsilon y)
   {\varepsilon^2\sin^2(y)\over \varepsilon^2y^2}dy
\\  =    {1\over\pi}
   \varphi(0)\int\limits_{-\infty}^{+\infty}{\sin^2(y)\over y^2}dy
  =    \varphi(0).
\end{split}
\end{equation}
Hence we can identify
\begin{equation}
   \lim_{\varepsilon \rightarrow 0}{\varepsilon\sin^2 \left({x\over\varepsilon}\right)\over\pi x^2}=\delta(x).
\end{equation}

 \item
In order to prove that $\frac{1}{\pi} \frac{n  e^{-x^2}}{1+n^2x^2} $ is  a  $\delta$-sequence
we proceed again by integrating over a good test function $\varphi$,
and with the hint that $\int\limits_{-\infty}^{+\infty} dx/ (1+x^2) =\pi$ we obtain
 \begin{equation}
\begin{split}
\lim_{n \rightarrow \infty}  \frac{1}{\pi}
\int\limits_{-\infty}^{+\infty}
\frac{n  e^{-x^2}}{1+n^2x^2}
   \varphi(x)
dx
\\  \textrm{[variable substitution }    y={xn} , x={y\over n}, {dy\over dx}={n}, dx={dy\over n}
\textrm{]}
\\   =
\lim_{n \rightarrow \infty}   \frac{1}{\pi}
\int\limits_{-\infty}^{+\infty}
\frac{n  e^{-\left( \frac{y}{n} \right)^2}}{1+y^2}
   \varphi \left( \frac{y}{n} \right)
\frac{dy}{n}
\\  =
\frac{1}{\pi}
\int\limits_{-\infty}^{+\infty}
\lim_{n \rightarrow \infty}  \left[     e^{-\left( \frac{y}{n} \right)^2}  \varphi \left( \frac{y}{n} \right) \right]
\frac{1}{1+y^2}
dy
\\   =       \frac{1}{\pi}
\int\limits_{-\infty}^{+\infty}
\left[     e^{0}  \varphi \left( 0\right) \right]
\frac{1}{1+y^2}
dy
\\   =    \frac{\varphi \left( 0\right)}{\pi}
\int\limits_{-\infty}^{+\infty}
\frac{1}{1+y^2}
dy   =
\frac{\varphi \left( 0\right)}{\pi} \pi
   =
\varphi \left( 0\right).
\end{split}
\end{equation}
Hence we can identify
\begin{equation}
   \lim_{n \rightarrow \infty}{\frac{1}{\pi} \frac{n  e^{-x^2}}{1+n^2x^2}}=\delta(x).
\end{equation}

\item
Let us prove that
$x^n\delta^{(n)}[\varphi]=C\delta [\varphi]$    and determine the constant $C$.
We proceed again by integrating over a good test function $\varphi$.
First note that if
$\varphi (x)$ is a good test function, then so is
$x^n\varphi (x)$.
 \begin{equation}
\begin{split}
    x^n\delta^{(n)}[\varphi] =
   \int dx x^n\delta^{(n)}(x)\varphi(x) \\
   = \int dx \delta^{(n)}(x)\bigl[x^n\varphi(x)\bigr]
   =(-1)^n \int dx \delta(x)\bigl[x^n\varphi(x)\bigr]^{(n)}\\
  =(-1)^n \int dx \delta(x)\bigl[nx^{n-1}\varphi(x)+x^n\varphi'(x)
   \bigr]^{(n-1)}=
 \cdots \\
   =(-1)^n \int dx \delta(x)\left[\sum_{k=0}^n
           \left(
          \begin{array}{c}
          n\\ k
           \end{array}\right) (x^n)^{(n-k)}\varphi ^{(k)}(x)\right]
    \\
   =(-1)^n \int dx
\delta(x)\bigl[n!\varphi(x)+n\cdot n!x\varphi'(x) +
\cdots +x^n\varphi^{(n)}(x)
   \bigr] \\
   =(-1)^n n! \int dx \delta(x)\varphi(x)
   =(-1)^n n!  \delta[\varphi ]
,
\end{split}
\end{equation}
and hence  $ C=(-1)^n n!$.

\item
Let us simplify $\int_{-\infty}^\infty \delta (x^2-a^2)g(x)\; dx$.
First recall Equation~(\ref{2011-m-distdp1})  stating that
$$
   \delta(f(x))=\sum_{i}{\delta(x-x_i)\over|f'(x_i)|},
$$
whenever $x_i$ are simple roots of  $f(x)$, and $f'(x_i)\neq 0$.
In our case, $
   f(x)=x^2-a^2=(x-a)(x+a)
$, and the roots are  $x=\pm a$.
Furthermore,
$
   f'(x)=(x-a)+(x+a)= 2x
$; therefore
$
   |f'(a)|=|f'(-a)|=2|a|$.
As a result,
$$
    \delta(x^2-a^2)=\delta\bigl((x-a)(x+a)\bigr)={1\over|2a|}
   \bigl(\delta(x-a)+\delta(x+a)\bigr).
$$
Taking this into account we finally obtain
 \begin{equation}
\begin{split}
  \int\limits_{-\infty}^{+\infty}\delta(x^2-a^2)
   g(x)dx\\
  =
\int\limits_{-\infty}^{+\infty}{\delta(x-a)+\delta(x+a)\over
   2|a|}g(x)dx\\
  =    {g(a)+g(-a)\over2|a|}.
\end{split}
\end{equation}

\item
Let us evaluate
\begin{equation}
I=
\int_{-\infty}^\infty
\int_{-\infty}^\infty
\int_{-\infty}^\infty
\delta (x_1^2+x_2^2+x_3^2-R^2) d^3x
\end{equation}
for  $R \in {\Bbb R}, R >0 $.
We may, of course, retain the standard Cartesian coordinate system and evaluate the integral by ``brute force.''
Alternatively,
a more elegant way is to use the spherical symmetry of the problem and use spherical coordinates $r, \Omega (\theta ,\varphi )$ by rewriting $I$
into
\begin{equation}
I=  \int_{r,\Omega} r^2  \delta (r^2-R^2) d\Omega dr.
\end{equation}
As the integral kernel $\delta (r^2-R^2)$ just depends on the radial coordinate $r$
the angular coordinates just integrate to $4\pi$.
Next we make use of Equation~(\ref{2011-m-distdp1}), eliminate the solution for $r=-R$, and obtain
\begin{equation}
\begin{split}
I= 4\pi   \int_0^\infty r^2  \delta (r^2-R^2)  dr \\
 =  4\pi   \int_0^\infty  r^2 \frac{\delta (r+R) + \delta (r-R)}{2R}  dr \\
  =   4\pi   \int_0^\infty  r^2 \frac{\delta (r-R)}{2R}  dr \\
  =   2 \pi R.
\end{split}
\end{equation}

\item
Let us compute
\begin{equation}
\int_{-\infty}^\infty \int_{-\infty}^\infty  \delta
 (x^3-y^2+2y)\delta (x+y)H (y-x-6)f(x,y) \,dx\,dy.
\end{equation}

First, in dealing with
$\delta(x+y)$, we evaluate the $y$ integration at $x=-y$ or $y=-x$:
$$
   \int_{-\infty}^\infty \delta(x^3-x^2-2x)H(-2x-6)f(x,-x)  dx
$$
Use of Equation~(\ref{2011-m-distdp1})
$$
   \delta(f(x))=\sum_{i}{1\over|f'(x_i)|}\delta(x-x_i),
$$
at the roots
\begin{equation}
\begin{split}
   x_1=0\\
   x_{2,3}={1\pm\sqrt{1+8}\over 2}={1\pm3\over2}=\left\{{2\atop-1}\right.
\end{split}
\end{equation}
of the argument $f(x)=x^3-x^2-2x=x(x^2-x-2)=x(x-2)(x+1)$ of the remaining $\delta$-function,
together with
$$
 f'(x)=  {d\over dx}(x^3-x^2-2x)=3x^2-2x-2 ;
$$
yields
\begin{equation}
\begin{split}
   \int\limits_{-\infty}^\infty dx
         {\delta(x)+\delta(x-2)+\delta(x+1)\over|3x^2-2x-2|}
         H(-2x-6)f(x,-x) \\
   ={1\over|-2|}\underbrace{H(-6)}_{ =0 }f(0,-0)+
      {1\over|12-4-2|}\underbrace{H(-4-6)}_{ =0 }f(2,-2)  +\\
   +\ {1\over|3+2-2|}\underbrace{H(2-6)}_{ =0 }f(-1,1)
=0
\end{split}
\end{equation}

\item
When simplifying derivatives of generalized functions it is always useful to evaluate their properties
--
such as $x\delta(x)=0$, $f(x)\delta(x-x_0)=f(x_0)\delta(x-x_0)$, or $\delta (-x)=\delta (x)$
--
first and  before proceeding with the next differentiation or evaluation.
We shall present some applications of this ``rule'' next.

First, simplify
\begin{equation}
\left({d\over dx}-\omega \right)H (x)e^{\omega x}
\end{equation}
as follows
\begin{equation}
\begin{split}
 {d\over dx}\left[H(x)e^{\omega
x}\right]-\omega H(x)e^{\omega x}\\
  =
   \delta(x)e^{\omega x}+\omega H(x)e^{\omega x}-\omega H(x)
   e^{\omega x}\\
  =
   \delta(x)e^{0}
   \\
  =  \delta(x)
\end{split}
\end{equation}

\item
Next, simplify
\begin{equation}
\left({d^2\over dx^2}+\omega^2 \right){1\over \omega }H
 (x)\sin (\omega x)
\end{equation}
as follows
\begin{equation}
\begin{split}
{d^2\over dx^2}\left[{1\over\omega}H(x)\sin(\omega
x)\right]+\omega H(x)
      \sin(\omega x)\\
     =   {1\over\omega}{d\over dx}\Bigl[\underbrace{\delta(x)\sin(\omega x)}_
      {\mbox{$=0$}}+\omega H(x)\cos(\omega x)\Bigr]+\omega H(x)
      \sin(\omega x)\\
   =   {1\over\omega}\Bigl[\omega\underbrace{\delta(x)\cos(\omega x)}_
      {\mbox{$\delta(x)$}}-\omega^2H(x)\sin(\omega x)\Bigr]+
      \omega H(x)\sin(\omega x)=\delta(x)
\end{split}
\end{equation}

\begin{marginfigure}
{\color{black}
\begin{center}
\begin{tabular}{c}

\begin{tikzpicture}[ scale=0.4,
 declare function={
    func(\x)= (\x )
   ;
                  } ]

\tikzstyle{every path}=[line width=3pt]

\begin{axis}[axis lines=middle, draw=gray!80,
axis equal,
xtick={-1,0,1},
ytick={-1,0,1},
ticklabel style = {font=\huge },
every axis x label/.style={
    at={(ticklabel* cs:1)},
    anchor=west,
    font=\huge ,
},
every axis y label/.style={
    at={(ticklabel* cs:1)},
    anchor=south,
    font=\huge ,
},
xlabel={$x$},
ylabel={$f(x)=f_1(x)f_2(x)$}
]

\addplot[orange,line width=2pt,smooth,domain=-1.5:0] {0};
\addplot [
orange,
domain=0:1,
line width=3pt
]  {func(x)};
\addplot[orange,line width=2pt,smooth,domain=1:1.5] {0};

\end{axis}
\end{tikzpicture}
\\
(a) $\,$\\
\\
\begin{tikzpicture}[ scale=0.4,
 declare function={
    func(\x)= (\x <= 0) * (0)   +
              and(\x >= 0,\x <= 1) * ( 1 )     +
              (\x > 1) * (0 )
   ;
                  } ]

\tikzstyle{every path}=[line width=3pt]

\begin{axis}[axis lines=middle, draw=gray!80,
axis equal,
xtick={-1,0,1},
ytick={-1,0,1},
ticklabel style = {font=\huge },
every axis x label/.style={
    at={(ticklabel* cs:1)},
    anchor=west,
    font=\huge ,
},
every axis y label/.style={
    at={(ticklabel* cs:1)},
    anchor=south,
    font=\huge ,
},
xlabel={$x$},
ylabel={\begin{tabular}{c}$f_1(x)=H(x)H(1-x)$\\$=H(x)- H(x-1)$\end{tabular}}
]

\addplot[blue,line width=2pt,smooth,domain=-1.5:0] {0};
\addplot[blue,line width=2pt,smooth,domain=0:1] {1};
\addplot[blue,line width=2pt,smooth,domain=1:1.5] {0};

\end{axis}
\end{tikzpicture}
\\
(b) $\,$\\
\\
\begin{tikzpicture}[ scale=0.4,
 declare function={
    func(\x)= (\x <= 0) * (\x)   +
              (\x > 0) * (\x)
   ;
                  } ]

\tikzstyle{every path}=[line width=3pt]

\begin{axis}[axis lines=middle, draw=gray!80,axis equal,
xtick={-1,0,1},
ytick={-1,0,1},
ticklabel style = {font=\huge },
every axis x label/.style={
    at={(ticklabel* cs:1)},
    anchor=west,
    font=\huge ,
},
every axis y label/.style={
    at={(ticklabel* cs:1)},
    anchor=south,
    font=\huge ,
},
xlabel={$x$},
ylabel={$f_2(x)=x$}
]
\addplot [
blue,
domain=-1.5:1.5,
samples=201,
line width=3pt
]  {func(x)};
\end{axis}
\end{tikzpicture}
\\
(c)
\end{tabular}
\end{center}
}
\caption{Composition of
$f (x)=f_1(x)f_2(x)$.
}
\label{2011-m-fcof1}
\end{marginfigure}

\item
Let us compute the $n$th derivative of
\begin{equation}
f (x)
=
\begin{cases}
0  & \textrm{ for }    x< 0 ,\\
x  & \textrm{ for }   0\le x\le 1, \\
0  &\textrm{ for }  x>1.
\end{cases}
\end{equation}

As depicted in Figure~\ref{2011-m-fcof1},
$f$ can be composed from two functions $f(x)=f_2(x)\cdot f_1(x)$;
and this composition can be done in at least two ways.

Decomposition {(i)} yields
\begin{equation}
\begin{split}
   f(x)=x\bigl[H(x)-H(x-1)\bigr]=xH(x)-xH(x-1)\\
   f'(x)=H(x)+\underbrace{x\delta(x)}_{=0}-H(x-1)-x\delta(x-1)
\end{split}
\end{equation}
Because of $x\delta(x-a)=a\delta(x-a)$,
\begin{equation}
\begin{split}
   f'(x)=H(x)-H(x-1)-\delta(x-1)\\
   f''(x)=\delta(x)-\delta(x-1)-\delta'(x-1)
\end{split}
\end{equation}
and hence   by induction, for $n>1$,
\begin{equation}
f^{(n)}(x)=\delta^{(n-2)}(x)-\delta^{(n-2)}(x-1)-   \delta^{(n-1)}(x-1)
.
\end{equation}

Decomposition {(ii)}  yields the same result as  decomposition {(i)}, namely
\begin{equation}
\begin{split}
   f(x)=xH(x)H(1-x)\\
   f'(x)=H(x)H(1-x)+         \underbrace{x\delta(x)}_{ =0 } H(1-x)+\underbrace{xH(x)(-1)\delta(1-x)}_{=-H(x)\delta(1-x)}\\
   =H(x)H(1-x) - \delta(1-x)\\
\text{[with $\delta(x)=\delta(-x)$]}  =        H(x)H(1-x) - \delta(x-1)\\
   f''(x)=\underbrace{\delta(x)H(1-x)}_{ =\delta(x) }         +\underbrace{(-1)H(x)\delta(1-x)}_{ -\delta(1-x) }        -\delta'(x-1) \\
        =\delta(x)-\delta(x-1)-\delta'(x-1);
\end{split}
\end{equation}
and hence   by induction, for $n>1$,
\begin{equation}
f^{(n)}(x)=\delta^{(n-2)}(x)-\delta^{(n-2)}(x-1)-
   \delta^{(n-1)}(x-1)
.
\end{equation}

\item
Let us compute the $n$th derivative of
 \begin{equation}
f (x)
=
\begin{cases}
\vert \sin x\vert & \textrm{ for } -\pi \le x\le \pi ,\\
 0  & \textrm{ for }  \vert x\vert >\pi .
\end{cases}
\end{equation}

$$
   f(x)=|\sin x|H(\pi+x)H(\pi-x)
$$
$$
   |\sin x|=\sin x\mbox{ sgn}(\sin x)=\sin x\mbox{ sgn\,}x \text{ for }
     -\pi<x<\pi ;
$$
hence we start from
$$
  f(x)=\sin x\mbox{ sgn\,}xH(\pi+x)H(\pi-x),
$$
Note that
\begin{eqnarray*}
   \mbox{ sgn\,}x&=&H(x)-H(-x),\\
   (\mbox{ sgn\,}x)'&=&H'(x)-H'(-x)(-1)=\delta(x)+\delta(-x)=
                   \delta(x)+\delta(x)=2\delta(x).
\end{eqnarray*}
\begin{eqnarray*}
   f'(x)   &=   &\cos x\mbox{ sgn\,}xH(\pi+x)
           H(\pi-x)+\sin x2\delta(x)H(\pi+x)H(\pi-x)+\\
           &    &+\sin x\mbox{ sgn\,}x\delta(\pi+x)H(\pi-x) +
           \sin x\mbox{ sgn\,}x H(\pi+x)\delta(\pi-x)(-1)=\\
           &=   &\cos x\mbox{ sgn\,}xH(\pi+x)H(\pi-x)\\
   f''(x)   &=   & -\sin x\mbox{ sgn\,}xH(\pi+x)H(\pi-x)+
             \cos x2\delta(x)H(\pi+x)H(\pi-x)+\\
           &    &+\cos x\mbox{ sgn\,}x\delta(\pi + x)H(\pi - x)
              + \cos x\mbox{ sgn\,}x H(\pi + x)\delta(\pi - x)(-1)=\\
           &=   & -\sin x\mbox{ sgn\,}xH(\pi+x)H(\pi-x)+
             2\delta(x)+\delta(\pi+x)+\delta(\pi-x)\\
   f'''(x)   &=   & -\cos x\mbox{ sgn\,}xH(\pi+x)H(\pi-x)-
             \sin x2\delta(x)H(\pi+x)H(\pi-x)-\\
           &    &-\sin x\mbox{ sgn\,}x\delta(\pi+x)H(\pi-x) -
             \sin x \mbox{ sgn\,}x H(\pi+x)\delta(\pi-x)(-1)+\\
           &    &+2\delta'(x)+\delta'(\pi+x)-\delta'(\pi-x)=\\
           &=   & -\cos x\mbox{ sgn\,}xH(\pi+x)H(\pi-x)+
             2\delta'(x)+\delta'(\pi+x)-\delta'(\pi-x)\\
   f^{(4)}(x)   &=   & \sin x\mbox{ sgn\,}xH(\pi+x)H(\pi-x)-
             \cos x2\delta(x)H(\pi+x)H(\pi-x)-\\
           &    &-\cos x\mbox{ sgn\,}x\delta(\pi+x)H(\pi-x) -
             \cos x \mbox{ sgn\,}x H(\pi+x)\delta(\pi - x)(-1)+\\
           &    &+2\delta''(x)+\delta''(\pi+x)+\delta''(\pi-x)=\\
           &=   & \sin x\mbox{ sgn\,}xH(\pi+x)H(\pi-x)-
             2\delta(x)-\delta(\pi+x)-\delta(\pi-x)+\\
           &    &+2\delta''(x)+\delta''(\pi+x)+\delta''(\pi-x);
\end{eqnarray*}
hence
\begin{eqnarray*}
   f^{(4)}   &=   &f(x)-2\delta(x)+2\delta''(x)-
             \delta(\pi+x)+\delta''(\pi+x)-\delta(\pi-x)+\delta''(\pi-x), \\
   f^{(5)}   &=   &f'(x)-2\delta'(x)+2\delta'''(x)-
             \delta'(\pi + x)+\delta'''(\pi + x)+\delta'(\pi - x)-
             \delta'''(\pi - x);
\end{eqnarray*}
and thus  by induction
\begin{eqnarray*}
 f^{(n)}&=&f^{(n-4)}(x)-2\delta^{(n-4)}(x)+
             2\delta^{(n-2)}(x)-\delta^{(n-4)}(\pi+x)+\\
   &&+\delta^{(n-2)}(\pi+x)+(-1)^{n-1}
             \delta^{(n-4)}(\pi-x)+(-1)^n\delta^{(n-2)}(\pi-x)\\
   &&(n=4,5,6,\dots)
\end{eqnarray*}

\end{enumerate}

\eexample
}

\begin{center}
{\color{olive}   \Huge
\floweroneleft
}
\end{center}

\chapter*{\color{BurntOrange} \thispagestyle{empty} {\fontsize{40}{168} \selectfont Part III} \\ {\fontsize{30}{68} \selectfont Differential equations}
\addcontentsline{toc}{part}{Part III:  Differential equations}
{\newpage \thispagestyle{empty}   $\;$ \vskip 9 true cm \begin{center}\includegraphics[width=0.7\textwidth, angle=-20]{2019-mm-swimmer.png}\end{center}}}
\chapter{Green's function}
\label{2011-m-gf}

This chapter is the beginning of a series of chapters dealing with the solution of differential equations related to theoretical physics.
\index{differential equation}
These differential equations are {\em linear}; that is, the ``sought after'' function $\Psi (x), y(x), \phi (t)$ {\it et cetera}
occur only as a polynomial of degree zero (the inhomogeneous term) and one, and {\em not} of any higher degree, such as, for instance, $[y(x)]^2$.
\marginnote{The order of an ordinary differential equation is the order of the highest derivative that appears in the equation.
The term ``degree'' is used in a variety of ways. Often it is defined as the power of its highest derivative,
after the ordinary differential equation has been made rational and integral in all of its derivatives.
Note that some ordinary differential equations have no degree according to this definition; for instance,
$ y'' + \cos y' = 0 $.}

\section{Elegant way to solve linear differential equations}

Green's functions present a very elegant way of solving linear differential equations
of the form
\begin{equation}
\begin{split}
{\cal L}_x y(x)=   f(x) \textrm{, with the differential operator}\\
{\cal L}_x =
a_n (x)\frac{d^n}{dx^n} +
a_{n-1} (x)\frac{d^{n-1}}{dx^{n-1}} +
\ldots
+
a_1 (x)\frac{d}{dx}
+
a_0 (x) \\
\qquad = \sum_{j=0}^n  a_j(x) \frac{d^j}{dx^j} ,
\end{split}
\label{2011-m-egfp}
\end{equation}
where $a_i(x)$, $0\le i \le n$ are functions of $x$.
The idea of the Green's function method is quite straightforward:
if we are able to obtain the ``inverse'' $G$ of the differential operator  ${\cal L}$ defined by
\begin{equation}
{\cal L}_x G(x,    x'   )=   \delta (x-    x'   ),
\label{2011-m-egfp01}
\end{equation}
with  $\delta$ representing Dirac's delta function, then the solution to the inhomogeneous differential
equation   (\ref{2011-m-egfp}) can be obtained by integrating  $G(x,x')$ alongside with
\index{inhomogeneous differential equation}
the inhomogeneous term $f(    x'   )$; that is, by forming
\begin{equation}
\begin{split}
y(x) = \int_{-\infty}^\infty
G(x,    x'   )
f(    x'   )
d     x'   .
\end{split}
\label{2011-m-egfp1}
\end{equation}
This claim, as posted in Equation (\ref{2011-m-egfp1}),
can be verified by explicitly applying the differential operator  ${\cal L}_x$
to the solution $y(x)$,
\begin{equation}
\begin{split}
{\cal L}_x y(x)\\
\qquad  =
{\cal L}_x \int_{-\infty}^\infty
G(x,    x'   )
f(    x'   )
d     x'        \\
\qquad  =
\int_{-\infty}^\infty
{\cal L}_x G(x,    x'   )
f(    x'   )
d     x'      \\
\qquad  =
\int_{-\infty}^\infty
\delta (x-    x'   )
f(    x'   )
d     x'      \\
\qquad  =
f(x)  .
\end{split}
\label{2011-m-egfp12}
\end{equation}

{
\color{blue}
\bexample

Let us check whether
$G(x,x')=H (x-x')\sinh (x-x')$
is a Green's function of the differential operator ${\cal L}_x={d^2\over dx^2}-1$.
In this case, all we have to do is to verify that   ${\cal L}_x$, applied to $G(x,x')$, actually renders $\delta (x-x')$,
as required by Equation (\ref{2011-m-egfp01}).

\begin{equation}
\begin{split}
   {\cal L}_xG(x,x')=\delta(x-x')\\
   \left({d^2\over dx^2}-1\right)H(x-x')\sinh(x-x')
   \stackrel{?}{=} \delta(x-x')
\end{split}
\end{equation}
Note that $\frac{d}{dx}\sinh x=\cosh x$
and ${d\over dx}\cosh x=\sinh x$  and, therefore,
\begin{equation}
\begin{split}
   {d\over dx}\left(\underbrace{\delta(x-x')\sinh(x-x')}_{\mbox{$=0$}}+H(x-x')\cosh(x-x')\right) \qquad\qquad \qquad \\
     -H(x-x')\sinh(x-x')\\
=
   \underbrace{\delta(x-x')\cosh(x-x')}_{=\delta(x-x')} + H (x-x')\sinh(x-x')\qquad \qquad\qquad  \\
-H (x-x')\sinh(x-x')=
   \delta(x-x').
\end{split}
\end{equation}
\eexample
}

\section{Nonuniqueness of solution}

The solution (\ref{2011-m-egfp12}) so obtained is {\em not unique}, as it is only a special solution to the inhomogeneous
equation (\ref{2011-m-egfp}).
The general solution to (\ref{2011-m-egfp}) can be found
by adding the general solution $y_0(x)$
of the corresponding {\em homogeneous} differential equation
\index{homogeneous differential equation}
\begin{equation}
\begin{split}
{\cal L}_x y_0(x)=   0
\end{split}
\label{2011-m-egfp0}
\end{equation}
to one special solution -- say, the one obtained in Equation (\ref{2011-m-egfp12}) through Green's function techniques.

{\color{OliveGreen}
\bproof
Indeed,  the most general solution
\begin{equation}
Y(x)  = y(x) + y_0(x)
\label{2011-m-egfpgs}
\end{equation}
clearly is a solution of the inhomogeneous differential equation (\ref{2011-m-egfp12}),
as
\begin{equation}
{\cal L}_x Y(x)  ={\cal L}_x y(x) + {\cal L}_x y_0(x) =f(x)+0 =f(x).
\label{2011-m-egfpgs2}
\end{equation}

Conversely, any two distinct special solutions $y_1(x)$ and $y_2(x)$ of the inhomogeneous differential equation (\ref{2011-m-egfp12})
differ only by a function which is a solution of the homogeneous differential equation (\ref{2011-m-egfp0}), because
due to linearity of ${\cal L}_x$, their difference
$y_1(x) - y_2(x)$ can
be parameterized by some function $y_0$   which is the solution of the homogeneous differential equation:
\begin{equation}
{\cal L}_x [y_1(x) - y_2(x)]  ={\cal L}_x y_1(x) + {\cal L}_x y_2(x) =f(x) -f(x) =0.
\label{2011-m-egfpgs3}
\end{equation}

\eproof
}

\section{Green's functions of translational invariant differential operators}

From now on, we assume that the coefficients $a_j(x) =a_j$
in Equation (\ref{2011-m-egfp}) are constants, and thus are {\em translational invariant};
that is, ${\cal L}_{x-x'} = {\cal L}_x$.
Then the differential operator ${\cal L}_x$, as well as the entire {\it Ansatz} (\ref{2011-m-egfp01})
for   $G(x,x')$, is translation invariant,
because derivatives are defined only by relative distances, and $\delta (x-x')$
is translation invariant for the same reason.
Hence we may rewrite
\begin{equation}
G(x,x')\rightarrow  G(x-x').
\end{equation}
For such translation invariant systems, the Fourier analysis \index{Fourier analysis}
presents an excellent way of
analyzing the situation.

{\color{OliveGreen}
\bproof
Let us see why translation invariance of the coefficients
$a_j (x)=a_j(x+\xi )=a_j$
under the translation $x\rightarrow x + \xi $ with arbitrary $\xi $ --
that is, independence of the coefficients $a_j$ on the ``coordinate''
or ``parameter'' $x$ -- and thus of
the Green's function, implies a simple form of the latter.
Translanslation invariance of the Green's function really means
\begin{equation}
G(x+\xi ,x'+\xi )= G(x,x').
\end{equation}
Now set $\xi = - x'$; then we can define a new Green's function that just depends on
one argument (instead of previously two), which is the difference of the old arguments
\begin{equation}
G(x - x',x' - x')= G(x - x',0)\rightarrow G(x-x').
\end{equation}
\eproof
}

\section{Solutions with fixed boundary or initial values}

For applications, it is important to adapt the solutions of some inhomogeneous differential equation
to boundary and initial value problems.
\index{initial value problem}
\index{boundary value problem}
In particular, a properly chosen $G(x-    x'   )$, in its dependence on the parameter $x$, ``inherits''
some behavior of the solution $y(x)$.
Suppose, for instance, we would like to find solutions with $y(x_i)=0$
for some parameter values $x_i$, $i=1,\ldots ,k$.
Then, the Green's function $G$ must vanish there also
\begin{equation}
G(x_i-    x'   ) = 0  \textrm{ for } i=1,\ldots ,k.
\end{equation}

\section{Finding Green's functions by spectral decompositions}

It has been mentioned earlier
(cf. Section  \ref{2012-m-efed1} on page \pageref{2012-m-efed1})
that the  $\delta$-function
can be expressed in terms of various
{\em eigenfunction expansions}.
\index{eigenfunction expansion}
We shall make use of these expansions here.\cite[-10mm]{duffy2001}

Suppose $\psi_i(x)$ are
{\em eigenfunctions}
\index{eigenfunction}
of the differential operator ${\cal L}_x $,
and $\lambda_i$ are the associated {\em eigenvalues}; that is,
\index{eigenvalue}
\begin{equation}
{\cal L}_x \psi_i(x) =\lambda_i \psi_i(x).
\end{equation}
Suppose further that ${\cal L}_x$ is of degree $n$,
and therefore (we assume without proof)
that we know all (a complete set of) the $n$ eigenfunctions
$\psi_1(x), \psi_2(x), \ldots ,\psi_n(x)$ of ${\cal L}_x$.
In this case, orthogonality  of the system of eigenfunctions holds, such that
\begin{equation}
\int_{-\infty}^\infty \overline{\psi_i(    x    )}\psi_j(x)  dx =\delta_{ij} ,
\label{2011-m-egfcr1orthogon}
\end{equation}
as well as completeness, such that
\index{resolution of the identity}
\index{completeness}
\begin{equation}
\sum_{i=1}^n \overline{\psi_i( x' )}\psi_i(x) =\delta ( x - x')
.
\label{2011-m-egfcr1}
\end{equation}
$\overline{\psi_i(    x'   )}$ stands for the complex conjugate of ${\psi_i(    x'   )}$.
The sum in Equation (\ref{2011-m-egfcr1}) stands for an integral in the case of continuous spectrum of  ${\cal L}_x $.
In this case, the Kronecker $\delta_{ij}$ in  (\ref{2011-m-egfcr1orthogon}) is replaced by the Dirac delta function $\delta (k-k')$.
It has been mentioned earlier that the  $\delta$-function
can be expressed in terms of various
{\em eigenfunction expansions}.
\index{eigenfunction expansion}

The Green's function of ${\cal L}_x$
can be written as the spectral sum of the product of the (conjugate)  eigenfunctions,
divided by the eigenvalues $\lambda_j$; that is,
\begin{equation}
G(x-    x'   ) =\sum_{j=1}^n \frac{\overline{\psi_j(    x'   )}\psi_j(x) }{\lambda_j}.
\label{2011-m-egfpgsst}
\end{equation}

{\color{OliveGreen}
\bproof
For the sake of proof, apply  the differential operator  ${\cal L}_x$  to the Green's function {\em Ansatz}
 $G$ of Equation (\ref{2011-m-egfpgsst}) and verify that it satisfies Equation (\ref{2011-m-egfp01}):
\begin{equation}
\begin{split}
{\cal L}_x G(x-    x'   ) \\
\qquad ={\cal L}_x \sum_{j=1}^n \frac{ \overline{\psi_j(    x'   )}\psi_j(x)}{\lambda_j}
\\
\qquad =\sum_{j=1}^n \frac{\overline{\psi_j(    x'   )}[{\cal L}_x \psi_j(x)] }{\lambda_j}
\\
\qquad =\sum_{j=1}^n \frac{\overline{\psi_j(    x'   )}[\lambda_j \psi_j(x)] }{\lambda_j}
\\
\qquad =\sum_{j=1}^n   \overline{\psi_j(    x'   )}  \psi_j(x)
\\
\qquad =\delta (x-    x'   ).
\end{split}
\end{equation}

\eproof
}

{
\color{blue}
\bexample
\begin{enumerate}
\item
For a demonstration of completeness of systems of eigenfunctions,
consider, for instance, the differential equation corresponding to the harmonic vibration
[please do not confuse this with the harmonic oscillator (\ref{2012-m-ch-fa-hphoe})]
\begin{equation}
{\cal L}_t \phi (t) = \frac{d^2}{dt^2} \phi (t)= k^2  ,
\end{equation}
with $k  \in {\Bbb R}$.

Without any boundary conditions the associated eigenfunctions are
\begin{equation}
\psi_\omega (t) = e^{\pm i\omega t},
\end{equation}
with $0 \le \omega \le \infty$, and with eigenvalue $-\omega^2$.
Taking the complex conjugate $\overline{\psi_\omega (t')}$
of $\psi_\omega (t')$
and integrating the product $ \psi_\omega (t)\overline{\psi_\omega (t')}$
over $\omega$ yields
[modulo a constant factor which depends on the choice of Fourier transform parameters;
see also Equation (\ref{2011-m-eftdelta1})]
\begin{equation}
\begin{split}
\int_{-\infty}^{\infty} \overline{\psi_\omega (t')}\psi_\omega (t) d\omega \\
  =
\int_{-\infty}^{\infty} e^{ i\omega t'}  e^{-i\omega t} d\omega \\
  =
\int_{-\infty}^{\infty}  e^{-i\omega (t-t')} d\omega \\
 =
2 \pi \delta (t-t').
\end{split}
\end{equation}
The associated Green's function -- together with a prescription to circumvent the pole at the origin -- is defined by
\begin{equation}
G(t - t') = \int_{-\infty}^{\infty}  \frac{e^{\pm i\omega (t-t')}}{(-\omega^2)} d\omega
.
\end{equation}
The solution is obtained by multiplication with the constant $k^2$, and by integration
 over $t'$;
that is,
\begin{equation}
\phi (t)=\int_{-\infty}^{\infty}  G(t - t') k^2 dt'
=-\int_{-\infty}^{\infty} \left(\frac{k}{\omega}\right)^2  e^{\pm i\omega (t-t')} d\omega \;dt'
.
\end{equation}

Suppose that, additionally, we impose boundary conditions; e.g.,
$\phi (0) = \phi ( L ) =0$,
representing a string ``fastened'' at positions $0$ and $L$.
In this case
the eigenfunctions change to
\begin{equation}
\psi_n (t) = \sin ( \omega_n t)= \sin \left( \frac{n \pi }{L} t\right),
\end{equation}
with $\omega_n= \frac{n \pi }{L}$ and $n \in {\Bbb Z}$.
We can deduce orthogonality and completeness from the
orthogonality relations for sines
(\ref{2012-m-ch-orsc}).

\item

For the sake of another example suppose,
from the Euler-Bernoulli bending theory, we know (no proof is given here)
that the equation for the quasistatic bending of slender,
isotropic, homogeneous beams of constant cross-section under an applied transverse load $q(x)$
is given by
\begin{equation}
{\cal L}_x y(x)= \frac{d^4}{dx^4} y(x)=  q(x)\approx c,
\label{2011-m-eebbe}
\end{equation}
with constant $c\in {\Bbb R}$.
Let us further assume the boundary conditions
\begin{equation}
y(0)= y(L)=\frac{d^2}{dx^2}y(0)= \frac{d^2}{dx^2}y(L) = 0.
\end{equation}
Also, we require that $y$(x) vanishes everywhere except inbetween $0$ and $L$; that is, $y(x)=0$ for $x=(-\infty,0)$ and for $x=(L,\infty)$.
Then in accordance with these boundary conditions,
the system of eigenfunctions $\{ \psi_j (x) \}$
of  ${\cal L}_x $ can be written as
\begin{equation}
 \psi_j (x) = \sqrt{\frac{2}{L}} \sin \left( \frac{\pi j x}{L}  \right)
\end{equation}
for $j=1,2,\ldots$.
The associated eigenvalues
 $$\lambda_j =\left( \frac{\pi j }{L}\right)^4$$
can be verified through explicit differentiation
\begin{equation}
\begin{split}
{\cal L}_x  \psi_j (x) = {\cal L}_x  \sqrt{\frac{2}{L}} \sin \left( \frac{\pi j x}{L}  \right)
\\   =  \left( \frac{\pi j }{L}\right)^4  \sqrt{\frac{2}{L}} \sin \left( \frac{\pi j x}{L}  \right)
 =  \left( \frac{\pi j }{L}\right)^4   \psi_j (x) .
\end{split}
\end{equation}
The cosine functions which are also solutions of the
Euler-Bernoulli equations  (\ref{2011-m-eebbe}) do not vanish at the origin $x=0$.

Hence,
\begin{equation}
\begin{split}
G(x-x')  = {\frac{2}{L}} \sum_{j=1}^\infty
\frac{\sin \left( \frac{\pi j x}{L}  \right)   \sin \left( \frac{\pi j x'}{L}  \right) }
{\left( \frac{\pi j }{L}\right)^4}
\\ \qquad =  {\frac{2L^3}{\pi^4}} \sum_{j=1}^\infty
\frac{1} {j^4} \sin \left( \frac{\pi j x}{L}  \right)   \sin \left( \frac{\pi j x'}{L}  \right)
\end{split}
\end{equation}

Finally the solution can be calculated explicitly by
\begin{equation}
\begin{split}
y(x) = \int_0^L G(x-x')  g(x') dx'
\\
\approx
\int_0^L  c \left[ {\frac{2L^3}{\pi^4}} \sum_{j=1}^\infty
\frac{1} {j^4} \sin \left( \frac{\pi j x}{L}  \right)   \sin \left( \frac{\pi j x'}{L}  \right) \right] dx'
\\ =
{\frac{2c L^3}{\pi^4}}
\sum_{j=1}^\infty
\frac{1} {j^4} \sin \left( \frac{\pi j x}{L}  \right)   \left[ \int_0^L    \sin \left( \frac{\pi j x'}{L}  \right)  dx'  \right]
\\ =
{\frac{4c L^4}{\pi^5}}
\sum_{j=1}^\infty
\frac{1} {j^5} \sin \left( \frac{\pi j x}{L}  \right)   \sin^2 \left( \frac{\pi j }{2}  \right)
\end{split}
\end{equation}

\end{enumerate}

\eexample
}

\section{Finding Green's functions by Fourier analysis}
\index{Fourier analysis}

If one is dealing with translation invariant systems of the form
\begin{equation}
\begin{split}
{\cal L}_x y(x)=   f(x) \textrm{, with the differential operator}\\
{\cal L}_x =
a_n  \frac{d^n}{dx^n} +
a_{n-1}  \frac{d^{n-1}}{dx^{n-1}} +
\ldots
+
a_1  \frac{d}{dx}
+
a_0   \\
\qquad = \sum_{j=0}^n  a_j \frac{d^j}{dx^j} ,
\end{split}
\label{2011-m-egfpti}
\end{equation}
with constant coefficients $a_j$,
then one can apply the following strategy using Fourier analysis to obtain the Green's function.

First,  recall that, by Equation~(\ref{2011-m-eftdelta}) on page~\pageref{2011-m-eftdelta}
the Fourier transform  $ \widetilde{\delta}(k) $ of the delta function $ \delta (x) $,
as defined by the conventions  $A=B=1$  in Equation~(\ref{2011-m-efta1mg}),
\marginnote{$A$ and $B$ refer to Equation~(\ref{2011-m-efta1mg}) on page~\pageref{2011-m-efta1mg}.}
is just a constant~$1$.
Therefore, $\delta$ can be written as
\begin{equation}
\delta (x-x')=
\frac{1}{2\pi}
\int_{-\infty}^\infty  e^{i{k(x-x')}} dk
\end{equation}

Next, consider the Fourier transform of the Green's function
\begin{equation}
 \widetilde{G}(k)=
 \int_{-\infty}^\infty  G(x) e^{-i{kx}} dx
\end{equation}
and its inverse transform
\begin{equation}
G(x)=
\frac{1}{2\pi}
\int_{-\infty}^\infty  \widetilde{G}(k) e^{i{kx}} dk
.
\label{2011-m-egfptift}
\end{equation}

Insertion of Equation (\ref{2011-m-egfptift})
into   the {\it Ansatz}
${\cal L}_x G(x-x') = \delta (x-x')$
yields
\begin{equation}
\begin{split}
{\cal L}_x G(x)
 =
{\cal L}_x
\frac{1}{2\pi}
\int_{-\infty}^\infty  \widetilde{G}(k) e^{i{kx}} dk
 =
\frac{1}{2\pi}
\int_{-\infty}^\infty  \widetilde{G}(k) \left({\cal L}_x   e^{i{kx}}\right) dk  \\
= \delta (x) =
\frac{1}{2\pi}
\int_{-\infty}^\infty  e^{i{kx}} dk
.
\end{split}
\label{2011-m-egfptift1}
\end{equation}
and thus,
if ${\cal L}_x   e^{i{kx}} = {\cal P}(k)   e^{i{kx}}$, where ${\cal P}(k) $ is a polynomial in $k$,
\begin{equation}
\frac{1}{2\pi}
\int_{-\infty}^\infty
\left[\widetilde{G}(k) {\cal P}(k)    -1 \right] e^{i{kx}}
dk
=
0.
\end{equation}
Therefore,  the bracketed part of the integral kernel needs to vanish;
\marginnote{Note that
$
\int_{-\infty}^{\infty} f(x)\cos (kx) dk
=
-i \int_{-\infty}^{\infty} f(x)\sin (kx) dk
$
cannot be satisfied for arbitrary $x$ unless $f(x)=0$.
}
and we obtain
\begin{equation}
\begin{split}
\widetilde{G}(k) {\cal P}(k)    -1 \equiv  0 \textrm{, or }
\widetilde{G}(k) \equiv \text{``}\left( {\cal L}_k \right)^{-1}\text{''},
\end{split}
\label{2011-m-egfptift2}
\end{equation}
where ${\cal L}_k$ is obtained from ${\cal L}_x$
by substituting every derivative $\frac{d}{dx}$ in the latter
by $ik$ in the former.
As a result, the Fourier transform   is obtained through $\widetilde{G}(k)   = 1/{\cal P}(k)$; that is, as one divided by a polynomial ${\cal P}(k)$ of degree $n$,
the same degree as the highest order of derivative in ${\cal L}_x$.

In order to obtain the Green's function $G(x)$,
and to be able to integrate over it with the inhomogeneous term $f(x)$,
we have to Fourier transform $\widetilde{G}(k)$ back to ${G}(x)$.
This often turns out the most difficult part of the computation.
It can be mastered with the help of Jordan's Lemma, as outlined in Section~\ref{2021-mm-ch-ca-jl}
 \index{Jordan's lemma}
and the residue theorem, as outlined in Section~\ref{2021-mm-ch-ca-rt}.
 \index{residue theorem}

Then we have to make sure that the solution obeys the initial conditions,
and,  if necessary, we have to add solutions of the homogeneous equation
${\cal L}_x G(x-x')=0$. That is all.

\section{Advanced, retarded, causal and anti-causal Green's functions}
\index{advanced Green's function}
\index{retarded Green's function}
\index{causal Green's function}
\index{anti-causal Green's function}

 \begin{marginfigure}
        \begin{center}
        \begin{tabular}{cc}
\resizebox{.50\textwidth}{!}{
\begin{tikzpicture}  [scale=0.3]

\tikzstyle{every path}=[line width=2pt]


\draw[draw=gray!80,->] (0,-5) + (0,-0.5cm)  -- (0,5) -- +(0,2cm) node[above right] {$\Im k$};
\draw[draw=gray!80,->] (-5,0) +(-0.5cm,0) -- (5,0) -- +(1cm,0) node[below right] {$\Re k$};

\draw[orange,->] (-5,0.8) -- (5,0.8);
\draw[blue,dashed,->] (-5,0.8) -- (4.8,0.8);


\draw [color=orange,->] (5,0.8) arc [start angle=0,end angle=180,x radius=5cm, y radius=5cm];
\node[orange] at (5.8,5.1) {$x > 0$};
\draw [color=blue,dashed,->] ( 5,0.8) arc [start angle=360,end angle=180,x radius=5cm, y radius=5cm];
\node[blue] at (5,-4.2) {$x  < 0$};


\end{tikzpicture}
}
&
\resizebox{.50\textwidth}{!}{
\begin{tikzpicture}  [scale=0.3]

\tikzstyle{every path}=[line width=2pt]


\draw[draw=gray!80,->] (0,-5.8) + (0,-0.5cm)  -- (0,5) -- +(0,0.5cm) node[above right] {$\Im k$};
\draw[draw=gray!80,->] (-5,0) +(-0.5cm,0) -- (5,0) -- +(1cm,0) node[below right] {$\Re k$};

\draw[orange,->] (-5,-0.8) -- (5,-0.8);
\draw[blue,dashed,->] (-5,-0.8) -- (4.8,-0.8);


\draw [color=orange,->] (5,-0.8) arc [start angle=0,end angle=180,x radius=5cm, y radius=5cm];
\node[orange] at (5,4.2) {$x > 0$};
\draw [color=blue,dashed,->] ( 5,-0.8) arc [start angle=360,end angle=180,x radius=5cm, y radius=5cm];
\node[blue] at (5.8,-5.1) {$x  < 0$};


\end{tikzpicture}
}
\\
retarded&advanced\\
\resizebox{.50\textwidth}{!}{
\begin{tikzpicture}  [scale=0.3]

\tikzstyle{every path}=[line width=2pt]


\draw[draw=gray!80,->] (0,-5) + (0,-0.5cm)  -- (0,5) -- +(0,1cm) node[above right] {$\Im k$};
\draw[draw=gray!80,->] (-5,0) +(-0.5cm,0) -- (5,0) -- +(1cm,0) node[below right] {$\Re k$};

\draw[orange,->] (-5,-0.8) -- (-0.8,-0.8) -- (0.8,0.8) -- (5,0.8);
\draw[blue,dashed,->] (-5,-0.8) -- (-0.8,-0.8) -- (0.8,0.8) -- (4.8,0.8);


\draw [color=orange,->] (0,0) ++(10:5) arc  (10:190:5);
\node[orange] at (5.8,5.1) {$x > 0$};
\draw [color=blue,dashed,->] (0,0) ++(10:5) arc  (370:190:5);
\node[blue] at (5,-5.2) {$x  < 0$};


\end{tikzpicture}
}
&
\resizebox{.50\textwidth}{!}{
\begin{tikzpicture}  [scale=0.3]

\tikzstyle{every path}=[line width=2pt]


\draw[draw=gray!80,->] (0,-5) + (0,-0.5cm)  -- (0,5) -- +(0,1cm) node[above right] {$\Im k$};
\draw[draw=gray!80,->] (-5,0) +(-0.5cm,0) -- (5,0) -- +(1cm,0) node[below right] {$\Re k$};

\draw[orange,->] (-5,0.8) -- (-0.8,0.8) -- (0.8,-0.8) -- (5,-0.8);
\draw[blue,dashed,->] (-5,0.8) -- (-0.8,0.8) -- (0.8,-0.8) -- (4.8,-0.8);


\draw [color=orange,->] (0,0) ++(-10:5) arc  (-10:170:5);
\node[orange] at (5.8,5.1) {$x > 0$};
\draw [color=blue,dashed,->] (0,0) ++(-10:5) arc  (350:170:5);
\node[blue] at (5,-5.2) {$x  < 0$};


\end{tikzpicture}
}
\\
causal&anti-causal
\end{tabular}
        \end{center}
        \caption{\label{2021-mm-ch-gf-types-r-a-c-ac}
                Four types of Green's functions; their differences being solutions of the homogeneous differential equation.
}
        \end{marginfigure}

Note that if one solves the Fourier integration by analytic continuation into the $k$-plane, different integration paths
lead to special solutions.
There are four types of Green's functions defined by the pathes across the real axis $\Re k$
in the complex $k$-plane, as drawn in Figure~\ref{2021-mm-ch-gf-types-r-a-c-ac}:
\begin{itemize}
\item[(i)] the path of the retarded Green's function along the real $k$-axis  is ``slightly shifted'' by a ``very small'' positive imaginary number $i\varepsilon$, with $\varepsilon \in \mathbb{R}$ and $\varepsilon \ll 1$;
\item[(ii)] the path of the advanced Green's function along the real $k$-axis is ``slightly shifted'' by a ``very small'' negative imaginary number $-i\varepsilon$, with $\varepsilon \in \mathbb{R}$ and $\varepsilon \ll 1$;
\item[(iii)] the path of the causal Green's function along the real $k$-axis
is ``slightly shifted'' by a ``very small'' negative imaginary number $-i\varepsilon$ until $k=0$,
and then is ``slightly shifted'' by a ``very small'' positive imaginary number $i\varepsilon$;
\item[(iv)] the path of the anti-causal Green's function along the real $k$-axis
is ``slightly shifted'' by a ``very small'' positive imaginary number $i\varepsilon$ until $k=0$,
and then is ``slightly shifted'' by a ``very small'' negative imaginary number $-i\varepsilon$.
\end{itemize}

The resulting Green's functions all yield solutions of the same inhomogeneous differential equation.
Therefore, they differ only by some particular solutions of the respective homogeneous differential equation.

{
\color{blue}
\bexample

Let us consider a few examples for this procedure.

\begin{enumerate}

\item
First, let us solve the differential equation $y' -y=t$
on the interval $[0,\infty )$ with the boundary conditions $y(0)=0$.

We observe that the associated differential operator is given by
$$
 {\cal L}_t  = \frac{d}{dt} -1,
$$
and the inhomogeneous term can be identified with $f(t)=t$.

We use the {\it Ansatz} $G_1(t,t')={1\over2\pi}\int\limits_{-\infty}^{+\infty}
\tilde G_1(k)e^{ik(t-t')}dk$; hence
\begin{equation}
\begin{split}
   {\cal L}_t G_1(t,t')={1\over2\pi}\int\limits_{-\infty}^{+\infty}
      \tilde G_1(k)\underbrace{\left({d\over dt}-1\right)e^{ik(t-t')}}_
      {\mbox{$=(ik-1)e^{ik(t-t')}$}}dk \\
   =\delta(t-t')={1\over2\pi}\int\limits_{-\infty}^{+\infty}e^{ik(t-t')}dk
\end{split}
\end{equation}
Now compare the kernels of the Fourier integrals of ${\cal L}_tG_1$ and $\delta$:
\begin{equation}
\begin{split}
   \tilde G_1(k)(ik-1)=1\Longrightarrow \tilde G_1(k)={1\over ik-1}
   ={1\over i(k+i)}
\\
   G_1(t,t')={1\over2\pi}\int\limits_{-\infty}^{+\infty}
   {e^{ik(t-t')}\over i(k+i)}dk
\end{split}
\label{2018-m-ch-gf-ip1}
\end{equation}
This integral can be evaluated by analytic continuation of the kernel to the imaginary $k$-plane,
by ``closing'' the integral contour ``far above the origin,''
and
by using the Cauchy integral and residue theorems of complex analysis.
The paths in the upper and lower integration plane are drawn in Fig.~\ref{2011-m-ch-gf-fe}.
{\color{black}
\begin{marginfigure}
\begin{center}
\begin{tikzpicture}  [scale=0.4]

\tikzstyle{every path}=[line width=2pt]


\draw[draw=gray!80,->] (0,-5) + (0,-0.5cm)  -- (0,5) -- +(0,1cm) node[above right] {$\Im k$};
\draw[draw=gray!80,->] (-5,0) +(-0.5cm,0) -- (5,0) -- +(1cm,0) node[below right] {$\Re k$};

\draw[orange,->] (-5,0) -- (5,0);
\draw[blue,dashed,->] (-5,0) -- (4.8,0);


\draw [color=orange,->] (5,0) arc [start angle=0,end angle=180,x radius=5cm, y radius=5cm];
\node[orange] at (5,4.5) {$t-t'> 0$};
\draw [color=blue,dashed,->] ( 5,0) arc [start angle=360,end angle=180,x radius=5cm, y radius=5cm];
\node[blue] at (5,-4.5) {$t-t' < 0$};

\filldraw [color=blue] (0,-2) circle (3pt) node[above right] {$-i$};

\end{tikzpicture}
\caption{Plot of the two paths reqired for solving the Fourier integral~(\ref{2018-m-ch-gf-ip1}).}
\label{2011-m-ch-gf-fe}
\end{center}
\end{marginfigure}
}
Note that, by  Jordan's Lemma, as outlined in Section~\ref{2021-mm-ch-ca-jl},
 \index{Jordan's lemma}
the line integral ``closures'' through the respective half-circle paths---the Jordan arcs---vanish.
\index{Jordan arc}
Thereby, the only nonzero contribution to the ``upper'' and ``lower'' contour integrals
comes from the integral along the entire real $k$-axis.
On the other hand we can apply
the residue theorem, as outlined in Section~\ref{2021-mm-ch-ca-rt},  \index{residue theorem} which yields
\begin{equation}
G_1(t,t') =
\begin{cases}
0
&
\textrm{ for } t>t'\\
-2\pi i\,{\rm Res}\,\left({1\over2 \pi i} {e^{ik(t-t')}\over k+i };-i\right)= -e^{t-t'}
&
\textrm{ for } t<t'.
\end{cases}
\end{equation}
Hence we obtain a Green's function for the inhomogeneous differential equation
$$
G_1(t,t')=-H (t'-t)e^{t-t'}
$$
However, this Green's function and its associated (special) solution does
not obey the boundary conditions
$G_1(0,t')=-H (t')e^{-t'}\ne0$ for $t'\in[0,\infty)$.

Therefore, we have to fit the Green's function by adding an appropriately weighted solution to the homogeneous differential equation.
The homogeneous Green's function is found by
$
   {\cal L}_tG_0(t,t')=0$,
and thus, in particular,
${d\over dt}
   G_0=G_0\Longrightarrow G_0=ae^{t-t'}
$.
with the {\em Ansatz}
$$
G(0,t')=G_1(0,t')+G_0(0,t';a)=-H (t')e^{-t'}+ae^{-t'}  $$
for the general solution we can choose the constant coefficient $a$ so that
$$
G(0,t')=0
$$
For $a=1$,
 the Green's function and thus the solution obeys the boundary value conditions;
that is,
$$
G(t,t')=\bigl[1-H (t'-t)\bigr]    e^{t-t'}.
$$
Since $H (-x)=1-H (x)$, $G(t,t')$ can be rewritten as
$$
   G(t,t')=H (t-t')e^{t-t'}.
$$

In the final step we obtain the solution through integration of $G$ over the inhomogeneous term  $t$:
\begin{equation}
\begin{split}
   y(t)=
\int_0^\infty G(t,t')t'dt'=
\int_0^\infty \underbrace{H (t-t')}_{=1\text{ for }t'<t} e^{t-t'}t'dt'
       =\int\limits_0^t e^{t-t'}t'dt'\\=e^t\int\limits_0^t t'e^{-t'}dt'
       =e^t\left(-t'e^{-t'}\Bigr|_0^t-\int\limits_0^t(-e^{-t'})dt'
          \right) \\
       =e^t\left[(-te^{-t})-e^{-t'}\Bigr|_0^t\right]=e^t\left(
          -te^{-t}-e^{-t}+1\right)=e^t-1-t.
\end{split}
\end{equation}

It is prudent to check whether this is indeed a solution of the differential equation
satisfying the boundary conditions:
\begin{equation}
\begin{split}
{\cal L}_t  y(t)=  \left(\frac{d}{dt} -1\right) \left(e^t-1-t\right)
=e^t-1 - \left(e^t-1-t\right) =t,\\
 \textrm{and }y(0)= e^0-1-0=0
.
\end{split}
\end{equation}


\item
Next, let us solve the differential equation
 ${d^2y\over dt^2}+y=\cos t$
on the intervall $t\in [0,\infty )$ with the boundary conditions  $y(0)=y'
 (0)=0$.

First, observe that
$
{\cal L} = {d^2 \over dt^2}+1.
$
The Fourier {\em Ansatz} for the Green's function is
\begin{equation}
\begin{split}
   G_1(t,t')={1\over2\pi}\int\limits_{-\infty}^{+\infty}
               \tilde G(k)e^{ik(t-t')}dk\\
   {\cal L}G_1={1\over2\pi}\int\limits_{-\infty}^{+\infty}
          \tilde G(k)\left({d^2\over dt^2}+1\right)e^{ik(t-t')}dk\\
       ={1\over2\pi}\int\limits_{-\infty}^{+\infty}\tilde G(k)((ik)^2+1)
          e^{ik(t-t')}dk\\
       =\delta(t-t')={1\over2\pi}\int\limits_{-\infty}^{+\infty}
          e^{ik(t-t')}dk
\end{split}
\end{equation}
Hence
$ \tilde G(k)(1-k^2)=1$
and thus
$
\tilde G(k)= \frac{1}{1-k^2}= -\frac{1}{(k+1)(k-1)}
$.
The Fourier transformation is
\begin{equation}
\begin{split}
    G_1(t,t')=-{1\over2\pi}\int\limits_{-\infty}^{+\infty}
                    {e^{ik(t-t')}\over(k+1)(k-1)}dk\\
                 =-{1\over 2\pi}2\pi i\left[\,{\rm Res}\left(
                    {e^{ik(t-t')}\over(k+1)(k-1)};k=1\right)\right.\\
\left.+{\rm Res}\left({e^{ik(t-t')}\over(k+1)(k-1)};k=-1\right)\right]
H (t-t')
\end{split}
\end{equation}
Note that, by  Jordan's Lemma, as outlined in Section~\ref{2021-mm-ch-ca-jl},
 \index{Jordan's lemma}
the line integral ``closures'' through the respective half-circle paths vanish.
Thereby, the only nonzero contribution to the ``upper'' and ``lower'' contour integrals
comes from the integral along the entire real $k$-axis.
On the other hand we can apply
the residue theorem, as outlined in Section~\ref{2021-mm-ch-ca-rt}.  \index{residue theorem}
The path in the upper  integration plain, corresponding to the advanced Green's function, is drawn in Fig. \ref{2011-m-ch-gf-fe2}.
\index{advanced Green's function}
Any other integration path---retarded, causal or anti-causal---would also be good; but in this case the solution would differ
by a particular solution of the inhomogenuous differential equation, corresponding to different boundary values.
(Here we take the ``most convenient one'' from a hindsight perspective.)
{\color{black}
\begin{marginfigure}
\begin{center}
\begin{tikzpicture}  [scale=0.4]

\tikzstyle{every path}=[line width=2pt]


\draw[draw=gray!80,->] (0,-5) + (0,-0.5cm)  -- (0,5) -- +(0,1cm) node[above right] {$\Im k$};
\draw[draw=gray!80,->] (-5,0) +(-0.5cm,0) -- (5,0) -- +(1cm,0) node[below right] {$\Re k$};

\draw[orange,->] (-5,0) -- (5,0);
\draw[blue,dashed,->] (-5,0) -- (4.8,0);


\draw [color=orange,->] (5,0) arc [start angle=0,end angle=180,x radius=5cm, y radius=5cm];
\node[orange] at (5,4.5) {$t-t'> 0$};
\draw [color=blue,dashed,->] ( 5,0) arc [start angle=360,end angle=180,x radius=5cm, y radius=5cm];
\node[blue] at (5,-4.5) {$t-t' < 0$};

\draw[line width=1pt,blue,dotted,->] (-2,0) -- (-2,1);
\draw[line width=1pt,blue,dotted,->] ( 2,0) -- ( 2,1);
\filldraw [color=blue] (-2,1.2) circle (3pt) node[above] {$-1+i\varepsilon$};
\filldraw [color=blue] ( 2,1.2) circle (3pt) node[above] {$ 1+i\varepsilon$};

\end{tikzpicture}
\end{center}
\caption{Plot of the   path  required for solving the Fourier integral, with the pole description of ``pushed up`` poles.}
\label{2011-m-ch-gf-fe2}
\end{marginfigure}
}
\begin{equation}
\begin{split}
   G_1(t,t')=-{i\over2}\left(e^{i(t-t')}-e^{-i(t-t')}\right)H (t-t')\\
            ={e^{i(t-t')}-e^{-i(t-t')}\over2i}H (t-t')=
               \sin(t-t')H (t-t')\\
   G_1(0,t')=\sin(-t')H (-t')=0\textrm{  since  }\quad t'>0\\
   G_1'(t,t')=\cos(t-t')H (t-t')+\underbrace{\sin(t-t')\delta(t-t')}_
                {\mbox{$=0$}}\\
   G_1'(0,t')=\cos(-t')H (-t')=0
.
\end{split}
\end{equation}
$G_1$ already satisfies the boundary conditions; hence we do not need to find the Green's function $G_0$ of the homogeneous equation.
\begin{equation}
\begin{split}
   y(t)=\int\limits_0^\infty G(t,t')f(t')dt'=
          \int\limits_0^\infty \sin(t-t')\underbrace{H  (t-t')}_
          {\mbox{$=1$ for $t>t'$}}\cos t' dt' \\
       =\int\limits_0^t\sin(t-t')\cos t' dt'=
          \int\limits_0^t(\sin t\cos t'-\cos t\sin t')\cos t' dt' \\
       =\int\limits_0^t\bigl[\sin t(\cos t')^2-\cos t\sin t'\cos t'
          \bigr]dt'=\\
       =\sin t\int\limits_0^t(\cos t')^2dt'-\cos t\int\limits_0^t
          sin t'\cos t'dt' \\
       =\sin t\left.\left[{1\over2}(t'+\sin t'\cos t')\right]\right|_0^t-
          \cos t\left.\left[{\sin^2 t'\over2}\right]\right|_0^t \\
       ={t\sin t\over2}+{\sin^2 t\cos t\over 2}-{\cos t\sin^2 t\over2}=
          {t\sin t\over2}.
\end{split}
\end{equation}

Again it is prudent to check whether this is indeed a solution of the differential equation
with
\begin{equation}
\begin{split}
y'(t) = \frac{1}{2}
\left(
\sin t + t \cos t
\right),
\\
y''(t) = \frac{1}{2}
\left(
\cos t +  \cos t - t \sin t
\right)
=
\cos t - \frac{1}{2} t \sin t
,
\\
\text{and thus}\\
y''(t) + y(t) =
\cos t - \frac{1}{2} t \sin t
+
\frac{1}{2} t \sin t
=
\cos t
,
\end{split}
\end{equation}
satisfying the boundary conditions
$y(0)= \frac{1}{2} 0 \sin 0 =0$
and
$y'(0) = \frac{1}{2}
\left(
\sin 0 + 0 \cos 0
\right)=0$.

\end{enumerate}

\eexample
}

\begin{center}
{\color{olive}   \Huge
\decofourright
}
\end{center}

\chapter{Sturm-Liouville theory}
\label{2011-m-ch-sl}

This is only a very brief ``dive into Sturm-Liouville theory,''
which has many fascinating aspects and connections to Fourier analysis, the
special functions of mathematical physics, operator theory, and linear algebra.\cite{birkhoff-Rota-48,Al-Gwaiz,everitt-handbook-sl}
In physics, many formalizations involve second order linear ordinary differential equations (ODEs),
\marginnote{Here the term
{\em ordinary} -- in contrast with  {\em partial} --
is used to indicate that its terms and solutions just depend on a single one independent variable.
Typical examples of a partial differential equations are the Laplace or the wave equation in three spatial dimensions.}
\index{ordinary differential equation}
\index{ODE}
which, in their most general form, can be written as\cite{herman-sc}
\begin{equation}
{\cal L}_x y(x) =
a_0(x) y(x)  +
a_1(x) \frac{d}{dx}y(x)+
a_2(x) \frac{d^2}{dx^2}y(x) =
f(x).
\label{2011-m-ch-sl1}
\end{equation}
The differential operator associated with this differential equation is defined by
\begin{equation}
{\cal L}_x  = a_0(x)
+
a_1(x) \frac{d}{dx}+
a_2(x) \frac{d^2}{dx^2}.
\label{2011-m-ch-sl1do}
\end{equation}
The solutions $y(x)$ are often subject to boundary conditions of various forms:
\begin{itemize}
\item
{\em Dirichlet boundary conditions}
\index{Dirichlet boundary conditions}
are of the form $y(a)=y(b)=0$ for some $a,b$.

\item
(Carl Gottfried) {\em  Neumann boundary conditions}
\index{Neumann boundary conditions}
are of the form $y'(a)=y'(b)=0$ for some $a,b$.

\item
{\em Periodic boundary conditions}
\index{periodic boundary conditions}
are of the form $y(a)=y(b)$ and $y'(a)=y'(b)$ for some $a,b$.
\end{itemize}

\section{Sturm-Liouville form}

Any second order differential equation of the general form (\ref{2011-m-ch-sl1})
can be rewritten into a differential equation of the {\em Sturm-Liouville form}
\index{Sturm-Liouville form}
\begin{equation}
\begin{split}
{\cal S}_{x} y(x) =
\frac{d}{dx}
\left[
p(x)
\frac{d}{dx}
\right]  y(x)
+
q(x) y(x)
=
F(x), \\ \qquad
\textrm{with }  p(x)=e^{\int \frac{a_1(x)}{a_2(x)} dx},\\  \qquad
q(x)=p(x) \frac{a_0(x)}{a_2(x)}=  \frac{a_0(x)}{a_2(x)}e^{\int \frac{a_1(x)}{a_2(x)} dx},\\   \qquad
F(x)=p(x) \frac{f(x)}{a_2(x)}= \frac{f(x)}{a_2(x)} e^{\int \frac{a_1(x)}{a_2(x)} dx}
\end{split}
\label{2011-m-ch-sl2}
\end{equation}
The associated differential operator
\begin{equation}
\begin{split}
{\cal S}_{x}  =
\frac{d}{dx}
\left[
p(x)
\frac{d}{dx}
\right]
+
q(x)     \\
\qquad =
p(x)\frac{d^2}{dx^2}
+
p'(x)
\frac{d}{dx}
+
q(x)
\end{split}
\label{2011-m-ch-sl3}
\end{equation}
is called
{\em Sturm-Liouville differential operator}.
\index{Sturm-Liouville differential operator}
It is very special: compared to the general form
(\ref{2011-m-ch-sl1}) the transformation (\ref{2011-m-ch-sl2})
yields
\begin{equation}
a_1(x) = a_2'(x)
.
\label{2018-mm-ch-sl3b}
\end{equation}

{\color{OliveGreen}
\bproof
For a proof, we insert $p(x)$, $q(x)$ and $F(x)$
into the Sturm-Liouville form of Equation (\ref{2011-m-ch-sl2}) and compare it with
Equation (\ref{2011-m-ch-sl1}).

\begin{equation}
\begin{split}
\left\{  \frac{d}{dx}
\left[
e^{\int \frac{a_1(x)}{a_2(x)} dx}
\frac{d}{dx}
\right]
+
\frac{a_0(x)}{a_2(x)}e^{\int \frac{a_1(x)}{a_2(x)} dx} \right\} y(x)
=
\frac{f(x)}{a_2(x)}e^{\int \frac{a_1(x)}{a_2(x)} dx}\\
e^{\int \frac{a_1(x)}{a_2(x)} dx} \left\{
\frac{d^2}{dx^2} +
\frac{a_1(x)}{a_2(x)}
\frac{d}{dx}
+
\frac{a_0(x)}{a_2(x)}\right\} y(x)
=
\frac{f(x)}{a_2(x)}e^{\int \frac{a_1(x)}{a_2(x)} dx}\\
\left\{
a_2(x)\frac{d^2}{dx^2} +
a_1(x)\frac{d}{dx}
+
a_0(x)\right\} y(x)
=
f(x).
\end{split}
\label{2011-m-ch-sl21}
\end{equation}

\eproof
}

\section{Adjoint and self-adjoint operators}
\index{adjoint  operator}

In operator theory,
just as in matrix theory,
we can define an
{\em adjoint operator}
(for finite dimensional Hilbert space, see Section~\ref{2014-m-fdvs-adjoint} on page \pageref{2014-m-fdvs-adjoint})
{\it via} the scalar product
defined in Equation (\ref{2011-m-ch-slspbtf}).
In this formalization,
the Sturm-Liouville differential operator ${\cal S}$
is self-adjoint.

Let us first define the
{\em domain}  of a differential operator ${\cal L}$ as the set of all  square integrable
(with respect to the weight  $\rho  (x)$)
\index{domain}
functions $\varphi$ satisfying boundary conditions.
\begin{equation}
\int_a^b \vert \varphi (x) \vert^2 \rho  (x) dx < \infty
.
\end{equation}

Then, the adjoint operator  ${\cal L}^\dagger$ is defined by  satisfying
\begin{equation}
\begin{split}
\langle \psi \mid {\cal L} \varphi \rangle
=
\int_a^b
\psi (x) [{\cal L}\varphi (x)]
\rho  (x)         dx
\\
\quad =\langle {\cal L}^\dagger \psi \mid \varphi \rangle
=
\int_a^b
[{\cal L}^\dagger \psi (x)] \varphi (x)
\rho  (x)         dx
\end{split}
\label{2011-m-ch-slspbtfao}
\end{equation}
for all $\psi (x)$ in the domain of ${\cal L}^\dagger$ and $\varphi (x)$ in the domain of ${\cal L}$.

Note that  in the case of second order differential operators
in the standard form (\ref{2011-m-ch-sl1do})  and with $\rho  (x) = 1$,
we can move the differential quotients and the entire differential operator in
\begin{equation}
\begin{split}
\langle \psi \mid {\cal L} \varphi \rangle
=
\int_a^b
\psi (x) [{\cal L}_x\varphi (x)]
\rho  (x)         dx   \\
\qquad =
\int_a^b
\psi (x)
[a_2(x) \varphi''(x) + a_1(x) \varphi'(x) + a_0(x) \varphi (x)]
dx
\end{split}
\label{2011-m-ch-slspbtfao1a}
\end{equation}
from
$\varphi$ to $\psi$
by one and two partial integrations.

Integrating the  kernel $a_1(x) \varphi'(x)$ by parts yields
\begin{equation}
\int_a^b
\psi (x)
  a_1(x) \varphi'(x)
dx
=
\left.
\psi (x)
  a_1(x) \varphi(x)
\right|_a^b -  \int_a^b
(\psi (x)
  a_1(x) )'\varphi(x)
dx
.
\end{equation}

Integrating the  kernel $a_2(x) \varphi''(x)$ by parts twice yields
\begin{equation}
\begin{split}
\int_a^b
\psi (x)
  a_2(x) \varphi''(x)
dx
=
\left.
\psi (x)
  a_2(x) \varphi'(x)
\right|_a^b -  \int_a^b
(\psi (x)
  a_2(x) )'\varphi'(x)
dx
\\
\qquad =
\left.
\psi (x)
  a_2(x) \varphi'(x)
\right|_a^b -
\left.
(\psi (x)
  a_2(x))' \varphi(x)
\right|_a^b
+
  \int_a^b
(\psi (x)
  a_2(x) )''\varphi(x)
dx    \\
\qquad =
\left.
\psi (x)
  a_2(x) \varphi'(x)
 -
(\psi (x)
  a_2(x))' \varphi(x)
\right|_a^b
+
  \int_a^b
(\psi (x)
  a_2(x) )''\varphi(x)
dx
.
\end{split}
\end{equation}
Combining these two calculations yields
\begin{equation}
\begin{split}
\langle \psi \mid {\cal L} \varphi \rangle
=
\int_a^b
\psi (x) [{\cal L}_x\varphi (x)]
\rho  (x)         dx   \\
\qquad =
\int_a^b
\psi (x)
[a_2(x) \varphi''(x) + a_1(x) \varphi'(x) + a_0(x) \varphi (x)]
dx      \\
\qquad =
\left.
\psi (x)
  a_1(x) \varphi(x)
+
\psi (x)
  a_2(x) \varphi'(x)
 -
(\psi (x)
  a_2(x))' \varphi(x)
\right|_a^b   \\
\qquad
\qquad
+
  \int_a^b
[( a_2(x)\psi (x) )'' - ( a_1(x)\psi (x)  )'    + a_0(x) \psi (x)] \varphi(x)
dx
.
\end{split}
\label{2011-m-ch-slspbtfao1aaa}
\end{equation}
If the sum of the``surface'' terms vanish for some reason
-- such that, for instance, because of boundary conditions on
$\psi $,  $\varphi$, $\psi'$, $\varphi'$ or other conditions
like
$\psi (a)= \varphi (a) =0$  or
$\psi' (a) = \varphi' (a)=0$
or $\psi (b)= \varphi (b) =0$  or
$\psi' (b) = \varphi' (b)=0$
 and
$a_1(x)=a_2'(x)$ in the case of the Sturm-Liouville
operator ${\cal S}_x$ -- then
\begin{equation}
\begin{split}
\left.
\psi (x)
  a_1(x) \varphi(x)
+
\psi (x)
  a_2(x) \varphi'(x)
 -
(\psi (x)
  a_2(x))' \varphi(x)
\right|_a^b
\\=
\left.
\psi (x) a_1(x) \varphi(x)
+
\psi (x)
  a_2(x) \varphi'(x)
 -
 \psi' (x)
  a_2(x)  \varphi(x)
-  \psi (x)
  a_2'(x)  \varphi(x)
\right|_a^b
\\=
\psi (x)
\left[
  a_1(x) - a_2'(x)
\right]
\varphi(x)
+ a_2(x)
\left[
\psi (x)
  \varphi'(x)
 -
 \psi' (x)
   \varphi(x)
\right]
\Big|_a^b
=0.
\label{2016-m-ch-sl-bc}
\end{split}
\end{equation}
Therefore, Equation (\ref{2011-m-ch-slspbtfao1aaa}) reduces to, and
Equation (\ref{2011-m-ch-slspbtfao}) results in,
\begin{equation}
\begin{split}
  \int_a^b
[( a_2(x)\psi (x) )'' - ( a_1(x)\psi (x)  )'    + a_0(x) \psi (x)] \varphi(x)
dx  \\
=  \langle  \psi \mid {\cal L}_x \varphi \rangle
=  \langle {\cal L}_x^\dagger \psi \mid  \varphi \rangle
,
\label{2011-m-ch-slspbtfao1aaa1}
\end{split}
\end{equation}
and we can identify the adjoint differential operator of ${\cal L}_x$ with
\begin{equation}
\begin{split}
{\cal L}_x^\dagger
=\frac{d^2}{dx^2}  a_2(x)  - \frac{d }{dx } a_1(x)    + a_0(x) \\
\qquad =
\frac{d}{dx} \left[ a_2(x) \frac{d}{dx} + a'_2(x)\right]  - a'_1(x) - a_1(x)\frac{d }{dx }    + a_0(x) \\
\qquad =
 a'_2(x) \frac{d}{dx} + a_2(x) \frac{d^2}{dx^2} + a''_2(x) + a'_2(x)\frac{d}{dx}
  - a'_1(x) - a_1(x)\frac{d }{dx }    + a_0(x)  \\
\qquad =
 \underbrace{a_2(x)}_{\tilde{a}_2} \frac{d^2}{dx^2}
+
\underbrace{[2 a'_2(x)- a_1(x) ]}_{\tilde{a}_1}  \frac{d}{dx}
+ \underbrace{a''_2(x)   - a'_1(x)   + a_0(x)}_{\tilde{a}_0}
.
\end{split}
\label{2011-m-ch-slspbtfaob}
\end{equation}

The operator ${\cal L}_x$ is called self-adjoint if
\index{self-adjoint transformation}
\begin{equation}
{\cal L}_x^\dagger
=  {\cal L}_x \;
,
\label{2011-m-ch-slspbtfaob1}
\end{equation}
that is, if
$a_1 = \tilde{a}_1$,
$a_2 = \tilde{a}_2$, and
$a_3 = \tilde{a}_3$.

Next we shall show that, in particular, the Sturm-Liouville differential operator~(\ref{2011-m-ch-sl3}) is self-adjoint,
and that all second order differential operators [with the boundary condition~(\ref{2016-m-ch-sl-bc})] which are  self-adjoint
are of the Sturm-Liouville form.

{\color{OliveGreen}
\bproof

In order to prove that the  Sturm-Liouville differential operator
\begin{equation}
\begin{split}
{\cal S}
=       \frac{d}{dx}
\left[
p(x)
\frac{d}{dx}
\right]
+
q(x)
= p(x) \frac{d^2}{dx^2}
+p'(x) \frac{d}{dx}
+q(x)
\end{split}
\label{2011-m-ch-slspbtfaoc}
\end{equation}
from Equation~(\ref{2011-m-ch-sl3})
is self-adjoint, we verify
Equation~(\ref{2011-m-ch-slspbtfaob1})
with
${\cal S}^\dagger$
taken from
Equation~(\ref{2011-m-ch-slspbtfaob}).
Thereby, we identify
$a_2(x) = p(x)$,
$a_1(x) = p'(x)$,
and
$a_0(x) = q(x)$; hence

\begin{equation}
\begin{split}
{\cal S}_x^\dagger
=a_2(x) \frac{d^2}{dx^2}
+
[2 a'_2(x)- a_1(x) ]  \frac{d}{dx}
+ a''_2(x)   - a'_1(x)   + a_0(x) \\
  =
p(x)  \frac{d^2}{dx^2}
+ [2 p'(x)-p'(x)] \frac{d}{dx}
+p''(x)-p''(x)+q(x) \\
  =
p(x)  \frac{d^2}{dx^2}
+ p'(x) \frac{d}{dx}
+q(x)  = {\cal S}_x
.
\end{split}
\end{equation}

Alternatively we could argue from Eqs. (\ref{2011-m-ch-slspbtfaob}) and (\ref{2011-m-ch-slspbtfaob1}),
noting that
a differential operator is self-adjoint if and only if
\begin{equation}
\begin{split}
 {\cal L}_x
=a_2(x) \frac{d^2}{dx^2}   +  a_1(x)\frac{d }{dx }    + a_0(x) \\
=   {\cal L}_x^\dagger =
 a_2(x) \frac{d^2}{dx^2}
+
[2 a'_2(x)- a_1(x) ]  \frac{d}{dx}
+ a''_2(x)   - a'_1(x)   + a_0(x)
.
\end{split}
\end{equation}
By comparison of the coefficients,
\begin{equation}
\begin{split}
a_2(x)=a_2(x),\\
a_1(x) =
2 a'_2(x)- a_1(x)   ,\\
a_0(x) =
 a''_2(x)   - a'_1(x)   + a_0(x)
,
\end{split}
\end{equation}
and hence,
\begin{equation}
 a'_2(x)= a_1(x),
\end{equation}
which is exactly the form of the   Sturm-Liouville differential operator.

\eproof
}

\section{Sturm-Liouville eigenvalue problem}
\index{Sturm-Liouville eigenvalue problem}

The Sturm-Liouville eigenvalue problem is given by the differential equation
\marginnote{The minus sign ``$-\lambda $'' is here for purely convential reasons; to make the presentation compatible with other texts.}
\begin{equation}
\begin{split}
{\cal S}_{x}    \phi (x) = -\lambda \rho  (x) \phi (x)\textrm{, or } \\
\frac{d}{dx}
\left[
p(x)
\frac{d}{dx}
\right] \phi(x)
+
[q(x)  +   \lambda \rho  (x) ]\phi(x)=0
\end{split}
\label{2011-m-ch-slee}
\end{equation}
for $x\in(a,b)$ and continuous  $p'(x)$, $q(x)$ and $p(x)>0$, $\rho  (x)>0$.

It can be expected that, very similar to the spectral theory of linear algebra introduced in Section~\ref{2012-m-ch-Spectraltheorem}
on page~\pageref{2012-m-ch-Spectraltheorem}, self-adjoint operators have a spectral decomposition
involving real, ordered eigenvalues and complete sets of mutually orthogonal operators.
We mention without proof (for proofs, see, for instance, Ref.\cite{Al-Gwaiz}) that
we can formulate a spectral theorem as follows
\index{spectral theorem}
\begin{itemize}
\item
the eigenvalues $\lambda$ turn out to be real, countable, and ordered, and that there is a smallest eigenvalue $\lambda_1$
such that $\lambda_1<\lambda_2<\lambda_3< \cdots$;

\item
for each eigenvalue $\lambda_j$ there exists an eigenfunction
$\phi_j(x)$ with $j-1$ zeroes on $(a,b)$;

\item
eigenfunctions corresponding to different eigenvalues are {\em orthogonal}, and can be normalized, with respect
to the weight function \index{weight function}
$\rho  (x)$; that is,
\begin{equation}
\langle \phi_j \mid \phi_k \rangle
=
\int_a^b
\phi_j (x)\phi_k(x)
\rho  (x)         dx
= \delta_{jk}
\label{2011-m-ch-slonef}
\end{equation}

\item
the set of eigenfunctions   is {\em complete}; that is, any piecewise smooth function can be represented by
\begin{equation}
\begin{split}
f(x)=\sum_{k=1}^\infty c_k\phi_k(x), \\
\textrm{with }     \\
c_k=\frac{ \langle f \mid \phi_k\rangle }  { \langle \phi_k \mid \phi_k\rangle }= \langle f \mid \phi_k\rangle .
\end{split}
\label{2011-m-ch-sleecom}
\end{equation}

\item
the orthonormal (with respect to the weight $\rho $) set $\{\phi_j(x)\mid j\in {\Bbb N}\}$
is a {\em basis} of a Hilbert space with the inner product
\begin{equation}
\langle f \mid g\rangle
=
\int_a^b
f (x) g(x)
\rho  (x)         dx
.
\label{2011-m-ch-slspbtf}
\end{equation}

\end{itemize}

\section{Sturm-Liouville transformation into Liouville normal form}
 Let, for $x\in [a,b]$,
\begin{equation}
\begin{split}
[{\cal S}_x  +\lambda \rho (x)] y(x)=0,\\
\frac{d}{dx}\left[ p(x) \frac{d}{dx}\right] y(x) + [q(x) +\lambda \rho (x)]y(x)=0,\\
\left[p(x)\frac{d^2 }{dx^2} + p'(x) \frac{d}{dx} + q(x) +\lambda \rho (x)\right] y(x)=0,\\
\left[\frac{d^2 }{dx^2} + \frac{p'(x)}{p(x)} \frac{d}{dx} + \frac{q(x) +\lambda \rho (x)}{p(x)}\right] y(x)=0
\end{split}
\label{2011-m-ch-slspbtfaod}
\end{equation}
be a second order differential equation of the
Sturm-Liouville form.\cite{birkhoff-Rota-48}

This equation (\ref{2011-m-ch-slspbtfaod}) can be written in the
{\em Liouville normal form}
\index{Liouville normal form} containing no first order differentiation term
\begin{equation}
-\frac{d^2}{dt^2} w(t) + [\hat{q}(t) -\lambda ] w(t)=0  \textrm{, with }t\in [ t(a)  , t(b)] .
\label{2011-m-ch-slspbtfaolnf}
\end{equation}
It is obtained {\em via} the
{\em Sturm-Liouville transformation}
\index{Sturm-Liouville transformation} 
\begin{equation}
\begin{split}
\xi= t(x) =   \int_a^x \sqrt{\frac{\rho(s)}{p(s)}}  ds,   \\
w(t)= \sqrt[4]{p(x(t))\rho(x(t))} y (x(t)),
\end{split}
\end{equation}
where
\begin{equation}
\hat{q}(t)= \frac{1}{\rho }\left[-q -\sqrt[4]{p\rho }
\left(p\left( \frac{1}{\sqrt[4]{p\rho }}\right)'\right)'\right].
\end{equation}
The apostrophe represents derivation with respect to $x$.

{
\color{blue}
\bexample

For the sake of an example, suppose we want to know the normalized eigenfunctions of
\begin{equation}
x^2y'' + 3xy' + y =- \lambda y \textrm{, with } x\in [1,2]
\end{equation}
with the boundary conditions $y(1) = y(2) =0$.

The first thing we have to do is to  transform this differential equation
into its Sturm-Liouville form by identifying $a_2(x)=x^2$,
$a_1(x)=   3x$, $a_0 =1$, $\rho = 1$ such that $f (x)= - \lambda y (x)$; and hence
\begin{equation}
\begin{split}
p(x)=e^{\int \frac{ 3x }{ x^2 } dx}=e^{\int \frac{3}{x} dx}=e^{3\log{x}}=x^3,\\
q(x)=p(x) \frac{ 1 }{ x^2 }= x,\\
F(x)=p(x) \frac{\lambda y}{(-x^2)}= -\lambda x y\textrm{, and hence } \rho(x)= x
.
\end{split}
\end{equation}
As a result we obtain the  Sturm-Liouville form
\begin{equation}
{1\over x}((x^3y')' + xy)=-\lambda y .
\end{equation}

In the next step we apply the Sturm-Liouville transformation
\begin{equation}
\begin{split}
\xi= t (x)  =\int\sqrt{\rho (x)\over p(x)}dx=\int{dx\over x}=\log x,\\
w(t(x))=\sqrt[4]{p(x(t))\rho(x(t))} y (x(t))= \sqrt[4]{x^4} y (x(t))= x y,\\
\hat{q}(t)= \frac{1}{x}\left[-x -\sqrt[4]{x^4 }
\left(x^3\left( \frac{1}{\sqrt[4]{x^4 }}\right)'\right)'\right] =0
.
\end{split}
\end{equation}
We now take the {Ansatz} $y={1\over x} w ( t (x))={1\over x} w (\log x)$
and finally obtain  the Liouville normal form
\begin{equation}
- w '' (\xi)=\lambda  w (\xi) .
\end{equation}

As an {\em Ansatz}
for solving the Liouville normal form we use
\begin{equation}   w (\xi)=a\sin(\sqrt{\lambda}\xi)+b\cos(\sqrt{\lambda}\xi)
\end{equation}

The boundary conditions translate into $x=1\rightarrow\xi=0$,
and $x=2\rightarrow \xi=\log 2$.
From
$w (0)=0$ we obtain $b=0$.
From
$w (\log 2)=a\sin(\sqrt{\lambda}\log2)=0$
we obtain
$\sqrt{\lambda_n}\log2=n\pi$.

Thus the eigenvalues are
\begin{equation}
\lambda_n=
\left({n\pi\over\log2}\right)^2.
\end{equation}

The associated eigenfunctions are
\begin{equation}
w _n(\xi)=a\sin\left({n\pi\over\log2}
\xi\right),
\end{equation}
and thus
\begin{equation}
y_n={1\over x}a\sin\left({n\pi\over\log2}
\log x\right).
\end{equation}

We can check that they are orthonormal by inserting into Equation (\ref{2011-m-ch-slonef})
and verifying it; that is,
\begin{equation}
\int\limits_1^2 \rho (x)y_n(x)
y_m(x)dx=\delta_{nm};
\end{equation}
more explicitly,
\begin{equation}
\begin{split}
\int\limits_1^2 dx x\left({1\over x^2}\right)a^2
   \sin\left(n\pi{\log x\over\log2}\right)\sin\left(m\pi
   {\log x\over\log2}\right)\\
 \textrm{\huge [} \textrm{variable substitution } u={\log x\over\log2}\\
    {du\over dx}=
{1\over\log2}{1\over x},\; du={dx\over x\log2} \textrm{\huge ]}\\
  =
\int\limits_{u=0}^{u=1}
  du\log2a^2\sin(n\pi u)\sin(m\pi u)\\
  =
\underbrace{a^2\left({\log2\over2}\right)}_{\mbox{$=1$}}\,
\underbrace{2 \int_0^1du \sin(n\pi u)\sin(m\pi
u)}_{\mbox{$=\delta_{nm}$}}
 = \delta_{nm}.
\end{split}
\end{equation}

Finally, with $a=\sqrt{2\over\log 2}$
we obtain the solution
\begin{equation}
y_n=\sqrt{2\over\log2}{1\over x}
\sin\left(n\pi{\log x\over\log2}\right).
\end{equation}

\eexample
}

\section{Varieties of Sturm-Liouville differential equations}

A catalogue of Sturm-Liouville differential equations
comprises the following {\it species}, among many others.\cite[-15mm]{arfken05,Al-Gwaiz,everitt-handbook-sl}
Some of these cases are
tabelated as functions   $p$, $q$, $\lambda$ and $\rho$ appearing in the general form
of the Sturm-Liouville eigenvalue problem  (\ref{2011-m-ch-slee})
\begin{equation}
\begin{split}
{\cal S}_{x}    \phi (x) = -\lambda \rho  (x) \phi (x)\textrm{, or } \\
\frac{d}{dx}
\left[
p(x)
\frac{d}{dx}
\right] \phi(x)
+
[q(x)  +   \lambda \rho  (x) ]\phi(x)=0
\end{split}
\end{equation}
in Table \ref{2011-m-sl-t-varieties}.
\begin{table}
{\footnotesize
\begin{tabular}{lccccccccc}
\hline\hline
Equation & $ p(x)$ & $q(x)$ & $-\lambda$ & $\rho (x)$\\
\hline
Hypergeometric  & $x^{\alpha+1}(1-x)^{\beta +1} $ &   $0$ &  $ \mu $ &  $x^{\alpha}(1-x)^{\beta }$
\\
Legendre  & $1-x^2 $ &   $0$ &  $l(l+1) $ &  $1$
\\
Shifted Legendre    & $ x(1-x)$ &   $0$ &  $l(l+1) $ &  $1$
\\
Associated Legendre    & $1-x^2 $ &   $-\frac{m^2}{1-x^2}$ &  $l(l+1) $ &  $1$
\\
Chebyshev I   & $\sqrt{1-x^2} $ &   $0$ &  $n^2 $ &  $ \frac{1}{\sqrt{1-x^2}}$
\\
Shifted Chebyshev I   & $\sqrt{x(1-x)} $ &   $0$ &  $n^2 $ &  $ \frac{1}{\sqrt{x(1-x)}}$
\\
Chebyshev II   & $(1-x^2)^\frac{3}{2} $ &   $0$ &  $n(n+2) $ &  $  \sqrt{1-x^2} $
\\
Ultraspherical (Gegenbauer)   & $(1-x^2)^{\alpha + \frac{1}{2}} $ &   $0$ &  $n(n+2\alpha ) $ &  $  (1-x^2)^{\alpha - \frac{1}{2}}$
\\
Bessel   & $ x$ &   $-\frac{n^2}{x}$ &  $a^2 $ &  $ x$
\\
Laguerre   & $x e^{-x} $ &   $0$ &  $\alpha $ &  $e^{-x} $
\\
Associated Laguerre     & $x^{k+1} e^{-x} $ &   $0$ &  $\alpha -k$ &  $x^ke^{-x} $
\\
Hermite     & $x e^{-x^2} $ &   $0$ &  $2\alpha $ &  $e^{-x} $
\\
Fourier    & $1 $ &   $0$ &  $ k^2 $ &  $1$
\\
(harmonic oscillator)    &   &     &  $  $ &  $ $  \\
Schr\"odinger    & $1 $ &   $l(l+1)x^{-2}$ &  $ \mu $ &  $1$
\\
(hydrogen atom)    &   &     &  $  $ &  $ $  \\
\hline\hline
\end{tabular}
}
\caption{Some varieties of differential equations expressible as Sturm-Liouville differential equations}
\index{Hermite polynomial}
\index{Laguerre polynomial}
\index{Gegenbauer polynomial}
\index{Chebyshev polynomial}
\index{Legendre polynomial}
\label{2011-m-sl-t-varieties}
\end{table}

\begin{center}
{\color{olive}   \Huge
\decoone 
}
\end{center}

\chapter{Separation of variables}
\label{2011-m-ch-sv}

This chapter deals with the ancient alchemic suspicion of {\it ``solve et coagula''} that it is possible
to solve a problem by splitting it up into partial problems, solving these issues separately; and
consecutively joining together the partial solutions, thereby yielding the full answer to the problem
\marginnote{For a counterexample see the Kochen-Specker theorem on page~\pageref{2011-m-KST}.}
-- translated into the context of {\em partial differential equations;}
that is, equations with derivatives of more than one variable.
\index{partial differential equation}
Thereby, solving the separate partial problems is not dissimilar to
applying subprograms from some program library.

Already Descartes mentioned this sort of method in his
{\it Discours de la m{\'e}thode pour bien conduire sa raison et chercher la verit{\'e} dans les sciences}
(English translation:
{\em Discourse on the Method of Rightly Conducting One's Reason and of Seeking Truth})\cite{Descartes-Discourse}
stating that (in a newer translation\cite{Descartes-CW1})
\begin{quote}
{\em
[Rule Five:]
The whole method consists entirely in the ordering and arranging of the
objects on which we must concentrate our mind's eye if we are to
discover some truth. We shall be following this method exactly if we first
reduce complicated and obscure propositions step by step to simpler
ones, and then, starting with the intuition of the simplest ones of all, try
to ascend through the same steps to a knowledge of all the rest.
$\ldots$
[Rule Thirteen:]
If we perfectly understand a problem we must abstract it from every
superfluous conception, reduce it to its simplest terms and, by means of
an enumeration, divide it up into the smallest possible parts.
}
\end{quote}

The method of
separation of variables
is one among a couple of strategies to solve differential equations,\cite[-20mm]{Evans98,jaenich-an}
and it is a very important one in physics.

Separation of variables can be applied whenever we have no ``mixtures of derivatives and functional dependencies;''
more specifically,
whenever the  partial differential equation can be written as a sum
\begin{equation}
\begin{array}  {l}
{\cal L}_{x,y} \psi(x,y) = ({\cal L}_{x} + {\cal L}_y)\psi (x,y) =  0\textrm{, or}  \\
{\cal L}_{x} \psi (x,y) =  - {\cal L}_y \psi (x,y).
\end{array}
\label{2011-m-ch-sveq}
\end{equation}
Because in this case we may make an {\it ad hoc} {\em multiplicative}\sidenote[][-17mm]{Another possibility is an {\em additive} composition of the solution; cf.~\bibentry{Cherniavsky}.
}
{\it Ansatz}
\begin{equation}
\psi (x,y)= v(x)u(y) .
\label{2011-m-ch-sva}
\end{equation}
Inserting (\ref{2011-m-ch-sva}) into (\ref{2011-m-ch-sv}) effectively  separates the variable dependencies
\begin{equation}
\begin{array}  {l}
{\cal L}_{x} v(x)u(y) =   - {\cal L}_y v(x)u(y), \\
u(y)\left[ {\cal L}_{x} v(x)\right]
 =   -  v(x)\left[{\cal L}_y u(y)\right]
, \\
\frac{1}{v(x)}{\cal L}_{x} v(x) =   - \frac{1}{u(y)}{\cal L}_y u(y)=a,
\end{array}
\label{2011-m-ch-sv1}
\end{equation}
with constant $a$, because
$\frac{{\cal L}_{x} v(x)}{v(x)}$  does not depend on $x$,
and  $\frac{{\cal L}_y u(y)}{u(y)}$  does not depend on $y$.
Therefore,
neither side depends on $x$ or $y$; hence both sides are constants.

As a result, we can treat and integrate both sides separately; that is,
\begin{equation}
\begin{array}  {l}
\frac{1}{v(x)} {\cal L}_{x} v(x)= a,\\
\frac{1}{u(y)}{\cal L}_y u(y)=-a
,
\end{array}
\label{2011-m-ch-sv2}
\end{equation}
or
\begin{equation}
\begin{array}  {l}
{\cal L}_{x} v(x)- av(x)=0,\\
{\cal L}_y u(y)+a u(y) = 0
.
\end{array}
\label{2011-m-ch-sv2or}
\end{equation}

This separation of variable {\it Ansatz}
can be often used when the
{\em Laplace operator}
\index{Laplace operator}
$\Delta=\nabla  \cdot \nabla$
is involved, since there the partial derivatives with respect to different variables
occur in different summands.

The general solution
\marginnote{If we would just consider a single product of all general one parameter solutions we would run into the same problem as in the
entangled case on page \pageref{2012-m-ch-fdvs-dectp-gftp-fr} -- we could not cover all the solutions of the original equation.}
is a linear combination (superposition) of the products of all the linear independent solutions --
that is, the sum of the products of all separate (linear independent) solutions, weighted by an arbitrary scalar factor.

{
\color{blue}
\bexample

For the sake of demonstration, let us consider a few examples.

\begin{enumerate}
\item
Let us separate the homogeneous Laplace differential equation
\begin{equation}
\Delta \Phi =\frac{1}{u^2+v^2}
  \left(
    \frac{\partial^2\Phi}{\partial u^2}+
    \frac{\partial^2\Phi}{\partial v^2}
  \right)+
  \frac{\partial^2\Phi}{\partial z^2}
= 0
\label{2018-m-ch-sv-ldecc}
\end{equation}
in parabolic
cylinder coordinates $(u,v,z)$ with
${\bf x} = \left({1 \over 2} (u^2 - v^2), uv, z\right) $.

The separation of variables {\it Ansatz} is
\begin{equation}
\Phi(u,v,z)=\Phi_1(u)\Phi_2(v)\Phi_3(z).
\label{2018-m-ch-sv-ansatz}
\end{equation}
Inserting (\ref{2018-m-ch-sv-ansatz}) into (\ref{2018-m-ch-sv-ldecc}) and
division by $\Phi=\Phi_1\Phi_2\Phi_3$---that is, multiplication with $\frac{1}{\Phi_1 \Phi_2 \Phi_3 }$---yields

\begin{equation}
\begin{split}
  \frac{1}{u^2+v^2}
  \left(
    \Phi_2\Phi_3\frac{\partial^2\Phi_1}{\partial u^2}+
    \Phi_1\Phi_3\frac{\partial^2\Phi_2}{\partial v^2}
  \right)+
  \Phi_1\Phi_2\frac{\partial^2\Phi_3}{\partial z^2}=0
\\
  \frac{1}{u^2+v^2}
  \left(
    \Phi_2\Phi_3\frac{\partial^2\Phi_1}{\partial u^2}+
    \Phi_1\Phi_3\frac{\partial^2\Phi_2}{\partial v^2}
  \right)  =  -
  \Phi_1\Phi_2\frac{\partial^2\Phi_3}{\partial z^2}\\
\left[\text{multiplied with }
 \frac{1}{\Phi_1 \Phi_2 \Phi_3 }
\right]
\\
 \frac{1}{u^2+v^2}
  \left(
    \frac{\Phi_1''}{\Phi_1}+
    \frac{\Phi_2''}{\Phi_2}
  \right)=
  -\frac{\Phi_3''}{\Phi_3}=\lambda=\mbox{const.}
\end{split}
\end{equation}
$\lambda$ is constant because it does neither depend on $u,v$ [because of the right hand side
$ {\Phi_3'' (z)/\Phi_3 (z)}$],
nor on $z$ (because of the left hand side).
Furthermore,
$$
  \frac{\Phi_1''}{\Phi_1}- \lambda u^2 =
-  \frac{\Phi_2''}{\Phi_2}+   \lambda v^2=l^2=\mbox{const.}
$$
with constant $l$ for analogous reasons.
The three resulting differential equations are
\begin{eqnarray*}
  \Phi_1''-(\lambda u^2+l^2) \Phi_1 & = & 0, \\
  \Phi_2''-(\lambda v^2-l^2) \Phi_2 & = & 0, \\
  \Phi_3''+\lambda\Phi_3 & = & 0.
\end{eqnarray*}

\item
Let us separate the homogeneous
(i)  Laplace,
(ii) wave,
 and
(iii) diffusion    equations,
in
elliptic cylinder coordinates $(u,v,z)$ with
$\vec x = \left( a \cosh u \cos v, a \sinh u \sin v, z\right)$ and
\begin{eqnarray*}
  \Delta & = &  \frac{1}{a^2(\sinh^2u+\sin^2v)}
    \left[
      \frac{\partial^2}{\partial u^2}+
      \frac{\partial^2}{\partial v^2}
    \right]+\frac{\partial^2}{\partial z^2}.
\end{eqnarray*}

\end{enumerate}

\subsection*{ad (i):}
Again the separation of variables {\it Ansatz} is $\Phi(u,v,z)=\Phi_1(u)\Phi_2(v)\Phi_3(z)$.
Hence,
\begin{equation}
\begin{split}
  \frac{1}{a^2(\sinh^2u+\sin^2v)}
  \left(
    \Phi_2\Phi_3\frac{\partial^2\Phi_1}{\partial u^2}+
    \Phi_1\Phi_3\frac{\partial^2\Phi_2}{\partial v^2}
  \right)
  =-\Phi_1\Phi_2\frac{\partial^2\Phi_3}{\partial z^2},
\\
  \frac{1}{a^2(\sinh^2u+\sin^2v)}
  \left(
    \frac{\Phi_1''}{\Phi_1}+
    \frac{\Phi_2''}{\Phi_2}
  \right)=
  -\frac{\Phi_3''}{\Phi_3}=k^2=\mbox{const.}
  \Longrightarrow \Phi_3''+k^2\Phi_3=0
\\
  \frac{\Phi_1''}{\Phi_1}+
  \frac{\Phi_2''}{\Phi_2}=k^2a^2(\sinh^2u+\sin^2v),
\\
  \frac{\Phi_1''}{\Phi_1}-k^2a^2\sinh^2u=
  -\frac{\Phi_2''}{\Phi_2}+k^2a^2\sin^2v=l^2,
\end{split}
\end{equation}
and finally,
$$
  \begin{array}{rcccl}
    \Phi_1'' & - & (k^2a^2\sinh^2u+l^2)\Phi_1 & = & 0, \\
    \Phi_2'' & - & (k^2a^2\sin^2v-l^2)\Phi_2 & = & 0.
  \end{array}
$$

\subsection*{ad (ii):}
the wave equation is given by
$$
  \Delta\Phi=\frac{1}{c^2}\frac{\partial^2 \Phi}{\partial t^2}.
$$
Hence,
$$
  \frac{1}{a^2(\sinh^2u+\sin^2v)}
  \left(
    \frac{\partial^2}{\partial u^2}+\frac{\partial^2}{\partial v^2}
  \right)\Phi+
  \frac{\partial^2 \Phi}{\partial z^2}=
  \frac{1}{c^2}\frac{\partial^2 \Phi}{\partial t^2}.
$$
The separation of variables {\it Ansatz} is  $\Phi (u,v,z,t)=\Phi_1(u)\Phi_2(v)\Phi_3(z)T(t)$
\begin{equation}
\begin{split}
  \Longrightarrow
  \frac{1}{a^2(\sinh^2u+\sin^2v)}
  \left(
    \frac{\Phi_1''}{\Phi_1}+
    \frac{\Phi_2''}{\Phi_2}
  \right)+
  \frac{\Phi_3''}{\Phi_3}=\frac{1}{c^2}\frac{T''}{T}=-\omega^2=\mbox{const.},
\\
  \frac{1}{c^2}\frac{T''}{T}=-\omega^2 \Longrightarrow T''+c^2\omega^2T=0,
\\
  \frac{1}{a^2(\sinh^2u+\sin^2v)}
  \left(
    \frac{\Phi_1''}{\Phi_1}+
    \frac{\Phi_2''}{\Phi_2}
  \right)=
  -\frac{\Phi_3''}{\Phi_3}-\omega^2=k^2,
\\
  \Phi_3''+(\omega^2+k^2)\Phi_3=0
\\
  \frac{\Phi_1''}{\Phi_1}+
  \frac{\Phi_2''}{\Phi_2}=k^2a^2(\sinh^2u+\sin^2v)
\\
  \frac{\Phi_1''}{\Phi_1}-a^2k^2\sinh^2u=
  -\frac{\Phi_2''}{\Phi_2}+a^2k^2\sin^2v=l^2,
\end{split}
\end{equation}
and finally,
\begin{equation}
  \begin{split}
    \Phi_1''   -   (k^2a^2\sinh^2u+l^2)\Phi_1   =   0, \\
    \Phi_2''   -   (k^2a^2\sin^2v-l^2)\Phi_2   =   0.
  \end{split}
\end{equation}

\subsection*{ad (iii):}
The diffusion equation is
$\Delta\Phi=\frac{1}{D}\frac{\partial \Phi}{\partial t}$.

The separation of variables {\it Ansatz} is  $\Phi(u,v,z,t)=\Phi_1(u)\Phi_2(v)\Phi_3(z)T(t)$.
Let us take the result of (i), then
\begin{equation}
\begin{split}
  \frac{1}{a^2(\sinh^2u+\sin^2v)}
  \left(
    \frac{\Phi_1''}{\Phi_1}+
    \frac{\Phi_2''}{\Phi_2}
  \right)+\frac{\Phi_3''}{\Phi_3}=\frac{1}{D}\frac{T'}{T}=
    -\alpha^2=\mbox{const.}
\\
  T=Ae^{-\alpha^2Dt}
\\
  \Phi_3''+(\alpha^2+k^2)\Phi_3=0
  \Longrightarrow
  \Phi_3''=-(\alpha^2+k^2)\Phi_3
  \Longrightarrow
  \Phi_3=Be^{i\sqrt{\alpha^2+k^2} \, z}
\end{split}
\end{equation}
and finally,
\begin{equation}
  \begin{split}
    \Phi_1''   -   (\alpha^2k^2\sinh^2u+l^2)\Phi_1   =   0 \\
    \Phi_2''   -   (\alpha^2k^2\sin^2v-l^2)\Phi_2   =   0
.
  \end{split}
\end{equation}

\eexample
}

\begin{center}
{\color{olive}   \Huge
\leafleft
}
\end{center}

\chapter{Special functions of mathematical physics}
\label{2011-m-ch-sf}

Special functions\marginnote{This chapter follows several
approaches:~\bibentry{lebedev:1965:sft}, \bibentry{Wilf}, \bibentry{bell-specfun}, \bibentry{andrews:1999:sfu}, \bibentry{Kuznetsov} and \bibentry{Kisil}.}
\marginnote{For reference, consider~\bibentry{abramowitz:1964:hmf}, \bibentry{Brych-HBSF} and \bibentry{Gradshteyn}.}
often arise as solutions of differential equations; for instance as eigenfunctions
of differential operators in quantum mechanics.
Sometimes they occur  after several {\em separation of variables}
and substitution steps have transformed the physical problem into something manageable.
For instance, we might start out with some linear partial differential equation like the wave equation,
then separate the space from time coordinates,
then separate the radial from the angular components,
and finally, separate the two angular parameters.
After we have done that, we end up with several separate differential equations of the Liouville form;
among them the Legendre differential equation leading us to the Legendre polynomials.

In what follows, a particular class of special functions will be considered.
These functions are all special cases of the
{\em hypergeometric function},
\index{hypergeometric function}
which is the solution of the
{\em hypergeometric differential equation}.
\index{hypergeometric differential equation}
The hypergeometric function exhibits a high degree of
``plasticity,''
as many elementary analytic functions can be expressed by it.

First, as a prerequisite, let us define the gamma function.
Then we proceed to second order Fuchsian differential equations;
\index{Fuchsian equation}
followed by rewriting a Fuchsian  differential equation
into a hypergeometric differential equation.
Then we study the hypergeometric function as a solution to the
hypergeometric differential equation.
Finally, we mention some particular hypergeometric functions, such as the
Legendre orthogonal polynomials, and others.

Again, if not mentioned otherwise, we shall restrict our attention to
second order differential equations.
Sometimes -- such as for the Fuchsian class -- a generalization is possible but
not very relevant for physics.

\section{Gamma function}
\index{gamma function}

The gamma function $\Gamma (x)$ is an extension of the factorial (function)  $n!$ because
it generalizes the ``classical'' factorial, which is defined on the natural numbers,
to real or complex arguments (different from the negative integers and from zero); that is,
\begin{equation}
\Gamma (n+1) = n! \textrm { for } n \in {\Bbb N}
\textrm{, or }
\Gamma (n) = (n-1)! \textrm { for } n \in {\Bbb N}-0
.
\label{2011-m-ch-sfgamman}
\end{equation}

Let us first define the
{\em shifted factorial}
\index{shifted factorial}
or, by another naming,
the
{\em Pochhammer symbol}
\index{Pochhammer symbol}
\begin{equation}
\begin{split}
(a)_0\stackrel{{\tiny \textrm{ def }}}{=}1,
\\
(a)_n\stackrel{{\tiny \textrm{ def }}}{=}
a(a+1)\cdots (a+n-1)
,
\end{split}
\label{2011-m-ch-sfsf}
\end{equation}
where $n>0$ and $a$ can be any real or complex number.
If $a$ is a natural number greater than zero, $(a)_n=
\frac{\Gamma (a+n)}{\Gamma(a)}$.
Note that
$(a)_1=a$ and $(a)_2=a(a+1)$,
and so on.

With this definition of the shifted factorial,
\begin{equation}
\begin{split}
 z ! ( z +1)_ n
=1\cdot 2 \cdots  z  \cdot ( z +1)(( z +1)+1)\cdots (( z +1)+ n -1)  \\
\qquad
=1\cdot 2 \cdots  z  \cdot ( z +1)(  z +2)\cdots ( z + n ) \\
\qquad
= ( z + n )! ,
\\
\textrm{ or }
 z !
= \frac{( z + n )!}{  ( z +1)_ n }.
\end{split}
\label{2011-m-ch-sfsf1}
\end{equation}

Since
\begin{equation}
\begin{split}
( z + n )!
=( n + z )!     \\
\qquad
= 1\cdot 2 \cdots  n  \cdot ( n +1)( n +2)\cdots ( n + z )   \\
\qquad
=  n ! \cdot ( n +1)( n +2)\cdots ( n + z )   \\
\qquad
= n !( n +1)_ z  ,
\end{split}
\end{equation}
we can rewrite Equation~(\ref{2011-m-ch-sfsf1}) into
\begin{equation}
 z !
= \frac{ n !( n +1)_ z  }{  ( z +1)_ n }
=   \frac{ n !  n ^ z  }{  ( z +1)_ n }  \frac{( n +1)_ z  }{  n ^ z }
.
\label{2011-m-ch-sfsf2}
\end{equation}

The latter factor, for large $n$, converges as
\index{order of}
\marginnote{Again, just as on page \pageref{2018-mm-ch-di-otof}, ``$O(x)$'' means ``of the order of $x$'' or ``absolutely bound by'' in the following way:
if $g(x)$ is a positive function,  then $f(x)=O(g(x))$ implies that there exist a positive real number $m$
such that $\vert f(x) \vert < m g(x)$.}

\begin{equation}
\begin{split}
\frac{( n +1)_ z  }{  n ^ z }   =
\frac{( n +1)(( n +1)+1)\cdots (( n +1)+ z -1)}{  n ^ z }  \\
=
\underbrace{\frac{( n +1)}{n}
\frac{( n +2)}{n}
\cdots
\frac{( n +z)}{n}}_{z \text{ factors}}
  \\
=
\frac{ n ^ z  +O( n ^{ z -1})}{  n ^ z }
=
\frac{ n ^ z }{  n ^ z }  +
\frac{O( n ^{ z -1})}{ n ^ z }
=
1  + O( n ^{-1})
\stackrel{ n  \rightarrow \infty}{\longrightarrow} 1.
\end{split}
\end{equation}
In this limit, Equation~(\ref{2011-m-ch-sfsf2})  can be written as
\begin{equation}
 z !
= \lim_{ n \rightarrow \infty}  z !
=  \lim_{ n \rightarrow \infty} \frac{ n !  n ^ z  }{  ( z +1)_ n }
.
\label{2011-m-ch-sfsf3}
\end{equation}

Hence, for all $z\in {\Bbb C}$ which are not equal to a negative integer
--
that is,  $z \not\in\left\{-1,-2,\ldots\right\}$
--
we can, in analogy to the ``classical factorial,''
define   a ``factorial function shifted by one'' as
\begin{equation}
\Gamma ( z + 1)
\stackrel{{\tiny \textrm{ def }}}{=} \lim_{ n \rightarrow \infty} \frac{ n !  n ^ z  }{  ( z +1)_ n }.
\label{2011-m-ch-sfsfdga}
\end{equation}
That is,
$\Gamma (z+1)$ has been redefined to
allow  an analytic continuation of the ``classical'' factorial $z!$ for $z \in \mathbb{N}$:
in (\ref{2011-m-ch-sfsfdga}) $z$ just appears in an exponent and in the argument of a shifted factorial.

At the same time basic properties of the factorial are maintained:
because for very large $n$ and constant $z$ (i.e., $z\ll n$),
$(z+n)\approx n$, and
\begin{equation}
\begin{split}
\Gamma ( z  )
=  \lim_{ n \rightarrow \infty} \frac{ n !  n ^ { z -1}  }{  ( z )_n }
\\ \qquad
=  \lim_{ n \rightarrow \infty} \frac{ n !  n ^ { z -1}  }{  z(z+1)\cdots (z+n-1) }
\\
=
\lim_{ n \rightarrow \infty} \frac{ n !  n ^ { z -1}  }{  z(z+1)\cdots (z+n-1) }
\underbrace{\left(\frac{z+n}{z+n}\right)}_{1}
\\ \qquad
=
\lim_{ n \rightarrow \infty} \frac{ n !  n ^ { z -1} (z+n) }{  z(z+1)\cdots (z+n) }
\\
=
\frac{1}{z} \lim_{ n \rightarrow \infty} \frac{ n !  n ^ z  }{  ( z +1)_n }
=
\frac{1}{z} \Gamma ( z +1 )
.
\label{2011-m-ch-sfsfdga234}
\end{split}
\end{equation}
This implies that
\begin{equation}
\Gamma ( z + 1)
=  z
\Gamma ( z  ) .
\label{2011-m-ch-sfsfdgas1ind}
\end{equation}
Note that, since
\begin{equation}
(1)_n =1 (1+1) (1+2)\cdots (1+n-1)= n! ,
\end{equation}
Equation~(\ref{2011-m-ch-sfsfdga})  yields
\begin{equation}
\Gamma ( 1)
=  \lim_{ n \rightarrow \infty} \frac{ n !  n ^ 0  }{  (  1)_ n }
=  \lim_{ n \rightarrow \infty} \frac{ n !    }{  n !}  =1
.
\label{2011-m-ch-sfsfdgas1}
\end{equation}
By induction, Eqs.
(\ref{2011-m-ch-sfsfdgas1})
and
(\ref{2011-m-ch-sfsfdgas1ind})
yield
$\Gamma ( n +1 ) =n!$ for $n\in {\Bbb N}$.

We state without proof that, for complex numbers $z$
with positive real parts $\Re z >0 $, the gamma function $\Gamma (z)$ and similarly,
the
\index{beta function}
beta function
(\ref{2011-m-ch-sf-beta}),
can be defined by an integral
representation as the upper incomplete gamma function $\Gamma ( z, x )$
\index{incomplete gamma function}
\index{gamma function}
\begin{equation}
\begin{split}
\Gamma ( z ,x )
\stackrel{{\tiny \textrm{ def }}}{=}
\int_x^\infty t^{z-1}e^{-t}dt \text{, and }
\Gamma ( z )  \stackrel{{\tiny \textrm{ def }}}{=}  \Gamma ( z ,0 )  =
\int_0^\infty t^{z-1}e^{-t}dt
\label{2017-m-ch-sf-edgamma}
.
\end{split}
\end{equation}
{\color{OliveGreen}
\bproof
Note that Equation~(\ref{2011-m-ch-sfsfdgas1ind}) can be derived from this integral representation
of $\Gamma(z)$
by partial integration; that is [with $u=t^z$ and $v'=\exp (-t)$, respectively],
\begin{equation}
\begin{split}
\Gamma ( z+1 )
=
\int_0^\infty t^{z}e^{-t}dt
\\
=
\underbrace{\left. -t^{z}e^{-t}  \right|_0^\infty }_{=0}
-  \left[- \int_0^\infty \left( \frac{d }{dt}t^{z}\right) e^{-t} dt \right]
\\
=
\int_0^\infty z t^{z-1} e^{-t} dt
\\
=
z\int_0^\infty  t^{z-1} e^{-t} dt = z \Gamma ( z )
.
\end{split}
\end{equation}
Therefore, Equation~(\ref{2017-m-ch-sf-edgamma})
can be verified for $z\in \mathbb{N}$ by complete induction. The induction basis
$z=1$  can be directly evaluated:
\begin{equation}
\Gamma ( 1 )
=
\int_0^\infty \underbrace{t^{0}}_{=1} e^{-t}dt = \left. - e^{-t} \right|_0^\infty
= - \underbrace{e^{-\infty}}_{=0} - \underbrace{(- e^{0})}_{=-1} =1.
\end{equation}
\eproof
}

We also mention the following formul\ae:
\begin{equation}
\begin{split}
\Gamma \left( \frac{1}{2} \right)
=\int_0^\infty  \frac{1}{\sqrt{t}} e^{-t} dt \\
\text{[variable substitution: }
u=\sqrt{t}, t=u^2, dt=2u\,du\textrm{]}\\
=\int_0^\infty  \frac{1}{u} e^{-u^2} 2u \, du
=2\int_0^\infty   e^{-u^2} du
=\int_{-\infty}^\infty   e^{-u^2} du
=\sqrt{\pi } ,
\end{split}
\end{equation}
where the Gaussian integral~(\ref{2018-m-ch-di-gi}) on page~\pageref{2018-m-ch-di-gi}  has been used.
Furthermore, more generally, without proof\marginnote{See also Exercise~8.1.17, p.~509 of~\bibentry{arfken05}}
\begin{equation}
\Gamma \left( \frac{n}{2}\right)=\sqrt{\pi }\frac{( n-2)!!}{2^{(n-1)/2}} \textrm{, for  odd } n= 2k-1, k\in \mathbb{N}; \textrm {  and }
\end{equation}
\begin{equation}
\text{Euler's reflection formula }\; \Gamma ( x)\Gamma ( 1-x) =\frac{\pi}{\sin (\pi x)} .
\end{equation}

Here, the
{\em  double factorial}
\index{double factorial} is defined by
\begin{equation}
n!!=
\begin{cases}
1   & \text{if } n=-1,0,  \\
2\cdot 4\cdots (n-2)\cdot n&\text{for even }   n= 2k ,  k\in \mathbb{N} \\
1\cdot 3 \cdots (n-2) \cdot  n &\text{for odd }  n= 2k-1 ,  k\in \mathbb{N} .
\end{cases}
\end{equation}

Note that the even and odd cases can be respectively rewritten as
\begin{equation}
\begin{split}
\text{for even }   n= 2k ,  k\ge 1:
n!!= 2\cdot 4\cdots (n-2)\cdot n\\
[n= 2k, k\in\mathbb{N}] =  (2k)!!
  =  \prod_{i=1}^k (2i)= 2^{k} \prod_{i=1}^k i\\
  =  2^{k}\cdot  1\cdot 2\cdots (k-1)\cdot k = 2^k\, k!
\\
\text{for odd }  n= 2k-1 ,  k\ge 1:
n!!= 1\cdot 3 \cdots (n-2) \cdot  n \\
[n= 2k-1, k\in\mathbb{N}] =(2k-1)!!
 =  \prod_{i=1}^k (2i-1) \\
  = 1\cdot 3 \cdots  (2k-1)  \underbrace{\frac{(2k)!!}{(2k)!!}}_{=1}    \\
  = \frac{1\cdot 2 \cdots (2k-2) \cdot  (2k-1)\cdot  (2k)}{(2k)!!}   =  \frac{(2k)!}{2^k \,k!}\\
  = \frac{k! (k+1)(k+2) \cdots [(k+1)+k-2][(k+1)+k-1] }{2^k \,k!}
  =  \frac{(k+1)_k }{2^k }
\end{split}
\end{equation}


Stirling's formula\cite{Namias-1986}
\index{Stirling's formula}
[again, $O(x)$ means ``of the order of $x$'']
\begin{equation}
\begin{split}
\log n! = n \log n -n + O(\log (n))\textrm{, or }   \\
n! \stackrel{n\rightarrow \infty}{\longrightarrow} \sqrt{2\pi n} \left(\frac{n}{e}\right)^n
\textrm{, or, more generally, }\\
\Gamma (x) = \sqrt{\frac{2\pi }{x}}\left(\frac{x}{e}\right)^x \left( 1+ O\left(\frac{1}{x}\right) \right)
\end{split}
\end{equation}
is stated without proof.

\section{Beta function}
\index{beta function}
The  {\em beta function,}
also called the  {\em Euler integral of the first kind,} is a special function defined by
\index{Euler integral}
\begin{equation}
B(x,y)=\int_0^1 t^{x-1}(1-t)^{y-1} dt =\frac{\Gamma (x) \Gamma (y)}{\Gamma (x+y)} \textrm{ for } \Re x, \Re y >0
\label{2011-m-ch-sf-beta}
\end{equation}
No proof of the identity of the two representations in terms of an integral, and of $\Gamma$-functions is given.

\section{Fuchsian differential equations}
\index{Fuchsian equation}

Many differential equations of theoretical physics are Fuchsian equations.
We shall, therefore, study this class in some generality.

\subsection{Regular,  regular singular, and irregular singular point}
Consider the homogeneous differential equation   [Equation~(\ref{2011-m-ch-sl1}) on page \pageref{2011-m-ch-sl1} is inhomogeneous]
\begin{equation}
{\cal L}_x y(x) =   a_2(x) \frac{d^2}{dx^2}y(x)  +  a_1(x) \frac{d}{dx}y(x)+   a_0(x) y(x)
 =
0.
\label{2011-m-ch-sf-fc1}
\end{equation}
If $a_0(x)$, $a_1(x)$ and $a_2(x)$ are analytic at some point $x_0$ and in its neighborhood,
and if $a_2(x_0)\neq 0$
at $x_0$, then
 $x_0$
is called an {\em ordinary point}, or {\em regular point}.
\index{ordinary point}
\index{regular point}
We  state without proof that in this case the solutions around $x_0$ can be expanded as power series.
In this case we can  divide equation (\ref{2011-m-ch-sf-fc1}) by $a_2(x)$ and rewrite it
\begin{equation}
\frac{1}{a_2(x)}{\cal L}_x y(x) =   \frac{d^2}{dx^2}y(x) +   p_1  (x)    \frac{d}{dx}y(x)+    p_2  (x)   y(x)
 =
0,
\label{2011-m-ch-sf-fc2}
\end{equation}
with
$  p_1  (x)  =    a_1(x) / a_2(x)
$
and
$ p_2 (x) =     a_0(x) / a_2(x)
$.

If, however, $a_2(x_0)= 0$ and $a_1(x_0)$ or $a_0(x_0)$ are nonzero, then the $x_0$ is called
{\em singular point}
\index{singular point} of (\ref{2011-m-ch-sf-fc1}).
In the simplest case $a_2(x)$ has a {\em simple zero} at $x_0$:
then both $  p_1  (x)  $ and $  p_2  (x)  $
in (\ref{2011-m-ch-sf-fc2})
have at most simple poles.

Furthermore, for reasons disclosed later
-- mainly motivated by the possibility to write the solutions as power series --
a point $x_0$ is called a
{\em regular singular point}
\index{regular singular point}
of Equation~(\ref{2011-m-ch-sf-fc1})
if
\begin{equation}
\begin{split}
\lim_{x\rightarrow x_0} \left[ (x-x_0)\frac{ a_1(x)}{a_2(x)} \right]
=
\lim_{x\rightarrow x_0} \left[ (x-x_0) p_1(x) \right]
\textrm{, as well as }\\
\lim_{x\rightarrow x_0} \left[ (x-x_0)^2 \frac{a_0(x)}{a_2(x)} \right]
=
\lim_{x\rightarrow x_0} \left[ (x-x_0)^2 p_2(x) \right]
\end{split}
\end{equation}
both exist.
If anyone of these limits does not exist, the singular point is
an
{\em irregular singular point}.
\index{irregular singular point}

A linear ordinary differential equation is called {\em Fuchsian,}
or {\em Fuchsian differential equation}
\index{Fuchsian equation}
generalizable to arbitrary order $n$ of differentiation
\begin{equation}
\left[ \frac{d^n}{dx^n} + p_1  (x) \frac{d^{n-1}}{dx^{n-1}} + \cdots  + p_{n-1}  (x)  \frac{d}{dx}+    p_n  (x) \right]
y(x)
 =
0,
\label{2011-m-ch-sf-fc2gc}
\end{equation}
if every singular point, including infinity,
is regular, meaning that  $p_k (x)$ has at most poles of order $k$.

A  very important case is a Fuchsian of the second order
(up to second derivatives occur).
In this case,
we suppose that the coefficients in (\ref{2011-m-ch-sf-fc2})
satisfy the following conditions:
\begin{itemize}
\item
$p_1  (x)$ has at most {\em single poles}, and
\item
$p_2 (x)$ has at most  {\em double poles}.
\end{itemize}

The simplest realization of this case is for
$
a_2(x)= a(x-x_0)^2
$,
$
a_1(x)= b(x-x_0)
$,
$
a_0(x)= c
$
for some constant $a,b,c \in {\Bbb C}$.

Irregular singular points are a further ``escalation level above''
regular singular points, which are already an ``escalation level above'' regular points.
It might still be possible to cope with irregular singular points
by asymptotic (power) series
(cf. Section~\ref{2019-mm-ch-ca-Ritt} on page~\pageref{2019-mm-ch-ca-Ritt}).
Asymptotic series may be seen as a generalization of Frobenius series
\index{Frobenius series} for regular singular points,
which in turn can be perceived as a generalization of Taylor series for regular points;
but they require a much more careful analysis.\cite{Bender-Orszag}

\subsection{Behavior at infinity}
\label{2012-m-ch-feainfty}
In order to cope with infinity $z=\infty$
let us transform the Fuchsian equation $w'' +p_1(z)w' +p_2(z)w=0$
into the new variable $t={1\over z}$.

\begin{equation}
\begin{split}
t={1\over z},\ z={1\over t},\ u(t)\stackrel{{\tiny \textrm{ def }}}
=
w\left({1\over t}\right)=w(z)
\\
{dt\over dz}=-{1\over z^2}=-t^2\text{ and }
{dz\over dt}=-{1\over t^2}
\textrm{; therefore }
{d\over dz}
=   {dt\over dz} {d\over dt} =
-t^2{d\over dt}
\\
{d^2\over dz^2}=-t^2{d\over dt}\left(-t^2{d\over dt}\right)=
-t^2\left(-2t{d\over dt}-t^2{d^2\over dt^2}\right)=
2t^3{d\over dt}+t^4{d^2\over dt^2}
\\
w'(z)={d\over dz}w(z)=-t^2{d\over dt}u(t)=
-t^2u'(t)
\\
w''(z)={d^2\over dz^2}w(z)=
\left(2t^3{d\over dt}+t^4
{d^2\over dt^2}\right)u(t)=2t^3u'(t)+t^4u''(t)
\end{split}
\end{equation}

Insertion into the Fuchsian equation $w''+p_1(z)w'+p_2(z)w=0 $ yields
\begin{equation}
   2t^3u'+t^4u''+p_1\left({1\over t}\right)(-t^2u')+
   p_2\left({1\over t}\right)u=0,
\end{equation}
and hence,
\begin{equation}
    u''+\left[{2\over t}-{p_1\left({1\over t}\right)
   \over t^2}\right]u'+{p_2\left({1\over t}\right)\over t^4}u=0.
\end{equation}
From
\begin{equation}
\tilde p_1(t)\stackrel{{\tiny \textrm{ def }}}{=} {2\over t}-
{p_1\left({1\over t}\right)\over t^2}
\label{2016-m-ch-sf-idp1}
\end{equation}
and
\begin{equation}
\displaystyle \tilde p_2(t)\stackrel{{\tiny \textrm{ def }}}{=} {p_2\left({1\over t}\right)\over t^4}
\label{2016-m-ch-sf-idp2}
\end{equation}
follows the form of the rewritten differential equation
\begin{equation}
u''+\tilde p_1(t)u'+\tilde p_2(t)u=0.
\end{equation}
A necessary criterion for this equation to be Fuchsian is that $0$ is an ordinary,
or at least a regular singular, point.

Note that, for infinity to be a regular singular point,
$\tilde p_1(t)$ must have at most a pole of the order of $ t^{-1}$,
and
$\tilde p_2(t)$ must have at most a pole of the order of $t^{-2}$
at $t=0$.
Therefore,
$(1/t) p_1(1/t) = z p_1(z)$
as well as
$(1/t^2) p_2(1/t) = z^2 p_2(z)$
must both be analytic functions as $t\rightarrow 0$,
or $z\rightarrow \infty$.
This will be an important finding for the following arguments.

\subsection{Functional form of the coefficients in Fuchsian differential equations}

The functional form of the coefficients $p_1(x)$ and $p_2(x)$,
resulting from the assumption of merely regular singular points can be estimated as follows.

First, let us start with {\em poles at finite complex numbers}.
Suppose there are $k$ finite poles.
[The behavior of
$p_1(x)$ and $p_2(x)$ at
infinity will be treated later.]
Therefore, in Equation~(\ref{2011-m-ch-sf-fc2}),
the coefficients must be of the form
\begin{equation}
\begin{split}
p_1  (x)  = \frac{P_1(x)}{\prod_{j=1}^k (x-x_j)} ,\\
\textrm{and }
p_2 (x) =  \frac{P_2(x)}{\prod_{j=1}^k (x-x_j)^2} ,
\end{split}
\label{2011-m-ch-sf-eforp}
\end{equation}
where the $x_1,\ldots ,x_k$ are $k$ the (regular singular) points of the poles,
and $P_1(x)$ and $P_2(x)$ are {\em entire functions;}
that is, they are analytic (or, by another wording, holomorphic)
over the whole complex plane formed by $\{x \mid x \in {\Bbb C} \}$.

Second, consider possible {\em poles at infinity.}
Note that the requirement that infinity is regular singular will restrict the possible growth of
$p_1(x)$
as well as
$p_2(x)$
and thus, to a lesser degree, of
$P_1(x)$
as well as
$P_2(x)$.

As has been shown earlier,
because of the requirement that infinity is regular singular, as $x$ approaches infinity,
$p_1(x)x$
as well as
$p_2(x)x^2$
must both be analytic.
Therefore,
$p_1(x)$ cannot grow faster than $\vert x \vert^{-1}$,
and
$p_2(x)$ cannot grow faster than $\vert x \vert^{-2}$.

Consequently, by (\ref{2011-m-ch-sf-eforp}),
as $x$ approaches infinity,
$
P_1 (x) = p_1 (x) \prod_{j=1}^k (x-x_j)
$
does not grow faster than $\vert x \vert^{k-1}$ -- which in turn means that $P_1 (x)$ is bounded by some constant times $\vert x \vert^{k-1}$.
Furthmore,
$
P_2 (x) = p_2(x)\prod_{j=1}^k (x-x_j)^2
$
does not grow faster than $\vert x \vert^{2k-2}$
-- which in turn means that $P_2 (x)$ is bounded by some constant times $\vert x \vert^{2k-2}$.

Recall that both $P_1(x)$ and $P_2(x)$ are {\em entire functions.}
Therefore, because of the {\em generalized Liouville theorem}\cite[-50mm]{Greene}
(mentioned on page \pageref{2012-m-ch-ca-lt}),
\index{Liouville theorem}
\index{generalized Liouville theorem}
\label{2014-m-ch-sf-glt}
both
$P_1(x)$ and $P_2(x)$ must be polynomials
of degree of at most $k-1$ and $2k-2$, respectively.
\index{rational function}

Moreover, by using
{\em partial fraction decomposition}\sidenote[][-40mm]{See also, for instance,
Chapter~3, pp.~{29-42}, as well as Appendix C, p.~201 of~\bibentry{KristenssonC3},
and p.~146 of~\bibentry{Henrici-II}.

{\color{blue}
For a particular example, consider
$\frac{x^2+2x-18}{x^2+x-6}$, and first reduce the order of the polynomial $x^2+2x-18$
in the numerator by dividing it with the denominator $x^2+x-6$, resulting in $1+\frac{x-12}{x^2+x-6}$.
Now suppose that the following {\it Ansatz} could be made:
$\frac{x-12}{x^2+x-6}
= \frac{x-12}{(x-2)(x+3)}= \frac{A}{x-2} + \frac{B}{x+3}
= \frac{A(x+3)}{(x-2)(x+3)} + \frac{B(x-2)}{(x-2)(x+3)}$. Therefore,
$x-12 =  A(x+3) + B(x-2)$. By substituting $x=2$ and $x=-3$ one obtains $A=-2$ and $B=3$, respectively.
Hence $\frac{x^2+2x-18}{x^2+x-6} = 1 -\frac{2}{x-2} + \frac{3}{x+3}$.}}
\index{partial fraction decomposition}
of the rational functions
-- that is, the quotients $\frac{R(x)}{Q(x)}$ of polynomials $R(x)$
and nonzero $Q(x)$ --
in terms of their pole factors $x-x_j$,
we obtain from~(\ref{2011-m-ch-sf-eforp}) the general form of the coefficients
\begin{equation}
\begin{split}
p_1 (x)  = \sum_{j=1}^k \frac{A_j}{x-x_j},\\
\textrm{and }
p_2(x) = \sum_{j=1}^k \left[ \frac{B_j}{(x-x_j)^2}   +  \frac{C_j}{x-x_j} \right]
,
\end{split}
\label{2012-m-ch-sf-eforp12359}
\end{equation}
with constant $A_j, B_j, C_j \in {\Bbb C}$.
The resulting Fuchsian differential equation
is called
{\em Riemann differential equation}.
\index{Riemann differential equation}

Although we have considered an arbitrary finite number of poles,
for reasons that are unclear to this author,
physics is mainly concerned
with two poles (i.e., $k=2$)
at finite points, and one at infinity.

The {\em hypergeometric differential equation} is a {\em Fuchsian differential equation}
which has at most {\em three regular singularities}, including infinity,
at\cite{Kuznetsov}  $0$, $1$, and $\infty$.

\subsection{Frobenius method: Solution by power series}

\index{power series solution}
\index{Frobenius series}
Let us get more concrete about the solution of Fuchsian equations by power series expansions.
Thereby the general strategy is to transform an ordinary differential equation
into a system of (coupled) linear equations.
Because as it turns out the solutions of Fuchsian differential equations
\index{Fuchsian equation} can be expanded as {\em power series},
so that the differentiations can be performed explicitly.
The unknow coefficients of these power series  which ``encode the solutions'' are then
obtained by utilizing the linear independence of different powers in these series.
Thereby every factor multiplied by the powers in these series is enforced to vanish separately.

{
\color{blue}
\bexample
In order to obtain a feeling for power series solutions of differential equations,
consider the ``first order'' Fuchsian equation\cite{larson-edwards-calculus}
\begin{equation}
y'-\lambda  y=0.
\label{2011-m-ch-sf-pss1}
\end{equation}
Make the {\it Ansatz}, also known as
{\em Frobenius method},\cite{arfken05}
\index{Frobenius method}
 that the solution can be expanded into a power series of the form
\begin{equation}
y(x)=\sum_{j=0}^\infty a_j x^j.
\label{2019-m-ch-sf-pss12}
\end{equation}
Then,  the second term of Equation~(\ref{2011-m-ch-sf-pss1}) is
$-\lambda   \sum_{j=0}^\infty a_j x^j$, whereas the first term   can be written as
\begin{equation}
\begin{split}
\left(\frac{d}{dx}  \sum_{j=0}^\infty a_j x^j\right) =
 \sum_{j=0}^\infty ja_j x^{j-1}=
 \sum_{j=1}^\infty ja_j x^{j-1}\\
=
 \sum_{m=j-1=0}^\infty (m+1)a_{m+1} x^{m}=
 \sum_{ j =0}^\infty (j+1)a_{j+1} x^{j}.
\end{split}
\label{2011-m-ch-sf-pss2iuzuiz}
\end{equation}
As a result the differential equation~(\ref{2011-m-ch-sf-pss1}) can be written
in terms of the sums in~(\ref{2011-m-ch-sf-pss2iuzuiz}) and~(\ref{2019-m-ch-sf-pss12}):
\begin{equation}
 \sum_{ j =0}^\infty (j+1)a_{j+1} x^{j}
-\lambda  \left(\sum_{j=0}^\infty a_j x^j\right)=
 \sum_{ j =0}^\infty x^j\big[(j+1)a_{j+1}
-\lambda a_j \big]=
0.
\label{2019-m-ch-sf-pss1}
\end{equation}
Note that polynomials $x^i$ and $x^j$ of different degrees $i\neq j$ are linearly independent of each other,
so the differences $(j+1)a_{j+1} -\lambda a_j$
in (\ref{2019-m-ch-sf-pss1}) have to be zero for all $j\ge 0$.
Thus by comparing the coefficients of $x^j$,  for $n\ge 0$,
in~(\ref{2011-m-ch-sf-pss2iuzuiz}) and in $\lambda$ times the sum~(\ref{2019-m-ch-sf-pss12}) one obtains
\begin{equation}
\begin{split}
(j+1)a_{j+1}= \lambda   a_j\textrm{, or }\\
a_{j+1}= \frac{\lambda   a_j}{j+1} =a_0 \frac{\lambda ^{j+1}}{(j+1)!}\textrm{; that is, }
a_{j }= a_0 \frac{\lambda ^{j }}{j!}
 .
\end{split}
\end{equation}
Therefore,
\begin{equation}
y(x)=\sum_{j=0}^\infty a_0 \frac{\lambda ^{j }}{j!} x^j=a_0 \sum_{j=0}^\infty \frac{(\lambda  x)^{j }}{j!} =a_0 e^{\lambda  x}.
\end{equation}

\eexample
}

In the Fuchsian case let us consider the following {\it Frobenius Ansatz}
to expand the solution as a {\em generalized power series} around a regular singular point $x_0$,
which can be  motivated by Equation~(\ref{2011-m-ch-sf-eforp}), and by the {\em Laurent series expansion}
\index{Laurent series}
(\ref{011-m-ch-ca-else1})--(\ref{011-m-ch-ca-else2}) on
page \pageref{011-m-ch-ca-else1}:
\begin{equation}
\begin{split}
  p_1  (x)  = \frac{A_1(x)}{x-x_0}=\sum_{j=0}^\infty \alpha_j (x-x_0)^{j-1} \textrm{  for } 0 < \vert x-x_0 \vert < r_1,\\
 p_2 (x) = \frac{A_2(x)}{(x-x_0)^2}=\sum_{j=0}^\infty \beta_j (x-x_0)^{j-2} \textrm{  for } 0 < \vert x-x_0 \vert < r_2,\\
y(x)=  (x-x_0)^{\sigma} \sum_{l=0}^\infty  (x-x_0)^{l} w_l
=  \sum_{l=0}^\infty (x-x_0)^{l + \sigma} w_l \textrm{, with } w_0\neq 0
,
\end{split}
\label{2011-m-ch-sf-pss2}
\end{equation}
where $A_1(x)= [(x-x_0) a_1(x)]/a_2(x)$
and $A_2(x)= [(x-x_0)^2 a_0(x)]/a_2(x)$.
Equation~(\ref{2011-m-ch-sf-fc2})
then becomes
\begin{equation*}
\begin{split}
\frac{d^2}{dx^2}y(x) +   p_1  (x)    \frac{d}{dx}y(x)+    p_2  (x)   y(x)     =   0,    \\
\left[\frac{d^2}{dx^2}  + \sum_{j=0}^\infty \alpha_j (x-x_0)^{j-1}  \frac{d}{dx} + \sum_{j=0}^\infty \beta_j (x-x_0)^{j-2}\right]
\sum_{l=0}^\infty w_l (x-x_0)^{l + \sigma}     =   0,    \\
\sum_{l=0}^\infty ({l + \sigma})({l + \sigma-1}) w_l (x-x_0)^{l + \sigma -2}\qquad \qquad \\
\qquad + \left[  \sum_{l=0}^\infty ({l + \sigma}) w_l (x-x_0)^{l + \sigma -1}\right]  \sum_{j=0}^\infty \alpha_j (x-x_0)^{j-1} \qquad \qquad
\\
\qquad + \left[\sum_{l=0}^\infty w_l (x-x_0)^{l + \sigma}\right] \sum_{j=0}^\infty \beta_j (x-x_0)^{j-2}
    =   0,    \\
(x-x_0)^{\sigma-2}  \sum_{l=0}^\infty (x-x_0)^{l}
\Bigg[
({l + \sigma})({l + \sigma-1})   w_l  \qquad \qquad \\
\qquad + ({l + \sigma})   w_l    \sum_{j=0}^\infty \alpha_j (x-x_0)^{j}
+  w_l  \sum_{j=0}^\infty \beta_j (x-x_0)^{j}
\Bigg]
    =   0,
\\
(x-x_0)^{\sigma-2} \left[ \sum_{l=0}^\infty
({l + \sigma})({l + \sigma-1})   w_l (x-x_0)^{l}\right. \qquad \qquad \\
\qquad +  \sum_{l=0}^\infty ({l + \sigma})   w_l    \sum_{j=0}^\infty \alpha_j (x-x_0)^{l+j}
\left.
 +  \sum_{l=0}^\infty  w_l  \sum_{j=0}^\infty \beta_j (x-x_0)^{l+j}  \right]
    =   0.
\end{split}
\end{equation*}
Next, in order  to reach a  common power of $(x-x_0)$,
we perform an index identification  in the second and third summands (where the order of the sums change):
$l=m$ in the first summand, as well as an index shift
$
l+j =m
$, and thus
$j= m-l$.
Since
$l\ge 0$  and  $j\ge 0$, also $m=l+j$ cannot be negative.
Furthermore,
$0\le j = m-l$, so that  $l\le m$.
\begin{equation}
\begin{split}
(x-x_0)^{\sigma-2} \left[  \sum_{l=0}^\infty
({l + \sigma})({l + \sigma-1})   w_l (x-x_0)^{l} \right.    \qquad \qquad
\\
\qquad + \sum_{j=0}^\infty  \sum_{l=0}^\infty ({l + \sigma})   w_l    \alpha_j (x-x_0)^{l+j}  \qquad \qquad
\\
\left.
\qquad + \sum_{j=0}^\infty  \sum_{l=0}^\infty  w_l  \beta_j (x-x_0)^{l+j}  \right]
    =   0,\\
(x-x_0)^{\sigma-2} \left[   \sum_{m=0}^\infty
({m + \sigma})({m + \sigma-1})   w_m (x-x_0)^{m} \right. \qquad \qquad \\
\qquad + \sum_{m=0}^\infty  \sum_{l=0}^m ({l + \sigma})   w_l    \alpha_{m-l} (x-x_0)^{l+m-l}  \qquad \qquad
\\
\left.
\qquad + \sum_{m=0}^\infty  \sum_{l=0}^m  w_l  \beta_{m-l} (x-x_0)^{l+m-l}     \right]
    =   0,\\
(x-x_0)^{\sigma-2}\left\{     \sum_{m=0}^\infty
(x-x_0)^{m} \left[
({m + \sigma})({m + \sigma-1})   w_m \right.\right. \qquad \qquad \\
   \left.  \left.  \qquad + \sum_{l=0}^m  ({l + \sigma})   w_l   \alpha_{m-l}
 +  \sum_{l=0}^m  w_l \beta_{m-l}
\right]
\right\}
    =   0,    \\
(x-x_0)^{\sigma-2}\left\{     \sum_{m=0}^\infty
(x-x_0)^{m} \left[
({m + \sigma})({m + \sigma-1})   w_m
\right.
\right.   \qquad \qquad
 \\
\qquad \qquad +
\left.
\left.
 \sum_{l=0}^m w_l  \left( ({l + \sigma}) \alpha_{m-l}
 + \beta_{m-l}
\right)
\right]
\right\}
    =   0.
\end{split}
\label{2011-m-ch-sf-pss646465}
\end{equation}

If we can divide this equation through  $(x-x_0)^{\sigma-2}$
and exploit the linear independence of the polynomials $(x-x_0)^{m}$,
we obtain an infinite number of equations for the infinite number of coefficients $w_m$
by requiring that all the terms ``inbetween'' the $[\cdots ]$--brackets in Equation~(\ref{2011-m-ch-sf-pss646465})
vanish {\em individually.}
In particular, for $m=0$ and $w_0\neq 0$,
\begin{equation}
\begin{split}
({0+ \sigma})({0 + \sigma-1})   w_0
+
 w_0  \left( ({0 + \sigma}) \alpha_{0} + \beta_{0}\right)
    =   0\\
 f_0(\sigma ) \stackrel{{\tiny \textrm{ def }}}{=} \sigma({\sigma-1}) +  \sigma \alpha_{0} + \beta_{0}           =   0
.
\end{split}
\label{2011-m-ch-sf-pss2sigma}
\end{equation}
The {\em radius of convergence} of the solution will,
\index{radius of convergence}
in accordance with the Laurent series expansion, extend to the next singularity.

Note that in  Equation~(\ref{2011-m-ch-sf-pss2sigma}) we have defined $f_0(\sigma )$ which we will use now.
Furthermore, for successive $m>0$, and with the definition of
\begin{equation}
f_m(\sigma ) \stackrel{{\tiny \textrm{ def }}}{=} \alpha_m \sigma +\beta_m,
\end{equation}
we obtain the sequence of linear equations
\begin{equation}
\begin{split}
w_0f_0(\sigma ) =0,\\
w_1f_0(\sigma +1)+w_0f_1(\sigma )  =0,\\
w_2f_0(\sigma +2)+w_1f_1(\sigma +1) +w_0f_2(\sigma )  =0,\\
\vdots   \quad    \\
w_nf_0(\sigma+ n)   +w_{n-1}f_1(\sigma+ n-1)+ \cdots +w_0f_n(\sigma )  =0, \\
\vdots \quad
\end{split}
\label{2011-m-ch-sf-pss2sigmaiter}
\end{equation}
which can be used for an inductive determination of the coefficients $w_m$.

Equation~(\ref{2011-m-ch-sf-pss2sigma}) is a quadratic equation
$   \sigma^2 +\sigma (\alpha_{0}-1 ) +  \beta_{0}  =0$
for the
{\em characteristic exponents}
\index{characteristic exponents}
\begin{equation}
\sigma_{1,2} = \frac{1}{2} \left[1 - \alpha_{0} \pm \sqrt{(1 - \alpha_{0})^2-4  \beta_{0}}\right]
\end{equation}
We state without proof that, if the difference of the characteristic exponents
\begin{equation}
\sigma_1 - \sigma_2  =   \sqrt{(1 - \alpha_{0})^2-4  \beta_{0}}
\end{equation}
is  {\em nonzero} and {\em not} an integer, then the two solutions found from   $\sigma_{1,2}$
through  the generalized series Ansatz  (\ref{2011-m-ch-sf-pss2}) are linear independent.

Intuitively speaking, the Frobenius method ``is in obvious trouble'' to find
the general solution of the Fuchsian equation
if the two characteristic exponents coincide (e.g., $\sigma_1 = \sigma_2$),
but it ``is also in trouble'' to find
the general solution
if  $\sigma_1 - \sigma_2 = m \in {\Bbb N}$;
that is, if, for some positive integer $m$,  $\sigma_1  = \sigma_2 + m > \sigma_2$.
Because in this case, ``eventually''
at $n=m$ in Equation~(\ref{2011-m-ch-sf-pss2sigmaiter}),
we obtain as iterative solution for the coefficient $w_m$ the term
\begin{equation}
\begin{split}
w_m  = - \frac{w_{m-1}f_1(\sigma_2 + m-1)+ \cdots +w_0f_m(\sigma_2  )}{f_0(\sigma_2 + m)}
\\
\qquad  = - \frac{w_{m-1}f_1(\sigma_1-1)+ \cdots +w_0f_m(\sigma_2  )}{\underbrace{f_0(\sigma_1)}_{ = 0}}
.
\end{split}
\end{equation}
That is,  the greater critical exponent $\sigma_1=\sigma_2 + m$   is a solution
of Equation~(\ref{2011-m-ch-sf-pss2sigma}) so that $f_0(\sigma_1)$ in the denominator vanishes.

In these cases the greater characteristic exponent  $\sigma_1 \ge \sigma_2$
can still be used to find a solution in terms of a power series,
but the smaller characteristic exponent $\sigma_2$ in general cannot.

\subsection{d'Alembert reduction of order}
\index{reduction of order}

If $\sigma_1=\sigma_2+n$ with $n\in {\Bbb Z}$,
then we find only a {\em single} solution of the Fuchsian equation in terms of the power series resulting
from inserting the {\em greater} (or equal) characteristic exponent.
In order to obtain another linear  independent solution we have to employ
a method based on the Wronskian,\cite[-40mm]{arfken05} or
the
d'Alembert reduction,\cite[-20mm]{Teschl-odr}
which is a general method to obtain another, linear independent solution
\index{d'Alembert reduction}
$y_2(x)$ from an existing particular solution  $y_1(x)$ by the {\it Ansatz}
(no proof is presented here)
\begin{equation}
y_2(x)=y_1(x)\int_x  v(s) ds.
\label{2011-m-ch-sf-dalambansatz}
\end{equation}
Inserting $y_2(x)$ from (\ref{2011-m-ch-sf-dalambansatz}) into the Fuchsian equation (\ref{2011-m-ch-sf-fc2}),
and using the fact that by assumption $y_1(x)$
is a solution of it,
yields
\begin{equation*}
\begin{split}
\frac{d^2}{dx^2}y_2(x) +   p_1  (x)    \frac{d}{dx}y_2(x)+    p_2  (x)   y_2(x)
 =
0, \\
\frac{d^2}{dx^2}y_1(x)\int_x  v(s) ds +   p_1  (x)    \frac{d}{dx}y_1(x)\int_x  v(s) ds+    p_2  (x)   y_1(x)\int_x  v(s) ds
 =
0, \\
\frac{d }{dx }\left\{
\left[\frac{d }{dx }y_1(x)\right]\int_x  v(s) ds
+y_1(x)   v(x)  \right\} \qquad  \\
+   p_1  (x)    \left[\frac{d}{dx}y_1(x)\right] \int_x  v(s) ds + p_1(x)   v(x)
+   p_2 (x)  y_1(x)\int_x  v(s) ds
 =
0, \\
\left[ \frac{d^2 }{dx^2 }y_1(x)\right] \int_x  v(s) ds
+\left[ \frac{d  }{dx  }y_1(x)\right]    v(x)
+\left[ \frac{d  }{dx  }y_1(x)\right]    v(x)
+y_1(x)  \left[ \frac{d  }{dx  } v(x) \right]    \\
+   p_1  (x)    \left[ \frac{d}{dx}y_1(x)\right] \int_x  v(s) ds + p_1  (x)   y_1(x)   v(x)
+   p_2 (x)  y_1(x)\int_x  v(s) ds
 =
0, \\
\left[ \frac{d^2 }{dx^2 }y_1(x)\right] \int_x  v(s) ds
+   p_1  (x)    \left[ \frac{d}{dx}y_1(x)\right] \int_x  v(s) ds
+   p_2 (x)  y_1(x)\int_x  v(s) ds    \quad   \\
+   p_1  (x)   y_1(x)   v(x)
+\left[ \frac{d  }{dx  }y_1(x)\right]    v(x)
+\left[ \frac{d  }{dx  }y_1(x)\right]    v(x)
+y_1(x)  \left[ \frac{d  }{dx  } v(x) \right]
 =
0, \\
\left[ \frac{d^2 }{dx^2 }y_1(x)\right] \int_x  v(s) ds
+   p_1  (x)    \left[ \frac{d}{dx}y_1(x)\right] \int_x  v(s) ds
+   p_2 (x)  y_1(x)\int_x  v(s) ds   \quad     \\
+y_1(x)  \left[ \frac{d  }{dx  } v(x) \right]
+2\left[ \frac{d  }{dx  }y_1(x)\right]    v(x)
 +   p_1  (x)   y_1(x)   v(x)
 =
0, \\
\underbrace{\left\{\left[ \frac{d^2 }{dx^2 }y_1(x)\right]
+   p_1  (x)    \left[ \frac{d}{dx}y_1(x)\right]
+   p_2 (x)  y_1(x)\right\}}_{=0}\int_x  v(s) ds     \qquad   \\
+y_1(x)  \left[ \frac{d  }{dx  } v(x) \right]
+\left\{ 2\left[ \frac{d  }{dx  }y_1(x)\right]
  +   p_1  (x)   y_1(x) \right\}  v(x)
 =
0, \\
 y_1(x)  \left[ \frac{d  }{dx  } v(x) \right]
+\left\{ 2\left[ \frac{d  }{dx  }y_1(x)\right]
  +   p_1  (x)   y_1(x) \right\}  v(x)
 =
0,
\end{split}
\end{equation*}
and finally,
\begin{equation}
      v'(x)  +  v(x)    \left\{ 2 \frac{y'_1(x)}{y_1(x)}    +   p_1  (x)   \right\} = 0.
\label{2011-m-ch-sf-fc2123}
\end{equation}

\subsection{Computation of the characteristic exponent}

 Let
 $w'' +p_1(z)w' +p_2(z)w=0$
be a Fuchsian equation.
From the Laurent series expansion of $p_1(z)$ and $p_2(z)$ in~(\ref{2011-m-ch-sf-pss2}) and Cauchy's integral formula  we can derive
the following equations, which are helpful in determining the characteristic exponent $\sigma$,
as defined in (\ref{2011-m-ch-sf-pss2sigma}) by
$
\sigma({\sigma-1}) +  \sigma \alpha_{0} + \beta_{0}           =   0
$:
\begin{equation}
\begin{split}
\alpha_0=\lim_{z\rightarrow z_0} (z-z_0)p_1(z),\\
\beta_0=\lim_{z\rightarrow z_0} (z-z_0)^2p_2(z),
\end{split}
\end{equation}
 where $z_0$ is a regular singular point.

{\color{OliveGreen}
\bproof

In order to find {$\alpha_0$}, consider the Laurent series for
\begin{equation}
\begin{split}
   p_1(z)=\sum_{k=-1}^\infty \tilde a_k(z-z_0)^k, \\
\textrm{ with }
   \tilde a_k={1\over 2\pi i}\oint p_1(s)(s-z_0)^{-(k+1)}ds.
\end{split}
\end{equation}
The summands  vanish  for $k<-1$, because $p_1(z)$ has at most a pole of order one at $z_0$.

An index change $n=k+1$, or  $k=n-1$,  as well as a redefinition
$\alpha_n \stackrel{{\tiny \textrm{ def }}}{=} \tilde a_{n-1}$
yields
\begin{equation}
p_1(z) =\sum_{n=0}^\infty \alpha_n(z-z_0)^{n-1},
\end{equation}
where
\begin{equation}
 \alpha_n={\tilde a}_{n-1}={1\over 2\pi i}
\oint p_1(s)(s-z_0)^{-n}ds;
\end{equation}
and, in particular,
\begin{equation}
\alpha_0={1\over2\pi i}\oint p_1(s)ds.
\label{2016-m-ch-sf-al0}
\end{equation}
Because the equation is Fuchsian,  $p_1(z)$ has at most a pole of order one at $z_0$.
Therefore, $(z-z_0) p_1(z)$  is analytic around $z_0$.
By multiplying  $p_1(z)$ with unity $1=(z-z_0)/(z-z_0)$
and insertion into~(\ref{2016-m-ch-sf-al0}) we obtain
\begin{equation}
   \alpha_0={1\over2\pi i}\oint{p_1(s)(s-z_0)\over(s-z_0)}ds.
\end{equation}
{\em Cauchy's integral formula}~(\ref{2012-m-ch-ca-cif}) on page \pageref{2012-m-ch-ca-cif}
yields
\begin{equation}
   \alpha_0=\lim_{s\to z_0}p_1(s)(s-z_0).
\end{equation}

Alternatively we may consider the
{\it Frobenius Ansatz}~(\ref{2011-m-ch-sf-pss2})
$  p_1(z)=\sum_{n=0}^\infty\alpha_n(z-z_0)^{n-1}$
which has again been motivated by the fact that $p_1(z)$ has at most a pole of order one at $z_0$.
Multiplication of this series by $(z-z_0)$ yields
\begin{equation}
   (z-z_0)p_1(z)=\sum_{n=0}^\infty\alpha_n(z-z_0)^n .
\end{equation}
In the limit $z\to z_0$,
\begin{equation}
\alpha_0  = \lim_{z\to z_0}(z-z_0)p_1(z)
.
\end{equation}

Likewise, let us find the expression for $\beta_0$  by considering the Laurent series for
\begin{equation}
\begin{split}
   p_2(z)=\sum_{k=-2}^\infty \tilde b_k(z-z_0)^k \\
\textrm{ with }
   \tilde b_k={1\over 2\pi i}\oint p_2(s)(s-z_0)^{-(k+1)}ds.
\end{split}
\end{equation}
The summands  vanish  for $k<-2$, because $p_2(z)$ has at most a pole of order two at $z_0$.

An index change  $n=k+2$, or  $k=n-2$,  as well as a redefinition
$\beta_n \stackrel{{\tiny \textrm{ def }}}{=} \tilde b_{n-2}$
yields
\begin{equation}
p_2(z) =\sum_{n=0}^\infty \beta_n(z-z_0)^{n-2},
\end{equation}
where
\begin{equation}
 \beta_n={\tilde b}_{n-2}={1\over 2\pi i}
\oint p_2(s)(s-z_0)^{-(n-1)}ds;
\end{equation}
and, in particular,
\begin{equation}
\beta_0={1\over2\pi i}\oint (s-z_0) p_2(s) ds.
\label{2016-m-ch-sf-bl0}
\end{equation}

Because the equation is Fuchsian,  $p_2(z)$ has at most a pole of order two at $z_0$.
Therefore, $(z-z_0)^2 p_2(z)$  is analytic around $z_0$.
By multiplying  $p_2(z)$ with unity $1=(z-z_0)^2/(z-z_0)^2$
and insertion into~(\ref{2016-m-ch-sf-bl0}) we obtain
\begin{equation}
   \beta_0={1\over2\pi i}\oint{p_2(s)(s-z_0)^2\over(s-z_0)}ds .
\end{equation}
{\em Cauchy's integral formula}~(\ref{2012-m-ch-ca-cif}) on page \pageref{2012-m-ch-ca-cif} yields
\begin{equation}
   \beta_0=\lim_{s\to z_0}p_2(s)(s-z_0)^2.
\end{equation}

Again another way to see this is with the
{\it Frobenius Ansatz}~(\ref{2011-m-ch-sf-pss2})   $p_2(z)=\sum_{n=0}^\infty\beta_n(z-z_0)^{n-2}$.
Multiplication with $(z-z_0)^2$, and taking the limit $z\to z_0$, yields
\begin{equation}
   \lim_{z\to z_0}(z-z_0)^2p_2(z)=\beta_n
.
\end{equation}

\eproof
}

{
\color{blue}
\bexample

\subsection{Examples}
Let us consider some examples involving Fuchsian equations of the second order.
\begin{enumerate}

\item
First, we shall prove that $z^2 y''(z) + z y'(z) - y(z) = 0$ is of the Fuchsian type, and compute the solutions with the Frobenius method.

Let us first locate the singularities of
\begin{equation}
y''(z) + \frac{y'(z)}{z} - \frac{y(z)}{z^2} = 0.
\label{2016-m-ch-sf-efmEu}
\end{equation}
One singularity is at the (finite) point  $z_0=0$.

In order to analyze the singularity at infinity, we have to transform the equation by $z=1/t$.
First observe that
$p_1(z)= 1/z$
and
$p_2(z)= -1/z^2$.
Therefore, after the transformation, the new coefficients, computed from
(\ref{2016-m-ch-sf-idp1}) and (\ref{2016-m-ch-sf-idp2}),
are
\begin{equation}
\begin{split}
\tilde p_1 (t) = \left( \frac{2}{t} - \frac{t}{t^2} \right) =   \frac{1}{t} , \textrm{ and }\\
\tilde p_2 (t) = \frac{- t^2}{t^4}  = -\frac{ 1}{t^2}
.
\end{split}
\end{equation}
Thereby we effectively regain the original type of equation~(\ref{2016-m-ch-sf-efmEu}).
We can thus treat both singularities at zero and infinity in the same way.

Both singularities are regular, as the coefficients
$p_1$ and $\tilde p_1$  have poles of order 1,
$p_2$ and $\tilde p_2$  have poles of order 2, respectively.
Therefore, the differential equation is Fuchsian.

In order to obtain solutions, let us first compute the  characteristic exponents
by
\begin{equation}
\begin{split}
\alpha_0 = \lim_{z \rightarrow 0} z p_1 (z) = \lim_{z \rightarrow 0} z\frac{1}{z} = 1
 , \textrm{ and }\\
\beta_0 = \lim_{z \rightarrow 0} z^2 p_2 (z) = \lim_{z \rightarrow 0} - z^2\frac{1}{z^2} = - 1
,
\end{split}
\end{equation}
so that, from
(\ref{2011-m-ch-sf-pss2sigma}),
\begin{equation}
\begin{split}
0= f_0(\sigma )  = \sigma({\sigma-1}) +  \sigma \alpha_{0} + \beta_{0}           =
\sigma^2 - \sigma +  \sigma  -1 = \\
= \sigma^2   -1
 , \textrm{ and  thus }
\sigma_{1,2} = \pm 1 .
\end{split}
\end{equation}

The first solution is obtained by insertion of the {\it Frobenius Ansatz}~(\ref{2011-m-ch-sf-pss2}),
in particular,
$y_1(x) =  \sum_{l=0}^\infty (x-x_0)^{l + \sigma} w_l$ with $\sigma_{1} =  1$ and  $x_0=0$
into (\ref{2016-m-ch-sf-efmEu}).
In this case,
\begin{equation}
\begin{split}
x^2 \sum_{l=0}^\infty (l+1)l x^{l - 1} w_l
+
x \sum_{l=0}^\infty (l+1) x^{l} w_l
-
\sum_{l=0}^\infty   x^{l+1} w_l = 0,
\\
\sum_{l=0}^\infty \left[ (l+1)l +  (l+1) - 1 \right] x^{l+1} w_l =
\sum_{l=0}^\infty w_l l (l+2)   x^{l+1} = 0.
\end{split}
\end{equation}
Since
the polynomials are linear independent, we obtain  $ w_l l (l+2) = 0  $ for all $l \ge 0$.
Therefore, for constant $A$,
\begin{equation}
 w_0  = A \textrm{, and }
 w_l  = 0 \textrm{ for } l > 0.
\end{equation}
So, the first solution is $y_1(z) = Az$.

The second solution, computed through the {\it Frobenius Ansatz}~(\ref{2011-m-ch-sf-pss2}),
is obtained by inserting
$y_2(x) =  \sum_{l=0}^\infty (x-x_0)^{l + \sigma} w_l$ with $\sigma_{2} =  -1$
into (\ref{2016-m-ch-sf-efmEu}).
This yields
\begin{equation}
\begin{split}
z^2 \sum_{l=0}^\infty (l-1)(l-2) (x-x_0)^{l - 3} w_l  +\\
+
z \sum_{l=0}^\infty (l-1) (x-x_0)^{l-2} w_l
-
\sum_{l=0}^\infty   (x-x_0)^{l-1} w_l = 0,
\\
\sum_{l=0}^\infty \left[ w_l (l-1)(l-2) + w_l (l-1) - w_l \right] (x-x_0)^{l-1} w_l = 0,
\\
\sum_{l=0}^\infty w_l l (l-2)   (x-x_0)^{l-1}  = 0.
\end{split}
\end{equation}
Since the polynomials are linear independent, we obtain  $ w_l l (l-2) = 0  $ for all $l \ge 0$.
Therefore, for constant $B,C$,
\begin{equation}
\begin{split}
 w_0  = B ,
 w_1  = 0,
 w_2  = C \textrm{, and }
 w_l  = 0 \textrm{ for } l > 2,
\end{split}
\end{equation}
So that the second solution is $y_2(z) = B\frac{1}{z}+ Cz$.

Note that $y_2(z)$ already represents the general solution of (\ref{2016-m-ch-sf-efmEu}).
\marginnote{Most of the coefficients are zero, so no iteration with a ``catastrophic divisions by zero'' occurs here.}
Alternatively we could have started from $y_1(z)$ and applied d'Alembert's {\it Ansatz}
(\ref{2011-m-ch-sf-dalambansatz})--(\ref{2011-m-ch-sf-fc2123}):
\begin{equation}
v' (z) + v (z)\left(\frac{2}{z} + \frac{1}{z}\right) = 0 \textrm{, or } v' (z) = - \frac{3 v (z)}{z}
\end{equation}
yields
\begin{equation}
\frac{d v}{v} = - 3\frac{d z}{z} \textrm{, and }\log v = -3 \log z \textrm{, or } v(z) = z^{-3}.
\end{equation}
Therefore, according to (\ref{2011-m-ch-sf-dalambansatz}),
\begin{equation}
y_2(z) = y_1(z) \int_z  v(s) ds  = A z \left(-\frac{1}{2z^2}\right) = \frac{A'}{z}  .
\end{equation}

\item
Find out whether the following differential equations are Fuchsian,
and enumerate the regular singular points:
\begin{equation}
\begin{split}
zw''+(1-z)w'=0 ,  \\
z^2w''+zw'-\nu ^2 w=0 ,  \\
z^2(1+z)^2w''+2z(z+1)(z+2)w'-4w=0 , \\
2z(z+2)w'' +w' -zw=0.
\end{split}
\end{equation}

{ ad 1:} $\displaystyle zw''+(1-z)w'=0\ \Longrightarrow
\ w''+{(1-z)\over z}w'=0$\\[2ex]
 {$z=0$:}
$$
   \alpha_0=\lim_{z\to0}z{(1-z)\over z}=1,\quad
   \beta_0=\lim_{z\to0}z^2\cdot0=0.
$$
The equation for the characteristic exponent is
$$
   \sigma(\sigma-1)+\sigma\alpha_0+\beta_0=0\Longrightarrow
   \sigma^2-\sigma+\sigma=0\Longrightarrow\sigma_{1,2}=0.
$$

\bigskip

\noindent  {$z=\infty$:} $  z={1\over t}$
$$
   \tilde p_1(t)={2\over t}-{{\left(1-{1\over t}\right)\over{1\over t}}\over
   t^2}={2\over t}-{\left(1-{1\over t}\right)\over t}={1\over t}
   +{1\over t^2}={t+1\over t^2}
$$
$\Longrightarrow$ not Fuchsian.

\bigskip

\noindent {  ad 2:} $\displaystyle z^2w''+zw'-v^2w=0\Longrightarrow
w''+{1\over z}w'-{v^2\over z^2}w=0$.\\[2ex]
 {$z=0$:}
$$
   \alpha_0=\lim_{z\to0}z{1\over z}=1,\quad
   \beta_0=\lim_{z\to0}z^2\left(-{v^2\over z^2}\right)=-v^2.
$$
$$
   \Longrightarrow \sigma^2-\sigma+\sigma-v^2=0\Longrightarrow\sigma_{1,2}=
   \pm v
$$

\bigskip

\noindent  {$z=\infty$:} $  z={1\over t}$
\begin{eqnarray*}
   \tilde p_1(t)&=&{2\over t}-{1\over t^2}t={1\over t}\\
   \tilde p_2(t)&=&{1\over t^4}\left(-t^2v^2\right)=-{v^2\over t^2}
\end{eqnarray*}
$$
   \Longrightarrow u''+{1\over t}u'-{v^2\over t^2}u=0 \Longrightarrow
   \sigma_{1,2}=\pm v
$$
$\Longrightarrow$ Fuchsian equation.

\bigskip

\noindent {  ad 3:}
$$
   z^2(1+z)^2w''+2z(z+1)(z+2)w'-4w=0\Longrightarrow
   w''+{2(z+2)\over z(z+1)}w'-{4\over z^2(1+z)^2}w=0
$$
 {$z=0$:}
$$
   \alpha_0=\lim_{z\to0}z{2(z+2)\over z(z+1)}=4,\quad
   \beta_0=\lim_{z\to0}z^2\left(-{4\over z^2(1+z)^2}\right)=-4.
$$
$$
   \Longrightarrow\sigma(\sigma-1)+4\sigma-4=\sigma^2+3\sigma-4=0
   \Longrightarrow\sigma_{1,2}=
   {-3\pm\sqrt{9+16}\over 2}=\left\{{-4\atop +1}\right.
$$
 {$z=-1$:}
$$
   \alpha_0=\lim_{z\to-1}(z+1){2(z+2)\over z(z+1)}=-2,\quad
   \beta_0=\lim_{z\to-1}(z+1)^2\left(-{4\over z^2(1+z)^2}\right)=-4.
$$
$$
   \Longrightarrow\sigma(\sigma-1)-2\sigma-4=\sigma^2-3\sigma-4=
   0\Longrightarrow\sigma_{1,2}=
   {3\pm\sqrt{9+16}\over 2}=\left\{{+4\atop -1}\right.
$$
 {$z=\infty$:}
\begin{eqnarray*}
   \tilde p_1(t)&=&{2\over t}-{1\over t^2}{2\left({1\over t}+2\right)
                   \over {1\over t}\left({1\over t}+1\right)}=
                   {2\over t}-{2\left({1\over t}+2\right)\over
                   1+t}={2\over t}\left(1-{1+2t\over 1+t}\right)\\
   \tilde p_2(t)&=&{1\over t^4}\left(-{4\over{1\over t^2}
                   \left(1+{1\over t}\right)^2}\right)=-{4\over t^2}
                   {t^2\over(t+1)^2}=-{4\over(t+1)^2}
\end{eqnarray*}
$$
   \Longrightarrow u''+{2\over t}\left(1-{1+2t\over1+t}\right)u'-
   {4\over(t+1)^2}u=0
$$
$$
   \alpha_0=\lim_{t\to0}t{2\over t}\left(1-{1+2t\over1+t}\right)=0,\quad
   \beta_0=\lim_{t\to0}t^2\left(-{4\over (t+1)^2}\right)=0.
$$
$$
   \Longrightarrow\sigma(\sigma-1)=0\Longrightarrow\sigma_{1,2}=
   \left\{{0\atop 1}\right.
$$
$\Longrightarrow$ Fuchsian equation.

\bigskip

\noindent {  ad 4:}
$$
   2z(z+2)w''+w'-zw=0\Longrightarrow w''+{1\over 2z(z+2)}w'-{1\over 2(z+2)}w=0
$$
 {$z=0$:}
$$
   \alpha_0=\lim_{z\to0}z{1\over 2z(z+2)}={1\over4},\quad
   \beta_0=\lim_{z\to0}z^2{-1\over 2(z+2)}=0.
$$
$$
   \Longrightarrow\sigma^2-\sigma+{1\over 4}\sigma=0\Longrightarrow
   \sigma^2-{3\over 4}\sigma=0\Longrightarrow\sigma_1=0,\sigma_2={3\over 4}.
$$
 {$z=-2$:}
$$
   \alpha_0=\lim_{z\to-2}(z+2){1\over 2z(z+2)}=-{1\over 4},\quad
   \beta_0=\lim_{z\to-2}(z+2)^2{-1\over 2(z+2)}=0.
$$
$$
   \Longrightarrow\sigma_1=0,\quad\sigma_2={5\over 4}.
$$
 {$z=\infty$:}
\begin{eqnarray*}
   \tilde p_1(t)&=&{2\over t}-{1\over t^2}\left({1\over 2{1\over t}
                   \left({1\over t}+2\right)}\right)={2\over t}-
                   {1\over 2(1+2t)}\\
   \tilde p_2(t)&=&{1\over t^4}{(-1)\over 2\left({1\over t}+2\right)}=
                   -{1\over 2t^3(1+2t)}
\end{eqnarray*}
$\Longrightarrow$ not a Fuchsian.

\item
Determine the solutions of
$$z^2w''+(3z+1)w'+w=0 $$  around the regular singular points.

The singularities are at $z=0$ and $z=\infty$.

\noindent  {Singularities   at $z=0$:}
$$
   p_1(z)={3z+1\over z^2}=
   {a_1(z)\over z} \textrm{ with }
   a_1(z)=3+{1\over z}
$$
  $p_1(z)$ has a pole of higher order than one; hence this is no Fuchsian equation;
and  $z=0$
is an irregular singular point.

\noindent  {Singularities   at $z=\infty$:}
\begin{itemize}
\item Transformation $\displaystyle z={1\over t}$, $w(z)\to u(t)$:
      $$
         u''(t)+\left[{2\over t}-{1\over t^2}p_1\left({1\over t}\right)
         \right]\cdot u'(t)+{1\over t^4}p_2\left({1\over t}\right)\cdot
         u(t)=0.
      $$
      The new coefficient functions are
      \begin{eqnarray*}
         \tilde p_1(t)&=&{2\over t}-{1\over t^2}p_1\left({1\over t}\right)=
                         {2\over t}-{1\over t^2}(3t+t^2)={2\over t}
                         -{3\over t}-1=-{1\over t}-1\\
         \tilde p_2(t)&=&{1\over t^4}p_2\left({1\over t}\right)=
                         {t^2\over t^4}={1\over t^2}
      \end{eqnarray*}
\item check whether this is a  regular singular point:
      $$
         \begin{array}{lll}
            \displaystyle \tilde p_1(t)=-{1+t\over t}={\tilde a_1(t)\over t} &
            ~~\mbox{ with }~~\tilde a_1(t)=-(1+t)&~\mbox{regular}\\
            \displaystyle \tilde p_2(t)={1\over t^2}={\tilde a_2(t)\over t^2} &
            ~~\mbox{ with }~~\tilde a_2(t)=1&~\mbox{regular}
         \end{array}
      $$
      $\tilde a_1$ and $\tilde a_2$ are regular at $t=0$, hence  this is a  regular singular point.
\item {\it Ansatz} around $t=0$:
the transformed equation is
      \begin{eqnarray*}
         u''(t)+\tilde p_1(t)u'(t)+\tilde p_2(t)u(t)&=&0\\
         u''(t)-\left({1\over t}+1\right)u'(t)+{1\over t^2}u(t)&=&0\\
         t^2 u''(t)-(t+t^2)u'(t)+u(t)&=&0
      \end{eqnarray*}
      The generalized power series is
      \begin{eqnarray*}
         u(t)  &=&\sum_{n=0}^\infty w_n t^{n+\sigma}\\
         u'(t) &=&\sum_{n=0}^\infty w_n (n+\sigma)t^{n+\sigma-1}\\
         u''(t)&=&\sum_{n=0}^\infty w_n (n+\sigma)(n+\sigma-1)
                  t^{n+\sigma-2}
      \end{eqnarray*}
      If we insert this into the transformed differential equation we obtain
      \begin{equation*}
         \begin{split}
           t^2\sum_{n=0}^\infty w_n(n+\sigma)
             (n+\sigma-1)t^{n+\sigma-2}-\\
           \qquad\quad -\ (t+t^2)\sum_{n=0}^\infty
             w_n(n+\sigma)t^{n+\sigma-1}+
             \sum_{n=0}^\infty w_n t^{n+\sigma}=0\\
          \sum_{n=0}^\infty w_n(n+\sigma)(n+\sigma-1)t^{n+\sigma}-
             \sum_{n=0}^\infty w_n(n+\sigma)t^{n+\sigma}-\\
           \qquad\quad -\ \sum_{n=0}^\infty w_n(n+\sigma)
             t^{n+\sigma+1}+\sum_{n=0}^\infty w_n t^{n+\sigma}=0
         \end{split}
      \end{equation*}
      Change of index: $m=n+1$, $n=m-1$ in the third sum yields
      $$
         \sum_{n=0}^\infty w_n\Bigl[(n+\sigma)(n+\sigma-2)+1\Bigr]
         t^{n+\sigma}-
         \sum_{m=1}^\infty w_{m-1}(m-1+\sigma)t^{m+\sigma}=0.
      $$
      In the second sum, substitute $m$ for $n$
      $$
         \sum_{n=0}^\infty w_n\Bigl[(n+\sigma)(n+\sigma-2)+1\Bigr]
         t^{n+\sigma}-
         \sum_{n=1}^\infty w_{n-1}(n+\sigma-1)t^{n+\sigma}=0.
      $$
      We write out explicitly the  $n=0$ term of the first sum
      \begin{equation*}
         \begin{split}
            \displaystyle w_0\Bigl[\sigma(\sigma-2)+1\Bigr]t^\sigma+
               \sum_{n=1}^\infty w_n\Bigl[(n+\sigma)(n+\sigma-2)+1\Bigr]
               t^{n+\sigma} \\
            \displaystyle \qquad -\ \sum_{n=1}^\infty
               w_{n-1}(n+\sigma-1)t^{n+\sigma}=0.
         \end{split}
      \end{equation*}
       The two sums can be combined
      \begin{equation*}
         \begin{split}
         \displaystyle w_0\Bigl[\sigma(\sigma-2)+1\Bigr]t^\sigma \\
         \displaystyle   +\, \sum_{n=1}^\infty
            \Bigl\{w_n\Bigl[(n+\sigma)(n+\sigma-2)+1\Bigr]
            -w_{n-1}(n+\sigma-1)\Bigr\}t^{n+\sigma}=0.
         \end{split}
     \end{equation*}
      The left hand side can only vanish for all  $t$ if the coefficients vanish; hence
      \begin{eqnarray}
         w_0\Bigl[\sigma(\sigma-2)+1\Bigr]&=&0, \label{eqn:5.3.1}\\
         w_n\Bigl[(n+\sigma)(n+\sigma-2)+1\Bigr]-
            w_{n-1}(n+\sigma-1)&=&0. \label{eqn:5.3.2}
      \end{eqnarray}
      ad (\ref{eqn:5.3.1}) for $w_0$:
      \begin{eqnarray*}
         \sigma(\sigma-2)+1&=&0\\
         \sigma^2-2\sigma+1&=&0\\
         (\sigma-1)^2&=&0\quad \Longrightarrow\ \sigma_\infty^{(1,2)}=1
      \end{eqnarray*}
      The characteristic exponent is $\sigma_\infty^{(1)}=
      \sigma_\infty^{(2)}=1$.\medskip\\
      ad (\ref{eqn:5.3.2}) for $w_n$:
      For the coefficients $w_n$ we obtain the recursion formula
      \begin{eqnarray*}
         w_n\Bigl[(n+\sigma)(n+\sigma-2)+1\Bigr]&=&w_{n-1}(n+\sigma-1)\\
         \Longrightarrow\ w_n&=&{n+\sigma-1\over (n+\sigma)(n+\sigma-2)+1}
         w_{n-1}.
      \end{eqnarray*}
      Let us insert $\sigma=1$:
      $$
         w_n={n\over (n+1)(n-1)+1}w_{n-1}={n\over n^2-1+1}w_{n-1}=
         {n\over n^2}w_{n-1}={1\over n}w_{n-1}.
      $$
      We can fix $w_0=1$, hence:
      \begin{eqnarray*}
         w_0&=&1={1\over 1}={1\over 0!}\\
         w_1&=&{1\over 1}={1\over 1!}\\
         w_2&=&{1\over 1\cdot 2}={1\over 2!}\\
         w_3&=&{1\over 1\cdot 2\cdot 3}={1\over 3!}\\
         &\vdots\\
         w_n&=&{1\over 1\cdot 2\cdot 3\cdot\,\cdots\,\cdot n}={1\over n!}
      \end{eqnarray*}
      And finally,
      $$
         u_1(t)=t^\sigma\sum_{n=0}^\infty w_n t^n=t\sum_{n=0}^\infty
         {t^n\over n!}=te^t .
      $$
\item Notice that both characteristic exponents are equal; hence we have to employ  the
d'Alembert reduction
      $$
         u_2(t)=u_1(t)\int\limits_0^t v(s)ds
      $$
      with
      $$
         v'(t)+v(t)\left[2{u_1'(t)\over u_1(t)}+\tilde p_1(t)\right]=0.
      $$
     Insertion of $u_1$ and $\tilde p_1$,
      \begin{eqnarray*}
         u_1(t)&=&te^t\\
         u_1'(t)&=&e^t(1+t)\\
         \tilde p_1(t)&=&-\left({1\over t}+1\right),
      \end{eqnarray*}
      yields
      \begin{eqnarray*}
         v'(t)+v(t)\left(2{e^t(1+t)\over te^t}-{1\over t}-1\right)&=&0\\
         v'(t)+v(t)\left(2{(1+t)\over t}-{1\over t}-1\right)&=&0\\
         v'(t)+v(t)\left({2\over t}+2-{1\over t}-1\right)&=&0\\
         v'(t)+v(t)\left({1\over t}+1\right)&=&0\\
         {dv\over dt}&=&-v\left(1+{1\over t}\right)\\
         {dv\over v}&=&-\left(1+{1\over t}\right)dt
      \end{eqnarray*}
      Upon integration of both sides we obtain
      \begin{eqnarray*}
         \int{dv\over v}&=&-\int\left(1+{1\over t}\right)dt\\
         \log v&=&-(t+\log t)=-t-\log t\\
         v&=&\exp(-t-\log t)=e^{-t}e^{-\log t}={e^{-t}\over t},
      \end{eqnarray*}
      and hence an explicit form of $v(t)$:
      $$
         v(t)={1\over t}e^{-t}.
      $$
      If we insert this into the equation for  $u_2$ we   obtain
      $$
         u_2(t)=te^t\int_0^t{1\over s}e^{-s}ds.
      $$

\item Therefore, with $  t={1\over z}$, $u(t)=w(z)$,
      the two linear independent solutions around the regular singular point at $z=\infty$ are
      \begin{equation}
      \begin{split}
         w_1(z)={1\over z}\exp\left({1\over z}\right)\textrm{, and}\\
         w_2(z)={1\over z}\exp\left({1\over z}\right)
                \int\limits_0^{1\over z}{1\over t}e^{-t}dt.
      \end{split}
      \end{equation}
\end{itemize}

\end{enumerate}
\eexample
}

\section{Hypergeometric function}
\index{hypergeometric function}

\subsection{Definition}
A
{\em hypergeometric series}
\index{hypergeometric series}
is a series
\begin{equation}
\sum_{j=0}^\infty c_j ,
\label{2011-m-ch-sfhserd}
\end{equation}
where the quotients $\frac{c_{j+1}}{c_j}$ are {\em rational functions}---that is, the quotient of two polynomials
$\frac{R(x)}{Q(x)}$, where $Q(x)$ is not identically zero---of $j$, so that they can be factorized:
\index{rational function}
\begin{equation}
\begin{split}
\frac{c_{j+1}}{c_j}
=
\frac{(j+a_1)(j+a_2)\cdots (j+a_p)}{(j+b_1)(j+b_2)\cdots (j+b_q)}
\left(\frac{x}{j+1}\right),
\\
 \textrm{ or } c_{j+1}
=  c_j
\frac{(j+a_1)(j+a_2)\cdots (j+a_p)}{(j+b_1)(j+b_2)\cdots (j+b_q)}
\left(\frac{x}{j+1}\right)
\\
\qquad =
 c_{j-1}
\frac{(j-1+a_1)(j-1+a_2)\cdots (j-1+a_p)}{(j-1+b_1)(j-1+b_2)\cdots (j-1+b_q)}
\\
\times
\frac{(j+a_1)(j+a_2)\cdots (j+a_p)}{(j+b_1)(j+b_2)\cdots (j+b_q)}
\left(\frac{x}{j}\right)
\left(\frac{x}{j+1}\right)
\\
\qquad =
 c_{0}
\frac{a_1 a_2\cdots a_p}{b_1 b_2\cdots b_q}
\cdots
\frac{(j-1+a_1)(j-1+a_2)\cdots (j-1+a_p)}{(j-1+b_1)(j-1+b_2)\cdots (j-1+b_q)}
\\
\times
\frac{(j+a_1)(j+a_2)\cdots (j+a_p)}{(j+b_1)(j+b_2)\cdots (j+b_q)}
\left(\frac{x}{1}\right)
\cdots
\left(\frac{x}{j}\right)
\left(\frac{x}{j+1}\right)
\\
\qquad =
 c_{0}
\frac{(a_1)_{j+1}(a_2)_{j+1}\cdots (a_p)_{j+1}}{(b_1)_{j+1}(b_2)_{j+1}\cdots (b_q)_{j+1}}
\left(\frac{x^{j+1}}{(j+1)!}\right)
.
\end{split}
\label{2011-m-ch-sfhser}
\end{equation}
The factor $j+1$ in the denominator of the first  line of (\ref{2011-m-ch-sfhser})
on the right
yields $(j+1)!$.
If it were not there ``naturally''
we may obtain it by compensation with a factor $j+1$ in the numerator.

With this iterated ratio~(\ref{2011-m-ch-sfhser}),
the hypergeometric series (\ref{2011-m-ch-sfhserd}) can be written in terms of
{\em shifted factorials},
\index{shifted factorial}
or, by another naming,
the
{\em Pochhammer symbol},
\index{Pochhammer symbol}
as
\begin{equation}
\begin{split}
\sum_{j=0}^\infty c_j
=
 c_0 \sum_{j=0}^\infty  \frac{( a_1)_j( a_2)_j\cdots ( a_p)_j}{( b_1)_j( b_2)_j\cdots ( b_q)_j}
\frac{x^j}{j!}
\\
\qquad =
c_0 {{}_pF_q} \left(
\begin{array}{cc}
a_1,\ldots ,a_p\\
b_1,\ldots ,b_q
\end{array} ; x
\right)     \textrm{, or }\\
\qquad =
c_0 {{}_pF_q} \left(
a_1,\ldots ,a_p;
b_1,\ldots ,b_q
 ; x
\right) .     \\
\end{split}
\label{2011-m-ch-sfhserd1}
\end{equation}

Apart from this definition {\it via}
hypergeometric series, the Gauss {\em hypergeometric function},
or, used synonymously,
the {\em Gauss series}
\index{Gauss series}
\index{Gauss hypergeometric function}
\begin{equation}
\begin{split}
{\;}_2F_1 \left(
\begin{array}{cc}
a ,b\\
c
\end{array} ; x
\right)
={\;}_2F_1 \left(
a ,b;c ; x
\right)
=   \sum_{j=0}^\infty  \frac{( a)_j( b)_j}{(c)_j} \frac{x^j}{j!}
\\
\qquad
=
1+ \frac{ab}{c} x   + \frac{1}{2!}\frac{a(a+1)b(b+1)}{c(c+1)} x^2
+ \cdots
\end{split}
\end{equation}
can be defined as a solution of a {\em Fuchsian differential equation}
which has at most {\em three regular singularities}
at $0$, $1$, and~$\infty$.

Indeed, any Fuchsian  equation
with
finite  {regular singularities} at $x_1$ and $x_2$ can be rewritten into the
{\em Riemann differential equation}
\index{Riemann differential equation}
(\ref{2012-m-ch-sf-eforp12359}),
which in turn can be rewritten into the
{\em Gaussian differential equation}
\index{Gaussian differential equation}
or
{\em hypergeometric differential equation}
\index{hypergeometric differential equation}
with
regular singularities
at $0$, $1$, and $\infty$.\cite[-0mm]{hille-69,birkhoff-Rota-48,KristenssonC3}
\marginnote{The Bessel equation has a regular singular point at $0$, and an irregular singular point at infinity.
\index{Bessel equation}}
This can be demonstrated by rewriting any such equation of the form
\begin{equation}
\begin{split}
 w''(x) + \left( \frac{A_1}{x-x_1}    + \frac{A_2}{x-x_2}
\right) w'(x)
\\ \qquad + \left(  \frac{B_1}{(x-x_1)^2}+\frac{B_2}{(x-x_2)^2} +\frac{C_1}{x-x_1} +\frac{C_2}{x-x_2}
\right)  w(x)  =0
\end{split}
\label{2011-m-ch-sfhserd12}
\end{equation}
through transforming Equation~(\ref{2011-m-ch-sfhserd12}) into the  {\em hypergeometric differential equation}
\index{hypergeometric differential equation}
\begin{equation}
\left[{}\frac{d^2}{dx^2}+ \frac{(a+b+1)x-c}{x(x-1)}\frac{d}{dx}+\frac{ab}{x(x-1)} \right]
{\;}_2F_1(a,b;c;x)=0,
\label{2011-m-ch-sfhserd121eq}
\end{equation}
where the solution is proportional to the Gauss hypergeometric function
\begin{equation}
w(x) \longrightarrow (x-x_1)^{\sigma^{(1)}_1} (x-x_2)^{ \sigma^{(2)}_2} {\;}_2F_1(a,b;c;x),
\end{equation}
 and the variable transform as
\begin{equation}
\begin{split}
x \longrightarrow x = \frac{x-x_1}{x_2-x_1}  \textrm{, with}\\
a=  {\sigma^{(1)}_1}+{\sigma^{(1)}_2}   +{\sigma^{(1)}_\infty},  \\
b=  {\sigma^{(1)}_1}+{\sigma^{(1)}_2}   +{\sigma^{(2)}_\infty} ,\\
c= 1+ {\sigma^{(1)}_1}   -{\sigma^{(2)}_1} .
\end{split}
\label{2011-m-ch-sfhserd121}
\end{equation}
where $\sigma^{(i)}_j$ stands for the $i$th characteristic exponent of the $j$th singularity.

{\color{OliveGreen}
\bproof

Whereas the full transformation from Equation~(\ref{2011-m-ch-sfhserd12})
to the hypergeometric differential equation  (\ref{2011-m-ch-sfhserd121eq}) will not been given, we shall show that
the Gauss hypergeometric function ${\;}_2F_1$ satisfies the hypergeometric differential equation (\ref{2011-m-ch-sfhserd121eq}).

First, define the differential operator
\begin{equation}
\vartheta = x \frac{d}{dx} ,
\label{2011-m-ch-sfhserddovt}
\end{equation}
and observe that
\begin{equation}
\begin{split}
\vartheta (\vartheta +c-1) x^n
=x \frac{d}{dx} \left( x \frac{d}{dx} +c-1\right) x^n\\  \qquad
=x \frac{d}{dx} \left( x n  x^{n-1}+c x^n- x^n\right)\\  \qquad
=x \frac{d}{dx} \left(   n  x^{n}+c x^n- x^n\right)\\     \qquad
=x \frac{d}{dx} \left(  n +c-1\right) x^n\\              \qquad
=n\left(  n +c-1\right) x^n.
\end{split}
\label{2011-m-ch-sfhserddovd1}
\end{equation}

Thus, if we apply  $\vartheta (\vartheta +c-1)$ to ${\;}_2F_1$, then
\begin{equation}
\begin{split}
\vartheta (\vartheta +c-1) {\;}_2F_1 (a,b;c;x)
=  \vartheta (\vartheta +c-1) \sum_{j=0}^\infty  \frac{( a)_j( b)_j}{(c)_j} \frac{x^j}{j!}
\\
=   \sum_{j=0}^\infty  \frac{( a)_j( b)_j}{(c)_j} \frac{j(j+c-1)x^j}{j!}
=   \sum_{j=1}^\infty  \frac{( a)_j( b)_j}{(c)_j} \frac{j(j+c-1)x^j}{j!}
\\
=   \sum_{j=1}^\infty  \frac{( a)_j( b)_j}{(c)_j} \frac{ (j+c-1)x^j}{(j-1)!}
\\
\textrm{[index shift: } j\rightarrow n+1, n=j-1, n\ge 0\textrm{]}
\\
=   \sum_{n=0}^\infty  \frac{( a)_{n+1}( b)_{n+1}}{(c)_{n+1}} \frac{ ({n+1}+c-1)x^{n+1}}{n!}
\\
=  x \sum_{n=0}^\infty  \frac{( a)_{n}(a+n)( b)_{n}(b+n)}{(c)_{n}(c+n)} \frac{ ({n }+c)x^{n}}{n!}
\\
=  x\sum_{n=0}^\infty  \frac{( a)_{n}( b)_{n}}{(c)_{n}} \frac{(a+n) (b+n)x^{n}}{n!}
\\
=   x(\vartheta +a)(\vartheta +b)  \sum_{n=0}^\infty  \frac{( a)_{n}( b)_{n}}{(c)_{n}} \frac{x^{n}}{n!}
=   x(\vartheta +a)(\vartheta +b)  {\;}_2F_1(a,b;c;x)
,
\end{split}
\label{2011-m-ch-sfhserddovd123}
\end{equation}
where we have used
\begin{equation}
\begin{split}
(a+n)x^n = (a+ \vartheta )x^n \text{, and}   \\
(a)_{n+1}=a(a+1)\cdots (a+n-1)(a+n) =  (a)_{n}(a+n).
\end{split}
\end{equation}
Writing out $\vartheta$ in  Equation~(\ref{2011-m-ch-sfhserddovd123}) explicitly yields
\begin{equation}
\begin{split}
\left\{\vartheta (\vartheta +c-1)  -   x(\vartheta +a)(\vartheta +b)\right\}  {\;}_2F_1(a,b;c;x) =0,
   \\
\left\{x \frac{d}{dx} \left(x \frac{d}{dx} +c-1\right)    -   x\left(x \frac{d}{dx}+a\right)\left(x \frac{d}{dx} +b\right) \right\}  {\;}_2F_1(a,b;c;x) =0,
   \\
\left\{ \frac{d}{dx} \left(x \frac{d}{dx} +c-1\right)    -    \left(x \frac{d}{dx}+a\right)\left(x \frac{d}{dx} +b\right) \right\}  {\;}_2F_1(a,b;c;x) =0,
   \\
\left\{ \frac{d}{dx} + x \frac{d^2}{dx^2} +(c-1)\frac{d}{dx}
 -    \left(x^2 \frac{d^2}{dx^2}+ x \frac{d}{dx}+ bx\frac{d}{dx}
\right.
\right.
  \qquad  \\
\left.
\left.
+  ax \frac{d}{dx} +ab\right) \right\}  {\;}_2F_1(a,b;c;x) =0,
   \\
\left\{  \left( x    -x^2 \right)\frac{d^2}{dx^2}
+
\left(
1+c-1-x-x(a+b)
\right)\frac{d}{dx}
+
ab  \right\}  {\;}_2F_1(a,b;c;x) =0
,   \\
\left\{ -   x(x-1)\frac{d^2}{dx^2}
-
\left(
c-x(1+a+b)
\right)\frac{d}{dx}
-
ab \right\}  {\;}_2F_1(a,b;c;x) =0,
   \\
\left\{ \frac{d^2}{dx^2}
+
\frac{x(1+a+b)-c}{x(x-1)}
\frac{d}{dx}
+
\frac{ab}{x(x-1)} \right\}  {\;}_2F_1(a,b;c;x) =0.
\end{split}
\label{2011-m-ch-sfhserddovd1234}
\end{equation}

\eproof
}

\subsection{Properties}

There exist many properties of the hypergeometric series.
In the following, we shall mention a few.

\begin{equation}
{d\over dz}{\;}_2F_1(a,b;c;z)={ab\over c}{\;}_2F_1(a+1,b+1;c+1;z) .
\end{equation}

{\color{OliveGreen}
\bproof

\begin{eqnarray*}
   {d\over dz}{\;}_2F_1(a,b;c;z)&=&{d\over dz}\sum_{n=0}^\infty
                           {(a)_n(b)_n\over(c)_n}{z^n\over n!}=\\
                        &=&\sum_{n=0}^\infty {(a)_n(b)_n\over(c)_n}
                           n{z^{n-1}\over n!}\\
                         &=&\sum_{n=1}^\infty
                           {(a)_n(b)_n\over(c)_n}{z^{n-1}\over(n-1)!}
\end{eqnarray*}
An index shift $n\to m+1$, $m=n-1$, and a subsequent renaming $m \to n$, yields
$$
   {d\over dz}{\;}_2F_1(a,b;c;z)=\sum_{n=0}^\infty
   {(a)_{n+1}(b)_{n+1}\over(c)_{n+1}}{z^n\over n!} .
$$
As
\begin{eqnarray*}
   (x)_{n+1}&=&x(x+1)(x+2)\cdots(x+n-1)(x+n)\\
   (x+1)_n  &=&\phantom{x}(x+1)(x+2)\cdots(x+n-1)(x+n)\\
   (x)_{n+1}&=&x(x+1)_n
\end{eqnarray*}
holds, we obtain
$$
   {d\over dz}{\;}_2F_1(a,b;c;z)=\sum_{n=0}^\infty{ab\over c}
   {(a+1)_n(b+1)_n\over(c+1)_n}{z^n\over n!}=
   {ab\over c}{\;}_2F_1(a+1,b+1;c+1;z).
$$

\eproof
}

We state {\em Euler's integral representation} for $\Re c>0$ and $\Re b>0$  without proof:
\begin{equation}
 {\;}_2F_1(a,b;c;x)=\frac{\Gamma(c)}{\Gamma(b)\Gamma(c-b)}
\int_0^1 t^{b-1}(1-t)^{c-b-1}(1-xt)^{-a} dt
.
\end{equation}

For  $\Re (c-a-b)>0$, we also state Gauss' theorem
\index{Gauss theorem}
\begin{equation}
 {\;}_2F_1(a,b;c;1)=\sum_{j=0}^\infty \frac{(a)_j(b)_j}{j! (c)_j} =\frac{\Gamma(c)\Gamma(c-a-b)}{\Gamma(c-a)\Gamma(c-b)}.
\end{equation}

{\color{OliveGreen}
\bproof
For a proof, we can set $x=1$ in Euler's integral representation, and the Beta function defined in Equation~
(\ref{2011-m-ch-sf-beta}).
\eproof
}

\subsection{Plasticity}

Some of the most important elementary functions can be expressed as hypergeometric series; most importantly the Gaussian one ${\;}_2F_1$,
which is
sometimes denoted by just $F$.
Let us enumerate a few.
\begin{eqnarray}
e^x
&=&
{\;}_0F_0\left(-;-;x\right)
\\
\cos x
&=&
{\;}_0F_1\left(-;\frac{1}{2};-\frac{x^2}{4}\right)
\\
\sin x
&=&
x
{\;}_0F_1\left(-;\frac{3}{2};-\frac{x^2}{4}\right)
\\
(1-x)^{-a}
&=&
{\;}_1F_0\left(a;-  ;x\right)
\\
\sin^{-1} x
&=&
x
{\;}_2F_1\left(\frac{1}{2},\frac{1}{2};\frac{3}{2};x^2\right)
\\
\tan^{-1} x
&=&
x
{\;}_2F_1\left(\frac{1}{2},1;\frac{3}{2};-x^2\right)
\\
\log (1 + x)
&=&
x
{\;}_2F_1\left(1,1;2;-x\right)
\\
H_{2n}(x)
&=&
\frac{(-1)^n(2n)!}{n!}
{\;}_1F_1\left(-n;\frac{1}{2}; x^2\right)
\\
H_{2n+1}(x)
&=&
2x
\frac{(-1)^n(2n+1)!}{n!}
{\;}_1F_1\left(-n;\frac{3}{2}; x^2\right)
\\
L_{n}^\alpha (x)
&=&
\left(
\begin{array}{c}
n+\alpha\\
n
\end{array}
\right)
{\;}_1F_1\left(-n;\alpha +1; x\right)
\\
P_{n}(x)
&=&   P^{(0, 0 )}_n (x)
=
{\;}_2F_1\left(-n,n+1; 1;\frac{1- x}{2}\right) ,
\\
C_{n}^\gamma (x)
&=& \frac{(2\gamma )_n}{\left( \gamma +\frac{1}{2}\right)_n}  P^{(\gamma -\frac{1}{2}, \gamma -\frac{1}{2} )}_n (x)
 ,
\\
T_{n}  (x)
&=& \frac{n!}{\left(  \frac{1}{2}\right)_n}  P^{( -\frac{1}{2},  -\frac{1}{2} )}_n (x)
,
\\
J_{\alpha }  (x)
&=& \frac{\left(\frac{x}{2}\right)^\alpha }{\Gamma (\alpha +1)}
{\;}_0F_1\left(- ; \alpha +1;-\frac{1 }{4}x^2\right) ,
\end{eqnarray}
where
$H$ stands for
{\em Hermite polynomials},
\index{Hermite polynomial}
$L$ for
{\em Laguerre polynomials},
\index{Laguerre polynomial}
\begin{equation}
P^{(\alpha, \beta )}_n (x)
=\frac{(\alpha +1)_n}{n!} {\;}_2F_1\left( -n,n+\alpha +\beta +1; \alpha +1;\frac{1- x}{2}\right)
\end{equation}
for
{\em Jacobi polynomials},
\index{Jacobi polynomial}
$C$ for
{\em Gegenbauer polynomials},
\index{Gegenbauer polynomial}
$T$ for
{\em Chebyshev polynomials},
\index{Chebyshev polynomial}
$P$  for
{\em Legendre polynomials},
\index{Legendre polynomial}
and
$J$  for   the
{\em Bessel functions of the first kind},
\index{Bessel function}
respectively.

{
\color{blue}
\bexample

\begin{enumerate}

\item
Let us prove that
 $$\log (1-z)=-z {\;}_2F_1(1,1,2;z) . $$
Consider
$$
   {\;}_2F_1(1,1,2;z)=\sum_{m=0}^\infty{[(1)_m]^2\over(2)_m}{z^m\over m!}=
   \sum_{m=0}^\infty{[1\cdot2\cdot\,\cdots\,\cdot m]^2\over
   2\cdot(2+1)\cdot\,\cdots\,\cdot(2+m-1)}{z^m\over m!}
$$
With
$$
   (1)_m=1\cdot2\cdot\,\cdots\,\cdot m=m!,\qquad
   (2)_m=2\cdot(2+1)\cdot\,\cdots\,\cdot (2+m-1)=(m+1)!
$$
follows
$$
   {\;}_2F_1(1,1,2;z)=\sum_{m=0}^\infty{[m!]^2\over(m+1)!}
   {z^m\over m!}=\sum_{m=0}^\infty {z^m\over m+1}.
$$
Index shift $k =m+1$
$$
   {\;}_2F_1(1,1,2;z)=\sum_{k=1}^\infty {z^{k-1}\over k}
$$
and hence
$$
   -z{\;}_2F_1(1,1,2;z)=-\sum_{k=1}^\infty{z^k\over k}.
$$
Compare with the series
$$
   \log(1+x)=\sum_{k=1}^\infty(-1)^{k+1}{x^k\over k}\qquad
   \mbox{for}\quad -1<x\leq1
$$
If one substitutes $-x$ for $x$, then
$$
   \log(1-x)=-\sum_{k=1}^\infty{x^k\over k}.
$$
The identity follows from the analytic continuation of $x$ to the complex $z$ plane.

\item
Let us prove that,  because of
$
(a+z)^n
=
\sum_{k=0}^n
\begin{pmatrix}
n\\
k
\end{pmatrix}
 z^ka^{n-k}$,
 $$(1-z)^n={\;}_2F_1(-n,1,1;z). $$

$$
   {\;}_2F_1(-n,1,1;z)=\sum_{i=0}^\infty{(-n)_i(1)_i\over(1)_i}{z^i\over i!}=
   \sum_{i=0}^\infty(-n)_i{z^i\over i!}.
$$
Consider $(-n)_i$
$$
   (-n)_i=(-n)(-n+1)\cdots(-n+i-1).
$$

For  $n\geq 0$
the series stops after a finite number of terms, because the factor
$-n+i-1=0$ for $i=n+1$ vanishes;
hence the sum of $i$ extends only
from $0$ to $n$.
Hence, if we collect the factors
 $(-1)$ which yield $(-1)^i$ we obtain
$$
   (-n)_i=(-1)^in(n-1)\cdots[n-(i-1)]=(-1)^i{n!\over(n-i)!}.
$$
Hence, insertion into the Gauss hypergeometric function yields
$$
   {\;}_2F_1(-n,1,1;z)=\sum_{i=0}^n(-1)^iz^i{n!\over i!(n-i)!}=
   \sum_{i=0}^n{n\choose i}(-z)^i.
$$
This is the binomial series
$$
   (1+x)^n=\sum_{k=0}^n{n\choose k}x^k
$$
with $x=-z$; and hence,
$$
   {\;}_2F_1(-n,1,1;z)=(1-z)^n.
$$

\item
 Let us prove that,  because of $\arcsin x=\sum_{k=0}^\infty
 {(2k)!x^{2k+1}\over 2^{2k}(k!)^2(2k+1)}$,
 $${\;}_2F_1\left({1\over 2},{1\over 2},{3\over 2}; \sin^2 z\right)
 ={z\over \sin z }  .$$

Consider
$$
   {\;}_2F_1\left({1\over 2},{1\over 2},{3\over 2};\sin^2z\right)=
   \sum_{m=0}^\infty{\left[\left({1\over 2}\right)_m\right]^2\over
   \left({3\over 2}\right)_m}{(\sin z)^{2m}\over m!}.
$$
We take
\begin{eqnarray*}
   (2n)!!&=& 2\cdot4\cdot\,\cdots\,\cdot(2n)=n!2^n\\
   (2n-1)!!&=& 1\cdot3\cdot\,\cdots\,\cdot(2n-1)={(2n)!\over2^n n!}
\end{eqnarray*}
Hence
\begin{eqnarray*}
   \left({1\over 2}\right)_m\!\!\!&=\!\!\!&{1\over 2}\cdot
   \left({1\over 2}+1\right)\cdots\left({1\over 2}+m-1\right)=
   {1\cdot3\cdot5\cdots(2m-1)\over2^m}={(2m-1)!!\over2^m}\\
   \left({3\over 2}\right)_m\!\!\!&=\!\!\!&{3\over 2}\cdot
   \left({3\over 2}+1\right)\cdots\left({3\over 2}+m-1\right)=
   {3\cdot5\cdot7\cdots(2m+1)\over2^m}={(2m+1)!!\over2^m}
\end{eqnarray*}
Therefore,
$$
   {\left({1\over 2}\right)_m\over\left({3\over 2}\right)_m}=
   {1\over 2m+1}.
$$
On the other hand,
\begin{eqnarray*}
   (2m)!&=&1\cdot2\cdot3\cdot\,\cdots\,\cdot(2m-1)(2m)=(2m-1)!!(2m)!!=\\
        &=&1\cdot3\cdot5\cdot\,\cdots\,\cdot(2m-1)\cdot
           2\cdot4\cdot6\cdot\,\cdots\,\cdot(2m)=\\
        &=&\left({1\over 2}\right)_m2^m\cdot2^m m!=2^{2m}m!
           \left({1\over 2}\right)_m
   \Longrightarrow\left({1\over 2}\right)_m={(2m)!\over 2^{2m}m!}
\end{eqnarray*}
Upon insertion one obtains
$$
   F\left({1\over 2},{1\over 2},{3\over 2};\sin^2z\right)=
   \sum_{m=0}^\infty{(2m)!(\sin z)^{2m}\over 2^{2m}(m!)^2(2m+1)}.
$$
Comparing with the series for arcsin one finally obtains
$$
   \sin z F\left({1\over 2},{1\over 2},{3\over 2};\sin^2z\right)=
   \arcsin(\sin z)=z.
$$
\end{enumerate}
\eexample
}

\subsection{Four forms}

We state without proof the four forms of the
Gauss hypergeometric function.\cite{macrobert:1967:she}
\begin{eqnarray}
{\;}_2F_1(a,b;c;x)
&=&(1-x)^{c-a-b}{\;}_2F_1(c-a,c-b;c;x)\\
&=&(1-x)^{ -a }{\;}_2F_1\left( a,c-b;c;\frac{x}{x-1}\right)\\
&=&(1-x)^{ -b }{\;}_2F_1\left( b,c-a;c;\frac{x}{x-1}\right).
\end{eqnarray}

\section{Orthogonal polynomials}
\label{2012-m-sf-fs}

Many systems or sequences of functions may serve as a {\em basis of linearly independent functions}
which are capable to ``cover'' -- that is, to approximate -- certain functional classes.\cite{herman-sc,Marcellan}
We have already encountered at least two such prospective bases [cf. Equation~(\ref{2011-m-fa-e1fc})]:
\begin{equation}
\{1,x,x^2,\ldots ,x^k,\ldots \} \textrm { with } f(x) =\sum_{k=0}^\infty c_k x^k,
\label{2011-m-ch-sfe1}
\end{equation}
and
\begin{equation}
\begin{split}
\left\{e^{ikx} \mid k\in {\Bbb Z} \right\} \quad \textrm{ for } f(x+2\pi)=f(x) \\
\qquad \textrm {  with }
f(x)= \sum _{k=-\infty}^\infty c_k e^{ikx},\\
\qquad  \textrm{  where }
c_k=\frac{1}{2\pi}\int_{-\pi}^\pi f(x) e^{-ikx} dx.
\end{split}
\end{equation}

In order to claim existence of such functional basis systems, let us first define what  {\em orthogonality} means in the  functional context.
\index{orthogonal functions}
Just as for linear vector spaces, we can define an
{\em inner product}
or {\em scalar product}
[cf. also Equation~(\ref{2012-m-ch-sfesp1})]
\index{inner product}
\index{scalar product}
of two real-valued functions $f(x)$ and $g(x)$ by the integral\cite{Wilf}
\begin{equation}
\langle   f \mid g\rangle
=
\int_a^b f(x)g(x) \rho(x) dx
\label{2011-m-ch-sfesp}
\end{equation}
for some suitable {\em weight function} $\rho (x)\ge 0$.  \index{weight function}
Very often, the weight function is set to the identity; that is,
$\rho(x) =\rho =1$.
We notice without proof that $\langle   f \vert g\rangle  $ satisfies all requirements of a scalar product.
A system of functions $\{\psi_0,\psi_1,\psi_2,\ldots ,\psi_k,\ldots \}$
is orthogonal if, for $j\neq k$,
\begin{equation}
\langle   \psi_j \mid \psi_k\rangle
=
\int_a^b \psi_j(x)\psi_k(x) \rho(x) dx
=0.
\label{2011-m-ch-sfespof}
\end{equation}

Suppose, in some generality,
that
 $\{f_0,f_1,f_2,\ldots ,f_k,\ldots \}$
is a sequence of nonorthogonal functions.
Then we can apply a {\em Gram-Schmidt orthogonalization process} to these functions
\index{Gram-Schmidt process}
and thereby obtain orthogonal functions
$\{\phi_0,\phi_1,\phi_2,\ldots ,\phi_k,\ldots \}$
by
\begin{equation}
\begin{split}
\phi_0 (x) =f_0(x), \\
\phi_k (x)
=
f_k(x)
-
\sum_{j=0}^{k-1}\frac{\langle f_k\mid \phi_j \rangle}{\langle \phi_j   \mid \phi_j \rangle}\phi_j (x).
\end{split}
\label{2011-m-ch-sfegs}
\end{equation}
Note that the proof of the {  Gram-Schmidt  process} in the functional
context is analogous to the one in the vector context.

\section{Legendre polynomials}
\label{2013-m-sf-lp}
\index{Legendre polynomial}

The polynomial functions in $\{1,x,x^2,\ldots ,x^k,\ldots \}$
are not mutually orthogonal because,
for instance,  with $\rho =1$ and $b=-a=1$,
\begin{equation}
\langle   1 \mid x^2\rangle
=
\int_{a=-1}^{b=1} x^2 dx
=
\left. \frac{x^3}{3} \right|_{x=-1}^{x=1}
=
\frac{2}{3}.
\end{equation}
Hence, by the   Gram-Schmidt  process we obtain
\begin{equation}
\begin{split}
\phi_0(x) =1,       \\
\phi_1(x) = x - \frac{\langle x \mid 1 \rangle}{\langle 1   \mid 1 \rangle}1
= x - 0 =x,   \\
\phi_2(x) = x^2 - \frac{\langle x^2 \mid 1 \rangle}{\langle 1   \mid 1 \rangle}1
 - \frac{\langle x^2 \mid x \rangle}{\langle x   \mid x \rangle}x
=  x^2 - \frac{2/3}{2}1 -0x    =    x^2 - \frac{1 }{ 3},\\
\vdots  \qquad\qquad
\end{split}
\end{equation}
If, on top of orthogonality,  we are ``forcing'' a type of ``normalization'' by defining
\begin{equation}
\begin{split}
P_l(x)\stackrel{{\tiny \textrm{ def }}}{=}\frac{\phi_l(x)}{\phi_l(1)},
\\
 \textrm{with }\; P_l(1)  =1,
\end{split}
\end{equation}
then the resulting orthogonal polynomials are the
{\em Legendre polynomials}  $P_l$; in particular,
\index{Legendre polynomials}
\begin{equation}
\begin{split}
P_0(x) =1,\\
P_1(x) =x,\\
P_2(x) =  \left. \left( x^2 - \frac{1 }{ 3}\right)\bigg/ \frac{2 }{ 3}  \right.
       =  \frac{1 }{ 2} \left( 3x^2 - 1\right)
,
\ldots
\end{split}
\label{2011-m-ch-sfelp}
\end{equation}
with $P_l(1)=1$, $l={\Bbb N}_0$.

Why should we be interested in orthonormal systems of functions?
Because, as pointed out earlier in the context of hypergeometric functions,
they could be alternatively defined as the eigenfunctions and solutions of certain differential equation,
such as, for instance, the Schr\"odinger equation, which may be subjected to a separation of variables.
For Legendre polynomials the associated differential equation is the  {\em Legendre equation}
\index{Legendre equation}
\begin{equation}
\begin{split}
\left\{
(1-x^2) \frac{d^2}{dx^2}
- 2x  \frac{d}{dx} +l(l+1)
\right\}P_l(x) = 0,
\\
\textrm{or } \left\{  \frac{d }{dx }\left[\left(1-x^2\right)\frac{d }{dx }\right]
+
 l(l+1) \right\}  P_l(x)= 0
\end{split}
\label{2011-m-ch-sfelpede}
\end{equation}
for $l\in {\Bbb N}_0$,
whose Sturm-Liouville form has been mentioned earlier in Table \ref{2011-m-sl-t-varieties}
on page \pageref{2011-m-sl-t-varieties}.
For a proof, we refer to the literature.

\subsection{Rodrigues formula}
\index{Rodrigues formula}

A third alternative definition  of Legendre polynomials
is by the  Rodrigues formula: for $l\ge 0$,
\begin{equation}
P_l(x)=\frac{1}{2^ll!}\frac{d^l}{dx^l}(x^2-1)^l \textrm{, for }l\in {\Bbb N}_0.
\label{2011-m-ch-sfrf}
\end{equation}
No proof of equivalence will be given.

For even $l$, $P_l(x)=P_l(-x)$ is an even function of $x$,
whereas
for odd $l$, $P_l(x)=-P_l(-x)$ is an odd function of $x$;
that is,
\begin{equation}
P_l(-x)=(-1)^lP_l(x).
\end{equation}
Moreover,
\begin{equation}
P_l(-1)=(-1)^l
\end{equation}
and, for $0 \le l \in {\mathbb N}$
\begin{equation}
P_l(0)
=\begin{cases}
0  &
\text{ for odd } l=2k+1, \; 0\le k\in {\mathbb N},   \\
\frac{(-1)^\frac{l}{2} (l)!}{2^{l} \left(\frac{l}{2}!\right)^2}  &
\text{ for even } l=2k, \; 0\le k\in {\mathbb N}.
\end{cases}
\label{2019-mm-ch-sf-lpaz}
\end{equation}

{\color{OliveGreen}
\bproof

Some of these equalities can be shown by  insertion into  the  Rodrigues formula, followed by a substitution:
\begin{eqnarray*}
   P_l(-x)&=&\left.{1\over 2^ll!}{d^l\over du^l}(u^2-1)^l\right|_{u=-x}
   =[u\to-u]=\\
   &=&\left.{1\over(-1)^l}{1\over2^ll!}{d^l\over du^l}(u^2-1)^l
   \right|_{u=x}=(-1)^lP_l(x).
\end{eqnarray*}

Because of the ``normalization'' $P_l(1)=1$ we obtain $P_l(-1)=(-1)^lP_l(1)=(-1)^l$.

And as $P_l(-0)=P_l(0)=(-1)^lP_l(0)$, we obtain $P_l(0)=0$ for odd $l$.
For even $l$ a proof of~(\ref{2019-mm-ch-sf-lpaz}) by the  Rodrigues formula
is rather lengthy and will not be given here.\sidenote[][]{See
\url{https://math.stackexchange.com/questions/1218068/proving-a-property-of-legendre-polynomials/1231213\#1231213}
for a derivation.}

\eproof
}

\subsection{Generating function}
\index{generating function}

For
$\vert x\vert<1$ and
$\vert t\vert<1$ the
Legendre polynomials $P_l(x)$ are the coefficients in the Taylor series expansion of the following generating function
\begin{equation}
g(x,t) =\frac{1}{\sqrt{1-2xt+t^2}}=\sum_{l=0}^\infty P_l(x)\, t^l
\end{equation}
 around $t=0$.
No proof is given here.

\subsection{The three term and other recursion formul\ae}
\index{three term recursion formula}

Among other things, generating functions are used for the derivation of certain recursion relations involving Legendre polynomials.

For instance, for $l=1,2,\ldots$, the   three term recursion formula
\begin{equation}
(2l+1)x P_l(x) = (l+1) P_{l+1}(x)+lP_{l-1}(x) ,
\label{2020-lp1}
\end{equation}
or, by substituting $l-1$ for $l$,  for $l=2,3\ldots$,
\begin{equation}
(2l-1)x P_{l-1}(x) = l P_{l}(x)+(l-1)P_{l-2}(x),
\end{equation}
can be proven as follows.

{\color{OliveGreen}
\bproof

$$
   g(x,t)={1\over\sqrt{1-2tx+t^2}}=\sum_{n=0}^\infty t^nP_n(x)
$$
$$
   {\partial\over\partial t}g(x,t)=-{1\over 2}(1-2tx+t^2)^{-{3\over 2}}
   (-2x+2t)={1\over\sqrt{1-2tx+t^2}}\,{x-t\over1-2tx+t^2}
$$
$$
   {\partial\over\partial t}g(x,t)={x-t\over1-2tx+t^2}
   \sum_{n=0}^\infty t^nP_n(x)=\sum_{n=0}^\infty nt^{n-1}P_n(x)
$$
$$
   (x-t)\sum_{n=0}^\infty t^nP_n(x)-(1-2tx+t^2)
   \sum_{n=0}^\infty nt^{n-1}P_n(x)=0
$$
$$
   \sum_{n=0}^\infty xt^nP_n(x)-
   \sum_{n=0}^\infty t^{n+1}P_n(x)-
   \sum_{n=1}^\infty nt^{n-1}P_n(x)+
$$
$$
   \hspace*{5cm}+\,\sum_{n=0}^\infty 2xnt^nP_n(x)-
   \sum_{n=0}^\infty nt^{n+1}P_n(x)=0
$$
$$
   \sum_{n=0}^\infty (2n+1)xt^nP_n(x)-
   \sum_{n=0}^\infty (n+1)t^{n+1}P_n(x)-
   \sum_{n=1}^\infty nt^{n-1}P_n(x)=0
$$
$$
   \sum_{n=0}^\infty (2n+1)xt^nP_n(x)-
   \sum_{n=1}^\infty nt^{n}P_{n-1}(x)-
   \sum_{n=0}^\infty (n+1)t^nP_{n+1}(x)=0,
$$
$$
   xP_0(x)-P_1(x)+\sum_{n=1}^\infty t^n\Bigl[(2n+1)xP_n(x)-nP_{n-1}(x)
   -(n+1)P_{n+1}(x)\Bigr]=0,
$$
hence
$$
    xP_0(x)-P_1(x)=0, \qquad (2n+1)xP_n(x)-nP_{n-1}(x)-
   (n+1)P_{n+1}(x)=0,
$$
hence
$$
  P_1(x)=xP_0(x), \qquad (n+1)P_{n+1}(x)=
   (2n+1)xP_n(x)-nP_{n-1}(x).
$$

\eproof
}

Let us prove
\begin{equation}
P_{l-1}(x)=P'_l(x)-2xP'_{l-1}(x)+P'_{l-2}(x) .
\label{2020-lp2}
\end{equation}

{\color{OliveGreen}
\bproof

$$
   g(x,t)={1\over\sqrt{1-2tx+t^2}}=\sum_{n=0}^\infty t^nP_n(x)
$$
$$
   {\partial\over\partial x}g(x,t)=-{1\over 2}(1-2tx+t^2)^{-{3\over 2}}
   (-2t)={1\over\sqrt{1-2tx+t^2}}\,{t\over1-2tx+t^2}
$$
$$
   {\partial\over\partial x}g(x,t)={t\over1-2tx+t^2}
   \sum_{n=0}^\infty t^nP_n(x)=\sum_{n=0}^\infty t^{n}P'_n(x)
$$
$$
   \sum_{n=0}^\infty t^{n+1}P_n(x)=\sum_{n=0}^\infty t^{n}P'_n(x)-
   \sum_{n=0}^\infty 2xt^{n+1}P'_n(x)+\sum_{n=0}^\infty t^{n+2}P'_n(x)
$$
$$
   \sum_{n=1}^\infty t^{n}P_{n-1}(x)=\sum_{n=0}^\infty t^{n}P'_n(x)-
   \sum_{n=1}^\infty 2xt^{n}P'_{n-1}(x)+\sum_{n=2}^\infty t^{n}P'_{n-2}(x)
$$
$$
   tP_0+\sum_{n=2}^\infty t^{n}P_{n-1}(x)=P'_0(x)+tP'_1(x)+
   \sum_{n=2}^\infty t^{n}P'_n(x)-
$$
$$
   \hspace*{5cm}-\,2xtP'_0-\sum_{n=2}^\infty 2xt^{n}P'_{n-1}(x)+
   \sum_{n=2}^\infty t^{n}P'_{n-2}(x)
$$
$$
   P'_0(x)+t\Bigl[P'_1(x)-P_0(x)-2xP'_0(x)\Bigr]+\hspace*{3cm}
$$
$$
   \hspace*{3cm}+\,\sum_{n=2}^\infty t^{n}[P'_n(x)-2xP'_{n-1}(x)+P'_{n-2}(x)-
   P_{n-1}(x)]=0
$$
$$
   P'_0(x)=0\textrm{, hence } P_0(x)={\rm const.}
$$
$$
   P'_1(x)-P_0(x)-2xP'_0(x)=0.
$$
Because of $P'_0(x)=0$ we obtain $P'_1(x)-P_0(x)=0$, hence $P'_1(x)=P_0(x)$, and
$$
   P'_n(x)-2xP'_{n-1}(x)+P'_{n-2}(x)-P_{n-1}(x)=0.
$$
Finally we substitute $n+1$ for $n$:
$$
   P'_{n+1}(x)-2xP'_{n}(x)+P'_{n-1}(x)-P_{n}(x)=0,
$$
hence
$$
    P_n(x)=P'_{n+1}(x)-2xP'_{n}(x)+P'_{n-1}(x).
$$

\eproof
}

Let us prove, from~(\ref{2020-lp1}) and~(\ref{2020-lp2}), that
\begin{equation}
P'_{l+1}(x)-P'_{l-1}(x)=(2l+1)P_l(x) .
\end{equation}

{\color{OliveGreen}
\bproof
Using~(\ref{2020-lp1}),
\begin{eqnarray*}
   (n+1)P_{n+1}(x)&=&(2n+1)xP_n(x)-nP_{n-1}(x)\quad\left|
   {d\over dx}\right.\\
   (n+1)P'_{n+1}(x)&=&(2n+1)P_n(x)+(2n+1)xP'_n(x)-nP'_{n-1}(x)
   \quad\Bigl|\cdot\,2\\
   \textrm{(i):}\quad(2n+2)P'_{n+1}(x)&=&2(2n+1)P_n(x)+2(2n+1)xP'_n(x)-2nP'_{n-1}(x)
\end{eqnarray*}
Using~(\ref{2020-lp2}),
$$
   P'_{n+1}(x)-2xP'_{n}(x)+P'_{n-1}(x)=P_{n}(x)\quad\Bigl|
   \cdot\,(2n+1)
$$
$$
   \textrm{(ii):}\quad (2n+1)P'_{n+1}(x)-2(2n+1)xP'_{n}(x)+(2n+1)P'_{n-1}(x)=
   (2n+1)P_{n}(x)
$$
We subtract (ii) from (i):
$$
   P'_{n+1}(x)+2(2n+1)xP'_n(x)-(2n+1)P'_{n-1}(x)=
$$
$$
   \hspace*{5cm}=\,(2n+1)P_n(x)+2(2n+1)xP'_n(x)-2nP'_{n-1}(x);
$$
hence
$$
   P'_{n+1}(x)-P'_{n-1}(x)=(2n+1)P_n(x).
$$
\eproof
}

\subsection{Expansion in Legendre polynomials}

We state without proof  that square integrable functions $f(x)$
can be written as series of Legendre polynomials
as
\begin{equation}
\begin{split}
 f(x)=\sum_{l=0}^\infty a_lP_l(x),\\
\textrm{with expansion coefficients } a_l={2l+1\over 2}
\int\limits_{-1}^{+1}f(x)P_l(x)dx.
 \end{split}
\end{equation}

{
\color{blue}
\bexample
Let us expand the Heaviside function defined in Equation~(\ref{2011-m-di-edhf})
\index{Heaviside function}
\begin{equation}
H(x)
=
\left\{
\begin{array}{rl}
1&\textrm{ for } x\ge  0\\
0&\textrm{ for } x <  0
\end{array}
\right.
\end{equation}
in terms of Legendre polynomials.

We shall use the recursion formula $(2l+1)P_l=P'_{l+1}-P'_{l-1}$ and rewrite
\begin{eqnarray*}
 a_l &=& {1\over 2}\int\limits_0^1\bigl(P'_{l+1}(x)-
              P'_{l-1}(x)\bigr)dx=
              {1\over 2}\bigl(P_{l+1}(x)-P_{l-1}(x)\bigr)
              \biggr|_{x=0}^1=\\
            &=& {1\over 2}\underbrace{\bigl[P_{l+1}(1)-P_{l-1}(1)\bigr]}
                _{\mbox{$=0$ because of}\atop \mbox{``normalization''}}-
                {1\over 2}\bigl[P_{l+1}(0)-P_{l-1}(0)\bigr]
.
\end{eqnarray*}
Note that $P_n(0)=0$ for {\em odd} $n$; hence $  a_l=0$ for
{\em even} $l\ne0$. We shall treat the case $l=0$ with $P_0(x)=1$ separately.
Upon substituting $2l+1$ for $l$ one obtains
$$
   a_{2l+1}=-{1\over 2}\biggl[P_{2l+2}(0)-P_{2l}(0)\biggr].
$$
Next, for even $l$, we shall use the formula~(\ref{2019-mm-ch-sf-lpaz})
$$
   P_l(0)=(-1)^{l\over 2}{l!\over
2^l\left(\left({l\over2}\right)!\right)^2},
$$
and, for {\em even} $l\geq 0$,  one obtains
\begin{eqnarray*}
   a_{2l+1}&=&-{1\over 2}\left[{(-1)^{l+1}(2l+2)!\over 2^{2l+2}((l+1)!)^2}
              -{(-1)^l(2l)!\over 2^{2l}(l!)^2}\right]=\\
   &=&(-1)^l{(2l)!\over 2^{2l+1}(l!)^2}\left[{(2l+1)(2l+2)\over 2^2(l+1)^2}
               +1\right]=\\
   &=&(-1)^l{(2l)!\over 2^{2l+1}(l!)^2}\left[{2(2l+1)(l+1)\over 2^2(l+1)^2}
               +1\right]=\\
   &=&(-1)^l{(2l)!\over 2^{2l+1}(l!)^2}\left[{2l+1+2l+2\over 2(l+1)}\right]=\\
   &=&(-1)^l{(2l)!\over 2^{2l+1}(l!)^2}\left[{4l+3\over2(l+1)}\right]=\\
   &=&(-1)^l{(2l)!(4l+3)\over 2^{2l+2}l!(l+1)!}\\
   a_0&=&{1\over 2}\int\limits_{-1}^{+1} H (x)\underbrace{P_0(x)}_
               {\mbox{$=1$}}dx={1\over 2}\int\limits_0^1dx={1\over 2};
\end{eqnarray*}
and finally
$$
  H (x)={1\over 2}+\sum_{l=0}^\infty
   (-1)^l{(2l)!(4l+3)\over2^{2l+2}l!(l+1)!}P_{2l+1}(x).
$$
\eexample
}

\section{Associated Legendre polynomial}
\index{associated Legendre polynomial}

Associated Legendre polynomials $P_l^m(x)$ are the solutions of the
{\em general Legendre equation}
\index{general Legendre equation}

\begin{equation}
\begin{split}
\left\{
(1-x^2)\frac{d^2}{dx^2}
-2x  \frac{d }{dx }
+
\left[ l(l+1)-\frac{m^2}{1-x^2} \right] \right\} P_l^m(x)= 0,
\\
\textrm{or } \left[  \frac{d }{dx }\left((1-x^2)\frac{d }{dx }\right)
+
 l(l+1)-\frac{m^2}{1-x^2} \right]  P_l^m(x)= 0
\end{split}
\label{2011-m-ch-sfgle}
\end{equation}
Equation~(\ref{2011-m-ch-sfgle})
reduces to the Legendre equation
 (\ref{2011-m-ch-sfelpede})
on page \pageref{2011-m-ch-sfelpede} for  $m=0$;
hence
\begin{equation}
P_l^0(x)=P_l(x).
\end{equation}
More generally,
by differentiating $m$ times the Legendre equation (\ref{2011-m-ch-sfelpede})
it can be shown that
\begin{equation}
P_l^m(x)=(-1)^m(1-x^2)^\frac{m}{2} \frac{d^m}{dx^m}P_l(x).
\end{equation}
By inserting $P_l(x)$ from the
{\em Rodrigues formula}
\index{Rodrigues formula}
for Legendre polynomials (\ref{2011-m-ch-sfrf})
we obtain
\begin{equation}
\begin{split}
P_l^m(x)=
(-1)^m(1-x^2)^\frac{m}{2} \frac{d^m}{dx^m} \frac{1}{2^ll!}\frac{d^l}{dx^l}(x^2-1)^l
\\
\qquad
=
\frac{(-1)^m(1-x^2)^\frac{m}{2}}{2^ll!} \frac{d^{m+l}}{dx^{m+l}}(x^2-1)^l
.
\end{split}
\end{equation}

In terms of the Gauss hypergeometric function
\index{hypergeometric function}
the associated Legendre polynomials can be generalized to
arbitrary complex indices $\mu$, $\lambda$ and argument $x$ by
\begin{equation}
P^\mu_\lambda (x) =
\frac{1}{\Gamma (1-\mu )}\left(\frac{1+x}{1-x}\right)^\frac{\mu}{2}
{}_2F_1
\left(
-\lambda , \lambda  +1; 1-\mu ;\frac{1-x}{2}
\right).
\end{equation}
No proof is given here.

\section{Spherical harmonics}
\index{spherical harmonics}
\label{2011-m-ch-sfshar}


Let us define the {\em spherical harmonics} $Y_l^m (\theta ,\varphi )$ by
\begin{equation}
Y_l^m (\theta ,\varphi ) =\sqrt{\frac{(2l+1)(l-m)!}{4\pi (l+m)!} }
P_l^m(\cos \theta )e^{im\varphi }\textrm{ for } -l\le m\le l.
.
\label{2014-m-ch-sf-sh}
\end{equation}

\marginnote{Twice continuously differentiable,
complex-valued solutions $u$ of the Laplace equation
$\Delta u =0$
are called  \index{harmonic function}
{\em harmonic functions}:~\bibentry{Axler:1994:HFT}.}
Spherical harmonics are solutions of the differential equation
\begin{equation}
\left\{
\Delta + l(l+1)
\right\}
Y_l^m (\theta ,\varphi ) =0
.
\end{equation}
This equation is what typically remains after separation and ``removal'' of the
radial part of the Laplace equation $\Delta \psi(r,\theta ,\varphi)=0$ in three dimensions
when the problem is invariant (symmetric) under rotations.

\section{Solution of the Schr\"odinger equation for a hydrogen atom}
\index{Schr\"odinger equation}

Suppose Schr\"odinger, in his 1926 {\it annus mirabilis}
--
a year
which seems to have been initiated by a trip to Arosa with `an old girlfriend from Vienna'
(apparently, it was neither his wife Anny who remained in Zurich, nor Lotte, nor Irene nor Felicie\cite{Moore-Schroedinger}),
--
came down from the mountains or from whatever realm he was in -- and
handed you over some partial differential equation  for the hydrogen atom
-- an equation
note that in the ``first quantization''  the quantum mechanical ``momentum operator'' ${\cal P}$ is identified with  $- i \hslash \nabla$)
\begin{equation}
\begin{split}
  \frac{1}{2\mu } {\cal P} ^2  \psi =\frac{1}{2\mu }\left({\cal P}_x^2 +{\cal P}_y^2 +{\cal P}_z^2\right) \psi = \left( E-V \right)\psi ,\\
\textrm{  or, with } V=-\frac{e^2}{4\pi \epsilon_0 r}, \\
 -\left( \frac{\hslash^2}{2\mu }\Delta +  \frac{e^2}{4\pi \epsilon_0r} \right)\psi ({\bf x})= E\psi, \\
 \textrm{ or }\left[ \Delta + \frac{2\mu }{\hslash^2} \left(\frac{e^2}{4\pi \epsilon_0r} + E \right)\right]\psi ({\bf x})=0,
\end{split}
\label{2011-m-ch-qae-sebf}
\end{equation}
which would later bear his name  -- and asked you if you could be so kind to please solve it for him.
Actually, by Schr\"odinger's own account\cite[-30mm]{ANDP:ANDP19263840404} this is exactly what he did:
\marginnote{In two-particle situations without external forces it is common to define the reduced mass
$\mu$ by $\frac{1}{\mu} = \frac{1}{m_1} +\frac{1}{m_2}= \frac{m_2+m_1}{m_1m_2}$, or $\mu= \frac{m_1m_2}{m_2+m_1}$,
where $m_1$ and $m_2$ are the masses of the constituent particles, respectively.
In this case, one can identify the electron mass $m_e$ with $m_1$, and the nucleon (proton) mass
$m_p\approx 1836 m_e \gg m_e$ with $m_2$, thereby allowing the approximation $\mu = \frac{m_e m_p}{m_e+m_p} \approx \frac{m_e m_p}{m_p} = m_e$.}
he handed  over this eigenwert equation to Hermann Klaus Hugo Weyl;
in this instance, he was not dissimilar from Einstein, who seemed to have employed a (human) computist on a very regular basis.
Schr\"odinger might also have hinted that $\mu$, $e$, and $\epsilon_0$ stand for some
\index{reduced mass}
(reduced) mass,
charge, and the  permittivity of the vacuum, respectively,
$\hslash$ is a constant of (the dimension of) action,
and $E$ is some eigenvalue which must be determined from
the solution of (\ref{2011-m-ch-qae-sebf}).

So, what could you do?
First, observe that the problem is spherical symmetric,
as the potential   just depends on the radius $r=\sqrt{{\bf x}\cdot {\bf x}}$,
and also the Laplace operator $\Delta =
\nabla \cdot \nabla $ allows spherical symmetry.
Thus we could write   the Schr\"odinger equation (\ref{2011-m-ch-qae-sebf})
in terms of spherical coordinates
$(r, \theta ,\varphi )$,
mentioned already as an example of orthogonal curvilinear coordinates in Equation~(\ref{2018-mm-ch-sc}),
with
\begin{equation}
\begin{split}
x  = r\sin \theta \cos \varphi \text{, }
y = r\sin \theta \sin \varphi \text{, }
z = r\cos \theta  \text{; and}\\
r=\sqrt{x^2+y^2+z^2}\text{, }
\theta = \arccos \left(\frac{z}{r}\right)\text{, }
\varphi=  \arctan \left(\frac{y}{x}\right).
\end{split}
\label{2018-mm-ch-sc1}
\end{equation}
$\theta$ is the polar angle in the $x$--$z$-plane measured
from the $z$-axis, with $0 \le \theta \le \pi$,
and $\varphi $ is  the azimuthal angle in the $x$--$y$-plane, measured
from the $x$-axis with $0 \le \varphi < 2 \pi$.
\index{spherical coordinates}
In terms of spherical coordinates the Laplace operator~(\ref{2018-mm-ch-laplaceocsc})
on page~\pageref{2018-mm-ch-laplaceocsc}
essentially ``decays into'' (that is, consists additively of)
a radial part and an angular part
\begin{equation}
\begin{split}
\Delta =
 \frac{\partial^2}{\partial x^2}
+
\frac{\partial^2}{\partial y^2}
+
\frac{\partial^2}{\partial z^2}
=
 \frac{\partial}{\partial x}
 \frac{\partial}{\partial x}
+
\frac{\partial}{\partial y}
 \frac{\partial}{\partial y}
+
\frac{\partial}{\partial z}
 \frac{\partial}{\partial z}
\\
=
\frac{1}{r^2} \left[ \frac{\partial}{\partial r}\left( r^2\frac{\partial}{\partial r}\right)
+
\frac{1}{\sin \theta}   \frac{\partial}{\partial \theta }
\sin \theta \frac{\partial}{\partial \theta }
+
\frac{1}{\sin^2 \theta} \frac{\partial^2}{\partial \varphi^2 }
\right].
\end{split}
\label{2011-m-ch-qae-losc}
\end{equation}

\newcommand{\comm}[1]{}
\comm{
=============================================================================

r[x_,y_,z_] := Sqrt[x^2 + y^2 + z^2];
\[Theta][x_,y_,z_] := ArcCos[z/Sqrt[x^2 + y^2 + z^2]];
\[Phi][x_,y_,z_] := ArcTan[ y / x ];

x[r,\[Theta]_,\[Phi]_] := r*Sin[\[Theta]]*Cos[\[Phi]];
y[r,\[Theta]_,\[Phi]_] :=  r*Sin[\[Theta]]*Sin[\[Phi]];
z[r,\[Theta]_,\[Phi]_] :=  r*Cos[\[Theta]];

(*
FullSimplify[
(
D[ r[x,y,z] , x] D[ f[r,\[Theta],\[Phi]] , r] + D[ \[Theta][x,y,z] , x] D[ f[r,\[Theta],\[Phi]] , \[Theta]] + D[ \[Phi][x,y,z] , x] D[ f[r,\[Theta],\[Phi]] , \[Phi]]
/. x -> r*Sin[\[Theta]]*Cos[\[Phi]]
/. y -> r*Sin[\[Theta]]*Sin[\[Phi]]
/. z -> r*Cos[\[Theta]]
+
D[ r[x,y,z] , y] D[ f[r,\[Theta],\[Phi]] , r] + D[ \[Theta][x,y,z] , y] D[ f[r,\[Theta],\[Phi]] , \[Theta]] + D[ \[Phi][x,y,z] , y] D[ f[r,\[Theta],\[Phi]] , \[Phi]]
/. x -> r*Sin[\[Theta]]*Cos[\[Phi]]
/. y -> r*Sin[\[Theta]]*Sin[\[Phi]]
/. z -> r*Cos[\[Theta]]
+
D[ r[x,y,z] , z] D[ f[r,\[Theta],\[Phi]] , r] + D[ \[Theta][x,y,z] , z] D[ f[r,\[Theta],\[Phi]] , \[Theta]] + D[ \[Phi][x,y,z] , z] D[ f[r,\[Theta],\[Phi]] , \[Phi]]
/. x -> r*Sin[\[Theta]]*Cos[\[Phi]]
/. y -> r*Sin[\[Theta]]*Sin[\[Phi]]
/. z -> r*Cos[\[Theta]]
)
]
*)

FullSimplify[ (FullSimplify[(
D[ r[x,y,z] , x] D[
FullSimplify[ (D[ r[x,y,z] , x] D[ f[r,\[Theta],\[Phi]] , r] + D[ \[Theta][x,y,z] , x] D[ f[r,\[Theta],\[Phi]] , \[Theta]] + D[ \[Phi][x,y,z], x] D[ f[r,\[Theta],\[Phi]] , \[Phi]]/. x -> r*Sin[\[Theta]]*Cos[\[Phi]]
/. y -> r*Sin[\[Theta]]*Sin[\[Phi]]
/. z -> r*Cos[\[Theta]]
),Element[r > 0 | 0 <= \[Theta] <= Pi | 0 <= \[Phi] <= 2 Pi  , Reals] ]
, r] +
D[\[Theta][x,y,z] , x] D[
FullSimplify[ (D[ r[x,y,z] , x] D[ f[r,\[Theta],\[Phi]] , r] + D[ \[Theta][x,y,z] , x] D[ f[r,\[Theta],\[Phi]] , \[Theta]] + D[ \[Phi][x,y,z] , x] D[ f[r,\[Theta],\[Phi]] , \[Phi]]/. x -> r*Sin[\[Theta]]*Cos[\[Phi]]
/. y -> r*Sin[\[Theta]]*Sin[\[Phi]]
/. z -> r*Cos[\[Theta]]
),Element[r > 0 | 0 <= \[Theta] <= Pi | 0 <= \[Phi] <= 2 Pi  , Reals] ]
, \[Theta]] +
D[ \[Phi][x,y,z] , x] D[
FullSimplify[ (D[ r[x,y,z] , x] D[ f[r,\[Theta],\[Phi]] , r] + D[ \[Theta][x,y,z] , x] D[ f[r,\[Theta],\[Phi]] , \[Theta]] + D[ \[Phi][x,y,z] , x] D[ f[r,\[Theta],\[Phi]] , \[Phi]]/. x -> r*Sin[\[Theta]]*Cos[\[Phi]]
/. y -> r*Sin[\[Theta]]*Sin[\[Phi]]
/. z -> r*Cos[\[Theta]]
),Element[r > 0 | 0 <= \[Theta] <= Pi | 0 <= \[Phi] <= 2 Pi  , Reals] ]
, \[Phi]]
+ (***********)
D[ r[x,y,z] , y] D[
FullSimplify[ (D[ r[x,y,z] , x] D[ f[r,\[Theta],\[Phi]] , r] + D[ \[Theta][x,y,z] , x] D[ f[r,\[Theta],\[Phi]] , \[Theta]] + D[ \[Phi][x,y,z] , x] D[ f[r,\[Theta],\[Phi]] , \[Phi]]/. x -> r*Sin[\[Theta]]*Cos[\[Phi]]
/. y -> r*Sin[\[Theta]]*Sin[\[Phi]]
/. z -> r*Cos[\[Theta]]
),Element[r > 0 | 0 <= \[Theta] <= Pi | 0 <= \[Phi] <= 2 Pi  , Reals] ]
, r] +
D[ \[Theta][x,y,z] , y] D[
FullSimplify[ (D[ r[x,y,z] , x] D[ f[r,\[Theta],\[Phi]] , r] + D[ \[Theta][x,y,z] , x] D[ f[r,\[Theta],\[Phi]] , \[Theta]] + D[ \[Phi][x,y,z] , x] D[ f[r,\[Theta],\[Phi]] , \[Phi]]/. x -> r*Sin[\[Theta]]*Cos[\[Phi]]
/. y -> r*Sin[\[Theta]]*Sin[\[Phi]]
/. z -> r*Cos[\[Theta]]
),Element[r > 0 | 0 <= \[Theta] <= Pi | 0 <= \[Phi] <= 2 Pi  , Reals] ]
, \[Theta]] +
D[ \[Phi][x,y,z] , y] D[
FullSimplify[ (D[ r[x,y,z] , x] D[ f[r,\[Theta],\[Phi]] , r] + D[ \[Theta][x,y,z] , x] D[ f[r,\[Theta],\[Phi]] , \[Theta]] + D[ \[Phi][x,y,z] , x] D[ f[r,\[Theta],\[Phi]] , \[Phi]]/. x -> r*Sin[\[Theta]]*Cos[\[Phi]]
/. y -> r*Sin[\[Theta]]*Sin[\[Phi]]
/. z -> r*Cos[\[Theta]]
),Element[r > 0 | 0 <= \[Theta] <= Pi | 0 <= \[Phi] <= 2 Pi  , Reals] ]
, \[Phi]]
+ (***********)
D[ r[x,y,z] , z] D[
FullSimplify[ (D[ r[x,y,z] , x] D[ f[r,\[Theta],\[Phi]] , r] + D[ \[Theta][x,y,z] , x] D[ f[r,\[Theta],\[Phi]] , \[Theta]] + D[ \[Phi][x,y,z] , x] D[ f[r,\[Theta],\[Phi]], \[Phi]]/. x -> r*Sin[\[Theta]]*Cos[\[Phi]]
/. y -> r*Sin[\[Theta]]*Sin[\[Phi]]
/. z -> r*Cos[\[Theta]]
),Element[r > 0 | 0 <= \[Theta] <= Pi | 0 <= \[Phi] <= 2 Pi  , Reals] ]
, r] +
D[ \[Theta][x,y,z] , z] D[
FullSimplify[ (D[ r[x,y,z] , x] D[ f[r,\[Theta],\[Phi]] , r] + D[ \[Theta][x,y,z] , x] D[ f[r,\[Theta],\[Phi]] , \[Theta]] + D[ \[Phi][x,y,z] , x] D[ f[r,\[Theta],\[Phi]] , \[Phi]]/. x -> r*Sin[\[Theta]]*Cos[\[Phi]]
/. y -> r*Sin[\[Theta]]*Sin[\[Phi]]
/. z -> r*Cos[\[Theta]]
),Element[r > 0 | 0 <= \[Theta] <= Pi | 0 <= \[Phi] <= 2 Pi  , Reals] ]
, \[Theta]] +
D[ \[Phi][x,y,z] , z] D[
FullSimplify[ (D[ r[x,y,z] , x] D[ f[r,\[Theta],\[Phi]] , r] + D[ \[Theta][x,y,z] , x] D[ f[r,\[Theta],\[Phi]] , \[Theta]] + D[ \[Phi][x,y,z] , x] D[ f[r,\[Theta],\[Phi]] , \[Phi]]/. x -> r*Sin[\[Theta]]*Cos[\[Phi]]
/. y -> r*Sin[\[Theta]]*Sin[\[Phi]]
/. z -> r*Cos[\[Theta]]
),Element[r > 0 | 0 <= \[Theta] <= Pi | 0 <= \[Phi] <= 2 Pi  , Reals] ]
, \[Phi]]
)]
/. x -> r*Sin[\[Theta]]*Cos[\[Phi]]
/. y -> r*Sin[\[Theta]]*Sin[\[Phi]]
/. z -> r*Cos[\[Theta]]
),Element[r > 0 | 0 <= \[Theta] <= Pi | 0 <= \[Phi] <= 2 Pi  , Reals] ]

JacoSp= FullSimplify[ {
 { D[Sqrt[x^2 + y^2 + z^2], x] /. x -> r*Sin[\[Theta]]*Cos[\[Phi]] /. y -> r*Sin[\[Theta]]*Sin[\[Phi]] /. z -> r*Cos[\[Theta]]             ,
  D[Sqrt[x^2 + y^2 + z^2], y] /. x -> r*Sin[\[Theta]]*Cos[\[Phi]] /. y -> r*Sin[\[Theta]]*Sin[\[Phi]] /. z -> r*Cos[\[Theta]]             ,
 D[Sqrt[x^2 + y^2 + z^2], z] /. x -> r*Sin[\[Theta]]*Cos[\[Phi]] /. y -> r*Sin[\[Theta]]*Sin[\[Phi]] /. z -> r*Cos[\[Theta]]              },
{  D[ArcCos[z/Sqrt[x^2 + y^2 + z^2]], x] /. x -> r*Sin[\[Theta]]*Cos[\[Phi]] /. y -> r*Sin[\[Theta]]*Sin[\[Phi]] /. z -> r*Cos[\[Theta]]      ,
  D[ArcCos[z/Sqrt[x^2 + y^2 + z^2]], y] /. x -> r*Sin[\[Theta]]*Cos[\[Phi]] /. y -> r*Sin[\[Theta]]*Sin[\[Phi]] /. z -> r*Cos[\[Theta]]      ,
 D[ArcCos[z/Sqrt[x^2 + y^2 + z^2]], z] /. x -> r*Sin[\[Theta]]*Cos[\[Phi]] /. y -> r*Sin[\[Theta]]*Sin[\[Phi]] /. z -> r*Cos[\[Theta]]     },
{  D[ ArcTan[ y / x ], x] /. x -> r*Sin[\[Theta]]*Cos[\[Phi]] /. y -> r*Sin[\[Theta]]*Sin[\[Phi]] /. z -> r*Cos[\[Theta]]                  ,
  D[ ArcTan[ y / x ], y] /. x -> r*Sin[\[Theta]]*Cos[\[Phi]] /. y -> r*Sin[\[Theta]]*Sin[\[Phi]] /. z -> r*Cos[\[Theta]]                  ,
 D[ ArcTan[ y / x ], z] /. x -> r*Sin[\[Theta]]*Cos[\[Phi]] /. y -> r*Sin[\[Theta]]*Sin[\[Phi]] /. z -> r*Cos[\[Theta]]                 }
       },Element[r > 0 | 0 <= \[Theta] <= Pi | 0 <= \[Phi] <= 2 Pi  , Reals] ];

MatrixForm[JacoSp]

=============================================================================
}

\subsection{Separation of variables {\it Ansatz}}

This can be exploited for a
{\em separation of variable} {\it Ansatz},
which, according to  Schr\"odinger, should be well known
(in German {\em sattsam bekannt})
by now (cf Chapter~\ref{2011-m-ch-sv}).
We thus write the solution $\psi$ as a product of functions
of separate variables
\begin{equation}
\psi (r, \theta ,\varphi )=R(r)\Theta(\theta)\Phi(\varphi) = R(r)Y_l^m ( \theta ,\varphi )
\label{2011-m-ch-qaesva}
\end{equation}
That the angular part $\Theta(\theta)\Phi(\varphi)$ of this product
will turn out to be
the spherical harmonics $Y_l^m ( \theta ,\varphi )$  introduced earlier
on page  \pageref{2011-m-ch-sfshar}
is nontrivial---at this point it is an {\em ad hoc} assumption
that may be motivated by the spherical symmetry of the electrostatic potential of a positive point charge representing the nucleus of the hydrogren atom.
Indeed, we may speculate that, once a spherical symmetry is established, the most important ``modulation''
is in the radial part of the solution, determined by the radial dependence of the (spherically symmetric) potential.
We will come back to its derivation in fuller detail later.

\subsection{Separation of the radial part from the angular one}

Let us first separate
the radial part $R(r)$ from the angular part of the Schr\"odinger equation (\ref{2011-m-ch-qae-sebf}),
written in terms of spherical coordinates, thereby reflecting the (hopefully) rotational invariance
\index{rotational invariance}
of the potential
or the configuration in general,
\begin{equation}
\begin{split}
\left\{
\frac{1}{r^2} \left[ \frac{\partial}{\partial r}\left( r^2\frac{\partial}{\partial r}\right)
+
\frac{1}{\sin \theta}   \frac{\partial}{\partial \theta }
\sin \theta \frac{\partial}{\partial \theta }
+
\frac{1}{\sin^2 \theta} \frac{\partial^2}{\partial \varphi^2 }
\right]
\right.    \\
+
\left.
\frac{2\mu }{\hslash^2} \left(\frac{e^2}{4\pi \epsilon_0 r} + E \right)\right\}
\psi (r, \theta ,\varphi  )=0
.
\end{split}
\label{2011-m-ch-qa1e}
\end{equation}
Multiplying~(\ref{2011-m-ch-qa1e}) with $r^2$ yields
\begin{equation}
\begin{split}
\left\{  \frac{\partial}{\partial r}\left( r^2\frac{\partial}{\partial r}\right) +
\frac{2\mu r^2}{\hslash^2} \left(\frac{e^2}{4\pi \epsilon_0 r} + E \right) \right. \qquad  \qquad \qquad \qquad \qquad\\
\qquad  \qquad +  \left.
\frac{1}{\sin \theta}   \frac{\partial}{\partial \theta }
\sin \theta \frac{\partial}{\partial \theta }
+
\frac{1}{\sin^2 \theta} \frac{\partial^2}{\partial \varphi^2 }
\right\}
\psi (r, \theta ,\varphi  )=0
.
\end{split}
\label{2011-m-ch-qae2}
\end{equation}
After division by $\psi (r, \theta ,\varphi  )= R(r)\Theta(\theta)\Phi(\varphi)$
and writing separate variables on separate sides of the equation one obtains
\begin{equation}
\begin{split}
\frac{1}{R( r )}
\left[  \frac{\partial}{\partial r}\left( r^2\frac{\partial}{\partial r}\right) +
\frac{2\mu r^2}{\hslash^2} \left(\frac{e^2}{4\pi \epsilon_0 r} + E \right) \right] R(r)
\qquad
\\  =
-\frac{1}{\Theta(\theta)\Phi(\varphi)} \left(
\frac{1}{\sin \theta}   \frac{\partial}{\partial \theta }
\sin \theta \frac{\partial}{\partial \theta }
+
\frac{1}{\sin^2 \theta} \frac{\partial^2}{\partial \varphi^2 }
\right)
\Theta(\theta)\Phi(\varphi)
.
\end{split}
\label{2011-m-ch-qae3}
\end{equation}
Because the left hand side  of this equation is independent of the angular variables
$\theta $ and $\varphi$, and its right hand side  is independent of the radial variable $r$,
both sides have to be independent with respect to variations of $r$, $\theta $ and $\varphi$, and
can thus be equated with a constant;
say, $\lambda$.
Therefore, we obtain two ordinary
differential equations: one for the radial part [after multiplication of (\ref{2011-m-ch-qae3}) with $R(r)$ from the left]
\begin{equation}
\left[  \frac{\partial}{\partial r} r^2\frac{\partial}{\partial r}  +
\frac{2\mu r^2}{\hslash^2} \left(\frac{e^2}{4\pi \epsilon_0 r} + E \right) \right] R(r)
 =  \lambda  R( r )  ,
\label{2011-m-ch-qae4a}
\end{equation}
and another one for the angular part [after multiplication of (\ref{2011-m-ch-qae3}) with $\Theta(\theta)\Phi(\varphi)$ from the left]
\begin{equation}
\left(
\frac{1}{\sin \theta}   \frac{\partial}{\partial \theta }
\sin \theta \frac{\partial}{\partial \theta }
+
\frac{1}{\sin^2 \theta} \frac{\partial^2}{\partial \varphi^2 }
\right)
\Theta(\theta)\Phi(\varphi)   =  -  \lambda  \Theta(\theta)\Phi(\varphi),
\label{2011-m-ch-qae4b}
\end{equation}
respectively.

\subsection{Separation of the polar angle $\theta$ from the azimuthal angle $\varphi $}

As already hinted in Equation~(\ref{2011-m-ch-qaesva})
the angular portion can still be separated into a polar and an azimuthal
part
because, when multiplied by $\sin^2 \theta /[\Theta(\theta)\Phi(\varphi)]$,
Equation~  (\ref{2011-m-ch-qae4b})
can be rewritten as
\begin{equation}
\left(
\frac{\sin \theta}{\Theta(\theta)}
\frac{\partial}{\partial \theta }
\sin \theta \frac{\partial \Theta(\theta)}{\partial \theta }
+  \lambda  \sin^2 \theta   \right)
+
\frac{1}{\Phi(\varphi)} \frac{\partial^2\Phi(\varphi)}{\partial \varphi^2 }
= 0,
\label{2011-m-ch-qae4bc}
\end{equation}
and hence
\begin{equation}
\begin{split}
\frac{\sin \theta}{\Theta(\theta)}
\frac{\partial}{\partial \theta }
\sin \theta \frac{\partial \Theta(\theta)}{\partial \theta }
+  \lambda  \sin^2 \theta
  = -
\frac{1}{\Phi(\varphi)} \frac{\partial^2\Phi(\varphi)}{\partial \varphi^2 }
=  m^2,
\end{split}
\label{2011-m-ch-qae8}
\end{equation}
where $m$ is some constant.

\subsection{Solution of the equation  for the azimuthal angle factor $\Phi(\varphi )$}

The  resulting differential equation  for $\Phi(\varphi )$
\begin{equation}
\frac{   d   ^2\Phi(\varphi )}{   d    \varphi^2 }
=  -m^2\Phi(\varphi),
\label{2011-m-ch-qaephi}
\end{equation}
has the general solution consisting of two linear independent parts
\begin{equation}
\Phi(\varphi) = A e^{im\varphi}+B e^{-im\varphi}.
\label{2011-m-ch-qae11}
\end{equation}
Because $\Phi$ must obey the periodic boundary conditions $\Phi(\varphi)=\Phi(\varphi  +2\pi)$,
$m$ must be an integer: let $B=0$; then
 \begin{equation}
\begin{split}
\Phi(\varphi)= Ae^{im\varphi} =\Phi(\varphi  +2\pi) = Ae^{im(\varphi  +2\pi)} = Ae^{im\varphi}e^{2i\pi m}  ,\\
1  =  e^{2i\pi m}  = \cos (2 \pi m) +i \sin (2 \pi m)  ,\\
\end{split}
\label{2018-mm-ch-sf-acin}
\end{equation}
which is only true for $m\in \mathbb{Z}$. A similar calculation yields the same result if $A=0$.

An integration shows that, if we require the system of functions $\{e^{im\varphi}\vert m \in \mathbb{Z}\}$
to be orthonormalized, then the two constants $A,B$ must be equal.
Indeed, if we define
\begin{equation}
\Phi_m(\varphi) = Ae^{im\varphi}
\label{2011-m-ch-qae11def}
\end{equation}
and require that it is normalized, it follows that
\begin{equation}
\begin{split}
\int_0^{2\pi} \overline {\Phi}_m(\varphi) \Phi_m(\varphi)d \varphi \\
\qquad     =
\int_0^{2\pi} \overline {A}e^{-im\varphi}Ae^{im\varphi} d \varphi \\
\qquad    =
\int_0^{2\pi}\vert A\vert^2 d \varphi \\
\qquad  = 2\pi  \vert A\vert^2  \\
\qquad
= 1,
\end{split}
\label{2011-m-ch-qae11normexpl}
\end{equation}
it is consistent to set
$
A= \frac{1} {\sqrt{2\pi } };
$
and hence,
\begin{equation}
\Phi_m(\varphi) = \frac{e^{im\varphi}} {\sqrt{2\pi }  }
\label{2011-m-ch-qae11normexplnendg}
\end{equation}
Note that, for different $m\neq n$, because $m-n\in \mathbb{Z}$,
\begin{equation}
\begin{split}
\int_0^{2\pi} \overline {\Phi}_n(\varphi) \Phi_m(\varphi)d \varphi
  =
\int_0^{2\pi} \frac{e^{-in\varphi}} {\sqrt{2\pi }  } \frac{e^{im\varphi}} {\sqrt{2\pi }}  d \varphi \\
  =
\int_0^{2\pi} \frac{e^{i(m-n)\varphi}} { 2\pi  }   d \varphi \
  =
\left. -\frac{i e^{i(m-n)\varphi}} { 2\pi (m-n) } \right|_{\varphi =0}^{\varphi = 2\pi}  = 0
.
\end{split}
\label{2011-m-ch-qae11normexplnoni}
\end{equation}

\subsection{Solution of the equation  for the  polar angle factor $\Theta (\theta )$}

The left-hand side of
Equation~(\ref{2011-m-ch-qae8}) contains only the polar coordinate.
Upon division by $\sin ^2 \theta$ we obtain
\begin{equation}
\begin{split}
\frac{1}{\Theta(\theta)\sin \theta}
\frac{   d   }{   d    \theta }
\sin \theta \frac{   d    \Theta(\theta)}{   d    \theta }
+  \lambda
= \frac{m^2}{\sin^2\theta}\textrm{, or }\\
\frac{1}{\Theta(\theta)\sin \theta}
\frac{   d   }{   d    \theta }
\sin \theta \frac{   d    \Theta(\theta)}{   d    \theta }   -\frac{m^2}{\sin^2\theta }
= -  \lambda    ,\\
\end{split}
\label{2011-m-ch-pcde}
\end{equation}

Now, first, let us consider the case $m=0$.
With the variable substitution
$x = \cos \theta$, and thus
$\frac{dx}{d\theta}= -\sin \theta$ and  $dx= -\sin \theta d\theta$, we obtain from (\ref{2011-m-ch-pcde})
\begin{equation}
\begin{split}
\frac{   d   }{   d    x }
\sin^2 \theta \frac{   d    \Theta(x)}{   d    x }
= -  \lambda  \Theta(x)   ,\\
\frac{   d   }{   d    x }
(1-x^2) \frac{   d    \Theta(x)}{   d    x } +  \lambda  \Theta(x)
=  0 ,\\
\left(x^2-1 \right)\frac{   d   ^2 \Theta(x)}{   d    x^2 } + 2 x \frac{   d    \Theta(x)}{   d    x } =  \lambda  \Theta(x)
,\\
\end{split}
\label{2011-m-ch-pcde1}
\end{equation}
which is of the same form as the {\em Legendre equation}
\index{Legendre equation}
(\ref{2011-m-ch-sfelpede})
mentioned on page \pageref{2011-m-ch-sfelpede}.

Consider the series {\it Ansatz}
\begin{equation}
\Theta(x) = \sum_{k=0}^\infty  a_k x^k
\label{2011-m-ch-pcde12}
\end{equation}
for solving (\ref{2011-m-ch-pcde1}).
\marginnote{This is actually a ``shortcut'' solution of the Fuchsian Equation mentioned earlier.}
Insertion into  (\ref{2011-m-ch-pcde1}) and comparing the coefficients of $x$
for equal degrees
yields the recursion relation
\begin{equation}
\begin{split}
\left(x^2-1 \right)\frac{   d   ^2  }{   d    x^2 }
 \sum_{k=0}^\infty  a_k x^k
   + 2 x \frac{   d    }{   d    x } \sum_{k=0}^\infty  a_k x^k
=  \lambda \sum_{k=0}^\infty  a_k x^k  ,\\
 \left(x^2-1 \right) \sum_{k=0}^\infty  k(k-1) a_k x^{k-2}
   + 2 x  \sum_{k=0}^\infty  k a_k x^{k-1}
=  \lambda \sum_{k=0}^\infty  a_k x^k  ,\\
\sum_{k=0}^\infty
\left[ \underbrace{k(k-1) + 2k}_{k(k+1)} -\lambda \right] a_k  x^k
-
\underbrace{\sum_{k=2}^\infty k(k-1) a_k x^{k-2}}_{\text{index shift }k-2=m\text{, }k=m+2}
=0
 ,\\
\sum_{k=0}^\infty
\left[ k(k+1) -\lambda \right] a_k  x^k
-
\sum_{m=0}^\infty (m+2)(m+1) a_{m+2} x^m
=0
 ,\\
\sum_{k=0}^\infty
\Big\{
\left[ k(k+1) -\lambda \right] a_k
-
 (k+1)(k+2) a_{k+2}
\Big\}
x^k
=0
 ,
\end{split}
\label{2011-m-ch-pcde123}
\end{equation}
and thus, by taking all polynomials of the order of $k$ and proportional to $x^k$,
so that, for $x^k\neq 0$ (and thus excluding the trivial solution),
\begin{equation}
\begin{split}
\left[ k(k+1) -\lambda \right] a_k -
 (k+1)(k+2) a_{k+2} =0,\\
a_{k+2} = a_k \frac{k(k+1) - \lambda}{(k+1)(k+2)}.
\end{split}
\label{2011-m-ch-pcde1236}
\end{equation}

In order to converge also for $x=\pm 1$, and hence for $\theta =0$ and $\theta= \pi$,
the sum in (\ref{2011-m-ch-pcde12})
has to have only {\em a finite number of terms}.
Because  if the sum would be infinite, the terms $a_k$, for large $k$,
would be dominated by $a_{k-2} O( k^2/k^2)= a_{k-2} O(1)$.
As a result  $a_k$
would converge to $a_k \stackrel{ k  \rightarrow \infty}{\longrightarrow} a_\infty$ with constant $a_\infty \neq 0$
Therefore, $\Theta$ would diverge as
$\Theta (1) \stackrel{ k  \rightarrow \infty}{\approx} k a_\infty  \stackrel{ k  \rightarrow \infty}{\longrightarrow}  \infty$.
That means that, in Equation~(\ref{2011-m-ch-pcde1236})
for some $k=l\in {\Bbb N}$, the coefficient  $a_{l+2}=0$ has to vanish; thus
\begin{equation}
\lambda = l(l+1).
\label{2011-m-ch-pcde12361}
\end{equation}
This results in {\it Legendre polynomials} $\Theta (x) \equiv P_l(x)$.
\index{Legendre polynomial}

Let us shortly mention the case $m\neq 0$.
With the same variable substitution  $x = \cos \theta$, and thus
$\frac{dx}{d\theta}= -\sin \theta$ and  $dx= -\sin \theta d\theta$ as before,
the equation for the polar angle dependent factor (\ref{2011-m-ch-pcde})
becomes
\begin{equation}
\begin{split}
\left[
\frac{   d   }{   d    x }
(1-x^2) \frac{   d    }{   d    x }  +  l(l+1)   -\frac{m^2}{ 1-x^2 }
\right]
\Theta(x)  =0,\\
\end{split}
\label{2011-m-ch-pcde9}
\end{equation}
This is exactly the form of the
general Legendre equation (\ref{2011-m-ch-sfgle}), whose solution is a multiple
of the associated Legendre polynomial   $P_l^m(x)$, with $\vert m\vert \le l$.

Note (without proof) that, for equal $m$, the $P_l^m(x)$ satisfy the orthogonality condition
\begin{equation}
\int_{-1}^{1} P_l^m(x) P_{l'}^m(x) dx =
{\frac{2 (l+m)!} {(2l+1)(l-m)!}}  \delta_{l l'}.
\end{equation}
Therefore  we obtain a normalized polar solution by dividing $P_l^m(x)$
by $\left\{\left[2 (l+m)!\right]/\left[(2l+1)(l-m)!\right]\right\}^{1/2}$.

In putting both normalized  polar and azimuthal angle
factors together we arrive at the spherical harmonics (\ref{2014-m-ch-sf-sh}); that is,
\begin{equation}
\Theta(\theta)\Phi(\varphi)
= \sqrt{\frac {(2l+1)(l-m)!}{2 (l+m)!}}  P_l^m(\cos \theta ) \frac {e^{im\varphi }}{\sqrt{2\pi}} =
 Y_l^m (\theta ,\varphi )
\end{equation}
for $-l\le m\le l$, $l\in {\Bbb N}_0$.
Note that the discreteness of these solutions
follows from physical requirements about their finite
existence.

\subsection{Solution of the equation  for radial factor $R(r)$}


The solution of the equation   (\ref{2011-m-ch-qae4a})
\begin{equation}
\begin{split}
\left[  \frac{   d   }{   d    r}  r^2\frac{   d   }{   d    r}   +
\frac{2\mu r^2}{\hslash^2} \left(\frac{e^2}{4\pi \epsilon_0r} + E \right) \right] R(r)
 =  l(l+1)  R( r ) \textrm{ , or}\\
-\frac{1}{R(r)} \frac{d}{   d    r}  r^2\frac{   d   }{   d    r} R( r ) +    l(l+1)
-
2
\frac{\mu e^2}{4\pi \epsilon_0\hslash^2} r
 = \frac{2\mu  }{ \hslash^2}  r^2 E
\end{split}
\label{2011-m-ch-qae4a19}
\end{equation}
for the radial factor $R(r)$
turned out to be the most difficult part for Schr\"odinger.\cite{Moore-Schroedinger}

Note that, since the additive term  $ l(l+1) $ in (\ref{2011-m-ch-qae4a19}) is non-dimensional,
so must be the other terms.
We can make this more explicit by the substitution of variables.

First, consider $y =\frac{r}{a_0}$ obtained by dividing $r$ by the
{\em Bohr radius}
\index{Bohr radius}
\begin{equation}
a_0= \frac{4\pi \epsilon_0 \hslash^2}{m_e e^2}\approx 5\; 10^{-11} m,
\label{2011-m-ch-qaebohrr}
\end{equation}
thereby assuming that the reduced mass is equal to the electron mass $\mu \approx m_e$.
More explicitly,
$r=ya_0= y  (4\pi \epsilon_0 \hslash^2)/(m_e e^2)$,
or $y= r/a_0= r  (m_e e^2)/(4\pi \epsilon_0 \hslash^2)$.
Furthermore, let us define $\varepsilon = E \frac{2\mu a_0^2}{\hslash^2}$.

These substitutions yield
\begin{equation}
\begin{split}
-\frac{1}{R(y)} \frac{d}{   d    y}  y^2\frac{   d   }{   d    y} R( y ) +    l(l+1)
-2 y
 = y^2 \varepsilon \textrm{, or}\\
-y^2 \frac{d^2}{   d    y^2} R( y )  - 2 y \frac{   d   }{   d    y} R( y )
+ \left[   l(l+1) -2 y   -\varepsilon   y^2  \right] R( y )
 = 0.
\end{split}
\label{2011-m-ch-qae4a191}
\end{equation}

Now we introduce a new function $\hat{R}$  {\it via}
\begin{equation}
R(  \xi )=  \xi ^l e^{-\frac{1}{2}  \xi }\hat{R}(  \xi ) ,
\label{2011-m-ch-qae4a19s}
\end{equation}
with $  \xi =\frac{2y}{n}$ and by replacing the energy variable with
$\varepsilon =-\frac{1}{n^2}$.
(It will later be argued that $\varepsilon$ must be discrete;  with $n\in {\Bbb N}-0$.)
This yields
\begin{equation}
\begin{split}
\xi  \frac{d^2}{   d    \xi^2}\hat{R}( \xi )  +[ 2 (l+1)-\xi] \frac{   d   }{   d    \xi } \hat{R}( \xi )
+ ( n-l-1  ) \hat{R}( \xi )
 = 0.
\end{split}
\label{2011-m-ch-qae4a191ssAe}
\end{equation}

The discretization of $n$ can again be motivated by requiring physical properties from
the solution; in particular, convergence.
Consider again a series solution {\it Ansatz}
\begin{equation}
\hat{R}(  \xi ) = \sum_{k=0}^\infty c_k\xi^k ,
\label{2011-m-ch-qae4a19sssA}
\end{equation}
which, when inserted into (\ref{2011-m-ch-qae4a191}),
yields
\begin{equation}
\begin{split}
\left\{
\xi  \frac{d^2}{   d    \xi^2}   + [ 2 (l+1)-\xi] \frac{   d   }{   d    \xi } + ( n-l-1  )
\right\}
\sum_{k=0}^\infty c_k\xi^k
=0,\\
\xi  \sum_{k=0}^\infty k(k-1) c_k\xi^{k-2}
+ [ 2 (l+1)-\xi] \sum_{k=0}^\infty k c_k\xi^{k-1}
+ ( n-l-1  )\sum_{k=0}^\infty c_k\xi^k
=0,\\
\underbrace{\sum_{k=1}^\infty \left[ \underbrace{k(k-1) + 2k(l+1)}_{=k(k+2l+1)}\right] c_k\xi^{k-1}}_{\text{index shift }k-1=m\text{, }k=m+1}
+
 \sum_{k=0}^\infty (-k + n-l-1  ) c_k\xi^k
=0,\\
\sum_{m=0}^\infty \left[ (m+1)(m+2l+2)\right] c_{m+1}\xi^m
+
 \sum_{k=0}^\infty (-k + n-l-1  ) c_k\xi^k
=0,\\
\sum_{k=0}^\infty
\Big\{
\left[ (k+1)(k+2l+2)\right] c_{k+1}
+
 (-k + n-l-1  ) c_k
\Big\}
\xi^k
=0,
\end{split}
\label{2011-m-ch-qae4a191ssA}
\end{equation}
so that, by comparing the coefficients of $\xi^k$, we obtain
\begin{equation}
\begin{split}
\left[ (k+1)(k+2l+2)\right] c_{k+1}
= -
 (-k + n-l-1  ) c_k  ,\\
c_{k+1}
=
c_k   \frac{k - n + l+ 1   }{(k+1) (k  + 2l + 2 )}  .\\
\end{split}
\label{2011-m-ch-qae4a191ssA5}
\end{equation}
Because of convergence of $\hat{R}$ and thus of $R$
--
note that, for large $\xi$ and $k$, the $k$'th term in
Equation~(\ref{2011-m-ch-qae4a19sssA}) determining $\hat{R}(  \xi )$
would behave as $\xi^k/k!$
and thus  $\hat{R}(  \xi )$ would roughly behave as the exponential function $e^\xi$
--
the series solution
(\ref{2011-m-ch-qae4a19sssA})
should terminate at some   $k =   n-l-1$,
or  $ n=  k+l+1$. Since $k$, $l$, and $1$ are all integers,
$n$ must be an integer as well.
And since $k\ge 0$,
and therefore $n-l-1 \ge 0$,
$n$ must at least be $l+1$, or
\begin{equation}
l\le n-1 .
\label{2011-m-ch-qae4a191ssA5hqz}
\end{equation}

Thus,  we end up with an {\em associated Laguerre equation}
\index{associated Laguerre equation} of the form
\begin{equation}
\left\{
  \xi    \frac{d^2}{   d     \xi  ^2 }
+[2(l+1)-  \xi  ]  \frac{d }{   d       \xi  }
+(n-l-1)
\right\}    \hat{R}(   \xi  ) =0
\textrm{, with } n \ge l+1\textrm{, and } n,l\in {\Bbb Z}.
\label{2011-m-ch-qaeale}
\end{equation}
Its   solutions are  the {\it associated Laguerre polynomials}
$L^{2l+1}_{n+l}$
which are the $(2l+1)$-th derivatives of the
Laguerre's polynomials  $L_{n+l}$; \index{Laguerre polynomial}
that is,
\begin{equation}
\begin{split}
 L_n(x)=e^x   \frac{d^n }{   d     x^n }  \left (x^ne^{-x}\right),\\
 L_n^m(x)=   \frac{d^m }{   d     x^m }  L_n(x).
\end{split}
\label{2011-m-ch-qaelp}
\end{equation}
This yields a normalized wave function
\begin{equation}
\begin{split}
R_n(r) =  {\cal N}
\left( \frac{2 {r}}{n a_0 }\right)^l e^{-\frac{r}{a_0 n}}
L^{2l+1}_{n+l} \left( \frac{2 {r}}{n a_0 }\right)\textrm{, with }\\
{\cal N}   =-\frac{2}{n^2}\sqrt{\frac{(n-l-1)!}{[(n+l)!a_0]^3}} ,
\end{split}
\label{2011-m-ch-qaecsrp}
\end{equation}
where
${\cal N}$ stands for the normalization factor.

\subsection{Composition of the general solution of the Schr\"odinger equation}

Now we shall coagulate
\marginnote{Always remember the alchemic  principle of {\it solve et coagula}!}
and combine the factorized solutions (\ref{2011-m-ch-qaesva})
into a complete solution of the Schr\"odinger equation
for $n+1,\, l,\, \vert m \vert  \in {\Bbb N}_0$, $0\le l\le n-1$, and $\vert m \vert \le l$,
\begin{equation}
\begin{split}
\psi_{n,l,m} (r, \theta ,\varphi )
 = R_n(r)Y_l^m ( \theta ,\varphi )\\
 = -\frac{2}{n^2}\sqrt{\frac{(n-l-1)!}{[(n+l)!a_0]^3}}
\left( \frac{2 {r}}{n a_0 }\right)^l e^{-\frac{r}{a_0 n}}
L^{2l+1}_{n+l} \left( \frac{2 {r}}{n a_0 }\right) \times \qquad \qquad  \\
\times \sqrt{\frac {(2l+1)(l-m)!}{2 (l+m)!}}  P_l^m(\cos \theta )
\frac {e^{im\varphi }}{\sqrt{2\pi}} .
\end{split}
\label{2011-m-ch-qaesva13444}
\end{equation}

\begin{center}
{\color{olive}   \Huge
\decofourleft
}
\end{center}

\chapter{Divergent series}
\index{divergent series}
\label{2011-m-ch-ds}

\newthought{Power series approximations} often occur
in physical situations in the context of solutions of ordinary differential equations;
for instance in celestial mechanics or in quantum field theory.\cite{Boyd99thedevil,PhysRev.85.631}
According to Abel\cite{Hardy:1949} they
appear to be the ``invention of the devil,'' even more so as\cite{rousseau-2004}  {\em ``for the most part,
it is true that the results are correct, which is very strange.''}

There appears to be another, complementary, more optimistic and less perplexed, view on diverging series,
a view that has been expressed by Berry as follows:\cite{berry-92}
{\em ``$\ldots$ an asymptotic series $\ldots$ is a compact encoding of a function,
and its divergence should be regarded not as a deficiency but as a source of information about the function.''}
In a similar spirit, Boyd quotes {\em Carrier's Rule}:
{\em ``divergent series converge faster than convergent series because they don't have to converge.''}

\newthought{The intuition behind such statements} is based on the observation that,
while convergent series representing some function
may converge very slowly and numerically intractably,
\sidenote{Contemplate on the feasibility
of computing the partial sum $\sum_{j=0}^n \frac{(-1)^j x^{2j+1}}{(2j+1)!} =\sin_n (x)$
of the sine funtion without ``shortcuts;''
that is, without computing the remainder of $\frac{x}{2\pi}$, subject to some finite machine precision, say, for $x=10^5$.
Or consider the convergence of the general series solution of the $N$-body problem\cite{Diacu96}}
asymptotical divergent series
representations of functions
may yield reasonable estimates in ``low'' order before they diverge ``fast''
later on (for higher polynomial order).
Ritt's theorem
\index{Ritt's theorem}
\index{theorem of Ritt}
mentioned in Section~\ref{2019-mm-ch-ca-Ritt}
provides a formal basis for this conjecture.

\section{Convergence, asymptotic divergence, and divergence: A zoo perspective}

Let us first define {\em convergence} in the context of series.
\index{convergence}
\index{convergent series}
A series
\begin{equation}
s=\sum_{j=0}^\infty a_j =a_0+a_1+a_2+\cdots
\end{equation}
is said to converge to the {\em sum}
$s$ if the {\em partial sum}
\begin{equation}
s_n\equiv s(n)=  \sum_{j=0}^n a_j =a_0+a_1+a_2+\cdots + a_n
\end{equation}
tends to a finite limit $s$ when $n\rightarrow \infty$;
otherwise it is said to diverge (it may remain finite but may alternate).

A power series about some number $c\in \mathbb{C}$
depends on some additional parameter $z$; it has partial sums of the form
\begin{equation}
s_n (z)\equiv s(n,z) =
\sum_{j=0}^n a_j (z - c)^j =a_0+a_1 (z-c) +a_2(z-c)^2+\cdots + a_n (z-c)^n.
\end{equation}
If $c=0$ then the partial sum of this series  $s_n (z) =  \sum_{j=0}^n a_j z^j$
is about the origin.
Power series are important because they are used for solving ordinary differential equations,
such as Frobenius series \index{Frobenius series} in the theory of
differential equations of the Fuchsian type.
\index{Fuchsian equation}

Power series have a rich enough structure to leave room
for some ``grey area'' in-between divergence and convergence.
\index{asymptotic series}
\index{semi-convergent series}
\index{convergently beginning series}
In Dingle's terms,\cite{Dingle-1973}
{\em
``the designation `asymptotic series' will be reserved
for those series in which for large values of the variable at all phases
the terms first progressively decrease in magnitude, then reach a minimum
and thereafter increase.''}
Those series could be useful in the case of irregular singularities
of an ordinary differential equation, for which the Frobenius method fails.
We shall come back to asymptotic series later in Section~\ref{2019-mm-ch-ds-aps}.

For a start consider a widely known diverging series: the
{\em harmonic series}
\index{harmonic series}
\begin{equation}
s=  \sum_{j=1}^\infty \frac{1}{j} = 1+\frac{1}{2}+\frac{1}{3}+\frac{1}{4}+\cdots .
\end{equation}
A medieval proof by Oresme (cf. p.~92 of Ref.\cite[10mm]{Edwards-1979})
uses approximations: Oresme points out that increasing numbers of summands in the series can be rearranged to yield numbers bigger than, say, $\frac{1}{2}$;
more explicitly,
$\frac{1}{3}+\frac{1}{4}> \frac{1}{4}+\frac{1}{4}  =\frac{1}{2}$,
$\frac{1}{5}+ \cdots +\frac{1}{8} > 4 \frac{1}{8}  =\frac{1}{2}$,
$\frac{1}{9}+ \cdots +\frac{1}{16} > 8 \frac{1}{16}  =\frac{1}{2}$,
and so on, such that the entire series must grow larger as $\frac{n}{2}$.
As $n$ approaches infinity, the series is unbounded and thus diverges.

One of the most prominent divergent series is Grandi's series,\cite{Sloane_oeis.org/A033999}
sometimes also referred to as
\index{Grandi's series}
\index{Leibniz's series}
Leibniz series\cite{leibnitz-1860,moore-1938,Hardy:1949,everest-2003}
\begin{equation}
s = \sum_{j=0}^\infty (-1)^j=
\lim_{n \rightarrow \infty} \left[\frac{1}{2}+\frac{1}{2}\left( -1\right)^n\right]
=1-1+1-1+1-\cdots ,
\label{2009-fiftyfifty-1s}
\end{equation}
whose summands may be -- inconsistently
-- ``rearranged,''
yielding
\begin{equation*}
\begin{split}
\textrm{ either }
1-1+1-1+1-1+\cdots = (1-1)+(1-1)+(1-1)-\cdots =0\\
\textrm{ or }
1-1+1-1+1-1+\cdots = 1+(-1+1)+ (-1+1) +\cdots =1.
\end{split}
\end{equation*}
One could tentatively associate the arithmetical average $1/2$ to represent ``the sum of Grandi's series.''

Another tentative approach  would be to first regularize this nonconverging expression by introducing a ``small entity''
$\varepsilon$ with $0 < \varepsilon <1$, such that $\vert \varepsilon - 1 \vert  <1$, which allows to formally sum up the geometric series
\begin{equation*}
s_\varepsilon \stackrel{\text{def}}{=} \sum_{j=0}^\infty (\varepsilon -1)^j=\frac{1}{1-(\varepsilon -1)}   =\frac{1}{2- \varepsilon  };
\end{equation*}
and then take the limit $s \stackrel{\text{def}}{=}  \lim_{\varepsilon \rightarrow 0^+} s_\varepsilon
= \lim_{\varepsilon \rightarrow 0^+}  1/(2- \varepsilon  ) =   1/2$.

Indeed, by {\em Riemann's rearrangement theorem},
\index{Riemann rearrangement theorem}
convergent series  which do not absolutely converge
(i.e., $\sum_{j=0}^n a_j$ converges but $\sum_{j=0}^n \left| a_j \right|$ diverges)
may be brought to  ``converge'' to arbitrary (even infinite) values
by permuting (rearranging) the (ratio of) positive and negative terms
(the series of which must both be divergent).

These manipulations
\marginnote{Every such strategy involving {\em finite means} fails miserably.}
could be perceived in terms of certain
paradoxes of infinity, such as Hilbert's hotel
\index{Hilbert's hotel}
which always has vacancies -- by ``shifting all of its guests one room further
down its infinite corridor''.\cite{rucker}

\section{Geometric series}
\index{geometric series}

As Grandi's series is a particular, ``pathologic,'' case of a {\em geometric series}
\index{geometric series}
we shall briefly review those in greater generality.
A finite geometric (power) series is defined by
(for convenience a multiplicative constant is ommitted)
\begin{equation}
\begin{split}
s_n(z)\equiv s(n,z) =
\sum_{j=0}^n z^j
=\underbrace{z^0}_{1}+z+z^2+  \cdots + z^n
.
\end{split}
\label{2019-mm-ch-ds-fgs}
\end{equation}
Multiplying both sides of
(\ref{2019-mm-ch-ds-fgs})
by $z$ gives
\begin{equation}
\begin{split}
z s_n(z)
= \sum_{j=0}^n z^{j+1}
=z+z^2+z^3+  \cdots + z^{n+1}
.
\end{split}
\label{2019-mm-ch-ds-fgsmbr}
\end{equation}
Subtracting
(\ref{2019-mm-ch-ds-fgsmbr})
from the original series
(\ref{2019-mm-ch-ds-fgs})
yields
\begin{equation}
\begin{split}
s_n(z) - z s_n(z)
= (1 - z)s_n(z)
= \sum_{j=0}^n z^{j} - \sum_{j=0}^n z^{j+1} \\
=
1+z+z^2+  \cdots + z^n
-
\left(z+z^2+z^3+  \cdots + z^{n+1}\right)
=1-z^{n+1},
\end{split}
\label{2019-mm-ch-ds-fgsmbrs}
\end{equation}
and
\begin{equation}
s_n(z)=  \frac{1-z^{n+1}}{1 - z}.
\label{2019-mm-ch-ds-fgsmbrss}
\end{equation}
Alternatively, by defining a ``remainder'' term
\begin{equation}
r_n(z)\stackrel{\text{def}}{=} \frac{z^{n+1}}{z-1},
\label{2019-fiftyfifty-1sgs-remainder}
\end{equation}
(\ref{2019-mm-ch-ds-fgsmbrss})
can be recasted into
\begin{equation}
\begin{split}
\frac{1}{1 - z}= s_n(z)+  \frac{z^{n+1}}{1 - z} = s_n(z) - r_n(z)\text{; and} \\
s_n(z) = \frac{1}{1-z} + r_n(z) = \frac{1}{z - 1} + \frac{z^{n+1}}{z-1}
.
\end{split}
\label{2019-mm-ch-ds-fgsmbrss1}
\end{equation}
\marginnote{Again the symbol ``$O$'' stands for ``of the order of'' or ``absolutely bound by'' in the following way:
if $g(x)$ is a positive function,  then $f(x)=O\left(g(x)\right)$
implies that there exist a positive real number $m$
such that $\vert f(x) \vert < m g(x)$. }.
\index{order of}
\index{big $O$ notation}
\index{Bachmann-Landau notation}
\index{asymptotic notation}
As $z \rightarrow 1$  the remainder diverges because the denominator tends to zero.
As $z \rightarrow -1$, again  the remainder diverges; but for a different reason: it does not converge
to a unique limit but alternates between $\pm \frac{1}{2}$.
If $\vert z\vert > 1$  the remainder  $r_n(z) = \frac{z^{n+1}}{1 - z} =O( z^n )$
grows without bounds; and
therefore the entire sum (\ref{2019-mm-ch-ds-fgs}) diverges in the $n\rightarrow \infty$ limit.

Only for $\vert z\vert <1$
the remainder  $r_n(z) = \frac{z^{n+1}}{1 - z} =O( z^n )$
vanishes in the $n\rightarrow \infty$ limit;
and, therefore, the infinite sum in the geometric series exists and
converges as a limit of~(\ref{2019-mm-ch-ds-fgs}):
\begin{equation}
\begin{split}
s(z) =
\lim_{n\rightarrow \infty}
s_n(z)=
\lim_{n\rightarrow \infty}
\sum_{j=0}^n z^j
=  z^j=1+z+z^2+  \cdots     \\
=1+ z(1+z+  \cdots)  =1+z s(z).
\end{split}
\label{2009-fiftyfifty-1sgs}
\end{equation}
Since $s(z)=1+z s(z)$
and
$s(z)-z s(z)=s(z)(1-z)=1$,
\begin{equation}
s(z)= \sum_{j=0}^\infty z^j= \frac{1}{1-z}.
\label{2012-fiftyfifty-1sgscont}
\end{equation}

\section{Abel summation  -- assessing paradoxes of infinity}
\index{Abel sum}

One ``Abelian'' way to ``sum up'' divergent series is by ``illegitimately continuing''
the argument
to  values for which the
infinite geometric series diverges;
thereby only taking its ``finite part''~(\ref{2012-fiftyfifty-1sgscont})
while at the same time neglecting or disregarding
the divergent remainder term~(\ref{2019-fiftyfifty-1sgs-remainder}).

For Grandi's series
this essentially amounts to substituting $z= - 1$
into~(\ref{2012-fiftyfifty-1sgscont}), thereby defining the
{\em Abel sum} (denoted by an ``A'' on top of equality sign)
\index{Abel sum}
\begin{equation}
s=\sum_{j=0}^\infty(-1)^j  = 1-1+1-1+1-1+\cdots \stackrel{{\rm A}}{=} \frac{1}{1-(-1)} = \frac{1}{2}.
\label{2012-fiftyfifty-1sgscont1}
\end{equation}

Another ``convergent value of a divergent series''
can, by a similar transgression of common syntatic rules,
be ``obtained'' by ``formally  expanding''  the square of the Abel sum of
Grandi's series
$
s^2 \stackrel{{\rm A}}{=} [1-(-x)]^{-2} = (1+x)^{-2}
$ for  $x=1$
into the Taylor series\cite{Kline-83}
around $t=0$, and using   $(-1)^{j-1}=(-1)^{j-1}(-1)^{2}=(-1)^{j+1}$:
\begin{equation}
\begin{split}
s^2   \stackrel{{\rm A}}{=} (1+x)^{-2}  \Big|_{x =1}=
\left.\left\{\left.\sum_{j=0}^\infty \frac{1}{j!}\left[\frac{d^j}{d{t}^j}  (1+t)^{-2}\right]   (x-t)^j\right|_{t=0}\right\} \right|_{x=1}
\\
=
\sum_{j=0}^\infty (-1)^j (j+1) =
\sum_{j=0}^\infty (-1)^{j+1} k =
1-2+3-4+5-\cdots
.
\label{2009-fiftyfifty-1s1}
\end{split}
\end{equation}
On the other hand, squaring the Grandi's series ``yields'' the Abel sum
\index{Grandi's series}
\begin{equation}
s^2 =
\left(\sum_{j=0}^\infty (-1)^j\right)
\left(\sum_{k=0}^\infty (-1)^k\right)
\stackrel{{\rm A}}{=} \left(\frac{1}{2}\right)^2=\frac{1}{4},
\end{equation}
so that, one could ``infer'' the Abel sum
\begin{equation}
s^2 =
1-2+3-4+5-\cdots  \stackrel{{\rm A}}{=} \frac{1}{4}.
\end{equation}

Once this identification is established, all of Abel's hell breaks loose:
One could, for instance, ``compute the finite sum\cite{Sloane_oeis.org/A000217} of all natural numbers\cite{Sloane_oeis.org/A000027}''
(a sum even mentioned on page~22 in a book on String Theory\cite{polchinski_1998}),
{\it via} formal analytic continuation as for the
Ramanujan summation~(\ref{2019-mm-ch-ds-Ramanujan}):  \index{Ramanujan summation}
\begin{equation}
S  =
\sum_{j=0}^\infty j = 1+2+3+4+5 + \cdots
\stackrel{{\rm A}}{=} \lim_{n\rightarrow \infty} \frac{n(n+1)}{2}
\stackrel{{\rm A}}{=} - \frac{1}{12}
\label{2016-m-ch-ds-natnumser}
\end{equation}
by sorting out
\begin{equation}
\begin{split}
S - \frac{1}{4}\stackrel{{\rm A}}{=}
S-s^2  = 1+2+3+4+5 + \cdots -
(1-2+3-4+5-\cdots )   \\
= 4 + 8+ 12 +  \cdots = 4 S
,
\label{2016-m-ch-ds-natnumser2}
\end{split}
\end{equation}
so that  $3 S
\stackrel{{\rm A}}{=}  - \frac{1}{4}$,
and, finally, $S
\stackrel{{\rm A}}{=}  - \frac{1}{12}$.

Note that the sequence of the partial sums $s^2_n=\sum_{j=0}^n (-1)^{j+1} j $
of $s^2$, as expanded in (\ref{2009-fiftyfifty-1s1}),
``appears to yield'' every integer once; that is,
$s^2_0 =0$,
$s^2_1 =0+1=1$,
$s^2_2 =0+1-2=-1$,
$s^2_3 =0+1-2+3=2$,
$s^2_4 =0+1-2+3-4=-2$,
$\ldots$,
$s^2_n =-\frac{n}{2}$ for even $n$,
and
$s^2_n =-\frac{n+1}{2}$ for odd $n$.
It thus establishes a strict one-to-one mapping
$s^2: {\Bbb N} \mapsto {\Bbb Z}$
of the natural numbers onto the integers.

These ``Abel sum'' type manipulations are outside of the radius of convergence
of the series and therefore cannot be expected to
result in any meaningful statement.
They could, in a strict sense, not even be perceived in terms of certain
paradoxes of infinity, such as Hilbert's hotel.
If they could quantify some sort of ``averaging'' remains questionable.
One could thus rightly consider any such
exploitations of infinities
as not only meaningless but outrightly wrong
-- even more so when committing to transgressions of convergence criteria.
Note nevertheless, that great minds have contemplated
geometric series for ever-decreasing ``Zeno squeezed'' computation cycle times,\cite[-30mm]{Russell-36,weyl:49}
or wondered in which state a (Thomson) lamp \index{Thomson lamp}   would
be after an infinite number of switching cycles
whose ever-decreasing switching times allow a geometric progression.\cite{thom:54}

\section{Riemann zeta function and Ramanujan  summation: Taming the beast}
\label{2019-mm-ds-zetafr}
\index{Ramanujan summation}

Can we make any sense\marginnote{For proofs and additional information see \S~3.7 in  \bibentry{Tao-2013}.}  of the seemingly absurd statement of the last section -- that
an infinite sum of all (positive) natural numbers appears to be both negative and ``small;'' that is, $-\frac{1}{12}$?
In order to set things up
let us introduce a generalization of the harmonic series:
the Riemann zeta function (sometimes also referred to as the Euler-Riemann zeta function)
\index{zeta function}
\index{Riemann zeta function}
\index{Euler-Riemann zeta function}
defined for $\Re t > 1$ by
\begin{equation}
\zeta (t) \stackrel{{\rm def}}{=}
\sum_{j=1}^\infty \frac{1}{j^t}
=
\prod_{p\;{\rm prime}} \left(\sum_{j=1}^\infty p^{-jt}\right)
=
\prod_{p\;{\rm prime}} \frac{1}{1-\frac{1}{p^t}}
\label{2016-m-ds-zeta}
\end{equation}
can be  continued analytically to all complex values $t\neq 1$.
Formally this analytic continuation yields the following
{\em Ramanujan summations}  (denoted by an ``R'' on top of equality sign)
for $t=0,-1,-2$ as follows\sidenote{For $t=-1$ this  has been ``derived''  earlier.}:
\begin{equation}
\begin{split}
1+1+1+1+1 + \cdots = \sum_{j=1}^\infty 1 \stackrel{R}{=} \zeta (0)  = -\frac{1}{2}
, \\
1+2+3+4+5 + \cdots = \sum_{j=1}^\infty j \stackrel{R}{=}  \zeta (-1)  = -\frac{1}{12}
, \\
1+4+9+16+25 + \cdots = \sum_{j=1}^\infty j^2 \stackrel{R}{=}  \zeta (-2)  = 0;
\end{split}
\label{2019-mm-ch-ds-Ramanujan}
\end{equation}
or, more generally,
for $s=1,2,\ldots$,
\begin{equation}
1+2^s+3^s+4^s+5^s + \cdots = \sum_{j=1}^\infty j^s \stackrel{R}{=} \zeta (-s) = -\frac{B_{s+1}}{ s+1}
,
\end{equation}
where $B_s$ are the
{\em Bernoulli numbers}.\cite{Sloane_oeis.org/A027642}
\index{Bernoulli numbers}

This scheme can be extended\cite{Masina-Grandi} to ``alternated''
zeta functions
\begin{equation}
1-2^s+3^s-4^s+5^s - \cdots =\sum_{j=1}^\infty \frac{(-1)^{j+1}}{j^s}
=      - \sum_{j=1}^\infty \frac{(-1)^j}{j^s} 
\end{equation}
by subtracting a similar series containing all even summands
twice:
\begin{equation}
\begin{split}
\sum_{j=1}^\infty \frac{(-1)^{j+1}}{j^s}
=
-\sum_{j=1}^\infty \frac{(-1)^j}{j^s}
=
\sum_{j=1}^\infty \frac{ 1 }{j^s}
-
2 \sum_{j=1}^\infty \frac{ 1 }{ (2j)^s}
\\
\stackrel{R}{=}
\zeta (s) - \frac{2}{2^s}\zeta (s)
=
\left(1  -2^{1-s} \right)   \zeta (s)
=\eta (s).
\end{split}
\label{2019-mm-ch-ds-Ramanujan-eta}
\end{equation}
$\eta (t) = \left(1  -2^{1-t} \right)   \zeta (t)  $
stands for the {\em Dirichlet eta function}.
\index{Dirichlet eta function}

By~(\ref{2019-mm-ch-ds-Ramanujan-eta}),  like in the Abel case, Grandi's series
corresponds to $s=0$, and sums up to
\begin{equation}
\begin{split}
1-1+1- \cdots = \sum_{j=1}^\infty \frac{(-1)^{j+1}}{j^0}  \stackrel{R}{=} \eta (0)  =  \left(1  - 2^1\right)  \zeta (0)
= -\zeta (0) = \frac{1}{2}
.
\end{split}
\label{2019-mm-ch-ds-Ramanujan-eta-Grandi}
\end{equation}

One way mathematicians cope with ``difficult entities''
such as generalized functions or divergent series is
to introduce suitable ``cutoffs'' in the form of multiplicative functions
and work with the resulting ``truncated'' objects instead.
We have encountered this both in Ritt's theorem
\index{Ritt's theorem}
\index{theorem of Ritt}
(cf. Section~\ref{2019-mm-ch-ca-Ritt}
on page~\pageref{2019-mm-ch-ca-Ritt})
and by inserting test functions associated with distributions
(cf. Chapter~\ref{2011-m-ch:gf}).

Therefore,
as Tao has pointed out,
if the divergent sums are multiplied with suitable ``smoothing'' functions\sidenote{An example of such smoothing function is
$\eta (x) = \theta\left(x^2-1\right) \exp \left(\frac{x^2}{x^2-1}\right)$, and, therefore,
$\eta \left(\frac{x}{N}\right) = \theta\left(x^2-N^2\right) \exp \left( \frac{x^2}{x^2-N^2}\right)$
defined in~(\ref{2019-m-ch-di-tf1-mod1})
on page~\pageref{2019-m-ch-di-tf1-mod1}.}
$\eta \left(\frac{j}{N}\right)$  which are
bounded, have a compact support,
and tend to $1$ at $0$ -- that is, $\eta (0) = 1$ for ``large $N$'' --
the respective smooth summations yield  smoothed asymptotics.
Then the divergent series
can be (somehow superficially\cite{Candelpergher-2017}) ``identified with''  their respective constant terms of their smoothed partial
sum asymptotics.

More explicitly, for the sum of natural numbers, and, more generally, for any fixed $s\in \mathbb{N}$ this yields\cite{Tao-2013}
\begin{equation}
\begin{split}
\sum_{j=1}^\infty j \eta \left(\frac{j}{N}\right)
= - \frac{1}{12} + C_{\eta ,1} N^2 + O\left(\frac{1}{N}\right)
,  \\
\sum_{j=1}^\infty j^s \eta \left(\frac{j}{N}\right)
= - \frac{B_{s+1}}{s+1} + C_{\eta ,s} N^{s+1} + O\left(\frac{1}{N}\right)
,
\end{split}
\label{2019-mm-ch-ds-Ramanujan-snn}
\end{equation}
where   $C_{\eta ,s}$ is the
{\em Archimedean factor}
\index{Archimedean factor}
\begin{equation}
\begin{split}
C_{\eta ,s}
\stackrel{\text{def}}{=}
\int_0^\infty x^s \eta (x) dx
.
\end{split}
\label{2019-mm-ch-ds-Ramanujan-af}
\end{equation}
Observe that~(\ref{2019-mm-ch-ds-Ramanujan-snn})
forces the Archimedean factor $C_{\eta ,1}$ to be positive and ``compensate for'' the constant factor
$-\frac{B_{s+1}}{s+1}$, which, for $s=1$, is negative and $-\frac{B_{2}}{2}=-\left(\frac{1}{6}\right)\left(\frac{1}{2}\right)=-\frac{1}{12}$.
In this case, as $N$ gets large, the sum diverges with $O\left(N^2\right)$, as can be expected from Gauss' summation formula $1+2+...+N= \frac{N(N+1)}{2}$
for the  the partial sum of the natural numbers up to $N$.

As can be expected both sides of (\ref{2019-mm-ch-ds-Ramanujan-snn}) diverge in the limit $N \rightarrow \infty$ and thus
$\eta \left(\frac{j}{N}\right)\rightarrow \eta (0)=1$.
For $s=1$ this could be interpreted as an instance of Ritt's theorem;
for arbitrary $s\in \mathbb{N}$ as a generalization thereof.

\section{Asymptotic power series}
\label{2019-mm-ch-ds-aps}

Divergent (power) series appear to be
living in the ``grey area'' in-between convergence and divergence, and, if
treated carefully,
may still turn out to be useful; in particular, when it comes to numerical approximations:
the first few terms of divergent series may (but not always do) ``converge''
to some ``useful functional'' value.
Alas, by taking into account more and more terms,
these series expansions eventually ``degrade'' through the rapidly increasing additional terms.
These cases have been termed asymptotic,\cite[-90mm]{Erdelyi-1956,Bender-Orszag,Balser-1994} semi-convergent, or convergently beginning series.
Asymptoticity has already been defined
in Section~\ref{2019-mm-ch-ca-Ritt} (on page~\pageref{2019-mm-ch-ca-Ritt}).
\index{asymptoticity}

Thereby the pragmatic emphasis is on a proper and ``useful'' (versus disadvantageous) {\em representation}
or {\em encoding} of entities such as functions and solutions of
ordinary differential equations by power series -- differential equations
with irregular singular points which are
not of the Fuchsian type, and not solvable by the Frobenius method.

The heuristic (not exact)
\index{optimal truncation rule}
{\em optimal truncation rule}
\sidenote{This pragmatic approach may cause some ``digestion problems;''
see Heaviside's remarks on page~\pageref{2013-m-ch-intro-cooking}.}
suggests that the best approximation to a function value from its divergent asymptotic
series expansion is often obtained by truncating the series (before or) at its smallest term.

{
\color{blue}
\bexample

To get a feeling for what can be expected in such scenarios
consider a ``canonical'' example:
With regard to convergence
the Stieltjes function\cite{Bleistein-Handelsman}
(formula~5.1.28, page~230 of Abramowitz and Stegun\cite[0mm]{abramowitz:1964:hmf} but with $x \mapsto \frac{1}{x}$)
\index{Stieltjes function}
\begin{equation}
\begin{split}
S(x)   = \int_0^\infty    \frac{e^{-t}}{1+tx} dt  = \frac{1}{x}  \int_0^\infty    \frac{e^{-t}}{\frac{1}{x}+t} dt
\end{split}
\label{2019-m-ch-ds-Stieltjes-function}
\end{equation}
can be represented by
power and inverse factorial series in three different ways:
\marginnote{In a metamatematical interpretation one might perceive mathematical entities such as functions
as ``ontologically'' existing in a ``Platonist universe of ideas''.
However, ontologic ``existence'' need not necessarily entail
concrete, operational, algorithmic ``epistemic access'' by formalizable means necessary for, say, physical prediction.
Therefore, pragmatic access to these functions presents an epistemic issue depending on our respective capacities and means to do so;
that is, it becomes means relative.}
\begin{itemize}
\item[(i)]
by the asymptotic Stieltjes series: for $n \in \mathbb{N}$,
\index{Stieltjes series}
\begin{equation}
\begin{split}
S(x) =
   \underbrace{\sum_{j=0}^n (-x)^j j!  }_{=S_n(x)}
+  \underbrace{(-x)^{n+1}(n+1)! \int_0^\infty   \frac{e^{-t}}{(1+tx)^{n+2}} dt}_{=R_n(x)},
\end{split}
\label{2019-m-ch-ds-asymptotic-Stieltjes-series}
\end{equation}
\item[(ii)]
by convergent Maclaurin series such as (Ramanujan found a series which converges even more rapidly)
\begin{equation}
\begin{split}
S(x)
=  \frac{e^\frac{1}{x}}{x}  \Gamma \left( 0, \frac{1}{x} \right)
=  -\frac{e^\frac{1}{x}}{x}  \left[  \gamma - \log x +\sum_{j=1}^\infty \frac{(-1)^j}{j!j x^j} \right]
,
\end{split}
\label{2019-m-ch-ds-Stieltjes-series}
\end{equation}
where

\begin{equation}
\gamma
=\lim_{n\rightarrow \infty}\left(
\sum_{j=1}^n \frac{1}{j}- \log n
\right) \approx 0.5772
\end{equation}
is the Euler-Mascheroni constant.\cite[-10mm]{Sloane_oeis.org/A001620}
\index{Euler-Mascheroni constant}
$\Gamma (z,x)$ represents the upper incomplete gamma function defined in~(\ref{2017-m-ch-sf-edgamma}).
\index{incomplete gamma function}

\item[(iii)]
by an inverse factorial series~\cite{Weniger2010}
\begin{equation}
S(x)= \sum_{j=0}^\infty \frac{(-1)^j}{\left(\frac{1}{x}\right)_{j+1}} \sum_{k=0}^j {S}_j^{(k)} k!,
\label{2022-afssf}
\end{equation}
where
$\left(x\right)_{j}=\Gamma \left(x+j\right)/\Gamma \left( x \right) = x \left(x +1\right)\cdots \left( x + j - 1 \right)$
and $(x)_0=1$
are Pochhammer symbols~(\ref{2011-m-ch-sfsf}) introduced on page~\pageref{2011-m-ch-sfsf},
\index{Pochhammer symbol}
and ${S}_j^{(k)}$
are Sterling numbers of the first kind
\index{Sterling numbers of the first kind}
that are the polynomial coefficients of the
Pochhammer symbol $(z-j+1)_j$
(Section~24.1.3, page~824 of Abramowitz and Stegun\cite[-15mm]{abramowitz:1964:hmf});
that is [cf. Equation(\ref{2022-m-ch-dsifps})],
\begin{equation}
\sum_{k=0}^j {S}_j^{(k)} z^k
= (z-j+1)_j
= (-1)^j  (-z)_j
\label{2022-ch-snfk}
\end{equation}
for $j \in \mathbb{N}\cup \{0\}$.
$(-1)^{j-k}{S}_j^{(k)}>0$
can also be identified with the (positive) number of permutations of $n$ symbols which have exactly $m$ cycles.
Because of the factor $(-1)^{j-k}$, as $j$ is fixed and $k$ varies, ${S}_j^{(k)}$ are strongly oscillating.
We conjecture without proof that in the inner sum of~(\ref{2022-afssf})
there are ``substantial'' cancellations.
The general method of conversion of a power series into an inverse factorial series will be discussed in the next section~\ref{2022-factser}.
\end{itemize}

A complete derivation\cite[-5mm]{sommer-Stieltjes}
of the Maclaurin series~(ii) is omitted; we just note that
the Stieltjes function $S(x)$ for real positive $x>0$ can be rewritten in terms of
the exponential integral
\index{exponential integral}
(e.g., formul\ae~5.1.1, 5.1.2, 5.1.4, page~227 of Abramowitz and Stegun\cite{abramowitz:1964:hmf})
\marginnote{See also \url{http://mathworld.wolfram.com/En-Function.html},
\url{http://functions.wolfram.com/GammaBetaErf/ExpIntegralEi/introductions/ExpIntegrals/ShowAll.html}
as well as  \bibentry{Masina-EIntegral}.}
\begin{equation}
\begin{split}
E_1(y) =   -\text{Ei}(-y)
= \Gamma \left( 0, y \right)
=
\int_{1}^\infty    \frac{e^{-uy}}{u} du
=
\int_{y}^\infty    \frac{e^{-u}}{u} du
\end{split}
\label{2019-m-ch-ds-Stieltjes-Ei}
\end{equation}
by first substituting $x=\frac{1}{y}$ in $S(x)$ as defined in~(\ref{2019-m-ch-ds-Stieltjes-function}), followed by the transformation of integration variable
$t = y(u-1)$, so that, for $y>0$,
\begin{equation}
\begin{split}
S\left(\frac{1}{y}\right)
=  \int_0^\infty    \frac{e^{-t}}{1+\frac{t}{y}} dt\\
[\text{substitution } t = y(u-1)\text{, }u=1+\frac{t}{y}\text{, }dt =y du]\\
=  \int_1^\infty    \frac{e^{-y(u-1)}}{1+\frac{y(u-1)}{y}} y du\\
=ye^{y}\int_1^\infty    \frac{e^{-yu}}{u}  du
=  ye^{y}E_1(y)  =  -ye^{y}\text{Ei}(-y)
\text{, or}\\
S(x)=  \frac{e^\frac{1}{x}}{x} E_1\left(\frac{1}{x}\right)
= \frac{e^\frac{1}{x}}{x} \Gamma \left( 0, \frac{1}{x} \right)
.
\end{split}
\label{2019-m-ch-ds-Stieltjes-function-Ei}
\end{equation}
The asymptotic Stieltjes
series~(\ref{2019-m-ch-ds-asymptotic-Stieltjes-series})
quoted in (i)
as well as the convergent
series~(\ref{2019-m-ch-ds-Stieltjes-series})
quoted in (ii)
can, for positive (real) arguments, be obtained by substituting the respective series for the exponential integral
(e.g., formul\ae~5.1.51, page~231 and~5.1.10,5.1.11, page~229  of Abramowitz and Stegun):
\begin{equation}
\begin{split}
E_1(y)
\sim
\frac{e^{-y}}{y}\sum_{j=0}^\infty (-1)^j j!  \frac{1}{y^j} =
\frac{e^{-y}}{y} \left( 1 - \frac{1}{y}  + 2 \frac{1}{y^2} + 6 \frac{1}{y^3} + \cdots
\right) \\
 = \Gamma \left( 0, y \right)=- \gamma - \log y - \sum_{j=1}^\infty \frac{ (-y)^j }{j (j!)}
,
\end{split}
\label{2019-m-ch-ds-Stieltjes-Ei-se}
\end{equation}
where again $\gamma$ stands for the Euler-Mascheroni constant
and $\Gamma (z,x)$ represents the upper incomplete gamma function
(cf. Formula~6.5.1, p~260   of Abramowitz and Stegun)
defined in~(\ref{2017-m-ch-sf-edgamma}).
\marginnote{It would be wrong but tempting
-- and would make the estimation of the remainder easier --
to treat the divergent series
very much like a geometric series outside its radius of convergence.}

The
divergent remainder of the asymptotic Stieltjes
series~(\ref{2019-m-ch-ds-asymptotic-Stieltjes-series})
can be estimated
by successive partial integrations of the Stieltjes function
and induction:
\begin{equation}
\begin{split}
S(x) =  \int_0^\infty    \frac{e^{-t}}{1+tx} dt
=
-\left. \frac{e^{-t}}{1+tx}\right \vert_{t=0}^{t=\infty}
-
x \int_0^\infty    \frac{xe^{-t}}{(1+tx)^2} dt  \\
= 1-      x \int_0^\infty    \frac{e^{-t}}{(1+tx)^2} dt\\
= 1-  x+    2x^2 \int_0^\infty    \frac{e^{-t}}{(1+tx)^2} dt\\
= 1-  x+    2x^2 \int_0^\infty    \frac{e^{-t}}{(1+tx)^3} dt\\
= 1-  x+    2x^2 -6 x^3\int_0^\infty    \frac{e^{-t}}{(1+tx)^4} dt\\
\vdots  \\
=\underbrace{\sum_{j=0}^n (-x)^j j!  }_{=S_n(x)}
+  \underbrace{(-x)^{n+1}(n+1)!   \int_0^\infty   \frac{e^{-t}}{(1+tx)^{n+2}} dt}_{=R_n(x)}
.
\end{split}
\label{2019-m-ch-ds-Stieltjes-function-pi}
\end{equation}
\begin{marginfigure}
{\color{black}
\begin{center}
\begin{tabular}{c}
\begin{tikzpicture} [ scale=0.53, every mark/.append style={mark size=2pt} ]
\begin{axis}[ 
    ticklabel style = {font=\Large },
    x label style={font=\Large ,at={(axis description cs:1.1,0.1)}},
    y label style={font=\Large ,at={(axis description cs:0.15,1.1)},rotate=270},
    xlabel={$n$},
    ylabel={$F_n(x)$},
    yticklabel style={
        /pgf/number format/fixed,
        /pgf/number format/precision=5
                     },
    scaled y ticks=false
]

\addplot  [
red, mark=*
]  table {   
1       2086.84
2       -2551.08
3       2602.16
4       -2229.
5       1635.93
6       -1048.05
7       595.202
8       -303.45
9       140.329
10      -59.3718
11      23.1491
12      -8.3693
13      2.82067
14      -0.890289
15      0.264232
16      -0.0740067
17      0.0196234
}
node [pos=0.9,label={[xshift=1.0cm, yshift=0.3cm, style={font=\Large}]$x=\frac{1}{5}$} ]{};

\end{axis}
\end{tikzpicture}
\\
\begin{tikzpicture} [ scale=0.53, every mark/.append style={mark size=2pt} ]
\begin{axis}[ 
    ticklabel style = {font=\Large },
    x label style={font=\Large ,at={(axis description cs:1.1,0.1)}},
    y label style={font=\Large ,at={(axis description cs:0.15,1.1)},rotate=270},
    xlabel={$n$},
    ylabel={$E_n(x)$},
    yticklabel style={
        /pgf/number format/fixed,
        /pgf/number format/precision=5
                     },
    scaled y ticks=false
]
\addplot  [
blue, mark=*
]  table {   
1       -0.0521109
2       0.0278891
3       -0.0201109
4       0.0182891
5       -0.0201109
6       0.0259691
7       -0.0385429
8       0.0646763
9       -0.121118
10      0.250471
}
node [pos=0.2,label={[xshift=1.0cm, yshift=0.3cm, style={font=\Large}]$x=\frac{1}{5}$} ]{};

\end{axis}
\end{tikzpicture}
\\
\begin{tikzpicture} [ scale=0.53, every mark/.append style={mark size=2pt} ]
\begin{axis}[ 
    ticklabel style = {font=\Large },
    x label style={font=\Large ,at={(axis description cs:1.1,0.1)}},
    y label style={font=\Large ,at={(axis description cs:0.15,1.1)},rotate=270},
    xlabel={$n$},
    ylabel={$E_n(x)$},
    yticklabel style={
        /pgf/number format/fixed,
        /pgf/number format/precision=5
                     },
    scaled y ticks=false
]
\addplot  [
orange, mark=*
]  table {   
1       -0.0156333
2       0.00436666
3       -0.00163334
4       0.000766661
5       -0.000433339
6       0.000286661
7       -0.000217339
8       0.000185861
9       -0.000177019
10      0.000185861
11      -0.000213307
12      0.000265694
13      -0.000357008
14      0.000514775
15      -0.000792899
16      0.00129938
17      -0.00225749
18      0.00414488
19      -0.00801963
20      0.0163094
21      -0.0347816
}
node [pos=0.2,label={[xshift=1.0cm, yshift=0.3cm, style={font=\Large}]$x=\frac{1}{10}$} ]{};

\end{axis}
\end{tikzpicture}
 \\
\begin{tikzpicture} [ scale=0.53, every mark/.append style={mark size=2pt} ]
\begin{axis}[ 
    ticklabel style = {font=\Large },
    x label style={font=\Large ,at={(axis description cs:1.1,0.1)}},
    y label style={font=\Large ,at={(axis description cs:0.15,1.1)},rotate=270},
    xlabel={$n$},
    ylabel={$E_n(x)$},
    yticklabel style={
        /pgf/number format/fixed,
        /pgf/number format/precision=5
                     },
    scaled y ticks=false
]
\addplot  [
green, mark=*
]  table {   
1       -0.0074708532778055
2       +0.001418035611083335
3       -0.0003597421666944323
4       +0.0001143319073796389
5       -0.00004369278397831078
6       +0.00001951709256486911
7       -0.00000998084982195557
8       +0.000005751386117691659
9       -0.000003687955446141089
10      +0.000002604938929784417
11      -0.000002009850279205416
12      +0.000001681981087964246
13      -0.000001517606096901325
14      +0.00000146867527561767
15      -0.000001517606096901325
16      +0.000001667760700430065
17      -0.000001942321669945457
18      +0.000002389777174482965
19      -0.00000309754802829687
20      +0.000004218885575557607
21      -0.000006024121470105115
22      +0.00000899895553019281
23      -0.00001403642920350112
24      +0.00002282018637023153
25      -0.0000386075062525082
26      +0.00006786716096018885
27      -0.0001237872400228213
28      +0.0002339676418122716
29      -0.0004576917964023153
30      +0.000925627080027081
31      -0.001933231931260382
32      +0.004165667292819641
}
node [pos=0.2,label={[xshift=1.0cm, yshift=0.3cm, style={font=\Large}]$x=\frac{1}{15}$} ]{};

\end{axis}
\end{tikzpicture}
\end{tabular}
\end{center}
\caption{The series approximation error   $F_n(x)= -\frac{e^\frac{1}{x}}{x}  \left[  \gamma - \log x +\sum_{j=1}^n \frac{(-1)^j}{j!j x^j} \right] -S(x)$
of the convergent Stieltjes series~(\ref{2019-m-ch-ds-Stieltjes-series})
for $x=\frac{1}{5}$, and
$E_n(x)=S_n(x) -S(x)$
of the Stieltjes series~(\ref{2019-m-ch-ds-asymptotic-Stieltjes-series})  as a function of increasing $n$
for $x\in \left\{ \frac{1}{5},\frac{1}{10},\frac{1}{15}\right\}$.}
  \label{2018-mm-ferrorS}
}
\end{marginfigure}
For $x>0$ the absolute value of the remainder $R_n(x)$
can be estimated to be bound from above by
\begin{equation}
\begin{split}
\vert R_n(x)\vert
=(n+1)! \, x^{n+1} \int_0^\infty   \frac{e^{-t}}{(1+xt)^{n+2}} dt
\le (n+1)! \, x^{n+1} \underbrace{\int_0^1    e^{-t} dt}_{=1}.
\end{split}
\label{2019-m-ch-ds-Stieltjes-function-pi-re}
\end{equation}
By examining\cite[9mm]{Erdelyi-1956} the partial series $\left| S_n(x) \right|=  \sum_{j=0}^n j! x^j$
with   the bound on the remainder $\left| R_n(x) \right| \le (n+1)!  \, x^{n+1}$
it can be inferred that  the bound on the remainder is of the same magnitude as the
first ``neglected''  term $(n+1)! x^{n+1}$.

A comparison of the argument $x$ of the Stieltjes series
with the number $n$ of terms contributing to $S_n(x)$ reveals three regions:
\begin{itemize}
\item[(i)] if $x=0$ the remainder vanishes for all $n$ and the series converges towards the constant $1$ (regardless of $n$).
\item[(ii)] if $x>1$ the series diverges; no matter what (but could be subjected to ``resummation procedures'' {\it \`a la} Borel,
cf Sections~\ref{2019-mm-ds-bresum}{\&}\ref{2019-mm-ch-ds-rsdg});
\item[(iii)] if $x = 1/y<1$ (and thus $y>1$)  the remainder $
\left| R_n\left(\frac{1}{y}\right) \right| \le \frac{(n+1)!}{y^{n+1}}   )$  is dominated by the $y$ term
until about $n=y$; at which point the factorial takes over and
the partial sum
$S_n(x)$ starts to become an increasingly worse approximation.

Therefore, although the
Stieltjes series is divergent for all $x>0$, in the domain $0<x<1$
it behaves very much like a convergent series until
about $n \approx x < 1$. In this $0< x < 1$ regime it makes sense to define an error estimate $E_k(x)=S_n(x)-S(x)$
as the difference between the partial sum $S_n(x)$, taken at $x$ and including terms up to the order of $x^n$,
and the exact value $S(x)$.

Figure~\ref{2018-mm-ferrorS}
depicts the asymptotic divergence of $S_n(x)$ for
$x\in \{\frac{1}{5},\frac{1}{10},\frac{1}{15}\}$
up to the respective adapted values $n \approx \frac{1}{x}$.
Since in the kernerls of the sums of the asymptotic Stieltjes series~(\ref{2019-m-ch-ds-asymptotic-Stieltjes-series})
$k_j(x)= (-1)^j j! x^j $
and the convergent Stieltjes series~(\ref{2019-m-ch-ds-Stieltjes-series})
$\frac{(-1)^j}{j!j x^j} = \frac{\left[k_j(x)\right]^{-1}}{ j }$
are ``almost inverse''   it can be expected that, for $0<x<1$, and if one is only willing to take ``the first view'' terms of these respective sums,  then the
former asymptotic Stieltjes series~(\ref{2019-m-ch-ds-asymptotic-Stieltjes-series}) will perform better than the latter
convergent Stieltjes series~(\ref{2019-m-ch-ds-Stieltjes-series})
the smaller $x\ll 1$   is.

\end{itemize}

 \eexample
 }

\section{Conversion of power series into inverse factorial series}
\index{factorial series}
\label{2022-factser}

\marginnote{The following recasting of power series into inverse factorial series closely follows \bibentry{Weniger2010}.}

We have already encountered a conversion of power series into inverse factorial series when discussing one ``epistemic access'' to, that is,
one representation of, the Stieltjes function in Equation~\ref{2022-afssf}.
In what follows  general power series
$f(z) = \sum_{n=0}^\infty a_n z^n$ will be rewritten
into (inverse) factorial series~\cite[-90mm]{Watson1912,Doetsch1972}
\begin{equation}
\begin{split}
f(z)
=
\alpha_0 \frac{1}{z}
+
\alpha_1 \frac{1!}{z(z+1)}
+
\alpha_2 \frac{2!}{z(z+1)(z+2)}+
\ldots
= \sum_{n=0}^\infty \alpha_n \frac{n!}{(z)_{n+1}}
,
\end{split}
\label{2022-m-ch-dsfs}
\end{equation}
where $(z)_{n+1}$ are Pochhammer symbols
\index{Pochhammer symbol}
which have been introduced in Equation~(\ref{2011-m-ch-sfsf}).
Thereby the main ``ingredient'' will be Sterling numbers
\index{Sterling numbers of the first kind}
which are defined and reviewed in Section~24.1.3, page~824 of Abramowitz and Stegun\cite[-20mm]{abramowitz:1964:hmf}.

To accomplish this task we first rewrite the power series in $z$ into an inverse power series in $\frac{1}{z}$
\begin{equation}
\begin{split}
f(z)
= \sum_{n=0}^\infty a_n z^n
= \frac{1}{z}  \sum_{n=0}^\infty \frac{a_n}{\left(\frac{1}{z}\right)^{n+1}}
.
\end{split}
\label{2022-m-ch-dsinv}
\end{equation}

\marginnote{Cf. Equation~(6), \S~30, p.~78 in  \bibentry{Nielsen-Gammafunktion}, as well as Equation~(A.14) in  \bibentry{Weniger2010}.}
Stirling numbers of the first kind~(\ref{2022-ch-snfk})
\index{Stirling numbers of the first kind}
${S}_j^{(n)}$
have infinite generating functions.
These serve as ``translations''---that is, as expansions from
an (inverse) power $\frac{1}{z^{n+1}}$ in terms of  inverse  factorial series $(z)_{n+j+1}$:
for $k  \in \mathbb{N} \cup \{0\}$,
\begin{equation}
\begin{split}
\frac{1}{z^{n+1}}  = \sum_{j=0}^\infty \frac{(-1)^j}{(z)_{n+j+1}} {S}_{n+j}^{(n)}
.
\end{split}
\label{2022-m-ch-dsngf}
\end{equation}
$(z)_{k+j+1}$ are Pochhammer symbols
\index{Pochhammer symbol}
introduced in Equation~(\ref{2011-m-ch-sfsf}).

The respective ``reverse'' expansion of a Pochhammer symbol $(z)_{k+1}$ in terms of  an inverse power series
$\frac{1}{z^{n+j+1}}$
for $k  \in \mathbb{N} \cup \{0\}$ and $\vert z \vert >0$
is given by
\begin{equation}
\begin{split}
\frac{1}{(z)_{n+1}}  = \sum_{j=0}^\infty \frac{(-1)^j}{z^{n+j+1}} {S}_{n+j}^{(n)}
.
\end{split}
\label{2022-m-ch-dsngfinverse}
\end{equation}
\marginnote{Cf. Equation~(9), \S~26, ~p.~68 in  \bibentry{Nielsen-Gammafunktion}, as well as Equation~(A.11) in  \bibentry{Weniger2010}.}

Insertion of~(\ref{2022-m-ch-dsngf})  into~(\ref{2022-m-ch-dsinv}), rearranging the order of the summations through an index shift $m = n+j$ yields
\begin{equation}
\begin{split}
f(z)
= \frac{1}{z}  \sum_{n=0}^\infty a_n \sum_{j=0}^\infty \frac{(-1)^j}{\left(\frac{1}{z}\right)_{n+j+1}} {S}_{n+j}^{(n)}
\\
= \frac{1}{z}  \sum_{j=0}^\infty \sum_{n=0}^\infty a_n \;  \frac{(-1)^j}{\left(\frac{1}{z}\right)_{n+j+1}} {S}_{n+j}^{(n)}
\\
[[ m = n+j \text{ with }  n \ge 0 \text{ and }  j \ge 0
\\
\Rightarrow  m \ge 0  \text{ and } j = m - n \ge 0 \Rightarrow m \ge n \text{ or }  n \le m]]
\\
= \frac{1}{z} \sum_{m=0}^\infty   \frac{(-1)^m}{\left(\frac{1}{z}\right)_{m+1}} \sum_{n=0}^m (-1)^{\pm n} \; {S}_{m}^{(n)} \; a_n
.
\end{split}
\label{2022-m-ch-dsinv2}
\end{equation}
So if we define the inverse power series
\begin{equation}
\begin{split}
f'(u)
= z f(z)
= \frac{1}{u} f\left( \frac{1}{u} \right)
= \sum_{m=0}^\infty \frac{a'_m}{u^{m+1}}
= \sum_{m=0}^\infty b_m' \, \frac{m!}{(u)_{m+1}}
,
\end{split}
\label{2022-m-ch-comp1}
\end{equation}
with $u=1/z$,
then, by comparison,
\begin{equation}
\begin{split}
b_m'= \frac{1}{m!} \frac{1}{\left(u \right)_{m+1}} \sum_{n=0}^m \underbrace{(-1)^{m\pm n}  \; {S}_{m}^{(n)}}_{>0} \; a_n'
.
\end{split}
\label{2022-m-ch-comp}
\end{equation}

{\color{OliveGreen}
\bproof

In what follows we turn to the proof of the Stieltjes factorial series~(\ref{2022-afssf}) in terms of the  Stirling's factorial series.
Stirling's factorial series,
\index{Stirling's factorial series}
also known as Waring's formula,
\index{Waring's formula}
can be derived by iteration for $\Re (z - w) > 0$
\marginnote{For a derivation of Stirling's factorial series see \S~30,~p.~77 in  \bibentry{Nielsen-Gammafunktion}.}\begin{equation}
\begin{split}
\frac{1}{z} \sum_{j=0}^\infty \left( \frac{w}{z} \right)^j=
\frac{1}{z}\cdot \frac{1}{1-\frac{w}{z}}  \\
=\frac{1}{z-w} =
\frac{1}{z} + \frac{w}{z(z-w)}=
\sum_{n=0}^\infty  \frac{(w)_n}{(z)_{n+1}}
,
\end{split}
\label{2022-m-ch-dswaring}
\end{equation}
where $\Re z$ stands for the real part of $z$.
\index{Pochhammer symbol}
$(w)_n$ and $(z)_{n+1}$ are Pochhammer symbols.

Note that $(z-n+1)_n$ in~(\ref{2022-ch-snfk}) can be rewritten as  $(-1)^j  (-z)_j$
since the following identity for Pochhammer symbols hold:
\begin{equation}
\begin{split}
(a-n+1)_n = \underbrace{(a-n+1)(a+1-n+1) \cdots (a-1)a}_{n\text{ times}} \\
= (-1)^n (-a+n-1)(-a-1+n-1) \cdots (-a+1)(-a)  \\
= (-1)^n  (-a)(-a+1) \cdots (-a-1+n-1)(-a+n-1) \\
= (-1)^n  (-a)_n
.
\end{split}
\label{2022-m-ch-dsifps}
\end{equation}
By replacing $z$ by $-z$ in  $(z-n+1)_n=(-1)^n  (-z)_n$  we obtain from
Equation~(\ref{2022-ch-snfk})---that is, from
$(z-j+1)_j = \sum_{k=0}^j z^k  {S}_j^{(k)}$,
\begin{equation}
\begin{split}
(z-n+1)_n = (-1)^n  (-z)_n  = \sum_{k=0}^n z^k  {S}_n^{(k)},
\\
(-z)_n  = (-1)^n  \sum_{k=0}^n z^k  {S}_n^{(k)}
,
\\
[[\text{or, with } z \mapsto -z, ]]
\\
(z)_n = (-1)^n  \sum_{k=0}^n (-1)^k  z^k {S}_n^{(k)}
.
\end{split}
\label{2022-m-ch-dsifps22}
\end{equation}

Insertion of~(\ref{2022-m-ch-dswaring})
with $w = -t$ and~(\ref{2022-m-ch-dsifps22})
with $-z = -t$
into~(\ref{2019-m-ch-ds-Stieltjes-function}), that is, into $S(x)   =  \frac{1}{x}  \int_0^\infty    \frac{e^{-t}}{\frac{1}{x}+t} dt$, yields~(\ref{2022-afssf}):
\begin{equation}
\begin{split}
S(z)
= \frac{1}{z}  \int_0^\infty    \frac{e^{-t}}{\frac{1}{z}+t} dt
= \frac{1}{z}  \int_0^\infty  \sum_{n=0}^\infty  \frac{(-t)_n}{\left(\frac{1}{z}\right )_{n+1}} e^{-t} dt
\\
= \frac{1}{z}  \sum_{n=0}^\infty  \frac{(-1)^n }{\left(\frac{1}{z}\right )_{n+1}}   \sum_{k=0}^n  {S}_n^{(k)} \underbrace{\int_0^\infty  t^k  \, e^{-t} dt}_{=\Gamma(k+1)=k!}
\\
=\frac{1}{z} \sum_{n=0}^\infty \frac{(-1)^n}{\left(\frac{1}{z}\right)_{n+1}} \sum_{k=0}^n {S}_n^{(k)} k!
.
\end{split}
\label{2022-m-ch-dsafssf}
\end{equation}

\bproof
}

\marginnote{For a discussion of convergence see  Section~3 of \bibentry{Weniger2010}, as well as \bibentry{Nielsen-Gammafunktion}, and \bibentry{landau1906uber}.}
Let us briefly consider the convergence of the
the  inverse factorial series (\ref{2022-m-ch-dsfs}), that is, of
$\sum_{n=0}^\infty  a_n \,{n!}/{(z)_{n+1}}= \sum_{n=0}^\infty  a_n \, {n!}/{\left[\Gamma(z+n+1)/\Gamma(z)\right]}$.
Note that its terms of the form $a_n {n!}/{(z)_{n+1}}$
can be estimated by considering the factor  ${n!}/{(z)_{n+1}}$, and with the help of
$\Gamma(z+a)/\Gamma(z+b)= z^{a-b}\left[1 + O\left(\frac{1}{z}\right) \right]$ for $z\rightarrow \infty$
({\S}~6, Formula~6.1.47  on p.~257  of Abramowitz and Stegun),
as follows:
\begin{equation}
\begin{split}
\frac{n!}{(z)_{n+1}}
= \frac{\Gamma(n+1)}{\left[\Gamma(z+n+1)/\Gamma(z)\right]}
= \frac{\Gamma(n+1)}{\Gamma(n+1+z)} \Gamma(z)  \\
= (n+1)^{-z}\left[1 + O\left(\frac{1}{n+1}\right) \right](z-1)!
= O \left( n^{-z} \right) \text{ for } n \rightarrow \infty.
\end{split}
\label{2022-m-ch-estimate}
\end{equation}

Therefore, the  inverse factorial series (\ref{2022-m-ch-dsfs})
converges with the possible
exception of the points $z =-m$ with $m \in   \mathbb{N} \cup \{0\}$
(where the Pochhammer symbols in the denominator might vanish)
if and only if the associated Dirichlet series
\index{Dirichlet series}
$\sum_{n=1}^\infty a_n \, n^{-z}$
converges.

A Dirichlet series has an abscissa of convergence
\index{abscissa of convergence}
$\Re (z) > \lambda$, that is, it converges on this half-plane.
$\lambda =- \infty$ in which case the Dirichlet series converges uniformly, or
$\lambda = \infty$ in which case the Dirichlet series diverges uniformly.
\marginnote{For a discussion of the convergence of Dirichlet series, see for instance \S~58,~255, page~456 of   \bibentry{Knoop1996}.}
Even if the inverse power series diverges factorially the respective inverse factorial series may converge; but this has to be
checked explicitly.

However, a convergence issue encountered in inverse factorial series is the Stokes phenomenon~\cite{Costin2016Aug,Costin_2017}:
the asymptotic behavior of functions need not be uniform in different regions of the complex plane, bounded by (anti-)Stokes lines.
In particular, inverse factorial series may not be suitable for the study of Stokes phenomena if Stokes lines
are present in the right complex half-plane $\Re ( \alpha ) > \lambda $ because of the singularities on these Stokes lines.
One may conjecture that inverse factorials might converge in regions where the associated power series are Borel summable;
yet convergence fails in the presence of Stokes lines.
This would mean that quantum field theories have convergent inverse factorial expansions only in less than four dimensions.

\section{Borel's resummation method -- ``the master forbids it''}
\index{Borel resummation}
\label{2019-mm-ds-bresum}

In what follows we shall review
a resummation method
invented by Borel\cite[-13mm]{Borel1899}
\marginnote{{\em ``The idea that a function could be determined by a divergent asymptotic series was a foreign one to the nineteenth century mind.
Borel, then an unknown young man, discovered that his summation method gave the ``right'' answer for many classical divergent series.
He decided to make a pilgrimage to Stockholm to see Mittag-Leffler, who was the recognized lord of complex analysis.
Mittag-Leffler listened politely to what Borel had to say and then,
 placing his hand upon the complete works by Weierstrass, his teacher, he said in Latin,
``The Master forbids it.''} quoted as {\em A tale of Mark Kac} on page 38  by  \bibentry{reed-sim4}.}
to obtain the exact convergent solution
(\ref{2011-m-ch-dseefasol})
of the differential equation  (\ref{2011-m-ch-dsee})
from the divergent series solution (\ref{2011-m-ch-dseess}).
First note that a suitable infinite series  can be rewritten as an integral,
thereby using the integral representation
(\ref{2011-m-ch-sfgamman}\&\ref{2017-m-ch-sf-edgamma})
$
n! = \Gamma ( n+1 ) =
\int_0^\infty t^{n}e^{-t}dt
$
of the factorial
as follows:
\begin{equation}
\begin{split}
\sum_{j=0}^\infty
a_j
=
\sum_{j=0}^\infty
a_j  \frac{j!}{j!}
=
\sum_{j=0}^\infty
  \frac{a_j}{j!}  j!
\\
=
\sum_{j=0}^\infty
  \frac{a_j}{j!}  \int_0^\infty t^j e^{-t} dt
\stackrel{{\rm B}}{=}
\int_0^\infty \left(\sum_{j=0}^\infty   \frac{a_j t^j}{j!}\right)   e^{-t} dt
.
\end{split}
\label{2012-m-ch-dsborel}
\end{equation}

A series  $\sum_{j=0}^\infty   a_j $
is {\em Borel summable}
\index{Borel summable}
if
$\sum_{j=0}^\infty   \frac{a_j t^j}{j!}$ has a non-zero radius of convergence,
if it can be extended along the positive real axis, and if the integral
(\ref{2012-m-ch-dsborel}) is convergent.
This integral is called the
{\em Borel sum}
\index{Borel sum}
of the series.
It can be obtained by taking $a_j$, computing the sum $\sigma (t) = \sum_{j=0}^\infty   \frac{a_j t^j}{j!}$,
and integrating $\sigma (t)$ along the positive real axis with a ``weight factor'' $e^{-t}$.

More generally, suppose
\begin{equation}
S(z)= z \sum_{j=0}^\infty
a_j  z^j  = \sum_{j=0}^\infty
a_j  z^{j+1}
\label{2018-m-ch-ds-fps}
\end{equation}
is some formal power series.
Then  its {\em Borel transformation}
\index{Borel transformation}
is defined by
\begin{equation}
\begin{split}
\sum_{j=0}^\infty
a_j z^{j+1}
=
\sum_{j=0}^\infty
a_j z^{j+1}  \frac{j!}{j!}
=
\sum_{j=0}^\infty
  \frac{a_j z^{j+1} }{j!}  \underbrace{j!}_{ \int_0^\infty t^j e^{-t}   dt }  \\
=
\sum_{j=0}^\infty
  \frac{a_jz^{j} }{j!}  \int_0^\infty t^j e^{-t} z dt
\stackrel{{\rm B}}{=}
\int_0^\infty \left(\sum_{j=0}^\infty   \frac{a_j (z t)^j}{j!}\right)  e^{-t} z dt \\
[\textrm{variable substitution }  y= z t, \; t = \frac{y}{z}  , \; dy = z \,dt, \; dt = \frac{dy}{z}]\\
\stackrel{{\rm B}}{=}
\int_0^\infty \left(\sum_{j=0}^\infty   \frac{a_j y^j}{j!}\right)  e^{-\frac{y}{z}}   dy  =
\int_0^\infty {\cal B}S (y)  e^{-\frac{y}{z}}   dy
.
\end{split}
\label{2012-m-ch-dsboreltrafo}
\end{equation}

Often, this is written with $z=1/t$, such that the  Borel transformation
\index{Borel transformation}
is defined by
\begin{equation}
\begin{split}
\sum_{j=0}^\infty
a_j t^{-(j+1)}
\stackrel{{\rm B}}{=}
\int_0^\infty {\cal B}S (y)  e^{- yt }   dy
.
\end{split}
\label{2012-m-ch-dsboreltrafo2}
\end{equation}

The {\em Borel transform}\sidenote[][0mm]{This definition differs from the standard definition of the Borel
transform  based on coefficients $a_j$ with $S(z)=   \sum_{j=0}^\infty
a_j  z^{j}$ introduced in
 \bibentry{Kleinert-Schulte},  \bibentry{Helling-2012},  \bibentry{Dorigoni-2014} and  \bibentry{Dunne-talk-ETH-2018}.}
\index{Borel transform}
of   $S(z)=   \sum_{j=0}^\infty
a_j  z^{j+1} =  \sum_{j=0}^\infty
a_j  t^{-(j+1)}$
is thereby defined as
\begin{equation}
{\cal B}S (y)
=
  \sum_{j=0}^\infty   \frac{a_j y^j}{j!}
.
\label{2012-m-ch-dsboreltransform}
\end{equation}


{
\color{blue}
\bexample

In the following, a few examples will be given.

\begin{itemize}
\item[(i)]
The Borel sum
of Grandi's series (\ref{2009-fiftyfifty-1s})
\index{Grandi's series}
is equal to its Abel sum:
\begin{equation}
\begin{split}
s= \sum_{j=0}^\infty (-1)^j
\stackrel{{\rm B}}{=}
\int_0^\infty \left(\sum_{j=0}^\infty   \frac{(-1)^j t^j}{j!}\right)   e^{-t} dt  \\
=
\int_0^\infty \underbrace{\left(\sum_{j=0}^\infty   \frac{(- t)^j}{j!}\right)}_{e^{-t}}   e^{-t} dt
=
\int_0^\infty    e^{-2t} dt  \\
[\textrm{variable substitution } 2t = \zeta, dt = \frac{1}{2} d \zeta ]\\
=
\frac{1}{2}
\int_0^\infty    e^{-\zeta } d\zeta      \\
=
\frac{1}{2}
\left.     \left(-e^{-\zeta }\right) \right|_{\zeta=0}^\infty
=
\frac{1}{2} \left(- \underbrace{e^{-\infty}}_{=0} + \underbrace{e^{-0}}_{=1}\right) = \frac{1}{2}
.
\end{split}
\end{equation}

\item[(ii)]
A similar calculation for $s^2$ defined in Equation~
(\ref{2009-fiftyfifty-1s1})
yields
\begin{equation}
\begin{split}
s^2= \sum_{j=0}^\infty (-1)^{j+1} j = (-1) \sum_{j=1}^\infty (-1)^j j
\stackrel{{\rm B}}{=}
-\int_0^\infty \left(\sum_{j=1}^\infty   \frac{(-1)^j j t^j}{j!}\right)   e^{-t} dt  \\
=
-\int_0^\infty \left(\sum_{j=1}^\infty   \frac{(-t)^j}{(j-1)!}\right)   e^{-t} dt
=
-\int_0^\infty \left(\sum_{j=0}^\infty   \frac{(-t)^{j+1}}{j!}\right)   e^{-t} dt  \\
=
-\int_0^\infty (-t) \underbrace{\left(\sum_{j=0}^\infty   \frac{(-t)^j}{j!}\right)}_{e^{-t}}   e^{-t} dt
=
-\int_0^\infty  (-t)  e^{-2t} dt  \\
[\textrm{variable substitution } 2t = \zeta, dt = \frac{1}{2} d \zeta ]\\
=
\frac{1}{4}
\int_0^\infty  \zeta  e^{-\zeta } d\zeta
=
\frac{1}{4}
\Gamma (2)
=
\frac{1}{4} 1! = \frac{1}{4}
,
\end{split}
\end{equation}
which is again equal to the Abel sum.
\index{Abel sum}

\item[(iii)]
The Borel transform
of a ``geometric'' series (\ref{2009-fiftyfifty-1sgs})
\index{geometric series}
$g(z)=a z \sum_{j=0}^\infty    z^{j}=a \sum_{j=0}^\infty    z^{j+1}$ with constant
coefficients $a$ and $0> z > 1$
is
\begin{equation}
{\cal B}g (y)
=
a \sum_{j=0}^\infty   \frac{y^j}{j!} = a e^y.
\end{equation}
The Borel transformation~(\ref{2012-m-ch-dsboreltrafo}) of this geometric series  is
\begin{equation}
\begin{split}
g(z) \stackrel{{\rm B}}{=} \int_0^\infty {\cal B}g (y)   e^{-\frac{y}{z}}   dy
=
\int_0^\infty a e^y  e^{-\frac{y}{z}}   dy
=
a \int_0^\infty e^{-\frac{y(1-z)}{z}}   dy  \\
\left[\textrm{variable substitution } x =  -y\frac{1-z}{z}, dy = -\frac{z}{1-z} dx \right]\\
=
\frac{-a z}{1-z} \int_0^{-\infty} e^{x}   dx
=
\frac{a z}{1-z} \int_{-\infty}^0 e^{x}   dx
=
a \frac{z}{1-z} (\underbrace{e^0}_{1}-\underbrace{e^{-\infty}}_{0})    =  \frac{a z}{1-z}.
\end{split}
\end{equation}

Likewise, the Borel transformation~(\ref{2012-m-ch-dsboreltrafo2})
of the geometric series
$g(t^{-1})=a   \sum_{j=0}^\infty    t^{-(j+1)}$ with constant $a$ and $t > 1$
is
\begin{equation}
\begin{split}
g(t^{-1}) \stackrel{{\rm B}}{=} \int_0^\infty {\cal B}g (y)   e^{-yt}   dy
=
\int_0^\infty a e^y  e^{-yt}   dy
=
a \int_0^\infty e^{- y(t-1) }   dy  \\
\left[\textrm{variable substitution } x =  - y(t-1) , dy = -\frac{1}{t-1} dx \right]\\
=
\frac{-a}{t-1} \int_0^{-\infty} e^{x}   dx
=
\frac{a }{t-1} \int_{-\infty}^0 e^{x}   dx
=
a \frac{1}{t-1} (\underbrace{e^0}_{1}-\underbrace{e^{-\infty}}_{0})    =  \frac{a }{t-1}.
\end{split}
\end{equation}

\end{itemize}
\eexample
}

\section{Asymptotic series as solutions of differential equations}
\label{2019-mm-ch-ds-rsdg}

Already in 1760 Euler observed\cite[-7mm]{Euler60} that what is today known as
the Stieltjes series
multiplied by  $x$; namely
the series
\begin{equation}
s(x) = x - x^2+2x^3-6x^4 + \ldots  = \sum_{j=0}^\infty (-1)^j j!  x^{j+1}   = x S(x)
,
\label{2011-m-ch-dseess}
\end{equation}
when differentiated, satisfies
\begin{equation}
\frac{d}{dx}s(x)= \frac{x-s(x)}{x^2},
\end{equation}
and thus in some way can be considered ``a solution'' of the  differential equation
\begin{equation}
\begin{split}
\left(x^2 \frac{d}{dx} +1\right) s(x) = {x},\;\text{ or }\;
\left(\frac{d}{dx} +\frac{1}{x^2}\right) s(x) = \frac{1}{x};
\end{split}
\label{2011-m-ch-dsee}
\end{equation}
resulting in a differential operator of the form ${\cal L}_x = \frac{d}{dx} +\frac{1}{x^2}$.

This equation has an irregular singularity at $x=0$ because
the coefficient of the zeroth derivative  $\frac{1}{x^2}$ has a pole of order $2$, which is greater than $1$.
Therefore,~(\ref{2011-m-ch-dsee}) is not of the Fuchsian type.
\index{Fuchsian equation}

Nevertheless, the differential equation~(\ref{2011-m-ch-dsee})
can be solved in five different ways:
\begin{itemize}
\item[(i)] by the convergent series solution~(\ref{2019-m-ch-ds-Stieltjes-seriesmwx})
based on the Stieltjes function~(\ref{2019-m-ch-ds-Stieltjes-function}), as pointed out earlier
(thereby putting in question speculations that it needs to be asymptotic divergent series to cope with irregular singularities
beyond the Frobenius {\it Ansatz});
\item[(ii)] by a proper (Borel) summation of Euler's divergent series~(\ref{2011-m-ch-dseess});
\item[(iii)] by quadrature, that is, direct integration of~(\ref{2011-m-ch-dsee});
\item[(iv)] by evaluating Euler's (asymptotic) divergent series~(\ref{2011-m-ch-dseess})
based on the Stieltjes series~(\ref{2019-m-ch-ds-Stieltjes-series})
to ``optimal order,''
and by comparing this approximation to the exact solution (by taking the difference); and
\item[(iv)] by evaluating the respective inverse factorial series~(\ref{2022-afssf}).
\end{itemize}

\subsection*{Solution by convergent series}
The differential equation~(\ref{2011-m-ch-dsee}) has a convergent series solution
which is inspired by the convergent series~(\ref{2019-m-ch-ds-Stieltjes-series})  for the Stieltjes function
multiplied by $x$; that is,
\begin{equation}
\begin{split}
s(x) =x S(x)
=   e^\frac{1}{x}   \Gamma \left( 0, \frac{1}{x} \right)
=  - e^\frac{1}{x}   \left[  \gamma - \log x +\sum_{n=1}^\infty \frac{(-1)^n}{n!n x^n} \right]
\end{split}
\label{2019-m-ch-ds-Stieltjes-seriesmwx}
\end{equation}
That~(\ref{2019-m-ch-ds-Stieltjes-seriesmwx}) is indeed a solution of~(\ref{2011-m-ch-dsee})
can be seen by direct insertion and a rather lengthy calculation.



\subsection*{Solution by asymptotic divergent series}
Without prior knowledge of $s(x)$ in~(\ref{2011-m-ch-dseess}) an
immediate way to solve~(\ref{2011-m-ch-dsee}) is a quasi {\em ad hoc} series {\it Ansatz}
similar to Frobenius' method; but allowing more general, and also diverging, series:
\begin{equation}
u(x)= \sum_{j=0}^\infty a_jx^j.
\end{equation}
When inserted into~(\ref{2011-m-ch-dsee}) $u(x)$ yields
\begin{equation}
\begin{split}
\left(x^2 \frac{d}{dx} +1\right) u(x) =
\left(x^2 \frac{d}{dx} +1\right) \sum_{j=0}^\infty a_jx^j = {x} \\
x^2 \sum_{j=0}^\infty a_j j x^{j-1} + \sum_{j=0}^\infty a_jx^j =
\sum_{j=0}^\infty a_j j x^{j+1} + \sum_{j=0}^\infty a_jx^j = {x} \\
  \left[\text{index substitution in first sum } i= j+1,\; j=i-1 \text{; then } i \rightarrow j\right]  \\
\sum_{i=1}^\infty a_{i-1} (i-1) x^i + \sum_{j=0}^\infty a_j x^j =
a_0 + \sum_{j=1}^\infty \left(a_{j-1} (j-1) +  a_j \right) x^j = x \\
a_0 + a_1 x + \sum_{j=2}^\infty \left(a_{j-1} (j-1) +  a_j \right) x^j = x
.
\end{split}
\label{2018-m-ch-adhocss}
\end{equation}
Since polynomials of different degrees are linear independent,
a comparison of coefficients appearing on the left hand side of~(\ref{2018-m-ch-adhocss}) with $x$
yields
\begin{equation}
\begin{split}
a_0 =0,
\;
a_1 = 1,
\\
a_j =  - a_{j-1} (j-1)  = - (-1)^j (j-1)!= (-1)^{j-1} (j-1)! \text{ for } j \ge 2
.
\end{split}
\label{2018-m-ch-adhocss1}
\end{equation}
This yields the sum~(\ref{2011-m-ch-dseess}) enumerated by Euler:
\begin{equation}
\begin{split}
u(x) = 0+ x +\sum_{j=2}^\infty (-1)^{j-1} (j-1)! x^j \\
=[j\rightarrow j+1] = x +\sum_{j=1}^\infty (-1)^{j} j! x^{j+1}
=  \sum_{j=0}^\infty (-1)^{j} j! x^{j+1} = s(x)
.
\end{split}
\label{2018-m-ch-adhocss12}
\end{equation}

Just as the Stieltjes series,
$s(x)$ is divergent for all $x\neq 0$:
for $j \ge 2$
its coefficients
$ a_j = (-1)^{j-1} (j-1)! $
have been enumerated in~(\ref{2018-m-ch-adhocss1}).
D'Alembert's criterion
yields
\begin{equation}
\lim_{j \rightarrow \infty}\left| \frac{ a_{j+1} }{ a_j } \right|
=
\lim_{j \rightarrow \infty}\left| \frac{ (-1)^{j} j! }{ (-1)^{j-1} (j-1)! } \right|
=\lim_{j \rightarrow \infty} j > 1 .
\end{equation}

\subsection*{Solution by Borel resummation of the asymptotic convergent series}
\index{Borel resummation}
In what follows the Borel summation will be used to formally sum up
the divergent series~(\ref{2011-m-ch-dseess}) enumerated by Euler.
A comparison
between~(\ref{2018-m-ch-ds-fps})
and~(\ref{2011-m-ch-dseess})
renders the coefficients
\begin{equation}
a_j =  (-1)^j \, j!,
\label{2019-mm-ch-dsboreltransform-euler}
\end{equation}
which can be used to compute the Borel transform~(\ref{2012-m-ch-dsboreltransform}) of Euler's divergent
series~(\ref{2011-m-ch-dseess})
\begin{equation}
{\cal B}S (y)
=
  \sum_{j=0}^\infty   \frac{a_j y^j}{j!}
  =   \sum_{j=0}^\infty   \frac{(-1)^j \, j!\, y^j}{j!}
  =   \sum_{j=0}^\infty    (-y)^j  = \frac{1}{1+y}
.
\label{2018-m-ch-dsboreltransform-euler}
\end{equation}
resulting in the Borel transformation~(\ref{2012-m-ch-dsboreltrafo} ) of Euler's divergent
series~(\ref{2011-m-ch-dseess})
\begin{equation}
\begin{split}
s(x) = \sum_{j=0}^\infty
a_j z^{j+1}
\stackrel{{\rm B}}{=}
\int_0^\infty {\cal B}S (y)  e^{-\frac{y}{x}}   dy
= \int_0^\infty \frac{ e^{-\frac{y}{x}} }{1+y}    dy \\
\left[\textrm{variable substitution } t =  \frac{y}{x} , dy =  z dt \right]\\
= \int_0^\infty \frac{ x e^{-t }}{1+xt}    dt
.
\end{split}
\label{2018-m-ch-dsboreltrafo2-Euler}
\end{equation}
Notice\cite{rousseau-2004} that the Borel transform~(\ref{2018-m-ch-dsboreltransform-euler})
``rescales'' or
``pushes'' the divergence of the series (\ref{2011-m-ch-dseess}) with zero radius of convergence
towards a ``disk'' or interval with finite radius of convergence and a singularity at $y=-1$.

\subsection*{Solution by integration}
An exact solution of~(\ref{2011-m-ch-dsee})
can also be found directly by {\em quadrature;}
that is, by direct integration (see, for instance, Chapter one of Ref.\cite[10mm]{birkhoff-Rota-48}).
It is not immediately obvious how to utilize direct integration in this case; the trick
is to make the following {\it Ansatz}:
\begin{equation}
s(x) = y (x) \exp \left( - \int \frac{ d x}{x^2} \right) = y(x)\exp \left[-\left(-\frac{1}{x}+C\right)\right]= k y(x)e^\frac{1}{x},
\label{201m-m-ch-dsans}
\end{equation}
with constant $k=e^{-C}$,
so that the ordinary differential equation~(\ref{2011-m-ch-dsee}) transforms into
\begin{equation}
\begin{split}
\left(x^2 \frac{d}{dx} +1\right)   s(x) =
\left(x^2 \frac{d}{dx} +1\right) y (x) \exp \left( - \int \frac{ d x}{x^2} \right) = x, \\
x^2 \frac{d  }{dx}\left[ y  \exp \left( - \int \frac{ d x}{x^2} \right)\right] + y  \exp \left( - \int \frac{ d x}{x^2} \right) = x, \\
x^2  \exp \left( - \int \frac{ d x}{x^2} \right)\frac{d y }{d x } + x^2 y \left(- \frac{1}{x^2}\right) \exp \left( - \int \frac{ d x}{x^2} \right) + y   \exp \left( - \int \frac{ d x}{x^2} \right) = x, \\
x^2  \exp \left( - \int \frac{ d x}{x^2} \right)\frac{d y }{d x }  = x, \\
  \exp \left( - \int \frac{ d x}{x^2} \right)\frac{d y }{d x } = \frac{1}{x}, \\
\frac{d y }{d x }  = \frac{ \exp \left( \int \frac{ d x}{x^2} \right)}{x}, \\
 y(x)  = \int \frac{1}{x} e^{ \int_x \frac{ d t}{t^2} } {d x }.
\end{split}
\end{equation}
More precisely, insertion into~(\ref{201m-m-ch-dsans}) yields, for some $a\neq 0$,
\begin{equation}
\begin{split}
s(x) =   e^{ - \int_a^x \frac{ dt}{t^2}}  y(x)=
-  e^{ - \int_a^x \frac{ dt}{t^2}} \int_0^x e^{ \int_a^t \frac{ ds}{s^2}} \left(-\frac{ 1}{t}\right) dt
\\
=
  e^{ - \left. \left(-\frac{1}{t}\right )\right|_a^x} \int_0^x e^{  \left. -\frac{1}{s}\right|_a^t} \left(\frac{ 1}{t}\right) dt
\\
=
  e^{ \frac{1}{x} -  \frac{1}{a} } \int_0^x e^{  -\frac{1}{t} +  \frac{1}{a} } \left(\frac{ 1}{t}\right) dt
\\
=
  e^{ \frac{1}{x}}\underbrace{e^{ -  \frac{1}{a} + \frac{1}{a}}}_{=e^0=1} \int_0^x e^{ -\frac{1}{t} } \left(\frac{ 1}{t}\right) dt
\\
=
e^{   \frac{1}{x}} \int_0^x  \frac{ e^{ - \frac{1}{t}}}{t} dt
\\
=
 \int_0^x  \frac{ e^{ \frac{1}{x} -\frac{1}{t}}}{t} dt.
\end{split}
\label{2011-m-ch-dseeeesola}
\end{equation}
With a change of the integration variable
\begin{equation}
\begin{split}
\frac{ z }{x} = \frac{1}{t}-\frac{1}{x}
, \; \textrm{ and thus }  \;
 z  = \frac{x}{t}-1
, \;  \textrm{ and } \;
t =  \frac{x}{1+  z }
, \\
\frac{dt}{d z  } =  -\frac{x}{(1+  z )^2}  , \;
\textrm{ and thus }  \;
dt =  -\frac{x}{(1+  z )^2} d z , \;    \\
\textrm{ and thus }  \;
\frac{dt}{t } =   \frac{-\frac{x}{(1+  z )^2}}{\frac{x}{1+  z }} d z
=   -\frac{ d z }{1+  z },
\end{split}
\label{2011-m-ch-dseeans}
\end{equation}
the integral (\ref{2011-m-ch-dseeeesola}) can be rewritten into the same form as Equation~(\ref{2018-m-ch-dsboreltrafo2-Euler}):
\begin{equation}
\begin{split}
s(x)= \int_{\infty}^0
\left(-\frac{e^{-\frac{ z }{x}}}{1+ z }\right) d z
= \int_0^\infty
\frac{e^{-\frac{ z }{x}}}{1+ z } d z  .
\end{split}
\label{2011-m-ch-dseefasol}
\end{equation}

Note that, whereas the series solution diverges for all nonzero $x$,
the solutions by quadrature~(\ref{2011-m-ch-dseefasol}) and by the Borel summation~(\ref{2018-m-ch-dsboreltrafo2-Euler})
are identical. They both converge and are well defined for all $x\ge 0$.

Let us now estimate the absolute difference between $s_{k}(x)$
which represents the partial sum of the Borel transform~(\ref{2018-m-ch-dsboreltransform-euler})
in the Borel transformation~(\ref{2018-m-ch-dsboreltrafo2-Euler}),
with $a_j =  (-1)^j \, j!$
from~(\ref{2019-mm-ch-dsboreltransform-euler}),
 ``truncated after the $k$th term'' and the exact solution $s(x)$; that is, let us consider
\begin{equation}
\begin{split}
  R_k(x)
\stackrel{{\tiny \textrm{ def }}}{=}
\Big\vert s(x) - s_{k}(x) \Big\vert=
\left\vert \int_0^\infty
\frac{e^{-\frac{ z }{x}}}{1+ z } d z
-
\sum_{j=0}^k (-1)^j j! x^{j+1} \right\vert .
\end{split}
\label{2011-m-ch-dseeanest}
\end{equation}

For any $x\ge 0$ this difference can be estimated\cite{rousseau-2004} by  a bound from above
\begin{equation}
  R_k(x)
\le
k! x^{k+1};
\label{2011-m-ch-dseeest}
\end{equation}
that is, this difference between the exact solution $s(x)$ and the diverging partial sum
$s_{k}(x)$ may become smaller than the first neglected term, and all subsequent ones.

{\color{OliveGreen}
\bproof

For a proof, observe that,
since
a partial {\em geometric series}
\index{geometric series}
is the sum of all the numbers in a geometric progression up to a certain power;
that is,
\begin{equation}
\sum_{j=0}^k r^j =   1+r+r^2+ \cdots +r^j+ \cdots +r^k .
\label{2011-m-ch-dsee124567}
\end{equation}
By multiplying both sides with $1-r$,
the sum (\ref{2011-m-ch-dsee124567}) can be rewritten as
\begin{equation}
\begin{split}
(1-r) \sum_{j=0}^k r^j=
(1-r) (1+ r+r^2+ \cdots +r^j+ \cdots +r^k)\\
=1+ r+r^2+ \cdots +r^j+ \cdots +r^k -
r(1+r+r^2+ \cdots +r^j+ \cdots +r^k +r^k) \\
=1+ r+r^2+ \cdots +r^j+ \cdots +r^k -
(r+r^2+ \cdots +r^j+ \cdots +r^k +r^{k+1}) \\
= 1-r^{k+1}
,
\end{split}
\end{equation}
and, since the middle terms all cancel out,
\begin{equation}
\sum_{j=0}^k r^j =  \frac{1-r^{k+1}}{1-r},
\;
\textrm{ or }
\;
\sum_{j=0}^{k-1} r^j =  \frac{1-r^{k}}{1-r}  =  \frac{1}{1-r} - \frac{r^{k}}{1-r}
.
\label{2011-m-ch-dsee12}
\end{equation}
Thus, for $r=-\zeta$, it is true that
\begin{equation}
\begin{split}
\frac{1}{1+\zeta}=\sum_{j=0}^{k-1} (-1)^j \zeta^j + (-1)^k \frac{\zeta^k}{1+\zeta},
\end{split}
\label{2011-m-ch-dsee13}
\end{equation}
and, therefore,
\begin{equation}
\begin{split}
f(x) =\int_0^\infty \frac{e^{-\frac{\zeta}{x}}}{1+\zeta}d\zeta \\
\qquad =
\int_0^\infty  e^{-\frac{\zeta}{x}}\left[
\sum_{j=0}^{k-1} (-1)^j \zeta^j + (-1)^k \frac{\zeta^k}{1+\zeta}
\right]
d\zeta \\
\qquad =
\sum_{j=0}^{k-1}(-1)^j\int_0^\infty  \zeta^j  e^{-\frac{\zeta}{x}}   d\zeta
 +
(-1)^k \int_0^\infty \frac{\zeta^ke^{-\frac{\zeta}{x}}}{1+\zeta}
d\zeta  .
\end{split}
\label{2011-m-ch-dsee14}
\end{equation}
Since [cf Equation~(\ref{2017-m-ch-sf-edgamma})]
\begin{equation}
k!= \Gamma(k+1)=    \int_0^\infty z^k e^{-z} dz,
\label{2011-m-ch-dsee15}
\end{equation}
one obtains
\begin{equation}
\begin{split}
\int_0^\infty  \zeta^j  e^{-\frac{\zeta}{x}}   d\zeta  \;
\textrm{ with substitution:  }z=\frac{\zeta}{x}, d \zeta =x dz    \\
= \int_0^\infty x^{j+1} z^j  e^{-z}   dz
= x^{j+1}\int_0^\infty  z^j  e^{-z}   dz
=  x^{j+1} k! ,
\end{split}
\label{2011-m-ch-dsee16}
\end{equation}
and hence
\begin{equation}
\begin{split}
f(x)  =
\sum_{j=0}^{k-1}(-1)^j\int_0^\infty  \zeta^j  e^{-\frac{\zeta}{x}}   d\zeta
 +
(-1)^k \int_0^\infty \frac{\zeta^ke^{-\frac{\zeta}{x}}}{1+\zeta}
d\zeta  \\
\qquad =
\sum_{j=0}^{k-1}  (-1)^j x^{j+1} k!
 +
\int_0^\infty (-1)^k \frac{\zeta^k e^{-\frac{\zeta}{x}}}{1+\zeta}
d\zeta  \\
\qquad =
f_k(x)  +R_k(x),
\end{split}
\label{2011-m-ch-dsee17}
\end{equation}
where $f_k(x)$ represents the partial sum of the power series, and $R_k(x)$ stands for the remainder,
the difference between $f(x)$ and $f_k(x)$.
The absolute of the remainder can be estimated by
\begin{equation}
\begin{split}
 R_k(x)
=
\int_0^\infty  \frac{\zeta^k e^{-\frac{\zeta}{x}}}{1+\zeta} d\zeta
\le
\int_0^\infty  \zeta^k e^{-\frac{\zeta}{x}} d\zeta
 = k! x^{k+1}.
\end{split}
\label{2011-m-ch-dsee18}
\end{equation}
\eproof
}

The functional form $k! x^k$ (times $x$) of the absolute error~(\ref{2011-m-ch-dseeanest})
suggests that, for $0<x<1$,
there is an ``optimal'' value $k \approx \frac{1}{x}$ with respect to convergence
of the partial sums $s(k)$ associated with Euler's asymptotic expansion of the solution~(\ref{2011-m-ch-dseess}):
up to this $k$-value the factor $x^k$ dominates the estimated absolute rest~(\ref{2011-m-ch-dseeest})
by suppressing it more than $k!$ grows.
\begin{marginfigure}
\begin{center}
\begin{tikzpicture} [ scale=0.53, every mark/.append style={mark size=2pt} ]
\begin{axis}[
    ymode=log,
    log ticks with fixed point,
    ticklabel style = {font=\Large },
    ymax=1,
    x label style={font=\Large ,at={(axis description cs:1.1,0.1)}},
    y label style={font=\Large ,at={(axis description cs:0.15,1.1)},rotate=270},
    xlabel={$k$},
    ylabel={$R_k(x)$}
]
\addplot  [
orange, mark=*
]  table {  
 0 0.10000000000000000000
 1 0.010000000000000000000
 2 0.0020000000000000000000
 3 0.00060000000000000000000
 4 0.00024000000000000000000
 5 0.00012000000000000000000
 6 0.000072000000000000000000
 7 0.000050400000000000000000
 8 0.000040320000000000000000
 9 0.000036288000000000000000
 10 0.000036288000000000000000
 11 0.000039916800000000000000
 12 0.000047900160000000000000
 13 0.000062270208000000000000
 14 0.000087178291200000000000
 15 0.00013076743680000000000
 16 0.00020922789888000000000
 17 0.00035568742809600000000
 18 0.00064023737057280000000
 19 0.0012164510040883200000
 20 0.0024329020081766400000
 21 0.0051090942171709440000
 22 0.011240007277776076800
 23 0.025852016738884976640
}
 node [pos=0.9 , above left, style={font=\Large}] {$x=\frac{1}{10}$};

\addplot  [
blue, mark=*
]  table {        
 0       0.0666667
1       0.00444444
2       0.000592593
3       0.000118519
4       0.0000316049
5       0.000010535
6       0.00000421399
7       0.00000196653
8       0.00000104882
9       0.000000629289
10      0.000000419526
11      0.000000307653
12      0.000000246122
13      0.000000213306
14      0.000000199085
15      0.000000199085
16      0.000000212358
17      0.000000240672
18      0.000000288807
19      0.000000365822
20      0.000000487762
21      0.000000682867
22      0.00000100154
23      0.00000153569
}
 node [pos=0.9 , above left, style={font=\Large}] {$x=\frac{1}{15}$};

\addplot  [
green, mark=*
]  table {        
 0       0.2
1       0.04
2       0.016
3       0.0096
4       0.00768
5       0.00768
6       0.009216
7       0.0129024
8       0.0206438
9       0.0371589
10      0.0743178
11      0.163499
12      0.392398
13      1.02024
14      2.85666
15      8.56997
16      27.4239
17      93.2413
18      335.669
19      1275.54
20      5102.17
21      21429.1
22      94288.0
23      433725.0
}
 node [pos=0.2 , below right, style={font=\Large}] {$x=\frac{1}{5}$};
\end{axis}
\end{tikzpicture}
\end{center}
\caption{The absolute error $R_k(x)$ as a function of increasing $k$
for $x\in \{\frac{1}{5},\frac{1}{10},\frac{1}{15}\}$.}
  \label{2018-mm-ferror}
\end{marginfigure}
However, this suppression of the absolute error as $k$ grows is eventually
-- that is, if $k > \frac{1}{x}$ -- compensated by the factorial function,
as depicted in Figure~\ref{2018-mm-ferror}: from $k \approx \frac{1}{x}$ the absolute error grows again,
so that the overall behavior of the absolute error $R_k(x)$  as a function of $k$ (at constant $x$)
is ``bathtub''-shaped; with a ``sink'' or minimum at $k \approx \frac{1}{x}$.

\section{Divergence of perturbation series in quantum field theory}

A formal entity such as the solution of an ordinary differential equation may have very
different representations and encodings; some of them with problematic issues.
The means available are often not a matter of choice but of pragmatism and even desperation.\cite{Boyd99thedevil}

This seems to apply also to field theories:
often one is restricted to perturbative solutions in terms of power series.
But these methods are problematic as they are applied in a situation where they are forbidden.

Presently there are two known reasons for the occurrence of asymptotically divergent power series in perturbative quantum field theories:
one is associated with expansion at an essential singularity,  such as $z=0$ for the function $e^\frac{1}{z}$
and the other with an exchange of the order of two limits,
such as exchanging an infinite sum with an integral if the domain of integration is not compact.

\newpage

\subsection{Expansion at an essential singularity}

The following argument is due to Dyson.\cite{PhysRev.85.631,LeGuillou-Zinn-Justin,Svozil-2023-axioms12010072}
Suppose
the overall energy of a system of a large number $N \gg 1$ of particles
of charge $q$ with mean kinetic energy (aka ``temperature'') $T$
and mean absolute potential $V$
consists of a kinetic and a potential part, like
\begin{equation}
E\sim T N + q^2 V \frac{N(N-1)}{2}\approx T N + \frac{q^2V}{2} N^2,
\label{2019-mm-ch-di-en}
\end{equation}
where ${N(N-1)}/{2}$ is the number of particle pairs.
Then the ground state energy is bound from below as long as the interaction is repulsive: that is, $q^2>0$.
However, for an attractive effective interaction $q^2<0$ and, in particular,
in the presence of (electron-positron) pair creation, the ground state may no longer be stable.
As a result of this instability of the ground state ``around'' $q^2=0$
one must expect that any physical quantity $F(q^2)$ which is calculated as a formal
power series in the coupling constant $q^2$ cannot be analytic around $q^2=0$.
Because, intuitively, even if $F(q^2)$ appears to be ``well behaved'' $F(-q^2)$ is not
if the theory is unstable for transitions from a repulsive to an attractive potential regime.

\marginnote{However, Dyson's argument does not apply to other series solutions~ \bibentry{Watson1912,Weniger2010}
which, for instance, converges on some open half-plane, such as the Dirichlet series.}
Therefore, it is strictly disallowed to develop $F(q^2)$ at $q^2=0$ into a Taylor series.
Insistence (or ignorance) in doing what is forbidden is penalized by an asymptotic divergent series at best.

To obtain a quantitative feeling for what is going on in such cases consider\cite{sommer-AR},
the functional integral
with a redefined exponential kernel from Equation~(\ref{2019-mm-ch-di-en}):
let $N=x^2$,   $T  = -\alpha$, and $g = - \frac{q^2V}{2}$, and
\begin{equation}
f\left(\alpha,g\right)=\int_{0}^{\infty}e^{-\alpha x^{2}-gx^{4}}dx  .
\label{2019-mm-ch-di-en2}
\end{equation}

For negative $g<0$ the term  $e^{-gx^{4}}=e^{|g|x^{4}}$ dominates the kernel, and the integral~(\ref{2019-mm-ch-di-en2}) diverges.
For $\alpha>0$ und $g>0$  this integral has a nonperturbative representation as
\begin{equation}
f\left(\alpha,g\right)=\frac{1}{4}\sqrt{\frac{\alpha}{g}}e^{\frac{\alpha^{2}}{8g}}K_{\frac{1}{4}}\left(\frac{\alpha^{2}}{8g}\right) ,
\label{2019-mm-ch-di-ennp}
\end{equation}
where $K_{\nu}\left(x\right)$ is the modified Bessel funktion of the second kind
(e.g., {\S}9.6, pp.~374-377  of Abramowitz and Stegun).
\marginnote{\url{http://mathworld.wolfram.com/ModifiedBesselFunctionoftheSecondKind.html}}

A divergent series is obtained by expanding $f\left(\alpha,g\right)$ from~(\ref{2019-mm-ch-di-en2})
in a Taylor series of the ``coupling constant'' $g\neq 0$
at $g=0$; and, in particular, by taking the limit $n\rightarrow \infty$ of the partial sum up to order $n$ of $g$:
\begin{equation}
\begin{split}
f_{n}\left(\alpha,g\right)
=
\frac{1}{2}\sum_{k=0}^{n}\frac{\left(-1\right)^{k}}{k!}\frac{\varGamma\left(2k+\frac{1}{2}\right)}{\alpha^{2k+\frac{1}{2}}}g^{k}
\qquad \qquad
\\=
\frac{1}{2}\left[
\sqrt{\pi}
+
\sum_{k=1}^{n}\frac{\left(-1\right)^{k}}{k!}\frac{\varGamma\left(2k+\frac{1}{2}\right)}{\alpha^{2k+\frac{1}{2}}}g^{k}
\right]
\\
=
\frac{1}{2 \sqrt{a}} \left(-\frac{g}{a^2}\right)^n \frac{\Gamma \left(\frac{1}{2} (4 n+1)\right)}{ \Gamma (n+1)}
{\;}_2F_2\left(1,-n;\frac{1}{4}-n,\frac{3}{4}-n;\frac{a^2}{4 g}\right)
.
\label{2019-mm-ch-di-en3}
\end{split}
\end{equation}




For fixed $\alpha=1$ the asymptotic divergence of~(\ref{2019-mm-ch-di-en3})
for $n \rightarrow \infty$ manifests itself differently for different values of $g>0$:
\begin{itemize}
\item For $g=1$   the nonperturbative expression~(\ref{2019-mm-ch-di-ennp}) yields
\begin{equation}
f\left(1,1\right)=\int_{0}^{\infty}e^{-x^{2}-x^{4}}dx=\frac{1}{4}  e^\frac{1}{8} K_{\frac{1}{4}}\left(\frac{1}{8}\right)\approx 0.684213  ,
\end{equation}
and
the series~(\ref{2019-mm-ch-di-en3})  starts diverging almost immediately
as the logarithm of the absolute error defined by
$R_n(1)= \log \left| f \left(1,1\right)-f_n \left(1,1\right)\right|$
and depicted in Figure~\ref{2019-mm-ferror1}, diverges.
\begin{marginfigure}
{\color{black}
\begin{center}
\begin{tabular}{c}
\begin{tikzpicture} [ scale=0.53, every mark/.append style={mark size=2pt} ]
\begin{axis}[
    log ticks with fixed point,
    ticklabel style = {font=\Large },
    x label style={font=\Large ,at={(axis description cs:1.1,0.1)}},
    y label style={font=\Large ,at={(axis description cs:0.15,1.1)},rotate=270},
    xlabel={$n$},
    ylabel={$R_n(1)$}
]
\addplot  [
orange, mark=*
]  table {
 0      -1.59942
 1      -0.77077
 2      0.894158
 3      3.07015
 4      5.60152
 5      8.40092
 6      11.414
 7      14.6041
 8      17.9452
 9      21.4178
 10     25.0067
 11     28.6999
 12     32.4877
 13     36.362
 14     40.3158
 15     44.3434
 16     48.4397
 17     52.6003
 18     56.8213
 19     61.0993
 20     65.4313
}
 node [pos=0.7 , above left, style={font=\Large}] {$q=1$};

\end{axis}
\end{tikzpicture}
\\
\begin{tikzpicture} [ scale=0.53, every mark/.append style={mark size=2pt} ]
\begin{axis}[
    log ticks with fixed point,
    ticklabel style = {font=\Large },
    x label style={font=\Large ,at={(axis description cs:1.1,0.1)}},
    y label style={font=\Large ,at={(axis description cs:0.15,1.1)},rotate=270},
    xlabel={$n$},
    ylabel={$R_n \left(\frac{1}{10}\right)$}
]
\addplot  [
blue, mark=*
]  table {
 0   -3.01219
 1   -4.05803
 2   -4.43997
 3   -4.4068
 4   -4.07194
 5   -3.50072
 6   -2.73561
 7   -1.80645
 8   -0.735293
 9   0.460913
 10  1.76876
 11  3.17737
 12  4.67777
 13  6.26244
 14  7.92495
 15  9.65979
 16  11.4622
 17  13.3279
 18  15.2532
 19  17.2348
 20  19.2698
}
 node [pos=0.7 , above left, style={font=\Large}] {$q=\frac{1}{10}$};

\end{axis}
\end{tikzpicture}
\\
\begin{tikzpicture} [ scale=0.53, every mark/.append style={mark size=2pt} ]
\begin{axis}[
    log ticks with fixed point,
    ticklabel style = {font=\Large },
    x label style={font=\Large ,at={(axis description cs:1.1,0.1)}},
    y label style={font=\Large ,at={(axis description cs:0.15,1.1)},rotate=270},
    xlabel={$n$},
    ylabel={$R_n \left(\frac{1}{100}\right)$}
]
\addplot  [
green, mark=*
]  table {
 0        -5.1
 1        -8.2
 2        -10.7
 3        -12.9
 4        -14.7
 5        -16.4
 6        -17.8
 7        -19.1
 8        -20.3
 9        -21.3
 10       -22.3
 11       -23.1
 12       -23.9
 13       -24.6
 14       -25.2
 15       -25.7
 16       -26.2
 17       -26.6
 18       -26.9
 19       -27.2
 20       -27.5
 21       -27.7
 22       -27.8
 23       -27.9
 24       -28.0
 25       -28.0
 26       -28.0
 27       -27.9
 28       -27.8
 29       -27.7
 30       -27.5
 31       -27.3
 32       -27.1
 33       -26.9
 34       -26.6
 35       -26.2
 36       -25.9
 37       -25.5
 38       -25.1
 39       -24.7
 40       -24.2
 41       -23.8
 42       -23.2
 43       -22.7
 44       -22.2
 45       -21.6
 46       -21.0
 47       -20.4
 48       -19.7
 49       -19.1
 50       -18.4
 51       -17.7
 52       -17.0
 53       -16.2
 54       -15.5
 55       -14.7
 56       -13.9
 57       -13.1
 58       -12.3
 59       -11.4
 60       -10.6
 61       -9.7
 62       -8.8
 63       -7.9
 64       -6.9
 65       -6.0
 66       -5.0
 67       -4.1
 68       -3.1
 69       -2.1
 70       -1.1
 71       -0.2
 72       1.0
 73       1.8
 74       3.2
}
 node [pos=0.7 , above left, style={font=\Large}] {$q=\frac{1}{100}$};

\end{axis}
\end{tikzpicture}
\end{tabular}
\end{center}
\caption{The logarithm of the absolute error $R_n$ as a function of increasing $n$
for $q\in \{1,\frac{1}{10},\frac{1}{100}\}$, respectively.\label{2019-mm-ferror1}     }
}
\end{marginfigure}

\item For $g=\frac{1}{10}$   the nonperturbative expression~(\ref{2019-mm-ch-di-ennp}) yields
\begin{equation}
f\left(1,\frac{1}{10}\right)=\int_{0}^{\infty}e^{-x^{2}-\frac{1}{10}x^{4}}dx =\frac{1}{2}\sqrt{\frac{5}{2}}e^{\frac{5}{4}}K_{\frac{1}{4}}\left(\frac{5}{4}\right)\approx 0.837043  ,
\end{equation}
and
the series~(\ref{2019-mm-ch-di-en3})  performs best at around $n=3$ or $4$ and then starts to deteriorate
as the logarithm of the absolute error defined by
$R_n \left(\frac{1}{10}\right)=  \log \left| f \left(1,\frac{1}{10}\right)-f_n \left(1,\frac{1}{10}\right)\right| $
and depicted in Figure~\ref{2019-mm-ferror1}, diverges.
%
%

\item For $g=\frac{1}{100}$ (a value which is almost as small as the coupling constant $g=\frac{1}{137}$ in quantum electrodynamics)
the nonperturbative expression~(\ref{2019-mm-ch-di-ennp}) yields
\begin{equation}
f\left(1,\frac{1}{100}\right)
=\int_{0}^{\infty}e^{-x^{2}-\frac{1}{100}x^{4}}dx
=\frac{5}{2}e^{\frac{25}{2}}K_{\frac{1}{4}}\left(\frac{25}{2}\right)\approx 0.879 849 554 945 695
.
\end{equation}
The series~(\ref{2019-mm-ch-di-en3}) performs best at around $n=25$ and then starts to deteriorate
as the logarithm of the absolute error defined by
$R_n \left(\frac{1}{100}\right) =  \log \left| f \left(1,\frac{1}{100}\right)-f_n \left(1,\frac{1}{100}\right)\right| $
and depicted in Figure~\ref{2019-mm-ferror1}, diverges.
\end{itemize}

\subsection{Forbidden interchange of limits}

A second ``source'' of divergence is the forbidden and thus incorrect interchange of limits --
in particular, an interchange between sums and integrals~\sidenote[][25mm]{See, for instance, the discussion in
Section~II.A of  \bibentry{PhysRevD.57.1144} based on  Lebesgue's dominated convergence theorem.} --
during the construction of the perturbation series.
Again one may perceive asymptotic divergence as a ``penalty'' for such manipulations.

For the sake of a demonstration, consider again the integral~(\ref{2019-mm-ch-di-en2})
\begin{equation}
f\left(1,g\right)
=\int_{0}^{\infty}e^{- x^{2}-gx^{4}}dx
=\int_{0}^{\infty}e^{- x^{2}}e^{-gx^{4}}dx
\label{2019-mm-ch-di-en3a}
\end{equation}
with $\alpha=1$. A Taylor expansion of the ``interaction part'' in the ``coupling constant'' $g$
of its kernel at $g=0$ yields
\begin{equation}
e^{-gx^{4}} =\sum_{k=0}^\infty \left(-x^4\right)^n \frac{1}{k!} g^k.
\label{2019-mm-ch-di-en3eip}
\end{equation}
This is perfectly legal; no harm done yet.
Consider the resulting kernel as a function of  the order $k$
of the Taylor series expansion, as well as of the ``coupling constant'' $g$
 and of the integration parameter $x$
for $\alpha=1$ in a similar notation as introduced in Equation~(\ref{2019-mm-ch-di-en3}):
\begin{equation}
F_k(g,x)= \frac{\left(-g\right)^k}{k!} e^{- x^{2}} x^{4k} .
\label{2019-mm-ch-di-en3eipke}
\end{equation}
Rather than  applying Lebesgue's dominated convergence theorem to $F_k(g,x)$
we directly show that an interchange of summation with integration yields a divergent series.


Indeed, the original order of limits in~(\ref{2019-mm-ch-di-en2})
yields a convergent expression~(\ref{2019-mm-ch-di-ennp}):
\begin{equation}
\begin{split}
\lim_{t\rightarrow \infty}
\lim_{n\rightarrow \infty}
\int_0^t dx
\sum_{k=0}^n
F_k(g,x) \qquad \qquad \\
=
\int_{0}^{\infty}e^{-x^{2}-gx^{4}}dx  = f\left(1,g\right)=
\frac{1}{4}\sqrt{\frac{1}{g}}e^{\frac{1}{8g}}K_{\frac{1}{4}}\left(\frac{1}{8g}\right)
.
\label{2019-mm-ch-di-en3eipke2c1}
\end{split}
\end{equation}

However, for $g\neq 0$ the interchange of limits results in a divergent series:
\begin{equation}
\begin{split}
\lim_{n\rightarrow \infty}
\lim_{t\rightarrow \infty}
\sum_{i=0}^n
\int_0^t
F_n(1,g,x)   dx =
\lim_{n\rightarrow \infty}
  f_n\left(1,g\right)   \\
=\lim_{n\rightarrow \infty}
\frac{1}{2}
\sum_{i=0}^n
\frac{(-g)^k}{k!} \Gamma\left(2k+\frac{1}{2}\right)
=
\frac{1}{2}\left[
\sqrt{\pi}
+
\lim_{n\rightarrow \infty}\sum_{i=1}^n
\frac{(-g)^k}{k!} \Gamma\left(2k+\frac{1}{2}\right)
\right].
\label{2019-mm-ch-di-en3eipke2c2}
\end{split}
\end{equation}

Notice that both the direct Taylor expansion of $f(\alpha,g)$ at the singular point $z=0$ as well as the
interchange of the summation from the legal Taylor expansion of $e^{-gx^{4}}$ with the integration in~(\ref{2019-mm-ch-di-en3eipke2c1})
yield the same (asymptotic) divergent expressions~(\ref{2019-mm-ch-di-en3})
and~(\ref{2019-mm-ch-di-en3eipke2c2}).


\subsection{On the usefulness of asymptotic expansions in quantum field theory}

It may come as a surprise that calculations involving asymptotic expansions in the coupling constants yield perturbation series
which perform well for many empirical predictions -- in some cases\cite[-30mm]{HAGIWARA2007173}
the differences between experiment and prediction as small as $10^{-9}$.
Depending on the temperament and personal inclinations to accept results from ``wrong'' evaluations
this may be perceived optimistically as well as pessimistically.

As we have seen the quality of such asymptotic expansions depends on the magnitude of the expansion parameter:
the higher it gets the worse is the quality of prediction in larger orders.
And the approximation will never be able to reach absolute accuracy.
However in regimes such as quantum electrodynamics,
for which the expansion parameter is of the order of 100, for all practical purposes\cite[-30mm]{bell-a}
and relative to our limited  means to compute the high order terms,
such an asymptotic divergent perturbative expansion
might be ``good enough'' anyway.
But what if this parameter is of the order of $1$?

Another question is whether resummation procedures can ``recover'' the ``right'' solution in terms of analytic functions.
This is an ongoing field of research. As long as low-dimensional toy models such as
the one covered in earlier sections are studied this might be possible, say, by (variants) of Borel summations.\cite[-45mm]{Sauzin-2014,Mas-2019}
However, for realistic, four-dimensional field theoretic models the situation may be very different and ``much harder.''\cite[-25mm]{ZINNJUSTIN20101454,neumaier-sum}
Let me finally quote Arthur M. Jaffe and Edward Witten:\cite{Jaffe-Witten-2000-QYMT}
{\em ``In most known examples, perturbation series,
i.e., power series in the coupling constant, are divergent expansions; even Borel and
other resummation methods have limited applicability.''}

\begin{center}
{\color{olive}   \Huge
 \floweroneright
}
\end{center}

\clearpage

\backmatter

 \bibliography{svozil}
 \bibliographystyle{plainnat}

\printindex

\end{document}